\documentclass[a4paper,12pt,twoside,dvips]{class_file/amthesis}  

\usepackage[square, numbers, comma, sort&compress]{natbib} 
\usepackage{verbatim}  
\usepackage{bm}
\usepackage{float}
\usepackage{multirow}

\usepackage{afterpage}
\hypersetup{urlcolor=blue, colorlinks=true} 
\usepackage{url}

\newcommand{\be}{\begin{eqnarray}}
\newcommand{\ee}{\end{eqnarray}}
\newcommand{\ben}{\begin{eqnarray*}}
\newcommand{\een}{\end{eqnarray*}}
\newcommand{\ba}{\begin{array}}
\newcommand{\ea}{\end{array}}

\newcommand{\bi}{\begin{itemize}}
\newcommand{\ei}{\end{itemize}}

\newcommand{\vect}[1]{\mathbf{#1}}

\input abstract/babarsym

\begin{document}

\frontmatter	  

\title { \bf{ \LARGE { \begin{center} Search for di-muon decays of a light scalar  Higgs boson in radiative {\boldmath $\Upsilon(1S)$} decays \end{center}}} } 
\authors  {Vindhyawasini Prasad}
\supervisors{Dr. Bipul Bhuyan  and  Dr. Poulose Poulose}
\degree{Doctor of Philosophy}
\faculty{Faculty of Science}
\institutelogo{\includegraphics[width=1.2in]{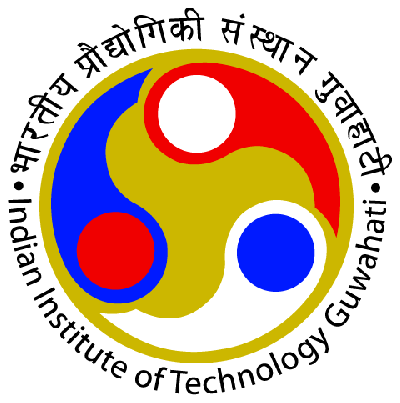}}
\department{Department of Physics}
\DEPARTMENT{Physics}
\institute{Indian Institute of Technology Guwahati}
\instituteadd{Guwahati 781039, India}
\date       {August 2012}
\subject    {}

\maketitle

\setstretch{1.3}  


\begin{certificate}
It is certified that the work contained in the thesis entitled ``{\em Search for di-muon decays of a light scalar  Higgs boson in radiative $\Upsilon(1S)$ decays}"
by Mr.\ Vindhyawasini Prasad, a Ph.D student of the Department of Physics, IIT Guwahati
was carried out under our joint supervision and has not 
been submitted elsewhere for award of any degree.

\vspace{10em}

Dr. Bipul Bhuyan \hfill Dr. Poulose Poulose

\setstretch{1.3}
\end{certificate}


\begin{dedication}
\Large\emph{This thesis is dedicated to the memory of my late beloved younger brother Sri Mata Prasad Mishra}
\end{dedication}


\begin{acknowledgements}

I would like to thank my thesis advisors, Dr. Bipul Bhuyan and Dr. Poulose Poulose for their guidance and constant support throughout my research work. They have  motivated me to take this work as a challenge and without their help it would have been impossible to finish the same. My other doctoral committee members: Dr. G. S. Setlur, Dr. T. N. Dey and Dr. K. Kapoor also deserve special thanks for their frank comments and encouragements.  I am thankful to all the faculty members and staff of the Department of Physics, Indian Institute of Technology Guwahati for their friendly behavior and help, whenever it was needed. My special thank extends to Mr. B. B. Purakayasthya  and Mr. Pallav Dutta for helping me to solve the computer related problems on many occasions.   

I wish to thank Prof. N. K. Mondal for providing me an opportunity to visit India Based Neutrino Observatory (INO) laboratory at Tata Institute of fundamental Research Center (TIFR), Mumbai, where I have learned different software techniques as well as worked on the LVDS-ECL-NIM translator for the INO data acquisition system (DAQ). These trainings gave me the first glimpse of a high energy physics experiment that kept me motivated throughout the period of this dissertation. I would also like to thank Dr. B. Satyanarayana and other INO colleagues for their kind help and suggestions during my visit of TIFR. 
 
   I wish to thank all the members of \babar\ Tau/QED working group, past and present, especially Dr. Yury Kolomensky, Dr. Randal Sobie, Dr. Bertrand Echenard and Dr. Albarto Lusiani. Their insight and suggestions proved to be  very important in finalizing this analysis. I would also like to extend my thanks to the review committee members: Dr. Bertrand Echenard, Dr. Andrew Manual Ruland, Dr. Randal Sobie, publication board chair, Dr. Bill Garry and the \babar\ Physics Analysis Coordinator, Dr. Abner Sofer for their valued suggestions.

I am thankful to my fellow research group members: Satendra, Deepanwita, Biswajit, Kamal, Deepanjali, Biswaranjan and Nitin for making a pleasant work atmosphere and their assistance time to time. My sincere thank also extends to Meera, Biswanath, Poulami, Jahir, Soumen, Sunita, Sangeetha, Supriya, Jharnali, Manirupa, Parvendra, Arindam, Rahul, Vipin, Niraj, Himanshu and many others colleagues for making my experience somewhat more intellectual and mostly for lots of fun.

Last but not least, I am highly grateful to my parents, Smt. Savitri Mishra and Sri Gulab Dhar Mishra, my brothers, sisters, nephews, relatives and all my well wishers for believing in me, for their constant love and mental support, and inspiring me not only to pursue my Ph.D but also to dedicate myself and my works in the development of basic science throughout the rest of my life.

\end{acknowledgements}


\begin{abstract}
We search for di-muon decays of a low-mass Higgs boson ($A^0$)  in the fully reconstructed decay chain of $\Upsilon(2S,3S) \to \pipi\Upsilon(1S)$, $\Upsilon(1S) \to \g A^0$, $A^0 \to \mumu$. The $A^0$ is predicted by several extensions of the Standard Model (SM), including the Next-to-Minimal Supersymmetric Standard Model (NMSSM). NMSSM introduces a \CP-odd light Higgs  boson whose mass could be  less than  10 \gevcc. The data samples used in this analysis contain $92.8 \times 10^6$  $\Upsilon(2S)$  and  $116.8 \times 10^6$ $\Upsilon(3S)$  events collected by the \babar\ detector. The $\Upsilon(1S)$ sample is selected by tagging the pion pair in the $\Upsilon(2S, 3S) \to \pi^+\pi^-\Upsilon(1S)$  transitions. We find no evidence for $A^0$ production and set $90\%$ confidence level (C.L.) upper limits on the product branching fraction  $\mathcal{B}(\Upsilon(1S) \to \g A^0) \times \mathcal{B}(A^0 \to \mumu)$ in the range of $(0.28 - 9.7)\times 10^{-6}$ for   $0.212 \le m_{A^0} \le 9.20$ \gevcc. We also combine our results with previous \babar\ results of $\Upsilon(2S,3S) \to \g A^0$, $A^0 \to \mumu$ to set limits on the effective coupling ($f_{\Upsilon}$) of the \b-quark to the $A^0$, $f_{\Upsilon}^2 \times \mathcal{B}(A^0 \to \mumu)$, at the level of  $(0.29 - 40)\times 10^{-6}$ for $0.212 \le m_{A^0} \le 9.2$ \gevcc.

\end{abstract}

\pagestyle{fancy}  
\tableofcontents  








\chapter{Preface}
 The Higgs boson is essential to explain the origin of mass of the elementary particles within the Standard Model (SM) via Higgs mechanism through spontaneous breaking of the electroweak symmetry. The Large Hadron Collider (LHC) experiment at CERN has found an evidence of a Higgs-like state which has a mass of $\approx 126$ \gevcc. However, a light Higgs boson  is also predicted by many extensions of the SM including the Next-to-Minimal Supersymmetric Standard Model (NMSSM).  The Higgs sector of the NMSSM contains a total three \CP-even, two \CP-odd and two charged Higgs bosons. The lightest \CP-odd Higgs boson ($A^0$)  could have a mass below the $b\overline{b}$ production threshold, avoiding the constraints of Large Electron-Positron (LEP) Collider experiment. Such low-mass Higgs boson can be detected at the \B-Factory via radiative $\Upsilon(nS) \to \g A^0$ ($n=1,2,3$) decays. These $\Upsilon$ resonances have narrow width and are produced below the $B\overline{B}$ threshold, providing a clean environment for new physics searches.          

In 2005, HyperCP experiment observed three anomalous events in the $\Sigma^+ \rightarrow p\mu^+\mu^-$ final state, that have been interpreted as candidates for CP-odd Higgs with the mass of $214.3\pm0.5$ MeV decaying into a pair of muons. In 2008, the CLEO experiment performed a search for $A^0$ production in the di-tau and di-muon in the final state in the radiative decays of $\Upsilon(1S)$ and ruled out the hyperCP prediction.  Similar searches have been performed by \babar\ experiment in several final states, including $\Upsilon(2S,3S) \to \g A^0$, $A^0 \to \mumu$, and more recently by  BESIII experiment in $\jpsi \to \g A^0$, $A^0 \to \mumu$, and by CMS experiment in $pp\to A^0$, $A^0 \to \mumu$. These results have ruled out the hyperCP prediction as well as a substantial fraction of the NMSSM parameter space.

 This thesis describes a search for the di-muon decays of the $A^0$ in the radiative decays of di-pion tagged $\Upsilon(1S)$   meson: $\Upsilon(2S,3S) \rightarrow \pi^+\pi^-\Upsilon(1S)$, $\Upsilon(1S) \rightarrow \gamma A^0$, $A^0 \rightarrow \mu^+\mu^-$. The data samples used in this analysis were collected at $\Upsilon(2S)$ and $\Upsilon(3S)$ resonances by \babar\ detector at the PEP-II asymmetric-energy $e^+e^-$ collider located at SLAC National Accelerator Laboratory. A clean $\Upsilon(1S)$ sample is selected by tagging the di-pions in the $\Upsilon(2S,3S) \rightarrow \pi^+\pi^- \Upsilon(1S)$ transition, resulting in a substantial background reduction compared to direct searches in $\Upsilon(2S,3S) \rightarrow \gamma A^0$ decays. We find no evidence for the $A^0$ production in the $\Upsilon(2S,3S)$ data samples, and set $90\%$ C.L. upper limits on the $\mathcal{B}(\Upsilon(1S) \rightarrow \gamma A^0) \times \mathcal{B}(A^0 \rightarrow \mu^+\mu^-)$ for $\Upsilon(2S)$, $\Upsilon(3S)$ and combined data of $\Upsilon(2S,3S)$ in the mass range of $0.212 \le m_{A^0} \le 9.20$ \gevcc. These results are combined with previous \babar\ measurements of $\Upsilon(2S,3S) \to \g A^0$, $A^0 \to \mumu$ to set limits on effective Yukawa coupling of bound \b-quark to the $A^0$. The results of this analysis have been published in Phys. Rev. D {\bf 87}, 031102 (R) (2013), [arXiv:1210.0287]. 

  This thesis is organized in six chapters as discussed bellow:

 {\bf Chapter 1} gives an overview of the SM and its limitations, and describes theoretically the most attractive replacement -- Supersymmetry. The Minimal Supersymmetric Standard Model (MSSM) solves the hierarchy problem of the SM, but fails to explain why $\mu$-parameter is of the order of electroweak scale which is so far from the next natural scale -- the Planck scale. The NMSSM solves this problem while generating a $\mu$-term and introduces an extra \CP-even and \CP-odd light Higgs bosons. Finally, this chapter reviews the phenomenology of the $A^0$.

{\bf Chapter 2} provides a short description of the PEP-II electron-positron collider and the \babar\ detector, which collected the $\Upsilon(2S,3S)$ datasets for this analysis. 

{\bf Chapter 3} describes the datasets used in this analysis, and the reconstruction of the $\Upsilon(2S,3S)$ decay chains: $\Upsilon(2S,3S) \rightarrow \pi^+\pi^- \Upsilon(1S)$, $\Upsilon(1S) \rightarrow \gamma A^0$, $A^0 \rightarrow \mu^+\mu^-$.  It describes the discriminative variables used to separate the signal from background. Monte Carlo (MC) simulated events are used to study the detector acceptance and optimize the event selection criteria. A blind analysis technique is used in this work, where the full data samples are kept blind until  all the selection criteria are finalized. A random forest (RF) classifier is used to improve the purity of $\Upsilon(1S)$ events from $\Upsilon(2S,3S) \to \pipi \Upsilon(1S)$ transitions. Finally, It estimates the remaining backgrounds after applying all the selection criteria.

{\bf Chapter 4}  discusses the signal and background probability density functions (PDFs), which are used to extract the signal from data. The fit procedure is validated by using a cocktail sample of $\Upsilon(2S,3S)$ generic and  $5\%$ of $\Upsilon(2S,3S)$ onpeak datasets, as well as a large number of Toy MC datasets. The full data sample of $\Upsilon(2S,3S)$ are unblinded after finalyzing all the selection criteria and the ML fitting procedure. The signal yields are extracted using the unblinded $\Upsilon(2S,3S)$ datasets. We also describe a trial factor study  used to compute the true significance, i.e. the probability for pure background sample to fluctuate up to a given value of the signal yield. 

{\bf Chapter 5}  describes the possible systematic uncertainties and their sources for this analysis.

{\bf Chapter 6}  presents the $90\%$ confidence level (CL) Bayesian upper limits on the product branching fraction of $\mathcal{B}(\Upsilon(1S) \rightarrow \gamma A^0) \times \mathcal{B}(A^0 \rightarrow \mu^+\mu^-)$ as a function of $m_{A^0}$,  including the systematic uncertainties. The combined upper limits of this result with previous \babar\ results of $\Upsilon(2S,3S) \to \g A^0$, $A^0 \to \mumu$ are also presented. Finally, we present a summary of the results and a brief conclusion.


\setstretch{1.3}  

\mainmatter	  
\pagestyle{fancy}  




















\chapter{Theoretical  $\&$  Phenomenological Framework}
\label{chapter1}
 This chapter begins with an overview of the Standard Model (SM) of particle physics,  including the Higgs mechanism which breaks the elctroweak symmetry spontaneously in the model and provides  masses to the $W^{\pm}$ and $Z^0$ gauge bosons and the fermions. Section~\ref{section:SMlimit}\ reviews some limitation of this model and  describes one of the possible theoretically attractive replacement -- supersymmetry. The Minimal  supersymmetric Standard Model (MSSM) solves the hierarchy problem of the SM, but fails to explain why the value of the $\mu$-parameter is of the order of electroweak scale, which is so far from the next natural scale -- the Planck scale. The Next-to-Minimal Supersymmetric Standard Model (NMSSM) cures this problem and predicts a \CP-odd light Higgs boson whose mass could be  less than twice the mass of the \b-quark.  Finally, section~\ref{section:constraints} reviews some phenomenology related to the light scalar Higgs boson.

\section{The Standard Model}
The SM of Particle Physics describes all the known fundamental particles and their interactions \cite{SM,SM2,SM1, BOOK, Sdawson}. It is a well established theory,  which has passed all the scrutiny by the high energy collider and precision experiments so far. Within this model, all the known matter is composed of spin-1/2 fermion constituents: the leptons and the quarks. There are six types of lepton flavors forming three generations, which are called  electron (e), muon ($\mu$) and tau ($\tau$) with electric charge $Q =-1$ (in the unit of the elementary charge of e), and the corresponding neutrinos $\nu_e$, $\nu_{\mu}$ and $\nu_{\tau}$ with $Q =0$. The quarks also comes in six different flavors: up (u),  down (d), charm (c), strange (s), top (t) and bottom (b), and have fractional charges $Q=+\frac{2}{3}, -\frac{1}{3}, +\frac{2}{3}, -\frac{1}{3}, +\frac{2}{3}$ and $-\frac{1}{3}$, respectively. These fermions interact with each other via exchange of gauge bosons of integral spin-1. The gauge fields in the SM describe the three interactions: the electromagnetic interaction,  the strong interaction and the weak interaction. The electromagnetic interaction is mediated by the photon ($\gamma$), the weak interaction is mediated by the weak vector bosons $W^{\pm}$ and $Z^0$, and strong interaction is mediated by the eight gluons ($g_i$). The gravity is not incorporated by the SM, because it is very weak compared to other interactions. 

The fermions and the gauge bosons acquire mass via Higgs mechanism \cite{HiggsMech, Englert, Guralnik, Kibble} through spontaneous breaking of electroweak symmetry, $SU(3)_C \times SU(2)_L \times U(1)_Y \rightarrow SU(3)_C \times U(1)_{em}$. In addition to providing the masses to the fermions and the $W^{\pm}$ and $Z^0$ gauge bosons, the Higgs mechanism predicts an additional electrically neutral scalar Higgs boson. A Higgs like state has recently been  discovered by the CMS and ATLAS experiments at CERN, and its mass is measured to be 126 \gevcc \cite{Higgs_discovery}. 

\subsection{Gauge Theories}
The gauge theory is a special class of quantum field theory that introduces an invariance principle used to describe the interaction among all the fundamental constituents of matter. The interactions between the fundamental particles are dictated by symmetry principles, which are intimately connected with the ideas of conserved physical quantities. The connection between symmetries and conservation laws is described in the framework of Lagrangian field theory. The gauge symmetry of a physical system is realized through the invariance of the Lagrangian under gauge transformations, which are characterized by Lie group. Global-invariance (phase invariance) under gauge transformation leads to a conserved  charge.  The local gauge invariance (space-times dependence of parameter of the system) of the Lagrangian introduces a vector field, called gauge field, which governs the interaction.  The quanta of the gauge fields are the gauge bosons mediating the interactions.

\subsubsection{Gauge theory of electromagnetic interaction}
The electromagnetic interaction is described by quantum electrodynamics \cite{QED}. The global invariance of  $U(1)$ in the QED introduces the conservation of the electric charge (Q). The local gauge invariance of the gauge theory gives rise to the gauge field corresponding to a  massless gauge boson (photon ($\gamma$)), which describes the interactions among the fundamental charged particles.  The coupling constant ($\alpha$) describes strength of the interaction between the photon and the fermions. However, $\alpha$ is a function of energy when quantum correction are considered. At low energy, the $\alpha$ is given by the fine structure constant, $\alpha = e^2/4\pi\hbar c = 1/137$. Due to the abelian nature of  the U(1) symmetry group, photon is charge-less, and do not interact with each other directly. The electromagnetic interaction is a long range interaction.

\subsubsection{Gauge theory of strong interaction}
The strong interaction is described by quantum chromodynamics (QCD) \cite{QCD}. The symmetry group of QCD is $SU(3)_C$, where $C$ refers to colour  and 3 refers to the three possible colour states of the quarks, normally termed as red, green and blue. Colour symmetry is exact, so QCD calculations are independent of the colour of the quarks. For example, probability of a red quark scattering off a green quark is the same as the probability of  a red quark scattering off a blue quark. The local gauge invariance of $SU(3)_C$ gives rise to eight types of the gluonic fields. QCD is a non-abelian theory, where the gluons carry both colour and anti-colour, in contrast to the photon in QED which does not carry the electric charge. Gluons interact with each other directly and as strongly as they do with quarks. Due to this gluon-gluon interaction, the strong force increases with distance resulting in confinement of quarks. This means, the quarks do not exist freely in Nature, but bind together by the strong force and form the mesons ($q\overline{q}$) and the baryons ($qqq$), where $q$ stands for a quark and $\overline{q}$ stands for an anti-quark.  

\subsubsection{Gauge theory of electroweak interaction}
  The electromagnetic and weak interactions are combined in an $SU(2)_L\times U(1)_Y$ gauge theory of electroweak interaction, developed by  Glashow, Weinberg and Salam \cite{SM, Electroweak}. The subscript \rm{\lq\lq $L$\rq\rq} indicates that only the left-handed (right-handed) components of the fermion (antifermion) fields take part in weak interactions. The fermions appear as left-handed doublets and right-handed singlets under the $SU(2)_L$. Global gauge invariance under the $SU(2)_L$ gauge transformation leads to the conservation of the weak-isospin, $T$. Requiring the local $SU(2)_L$ gauge invariance of the Lagrangian of the system introduces a weak-isospin triplet of the gauge fields, $W_{\mu}^i$, $i =1,2,3$. The $SU(2)_L$ is a non-abelian group which leads to the self-interactions of the gauge fields. The global gauge invariance under the $U(1)_Y$ transformation leads to the conservation of weak-hypercharge, $Y$. However, the local gauge invariance  of $U(1)_Y$ introduces  vector gauge field, $B_{\mu}$.  The weak-hypercharge, $Y$, third component of weak-isospin, $T_3$, and electric charge, $Q$ are related by the Gell-Mann-Nishijima relation:

\begin{equation}
Q = T_3 + \frac{1}{2}Y.
 \end{equation}

\subsection{Electroweak symmetry breaking in the SM: The Higgs Mechanism}

 The gauge invariance of  $SU(2)_L \times U(1)_Y$  requires  massless gauge bosons, since the presence of a mass term for the gauge boson  violates gauge invariance ($M^2A_{\mu}A^{\mu}$ is the not invariant under $A_{\mu} \rightarrow A_{\mu} - \partial_{\mu}\chi$,  where $\chi$ is a function of position in space time.  So $M^2$ must be zero in a gauge symmetric Lagrangian). This difficulty is circumvented by the Higgs mechanism through which electroweak symmetry breaking is achieved spontaneously \cite{HiggsMech}. The SM contains a  weak-isospin $SU(2)_L$ doublet of complex scalar Higgs fields (with weak-hypercharge Y=1),

\begin{equation}
\Phi = 
\begin{pmatrix}
\phi^+\\
\phi^0
\end{pmatrix}
\end{equation}

\noindent in the Lagrangian of the system. The most general renormalizable and $SU(2)_L \times U(1)_Y$ invariant Lagrangian allowed, involving only the gauge bosons and scalar fields is given by

\begin{equation}
\mathcal{L} = -\frac{1}{4}W_{\mu \nu}^i W^{\mu \nu i} - \frac{1}{4} B_{\mu \nu}B^{\mu \nu} + (D_{\mu}\Phi)^{\dagger}(D^\mu \Phi) - V(\Phi),
\end{equation}

\begin{figure}[htb]
\begin{center}
\includegraphics[width=0.5\textwidth]{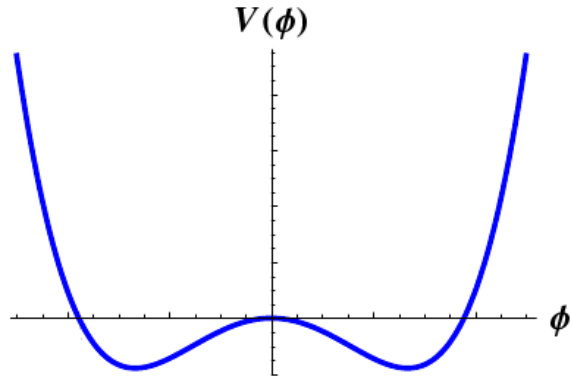}
\caption{One dimensional projection of Higgs potential (V($\phi$)) as a function of scalar field ($\phi$). The  (V($\phi$)) develops a vacuum expectation value (VEV) at $\phi = 0$ when $\mu^2 < 0$.}
\label{fig:VEV}
\end{center}
\end{figure}

\begin{center}
$W_{\mu\nu}^i = \partial_{\mu}W_{\nu}^i -\partial_{\nu}W_{\mu}^i - g \epsilon^{ijk}W_{\mu}^jW_{\nu}^k$,

$B_{\mu\nu} = \partial_{\mu}B_{\nu} - \partial_{\nu}B_{\mu} $,

\end{center}
\begin{equation}
D_{\mu} = \partial_{\mu} + \frac{1}{2}ig\tau^iW_{\mu}^i+\frac{1}{2}ig'YB_{\mu}, 
\end{equation} 

\noindent where $W_{\mu}^i$ ($i=1,2,3$) are the three massless $SU(2)_L$ gauge bosons, $B_{\mu}$ the massless $U(1)_Y$ gauge boson, and the scalar potential is given by

\begin{equation}
V(\Phi) = \mu^2|\Phi^{\dagger}\Phi| + \lambda|\Phi^{\dagger}\Phi|^2
\label{Phi},
\end{equation}

\noindent here $g$ and $g'$ are the gauge coupling constants of  $SU(2)_L$ and $U(1)_Y$, respectively. For a choice of $\lambda > 0$ and $\mu^2 < 0$, the state of minimum energy for the potential V is not at zero, but at $|\Phi^{\dagger}\Phi|=-\mu^2/2\lambda \equiv v^2/2$ (Figure~\ref{fig:VEV}). The scalar field thus develops a non-vanishing vacuum expectation value (VEV), which is degenerate. A single value of the VEV can be chosen, which is essentially a choice of a preferred \rm{\lq\lq direction\rq\rq} in the Higgs-doublet phase space. The usual choice is

\begin{equation}
\Phi(x) = \frac{1}{\sqrt{2}}
\begin{pmatrix}
0\\
v+H(x)
\end{pmatrix},
\label{Equation:vev}
\end{equation}

\noindent where $H(x)$ is a physical scalar filed. The choice of this new ground state \rm{\lq\lq spontaneously\rq\rq} breaks the $SU(2)_L \times U(1)_Y$ symmetries to $U(1)_{EM}$, while maintaining the renormalizability and unitarity of the theory. As the $U(1)_Y$ gauge symmetry remains unbroken in this transformation, the associated gauge boson, the photon, remains massless. However, three of the degrees of freedom of the scalar doublet (corresponding to Goldstone bosons) are \rm{\lq\lq eaten by\rq\rq} or transformed into the longitudinal polarization components of the weak-isospin triplet of bosons, giving the $W^{\pm}$ and $Z^0$ bosons their masses of $M_W = \frac{1}{2} v g$ and $M_{Z^0} = \frac{1}{2}v (g^2 +g'^2)^{1/2}$, respectively \cite{SM}. The mass eigenstates are expressed in terms of the gauge eigenstates as bellow: 

\begin{center}
$W_{\mu}^{\pm}= \frac{1}{\sqrt{2}}(W_{\mu}^1\mp iW_{\mu}^2)$, 

$Z_{\mu}^0=W_{\mu}^3cos\theta_W - B_{\mu}sin\theta_W$,
\begin{equation}
 A_{\mu} = W_{\mu}^3 sin \theta_W + B_{\mu}cos \theta_W,
 \end{equation}

\end{center}

\noindent where $A_{\mu}$ is the  gauge field of the electromagnetic interaction, and  $\theta_W$ is the Weinberg mixing angle. The remaining degree of freedom corresponds to a massive  neutral scalar particle, the Higgs boson, $H^0$.  The mass of this scalar is  given by $m_{H^0}^2 = 2v^2\lambda$.

\begin{figure}[htb]
\begin{center}
\includegraphics[width=0.45\textwidth]{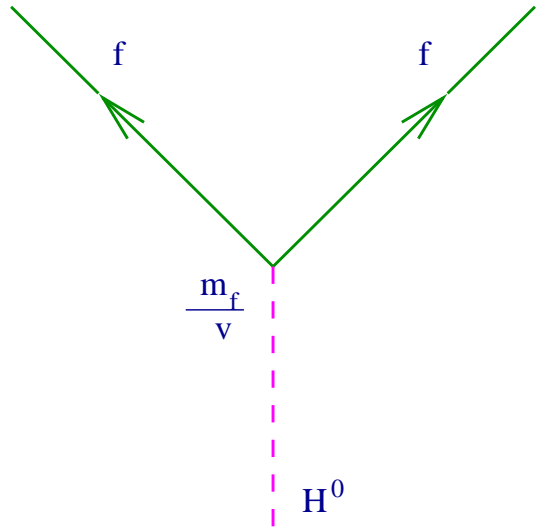}
\caption{The Yukawa coupling of the Standard Model Higgs boson to the fermions.}
\label{fig:Higgs}
\end{center}
\end{figure}

 The Higgs mechanism also provides masses to the quarks and leptons while including the following $SU(2)_L\times U(1)_Y$ gauge invariant terms for the first generation of leptons and quarks in the Lagrangian

\begin{center}
$\mathcal{L}_{\mathrm{Lepton}}= -g_e \begin{bmatrix}\begin{pmatrix}
\overline{\nu_e}~~~
\overline{e}
\end{pmatrix}_L \begin{pmatrix}
\phi^+ \\
\phi^0
\end{pmatrix}e_R + \overline{e}_R \begin{pmatrix}
\phi^- ~~~
\overline{\phi^0}
\end{pmatrix}\begin{pmatrix}
\nu_e \\
e
\end{pmatrix}_L \end{bmatrix},$
\end{center}

\begin{equation}
\mathcal{L}_{\mathrm{Quark}}= -g_d \begin{pmatrix}
\overline{u}~~~
\overline{d}
\end{pmatrix}_L \begin{pmatrix}
\phi^+ \\
\phi^0
\end{pmatrix}d_R - g_u\begin{pmatrix}
\overline{u}~~~
\overline{d}
\end{pmatrix}_L  \begin{pmatrix}
-\phi^0 \\ 
\phi^-
\end{pmatrix}u_R + hermitian~conjugate~ (h.c).
\label{Equation:Lang}
\end{equation}

\noindent Here $\begin{pmatrix}
-\phi^0 \\ 
\phi^-
\end{pmatrix} = -i\tau_2 \begin{pmatrix}
\phi^+ \\
\phi^0
\end{pmatrix}^*$, where $\tau_2= \begin{pmatrix}
~0 ~~
-i \\
i ~~~~~
0
\end{pmatrix}$  is the isospin version of the Pauli matrix. Second and third generations of leptons and quarks have similar expressions.  After breaking the symmetry spontaneously as discussed above, the Higgs scalar picks up a vacuum expectation value  given by equation~\ref{Equation:vev}. This will generate the mass term of the fermion, and an interaction term with the Higgs particle

\begin{center}
$\mathcal{L}_{\mathrm{Lepton}} = -m_e\overline{e}e -\frac{m_e}{v}\overline{e}e\mathrm{H}^0$,
\end{center}

\begin{equation}
\mathcal{L}_{\mathrm{Quark}} =- m_d \overline{d}d -m_u\overline{u}u -\frac{m_d}{v}\overline{d}d\mathrm{H}^0 - \frac{m_u}{v}\overline{u}u\mathrm{H}^0,
\end{equation}

\noindent with the identification $m_i = g_i v/\sqrt{2}$, where $m_i$ is the mass of each fermion $i$, $g_e$, $g_u$ and $g_d$ are the Yukawa coupling constants (Figure~\ref{fig:Higgs}). Thus, the strength of Higgs boson couplings to fermions  is proportional to the corresponding particle masses.

\section{Drawback of the SM}
\label{section:SMlimit}
The SM is the result of many experimental observations and progress in the theoretical understanding of Nature. Most of the theoretical results of the SM agree with the
 experimental data. However, the SM can not be quantified as a \rm{\lq\lq theory of everything\rq\rq}. There is no method to incorporate gravity which becomes important at energy scales approaching the Planck scale ($M_{Planck} = (8\pi G_N)^{-1/2} \sim 2.4 \times 10^{18}$ $\gevcc$) and so the SM must be considered as an effective theory at energies below this scale.  Some of the  important drawbacks of the SM and their possible solutions are described bellow: 

\subsection{Hierarchy problem of the SM}   
\begin{figure}[htb]
\begin{center}
\includegraphics[width=0.55\textwidth]{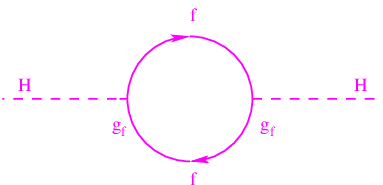}
\includegraphics[width=0.40\textwidth]{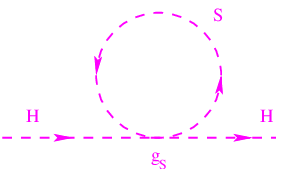}
\smallskip
\centerline{\hfill (a) \hfill \hfill (b) \hfill}
\smallskip
\caption{Loops affecting the squared Higgs mass from (a) fermions trilinear couplings and (b) scalar quartic couplings.}
\label{fig:div}
\end{center}
\end{figure}

  The mass of the SM Higgs boson is expected to be of the order of electroweak scale ($\sim {\cal O} (M_W)$). The self-coupling effects in the scalar Higgs field involving higher-order fermionic loops are quadratically divergent (Figure~\ref{fig:div}(a)). A cut-off scale $\Lambda_{cutoff}$ on the momentum integral can be introduced to prevent these radiative correction from going to infinity. The Higgs couples with fermion pair via an Yukawa interaction term of $-g_f\mathrm{H}\overline{f}f$ in the Lagrangian. At one loop each fermion contributes a correction of mass term, which is \cite{LucPape}

\begin{equation}
\Delta m_{H^0,f}^2 = \frac{g_f^2}{8\pi^2}\biggr[-\Lambda_{cutoff}^2 + 6m_f^2ln\biggr(\frac{\Lambda_{cutoff}}{m_f}\biggr)\biggr].
\label{Equ:SMQuad}
\end{equation}
 
\noindent These corrections blow up as $\Lambda_{cutoff} \to \infty$. To explain the $m_{H^0}^2 \sim {\cal O} (M_W)$ we need either $\Lambda_{cutoff} \lesssim 1$ \tev, or extreme fine tuning (adjusting the value of $g_f$  accordingly) so that the correction is of the electroweak scale. This difficulty is known as hierarchy problem of the SM.
   Supersymmetric extension of the SM solves the hierarchy problem of the SM while introducing the superpartners of each fundamental particles that differ by  half integral-spin \cite{martin}. The superpartners of the fermions also couple to the Higgs by a quartic interaction of the form $-g_S |H^0|^2|S|^2$, and thereby contribute to the Higgs mass corrections through loops as shown in Figure~\ref{fig:div}(b). The loop correction contributes to the Higgs mass  by:

\begin{equation}
\Delta m_{H^0,S}^2 = \frac{g_S}{16\pi^2}\biggr[\Lambda_{cutoff}^2 - 2m_S^2ln\biggr(\frac{\Lambda_{cutoff}}{m_S}\biggr)\biggr].
\label{Equ:SuperSym}
\end{equation}
 
\noindent It is seen from equation~\ref{Equ:SMQuad} and ~\ref{Equ:SuperSym} that if every fermion is accompanied by a scalars with coupling $g_S = 2g_f^2$, the quadratic divergences cancel exactly. After adding the equation ~\ref{Equ:SMQuad} and ~\ref{Equ:SuperSym}, the total correction is reduced to

 \begin{equation} 
 \Delta m_{H^0,Tot}^2 \simeq \frac{g_f^2}{4\pi^2}(m_S^2 - m_f^2)ln\biggr(\frac{\Lambda_{cutoff}}{m_S}\biggr).
\label{Equ:logarithmdiv}
\end{equation}

\subsection{Unification}
The SM unifies the electromagnetic and weak interactions at the electroweak scale. This motivation can also be extended to the grand unification scale \cite{GUT1,GUT2} where the electromagnetic, weak and strong interactions are unified together \cite{Unify} through the supersymmetric extensions of the SM.

\subsection{Dark Matter}
There is ample evidence from observation like the rotation curve of galaxies that luminous matter in the universe accounts for only a small fraction of the total matter-energy density. The unknown matter content of the universe is called the dark matter (DM) \cite{DM1, DM2, DM3}. It accounts for about $23\%$ of the total matter density of the observable universe, while the ordinary matter accounts for only $4.6\%$, with the remainder being attributted to dark energy. The SM does not have  viable candidate for DM  particles. However, there are extension of SM including  supersymmetric models which contain viable candidates for DM. We should mention that, such models do not explain the existence of dark energy, which contributes to about $73\%$  of the total energy of the universe.

\section{The Minimal Supersymmetric Standard Model}
As we see in the last section that one of the best motivated extensions of the SM of particle physics is the introduction of Supersymmetry (SUSY) \cite{martin,LucPape,MSSMSUSY}.  The SUSY solves the hierarchy problem of the SM and unifies the three forces of electromagnetic, weak and strong at the Grand unified scale.  The Minimal Supersymmtric Standard Model (MSSM) is a minimal supersymmetric extension of the SM \cite{Haber}. It is also based on the gauge symmetry of $SU(3)_C \times SU(2)_L \times U(1)_Y$. It transforms bosonic states into fermionic states and vice versa via an operator $Q$
\begin{equation}
Q|{\rm Boson}\rangle = |{\rm Fermion} \rangle ~~~~~~~~~~~~~~~~~~~~~~~Q|{\rm Fermion} \rangle = |{\rm Boson}\rangle.
\end{equation}
\noindent If the $Q$ and its hermition conjugate $Q^{\dagger}$ hold following commutation relations

~~~~~~~~~~~~~~~~~~~~~~~~~~~~~~~~~~~~~~~~~~~~~~~~~~~$\{Q,Q^{\dagger}\} = P^{\mu},$
\begin{equation}
\{Q,Q\} =  \{Q^{\dagger},Q^{\dagger}\} =  0,
\end{equation}
~~~~~~~~~~~~~~~~~~~~~~~~~~~~~~~~~~~~~~~~~~~~~~~~~~~ ~~~~~$[P^{\mu},Q] = [P^{\mu},Q^{\dagger}] = 0$,

\noindent  then the theory is able to describe the chiral fermions as they are observed in Nature \cite{martin}. Here, $P^{\mu}$ is the four-momentum, which is the generator of space-time translations. Irreducible representations of such types of algebra are called supermultiplets and describe the single particle state. A supermultiplet includes an equal number of fermionic ($n_F$) and bosonic ($n_B$) degrees of freedom, which means that  every SM particle has their own superpartner, which has the same quantum  numbers except their spin which differ by 1/2. The superpartner of fermions are scalar particles called sfermions, that of gauge boson  are  spin-1/2 particles called gauginos, and that of the Higgs bosons are spin-1/2 particles called Higgsinos.  

A gauge or vector supermultiplet  contains a massless vector boson ($n_B =2$) and the superpartner of this boson, a spin-1/2 Weyl fermion ($n_F =2$). The Weyl fermion does not have its own antiparticles in contrast to the Majorana fermion that is its own antiparticle. The known SM gauge bosons and  the corresponding gauginos are contained in  vector supermultiplet in the MSSM.  
   
A chiral supermultiplet contains a spin-1/2 Weyl fermion ($n_F =2$) and two real scalars (each $n_B =1$, spin 0), which can be described by a complex scalar field. The Higgs bosons, Higgsinos, and spin-1/2 fermions and sfermions are part of such  chiral superfields in the MSSM.  

 The MSSM postulates two complex $SU(2)_L$ doublet scalar superfields, denoted by $\hat{H}_u$  and  $\hat{H}_d$ with weak-hypercharges $Y=\pm 1$:

\begin{equation}
\hat{ H}_{ u} = 
\begin{pmatrix}
\hat{ H}_{ u}^+\\
\hat{ H}_{ u}^0
\end{pmatrix}
,
~~~~~~~
\hat{ H}_{ d} = 
\begin{pmatrix}
\hat{ H}_{ d}^0\\
\hat{ H}_{ d}^-
\end{pmatrix}.
\label{Eq:Higgsfield}
\end{equation}

\noindent The superpotential  of the MSSM  involving the Higgs fields  is given by \cite{martin}:
\begin{equation}
W = (g_u)^{ij}\hat{u}_i\hat{Q}_j.\hat{H}_u - (g_d)^{ij}\hat{d}_i\hat{Q}_j.\hat{H}_d - g_e^{ij}\hat{e}_i\hat{L}_j.\hat{H}_d + \mu \hat{H}_u.\hat{H}_d,
\label{Eq:superpot}
\end{equation}

\noindent where the labels $i,j$ are family indexes of quarks and leptons. The $g_u$, $g_d$ and $g_e$ are the Yukawa coupling constants of up-type quarks, down-type quarks and leptons, respectively. The $\mu$-term mixes the two Higgs superfields. 

 The gauge-invariant Higgs scalar potential built from the two Higgs doublets in Equation~\ref{Eq:Higgsfield}  is consistent with the electroweak sector of the SM and spontaneously breaks $SU(2)_L\times U(1)_Y$ down to $U(1)_{EM}$. It is given by:
\begin{equation}
V = \frac{1}{8}(g^2 +g'^2)(|H_d|^2 - |H_u|^2) + \frac{1}{2}g^2|H_d^{\dagger}H_u|^2 + \mu^2(|H_d|^2 +|H_u|^2) + m_{H_d}^2 |H_d|^2 + m_{Hu}^2 |H_u|^2 + \mu B(H_u.H_d +h.c.),
\label{Eq:muterm}
\end{equation}  
\noindent where $m_{H_d}^2$, $m_{H_u}^2$ and $B$ are soft supersymmetry breaking parameters. This scalar potential is minimized by the vacuum expectation values (VEV's) of the Higgs fields

\begin{equation}
\langle H_d \rangle = \frac{1}{\sqrt{2}}
\begin{pmatrix}
v_1\\
0
\end{pmatrix}
~
~
~
~
~
~
\langle H_u \rangle = \frac{1}{\sqrt{2}}
\begin{pmatrix}
0\\
v_2
\end{pmatrix},
\end{equation}

\noindent which spontaneously breaks the electroweak symmetry, $SU(2)_L\times U(1)_Y \to U(1)_{EM}$. A conventional notation is used to relate the two VEVs by $\rm{tan}\beta = v_1/v_2$. The two VEVs can then be defined as $v_1 = \langle H_d \rangle = v {\rm sin}\beta$ and $v_2 = \langle H_u \rangle = v {\rm cos}\beta$  (where $v = \sqrt{v_1^2 +v_2^2} = 2 m_W/g \simeq 246$ \gev). The physical MSSM Higgs sector consists of two neutral \CP-even ($H^0$ and $h^0$), a neutral \CP-odd (A) and a pair of charged ($H^{\pm}$) Higgs bosons. The MSSM also  contains four neutralinos ($\tilde{\chi}_{1,2,3,4}^0$), among which the $\tilde{\chi}_1^0$ is the lightest supersymmetric particle (LSP) in the R-parity conserving model, and is a viable candidate of DM.

\subsection{The $\mu$ problem in MSSM}
  The MSSM superpotential (Equation~\ref{Eq:superpot}) contains a $\mu$-term, which mixes the $\hat{H}_u$ and $\hat{H}_d$ chiral superfileds, is the only dimensional coupling in the superpotential. The value of $\mu$ is expected to be of the order of electroweak scale, which is many orders of magnitude smaller than the next natural scale, the Planck scale. A possible solution for this problem can be found in the framework of the Next-to-Minimal Supersymmetric Standard Model (NMSSM). 

\section{The Next-to-Minimal Supersymmetric Standard Model}
The NMSSM adds a singlet chiral superfield ($\hat{N}$)  to the MSSM \cite{Dermisek, Fine-tuning, Hiller}. The superpotential of the NMSSM contains a trilinear term along with an $\hat{N}^3$ term instead of the $\mu$-term of MSSM superpotential in Equation~\ref{Eq:superpot}, basically defined as 
\begin{equation}
W_{NMSSM} =  (.....) + \lambda \hat{N} \hat{H}_u.\hat{H}_d + \frac{\kappa}{3}\hat{N}^3,
\label{Eq:NMSSMSuperpot}
\end{equation}

\noindent where $\lambda$ and $\kappa$ are dimensionless Yukawa couplings, and $\hat{H}_u$ and $\hat{H}_d$ are up and down types of Higgs superfields. The associated soft terms, which break the supersymmetry explicitly, are $\lambda A_{\lambda}NH_uH_d +\frac{1}{3}A_{\kappa}N^3$. In the presence of these soft supersymmetry breaking terms, a vacuum expectation value (VEV) of $N$, which is of the order of electroweak scale generates an effective $\mu$-term with $\mu_{eff} = \lambda \langle N \rangle$, which solves the \rm{\lq $\mu$-problem\rq} of the MSSM \cite{Kim}.  As a result, the NMSSM Higgs sector contains a total of three \CP-even, two \CP-odd and two charged Higgs bosons.  This model also contains a total of five neutral fermionic states, $\tilde{\chi}_{1,2,3,4,5}^0$, which are LSP (in the R-parity model) and viable candidates of DM. The Higgs sector of the NMSSM contains  six independent parameters:
\begin{equation}
\lambda, ~~ \kappa, ~~ A_{\lambda}, ~~ A_{\kappa}, ~~ \mathrm{tan}\beta, ~~ \mu_{eff}, 
\end{equation}
\noindent where the sign conventions for the fields $\lambda$ and $\mathrm{tan}\beta$ should be always positive, while $\kappa$, $A_{\lambda}$, $A_{\kappa}$ and $\mu_{eff}$ may have either sign. 

The mass of the lightest \CP-odd Higgs boson ($A^0$) is controlled by the soft-trilinear coupling $A_{\lambda}$ and $A_{\kappa}$ and vanishes in the Peccei-Quinn symmetry limit,  $\kappa \rightarrow 0$ \cite{Peccei-Quinn}, or a global $U(1)_R$ symmetry in the limit of vanishing soft term, $A_{\lambda}, A_{\kappa} \rightarrow 0$, which is spontaneously broken by the VEVs, resulting in a Nambu-Goldstone boson in the spectrum \cite{UR}. This symmetry is explicitly broken by the trilinear soft terms so that the $A^0$ is naturally small.  In a generic case, the fermion coupling to the light pseudoscalar $A^0$ field can be defined by an interaction term:

\begin{equation}
\mathcal{L}^{f\overline{f}A^0} = -X_f \frac{m_f}{\mathrm{v}}A^0\overline{f}(i\gamma_5)f,
\end{equation}

\noindent where  $X_f$ is the coupling constant, which depends on the type of fermion with a mass $m_f$ \cite{Fullana, JHEP01}.   In the NMSSM, $X_d = \mathrm{cos}\theta_A\mathrm{tan}\beta$ for the down-type fermion pair and $X_u = \mathrm{cos}\theta_A\mathrm{cot}\beta$ for the up-type fermion pair, where  $\theta_A$ is the mixing angle between the singlet ($A_S$) component and MSSM like doublet component ($A_{MSSM}$) of the $A^0$.  With this mixing angle, the lighter \CP-odd state of the $A^0$ is defined as:
 \begin{equation}
 A^0 = \mathrm{cos}\theta_A A_{MSSM} + \mathrm{sin}\theta_A A_S.
\end{equation} 

  Such light state of the $A^0$ is  not excluded by the LEP constraints \cite{LEP-EXPT}, where the \CP-even Higgs boson, $h$, could decay dominantly into a pair of \CP-odd scalars \cite{Fine-tuning, UR, lep1,lep2,lep3,lep4,lep5}. The LEP experiment has also excluded  a SM-like $h$ decaying to $b\overline{b}$ for $m_h < 114$ \gevcc and  placed a strong constraints on  $e^+e^- \rightarrow Zh \rightarrow Zb\overline{b}$ as well as the effective coupling of $C_{eff}^2 \equiv [g_{ZZh}^2/g_{ZZh_{SM}}^2]\mathcal{B}(h\rightarrow b\overline{b})$ \cite{lep2}. The Large hadron collider (LHC) experiment  will also not be able to discover such scalar states if $h$ decays primarily into a pair of \CP-odd scalars with $m_A^0$ bellow the $B\overline{B}$ threshold \cite{lep1,lep4,lep5}. In this case, the $A^0$ can be accessible via the $\Upsilon$ decays \cite{Fullana,JHEP01, PRD41-1547, Mod-Phys-Lett17-2265, PRD72-103508, PRD76-051105, PLB665-219} while using the large datasets of the current generation of \B-Factories, such as  \babar,\  CLEO and Belle experiments.

\section{Phenomenology of the light scalar states}
\label{section:constraints}
 
The lightest state of the $A^0$ in the NMSSM is constrained to have the mass bellow the $b\overline{b}$ threshold, $2m_b$ \cite{Hiller}, to avoid the detection at LEP. A pseudo-scalar axion having a mass around 360--800 \mevcc and decaying into a lepton pair with a Higgs-like coupling is also predicted by models motivated by astrophysical observations \cite{nomura}.  The low mass Higgs boson could explain the origin of mass of the light elementary particles, the mystery related to the Dark matter and Dark energy which contributes more than $90\%$ matter density of the universe. In the framework of dark matter, the dark matter particles can annihilate into pairs of the dark photons, which subsequently decay to SM particles. In a minimal model \cite{Dark-photon}, the dark photon mass is generated via the Higgs mechanism, adding a dark Higgs boson in the theory. The mass hierarchy between dark photon and dark Higgs boson is not constrained  experimentally, so the dark Higgs boson could be light as well \cite{Dark-Higgs}. These light scalar states could be within the reach of present particle accelerators, such as the \B-Factory at SLAC.  

The branching fractions of $\mathcal{B}(\Upsilon(nS) \to \g A^0)$  ($n = 1,2,3$) are  related to the effective Yukawa coupling ($f_{\Upsilon}$) of the  \b-quark to the $A^0$ through \cite{Wilzek,Radiative_Quarkonium,P_Nason}: 

\begin{equation}
\frac{\mathcal{B}(\Upsilon(nS)\to \g A^0)}{\mathcal{B}(\Upsilon(nS) \to l^+l^-)} = \frac{f_{\Upsilon}^2}{2\pi \alpha}\biggl(1-\frac{m_{A^0}^2}{m_{\Upsilon(nS)}^2}\biggl), 
\label{Eq:fycoupling}
\end{equation}

\noindent  where $l \equiv e$ or $\mu$ and $\alpha$ is the running  fine structure constant. In the SM, the value of  $f_{\Upsilon}$ is defined as:
\begin{equation}
f_{\Upsilon, SM}^2 = \sqrt{2}G_F m_{b}^2 C_{QCD} \approx (2-3) \times 10^{-4}, 
\end{equation}

\noindent where $C_{QCD} \approx 0.7 -1.0$ \cite{Yury} includes the QCD loop corrections and relativistic corrections to $\mathcal{B}(\Upsilon(nS) \to \g A^0)$ \cite{P_Nason}, as well as the leptonic width of $\Upsilon(nS) \to l^+l^-$ \cite{Upsilonwidth}. However, the coupling of bound \b-quark to the $A^0$  in the NMSSM is $f_{\Upsilon, NMSSM}^2 = \sqrt{2}G_F m_{b}^2 X_d^2 C_{QCD}$. The Yukawa coupling also depends upon the axion constant $f_a$ in the axion model of Nomura and Thaler \cite{nomura}. A study of the NMSSM parameter space predicts the branching fraction of $\Upsilon(1S) \rightarrow \gamma A^0$ to be in the range of $10^{-6} - 10^{-4}$ depending upon the $A^0$ mass, $\mathrm{tan}\beta$ and $\mathrm{cos}\theta_A$  \cite{PRD76-051105}.

 In the SM, interactions between the leptons and gauge bosons are same for all the lepton flavors, and therefore the quantity $\mathrm{R}_{ll'} = \Gamma_{\Upsilon(1S) \rightarrow ll}/\Gamma_{\Upsilon(1S) \rightarrow l'l'}$ with $l$,  $l' =$  $e$, $\mu$, $\tau$ and $l' \ne l$, is expected to be close to one. In the NMSSM, any significant deviations of $\mathrm{R}_{ll'}$ from unity would violate lepton universality, which may arise due to presence of the $A^0$ that couples to the $\Upsilon(1S)$. \babar\ has measured the value of $\mathrm{R}_{\tau\mu}(\Upsilon(1S)) = 1.005 \pm 0.013(\mathrm{stat}) \pm 0.022(\mathrm{syst})$ using a sample of $(121.8 \pm 1.2)\times 10^6$ $\Upsilon(3S)$ events, which shows no significant deviation from the expected SM value \cite{PRL104-191801}. If the light \CP-odd Higgs boson $A^0$ has a mass in the range of $9.2 < m_{A^0} < 12$ \gevcc, the NMSSM can account for the anomalous muon magnetic moment \cite{arxiv:0803.2509}. 

For large value of $\mathrm{tan}\beta$, the $A^0$ will primarily decay to heavier down-type fermion that is kinematically available.  The branching fractions of $A^0 \rightarrow f\overline{f}$ as a function of tan$\beta$ and $m_{A^0}$ are summarized in \cite{NMSSM-constraind}. The same reference \cite{NMSSM-constraind} also summarizes the expected  $\mathcal{B}(\Upsilon(3S) \rightarrow \gamma A^0)$ for various SUSY model parameters, with the constraint that the model does not require the \rm{\lq\lq fine tuning\rq\rq} \cite{lep2}. \babar\ has previously searched for $A^0$ production in the radiative decays of $\Upsilon(nS) \rightarrow \gamma A^0$ with $n=1,2,3$, where the $A^0$ decays to muons \cite{Aubert:2009cp}, taus \cite{Aubert:2009cka}, invisible \cite{0808.0017,Sanchez:2010bm}, or hadrons \cite{Lees:2011wb}.  Similar searches have also been performed by CLEO in the di-muon and di-tau final states in radiative $\Upsilon(1S)$ decays \cite{CLEO_Higgs0}, and more recently by BESIII in the decay chain of $J/\psi \rightarrow \gamma A^0$, $A^0 \rightarrow \mu^+\mu^-$ \cite{BESIII_Higgs0}, and by CMS experiment in $pp\to A^0$, $A^0 \to \mumu$ \cite{CMS}.  \babar\ results \cite{Aubert:2009cp} for $A^0 \rightarrow \mu^+\mu^-$ decay  rules out approximately $80\%$ of the NMSSM parameter  space in the $m_{A^0} < 2m_{\tau}$ range at $\mathrm{tan}\beta = 3$. Reference \cite{Domingo} interprets the \babar\ \cite{Aubert:2009cp} and CLEO \cite{CLEO_Higgs0} results in terms of the limit of $X_d$ as a function of $m_{A^0}$ and predicts that these results fit with an approximate limit of $X_d < 0.5$ for $\mathrm{tan}\beta = 5$.

 This thesis describes  a search for a di-muon resonance in the fully reconstructed decay chain  of $\Upsilon(2S, 3S) \to \pi^+\pi^- \Upsilon(1S)$, 
 $\Upsilon(1S) \to \g A^0$, $A^0 \to \mumu$. This search is based on a sample of $(92.8 \pm 0.8) \times 10^6$  $\Upsilon(2S)$ and 
$(116.8 \pm 1.0) \times 10^6$ $\Upsilon(3S)$ mesons collected with the \babar\ detector at the PEP-II asymmetric-energy $\epem$ 
collider located at SLAC National Accelerator Laboratory. A sample of  $\Upsilon(1S)$ mesons is  selected by tagging the di-pion 
transition, which results in a substantial background reduction compared to direct searches of $A^0$ in $\Upsilon(2S,3S) \to \g A^0$ 
decays. We assume that the $A^0$ is a scalar or pseudo-scalar particle with a negligible  decay width  compared to the 
experimental resolution.
 






















\chapter{The \babar\ experiment}
\label{Chapter2}

\babar\ is a high luminosity $e^+e^-$ asymmetric energy collider experiment located at SLAC National Accelerator Laboratory, California, USA. It was primarily designed to study the \CP-violation in \B-meson decays, and therefore, for most of its run period, the experiment was operated at the \epem center of mass (CM) energy corresponding to the $\Upsilon(4S)$ resonance, which is just above the $B\overline{B}$ threshold. This has allowed the \babar\ to perform precision measurements of the \B meson decays, probing deeply into the phenomena of \CP-violation  and thereby establishing the CKM formalism \cite{CKMSM} of the SM. Despite its initial goal of the study of \CP-violation in \B-meson decays, the \babar\ experiment has also carried out significant studies in many other  fields of  high energy physics such as: $\tau$ physics, physics of the heavy quarks, decays of the D-mesons and physics beyond SM such as low mass Higgs searches.  To achieve the goal of some of these physics programs, the \babar\ has also collected the data at the CM energy corresponding to the $\Upsilon(2S)$ and $\Upsilon(3S)$ resonances in the last phase of the data acquisition period in 2008.  

This chapter outlines the design of the PEP-II \B-Factory and the \babar\ detector which enabled such a rich physics program from this experiment.

\section{The PEP-II accelerator}
\begin{figure}[htb]
\begin{center}
\includegraphics[width=6.6in]{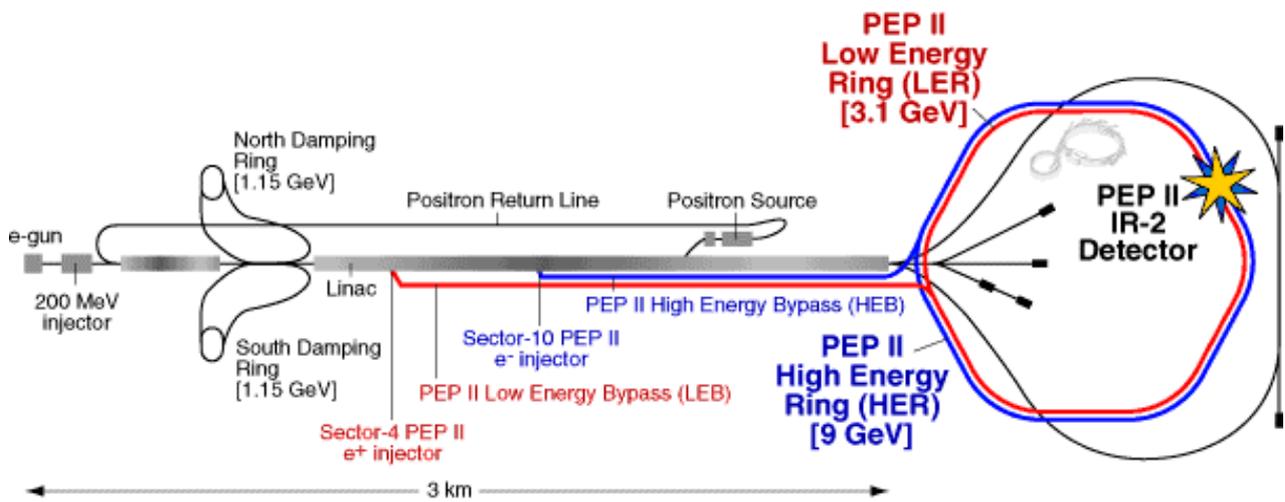}
\caption{The diagram of the PEP-II Accelerator.}
\end{center}
\label{fig:PEPII}
\end{figure}

The PEP-II  is an asymmetric energy \epem collider operating at the center-of-mass energy of 10.58 $\gevcc$ corresponding to the mass of the $\Upsilon(4S)$ resonance \cite{pepII}. This resonance subsequently  decays almost exclusively to both $\B^0\overline{B^0}$ and $\B^+\B^-$ pairs, which provide an ideal framework for studying the \CP-violation in the \B mesons decay.  A schematic of the overall layout of the PEP-II collider is shown in Figure~\ref{fig:PEPII}.

\noindent The \babar\ experiment uses two accelerators: the SLAC linear accelerator (linac) and the PEP-II storage ring facility. The SLAC linac accelerates the electron and positron beams to the required high energies, and then it injects them into the  PEP-II's storage rings. PEP-II consists of two storage rings, a high Energy Ring (HER) for the 9.0 GeV electron beam, and a low Energy Ring (LER) for the 3.1 GeV positron beam. The two beams move in opposite directions and collide at the interaction point, where the \babar\ detector is located. The asymmetric beam energies cause the $\Upsilon(4S)$ system to be Lorentz-boosted by a factor of $\beta \gamma = 0.56$ in the laboratory frame, which is important for  studying the \CP-violation in the \B-meson decays. This boost allows to reconstruct the decay vertices of the two \B-mesons with enough accuracy to determine the relative decay time needed for time dependent \CP-violation measurement.  

\begin{figure}[htb]
\centering
\includegraphics[width=5.5in]{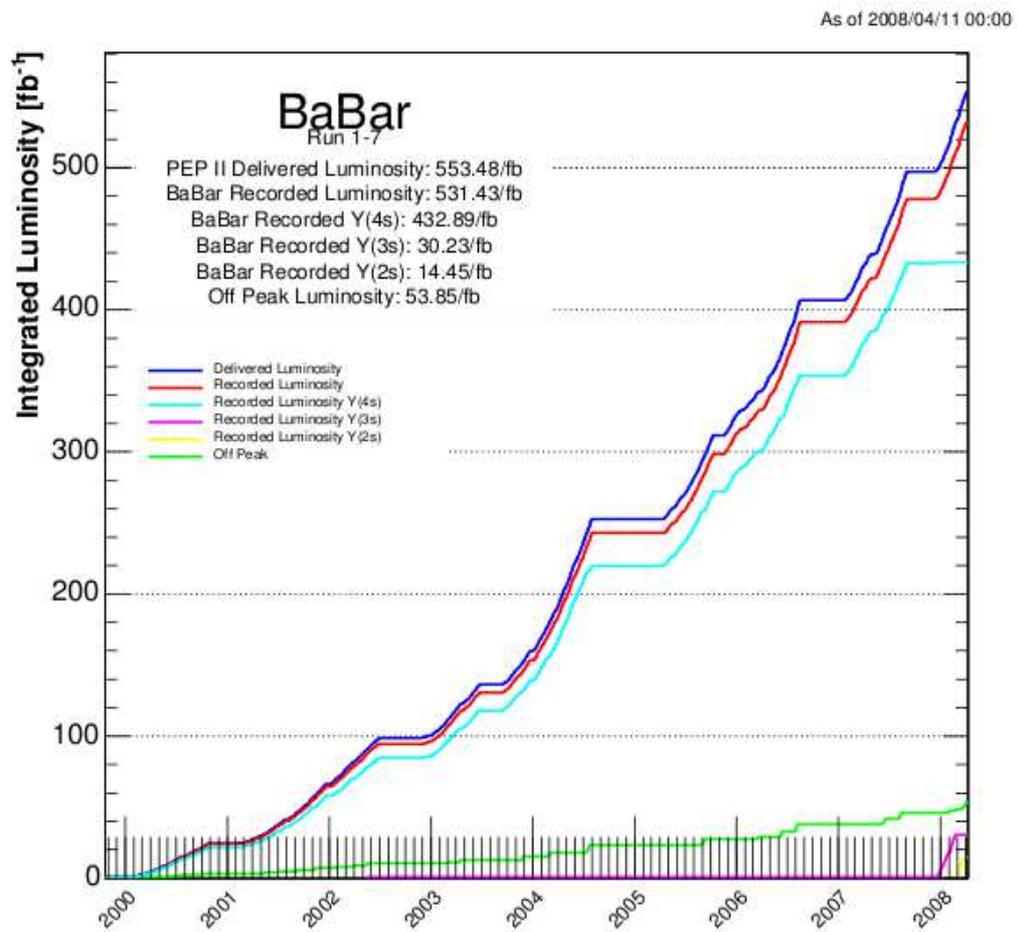}
\caption{Integrated luminosity delivered by PEP-II to the \babar\ experiment.}
\label{fig:lumi}
\end{figure}

PEP-II was operational from October 1999 to March 2008. During this period, the \babar\  experiment has collected about 476 million of $\Upsilon(4S)$ events with an integrated luminosity of 433 $fb^{-1}$, 120 million of $\Upsilon(3S)$ events with an integrated luminosity of 28.05 $fb^{-1}$, and 98 million of $\Upsilon(2S)$ events with an integrated luminosity of 14.4 $fb^{-1}$. \babar\  has also collected the data with an integrated luminosity of 53.85 $fb^{-1}$ outside these resonances (off-resonance), which are mostly used for continuum background study.  Figure~\ref{fig:lumi} shows the integrated luminosity of the experiment throughout its running period. 

\section{The \babar\ detector}
The \babar\ detector is located at the collision point of the PEP-II accelerator  \cite{BaBar-NIM}. To achieve the wide physics objective, it is necessary that the detector has a large acceptance, good vertexing, excellent reconstruction efficiencies for charged particles, good energy and momentum resolution, high lepton (particularly $e$  and $\mu$) and hadron identification efficiency and radiation hardness. 

\begin{figure}[!htb]
\centering
\includegraphics[width=4.8in]{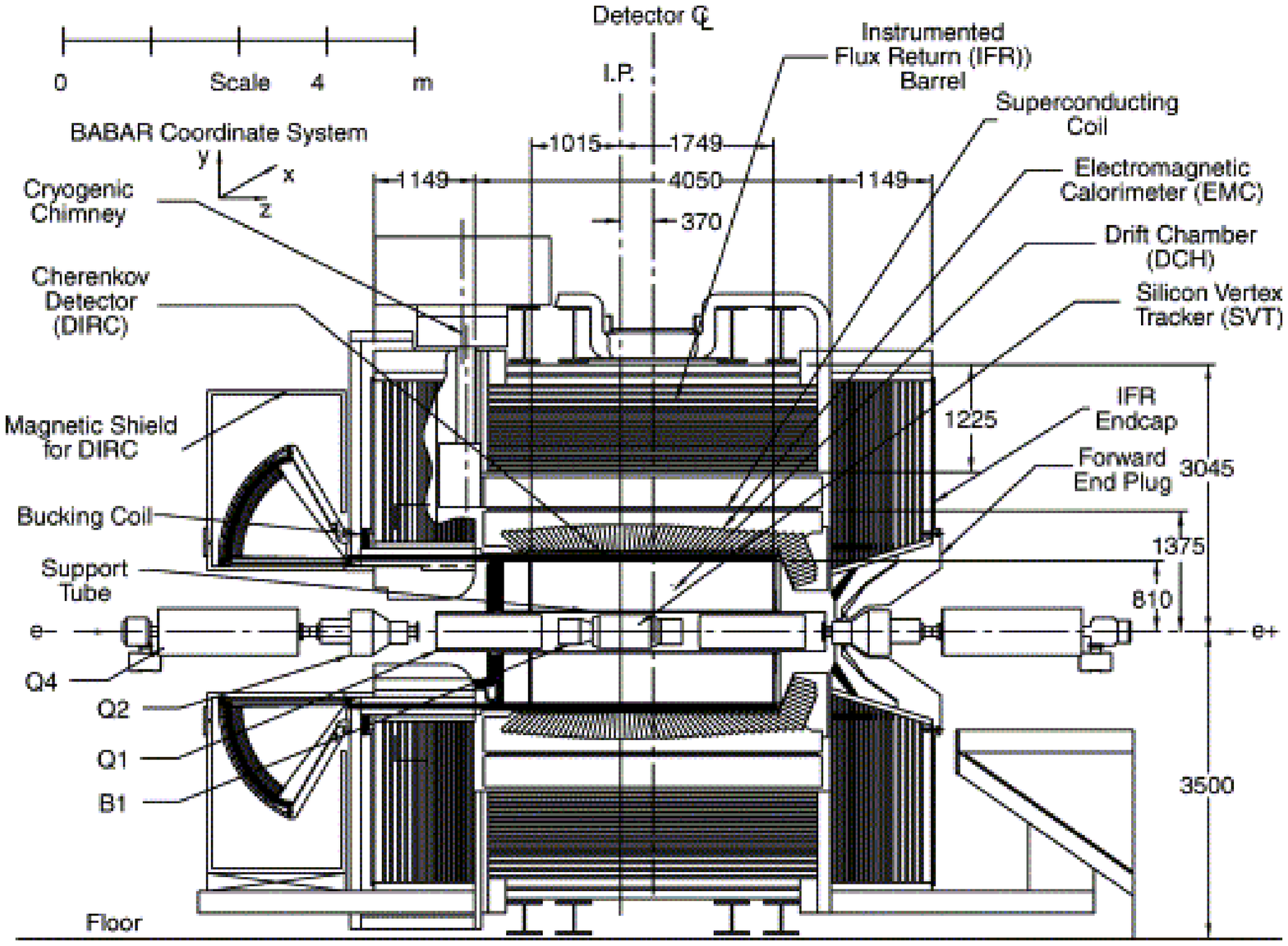}
\includegraphics[width=4.8in]{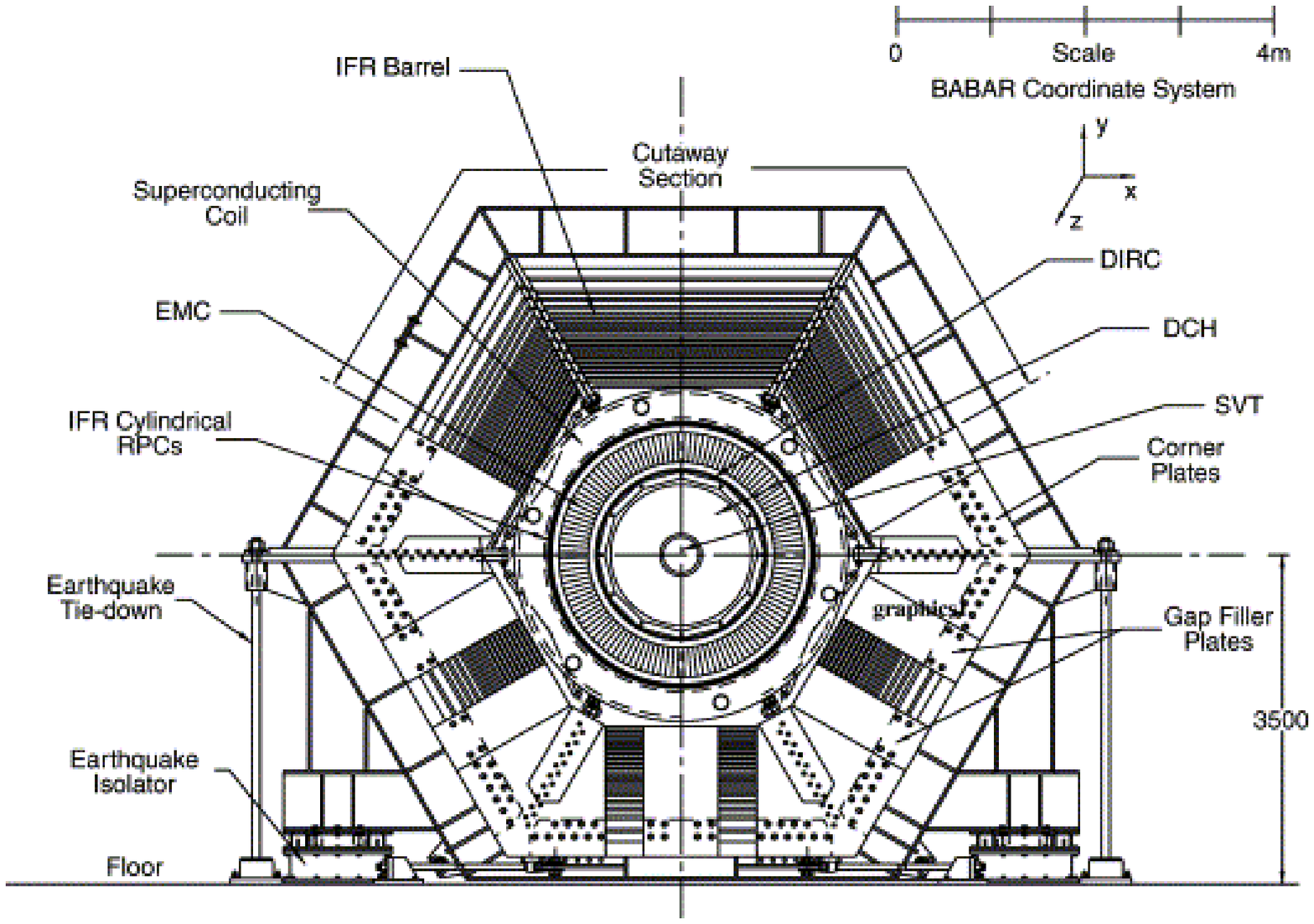}

\caption{Longitudinal section and front end  view of the \babar\ detector \cite{BaBar-NIM}.}
\label{figure:BaBardetector}
\end{figure}

 The \babar\ detector consists of five sub-detectors:  silicon vertex tracker (SVT) is positioned closest to the collision point and is responsible for measuring the decay vertices of the \B-mesons, a drift chamber (DCH) for charged particle tracking and momentum measurement, a ring-imaging Cerenkov detector for particle identification, and an electromagnetic calorimeter (EMC) for measuring the electromagnetic showers from electrons and photons. These detector subsystems are contained within a large solenoidal magnet capable of generating a 1.5 T magnetic field, and for which the steel flux return is instrumented with a muon detection system. The \babar\ detector is illustrated in Figure~\ref{figure:BaBardetector}, and the following subsections describe these sub-detectors in more detail.

\subsection{Silicon Vertex Tracker (SVT)}
The Silicon Vertex Tracker (SVT) \cite{BaBar-SVT} is a semiconductor based tracking sub-detector of the \babar\ experiment positioned very close to the collision point. It has been designed to provide the precise measurement of the reconstruction of the trajectory of the charged particles and decay vertices near the interaction region. It reconstructs the decay vertices of two primary \B-mesons at the $\Upsilon(4S)$ resonance to determine the time difference between two \B-mesons decay, which helps to study the time dependent \CP-asymmetries. It is also capable of reconstructing the low momentum  charged tracks bellow 120 \mevc that stop before reaching the DCH. 

\begin{figure}[!htb]
\centering
\includegraphics[width=3.2in]{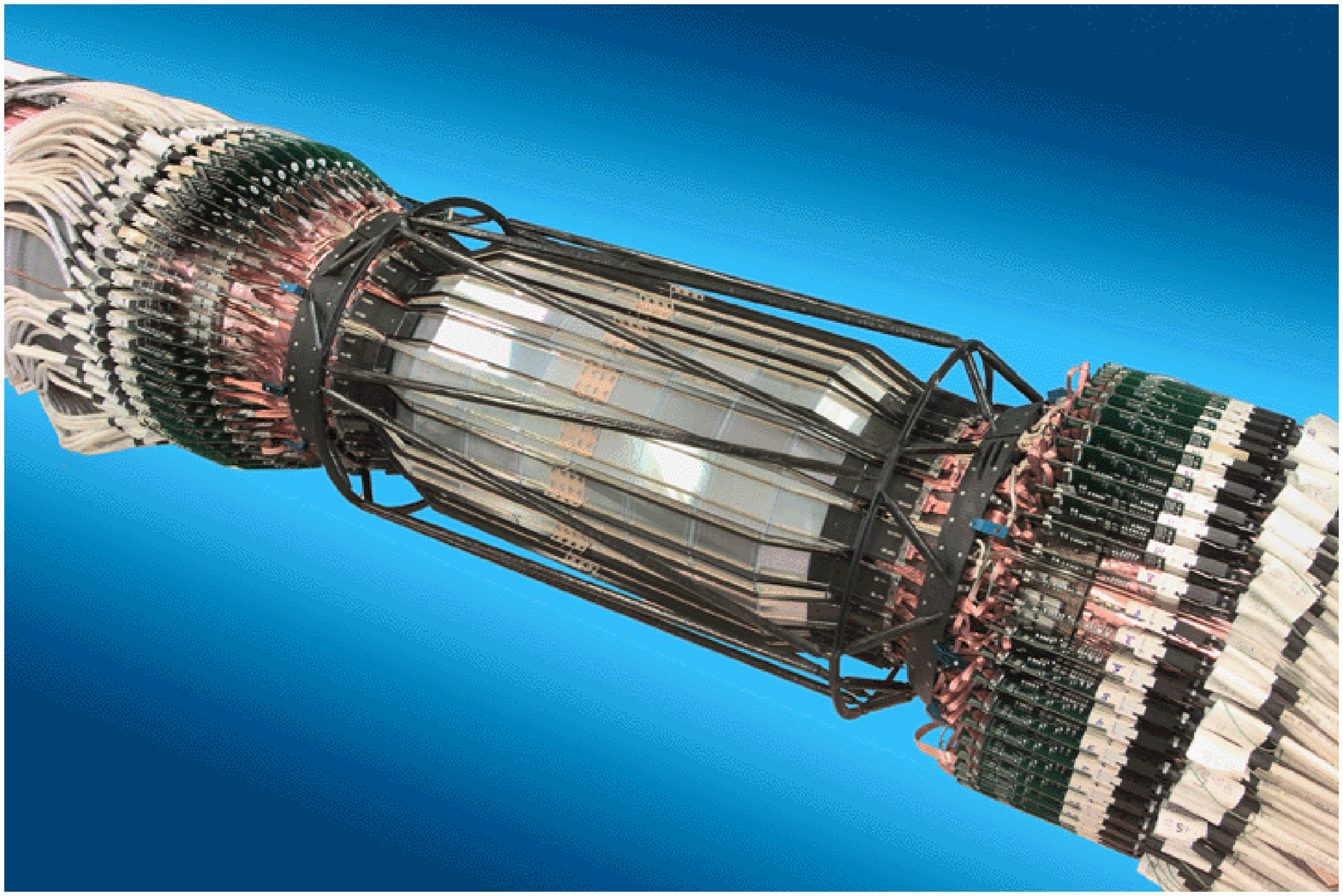}
\includegraphics[width=2.9in]{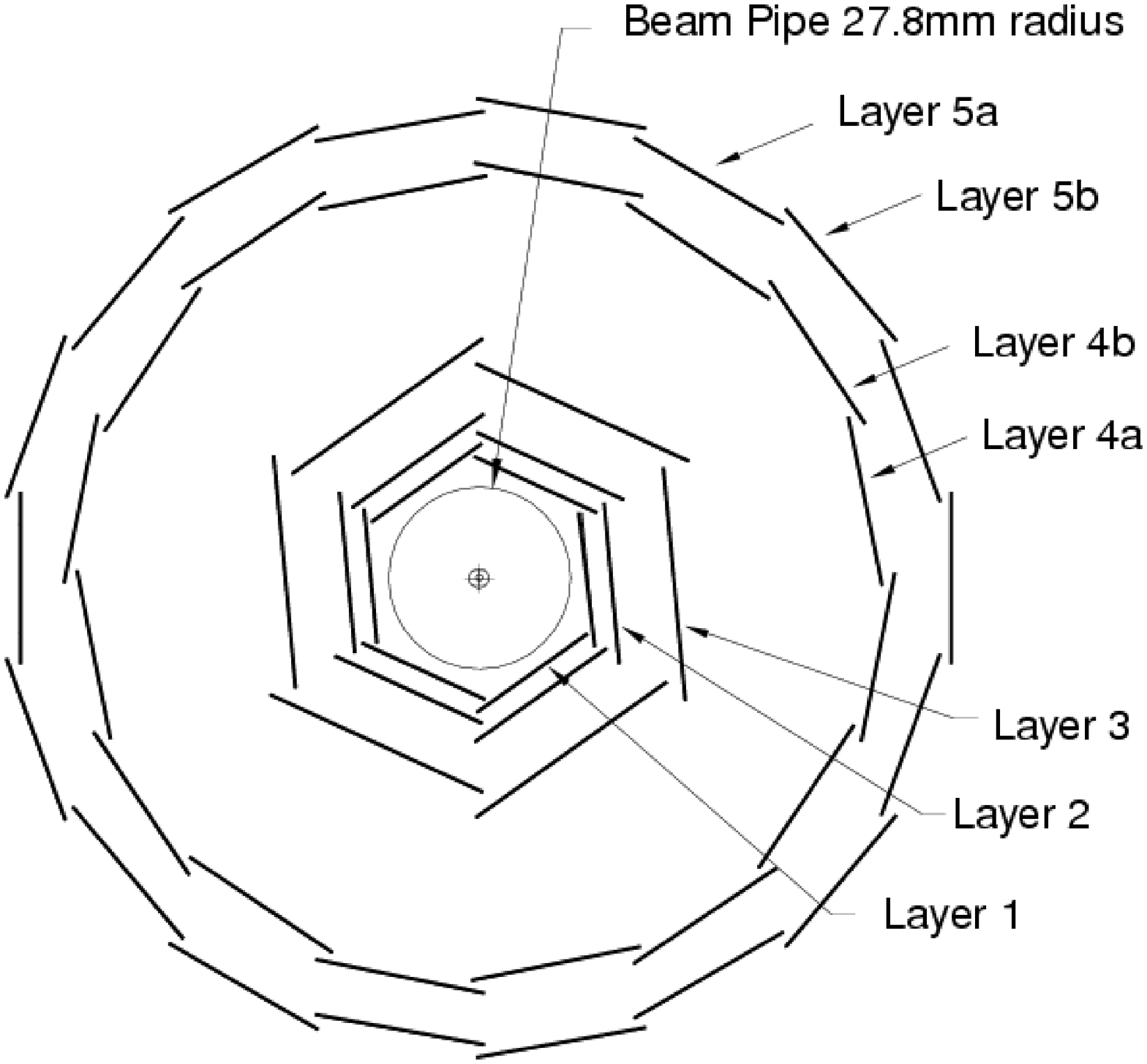}

\smallskip
\centerline{\hfill (a) \hfill \hfill (b) \hfill}
\smallskip

\caption{The Silicon Vertex Tracker (SVT) (a) fully assembled with visible outer layers and carbon fiber frame and (b) schematic view of the transverse section with the various layers around the beam pipe.}
\label{figure:SVT}
\end{figure}

The SVT consists of five concentric cylindrical layers of double-sided AC-coupled silicon micro-strip sensors. The strips on the one side of each sensor are oriented parallel to the beam direction and used to measure the azimuthal angle ($\phi$), while other side of the strips are perpendicular to the beam direction and used to measure the position of $z$. The inner 3 layers are barrel shaped and used to provide an accurate measurement of the impact parameters along $z$ direction and in the $x-y$ plane. However, the outer two layers are arch shaped and used to provide accurate polar angle measurement and can provide the standalone tracking for the low momentum particles that may not be capable of reaching the DCH. This arc design was chosen to minimize the amount of silicon required to cover the solid angle and to increase the crossing angle of the particles near the edges. These outer modules can not be tilted in $\phi$  like the inner modules because of their geometrical shape.  To avoid the gap in the $\phi$ coordinate, the two outer layers were divided into two sub-layers (4a, 4b, 5a, 5b as shown in Figure~\ref{figure:SVT} (b)), and placed at slightly different radii. Figure~\ref{figure:SVT} shows the fully assembled SVT with visible sensors of the outer layer and a transverse schematic view.

The SVT sensors are composed of a 300 $\mum$ thick n-type bulk silicon substrate with $p^+$ and $n^+$ strips on opposite sides. These sensors work in the reverse bias mode and are held at a voltage of about 10 V above the depletion voltage, where the typical depletion voltages are 25--35 V. When a charged particles passes through the SVT sensors, it ionizes the  materials  creating the electron-hole pairs. The electron drifts to the $n^+$ strips and hole drifts to the $p^+$ strips. This results in an electrical signal which is read-out via capacitive couplings between the strips and the electronics.

The alignment of the SVT is performed in the following two steps: the local alignment to determine the relative position of all the silicon sensors and the global alignment to correct the movement of the SVT with respect to the rest of the other \babar\ detectors. The local alignment of the 340 silicon sensors is performed by using a sample of $e^+e^- \rightarrow \mu^+\mu^-$ and the cosmic ray muons, and described by three translations, three rotations and a curvature. By using these parameters, it calculates the track residual  using the SVT only hit and performs a $\chi^2$ minimization to determine the best position for each sensor. The local alignment is stable and performed only rarely. Once the local alignment is done, the SVT also requires to align globally with respect to the DCH since it is not supported structurally by the rest of the other \babar\ detectors. The global alignment is performed by minimizing the difference between the track parameters fit with the SVT hit only as well as DCH hit only. Other monitoring systems such as temperature, humidity and electronic calibration  are also used regularly to ensure the successful SVT operation.

The SVT also includes a radiation protection system consisting of a PIN and a diamond diode sensors located very near to the collision point. The PIN diode consists of p and n-types of semiconductors which are separated by an intrinsic semiconductor located between the regions of these two semiconductors. These radiation systems  are used to protect the SVT by the colliding beams in the events of sudden high instantaneous or prolonged background levels that could damage the hardware components.    

The SVT performs with an efficiency of  $97\%$,  which is calculated for each half-module by comparing the number of associated hit to the number of tracks crossing the active area of the half-module. The spatial resolution of the SVT ranges from 10--15 $\mum$ for the inner layers and 30--40 $\mum$ for the outer layers. The spatial resolution of the SVT is determined by measuring the distance between the track trajectory and the hit for the high momentum tracks in the two-track events. The SVT is also used to measure the energy loss (\dedx) of the charged particles which passes through matter and deposit the energy in the sensor. The average \dedx is  used for the particle identification and gives a 2$\sigma$ separation between kaons and pions up to momentum of 500 \mevc and between kaons and protons up to 1 \gevc

\subsection {Drift Chamber (DCH)}

The DCH is designed to measure the charged particle momentum with minimum transverse momentum of $p_T > 100$ \mevc and the angular distribution with a high precision. It is the main tracking device of the \babar\ detector, and also enables the particle identification based on the \dedx measurement for the low momentum of particles where the DIRC is not effective. It is also crucial to  reconstruct the long lived particles such as $K_s^0$, which often decays outside or on the edge of the SVT, so the chamber should be able to measure the longitudinal positions of a tracks with a resolution of $\sim 1$ mm. Combined with SVT, the \babar\ tracking system provides excellent spatial and momentum resolution that enables the reconstruction of the exclusive \B and $D$-meson decays. The DCH complements the measurements of the impact parameter and the directions of the charged tracks provided by the SVT near the interaction point (IP), and it is also the key to the extrapolation of the  charged tracks to the DIRC, EMC and IFR.

\begin{figure}[htb]
\centering
\includegraphics[width=5.5in]{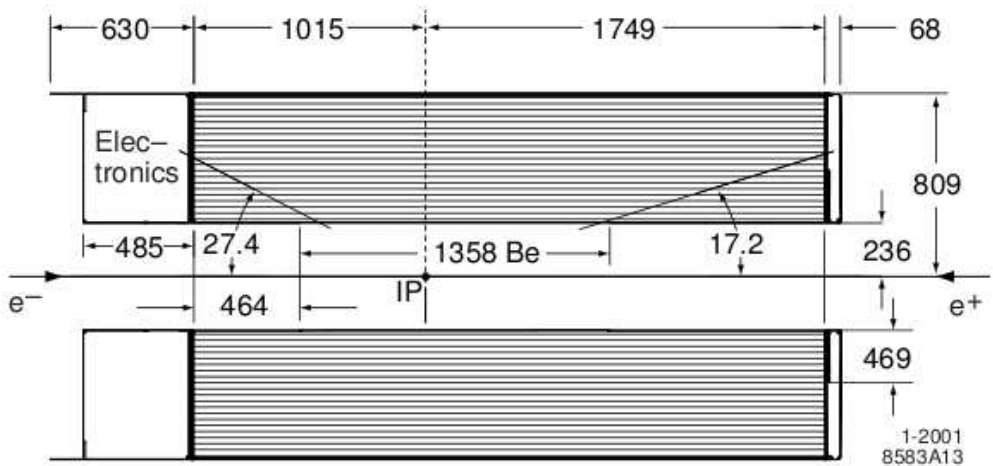}
\caption{Longitudinal cross-section of the drift chamber.}
\label{fig:drift}
\end{figure}

The DCH is a 280 cm long cylinder, with an inner radius of 23.6 cm and the outer radius of 81 cm (Figure~\ref{fig:drift}). Since the \babar\ events are boosted in the forward direction, its design is therefore optimized to reduce the material in the forward end in front of the endcap calorimeter, and offset by 37 cm from the IP to give greater coverage in the forward region. The forward endplate is made thinner (12 mm) in the acceptance region of the detector compared to the rear endplate (24 mm), and all the electronics are mounted on the rear backward endplate. The inner cylinder is made of 1 mm beryllium corresponding to $0.28\%$ of the radiation length ($X_0$), while the outer is made of 2 layes of carbon fiber of a honeycomb core correcponding to $1.5\%$ of the $X_0$.   

The DCH consists of 7104 drift cells, which are arranged in 10 super-layers of 4 layers each, for a total of 40 layers. The stereo angles of the super-layers alternate between axial (A) and stereo (U,V) in following order: AUVAUVAUVA. The stereo angles increase from 45 mrad in the innermost super-layer to 76 mrad in the outermost super-layer. The chamber is filled with a 80:20 gas mixture of helium:isobutane to provide good spatial separation and resolution for the \dedx measurement and reasonably short response time, where the helium is chosen to minimize the multiple scattering. 

\begin{figure}[!htb]
\centering
\includegraphics[width=2.9in]{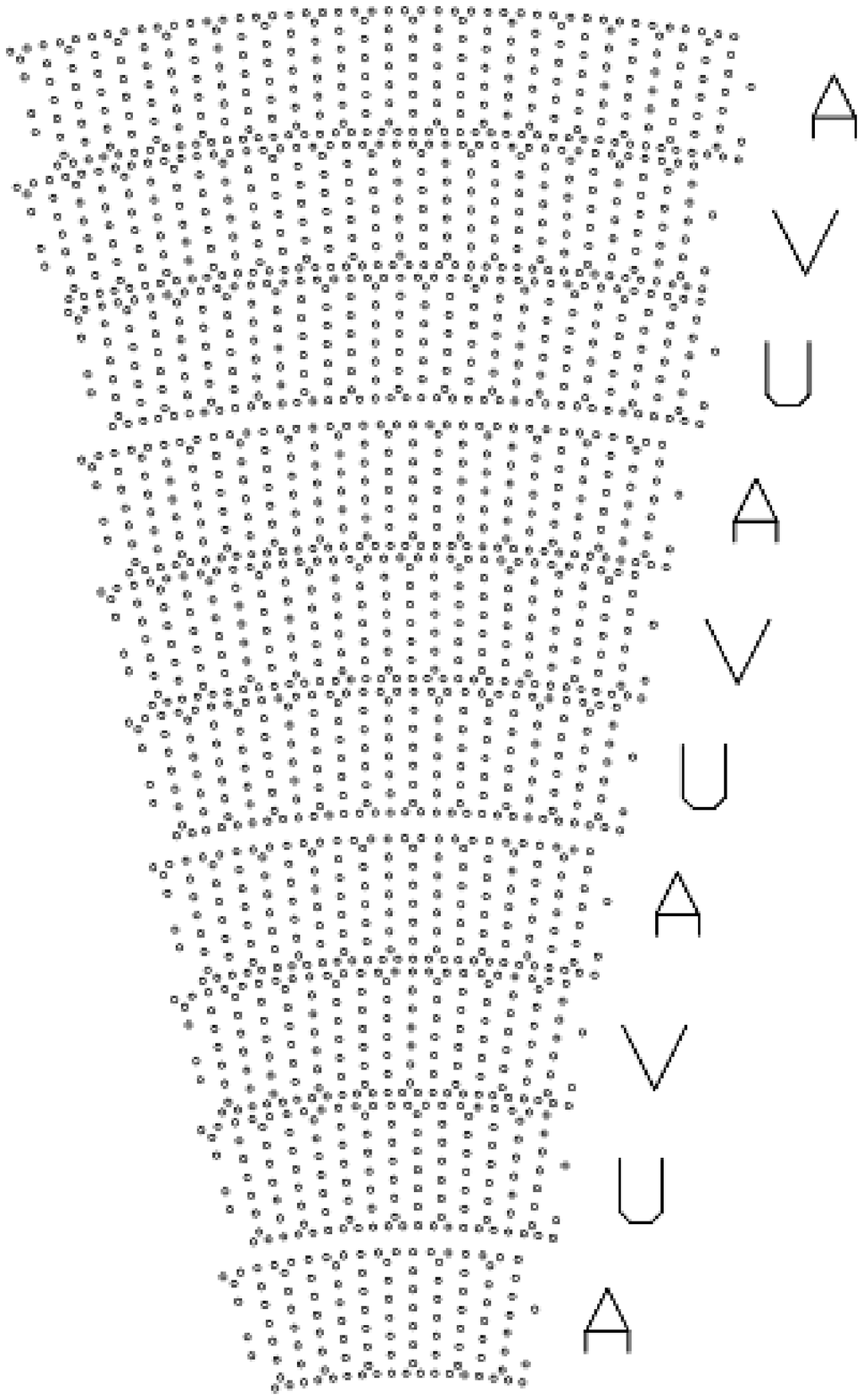}
\includegraphics[width=2.2in]{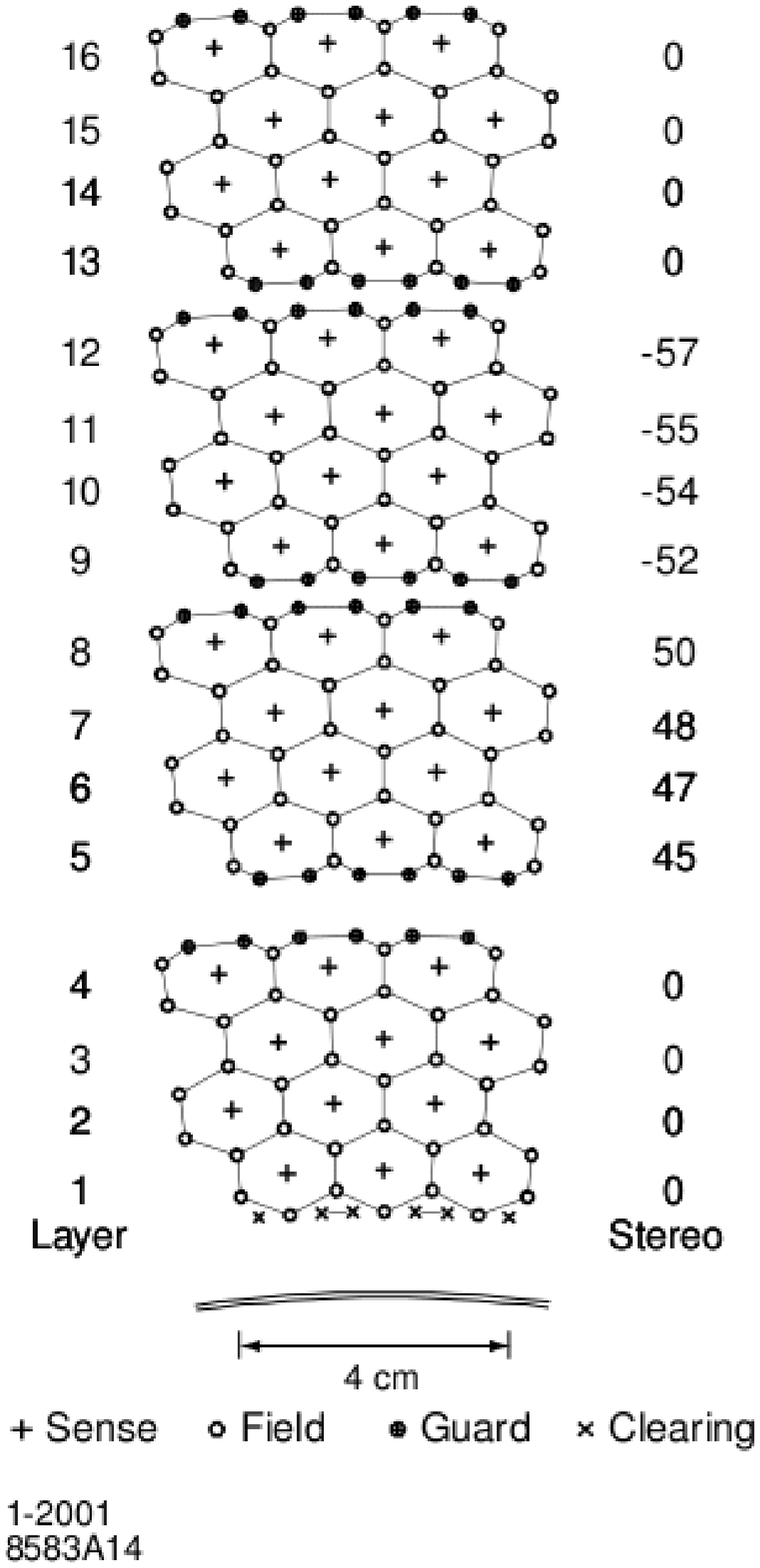}

\smallskip
\centerline{\hfill (a) \hfill \hfill (b) \hfill}
\smallskip

\caption{(a) Cell layout in the \babar\ drift chamber. (b) DCH drift cell configuration for the four innermost super-layers. The numbers on the right give the stereo angles in mrad of the sense wires in each layer. }
\label{figure:DCH}
\end{figure}

  Figure~\ref{figure:DCH} shows the design of the drift cells for the four innermost super-layers. The 7104 cells are hexagonal with a typical dimension of $1.2\times 1.8$ $\rm cm^2$, to minimize the drift time. The sense wires is a 25 $\mum$ gold-plated  tungsten-rhenium wire, while the field wires are  gold-plated aluminium with diameters of 120 $\mum$ and 80 $\mum$. A voltage of 1960 V is applied to the sense wires, while the field wires are held at ground.   

  The ionized charged particles in the gas produces the free electrons that are accelerated towards the sense wires by the applied electromagnetic field. These accelerated electrons are further ionized and result in an avalanche of the electric charge near the wire. The avalanche accumulates at the sense wire producing a measurable electrical signal, which is amplified and read-out to the electronics. The integrated charge and drift time (time required for the ionized electrons to reach  the sense wire) provide the ionization energy-loss and position information of the charged particles, respectively. 

The track of the charged particles is defined by five parameters ($d_0$, $\phi_0$, $\omega$, $z_0$, $\mathrm{tan}\lambda$), which are measured at the point of closest approach to the $z$-axis, and their associated error matrix. The $d_0$ and $z_0$ represent the distance of a track from the origin of the coordinate system in the $x-y$ plane and along the $z-$axis; The $\phi_0$ is the azimuthal angle of the track; $\lambda$ is the dip angle relative to the transverse plane, and $\omega = 1/p_T$ is the curvature of the track. Based upon the full width half maxima, the distributions of these variables have the following resolution values: $\sigma_{d_0}= 23~\mu \rm m$, $\sigma_{\phi_0}= 0.43$ mrad, $\sigma_{z_0}= 29~\mu \rm m$ and $\sigma_{\mathrm{tan}\lambda}= 0.53 \times 10^{-3}$. The DCH performs with a tracking efficiency of $(98 \pm 1)\%$ for  $p_T > 200$ \mevc and for polar angle $\theta > 500$ mrad at the voltage of 1960 V. The resolution of the measured $p_T$ can be written as a linear function of $\sigma_{p_T}/p_T = (0.13 \pm 0.01)\% p_T + (0.45 \pm 0.03)\%$.

\begin{figure}[!htb]
\centering
\includegraphics[width=5.0in]{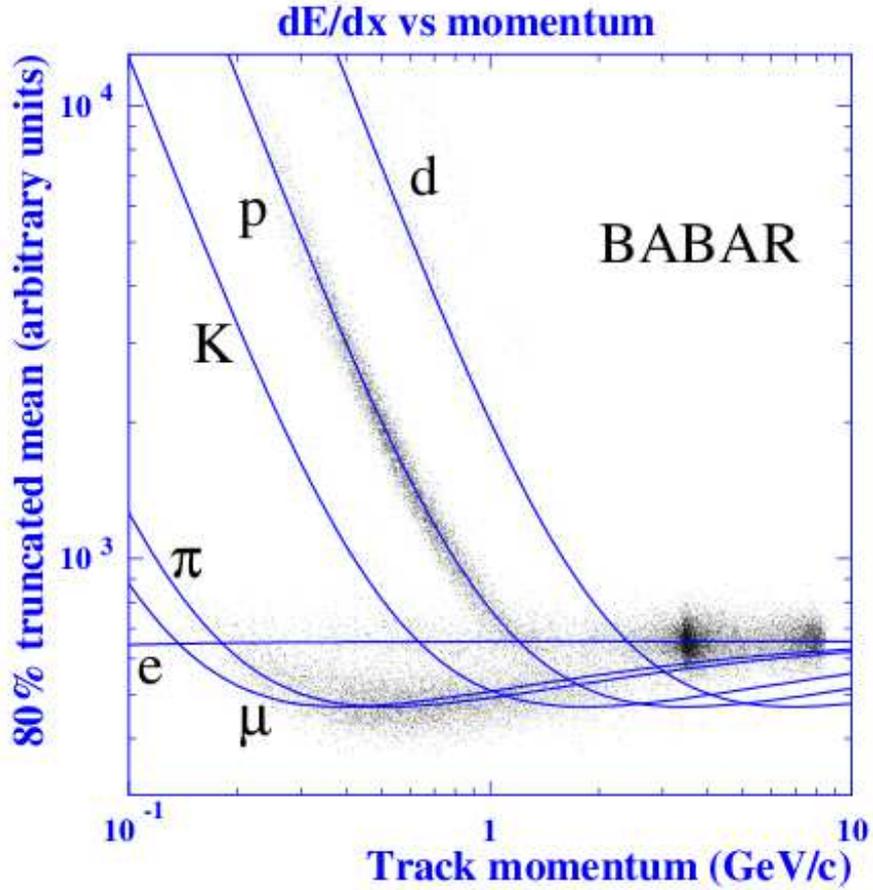}
\caption{\dedx in the DCH as a function of momentum for different particles.}
\label{fig:de/dx}
\end{figure}

The specific energy loss per track is computed as a truncated mean from  the lowest $80\%$ of the individual dE/dx measurements. This value is computed
 after incorporating all the corrections. The corrections are needed to account  for changes in gas  pressure and mixture; differences in cell geometry and  charge collection; signal saturation due to space charge buildup; non-linearties in the most probable energy loss at large track dip angles; and  changes in cell charge collection as a function of track entrance angle. The corrections are all done once for a given High-Voltage (HV) setting and a given gas mixture whilst the gain corrections must be updated run by run.  Corrections at the cell level can be large compared to the dE/dx resolution  for a  single cell, but have only a small impact on the average resolution of
  the ensemble of hits. The \dedx as a function momentum is shown in Figure~\ref{fig:de/dx}.  The DCH achieves good separation between K and $\pi$ upto 700 \mevc.    

\subsection {The Detector of Internally Reflected Cherenkov Light (DIRC)}
 The DIRC is a new type of ring-imaging Cherenkov detector used for hadronic particle identification in the \babar\ experiment. It provides the $\pi/K$ separation of greater than $4\sigma$ for all tracks from pion Cherenkov threshold up to 4.2 \gevc, and tags the flavor of a \B meson via the cascade decay of  $b \rightarrow c \rightarrow s$. Its imaging system is based upon the total internal reflection of Cherenkov photon produced in long quartz bar. When a  particle passes through the medium with a velocity  greater than the speed of light in that medium, it emits photons known as Cherenkov radiation. The angle of the Cherenkov radiation is defined as

\begin{equation}
\mathrm{cos}\theta_c = \frac{c}{nv},
\end{equation}     

\noindent where c is the velocity of light, n is the refractive index of the medium and v is the speed of the particle. 

\begin{figure}[!htb]
\centering
\includegraphics[width=6.5in]{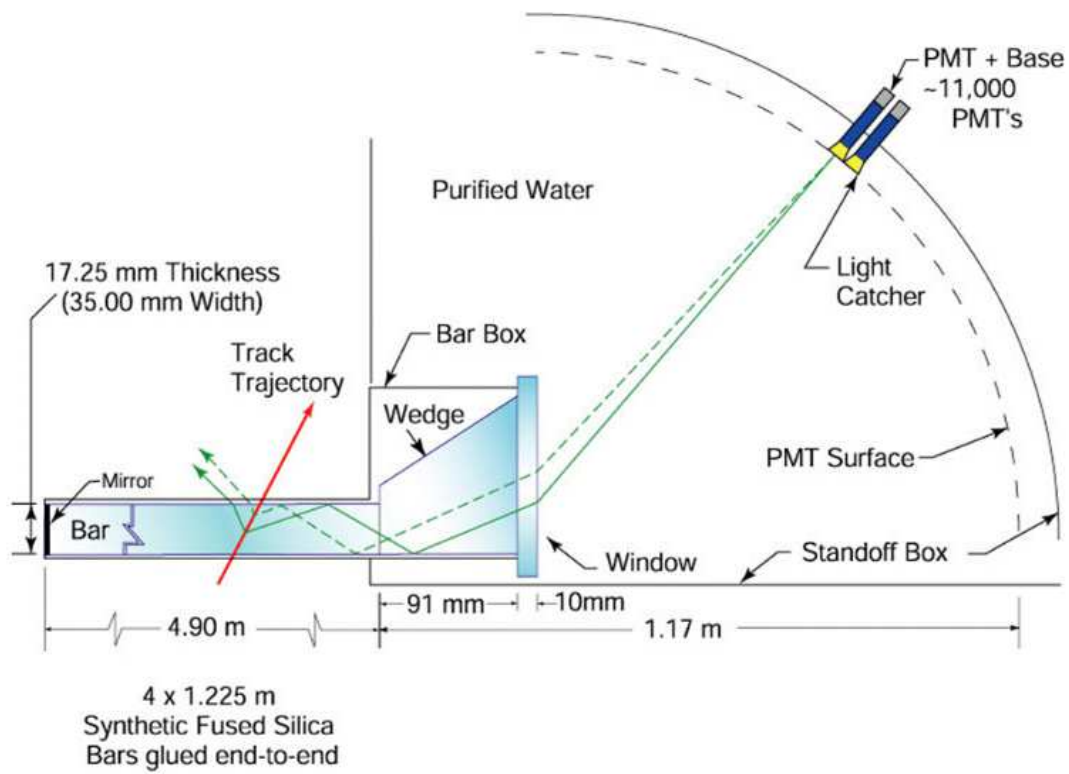}
\caption{Schematic of the DIRC fused silica radiator bar and imaging region.}
\label{fig:DIRC}
\end{figure}

The DIRC is a three-dimensional imaging device, used to sense the position and arrival time of the signal by using an array of densely packed photomultiplier tubes. It consists of 144 radiation-hard fused silica bars with an refractive index of n = 1.473. The bar serves both as radiators and as light pipes for the light trapped in the radiator by total internal reflection. A schematic of the DIRC geometry illustrating the principle of light production, imaging and transportation is shown in Figure~\ref{fig:DIRC}. Photons are generated by the particles above the Cherenkov threshold, trapped inside the bars and emerge into a water-filled expansion region, called a standoff box. A fused silica wedge is used to reflect photons at large angles to reduce the size of the required detection surface and hence recover those photons that would be lost due to internal reflection at the fused silica and water interface. Finally, the photo multiplier tubes (PMTs) detect the light and allow the Cherenkov angle and particle velocity to be measured. Once the velocity is known, the mass of the particle can be calculated using the momentum information from the DCH. 

\subsection{The Electromagnetic Calorimeter}
The EMC is designed to measure the photon showers with excellent efficiency, energy and angular resolution over the energy range from 20 \mev to 9 \gev. The capability of the EMC allows the detection of photons from $\pi^0$ and $\eta^0$ as well as from electromagnetic and radiative processes. Most of the photons are produced by neutral pion decays  with maximum energy of 200 \mev, hence the lower bound of energy is set to allow the reconstruction of \B-meson decays containing multiple $\pi^0$ mesons. However, the upper bound of the energy range is set by calibrating and monitoring the luminosity of the photons produced via the QED processes, like $e^+e^- \rightarrow e^+e^- (\gamma)$. The EMC is also used to identify the electrons which allows to study of semi-leptonic and rare decays of \B and $D$ mesons, and $\tau$ leptons, and  the reconstruction of vector mesons  like $J/\psi$.  

\begin{figure}[htb]
\centering
\includegraphics[width=6.5in]{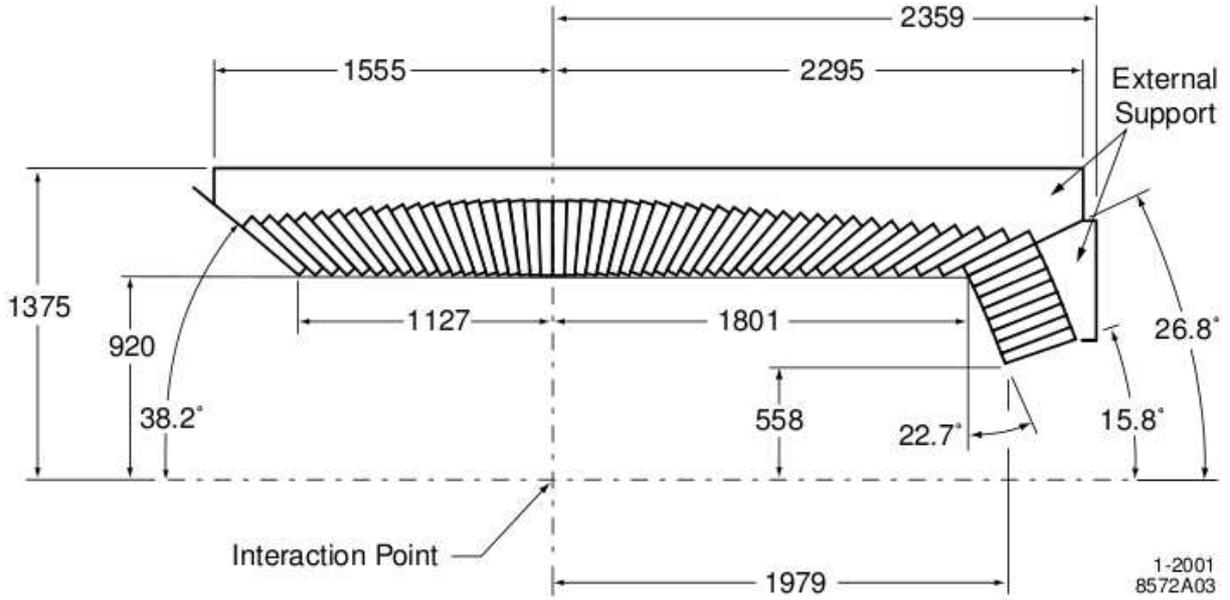}
\caption{A longitudinal cross-section of the EMC (only the top half is shown) indicating the arrangement of 56 crystal rings. The detector is axially symmetric around the $z$-axis. All dimensions are given in mm.}
\label{fig:EMC}
\end{figure}

The EMC consists of a cylindrical barrel and a conical forward endcap. It has a full coverage in the azimuth and extends in polar angle from $15.8^{\circ}$ to $141.8^{\circ}$ corresponding to a solid-angle coverage of $90\%$ in the CM system (Figure~\ref{fig:EMC}). The barrel contains 5,760 thallium-doped caesium iodide (CsI(Tl)) crystals arranged in 48 distinct rings with 120 identical crystals each. The endcap holds 820 crystals arranged in eight-rings, adding up to a total of 6,580 crystals. The crystals have a tapered trapezoidal cross-section and length of the crystals increases from 29.6 cm in the backward to 32.4 cm in the forward direction to limit the effects of shower leakage from increasingly higher energy particles.  Two silicon PIN diodes mounted on the rear face of each crystal are used to readout the scintillation light. 

A typical electromagnetic shower tends to spread over many adjacent crystals, forming a cluster of adjacent energy deposits. Pattern recognition algorithms are used to analyze the shower shape and to  check  whether cluster can be associated with the charged particles. Otherwise, the EMC cluster would be assumed to originate from a neutral particle.  The energy resolution of a homogeneous crystal calorimeter is empirically described by

\begin{equation}
\frac{\sigma_E}{E} = \frac{a}{4\sqrt{E(\gev)}}\oplus b,
\end{equation} 

\noindent where $\oplus$ signifies addition in quadrature, and  E and $\sigma_E$ are the energy and rms value of a photon.  The angular resolution is determined by the transverse crystal size and the distance from the interaction point, which is defined as
\begin{equation}
\sigma_{\theta} =  \sigma_{\phi} = \frac{c}{\sqrt{E(\gev)}} + d,
\end{equation}

\noindent The energy dependent terms $a$ and $c$ are dominant at low energy and arise due to the fluctuations in photon statistics and electronic noise in the readout chain. Furthermore, beam-generated background will lead  to a large numbers of additional photons that add additional noise. The constant terms  $b$ and $d$ are dominant at higher energies ($>1$  \gev) and arise due to non-uniformity in light collection and light absorption in the detector materials.

\subsection{The Instrument Flux Return (IFR)}
 The IFR was designed to identify the muons with high efficiency and good impurity, and to detect neutral hadrons (primarily $K_L^0$ and neutrons) over a wide range of momenta and angles. Muons are important for tagging the \B mesons via semileptonic decays, for the reconstruction of the vector mesons, like the J/$\psi$, and for the study of semi-leptonic and rare decays of \B and $D$ mesons and $\tau$ leptons. $K_L^0$ detection is important to study the exclusive \B decays, in particular \CP eigenstates. The IFR also helps in vetoing charm decays and improving the reconstruction of neutrinos.   

\begin{figure}[!htb]
\centering
\includegraphics[width=6.5in]{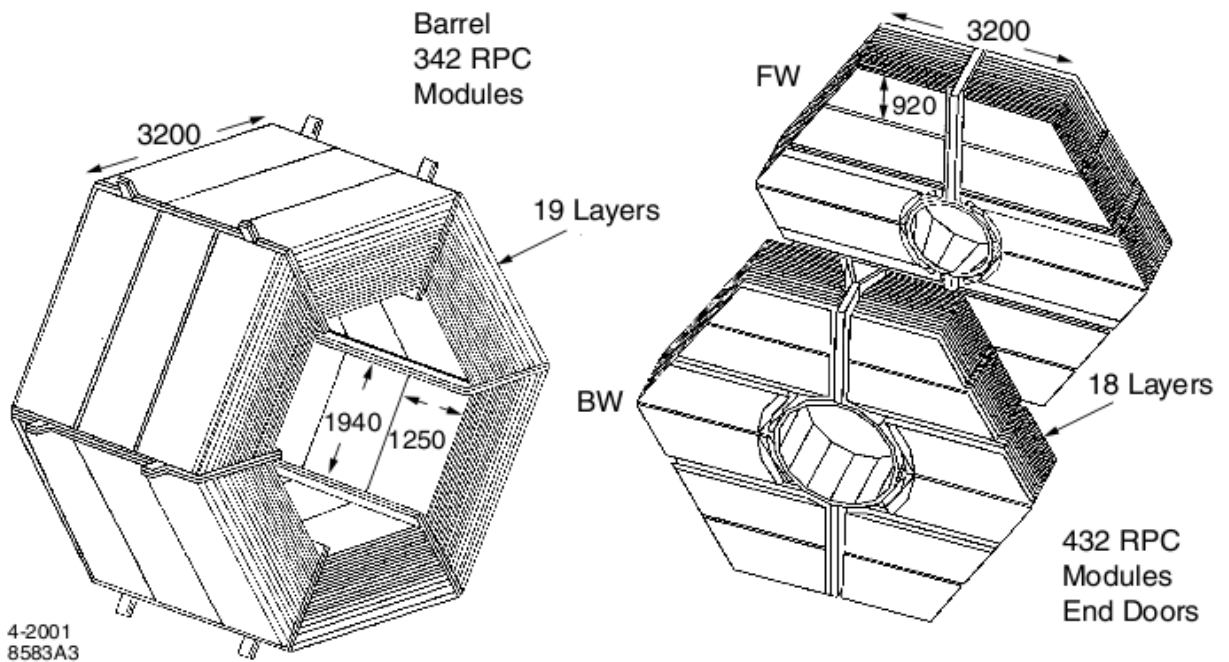}
\caption{Layout of IFR barrel and endcaps. All the units are given in mm.}
\label{fig:IFR}
\end{figure}

 The IFR consists of one barrel and two endcap and  uses the steel flux return of the magnet as a muon filter and hadron absorber. Single gap resistive plate chamber (RPC) with two-coordinate readout have also been chosen as an active detector. The IFR was originally equipped with 19 layers of RPC in the barrel and 18 in the endcaps. In addition, two layers of cylindrical RPCs are installed between the EMC and the magnet crystal to detect the particles existing the EMC. The RPC consists of two high resistivity Bakelite sheets coated with linseed oil separated by a 2 mm gap containing $56\%$ argon, $38.8\%$ Freon 134a, and $4.5\%$ isobutane. The RPCs operate in the limited streamer mode at $\sim 8$ kV, and  streamer signals readout by aluminum strips on the exterior of the plates.  An illustration of the layout of the IFR is shown Figure~\ref{fig:IFR}.    

Unfortunately. it was found that the RPC degraded rapidly. Therefore, the muon detection system was upgraded with  a Limited Streamer Tubes (LST) \cite{BaBar-LST1,BaBar-LST2} during the detector shutdown periods from 2004-2006. The inner 18 layers of the RPC detector were replaced with 12 layers of LST detectors and 6 layers are filled with brass absorber, to improve muon-hadron separation. The LST detector consists of 7--8 cells with a dimension of $380 \times 15 \times 17$ $\rm mm^3$. The cells are composed of PVC plastic coated with a graphite paint, which is maintained at a ground potential, a central high voltage gold plated anode which is held in place by 6 wires holders, and are filled with a (89:8:3) gas mixture of $\rm CO_2$, ISO-butane and Argon. The LST also operates in the streamer mode, and the signals are collected by the external read-out strips.

\subsection{Trigger Selection}
The trigger system \cite{Bad-194} was designed to select events of interest with a high, stable, and well-understood efficiency while rejecting background events and  keeping the total event rate under 120 Hz. \babar\ uses two types of the trigger systems: the hardware based level 1 Trigger (L1) and the software based Level 3 Trigger (L3).  The details of L1 and L3 trigger systems are described bellow:

\noindent {\bf Level 1 Trigger system}

The design of L1 trigger decision is based on the charged tracks in the DCH above a preset transverse momentum, showers in the EMC, and tracks detected in the IFR. The drift chamber trigger (DCT) processes the input  data consisting of one bit from each of the 7104 cells to identify tracks. The Electromagnetic Trigger (EMT) receives input from the 280 towers in the EMC, and identifies the energy deposits in the EMC. The IFR is divided  into ten sectors, namely the six barrel sextants and the four half end doors. The primary functions of the instrument flux return trigger (IFT) are to veto cosmic events and to identify muons from the interaction of $e^+e^- \rightarrow \mu^+\mu^-$, which can be used for measuring the detector parameters such as the luminosity. The output of the DCT, EMT and IFT are utilized to determine whether the signal event constraints a physics event by a Global Level Trigger. The frequency at which the events are accepted by the Level 1 Trigger is approximately 1 KHz.

\noindent {\bf Level 3 Trigger}

The L3 trigger is an online application that acts primarily as an event filter.  It implements in the framework of the Online Event Processing (OEP) and runs in parallel on a number of Unix processors. It is the first stage of the the DAQ system to select the events and is responsible to make a logging decision on the output of the L1 hardware trigger. Its design was required to reduce the L1 output of 1 kHz to a logging rate of about 120 Hz with a high efficiency in physics events of interest. It performs a partial reconstruction of the event based on the data from the drift chamber and drift chamber trigger as well as from the EMC. Its data are in part used by the offline reconstruction and its trigger decision records are input to the offline filters of DigiFilter and BGFilter.  

 The offline filters are based on prompt reconstruction (PR) used to provide a further selection of events before the full reconstruction. The selection is done using two levels of filters: the DigiFilter and BGFilter. The DigiFilter uses only  information available from the L1 and L3 triggers to make the selection. It is primarily used to remove the calibrated events, such as radiative Babha events. The BigiFilter first runs as a part of the offline reconstruction to find drift chamber tracks and EMC clusters. Based on these tracks and clusters information, an event classification is done, where the events classified as multi-hadron, $\tau$ or two-prong etc are identified.   

\section{Chapter Summary}
In this chapter, we describe the \babar\ detector, PEP-II accelerator complex and the trigger system which are used to collect the dataset analyzed in this thesis.
 
























\chapter{Event reconstruction and selection}
\label{Chapter3}
This chapter describes the event reconstruction and the selection criteria applied to select signal-like events for the decay  $\Upsilon(2S,3S) \to \pipi \Upsilon(1S)$, $\Upsilon(1S) \to \g A^0$, $A^0 \to \mu^+\mu^-$. A blind analysis \cite{blind} technique is used, where the $\Upsilon(2S,3S)$ datasets are blinded until all the selection criteria are finalized for an optimal value of signal-to-noise ratio. In this chapter, we describe the discriminative variables used to separate signal from background. A more advanced multivariate technique  based BumpHunter and  Random forest classifiers are also used to improve the purity of the $\Upsilon(1S)$ sample. We also discuss the $\Upsilon(2S)$ and $\Upsilon(3S)$ datasets used in this analysis along with the Monte-Carlo (MC) samples which are intended to model the data. The luminosities of these datasets are also documented.

\section{Data Sets}
 The data sample used in this analysis was collected during Run 7, specifically during a period between December 2007 and April 2008 by the \babar\ detector. The $\Upsilon(3S)$ dataset contains $(121.9 \pm 1.1)\times 10^6$ $\Upsilon(3S)$ events and  the $\Upsilon(2S)$ dataset contains $(98.3 \pm 0.9)\times 10^6$ $\Upsilon(2S)$ events. 

       The  $\Upsilon(3S)$ data set is divided into three sub samples: low, medium, and high which were collected in the beginning, middle and the end of Run7, respectively. The \rm{\lq\lq Low\rq\rq} data set corresponds to about $4.2\%$ of the total $\Upsilon(3S)$ on resonance data set and is used for checking the agreement between data and MC and finally, for validating the analysis procedure. For the $\Upsilon(2S)$ analyis, a similar \rm{\lq\lq Low\rq\rq} data set was generated which corresponds to $5.6\%$ of the total  $\Upsilon(2S)$ data set. The \rm{\lq\lq Low\rq\rq} samples are kept blinded untill all the selection criteria are finalized. We unblind these \rm{\lq\lq Low\rq\rq} samples later to validate the fit procedure after applying all the optimal selection cuts. The luminosities of these samples are shown in Table~\ref{table:YnS_data}. To avoid any bias, these samples are discarded from the final dataset.

\begin{table}
\centering
\begin{tabular}{|c |r|}
	\hline
Dataset Name   &Integrated Luminosity $(fb^{-1})$ \\[0.5ex]
\hline
\hline
\multicolumn{2}{|c|}{For $\Upsilon(3S)$ dataset} \\
\hline
\hline
AllEvents-Run7-R24b$\Upsilon(3S)$-OnPeak-Low  &1.173 \\
\hline
AllEvents-Run7-R24b$\Upsilon(3S)$-OnPeak-Medium (So far blind)  &25.594 \\
\hline
AllEvents-Run7-R24b$\Upsilon(3S)$-OnPeak-High (So far blind) &1.282 \\
\hline
\hline
\multicolumn{2}{|c|}{For $\Upsilon(2S)$ dataset} \\
\hline
\hline
AllEventsSkim-Run7-$\Upsilon(2S)$-OnPeak-R24d-LowOnpeak & 0.758 \\
\hline
AllEventsSkim-Run7-$\Upsilon(2S)$-OnPeak-R24d (So far blind) & 13.56  \\
\hline
\end{tabular} 
\caption{The luminosity of each data sample used in the analysis.}
\label{table:YnS_data}
\end{table} 
 
 MC simulated events  are used to study the detector acceptance and optimize the event selection procedure. The EvtGen package \cite{EVTGEV} is used to simulate the $\epem \to q\overline{q}$ $(q=u, d, s, c)$ and generic $\Upsilon(2S,3S)$ production, BHWIDE \cite{bhwide}   to simulate the Bhabha scattering and  KK2F  \cite{kk2f} to simulate the decay processes of  $\epem \to (\g) \mumu$ (radiative di-muon) and  $\epem \to (\g)\tautau$. Signal events are generated using a phase-space (P-wave) model for the $A^0 \to \mumu$ ($\Upsilon(1S) \to \g A^0$) decay and the hadronic matrix elements measured by the CLEO experiment  \cite{CLEO} are used for the $\Upsilon(2S,3S) \to \pi^+\pi^- \Upsilon(1S)$ transition. The detector response is simulated by GEANT4 \cite{GEANT4}, and time-dependent detector effects are  included in the simulation. The cross-sections for $e^+e^- \rightarrow q\overline{q}$  and lepton-pair productions are calculated from their values at the $\Upsilon(4S)$ assuming $1/s$ scaling, where $\sqrt{s}$ is the \epem CM energy at $\Upsilon(nS)$ ($n =2,3,4$) resonances. Table~\ref{table:SignalMC} summarizes the number of generated signal MC events at different masses for the decay chains of $\Upsilon(2S,3S) \to \pipi Y(1S)$, $\Upsilon(1S) \to \g A^0$, $A^0 \to \mumu$.  The cross-sections and luminosities of these  background decay processes are summarized in Table~\ref{table:YnS_MC}. We use these six types of background MCs  and a signal MC sample in the mass range of 0.212 - 9.46 GeV/$c^2$  to optimize the selection criteria.

\begin{table}
\begin{minipage}[b]{0.5\linewidth}\centering
\begin{tabular}{|c |r|}

	\hline

Mass of $A^0$ \gevcc  &  Number of events  \\[0.5ex]
\hline
\hline
\multicolumn{2}{|c|}{For $\Upsilon(3S)$ dataset} \\
\hline
\hline
 0.212    & 172k  \\
\hline
 0.214    & 172k \\
\hline
 0.216    & 172k  \\
\hline
 0.218    & 172k \\
\hline 
 0.220    & 172k \\
\hline 
 0.225    & 172k \\
\hline 
 0.300    & 172k \\
\hline 
 0.500    & 172k \\
\hline 
 0.750    & 172k  \\
\hline
 1.0    & 103k \\
\hline 
 1.5    & 172k \\
\hline
 2.0    & 103k \\
\hline 
 3.0    & 103k \\
\hline 
 4.0    & 103k \\
\hline
 5.0    & 103k \\
\hline
 6.0    & 103k \\
\hline 
 6.7    & 103k \\
\hline 
 7.0    & 95k \\
\hline 
 7.5    & 103k \\
\hline 
 8.0    & 103k  \\
\hline
 8.25    & 172k \\
\hline 
 8.5    & 172k \\
\hline
 8.75    & 172k \\
\hline
 9.0    & 103k  \\
\hline
 9.25    & 172k \\
\hline
 0.212 -- 9.46    & 204k \\ 
\hline
\end{tabular} 
\end{minipage}
\hspace{0.5cm}
\begin{minipage}[b]{0.05\linewidth}
\centering
\begin{tabular}{|c |r|}

	\hline

Mass of $A^0$ \gevcc  &  Number of events  \\[0.5ex]
\hline
\hline
\multicolumn{2}{|c|}{For $\Upsilon(2S)$ dataset} \\
\hline
\hline
0.212    & 126.2k \\
\hline
0.214    & 126.2k \\
\hline 
0.216    & 126.2k \\
\hline 
0.218    & 126.2k \\
\hline
0.220    & 126.2k \\
\hline 
0.225    & 126.2k \\
\hline
0.500    & 126.2k \\
\hline 
0.750    & 126.2k  \\
\hline
1.0    & 87k \\
\hline 
1.5    & 126.2k \\
\hline 
2.0    & 87k \\
\hline 
3.0    & 87k \\
\hline
4.0    & 87k \\
\hline
5.0    & 87k \\
\hline 
6.0    & 87k  \\
\hline
6.7    & 87k \\
\hline 
7.0    & 87k \\
\hline 
7.5    & 87k \\
\hline 
8.0    & 87k  \\
\hline
8.25    & 126.2k \\
\hline 
8.5    & 126.2k \\
\hline 
8.75    & 126.2k  \\
\hline
9.0    & 87k  \\
\hline
9.10    & 126.2k \\
\hline 
9.20    & 126.2k \\
\hline
9.25    & 126.2k \\
\hline 
0.212 -- 9.46    & 174k \\ 
\hline
\end{tabular} 
\end{minipage}
\caption{The number of  signal MC events generated at different masses for the decay chains of $\Upsilon(2S,3S)\rightarrow \pi^+\pi^-\Upsilon(1S),  \Upsilon(1S) \rightarrow \gamma A^0,  A^0 \rightarrow \mu^+\mu^-$.}
\label{table:SignalMC}
\end{table}

 \begin{table}
\centering
\begin{tabular}{|c |r|r|r|}
\hline
Decay Mode    &Generated Events  &Cross-section (nb)  &Luminosity $(fb^{-1})$ \\[0.5ex]
\hline
\hline
\multicolumn{4}{|c|}{For $\Upsilon(3S)$ dataset} \\
\hline
\hline 
 $\Upsilon(3S) \rightarrow anything $    & 215456000     & 4.19  & 51.42 \\
\hline
 $e^+e^- \rightarrow q\overline{q}~ (q=u,d,s)$         & 111576000     & 2.18  & 51.18 \\
\hline
 $e^+e^- \rightarrow c\overline{c}$        & 135224000     & 1.36  & 99.429 \\
\hline
 $e^+e^- \rightarrow  \tau^+\tau^-$        & 47632000     & 0.94  & 50.672 \\
\hline
 $e^+e^- \rightarrow \gamma e^+e^-$   & 283856000  & 25.79  & 11.01 \\
\hline
 $e^+e^- \rightarrow \gamma \mu^+\mu^-$     & 68744000  &1.1985  & 57.358 \\
\hline
\hline
\multicolumn{4}{|c|}{For $\Upsilon(2S)$ dataset} \\
\hline
 \hline
 $\Upsilon(2S) \rightarrow anything $   & 156400000     & 7.249  & 21.57 \\
\hline
 $e^+e^- \rightarrow q\overline{q}~ (q=u,d,s)$          & 91025000     & 2.31  & 39.41 \\
\hline
 $e^+e^- \rightarrow c\overline{c}$        & 51420000     & 1.44  & 35.71 \\
\hline
 $e^+e^- \rightarrow  \tau^+\tau^-$       & 20245000     & 1.04  & 19.47 \\
\hline
 $e^+e^- \rightarrow \gamma e^+e^-$   & 106268000  & 25.9  & 4.10 \\
\hline
 $e^+e^- \rightarrow \gamma \mu^+\mu^-$    & 26891000  & 1.30  & 20.69 \\
\hline
\end{tabular} 
\caption{Background MC samples for different decay processes, which are used in this analysis.}
\label{table:YnS_MC}
\end{table}

\section{Event Reconstruction and Event Pre-Selection}
The events of interest are reconstructed using \babar\ software packages designed for creating the lists of composite particles, automating the work of making combinations, performing the kinematic fits, making the pre-selection criteria and storing the events in an object-oriented based ROOT  ntuple files \cite{ROOT}. To streamline the decay processes of $\Upsilon(2S,3S) \rightarrow \pi^+\pi^- \Upsilon(1S)$, $\Upsilon(1S) \rightarrow \gamma A^0$, $A^0 \rightarrow \mu^+\mu^-$, the data and MC samples are filtered or \rm{\lq\lq skimmed\lq\lq}. We select events containing exactly four charged tracks and a single energetic photon with a center-of-mass (CM) energy  greater than 
$200 \mev$. The additional photons with CM energies below this threshold  are also allowed to be present in the events. The two highest momentum tracks in the CM frame are required  to have opposite charge, and are assumed to be muon candidates, combined to form the $A^0$ candidate. These tracks are required to have a distance of closest approach 
to the interaction point of less than 1.5 cm in the plane transverse to the beam and less than 10 cm along the beam-axis. The $\Upsilon(1S)$ candidate 
is reconstructed by combining the $A^0$ candidate with the energetic photon candidate and requiring the invariant mass of the $\Upsilon(1S)$ candidate to be between $9.0$ and 
$9.8 \gevcc$. The $\Upsilon(2S,3S)$ candidates are  formed by combining the $\Upsilon(1S)$ candidate with the two remaining tracks, 
assumed to be pions. The di-pion invariant mass must be in the range of [$2m_{\pi},(m_{\Upsilon(2S,3S)}-m_{\Upsilon(1S)})$],  compatible with the kinematic boundaries of 
the $\Upsilon(2S,3S) \rightarrow \pi^+\pi^- \Upsilon(1S)$ decay. Finally, we define the mass recoiling against the di-pion system to be:
\begin{equation}
       m^2_{\rm recoil}=s +m^2_{\pi\pi} - 2\surd s E_{\pi\pi}, 
 \end{equation}

\noindent where $\surd s$ is the collider CM energy (assumed to be $\surd s = M_{\Upsilon(3S, 2S)}$) and $E_{\pi\pi}$ is the energy of the di-pion system. We require that $m_{recoil}$ to be between 9.35 and 9.57 GeV/$c^2$. The $m_{\rm recoil}$ is used to identify the $\Upsilon(3S, 2S) \rightarrow \pi^+\pi^-\Upsilon(1S)$ transitions and it should be peaked at $\Upsilon(1S)$ mass for signal like events. The entire decay chain is fit imposing a mass constraint on the 
$\Upsilon(1S)$ and $\Upsilon(2S,3S)$ candidates, as well as requiring the energy of the $\Upsilon(2S,3S)$ candidate to be consistent 
with the $e^+e^-$ CM energy. 

To distinguish the signal from backgrounds, we calculate the reduced mass \cite{Yury} for an event which is defined as:
\begin{equation}
 m_{\rm  red} = \sqrt{m_{\mu^+\mu^-}^2 - 4 m_{\mu}^2}.
\end{equation}
 
\noindent  $m_{\rm red}$ is equal to twice the momentum of the muons in the rest frame of $A^0$, and has a smooth distribution in the region of the kinematic threshold $m_{\mumu} \approx 2m_{\mu}$ ($m_{\rm red} \approx 0$). It has a Gaussian-like distribution  for signal and a flat distribution for background.    

\begin{figure}[!htb]
\centering
\includegraphics[width=3.0in]{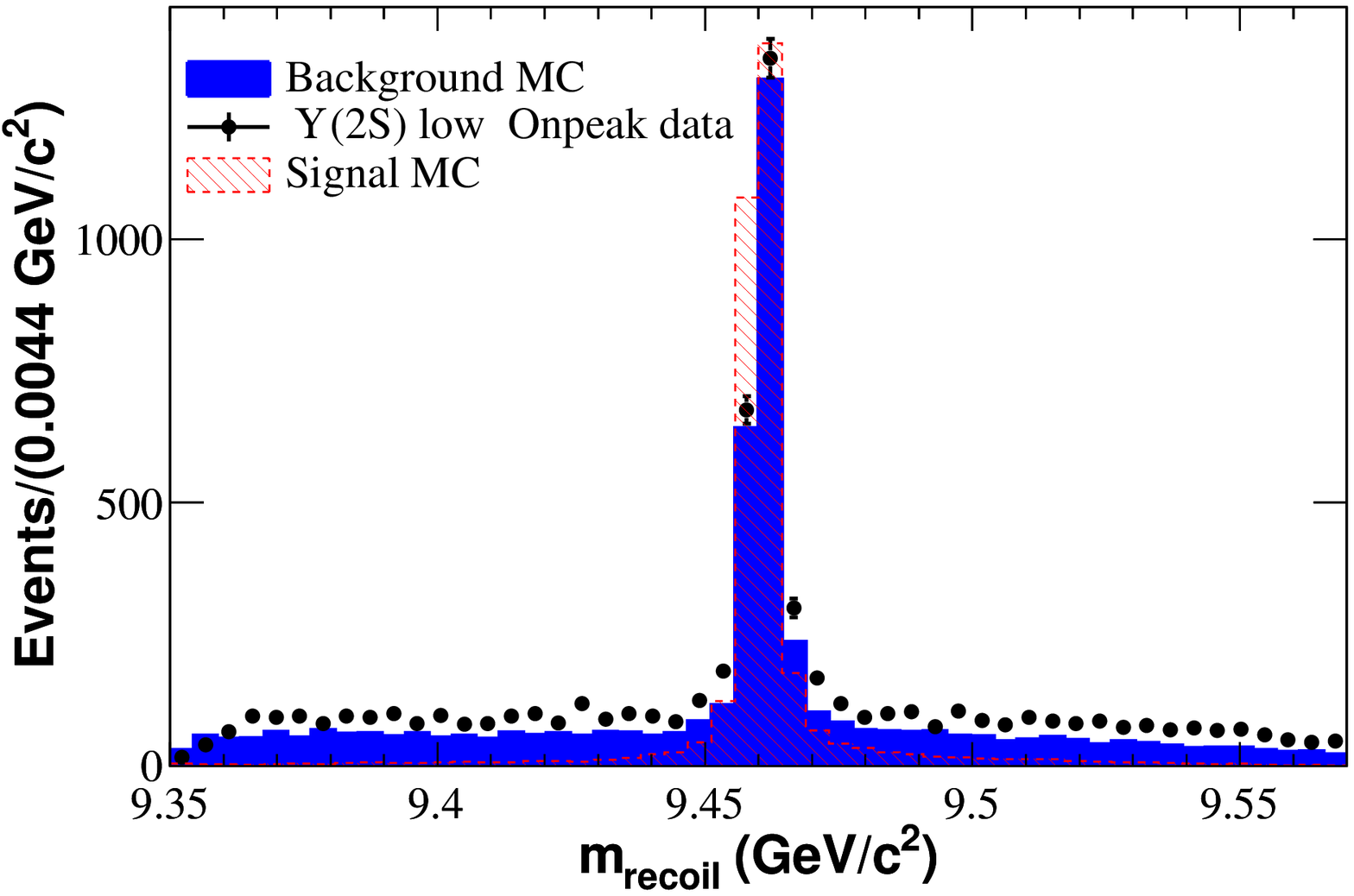}
\includegraphics[width=3.0in]{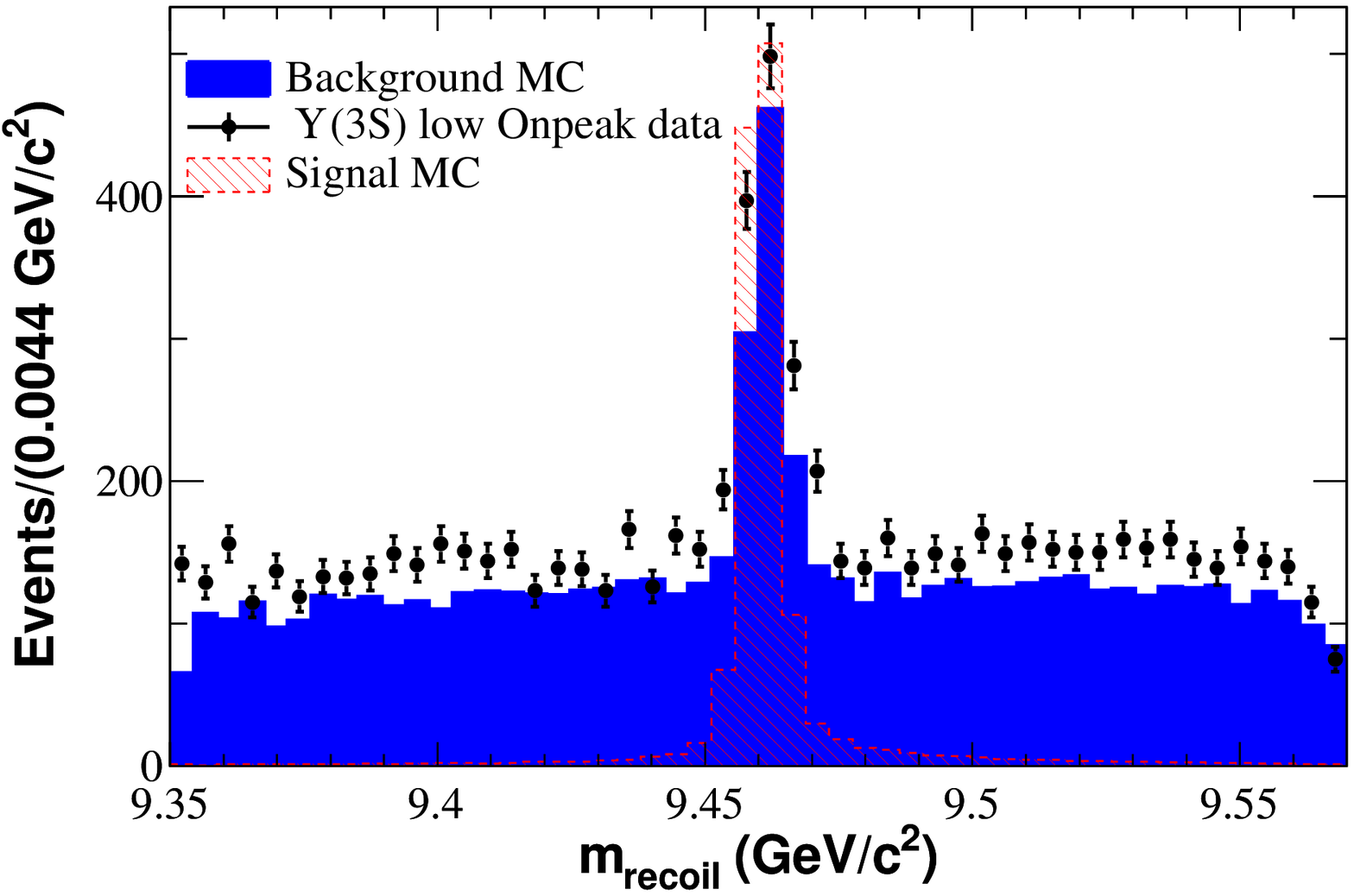}

\caption{The distribution of the $m_{\rm recoil}$ for the low onpeak data sample (dots), together with the production of the background and signal MC for $\Upsilon(2S)$ (left) and $\Upsilon(3S)$ (right). The mean of the recoil mass in background  MC has been corrected after comparing the recoil mass distributions in a control samples of data and MC, the details of which can be found  in section~\ref{section:mrec-study}.  The background MC is normalized to the $\Upsilon(2S,3S)$ low onpeak data samples.}
\label{fig:mrec-skim}
\end{figure}

 Further selection criteria are applied at the ntuple level. Events are required to satisfy L3 trigger (L3OutDch $|$$|$ L3OutEmc)  and filter (RecoBGFilter $\&$$\&$ DigiFilter) flags. The trigger selection criteria reduces significant amount of combinatorial backgrounds while maintaining the signal selection efficiencies up to $\sim 99.95\%$ for both the $\Upsilon(2S,3S)$ datasets. Further, we require that the momentum magnitude of most energetic charged particle to be less than 8.0 \gevc.  Figure~\ref{fig:mrec-skim} shows the $m_{\rm recoil}$ distribution for signal MC, combined background MC of  $\Upsilon(3S, 2S)$ generic, radiative bhabha, radiative di-muon, $\tau^+\tau^-$, $\c\overline{c}$ and uds. 

\section{Event Selection}
This section describes the variables used to discriminate between signal and  background processes. We also describe various multi-variate techniques used to discriminate signal events from the background events. The variables of interest can be split into three groups, which are pion, photon, and muon related variables. The pion related variables in the decay chains of $\Upsilon(3S, 2S) \rightarrow \pi^+\pi^- \Upsilon(1S)$ are identified by searching for two low momentum pions. The photon related variables in the decay chain of $\Upsilon(1S) \rightarrow \gamma A^0$ are identified by detecting a monochromatic photon. The muon related variables in the decay chain of $A^0 \rightarrow \mu^+\mu^-$  are identified by two high momentum muons. The kinematic variables related to these three groups are chosen as follows.

\subsection{Pion selection variables}

\begin{figure}[!htb]
\centering
\includegraphics[width=3.0in]{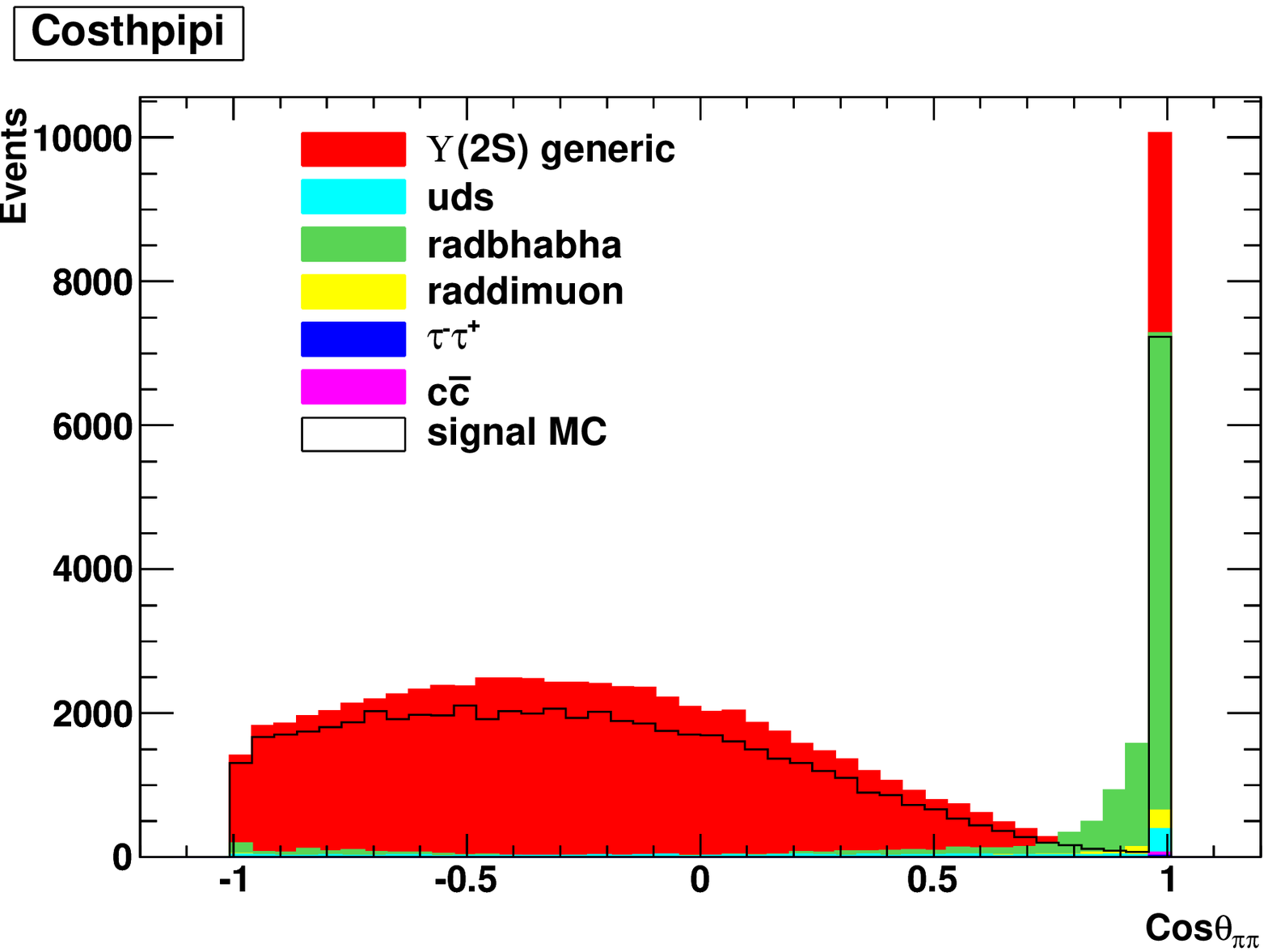}
\includegraphics[width=3.0in]{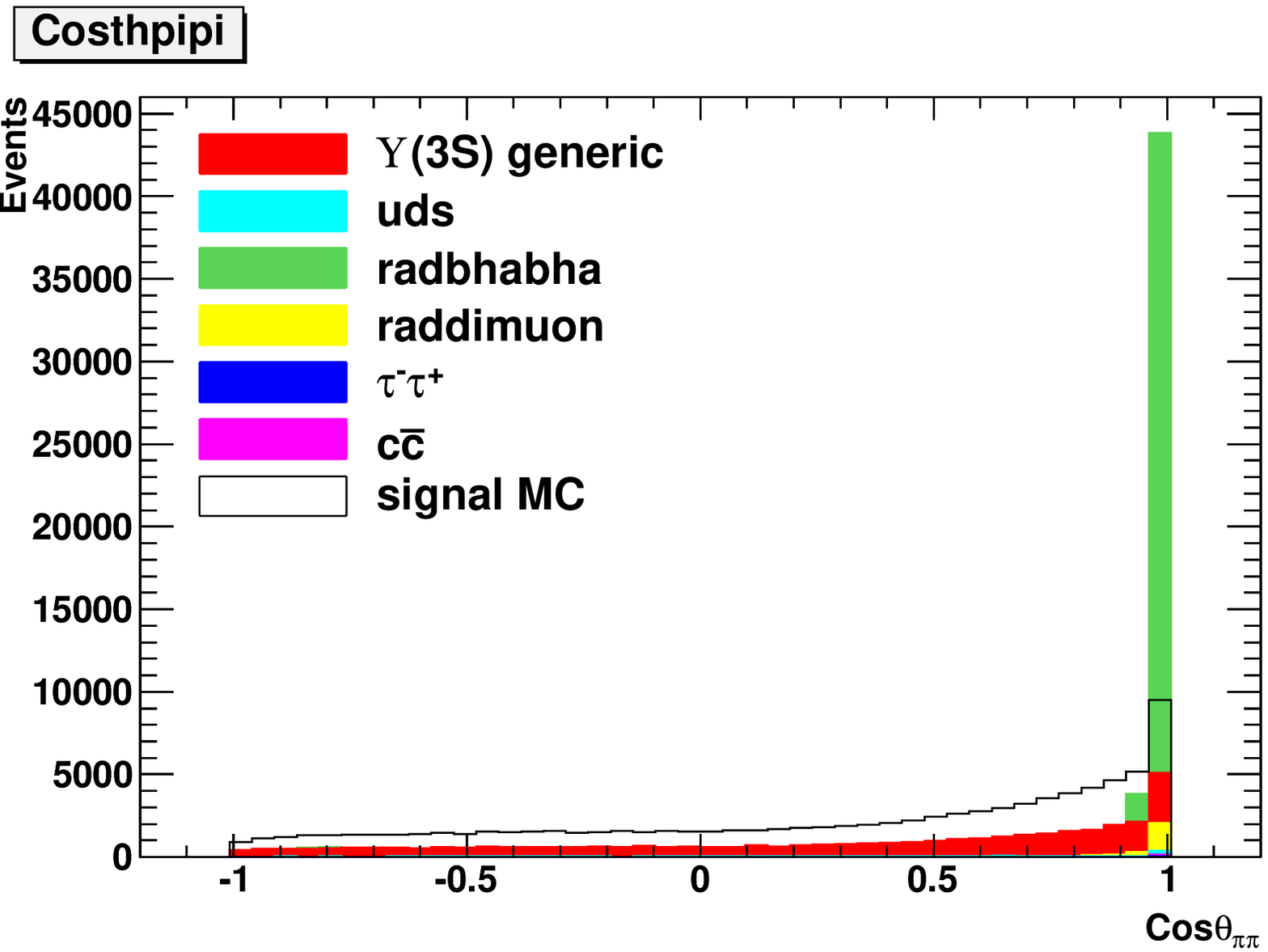}

\smallskip
\centerline{\hfill (a) \hfill \hfill (b) \hfill }
\smallskip

\includegraphics[width=3.0in]{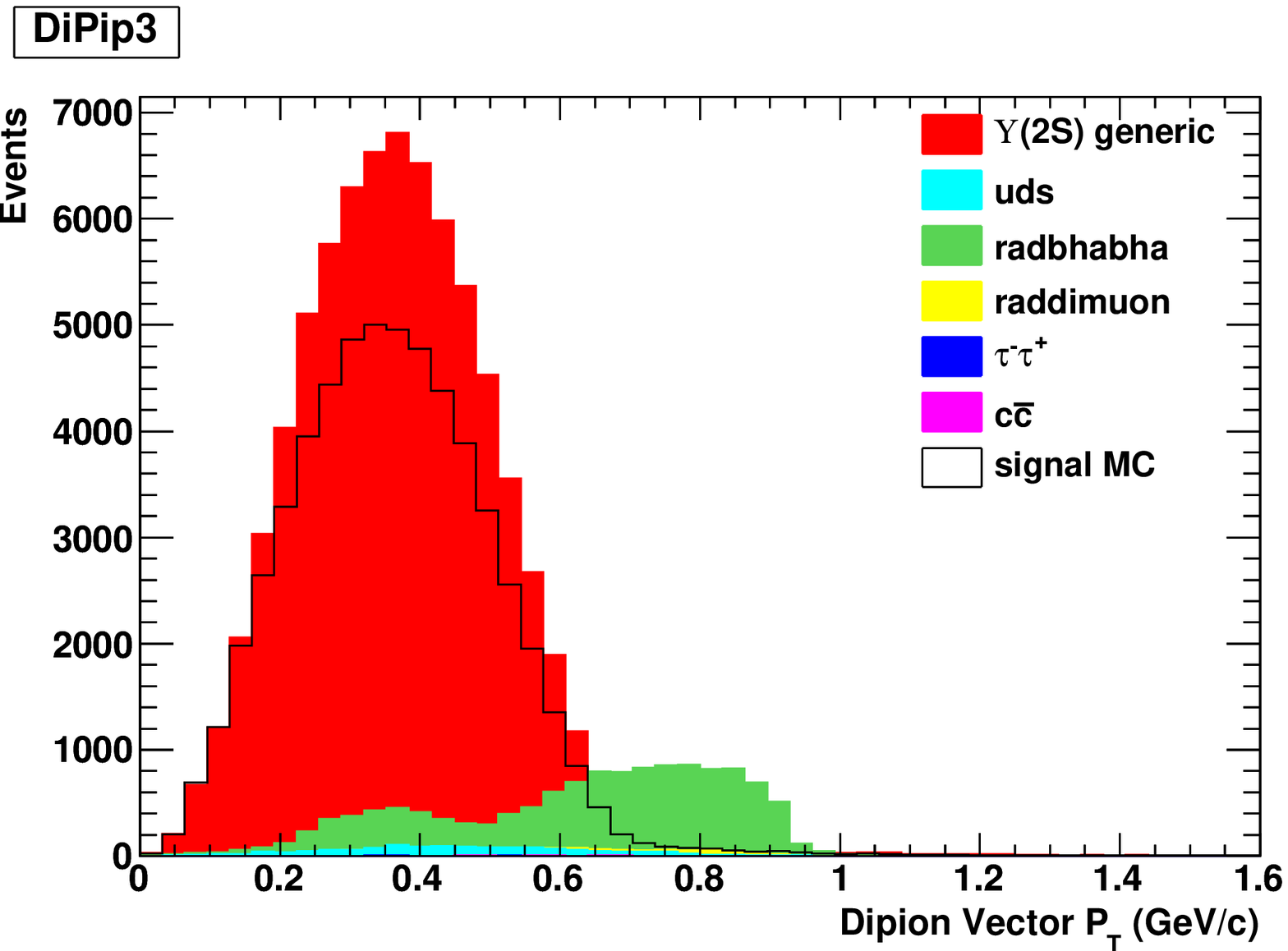}
\includegraphics[width=3.0in]{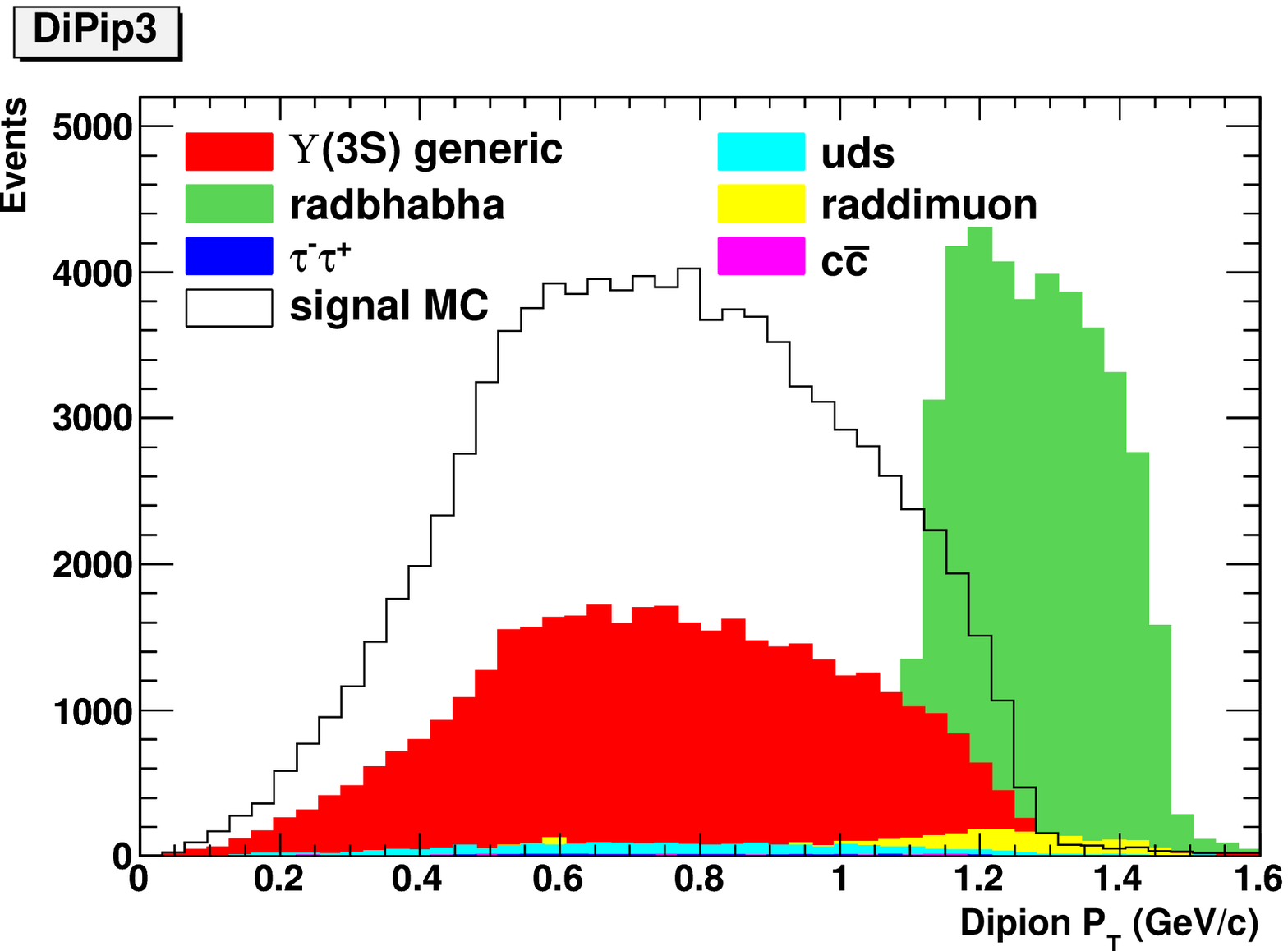}

\smallskip
\centerline{\hfill (c) \hfill \hfill (d) \hfill }
\smallskip

\includegraphics[width=3.0in]{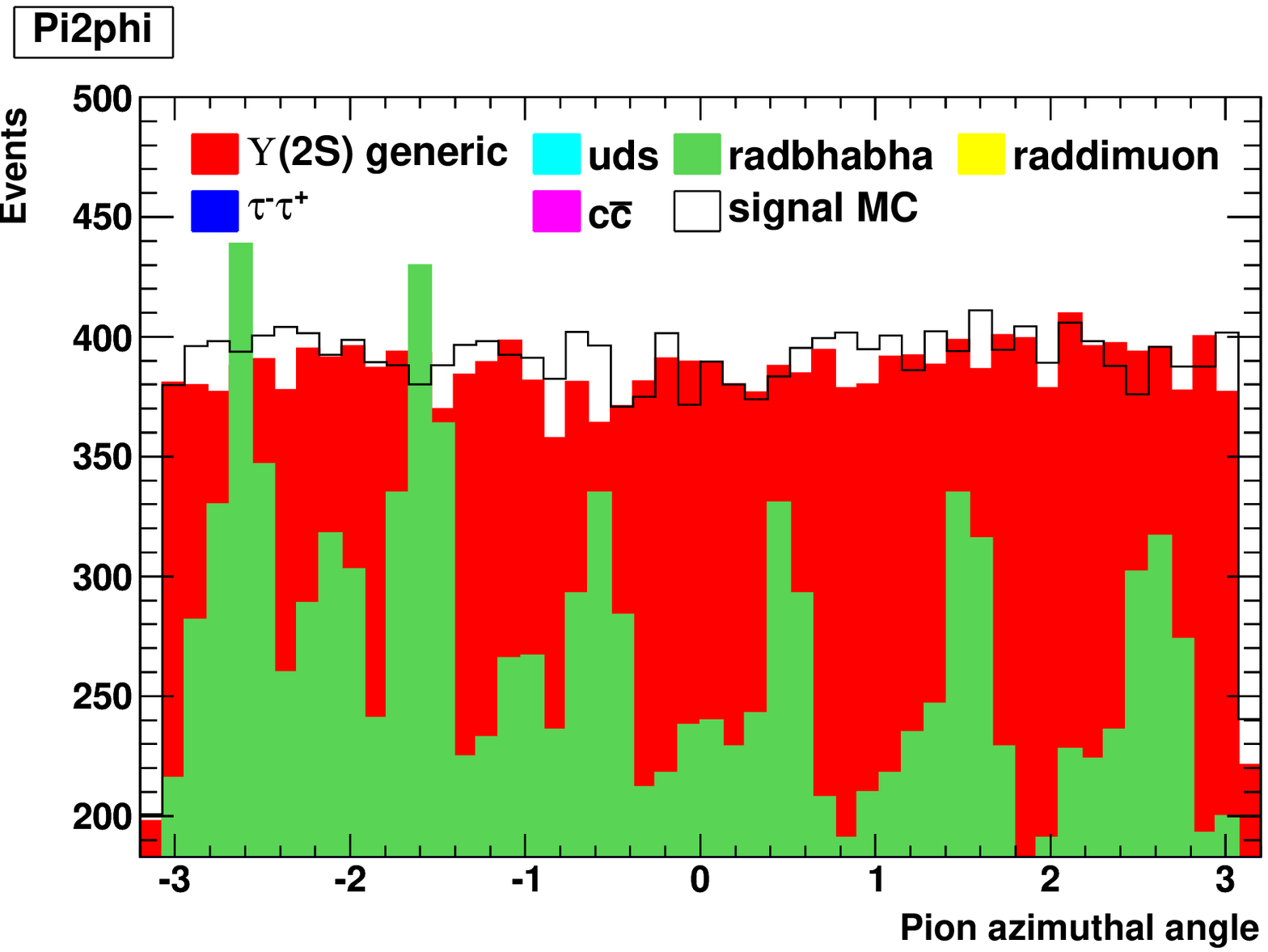}
\includegraphics[width=3.0in]{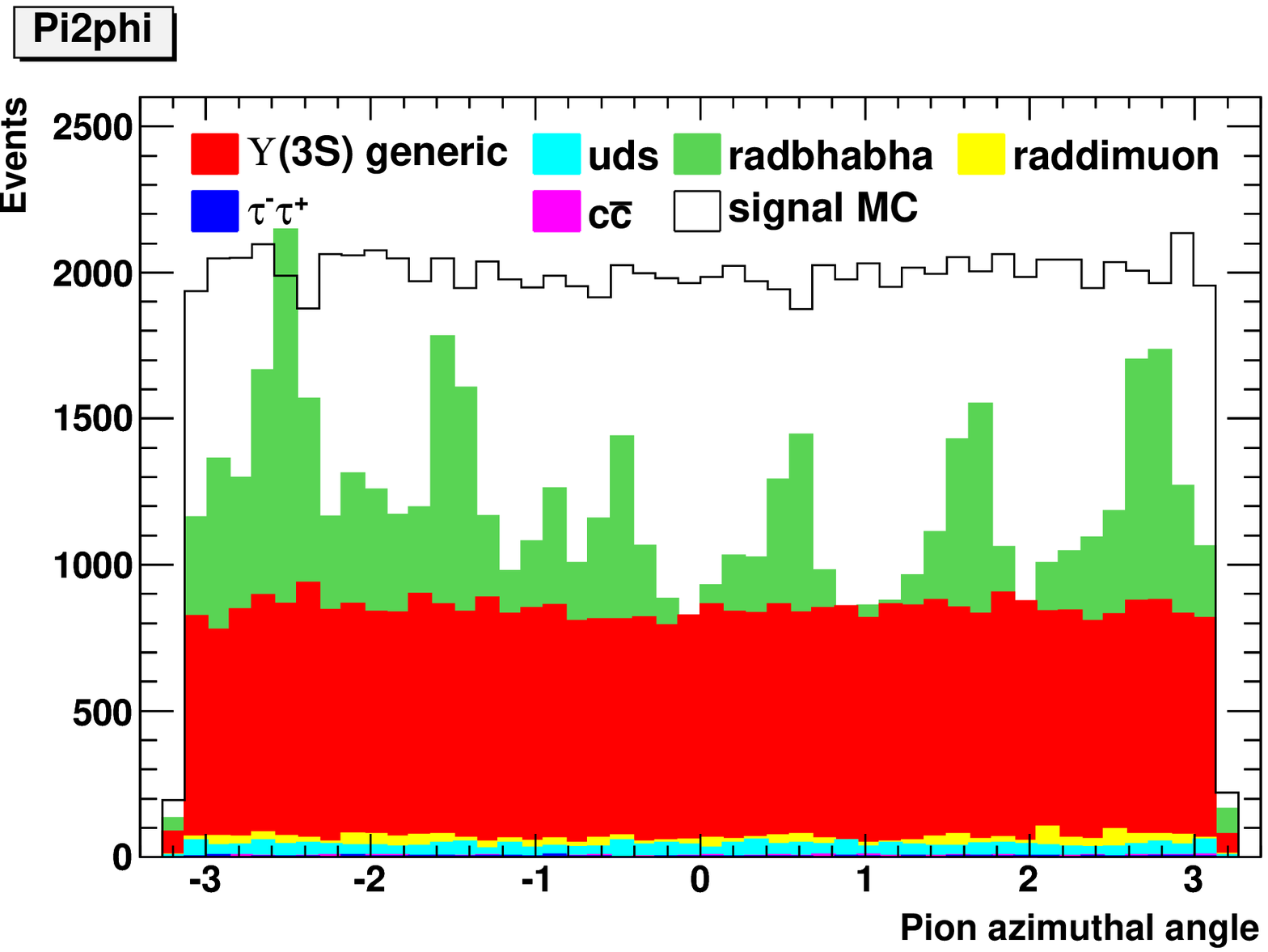}

\smallskip
\centerline{\hfill (e) \hfill \hfill (f) \hfill }
\smallskip

\caption {Di-pion related variables: (a, b) Cosine of angle between two pions in the laboratory frame (c, d) Transverse momentum of di-pion system in the laboratory frame and  (e, f) Azimuthal angle of pion. Left plots are for $\Upsilon(2S)$ and right plots are for $\Upsilon(3S)$.   All these variables are plotted  after applying the pre-selection criteria.} 

\label{fig:Di-PionVar1}
\end{figure}

\begin{figure}[!htb]
\centering 
\includegraphics[width=3.0in]{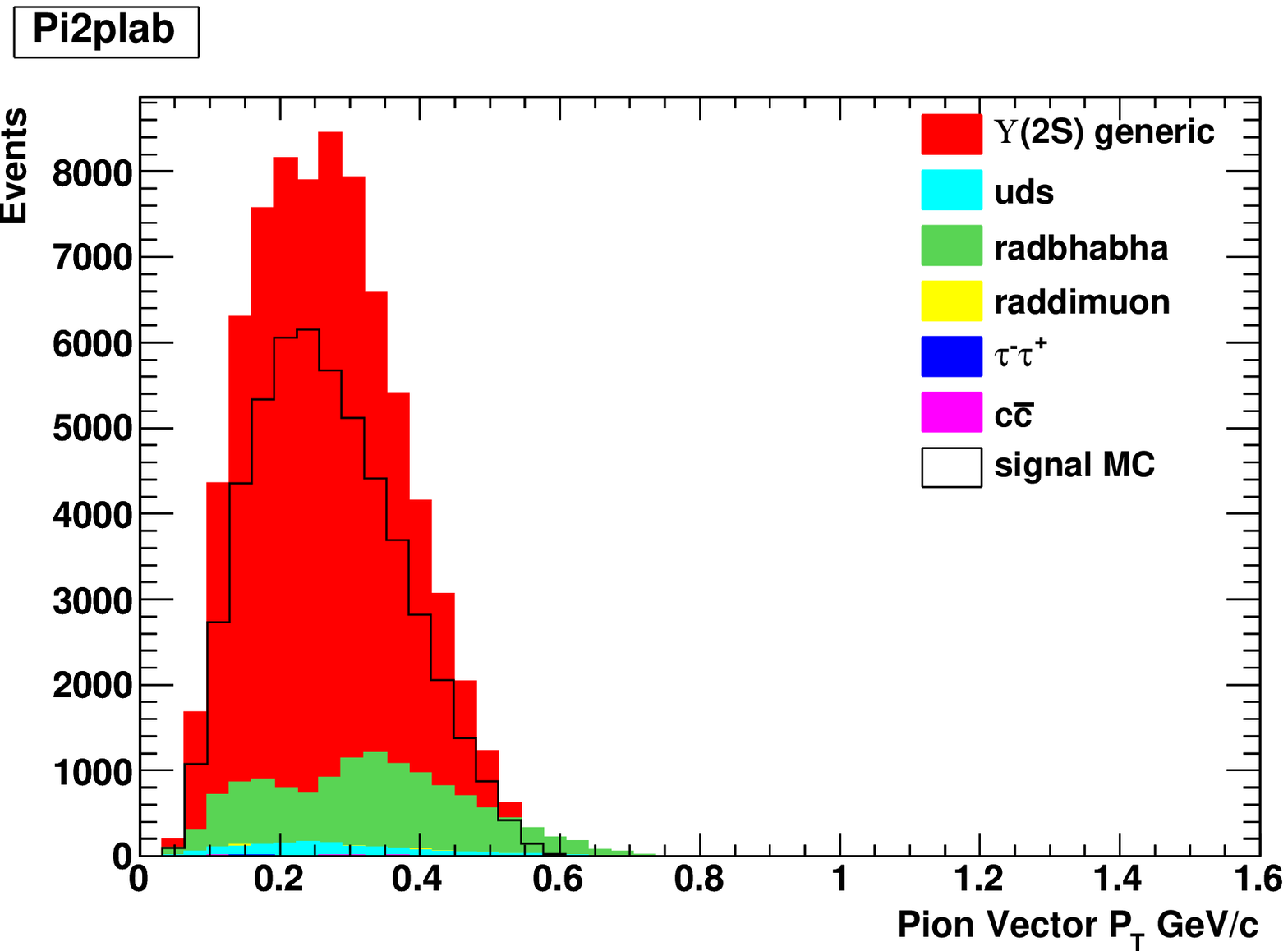}
\includegraphics[width=3.0in]{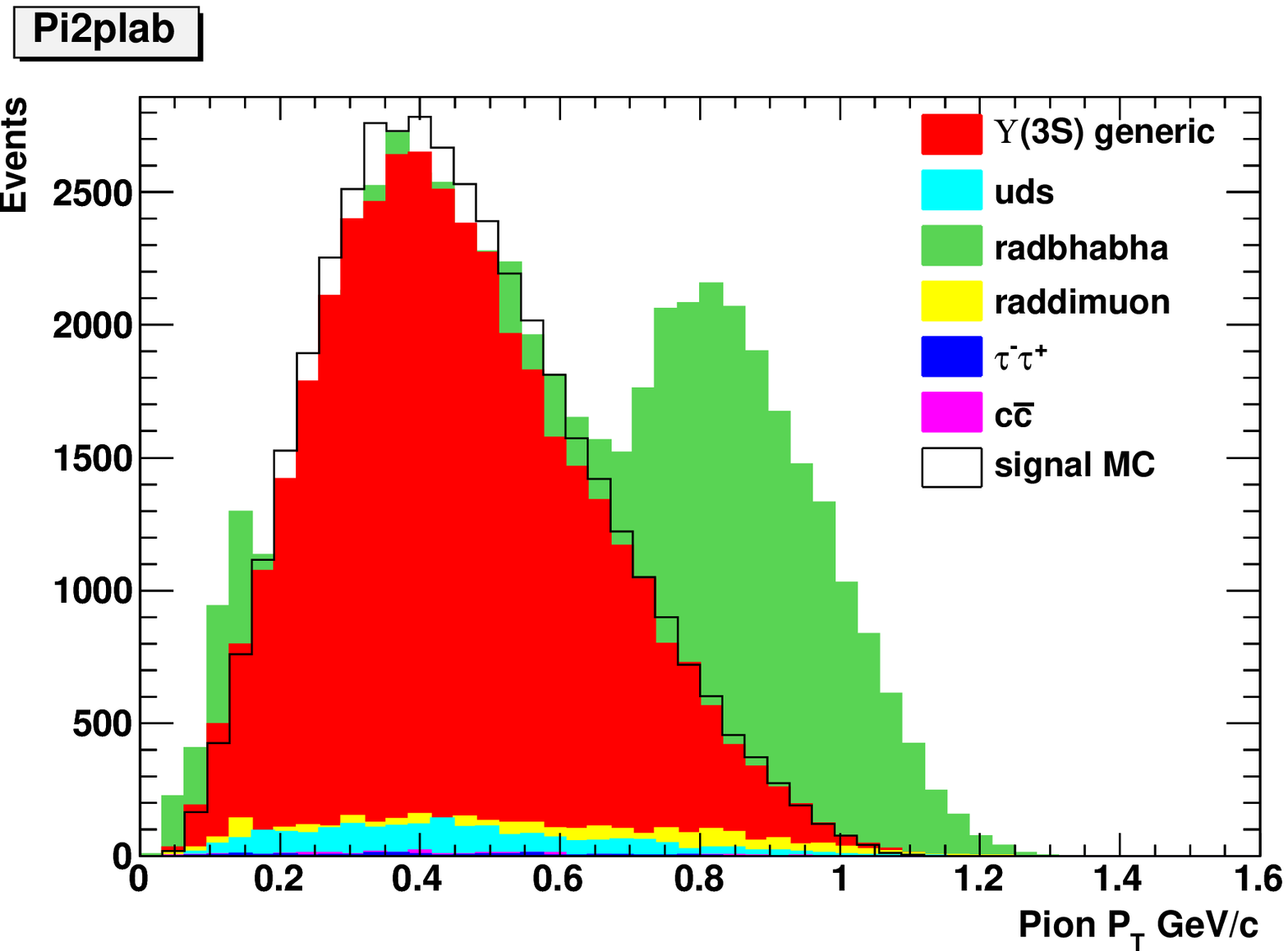}

\smallskip
\centerline{\hfill (a) \hfill \hfill (b) \hfill }
\smallskip

\includegraphics[width=3.0in]{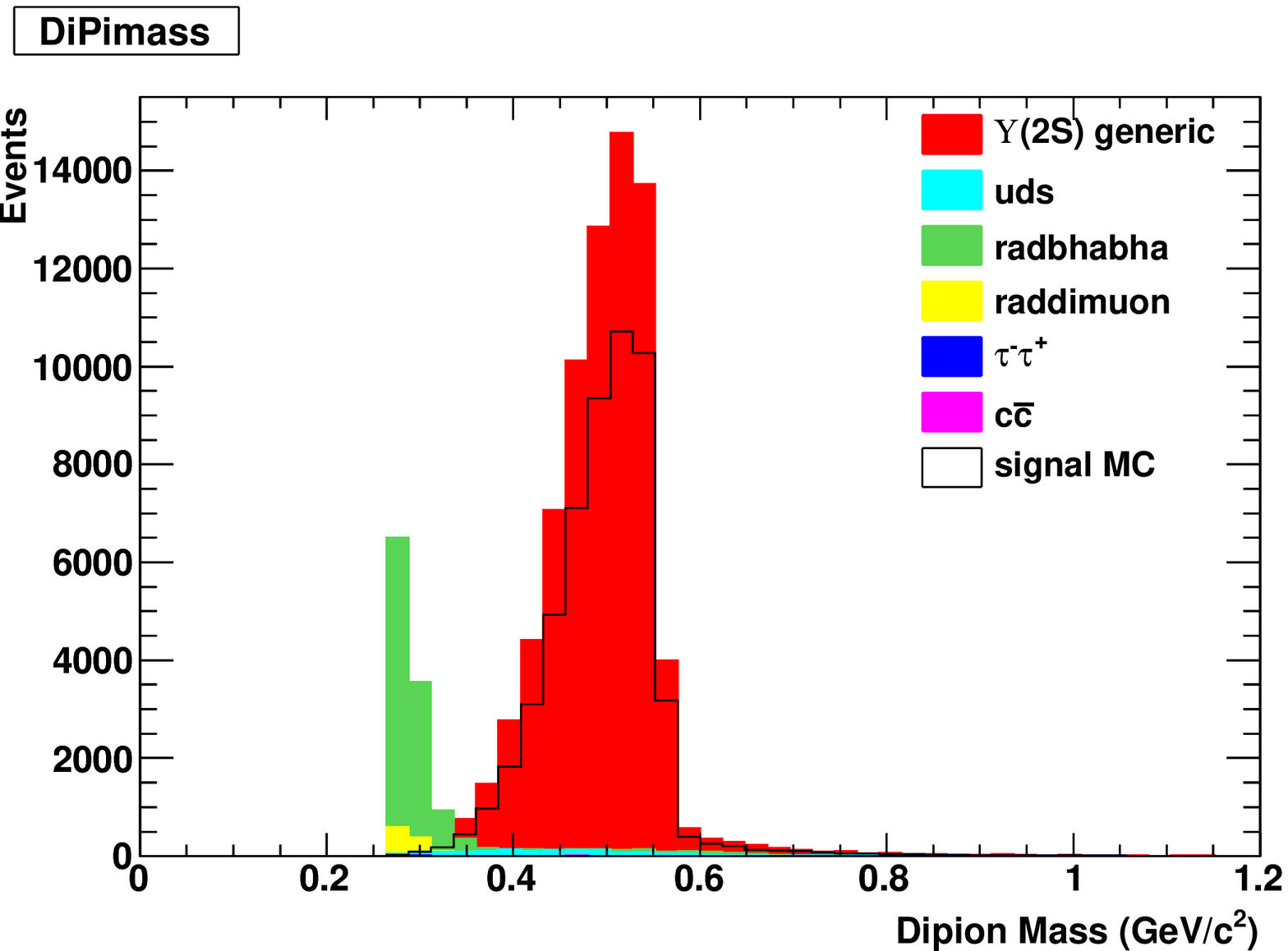}
\includegraphics[width=3.0in]{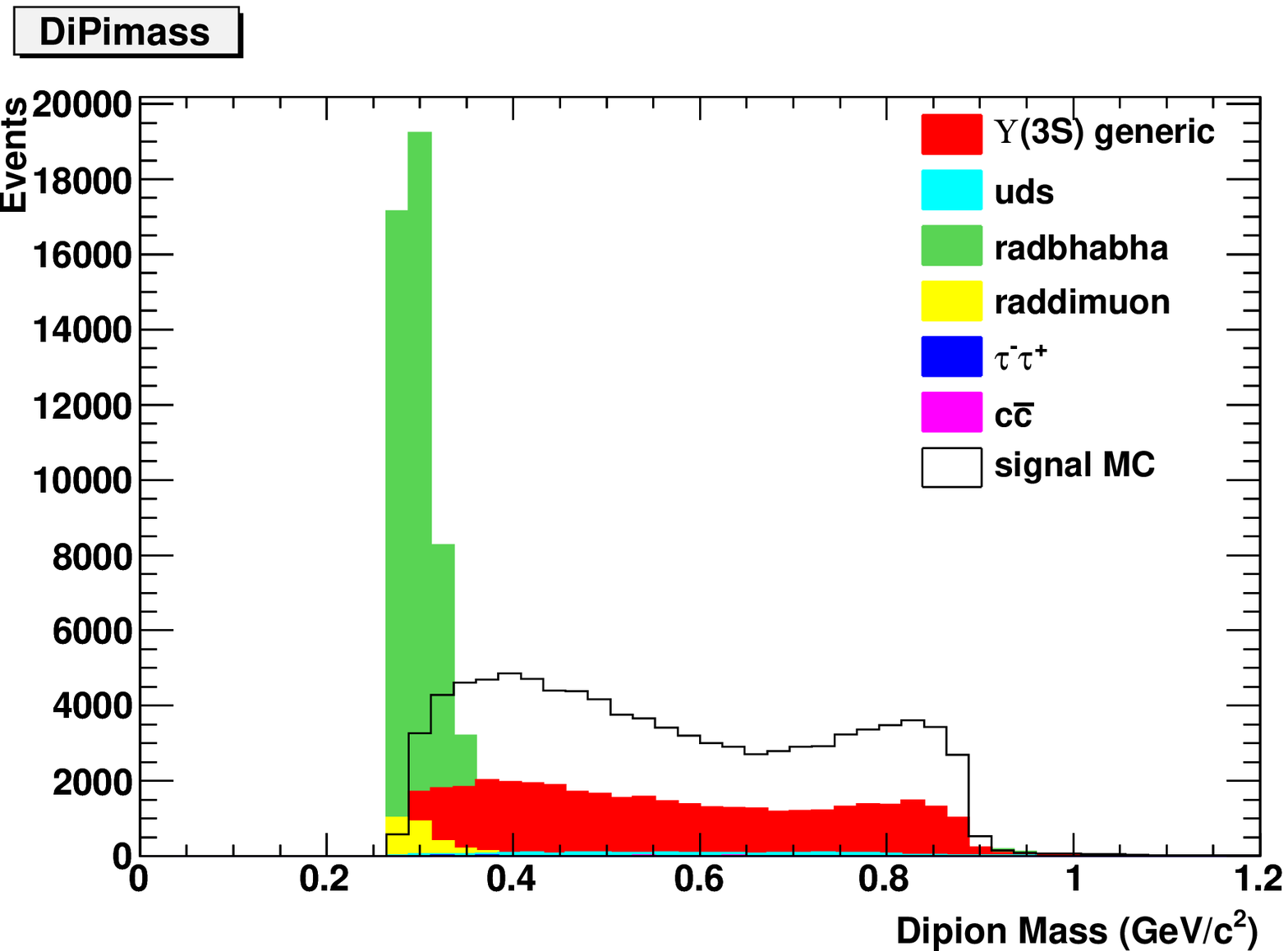}

\smallskip
\centerline{\hfill (c) \hfill \hfill (d) \hfill }
\smallskip

\includegraphics[width=3.0in]{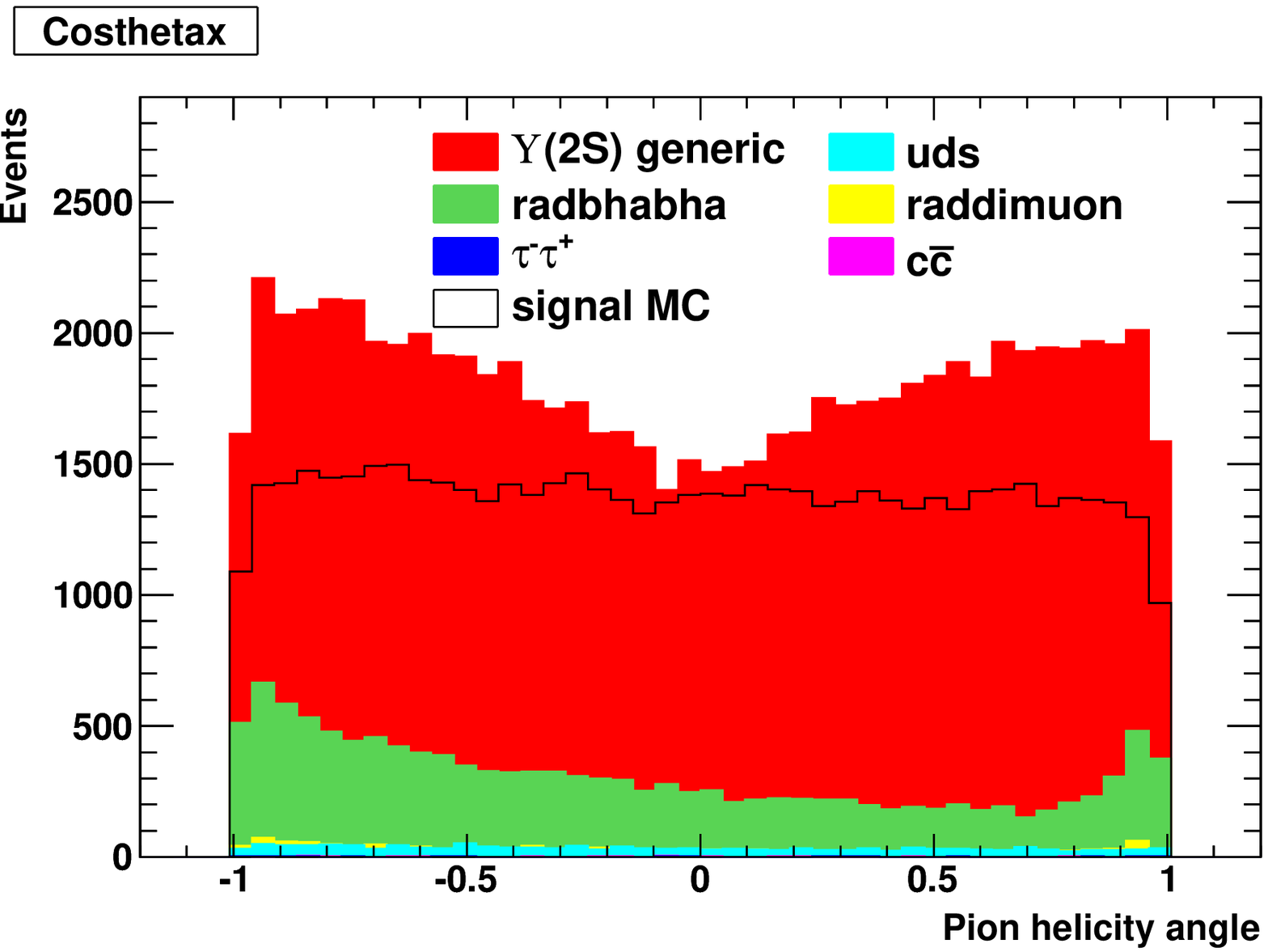}
\includegraphics[width=3.0in]{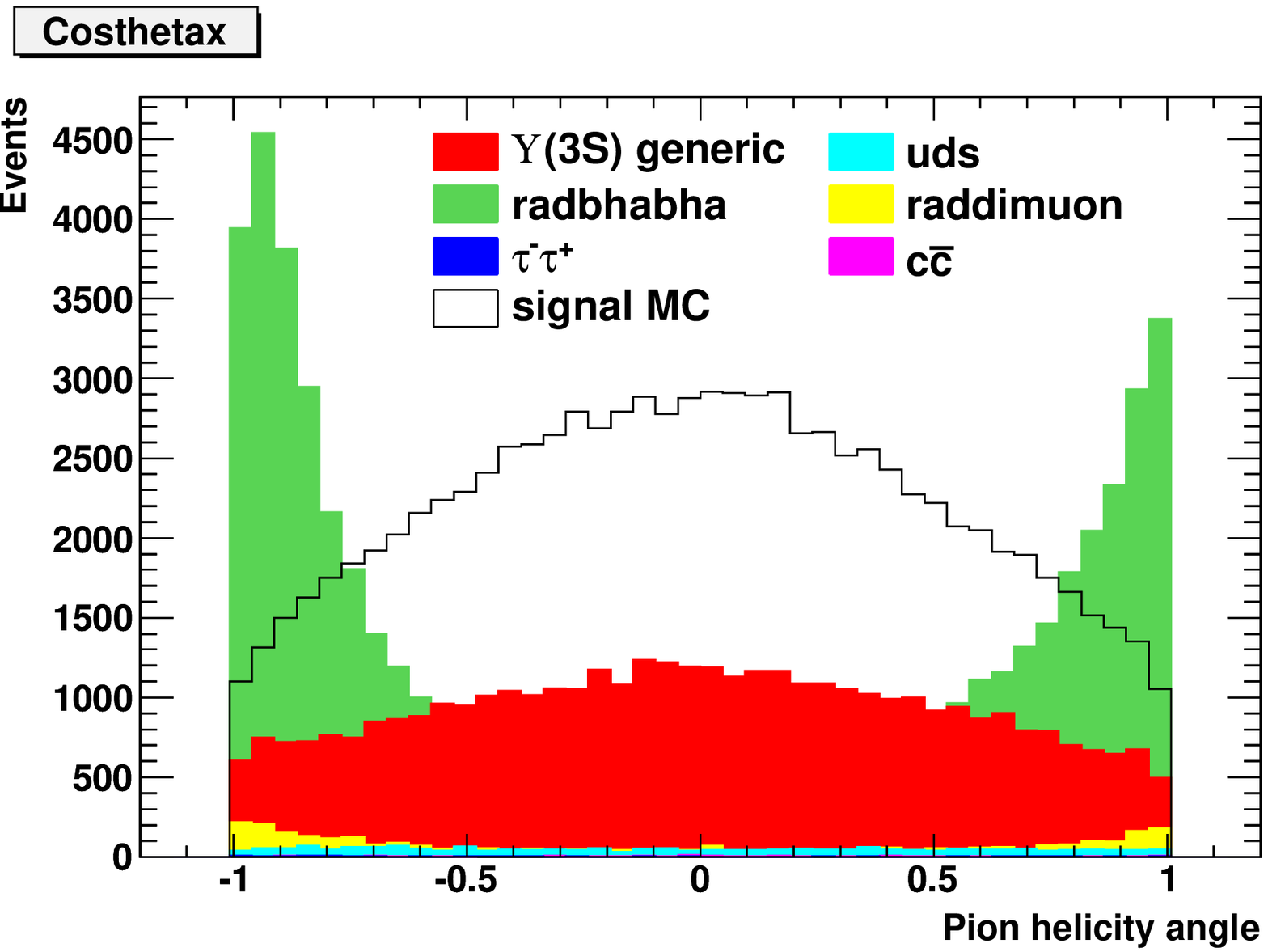}

\smallskip
\centerline{\hfill (e) \hfill \hfill (f) \hfill }
\smallskip

\caption {Di-pion related variables: (a, b) Transverse momentum of the pion  (c, d) Di-pion invariant mass and (e, f) Cosine of pion helicity angle. Left plots are for $\Upsilon(2S)$ and right plots are for $\Upsilon(3S)$. All these variables are plotted after applying the pre-selection criteria. } 

\label{fig:Di-PionVar2}
\end{figure}

\begin{figure}[!htb]
\centering 

\includegraphics[width=3.0in]{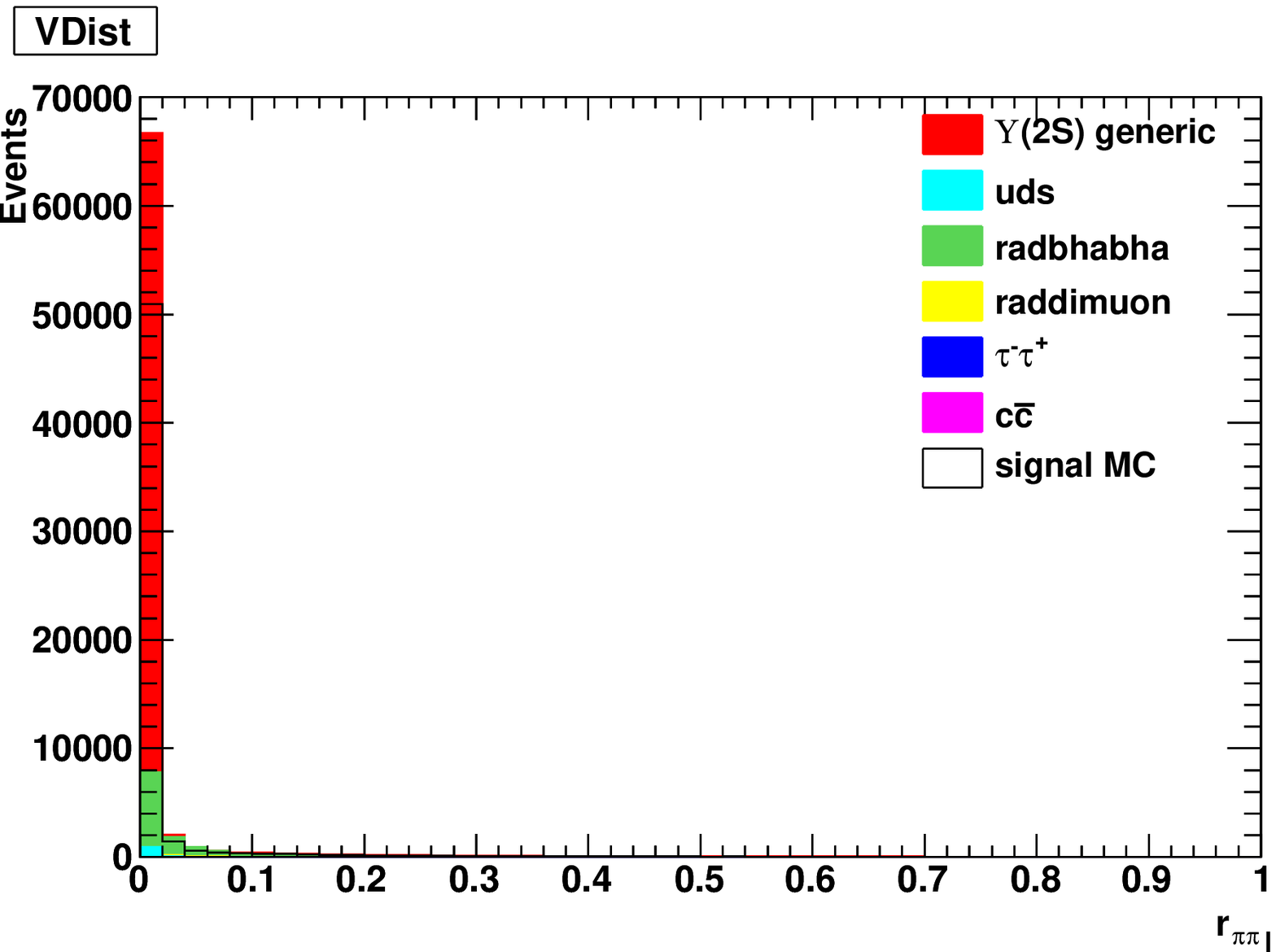}
\includegraphics[width=3.0in]{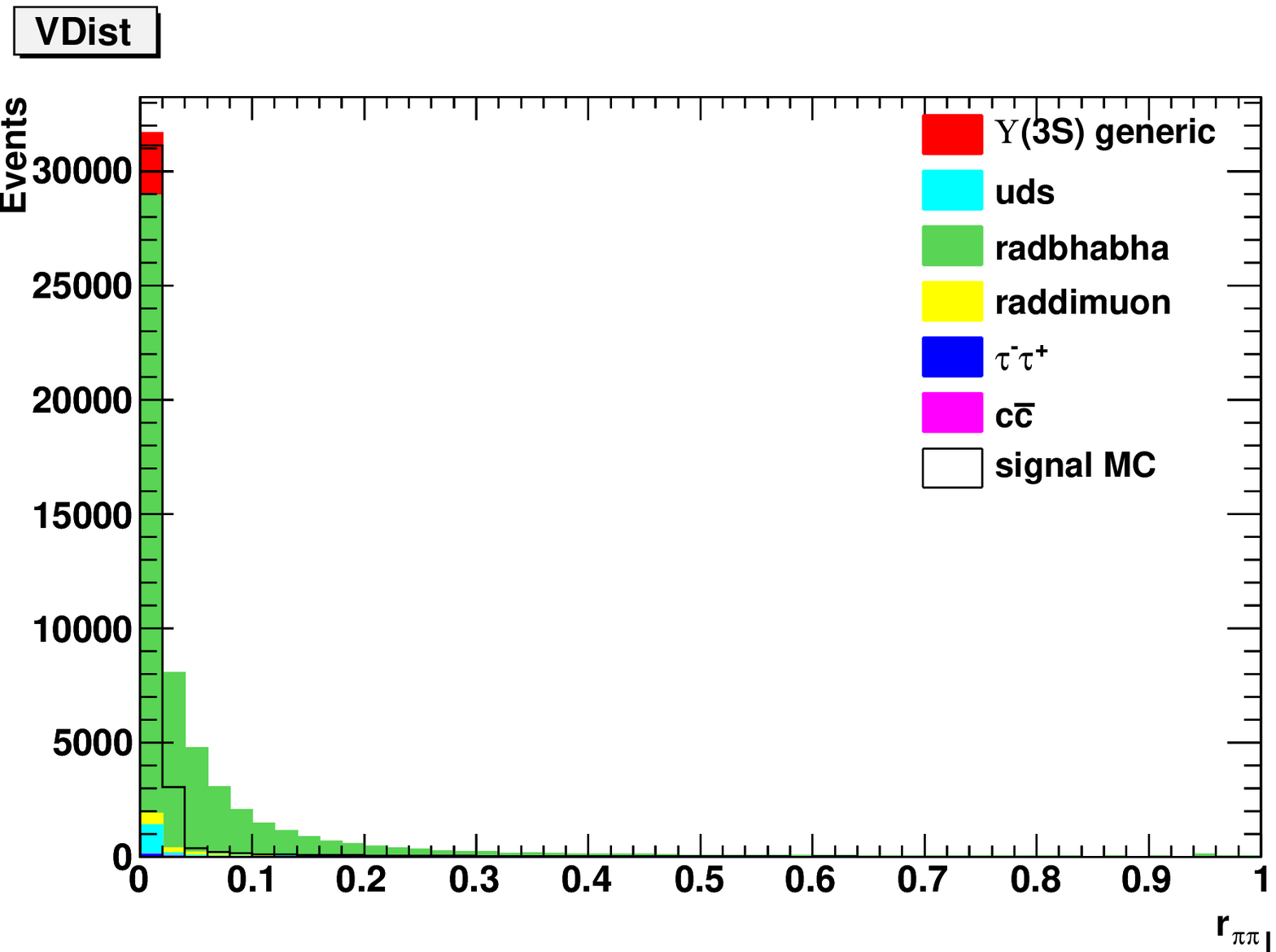}

\smallskip
\centerline{\hfill (a) \hfill \hfill (b) \hfill }
\smallskip

\includegraphics[width=3.0in]{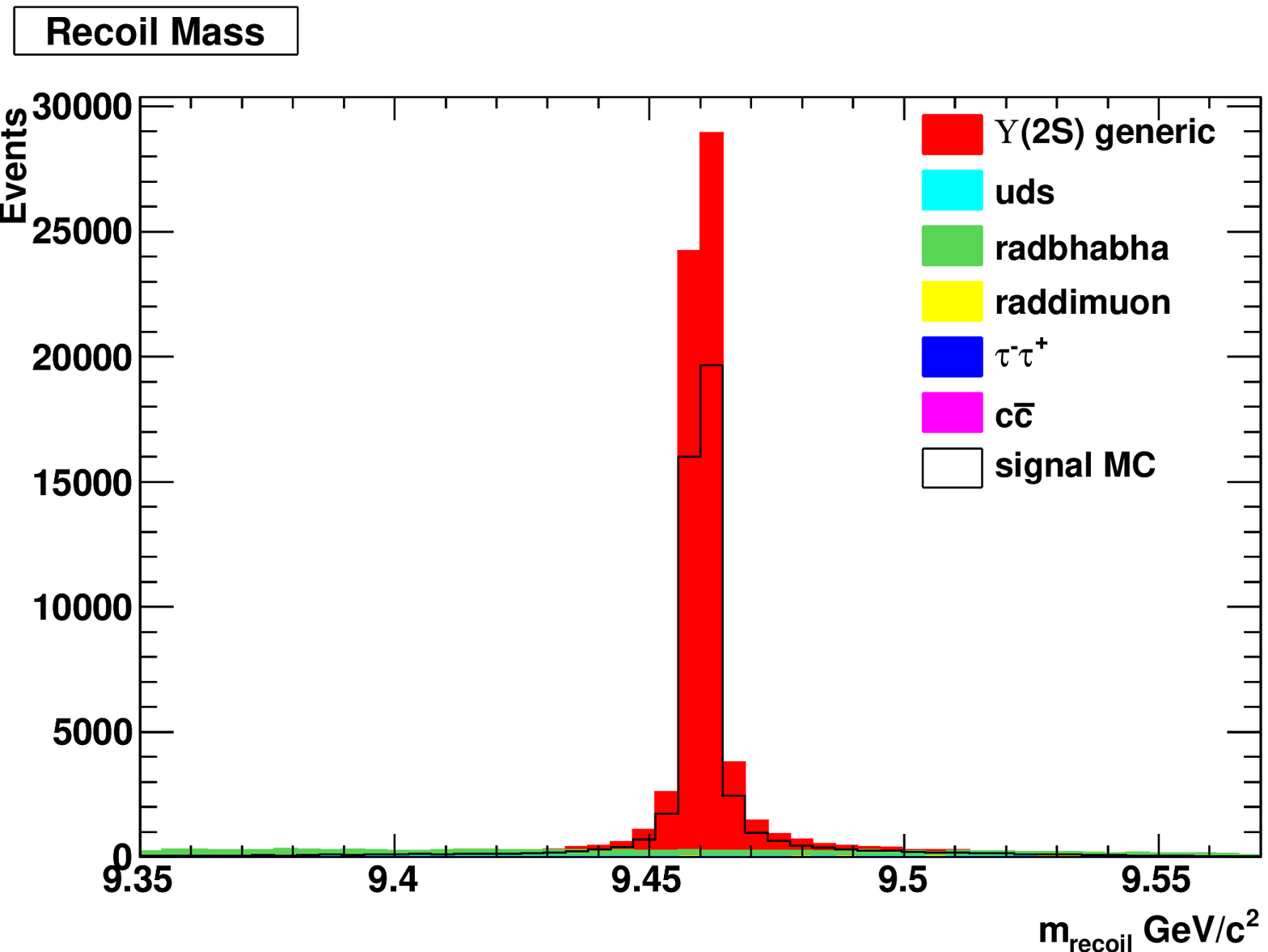}
\includegraphics[width=3.0in]{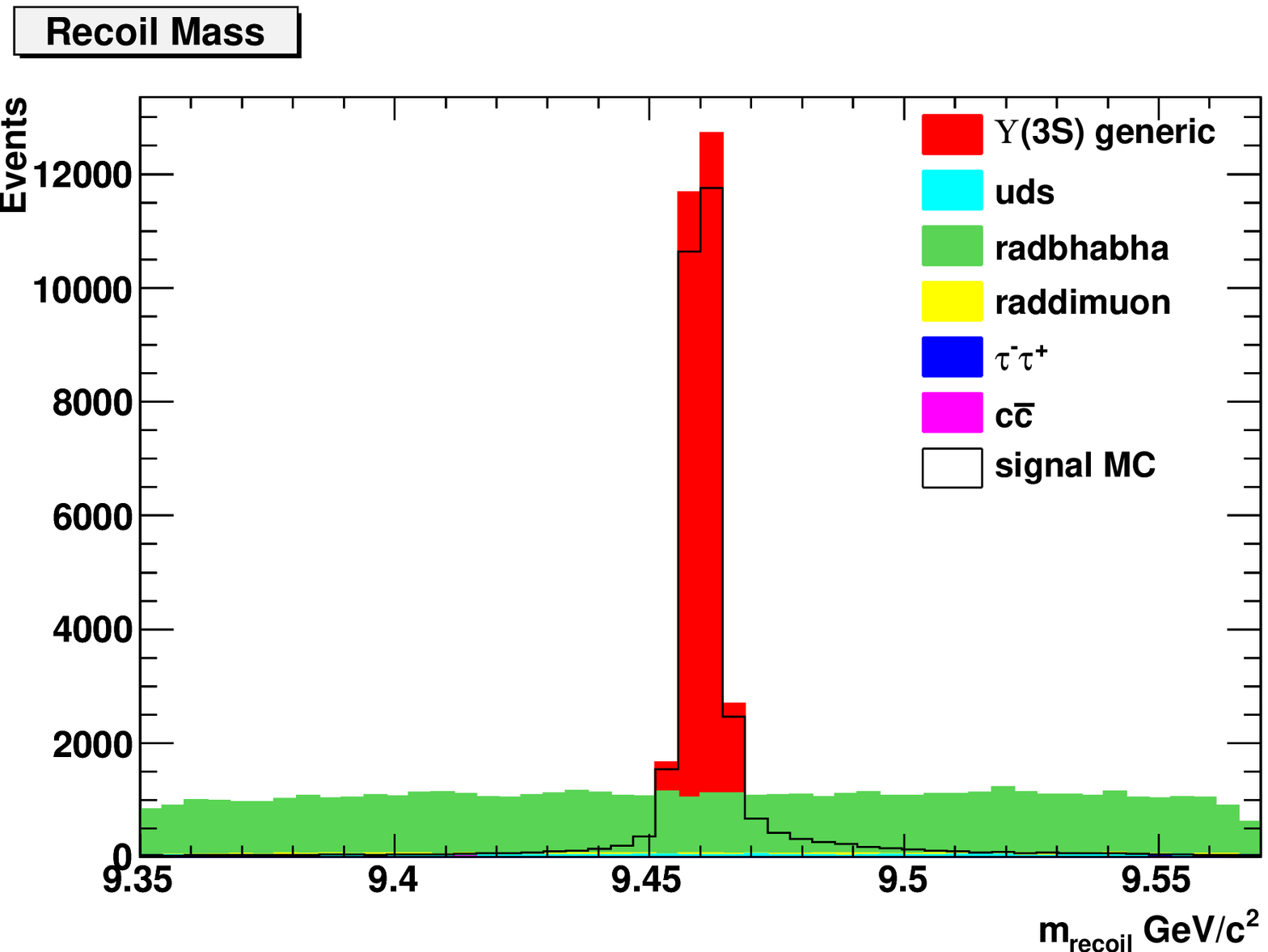}

\smallskip
\centerline{\hfill (c) \hfill \hfill (d) \hfill }
\smallskip

\caption {Di-pion related variables: (a, b)  Transverse position of the di-pion vertex and (g, h) Mass recoiling against the di-pion system. Left plots are for $\Upsilon(2S)$ and right plots are for $\Upsilon(3S)$. All these variables are plotted after applying the pre-selection criteria.} 

\label{fig:Di-PionVar3}
\end{figure}

\begin{itemize}
\item  {\bf Costhpipi:} The cosine of the angle between two pions in the laboratory frame, shown in  Figure~\ref{fig:Di-PionVar1}(a) for $\Upsilon(2S)$ and Figure~\ref{fig:Di-PionVar1}(b) for $\Upsilon(3S)$.

\item {\bf DiPip3:} The transverse momentum of the di-pion system  in the laboratory frame, shown in Figure~\ref{fig:Di-PionVar1}(c) for $\Upsilon(2S)$ and Figure~\ref{fig:Di-PionVar1}(d) for $\Upsilon(3S)$. 

\item  {\bf Pi2phi:} The azimuthal angle of each pion, shown in Figure~\ref{fig:Di-PionVar1}(e) for $\Upsilon(2S)$ and Figure~\ref{fig:Di-PionVar1}(f) for $\Upsilon(3S)$.

\item {\bf Pi2plab:} The transverse momentum of the  pions, shown in Figure~\ref{fig:Di-PionVar2}(a) for $\Upsilon(2S)$ and Figure~\ref{fig:Di-PionVar2}(b) for $\Upsilon(3S)$.

\item {\bf DiPimass:} The di-pion invariant mass, shown in Figure~\ref{fig:Di-PionVar2}(c) for $\Upsilon(2S)$ and Figure~\ref{fig:Di-PionVar2}(d) for $\Upsilon(3S)$.

\item  {\bf Costhetax:} The cosine of the angle formed between the $\pi^+$ in the di-pion frame and the direction of  the di-pion in the $\Upsilon(2S,3S)$ rest frame, shown in  Figure~\ref{fig:Di-PionVar2}(e) for $\Upsilon(2S)$ and Figure~\ref{fig:Di-PionVar2}(f) for $\Upsilon(3S)$.

\item {\bf VDist:} The transverse position of the di-pion vertex, shown in Figure~\ref{fig:Di-PionVar3}(a) for $\Upsilon(2S)$ and Figure~\ref{fig:Di-PionVar3}(b) for $\Upsilon(3S)$. 

\item  {\bf RecoilMass:} the mass recoiling against the di-pion system, shown in Figure~\ref{fig:Di-PionVar3}(c) for $\Upsilon(2S)$ and Figure~\ref{fig:Di-PionVar3}(d) for $\Upsilon(3S)$.

\end{itemize}

\begin{figure}
\centering 
\includegraphics[width=3.0in]{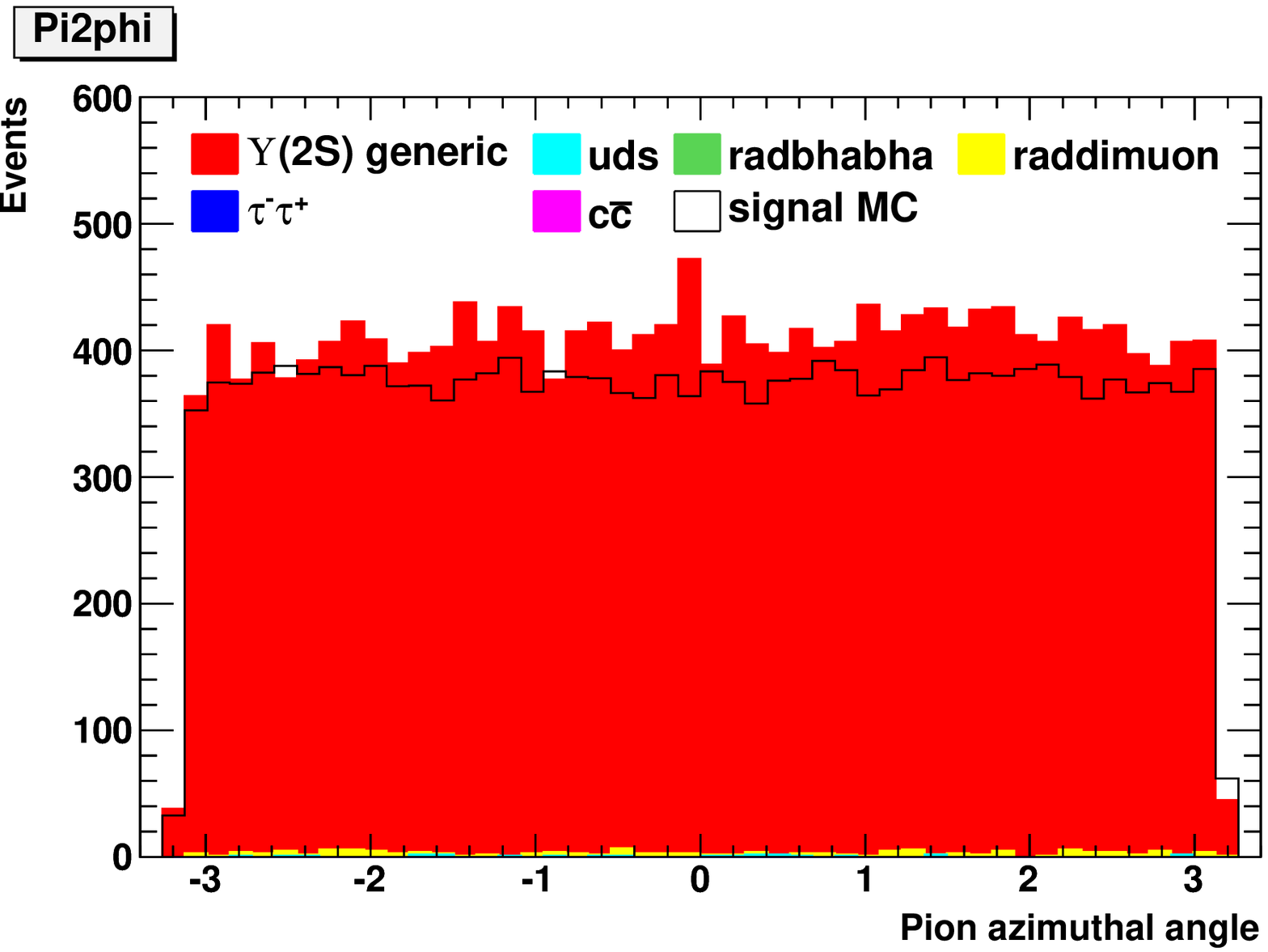}
\includegraphics[width=3.0in]{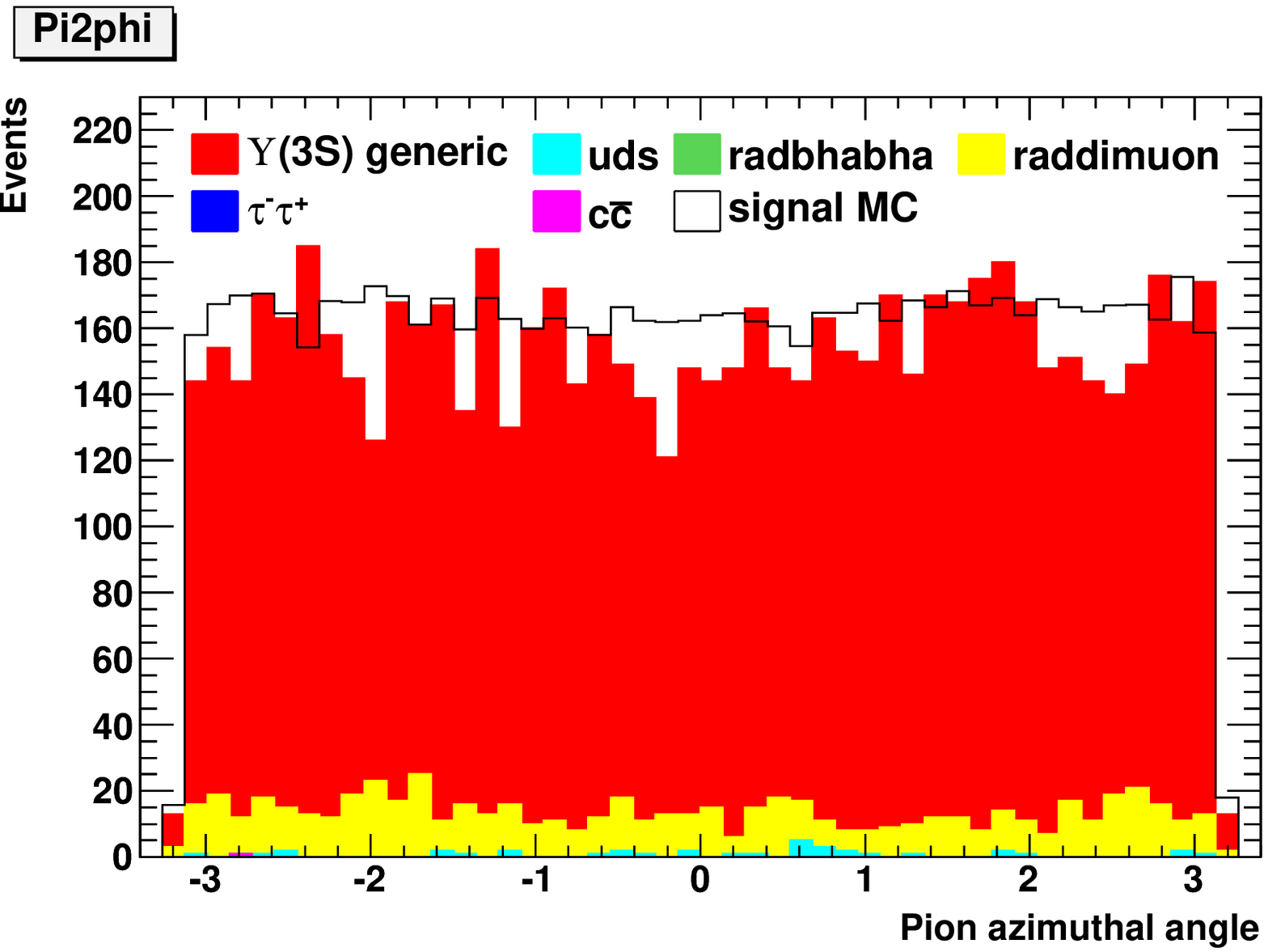}

\caption {Azimuthal angle of pion for signal, $\Upsilon(3S, 2S)$ generic, uds, radiative bhabha, radiative di-muon, $\tau^+\tau^-$ and $c\overline{c}$ events for $\Upsilon(2S)$ (left) and $\Upsilon(3S)$ (right). These variables are plotted after applying the pre-selection criteria as well as requiring that either one of the tracks of the $A^0$ reconstruction using the muon PID must be identified as muon.}
\label{fig:Pi2phi}
\end{figure}

The pion azimuthal angle in the radiative bhabha sample of both $\Upsilon(3S)$ and $\Upsilon(2S)$ datasets shows a multipeak structure, as shown in Figure~\ref{fig:Di-PionVar1}(e) and ~\ref{fig:Di-PionVar1}(f). The peak structure is understood to be due to the random tracks which is removed after requiring that either one of the charged tracks must be identified as muon for $A^0$ reconstruction using  muon Particle-ID (PID). Figure~\ref{fig:Pi2phi}  shows the azimuthal angle of pion after applying the muon ID cut. 

\subsection{Muon selection variables}
\begin{itemize}
\item {\bf BDTMuon[1,2]IDFakeRate:} We require either one of the charged tracks for the $A^0$ reconstruction must be identified as muon by a standard Muon particle-ID algorithm, where the $\mu$-to-$\pi$ misidentification rate is about $3\%$. Figure~\ref{fig:BDTMuon1IDFakeRate} shows the muon PID Boolean distribution of the $\Upsilon(2S,3S)$ datasets.

\begin{figure}
\centering 
\includegraphics[width=3.0in]{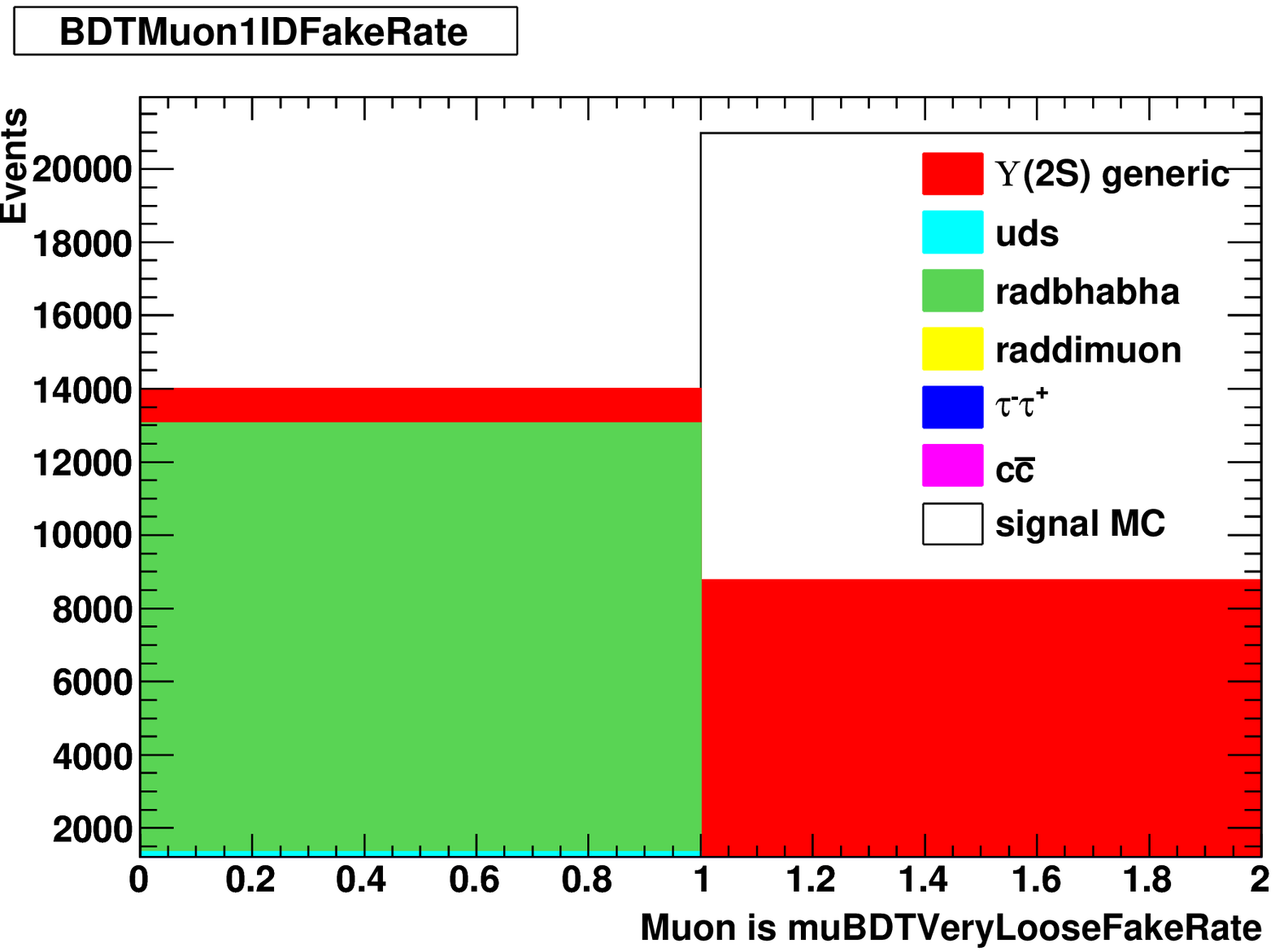}
\includegraphics[width=3.0in]{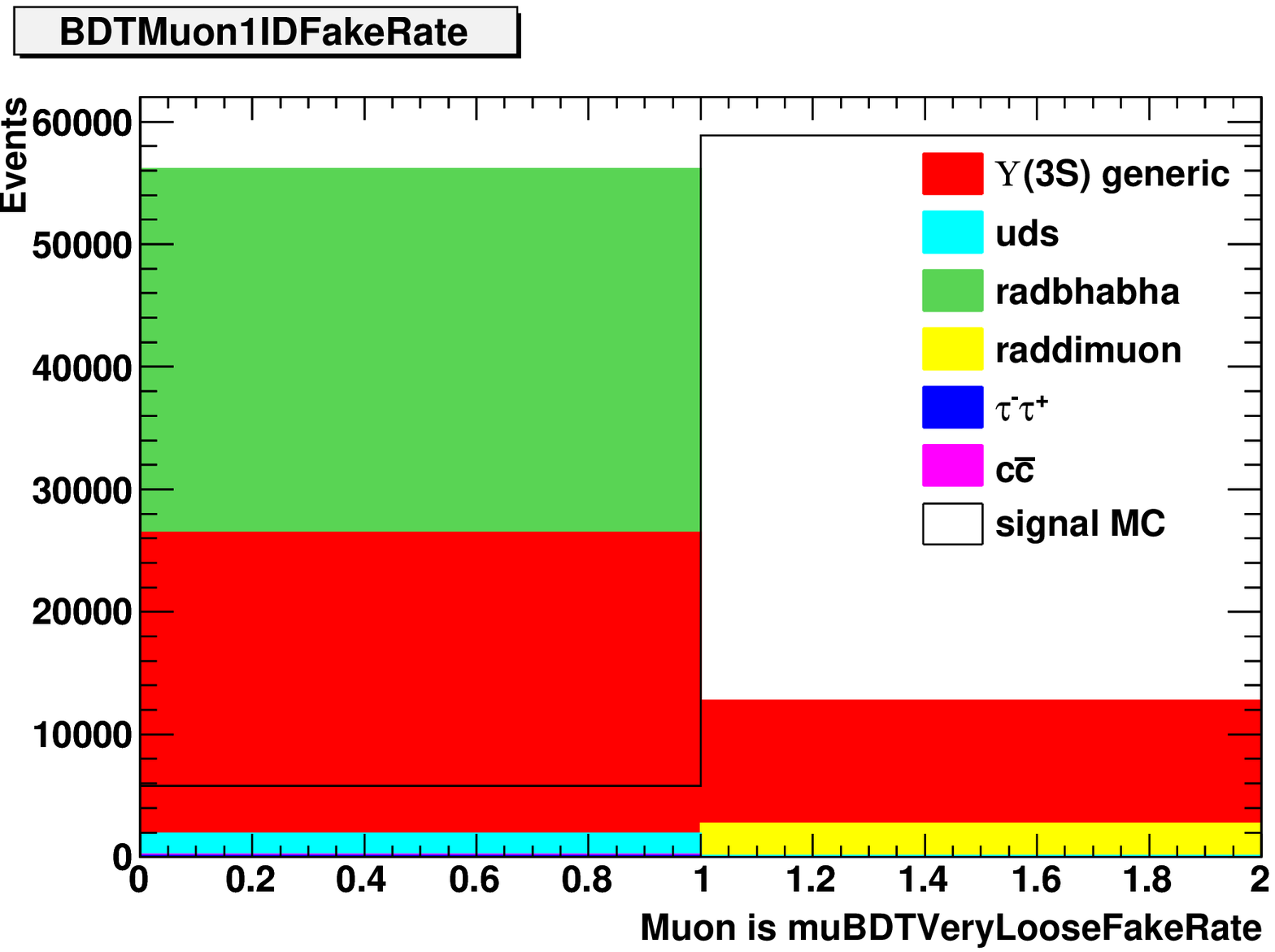}

\smallskip
\centerline{\hfill (a) \hfill \hfill (b) \hfill }
\smallskip

\includegraphics[width=3.0in]{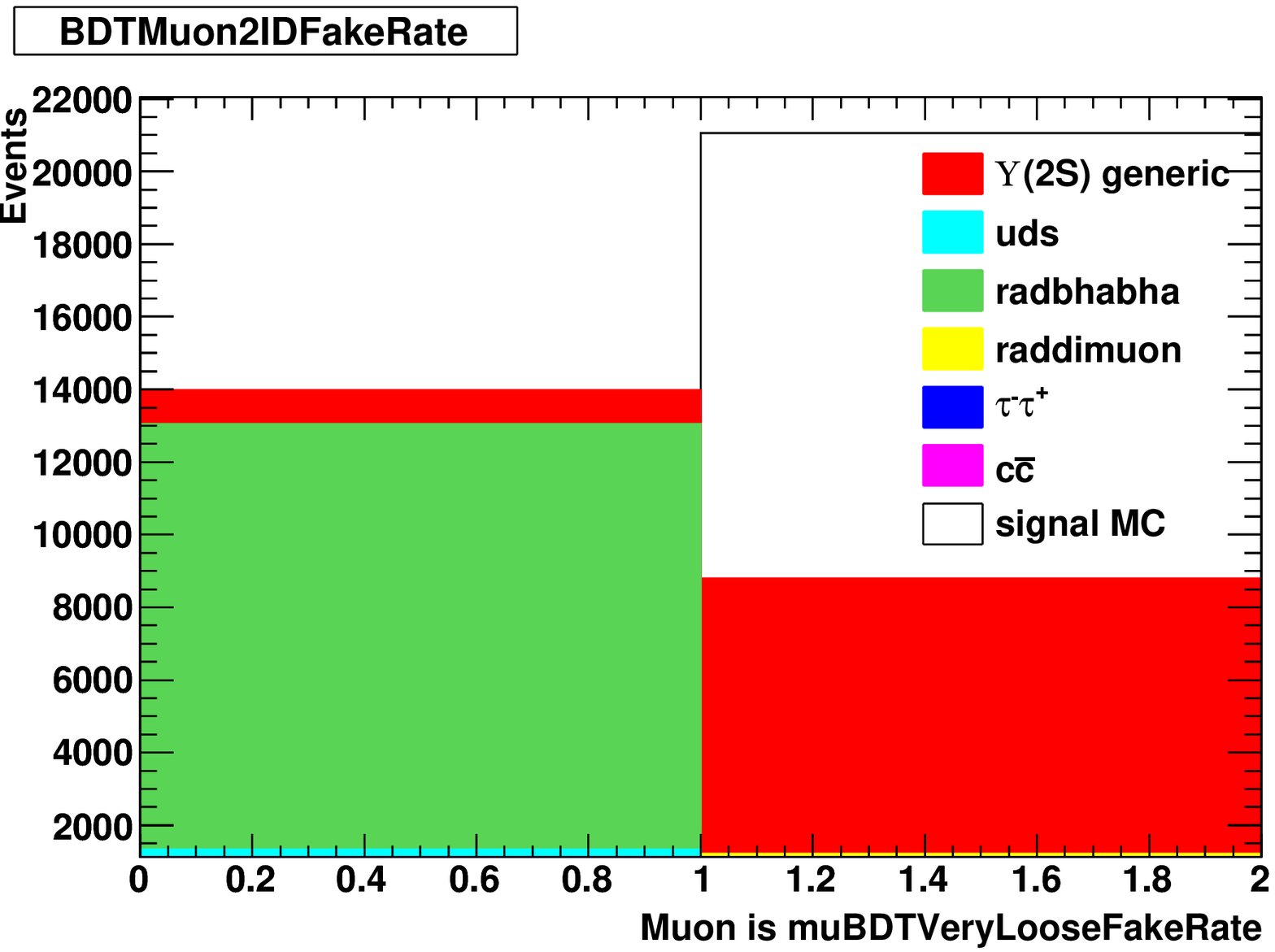}
\includegraphics[width=3.0in]{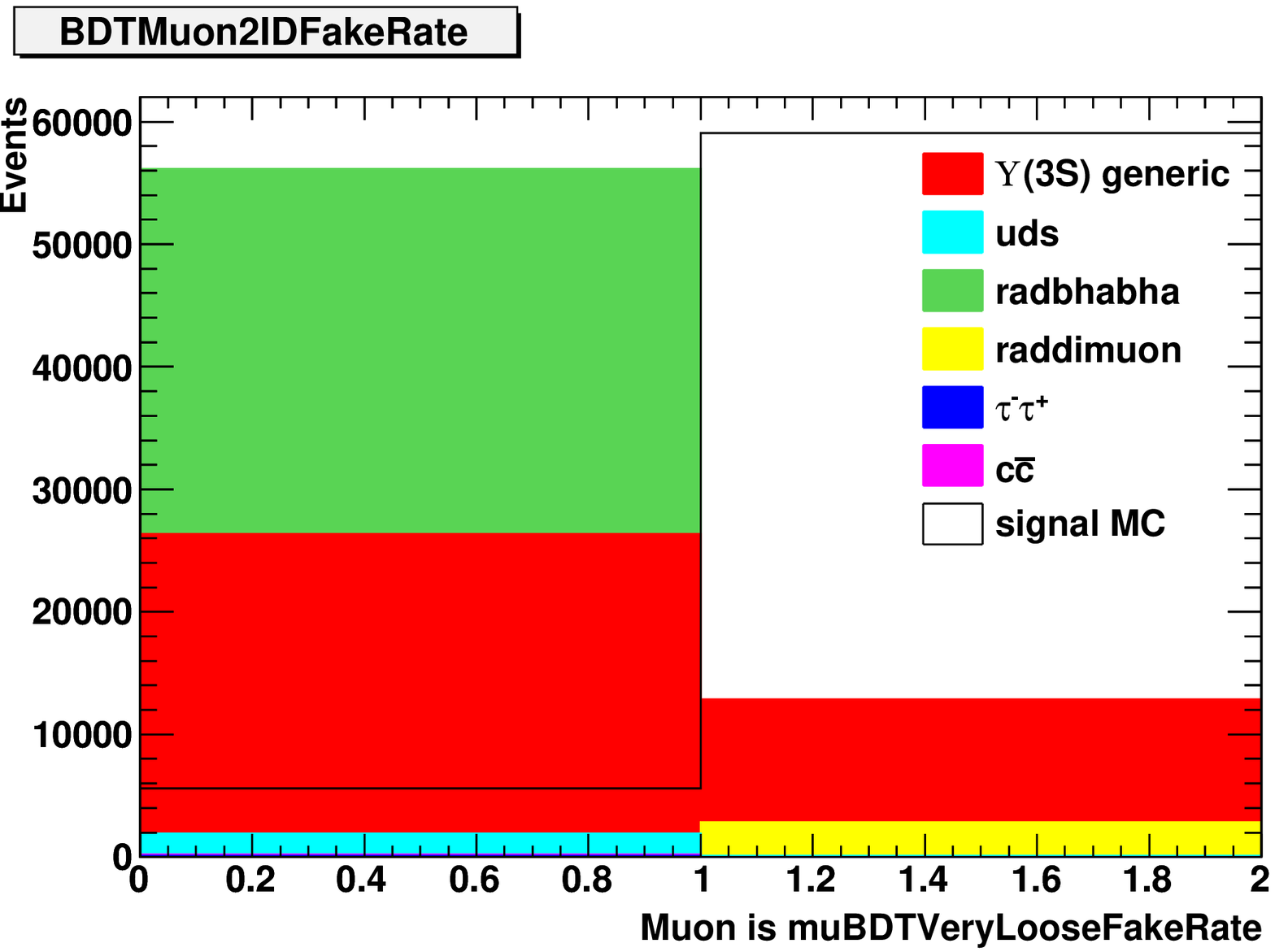}

\smallskip
\centerline{\hfill (c) \hfill \hfill (d) \hfill }
\smallskip

\caption {Muon particle-ID for $\Upsilon(3S, 2S)$ generic, uds, radiative bhabha, radiative di-muon, $\tau^+\tau^-$ and $c\overline{c}$  events for $\Upsilon(2S)$ (left) and $\Upsilon(3S)$ (right). We have plotted this variable at the pre-selection level.}

\label{fig:BDTMuon1IDFakeRate}
\end{figure}
\end{itemize}

\subsection{Track multiplicity and  photon selection variables}
\begin{itemize}
\item {\bf nTracks:}  We require that the number of charged tracks should be equal to four in the event. 

\begin{figure}[!htb]
\centering 
\includegraphics[width=3.0in]{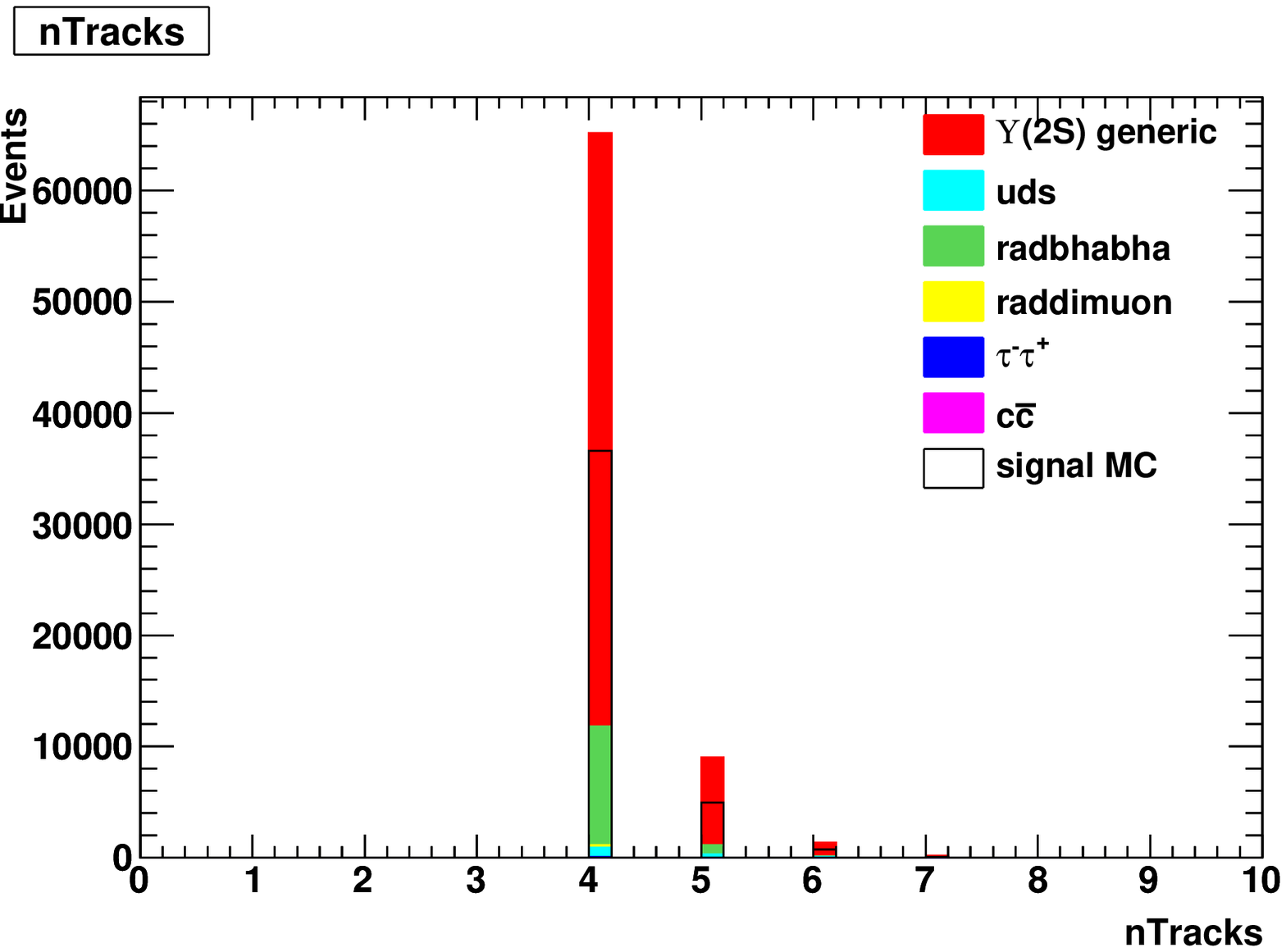}
\includegraphics[width=3.0in]{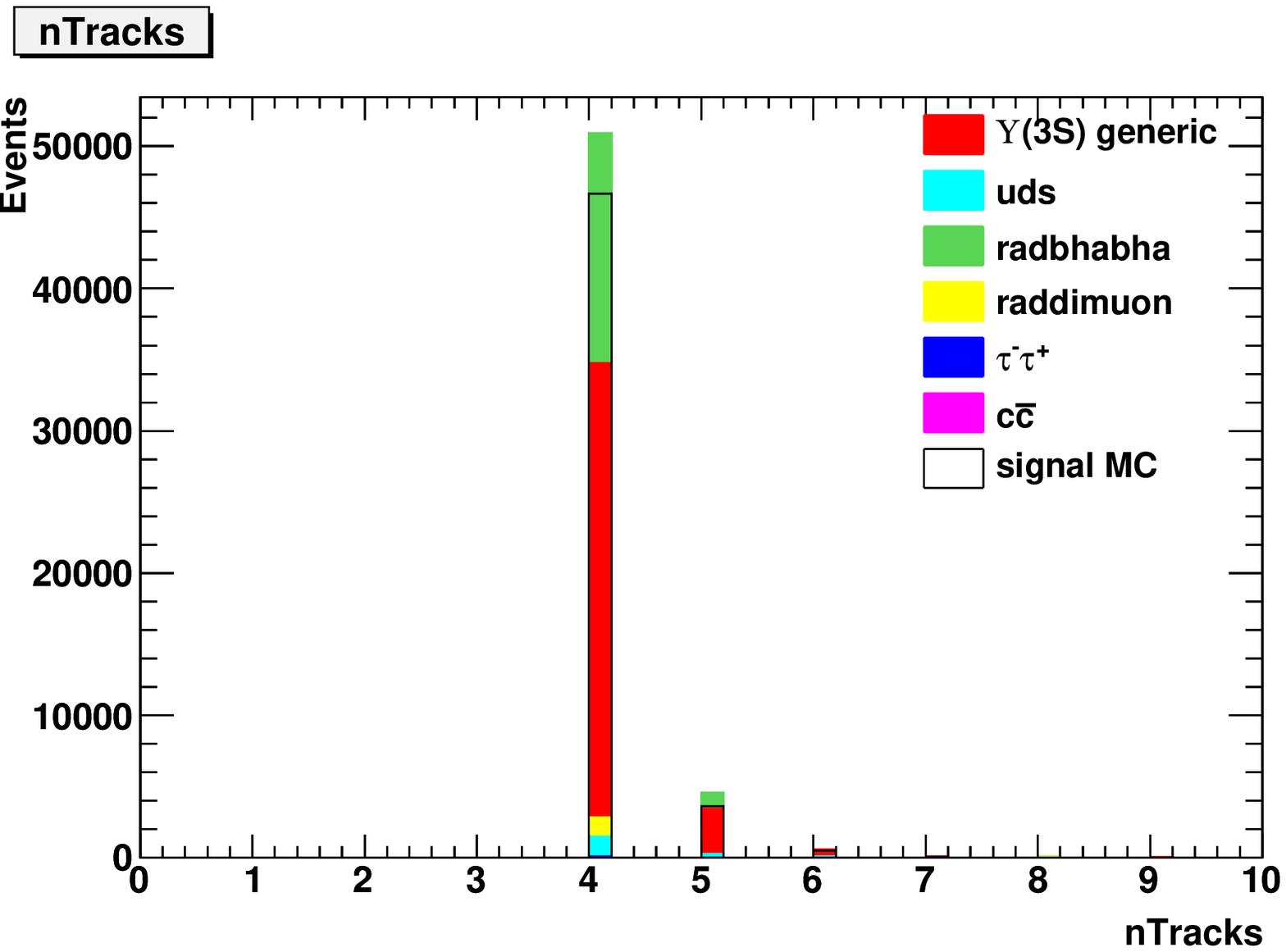}

\caption {Number of ChargedTracks in signal, $\Upsilon(3S, 2S)$ generic, uds, radiative bhabha, radiative di-muon, $\tau^+\tau^-$ and $c\overline{c}$ events for $\Upsilon(2S)$ (left) and $\Upsilon(3S)$ (right). We have plotted this variable t the pre-selection level.}

\label{fig:nTracks}
\end{figure}

\item {\bf xlmomgam:} The lateral moment  \cite{xlmomgam} of a photon candidate in the electromagnetic calorimeter  is defined as

\begin{equation}
  \mathrm{xlmomgam} = \frac{\sum_{i=3}^{N} E_i r_i^2}{\sum_{i=3}^N E_i r_i^2 +E_1r_0^2 +E_2r_0^2},
\end{equation}

\noindent where N is the number of crystals in the shower, $E_i$ is the energy deposited in the $i$th crystal, $r_i$ is the radius in the plane perpendicular to the line pointing from the interaction point to the shower center, and $r_0 = 5$ cm is the average distance between two crystals. The energies are ordered $E_1 > E_2 > ... > E_N$. The  xlmomgam quantity is used to differentiate the electromagnetic showers from the hadronic showers. The electromagnetic shower  typically deposits a large fraction of their energy  in one or two crystals, whereas the hadronic showers tend to be more spread out.  

\item {\bf Zmom42gam:} The Zernike $A_{42}$ moment  is defined as \cite{Zernike}:

\begin{equation}
A_{nm} = \sum_{k=1}^n(E_i/E)\cdot f_{nm}(\rho_i)e^{-im\phi_i},
\end{equation}
 
\noindent where $E_i$ is the energy deposited in the  $i^{th}$ crystal, E is the total energy deposited in the total crystals, $f_{nm}$ are the polynomials of degree n  and ($\rho_i$,$\phi_i$)  the location of the hit crystals in the EMC with respect to the center of the shower. The locations are defined in cylindrical coordinates with  z-axis running from the beam spot to the centroid, with $\rho_i = r_i/R_0$ where $R_0 = 15$ cm. $f_{nm}$ represents the Zernike function, 
\begin{equation}
f_{nm}(\rho) = \sum_{s =0}^{(n-m)/2} \frac{(-1)^s(n-s)!\rho^n-2s}{s!((n+m)/2-s)!((n-m)/2-s)!}
\end{equation}

\noindent with $m \le n$ and $(n-m)$ even. The Zmom42gam is used to characterize the azimuthal spread of the shower. It is also used to distinguish between electromagnetic  showers and hadronic showers, because hadronic showers tend to be more irregular than electromagnetic shower.

\begin{figure}
\centering 
\includegraphics[width=3.0in]{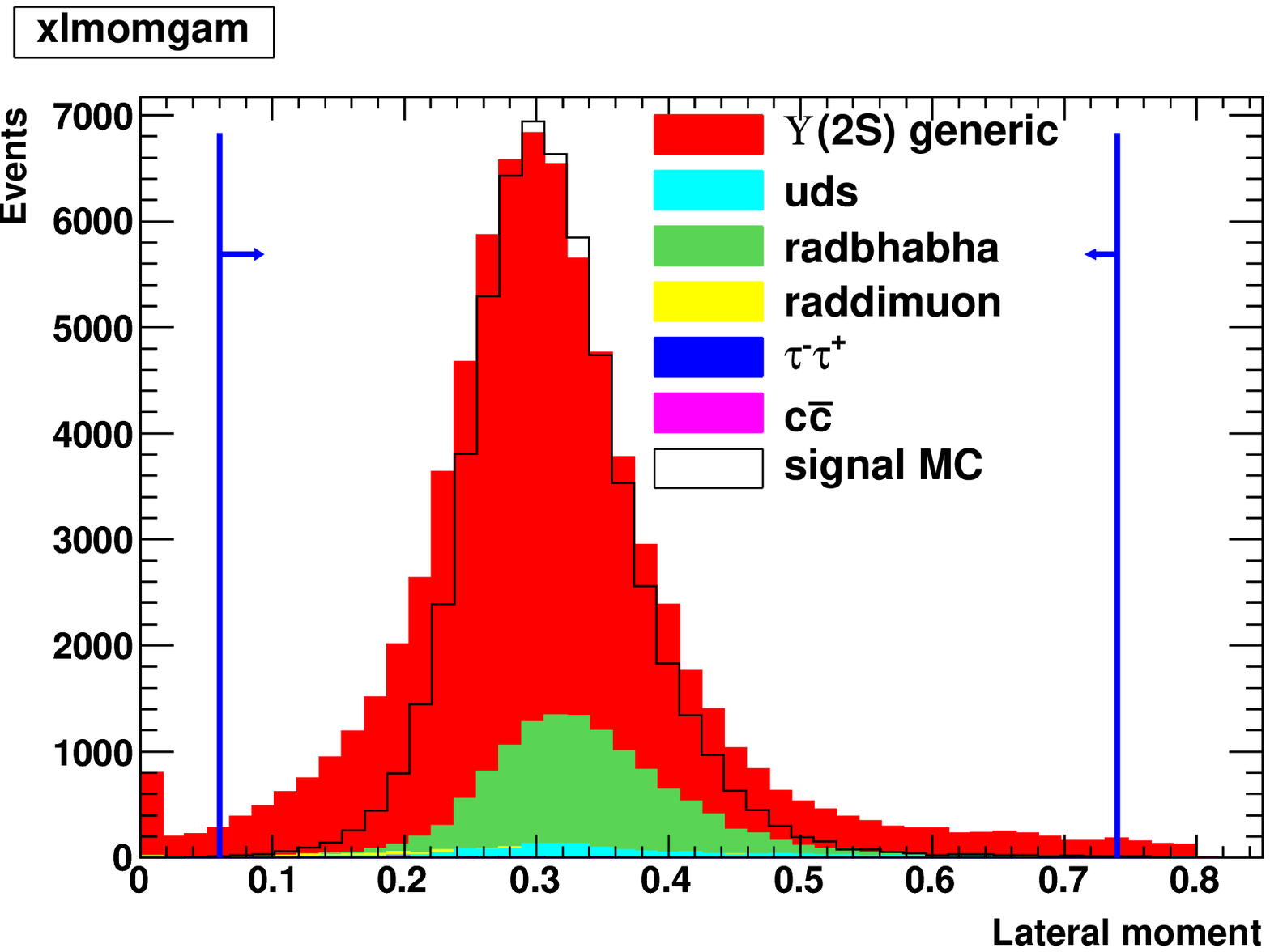}
\includegraphics[width=3.0in]{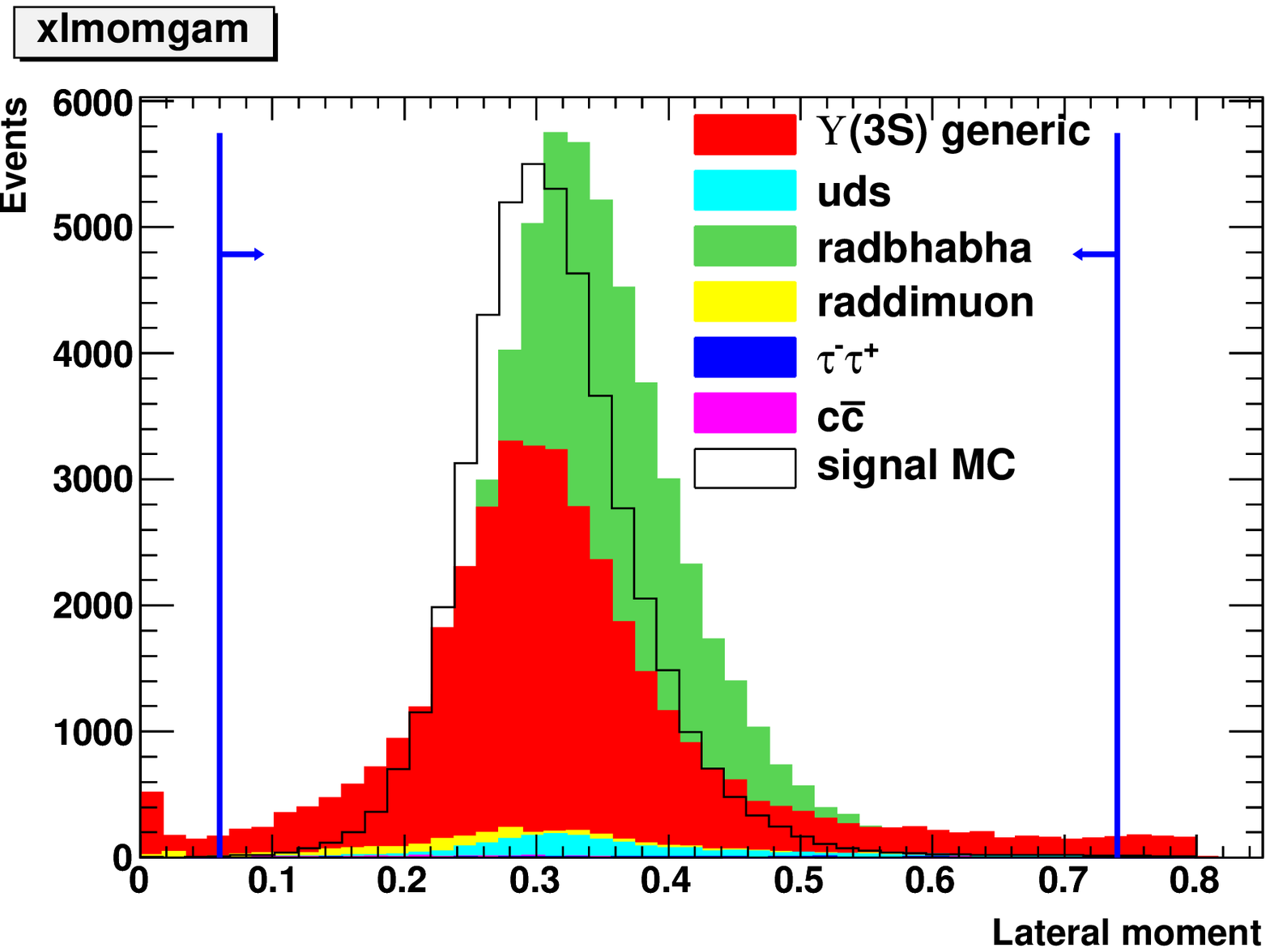}

\caption {Lateral moment associated with photon from signal, $\Upsilon(3S, 2S)$ generic, uds, radiative bhabha, radiative di-muon, $\tau^+\tau^-$ and $c\overline{c}$ events for $\Upsilon(2S)$ (left) and $\Upsilon(3S)$ (right). We have plotted this variable at the pre-selection level.}

\label{fig:xlmomgam}
\end{figure}

\begin{figure}
\centering 
\includegraphics[width=3.0in]{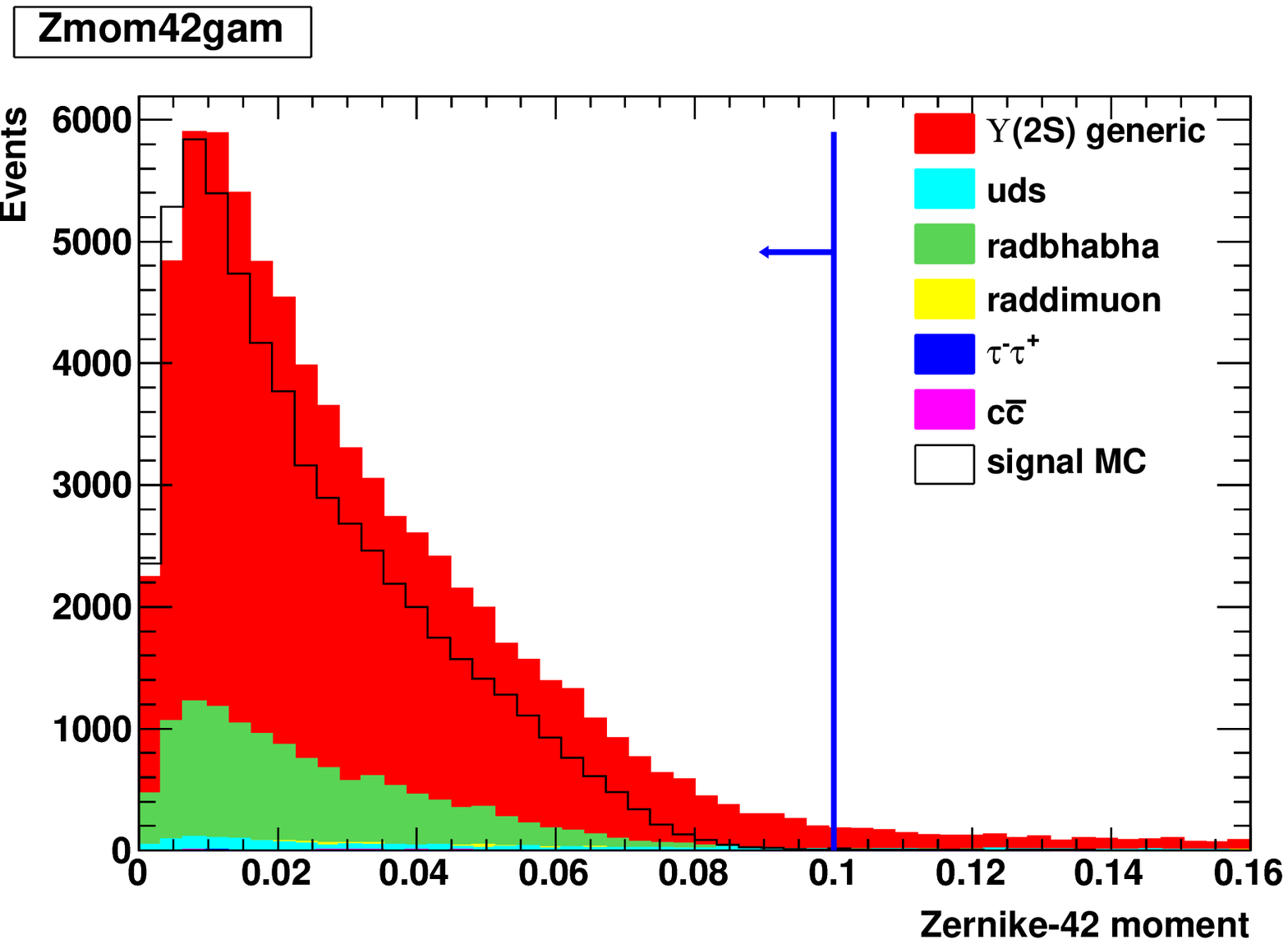}
\includegraphics[width=3.0in]{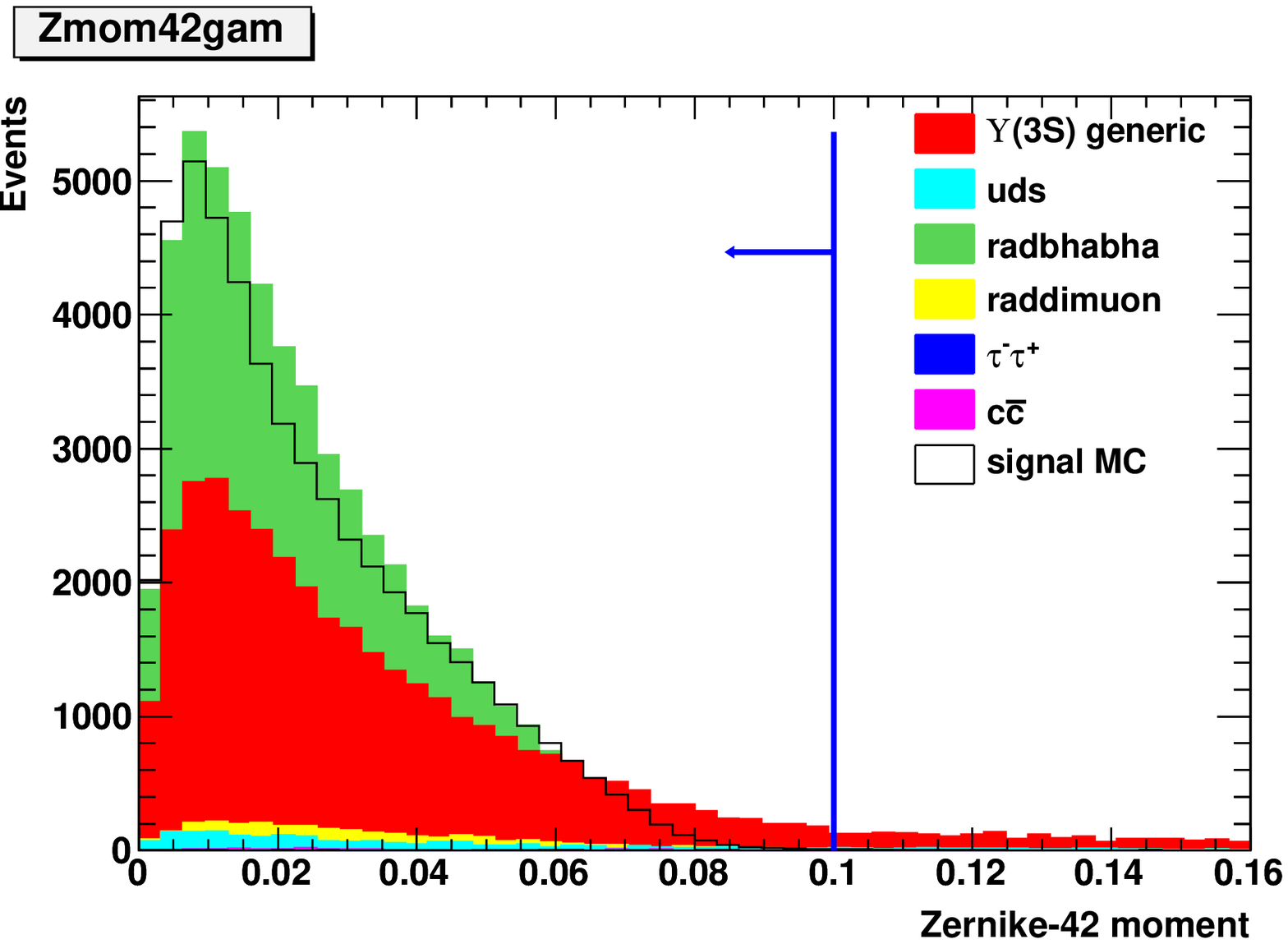}

\caption {Zernike-42 moment associated with photon from signal, $\Upsilon(3S, 2S)$ generic, uds, radiative bhabha, radiative di-muon, $\tau^+\tau^-$ and $c\overline{c}$ events for $\Upsilon(2S)$ (left) and $\Upsilon(3S)$ (right). We have plotted this variable at the pre-selection level.}

\label{fig:Zmom42gam}
\end{figure}

\end{itemize}

  We apply a loose selection cuts for the muon, track multiplicity and photon related variables. The selection criteria for  the muon, track multiplicity and photon related variables for $\Upsilon(2S)$ and $\Upsilon(3S)$ datasets are summarized in Table~\ref{table:track-photon}. Figure ~\ref{fig:BDTMuon1IDFakeRate}, ~\ref{fig:nTracks}, ~\ref{fig:xlmomgam} and ~\ref{fig:Zmom42gam} show the distributions of these variables.

\begin{table}  
\begin{center}
\begin{tabular}{|l|r|r|}
	\hline
&\multicolumn{2}{c|} {Selection} \\
\cline{2-3}
 \multicolumn{1}{|c|}{Variable name}  &  $\Upsilon(3S)$    & $\Upsilon(2S)$  \\
\hline

\hline
 Number of tracks &  = 4              & = 4    \\
\hline
 Lateral moment & [0.06, 0.75]      & [0.06, 0.75] \\
\hline
 Zernike-42 moment & $<$ 0.1      & $<$ 0.1   \\
\hline
Muon-ID & OR muon PID & OR muon PID \\
\hline
\end{tabular} 
\caption{Track multiplicity, photon and muon related selection variables.}
\label{table:track-photon}
\end{center}
\end{table}

\begin{figure}
\centering 
\includegraphics[width=3.0in]{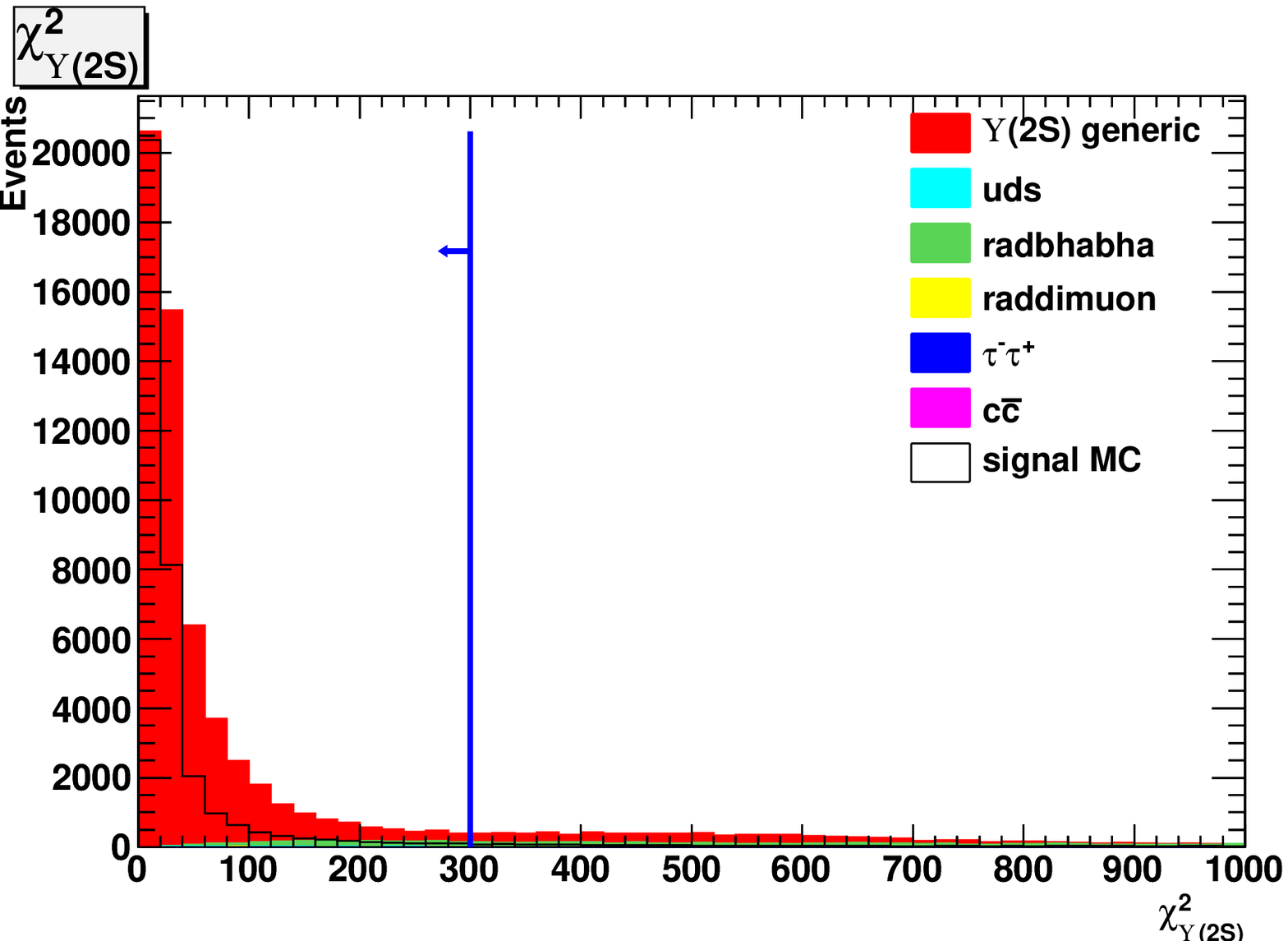}
\includegraphics[width=3.0in]{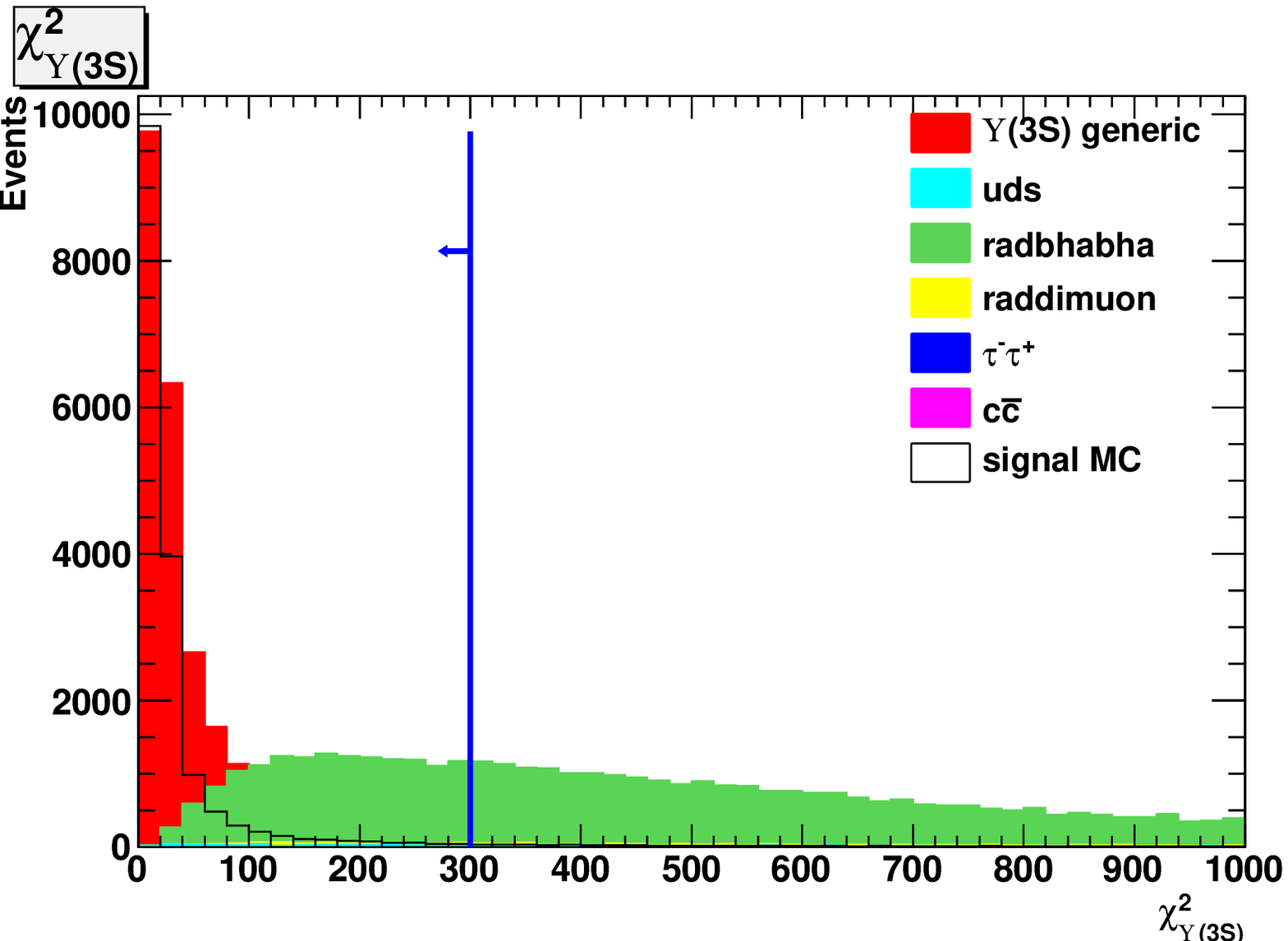}

\caption {The $\Upsilon(2S,3S)$ kinematic fit $\chi^2$ for signal, $\Upsilon(3S, 2S)$ generic, uds, radiative bhabha, radiative di-muon, $\tau^+\tau^-$  and $c\overline{c}$  events for $\Upsilon(2S)$ (left) and $\Upsilon(3S)$ (right). We have plotted this variable at the pre-selection level.}

\label{fig:Ynschi2}
\end{figure}

We also apply a selection cut on the $\Upsilon(3S, 2S)$ kinematic fit $\chi^2$ ($\chi_{\Upsilon(3S, 2S)}^2 < 300$), which is calculated after fitting the entire decay chain using the CM beam energy constraints on the $\Upsilon(3S, 2S)$ and mass constraints on the $\Upsilon(3S, 2S)$ and $\Upsilon(1S)$. Figure~\ref{fig:Ynschi2} shows the distribution of $\Upsilon(3S, 2S)$ kinematic fit $\chi^2$ variable.

\subsection{Multivariate Analysis}

\begin{figure}
\centering 
\includegraphics[width=3.0in]{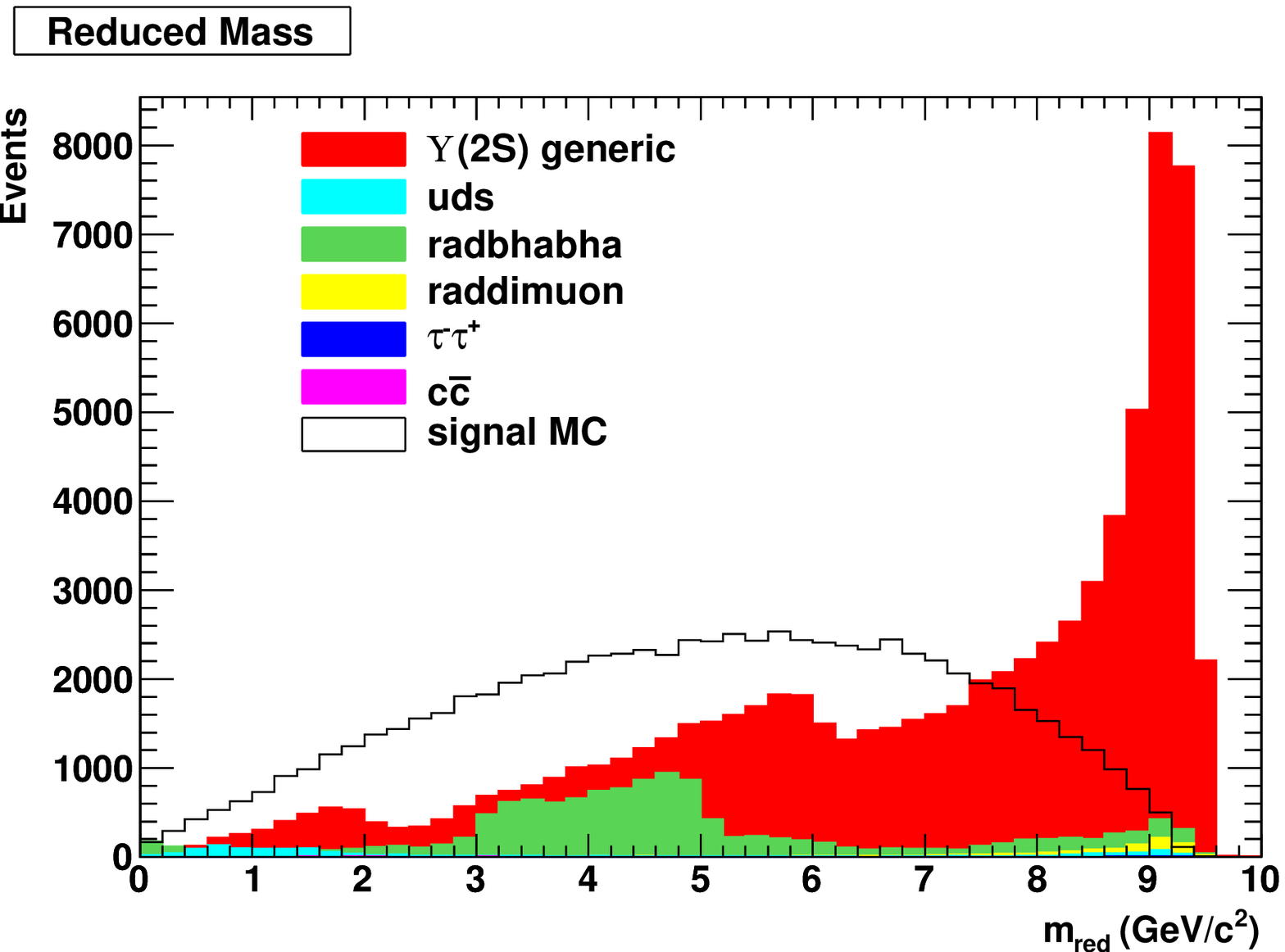}
\includegraphics[width=3.0in]{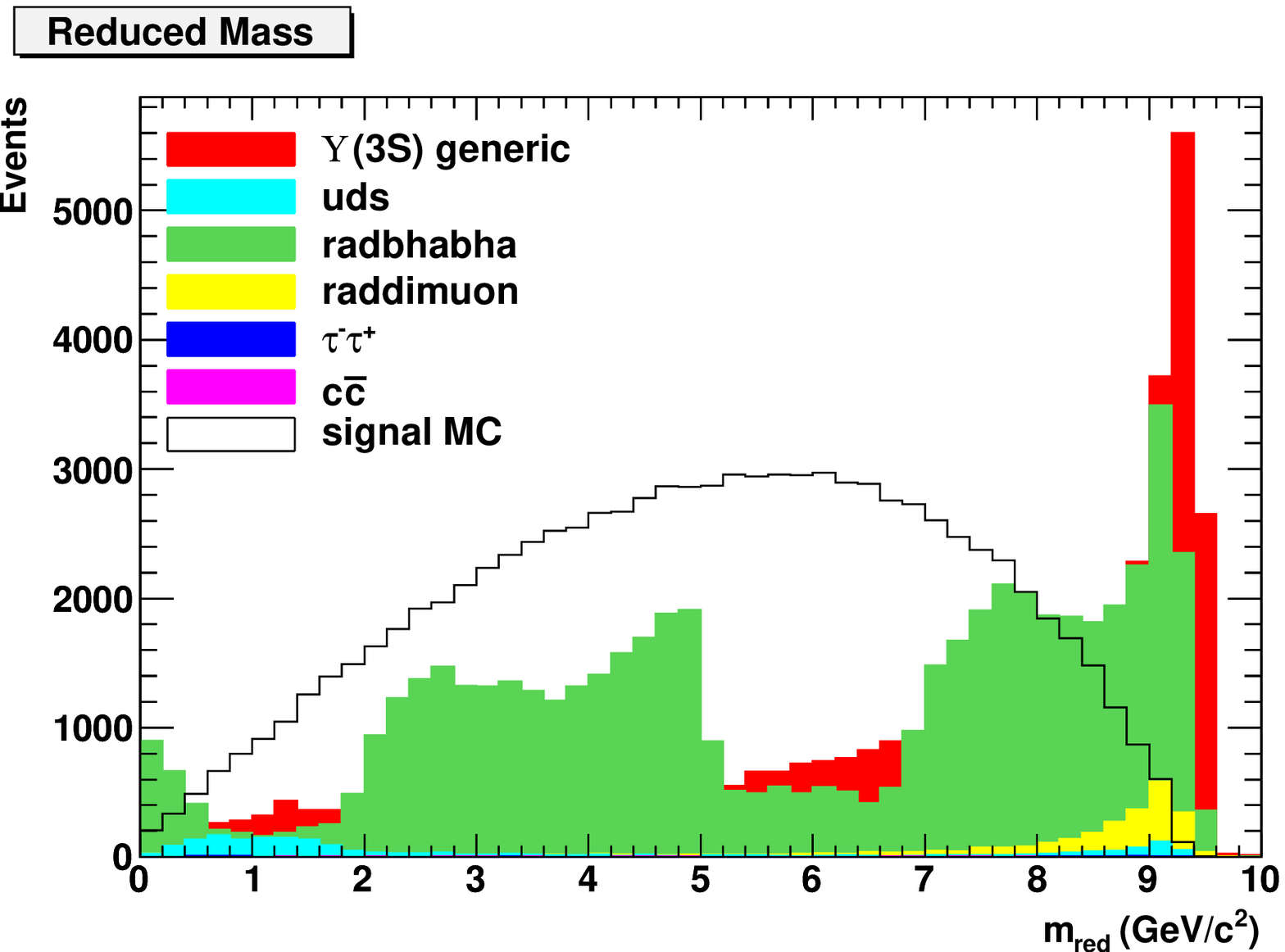}

\caption {Reduced mass distribution for signal, $\Upsilon(3S, 2S)$ generic, uds, radiative bhabha, radiative di-muon, $\tau^+\tau^-$  and $c\overline{c}$  events  for $\Upsilon(2S)$ (left) and $\Upsilon(3S)$ (right). We have plotted this variable at the pre-selection level.}

\label{fig:ReducedMass}
\end{figure}

   We use multivariate analysis (MVA) based BumpHunter algorithm and Random forest algorithm included in StatPatternRecognition \cite{SPR} to optimize pions related variables. The full $m_{\rm red}$ range is used to optimize the pion related variables for both the datasets which are shown in Figure~\ref{fig:ReducedMass}. 

\begin{figure}
\centering
\includegraphics[width=6.5in]{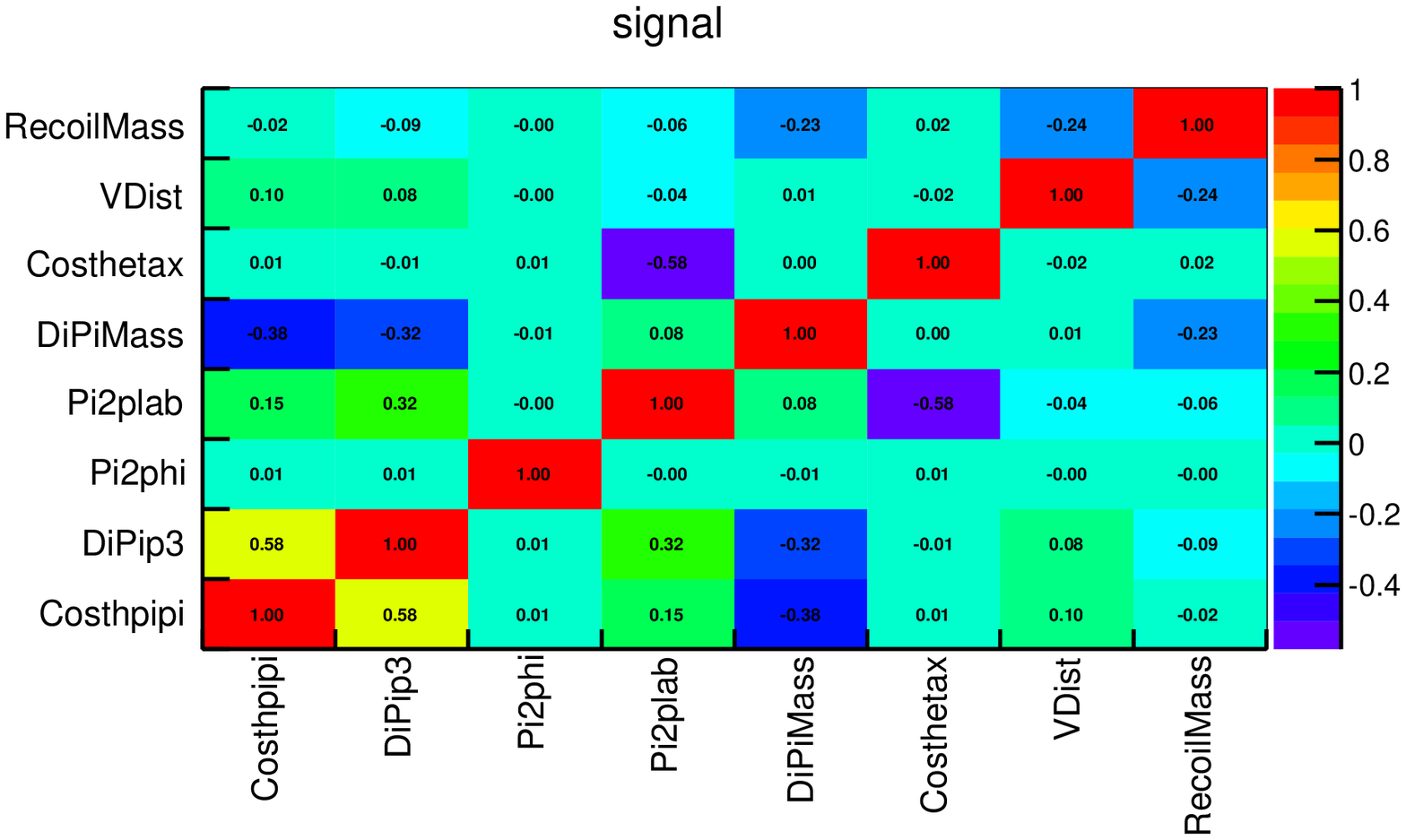}
\includegraphics[width=6.5in]{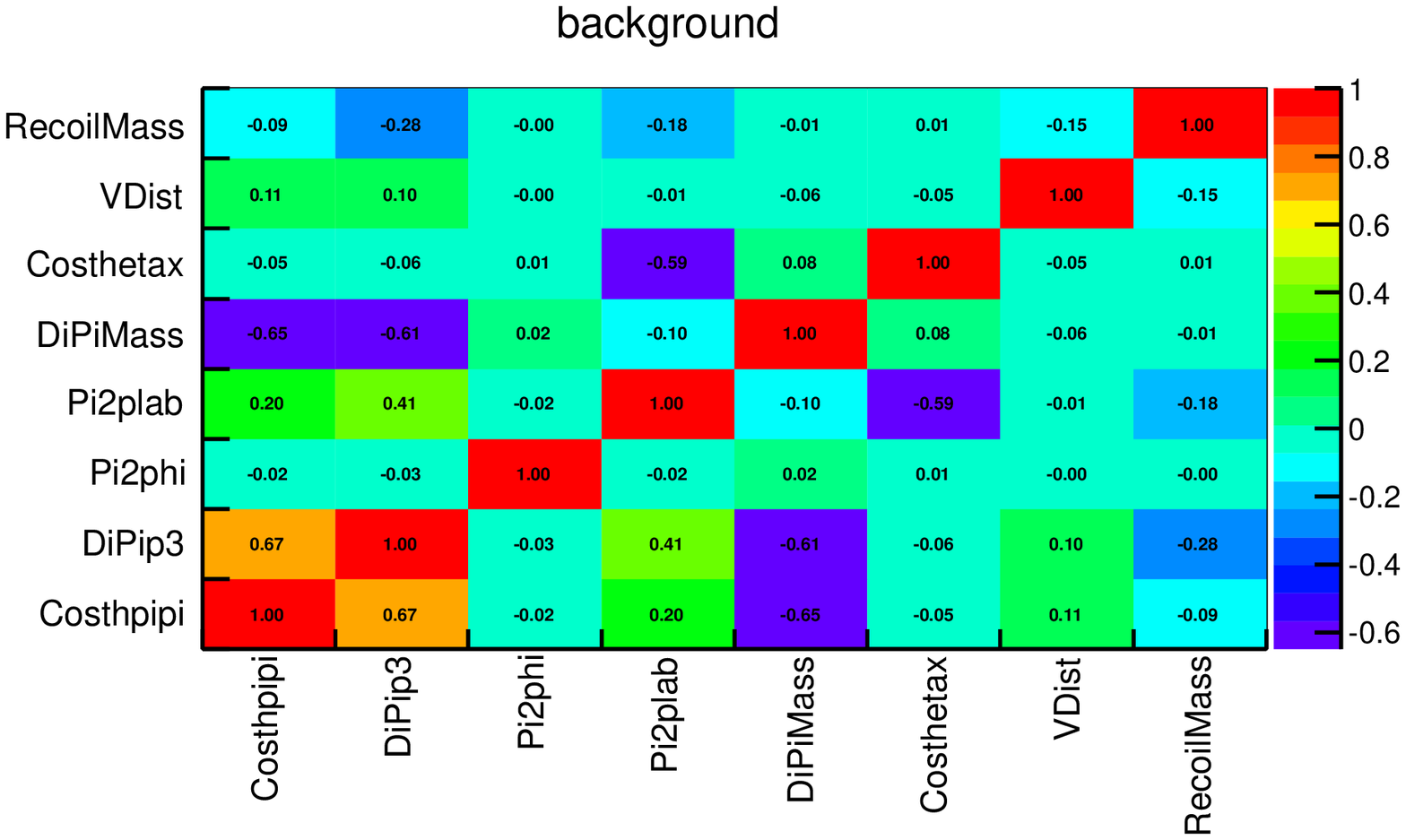}

\caption{Correlation between MVA input variables in $\Upsilon(2S)$ for signal and Background MCs.}
\label{fig:corrY2S} 
\end{figure}

\begin{figure}
\centering
\includegraphics[width=6.5in]{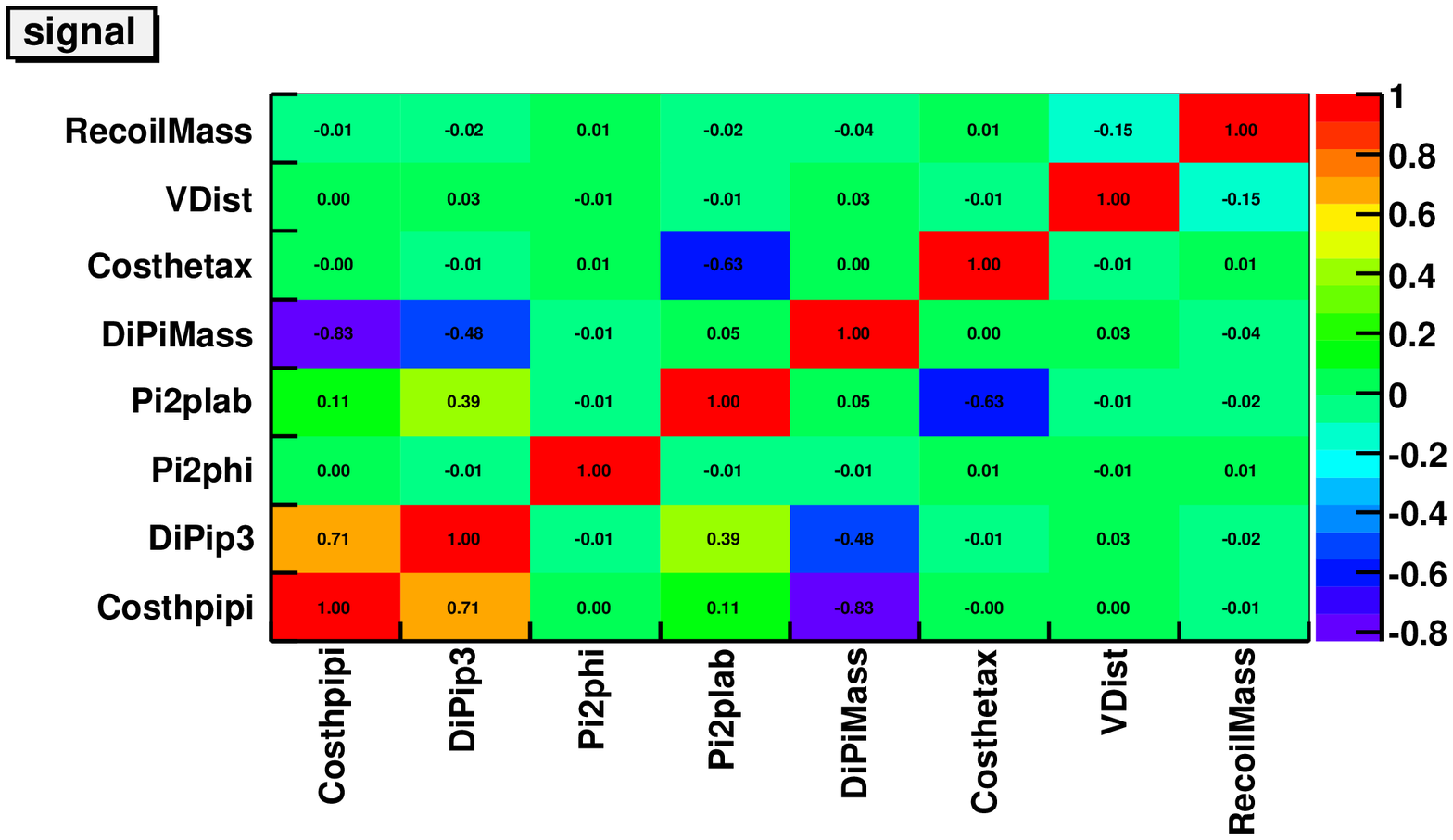}
\includegraphics[width=6.5in]{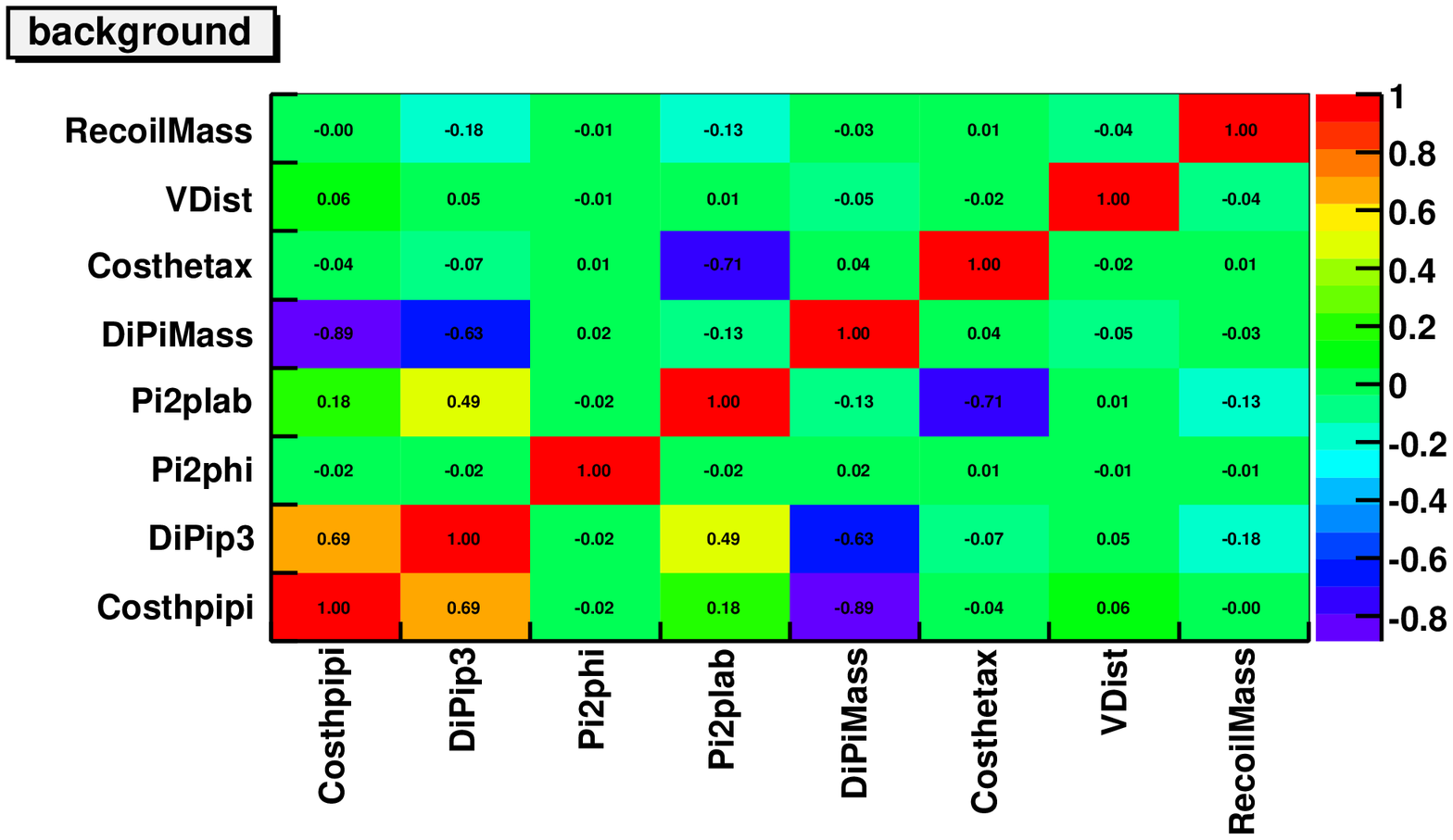}

\caption{Correlation between MVA input variables in $\Upsilon(3S)$ for signal and Background MCs.}
\label{fig:corrY3S} 
\end{figure}

We split the data sample into 3 sub-samples, one for training set, one for validation set, and one for test set. The training and validation samples are used to train the MVAs. The test sample is used to check the performance of the MVAs after applying the selection criteria. Figure~\ref{fig:corrY2S} and ~\ref{fig:corrY3S} show the the correlation between the input variables for signal and background in $\Upsilon(2S)$ and $\Upsilon(3S)$ datasets, respectively, which are used to train the BumpHunter and Random forest classifiers.

\subsubsection{Variable selection optimization using BumpHunter classifier}
            The BumpHunter classifier is based on PRIM algorithm \cite{Prim}.  This classifier  searches for a series of selection criteria that define an n-dimensional cuboid in n-dimensional variable space. Once a suitable region is found, the selection criteria are adjusted to optimize the figure of merit (FOM), such that the proportion of the number of events excluded by this adjustment does not exceed a fixed amount. This amount is known as the \rm{\lq\lq peel\rq\rq}  parameter. The process is repeated until a cuboid is found which maximizes the FOM. In this analysis, we use  Punzi FOM \cite{Punzi} for optimization which is defined as:

\begin{equation}
   \frac{\epsilon}{0.5N_{\sigma} +\sqrt {B}},
 \end{equation}

\noindent where $N_{\sigma}$ is the number of standard deviations desired from the result, and $\epsilon$ and $B$ are the average efficiency and background yield over a broad $m_{A^0}$ range, respectively.  

  To train the BumpHunter classifier, we weigh the background MC by Run7 $\Upsilon(3S)$ onpeak luminosity (28.049 $fb^{-1}$) and weigh the signal MC by determining the number of expected signal event in our data while assuming a branching ratio of $10^{-6}$. We train the BumpHunter MVA using training and validation sample of $\Upsilon(3S)$ to optimize the selection cuts. The peel parameter is varied between $1\%$ and $95\%$. The optimal peel parameter (maximizing the FOM) is found to be $20\%$. The cuts determined by the algorithm are shown in Table~\ref{table:BumpHunter}.     

\begin{table}

\begin{center}
\begin{tabular}{|c|r|}

\hline
Variable name  &Selection \\
\hline
 Cosine of angle between two pions & $>$ -0.999 \\
\hline
 Di-Pion transverse momentum & $<$ 1.239 GeV/c \\
\hline
 Pion transverse momentum & [0.070, 1.021] GeV/c \\
\hline
Pion helicity angle   & [ -0.966, 0.947] \\
\hline
Di-Pion mass &  [0.293, 0.894]GeV/$c^2$ \\
\hline
Transverse position of di-pion vertex & [$5.50\times 10^{-6}$, 0.041] \\
\hline
RecoilMass & [9.451, 9.470] GeV/$c^2$ \\
\hline
\end{tabular}
\caption{Optimal set of cuts obtained from the BumpHunter with a peel parameter of 0.2 for $\Upsilon(3S)$.}
\label{table:BumpHunter}
\end{center}
\end{table}

We then apply these optimal cuts to the test sample and check the performance. We find 23925 signal MC events and 10009 background MC events for $\Upsilon(3S)$. This will be our benchmark numbers for a more complex multivariate analysis.

\subsubsection{Variable selection optimization using Random Forest classifier}
   We use another advanced tool, the Random Forest (RF) classifier which was proposed by Breiman in 2001 \cite{RF}. RF is a method by which a number of decision tress are trained and the output of the algorithm is taken as the weighted vote of the output of each decision trees. Unlike the BumpHunter, a decision tree recursively splits training data into rectangular region (nodes). For each node, the tree examines all possible binary splits in each dimension and selects the one with the optimized FOM. In our case, the decision tree sets the weights for the vote, to maximize the Gini index (the FOM for this approach). StatPatternRecognition uses negative Gini-index (= $-2p.q$),  where $p$ and $q=1-p$ are fractions of correctly and incorrectly classified events in each node. The Gini index is related to the minimization of the loss of events from each category. Each training cycle grows a decision tree from a random set of input variables - thus the name, random forest.

\begin{figure}
\centering
\includegraphics[width=3.0in]{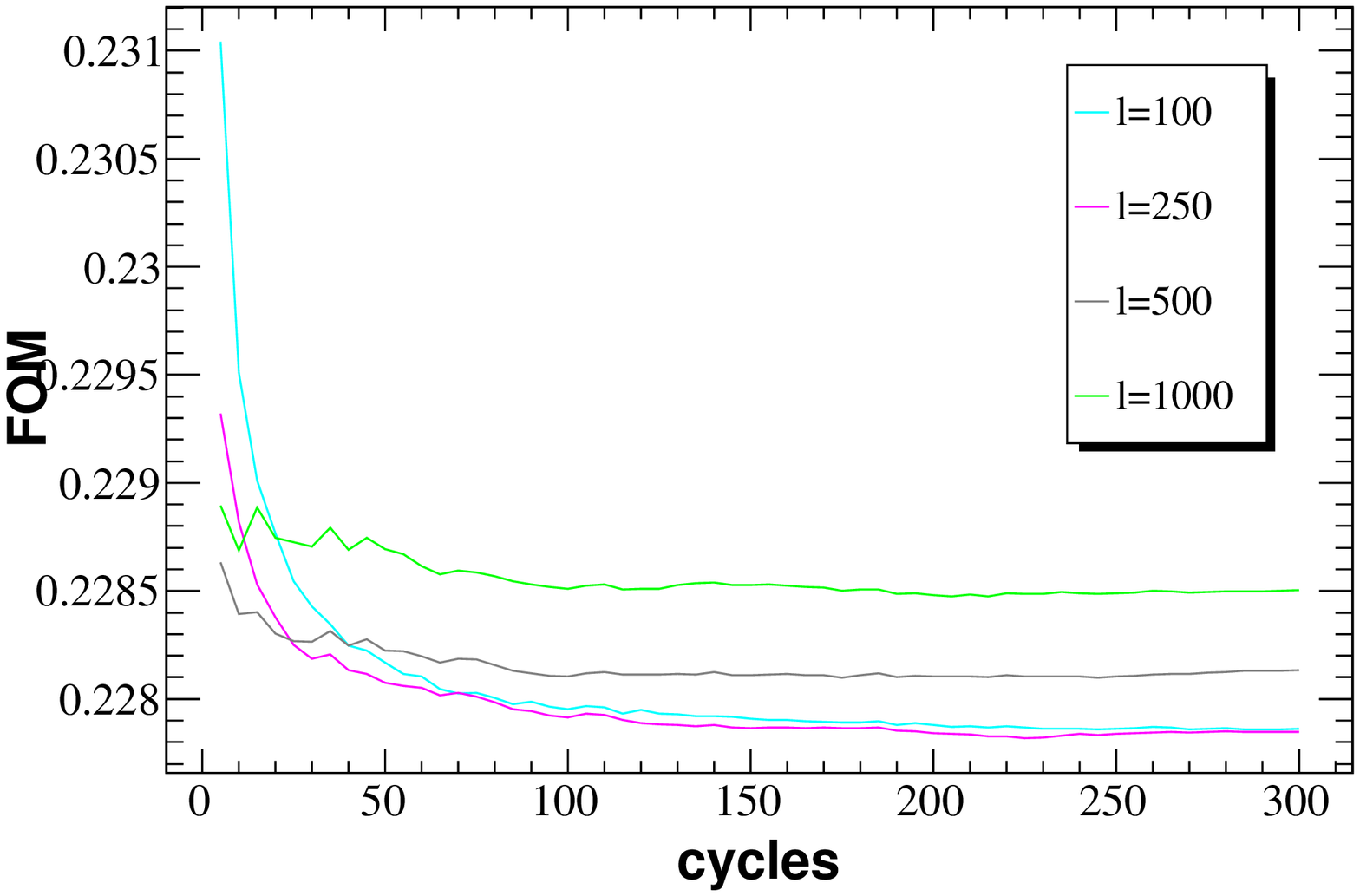}
\includegraphics[width=3.0in]{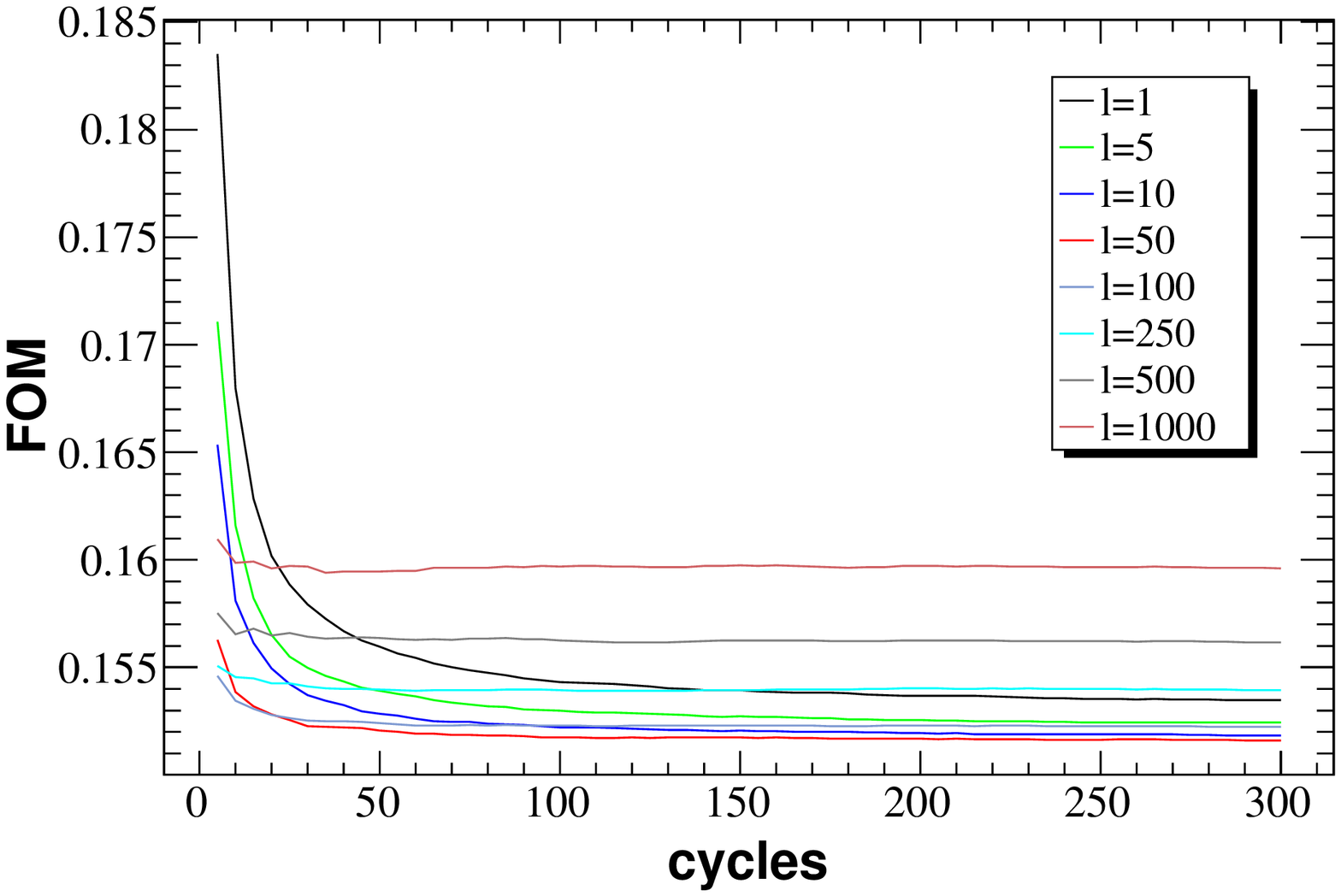}

\caption{Figure of merit (FOM) of the RF vs. the number of trees grown (training cycles), for a variety of minimal events allowed per terminal node {\bf l} for $\Upsilon(2S)$ (left) and  $\Upsilon(3S)$ (right). We find the lowest FOM: l=250 for $\Upsilon(2S)$ and l=50 for $\Upsilon(3S)$.}

\label{fig:FOM} 
\end{figure}

\begin{figure}
\centering
\includegraphics[width=3.0in]{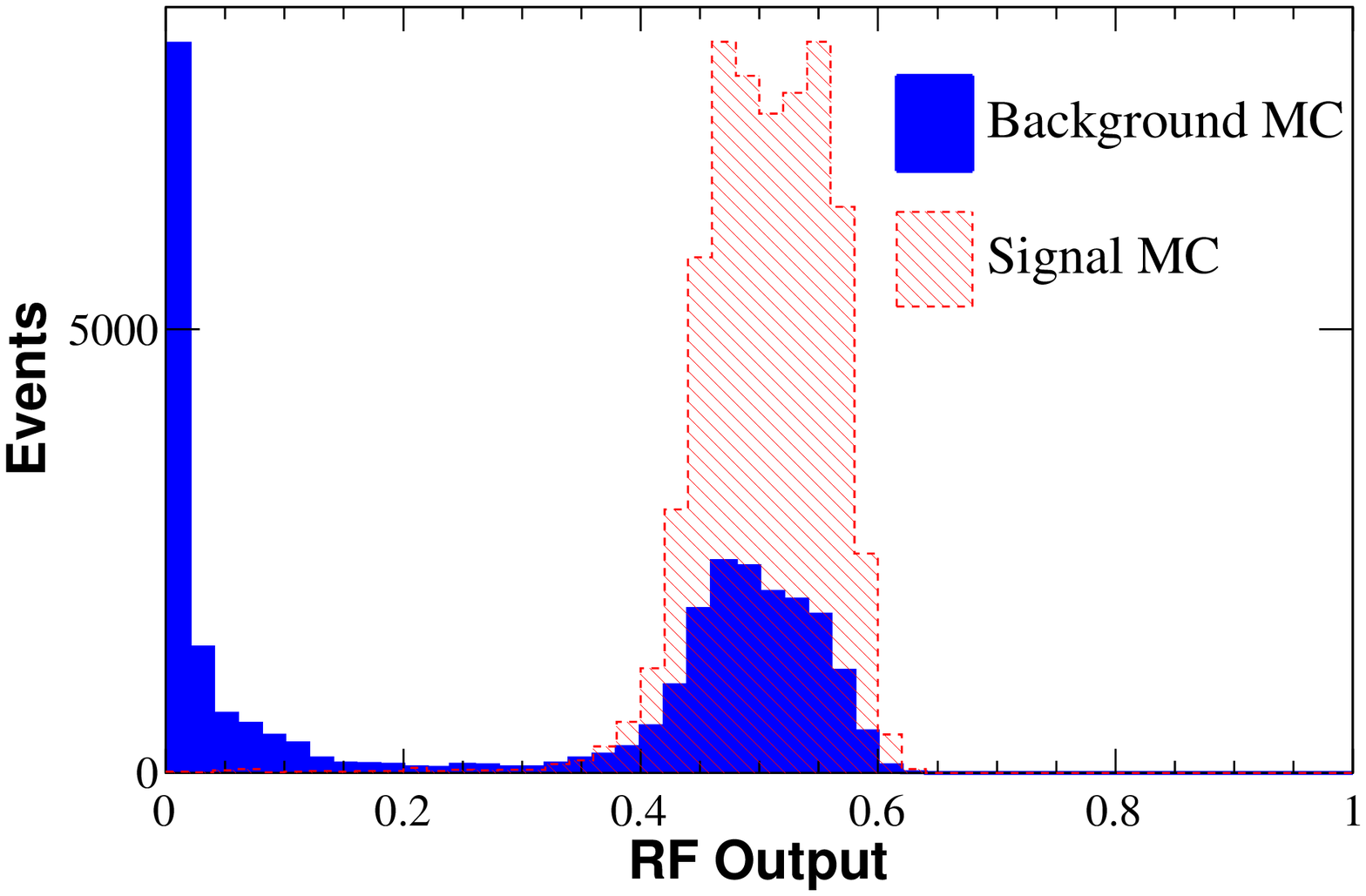}
\includegraphics[width=3.0in]{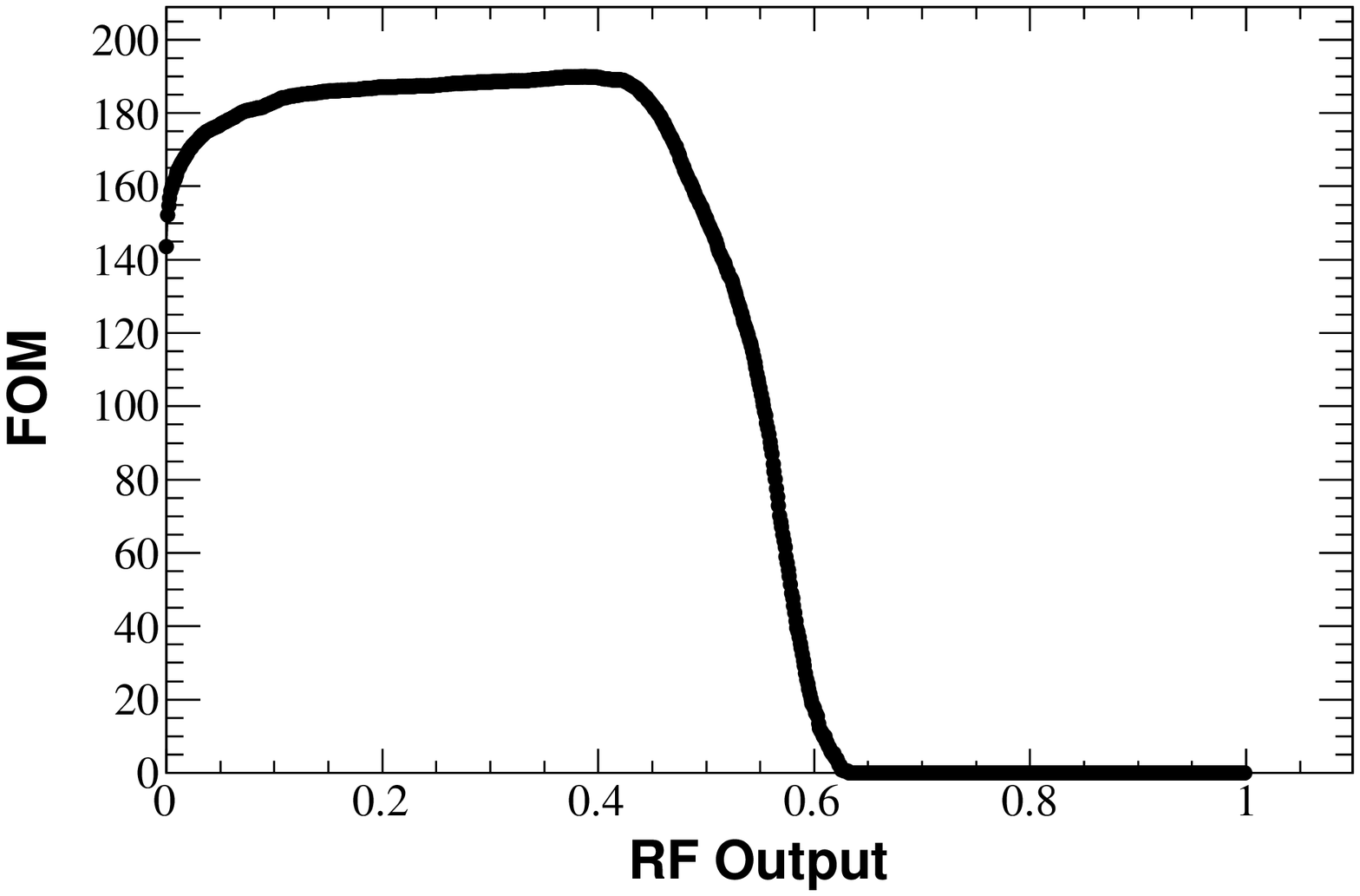}

\smallskip
\centerline{\hfill (a) \hfill \hfill (b) \hfill }
\smallskip

\includegraphics[width=3.0in]{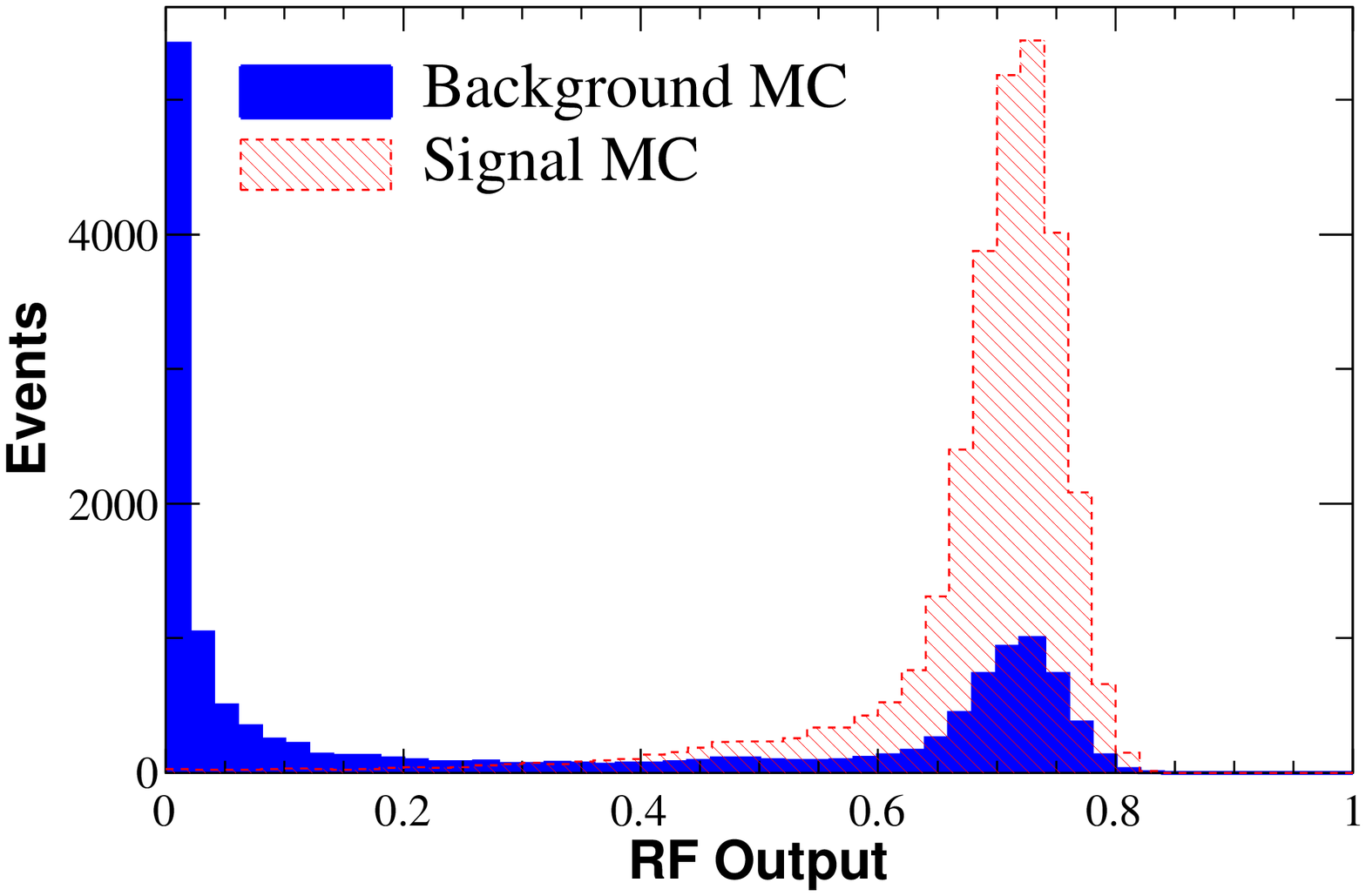}
\includegraphics[width=3.0in]{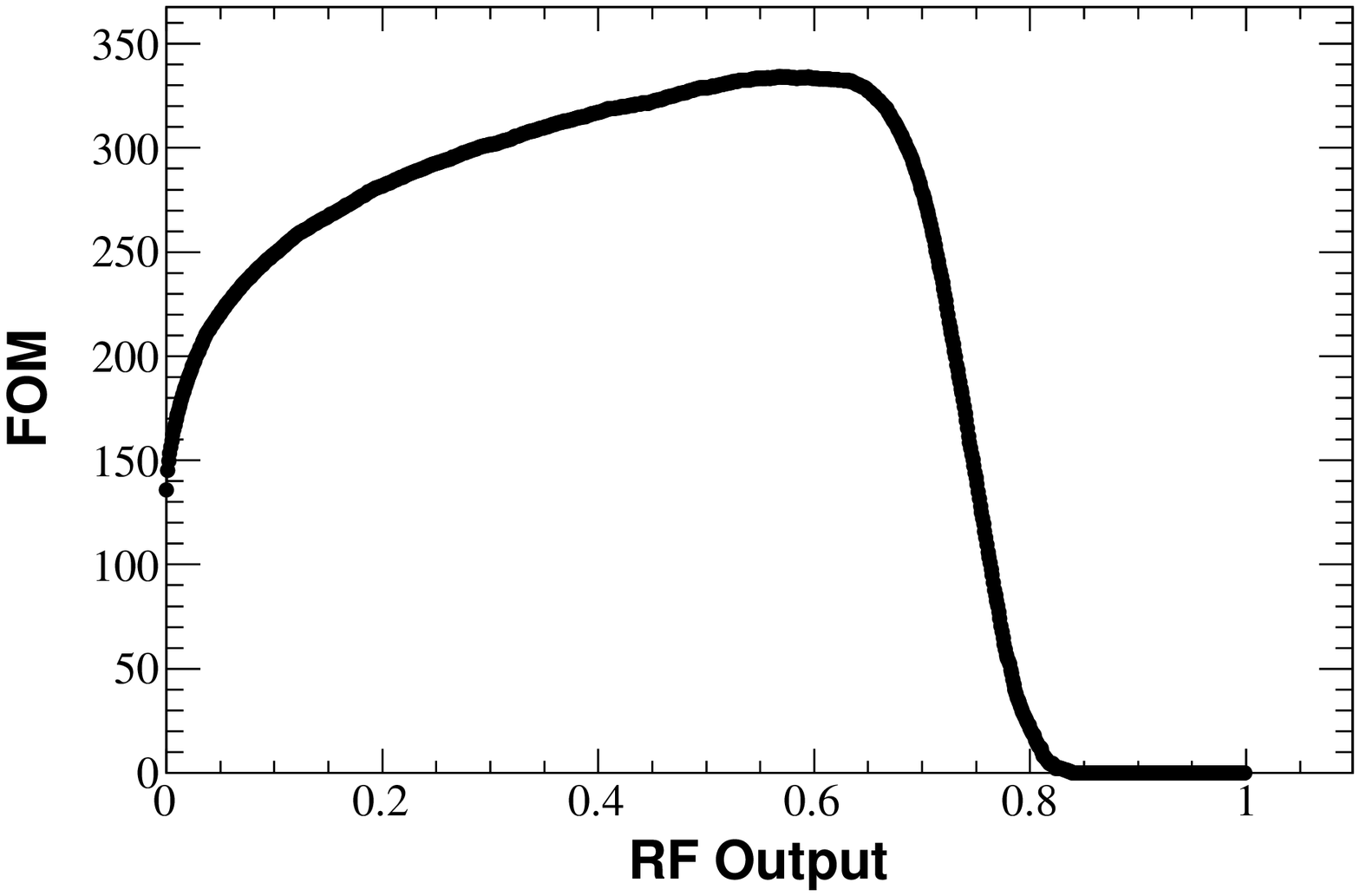}

\smallskip
\centerline{\hfill (c) \hfill \hfill (d) \hfill }
\smallskip

\caption{ (a,c) The output of the RF for both signal and combined background MC for $\Upsilon(2S,3S)$  and  (b,C) Punzi's FOM as a function of RF cut. The backgrounds are normalized by $\Upsilon(3S, 2S)$ onpeak data sets. Top plots are for $\Upsilon(2S)$ and bottom plots are for $\Upsilon(3S)$.   The optimized cut is RF $>$ 0.388 (0.568) for the $\Upsilon(2S)$ ($\Upsilon(3S)$) dataset. }

\label{fig:MVARF} 
\end{figure}

We can control two parameters during the training process: the number of tress grown (training cycles) and the minimum number of events which are allowed to populate a terminal node of the tree (a node with no further splits). We fix the number of trees to 300 and try a variety of minimal events per terminal node, which we denote by \rm{\lq\lq  {\bf l}\rq\rq}. Figure~\ref{fig:FOM}  shows the resulting training curves for the FOM vs. training cycle. We find the best performance (lowest FOM) for l = 50 for $\Upsilon(3S)$ and l = 250 for $\Upsilon(2S)$. The output of the RF,  for both signal and combined background MC is shown in Figure~\ref{fig:MVARF}. We use these RF outputs to calculate the survived signal and background events.   

 We cross-check the performance of RF algorithm against the BumpHunter algorithm for $\Upsilon(3S)$. We fix the cut on the RF output to a particular value to achieve the same background yield as BumpHunter and compare the signal MC yield and we find that the RF returns 5.88$\%$ more signal MC events (25420 events) for the same background for $\Upsilon(3S)$. We shall use RF classifier for further analysis for both $\Upsilon(3S)$ and $\Upsilon(2S)$ datasets.
  
 We optimize the cut on the RF  discriminant using the Punzi figure of merit (FOM):
\begin{equation}
   \frac{\epsilon}{0.5N_{\sigma} +\sqrt { \sum_{i}(B_{i} \times w_{i}})},   
 \end{equation}

\noindent where $N_{\sigma} =3$, $\epsilon$ is the average efficiency, $B_i$ and $w_i$ are the number of  background events and background weights ( for different i= $\Upsilon(3S, 2S)$ generic, uds, radiative bhabha, radiative di-muon, $\tau^+\tau^-$  and $c\overline{c}$) respectively. The weight of each data-set is defined as the ratio of two quantities a/b, where a is Run7 $\Upsilon(3S, 2S)$ onpeak luminosity and b is background sample luminosity. The RF  output for signal and background MCs in the test sample is shown in Figure~\ref{fig:MVARF}  for both  $\Upsilon(3S)$ and $\Upsilon(2S)$ datasets. The optimized plot for the Punzi's FOM vs. RF  discriminant is also  shown in Figure~\ref{fig:MVARF}(b) and Figure~\ref{fig:MVARF}(d) for $\Upsilon(2S)$ and $\Upsilon(3S)$ sample, respectively.

 \subsection{Final selection}
\label{section:selection}
 The final selection  criteria for the $\Upsilon(3S, 2S) \rightarrow \pi^+\pi^- \Upsilon(1S)$; $\Upsilon(1S) \rightarrow \gamma A^0$; $A^0 \rightarrow \mu^+\mu^-$ analysis includes the following:

\begin {itemize}
\item Track multiplicity, photon and muon related cuts as described in Table~\ref{table:track-photon}.
\item Pion related variables using RF classifier : $RF > 0.568$ for $\Upsilon(3S)$ and $RF > 0.388$ for $\Upsilon(2S)$.
\item $\Upsilon(3S, 2S)$ kinematic fit $\chi^2$ : $\chi_{\Upsilon(3S, 2S)}^2 < 300$.

\end {itemize}

\begin{figure}
\centering
\includegraphics[width=3.0in]{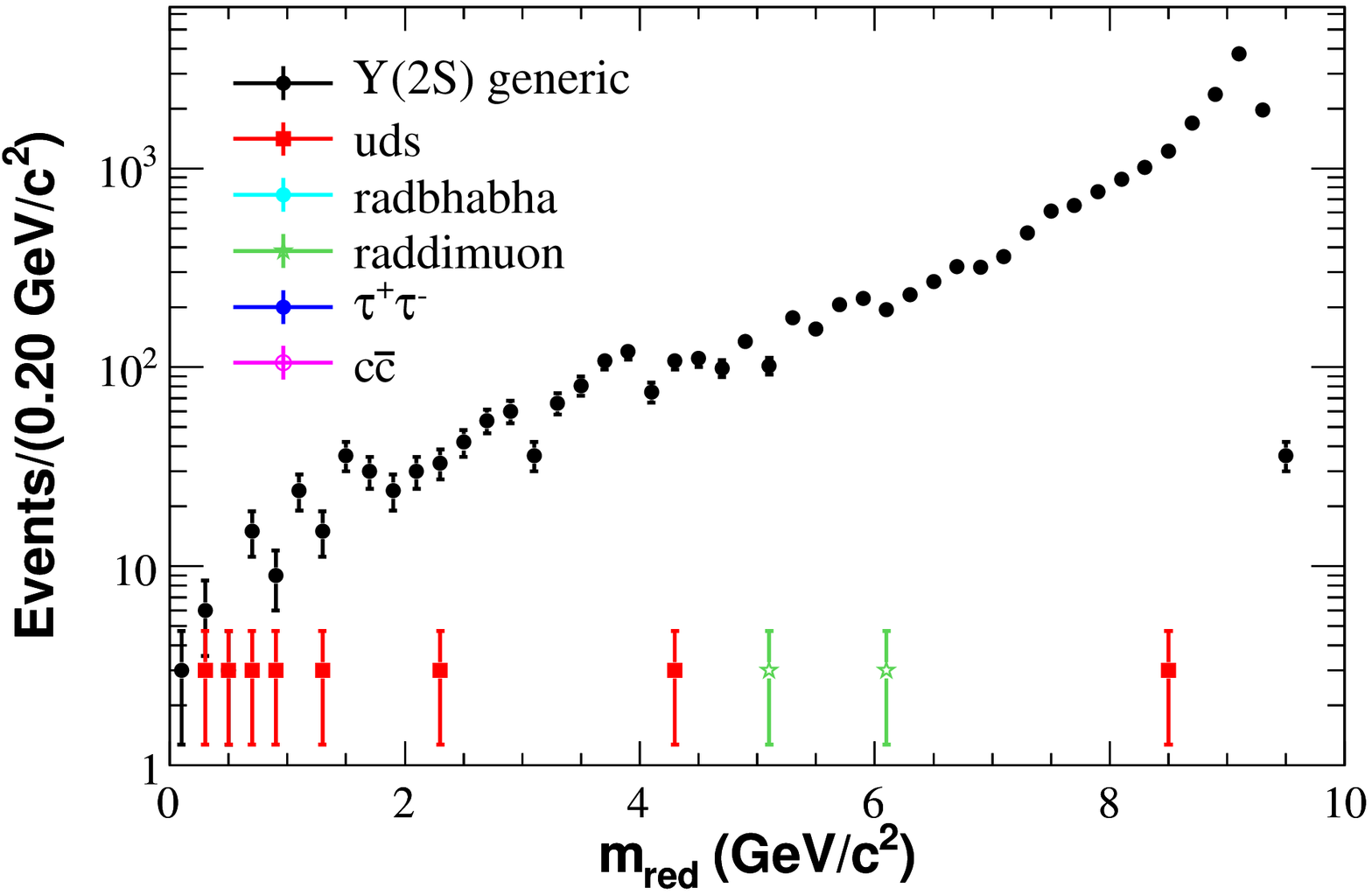}
\includegraphics[width=3.0in]{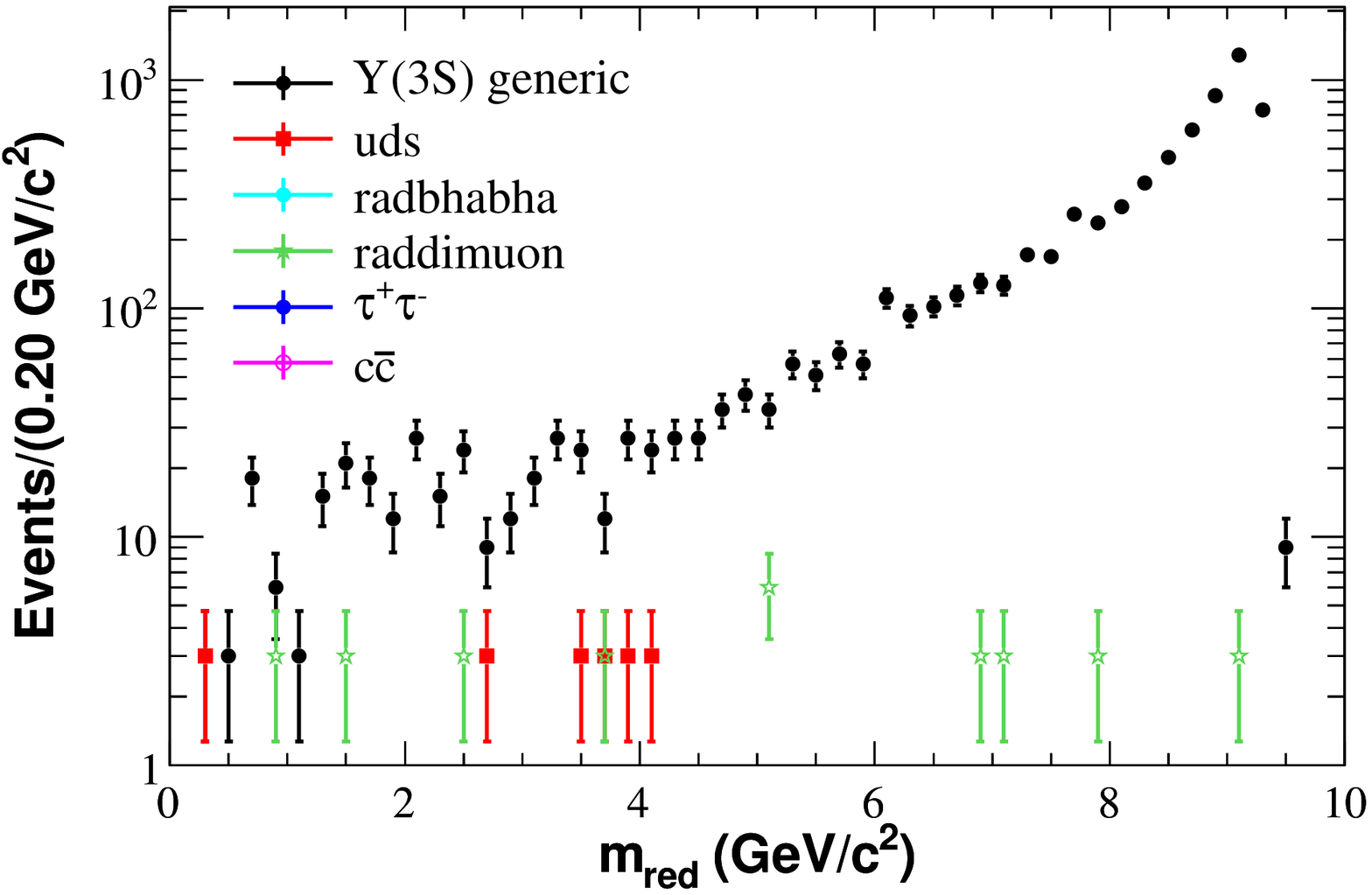}

\caption{$m_{\rm red}$ distribution of the remaining background MCs events in the $\Upsilon(2S,3S)$ datasets after applying all the optimal selection cuts. The left plot is for $\Upsilon(2S)$ and right plot is for $\Upsilon(3S)$. The test sample is scaled up by three to represent the full data samples in both $\Upsilon(2S)$ and $\Upsilon(3S)$. The most dominant remaining background is $\Upsilon(2S,3S)$ generic decays in both the datasets. Contributions from other backgrounds are negligible.}. 

\label{fig:mredY3S} 
\end{figure}

We then apply the optimal selection cuts to the test samples for both signal and background MCs.  The signal MC sample is used to compute the signal selection efficiency as a function of $m_{A^0}$  after applying all the selection cuts. We use $m_{\rm red}$ distribution to perform the maximum likelihood (ML) fit for the signal yield extraction from data, the result of which will be presented in the next chapters. The signal efficiency varies between 38.3\% (40.4\%) 
and 31.7\% (31.6\%) for $\Upsilon(2S)$ ($\Upsilon(3S)$), and decreases monotonically with $m_{A^0}$. Figure~\ref {fig:mredY3S}  shows the remaining background events in the   $\Upsilon(3S)$ and $\Upsilon(2S)$ data samples, after scalling up the number of events by 3 to represent the full sample. The background is dominated by $\Upsilon(2S,3S)$ generic decays, rest of the other sources are negligible \cite{Bad-2304}.

\begin{figure}
\centering
\includegraphics[width=3.0in]{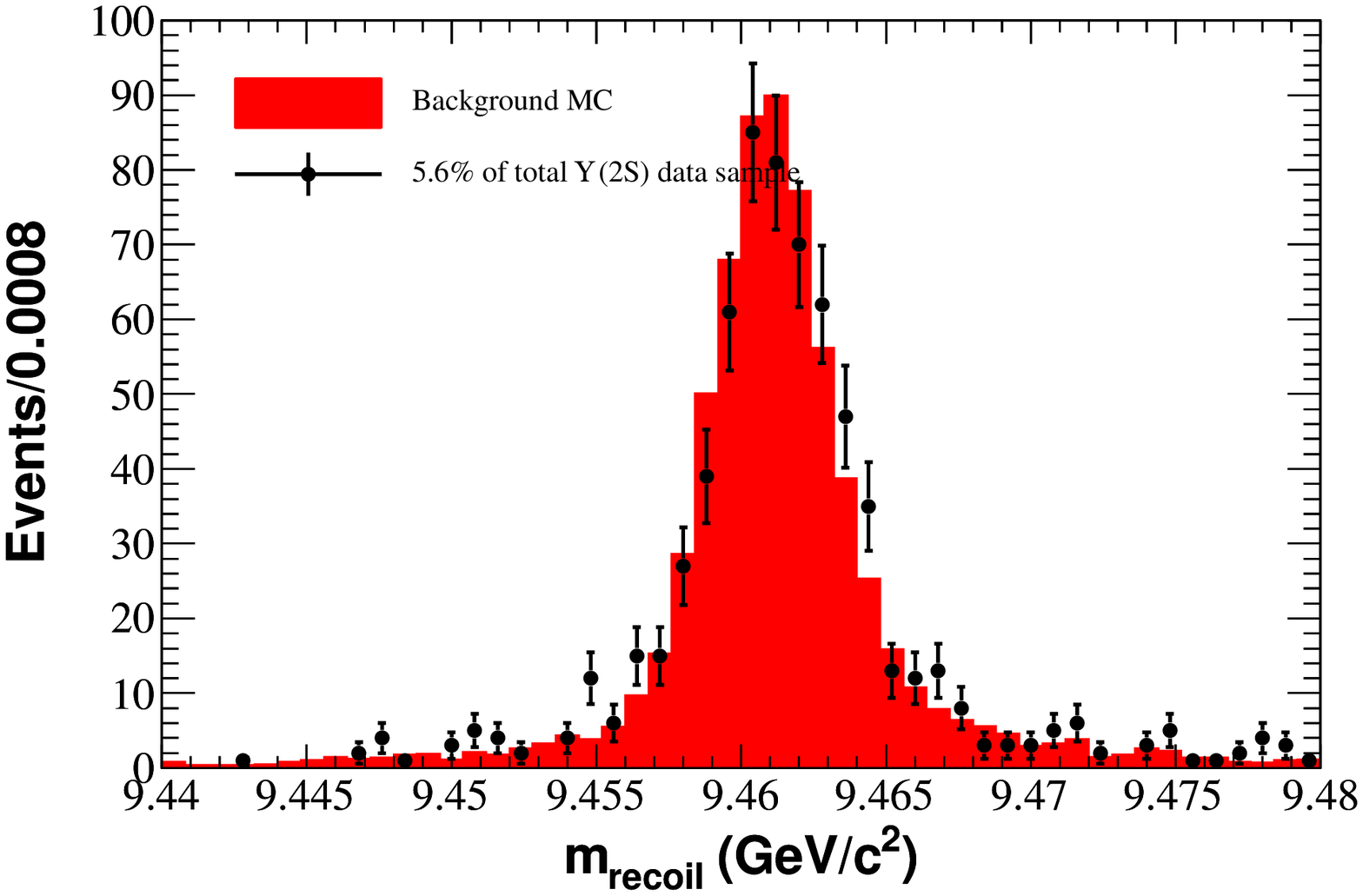}
\includegraphics[width=3.0in]{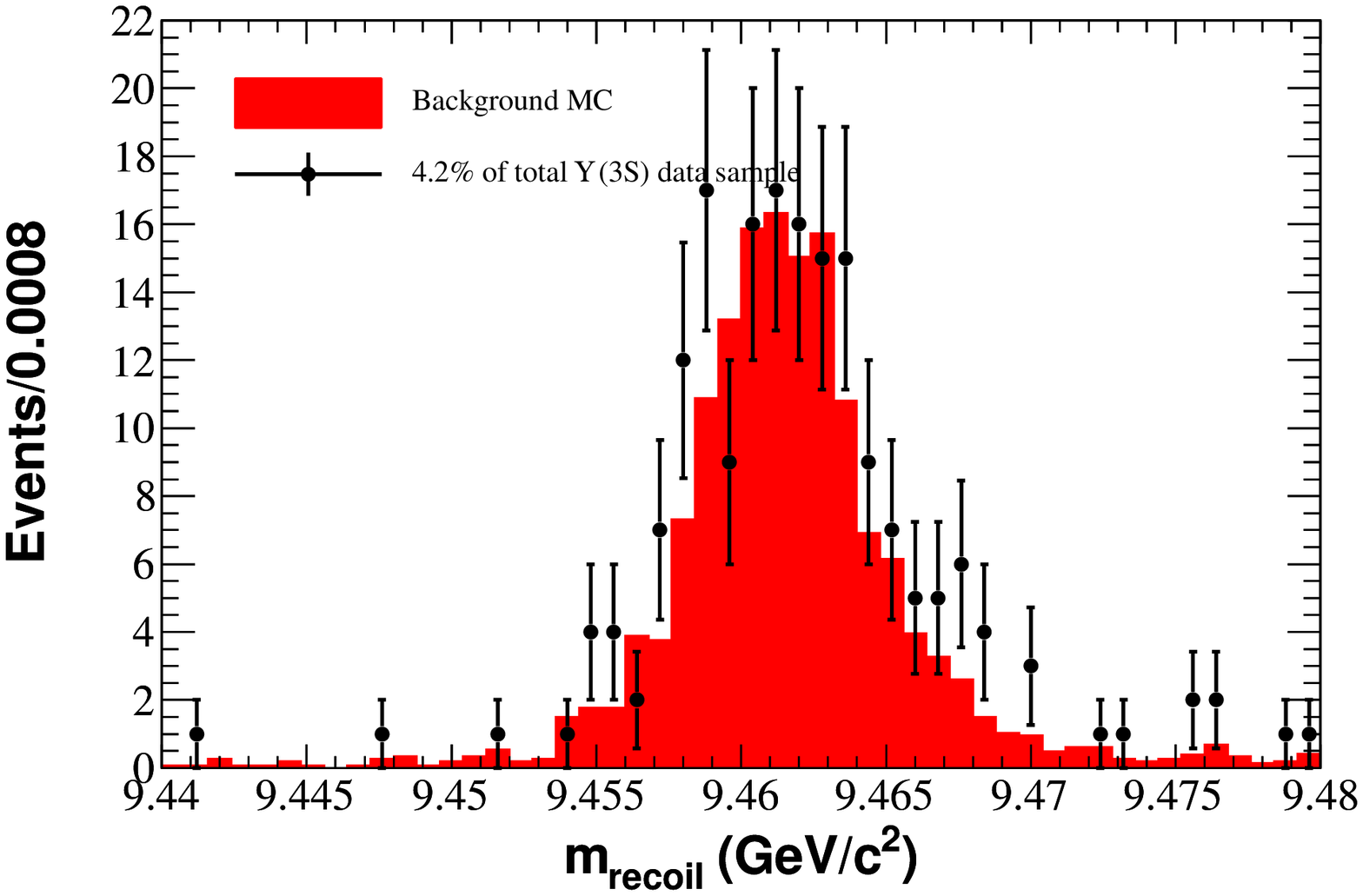}

\caption {$m_{recoil}$ distributions for generic and low onpeak datasets after applying all the selection criteria. Left plot is for $\Upsilon(2S)$ and right plot is for $\Upsilon(3S)$.  The mean of the recoil mass in MC has been corrected after comparing the recoil mass distributions in a control samples of data and MC, the details of which can be found  in section~\ref{section:mrec-study}.}

\label{fig:mrecY2-3S}
\end{figure}

\begin{figure}
\centering
\includegraphics[width=3.0in]{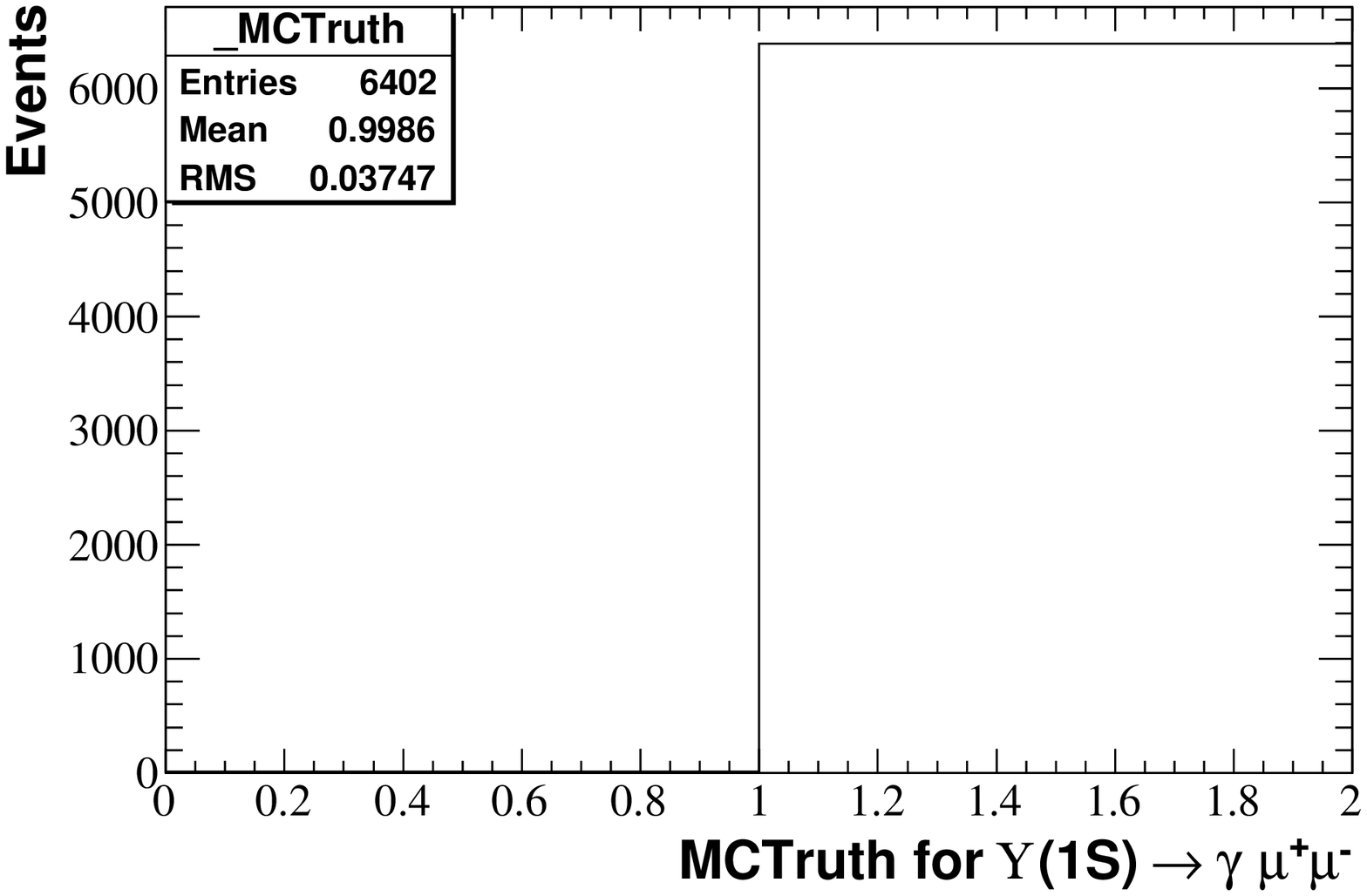}
\includegraphics[width=3.0in]{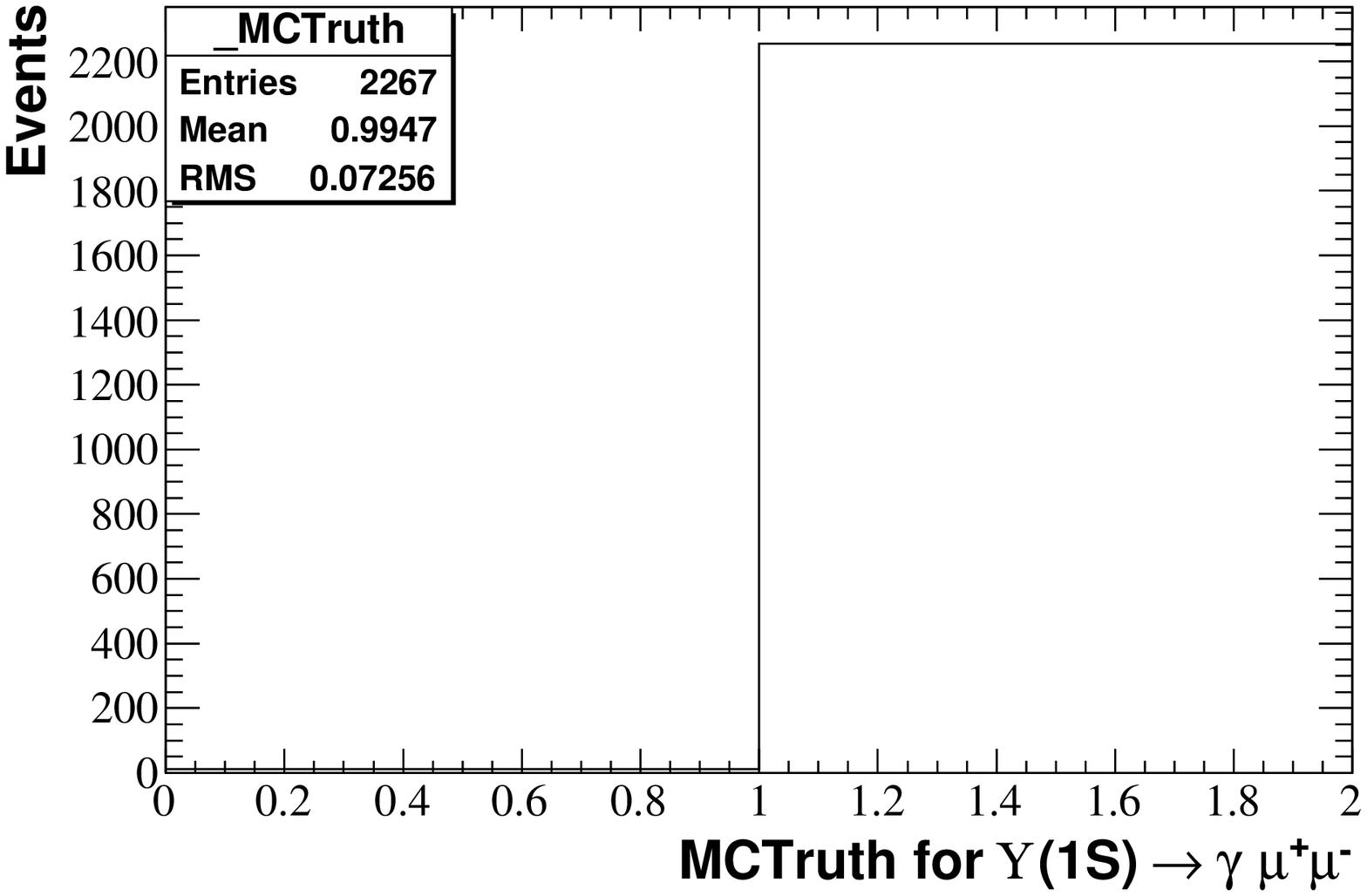}

\caption {MC-Truth Boolean distributions for $\Upsilon(1S) \rightarrow \gamma \mu^+ \mu^-$ decays. Left plot is for $\Upsilon(2S)$ and right plot is for $\Upsilon(3S)$.}

\label{fig:MC-Truth}
\end{figure}

The $m_{\rm recoil}$ distributions of generic events and low onpeak datasets show that  about $93\%$ of the $\Upsilon(3S,2S)$ generic events decay via  $\Upsilon(3S,2S) \rightarrow \pi^+\pi^-\Upsilon(1S)$,$\Upsilon(1S) \rightarrow anything$ (Figure~\ref {fig:mrecY2-3S}).  Using MC-Truth information of the survived background events, it is found that about $99\%$ of the events decay via $\Upsilon(1S) \rightarrow \gamma \mu^+ \mu^-$.  Figure~\ref {fig:MC-Truth} shows a MC-Truth Boolean distributions for $\Upsilon(1S) \rightarrow \gamma \mu^+ \mu^-$ decays for both the datasets after applying all the selection cuts.

\section{Corrections of mean and width of  $m_{\rm recoil}$}
\label{section:mrec-study}
After the event reconstruction, it was observed that the mean and sigma of the $m_{\rm recoil}$ distribution is shifted by 1.0 \mevcc in MC, while compared to data. We use a control sample of $\Upsilon(2S,3S) \rightarrow \pi^+\pi^-\Upsilon(1S)$, $\Upsilon(1S) \rightarrow \mu^+\mu^-$ in data and MC to study the mean and width value of $m_{\rm recoil}$.  We apply the following selection criteria to both data and MC after  reconstructing the events: 
\begin{itemize}
\item Two pions must not be misidentified as electron using a particle-ID algorithm where the $\pi$-to-$e$ mis-identification rate is about $0.1\%$.
\item Both leptons must be identified as muons by a muon particle-ID  algorithms.
\item CM energy and momentum are within $|\Delta E| < 0.2$ GeV and $|\Delta P| < 0.2$ GeV/c.
\item The number of the charged tracks must be equal to four.
\item RF selection cuts of the $\Upsilon(2S,3S)$ datasets.
\end {itemize}   

\begin{figure}
\centering
\includegraphics[width=3.0in]{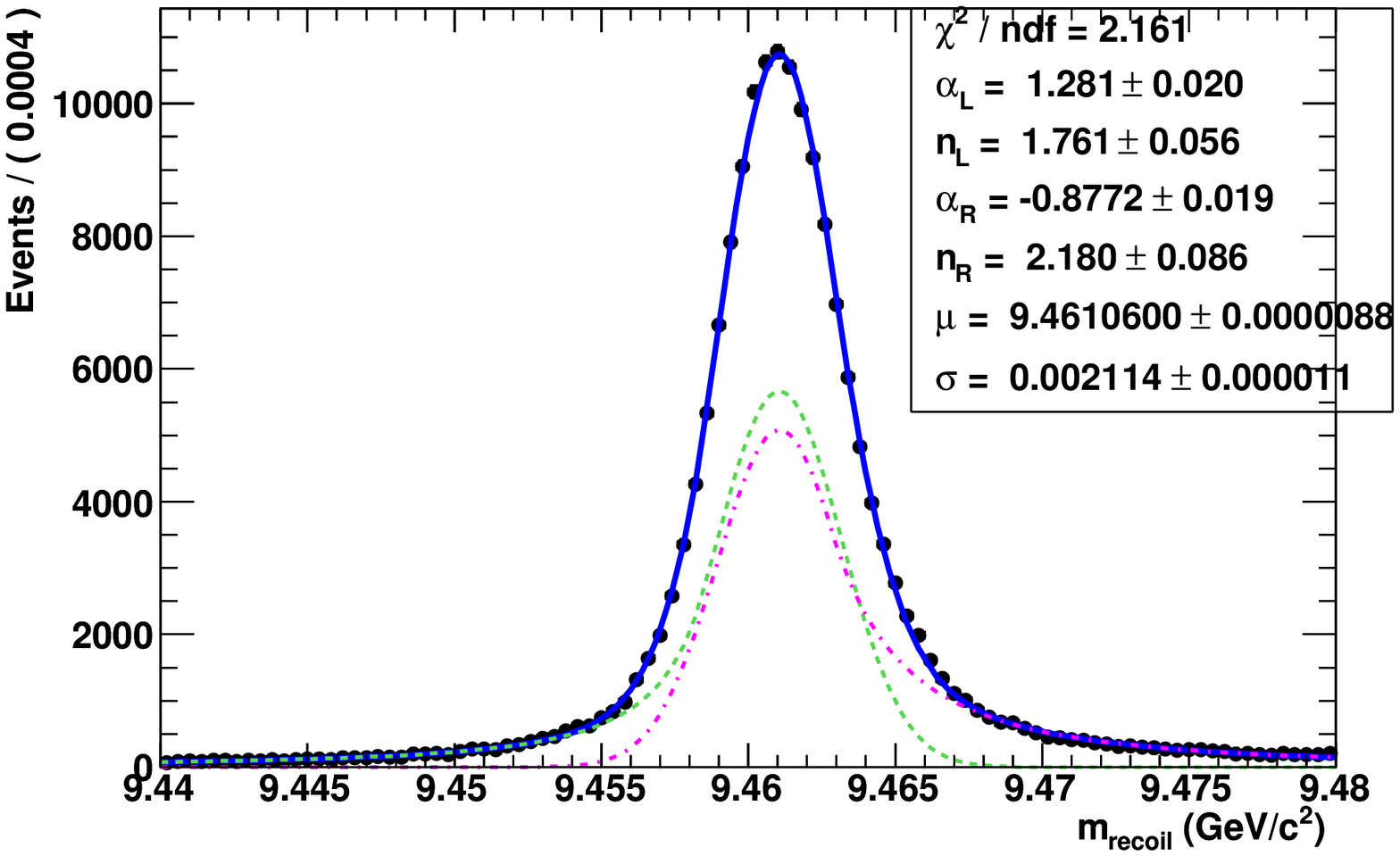}
\includegraphics[width=3.0in]{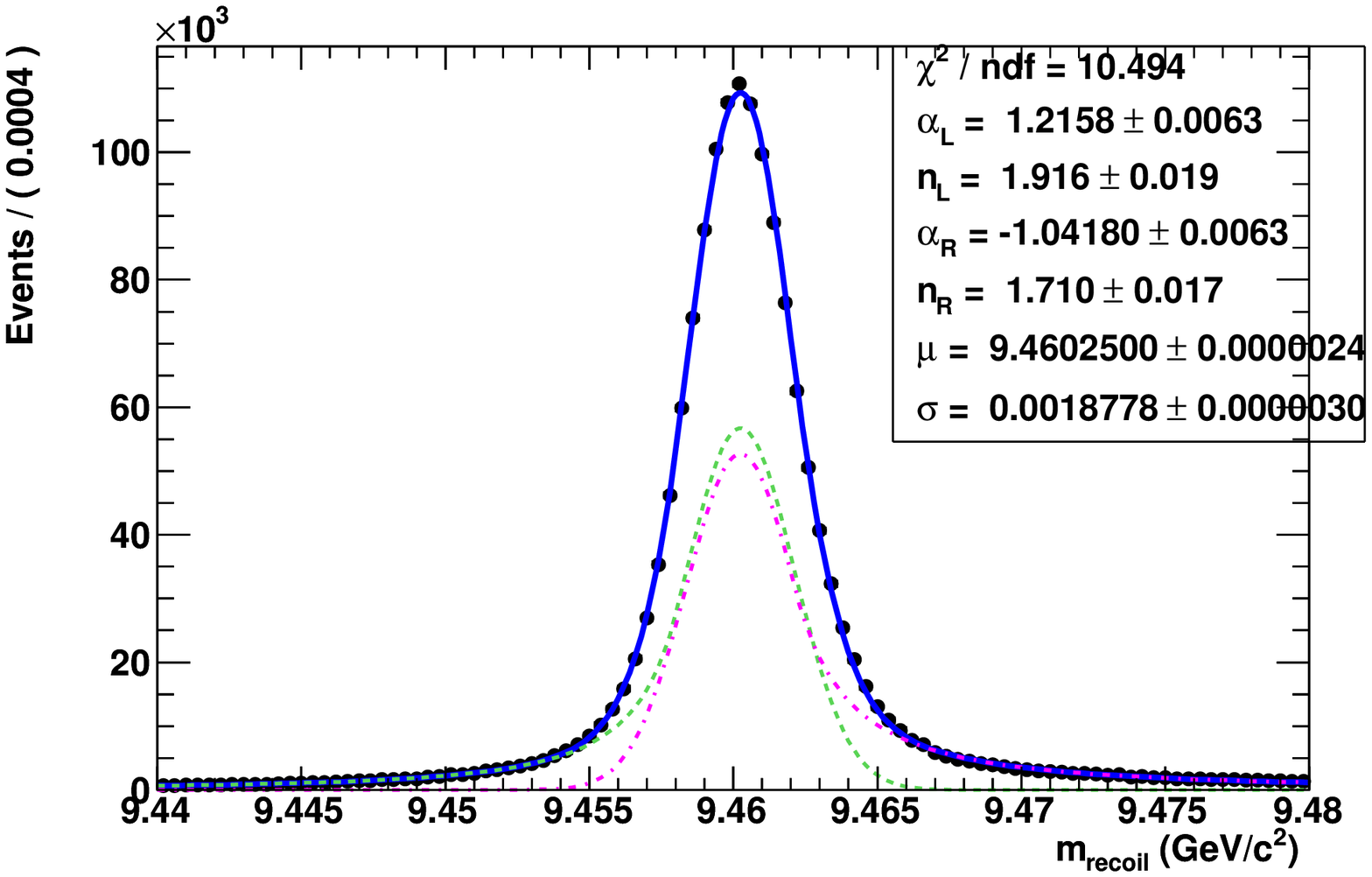}

\caption {$\mathrm{m_{recoil}}$ distribution in $\Upsilon(2S) \rightarrow \pi^+\pi^-\Upsilon(1S)$, $\Upsilon(1S) \rightarrow \mu^+\mu^-$ events after applying all the selection cuts as mentioned in the section~\ref{section:mrec-study} including the RF selection cuts of $\Upsilon(2S)$. Left plot is for the data and right plot is for MC.}

\label{Fig:MrecoilY2S} 
\end{figure}

\begin{figure}
\centering
\includegraphics[width=3.0in]{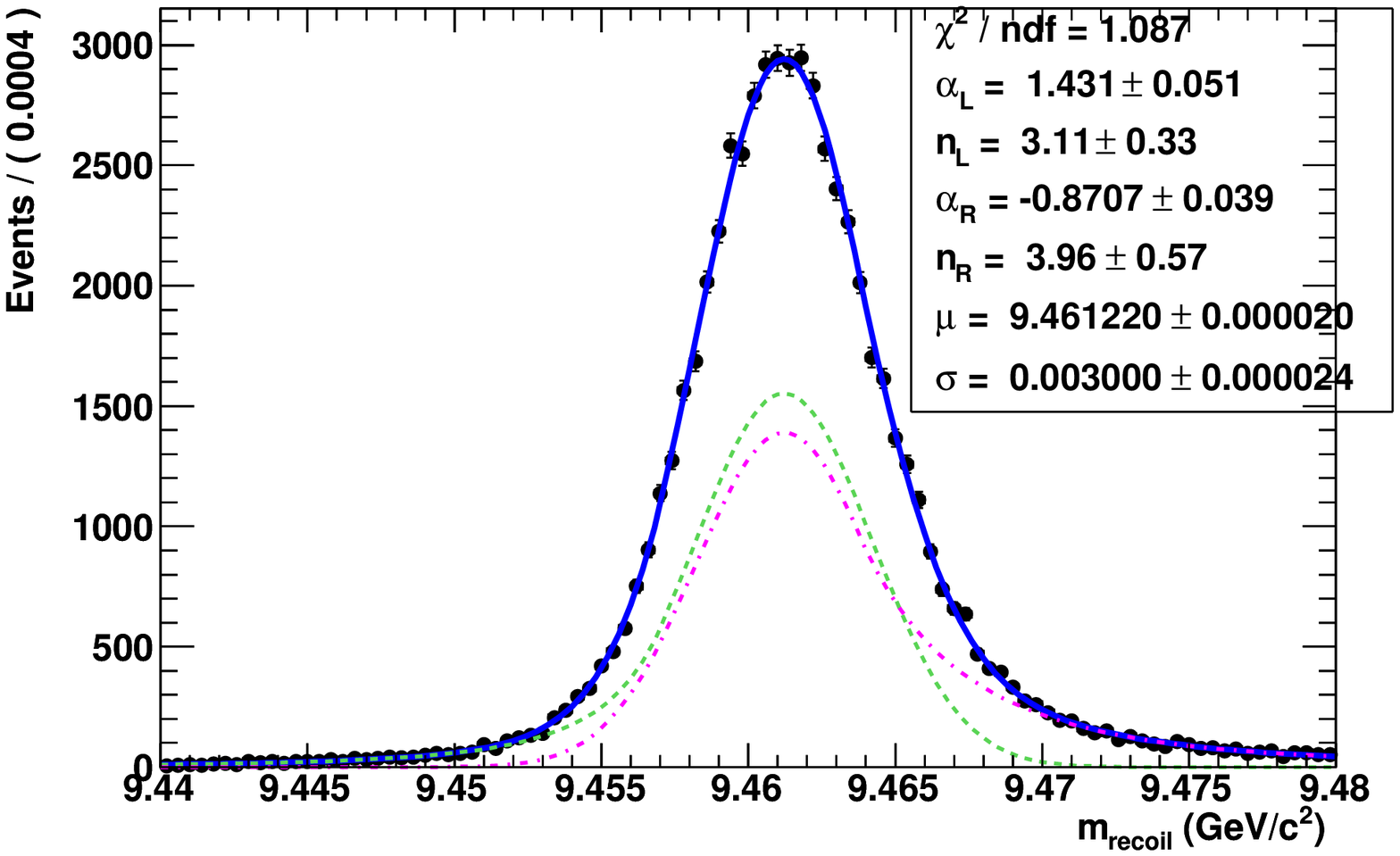}
\includegraphics[width=3.0in]{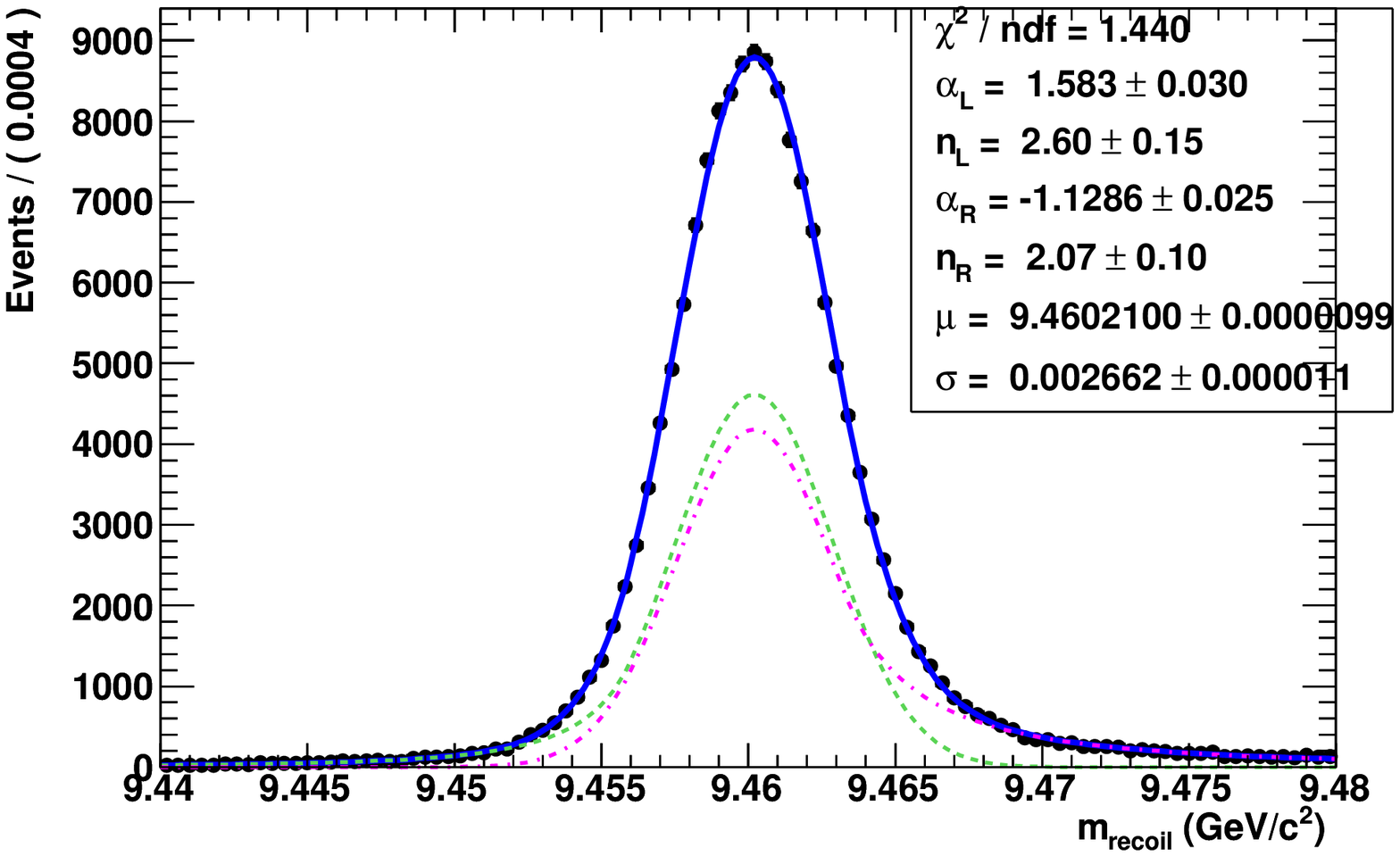}
\caption {$\mathrm{m_{recoil}}$ distribution in $\Upsilon(3S) \rightarrow \pi^+\pi^-\Upsilon(1S)$, $\Upsilon(1S) \rightarrow \mu^+\mu^-$ events after applying all the selection cuts as mentioned in the section~\ref{section:mrec-study} including the RF selection cuts of $\Upsilon(3S)$. Left plot is for the data and right plot is for MC.}

\label{Fig:MrecoilY3S} 
\end{figure}

\noindent We use a sum of two Crystal Ball (CB) functions  \cite{CB1} with opposite side tails to model the $m_{\rm recoil}$. The detail description about the CB function has been presented in the section~\ref{section:PDF} in chapter 4. The fit to the $\mathrm{m_{recoil}}$ distributions in both data and MC samples for both $\Upsilon(2S)$ and $\Upsilon(3S)$ datasets are shown in Figure~\ref{Fig:MrecoilY2S} and ~\ref{Fig:MrecoilY3S}, respectively. The mean of the recoil mass distribution in data appears to be shifted by less than 1 MeV/$c^2$ and is also  wider than MC, for both the $\Upsilon(2S,3S)$ datasets. We correct the mean and width of the recoil mass distribution in MC by the observed difference in data and MC. 

\section{Chapter Summary}
  In this chapter, we have described the event reconstruction and the methods of event selection using the different multivariate techniques. The datasets used for this analysis are also presented. Finally, we have discussed the remaining backgrounds after applying all the selection criteria. In the following chapter, we will describe the signal and background probability density functions (PDFs), which are used to extract the signal events from data.

 



















\chapter{Maximum Likelihood Fit}
\label{Chapter4}
This chapter begins with an overview of the maximum likelihood (ML)
fit used to extract the signal events from the data \cite{MLFit1,
MLFit2}. The RooFit \cite{roofit} and RooRarFit \cite{Roorarfit} packages are used to conduct the 1d unbinned ML fit to the $m_{\rm red}$ distributions 
 in the data samples. The signal and background probability density functions (PDFs) are developed using  signal MC samples generated at 26 $m_{A^0}$ points and the
combined background MC, respectively. The fit validations are done using a cocktail
samples of $\Upsilon(2S,3S)$ low onpeak datasets and $\Upsilon(2S,3S)$
generic MCs, as well as a large number of the Toy MC experiments with
different embedded signal events at selected $m_{A^0}$ points. The
bias of the fit is considered as an additive systematic
uncertainty. Finally, this chapter describes the trial factor study
used to compute the true significance i.e., the probability for pure
background event to fluctuate up to a given value of the signal
yield.  

 \section{Theoretical overview of the ML fit} 
The ML fit is a technique used to estimate the values of the parameters
for a given finite sample of the data. Suppose a measurement of the
random variable x is repeated several times for a finite values of
$x_1,....x_n$, where each $x_i$  follow a probability density function (PDF) of
$f(x_i;\theta)$ for a particular value of $\theta$.  Then the likelihood
function in the interval of $[x_i,x_i + dx_i]$ is defined as:

\begin{equation}
\mathcal{L}({\theta}) = \prod_{i=1}^n f(x_i;{\theta}).
\label{Eq:MLFit}
\end{equation}

\noindent The likelihood defined by equation  equation~\ref{Eq:MLFit} is called an unbinned likelihood, which is evaluated at each data point and no binning of the data is needed. In practice one often uses the negative log-likelihood  (NLL)

 \begin{equation}
-log \mathcal{L}({\theta}) = -\sum_{i=1}^n log f(x_i;{\theta}),
\end{equation}
   
\noindent that makes easier to estimate a parameter value while minimizing the NLL function. The unbinned ML estimator $\hat{\theta}$ for a parameter
vector $\vect{\theta}$ is defined as the value of $\vect{\theta}$ for
which the likelihood is maximal, or equivalently the negative
log-likelihood is minimal.

The statistical uncertainty on a parameter $\theta$ is defined as the
square-root of the variance. The ML estimator for the variance on
$\theta$ is given by the second derivative of the log-likelihood at
$\theta = \hat{\theta}$.

\begin{equation}
\hat{\sigma}(\theta)^2 = \hat{V}(\theta) = \biggl(\frac{d^2 log(\mathcal{L}(\theta))}{d^2\theta}\biggl)^{-1},
\end{equation}

\noindent In case there are multiple parameters, the variance of the
ensemble of parameters is represented by the covariance matrix, which
is defined as:

\begin{equation}
V(\theta,\theta') = \langle\theta\theta'\rangle - \langle \theta \rangle \langle \theta' \rangle = \biggl(\frac{ \partial^2 log (\mathcal{L}({\theta},{\theta'}))}{\partial\theta \partial \theta'}\biggl)^{-1},
\end{equation}

\noindent which can also be expressed in terms of variance and a
correlation matrix
\begin{equation}
 V(\theta,\theta') = \sqrt{V(\theta) V(\theta')} \cdot \rho(\theta,\theta'), 
\end{equation} 

\noindent Here $\rho(\theta,\theta')$ expresses the correlation
between the parameters of $\theta$ and $\theta'$ and have their values in the range
of [-1,1].

\subsection{Extended ML Fit}
The extended ML function includes an extra factor for the probability of obtaining a sample of size N from a Poisson distribution of a mean $\nu$ 
\begin{equation}
\mathcal{L}(\nu,\theta) = \prod_{i=1}^n f(x_i;{\theta})\cdot e^{-\nu}\frac{\nu^N}{N!},
\end{equation}

\noindent where the $\nu$ describes the expected rate at which the total number of events are produced. The extended ML function is used to determine the number of signal  and background  events in a given data sample through a fit. The most straightforward approach to such an analysis is to define  a composite probability density function (PDF) of $\mathcal{L}(x,\theta,\theta')$ as follows:

\begin{equation}
\mathcal{L}(x,\theta,\theta') = \frac{N_S}{N_S+N_B}\cdot S(x;\theta) + \frac{N_B}{N_S+N_B}\cdot B(x;\theta'),
\end{equation}
\noindent where $N_S$ and $N_B$ are the number of signal and background events, respectively,  $N = N_S + N_B$  the total number of events in the data sample, and $S(x;\theta)$ and $B(x;\theta')$  the
PDFs of signal and background, respectively. A minimization of the extended ML fit estimates the yield of the $N_S$ and $N_B$.

\section{Signal PDF}
\label{section:PDF}
In this analysis we perform an one-dimensional extended ML fit to the $m_{\rm red}$ distribution to extract the number of signal events. 
The $m_{\rm red}$ distributions of the signal are parametrized by a sum of two Crystal Ball (CB) \cite{CB1} functions with opposite-side tails. The CB function is given by,
\begin{equation}
 f(x|\mu,\sigma,\alpha,n)=C.\begin{cases}exp(\frac{-(x-\mu)^2}{2\sigma^2}) , & \frac{x - \mu}{\sigma} > -\alpha \\ 
(\frac{n}{|\alpha|})^n exp(-\frac{\alpha^2}{2})\cdot(\frac{n}{|\alpha|}-|\alpha|+\frac{x-\mu}{\sigma})^{-n} , & \frac{x - \mu}{\sigma} \le -\alpha
\end{cases} 
 \end{equation}
\noindent where $\alpha$ determines where the usual Gaussian turns
into a power function with the tail parameter n, and C is overall
normalization. We constrain the mean ($\mu$) parameters of the two CB
functions to be the same, and for $m_{A^0} > 0.5$ GeV/$c^2$ we also
fix the relative weight of each CB to $\rm frac = 0.5$.  For $m_{A^0}
> 0.5$ GeV/$c^2$ we also constrain the width $(\sigma)$ parameters of
the two CB functions to be same. Thus, in this mass range there are
six floated parameters: mean ($\mu$), sigma ($\sigma$), two tail
cutoffs ($\alpha_L,\alpha_R$), and two powers ($n_L,n_R$). For
$m_{A^0} \le 0.5$, we float the two widths $\sigma_L$ and $\sigma_R$
separately, for a total of seven free parameters. We fit over fixed
intervals in the mass regions: $0.002 \le m_{\rm red} \le 1.85$
\gevcc for $0.212 \le m_{A^0} \le 1.50$ \gevcc, $1.40 \le m_{\rm red}
\le 5.6$ \gevcc for $1.502 \le m_{A^0} < 5.36$ \gevcc and $5.25 \le
m_{\rm red} \le 7.3$ \gevcc for $5.36 \le m_{A^0} \le 7.10$
\gevcc. Above this range, we use sliding intervals $\mu-0.2 < m_{\rm
red}< \mu+ 0.15$ \gevcc.
    
\begin{figure}
\centering
 \includegraphics[width=2.0in]{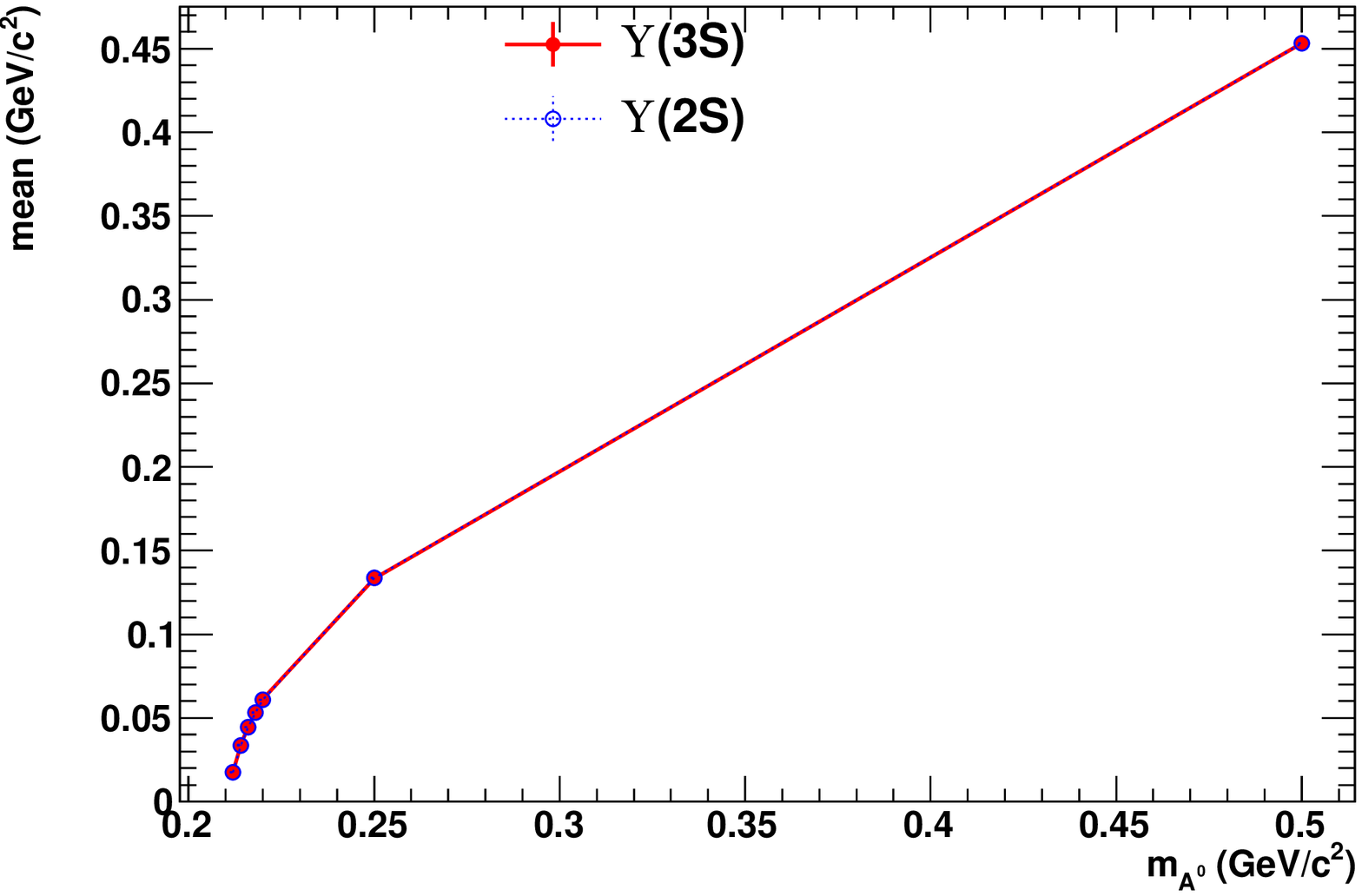}
 \includegraphics[width=2.0in]{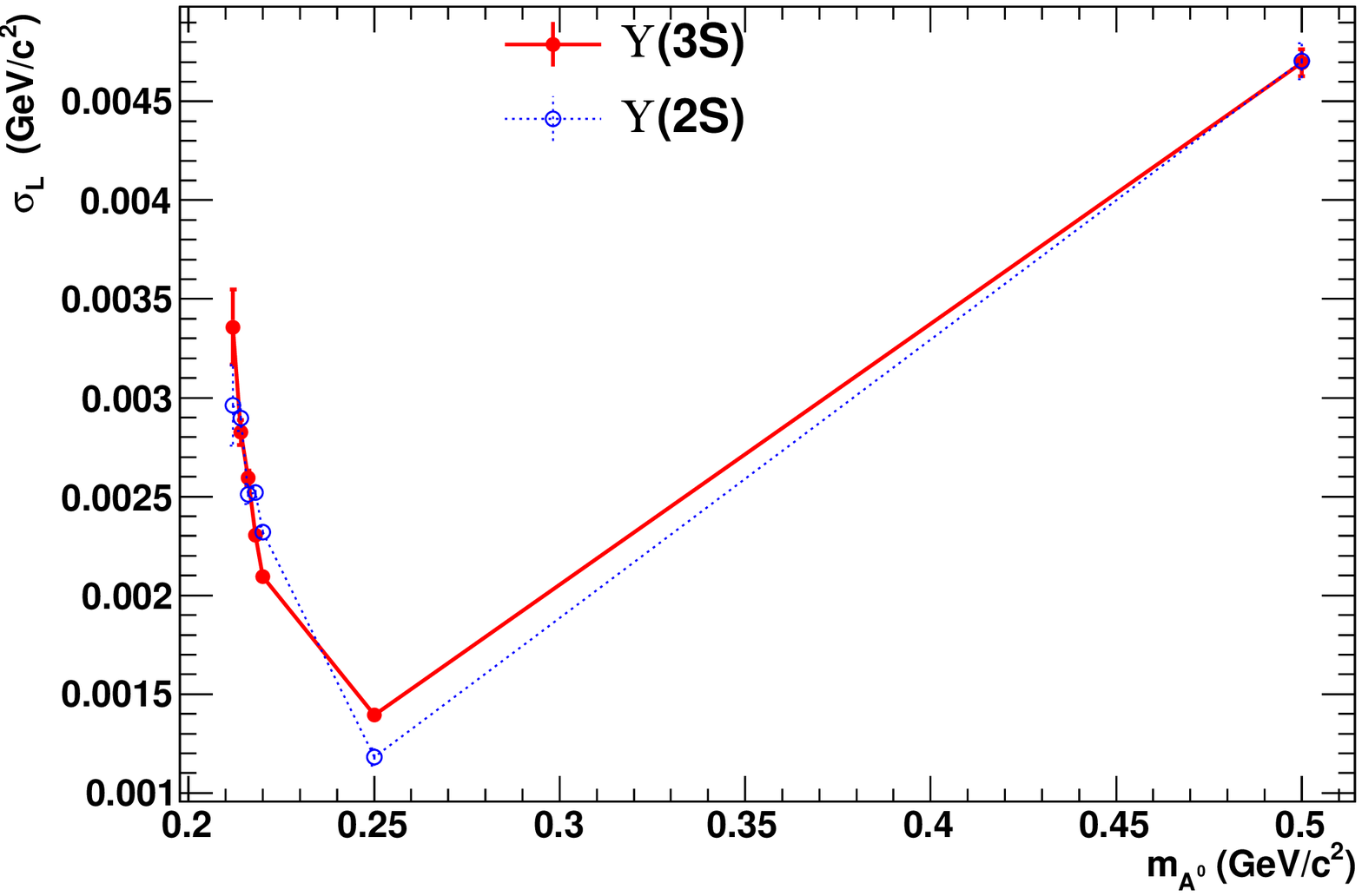}
 \includegraphics[width=2.0in]{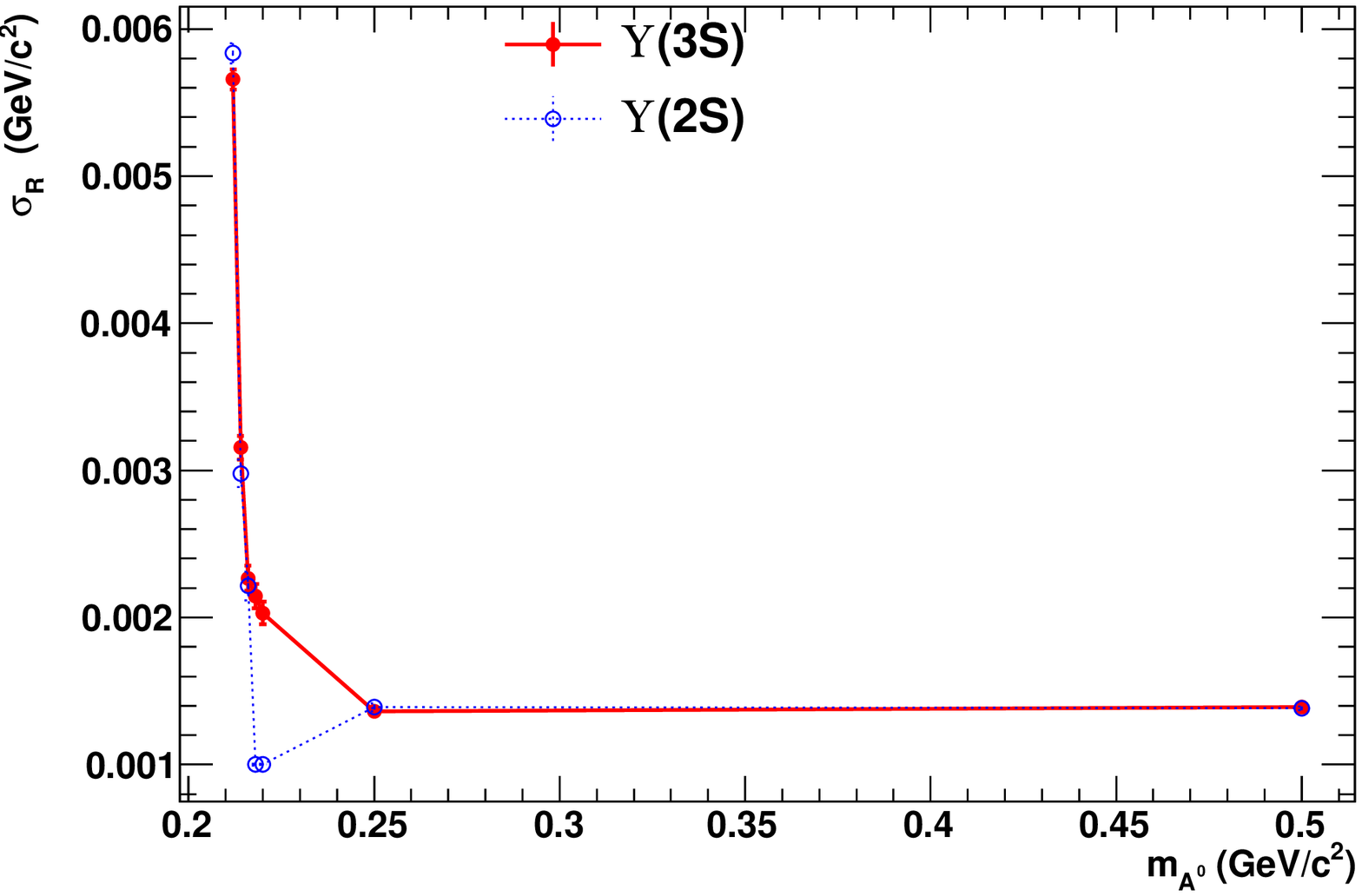}

\smallskip
\centerline{\hfill (a) \hfill \hfill (b) \hfill \hfill (c) \hfill}
\smallskip

 \includegraphics[width=3.0in]{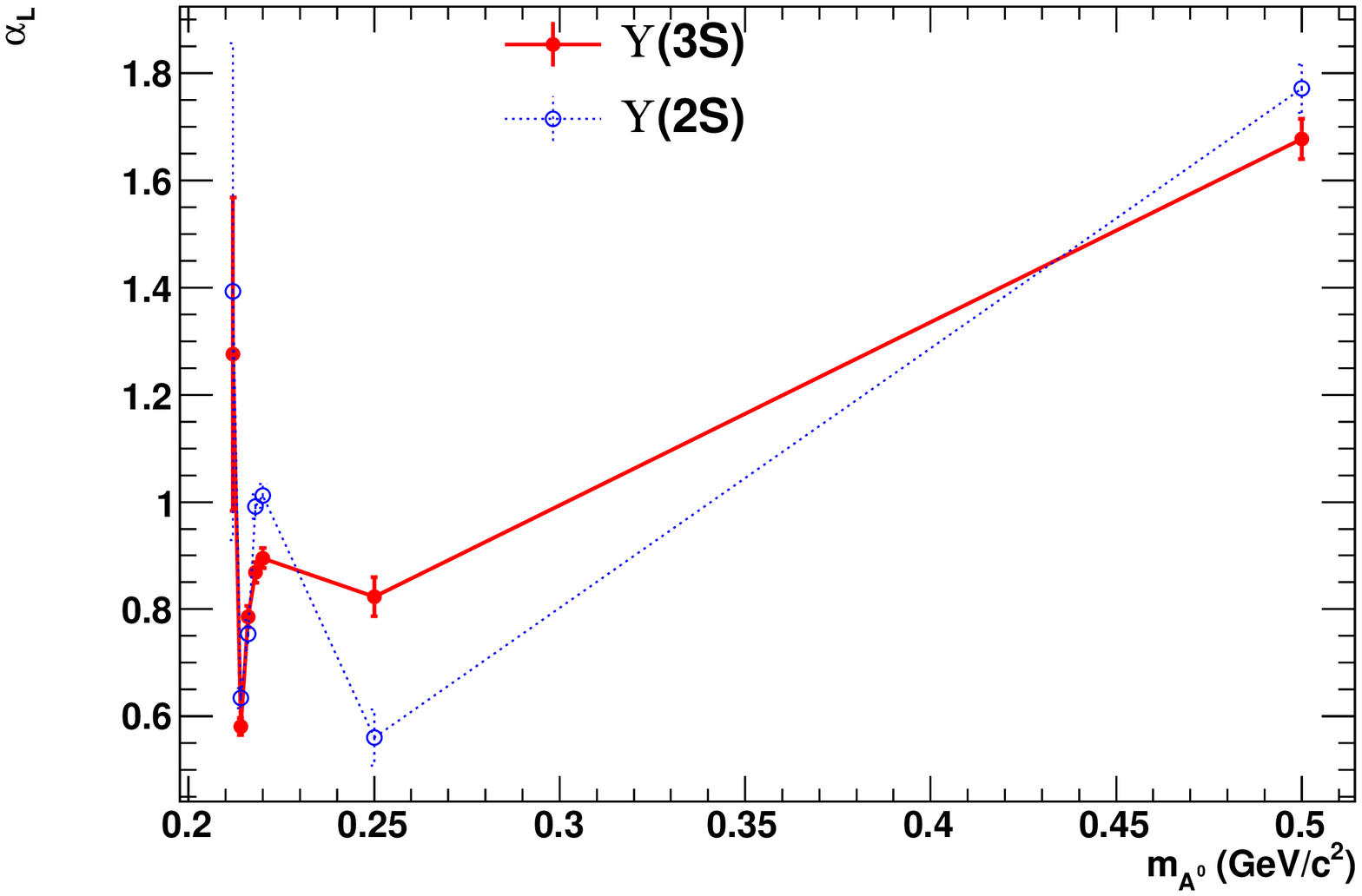}
 \includegraphics[width=3.0in]{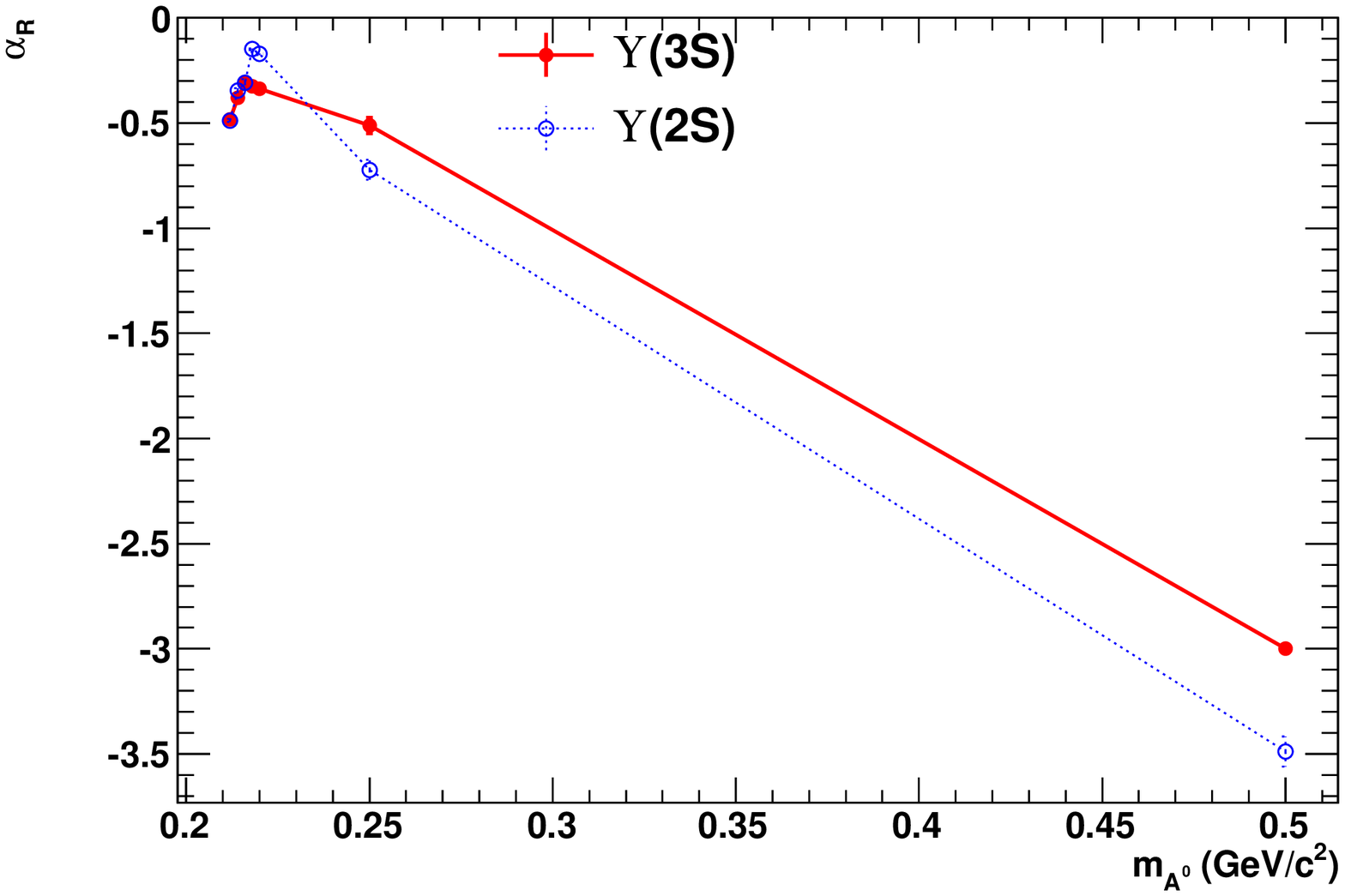}

\smallskip
\centerline{\hfill (d) \hfill \hfill (e) \hfill}
\smallskip

 \includegraphics[width=3.0in]{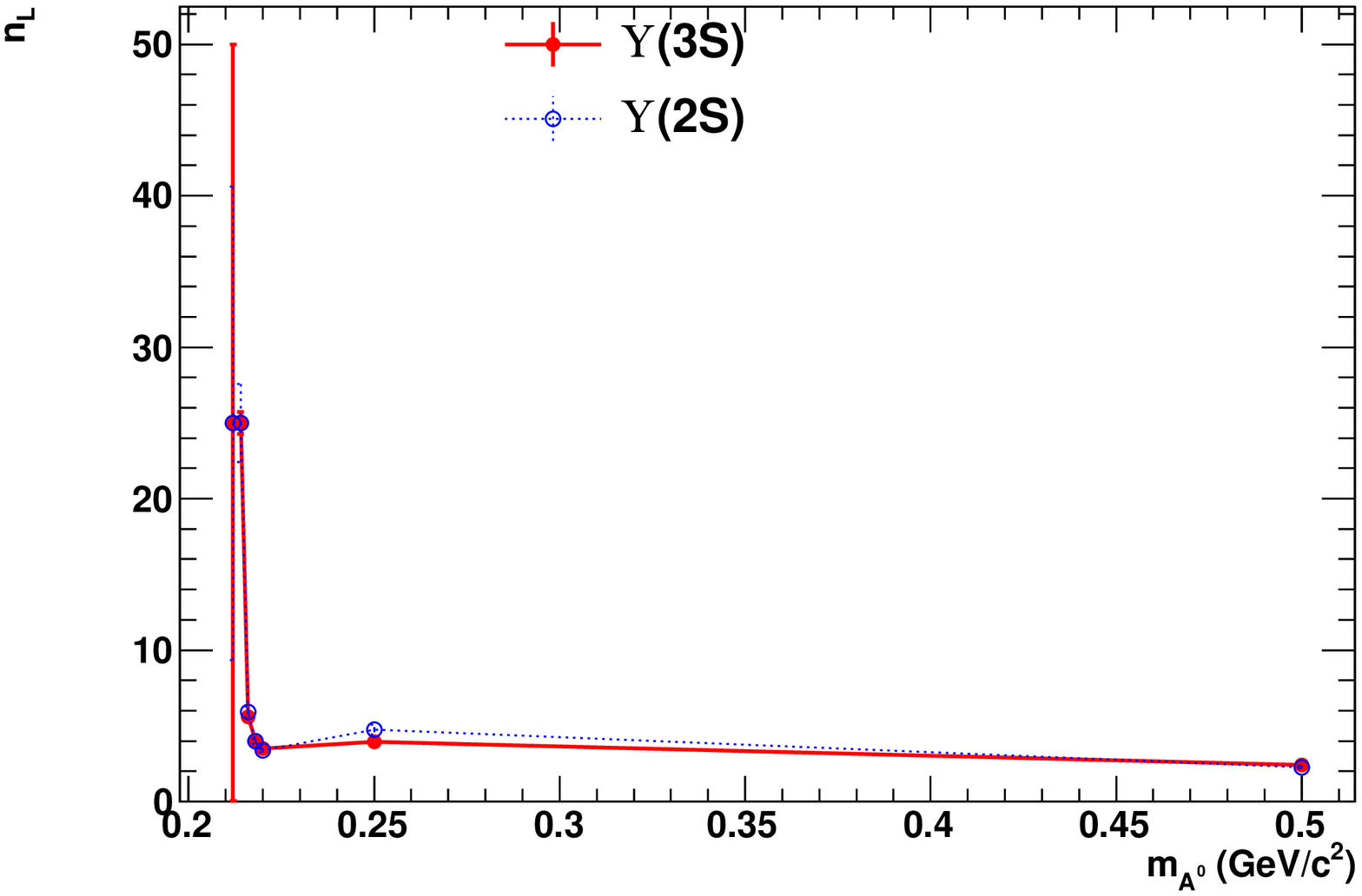}
 \includegraphics[width=3.0in]{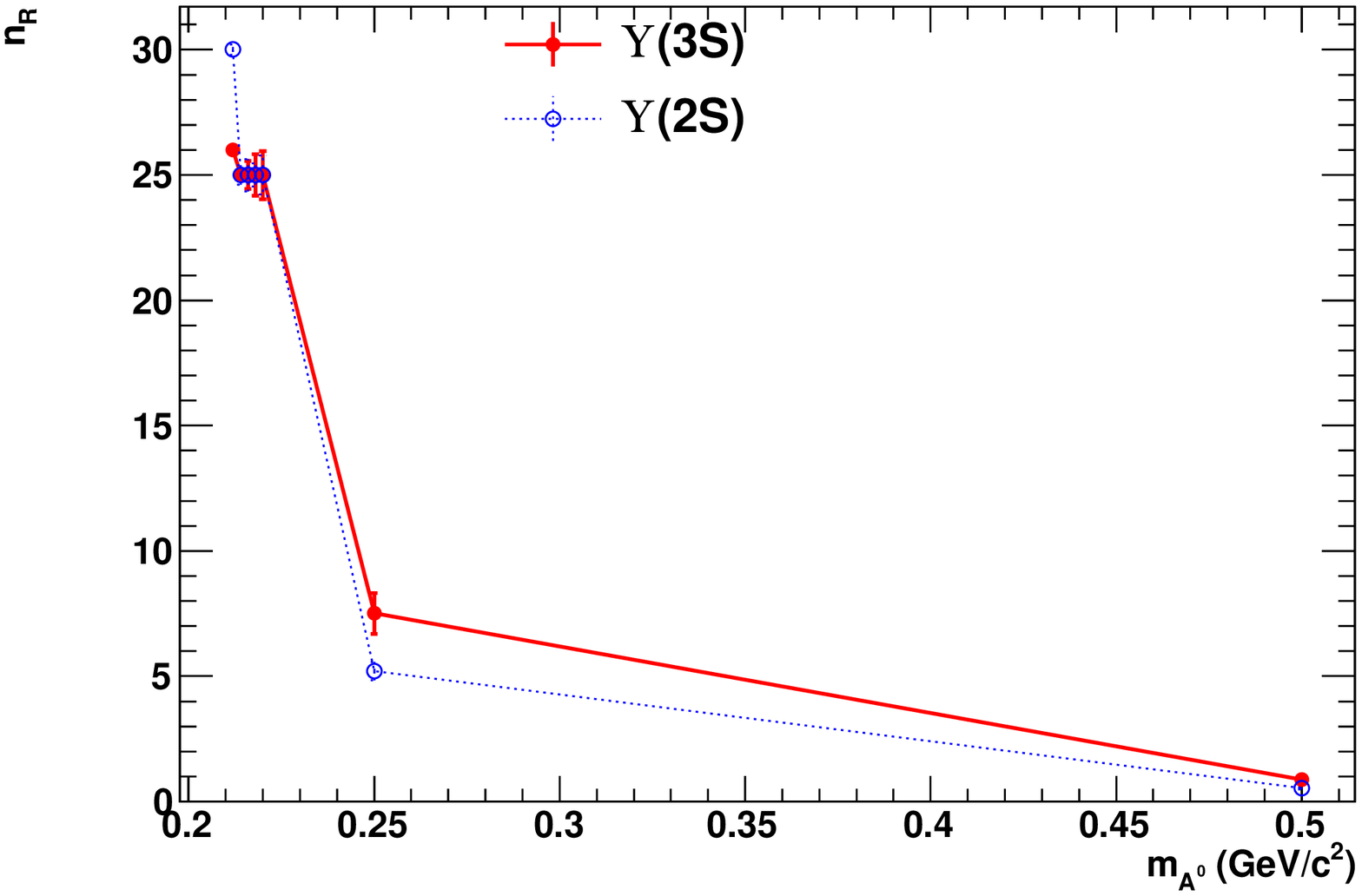}

\smallskip
\centerline{\hfill (f) \hfill \hfill (g) \hfill}
\smallskip

\caption {Parameters of the 1d ML fit to $m_{\rm red}$ distributions for
signal MC for $m_{A^0} \le 0.5$ GeV/$c^2$: (a) mean of both CB
functions (b) width of the \rm{\lq\lq left\rq\rq} CB shapes (c) width
of the \rm{\lq\lq right\rq\rq} CB shapes (d) cutoff of the \rm{\lq\lq
left\rq\rq} CB (e) cutoff of the \rm{\lq\lq right \rq\rq} CB (f) power
of the \rm{\lq\lq left\rq\rq} CB and (g) power of the \rm{\lq\lq
right\rq\rq} CB. }

\label{fig:Pdfparles0.5}
\end{figure}

\begin{figure}
\centering
 \includegraphics[width=3.0in]{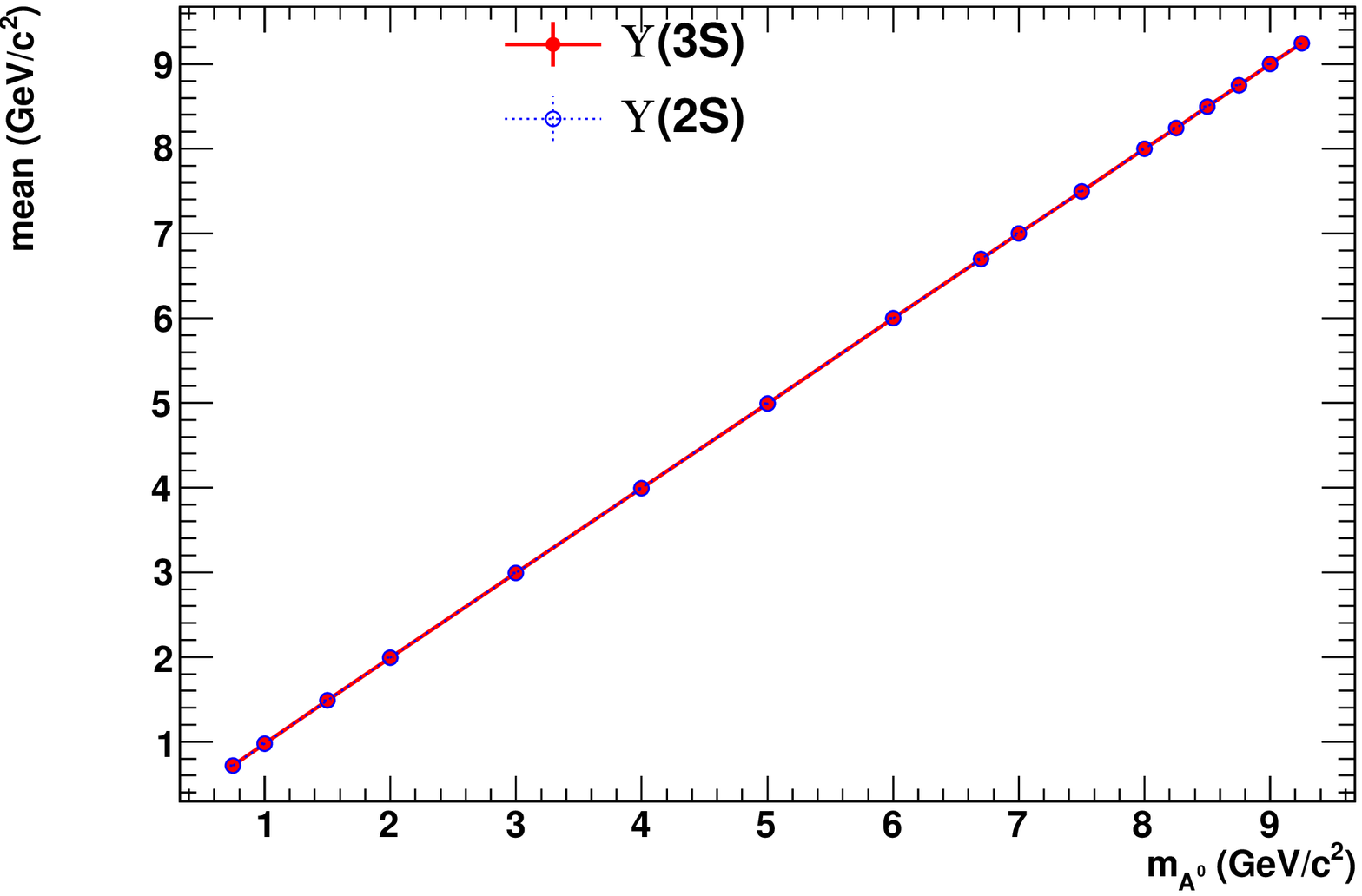}
 \includegraphics[width=3.0in]{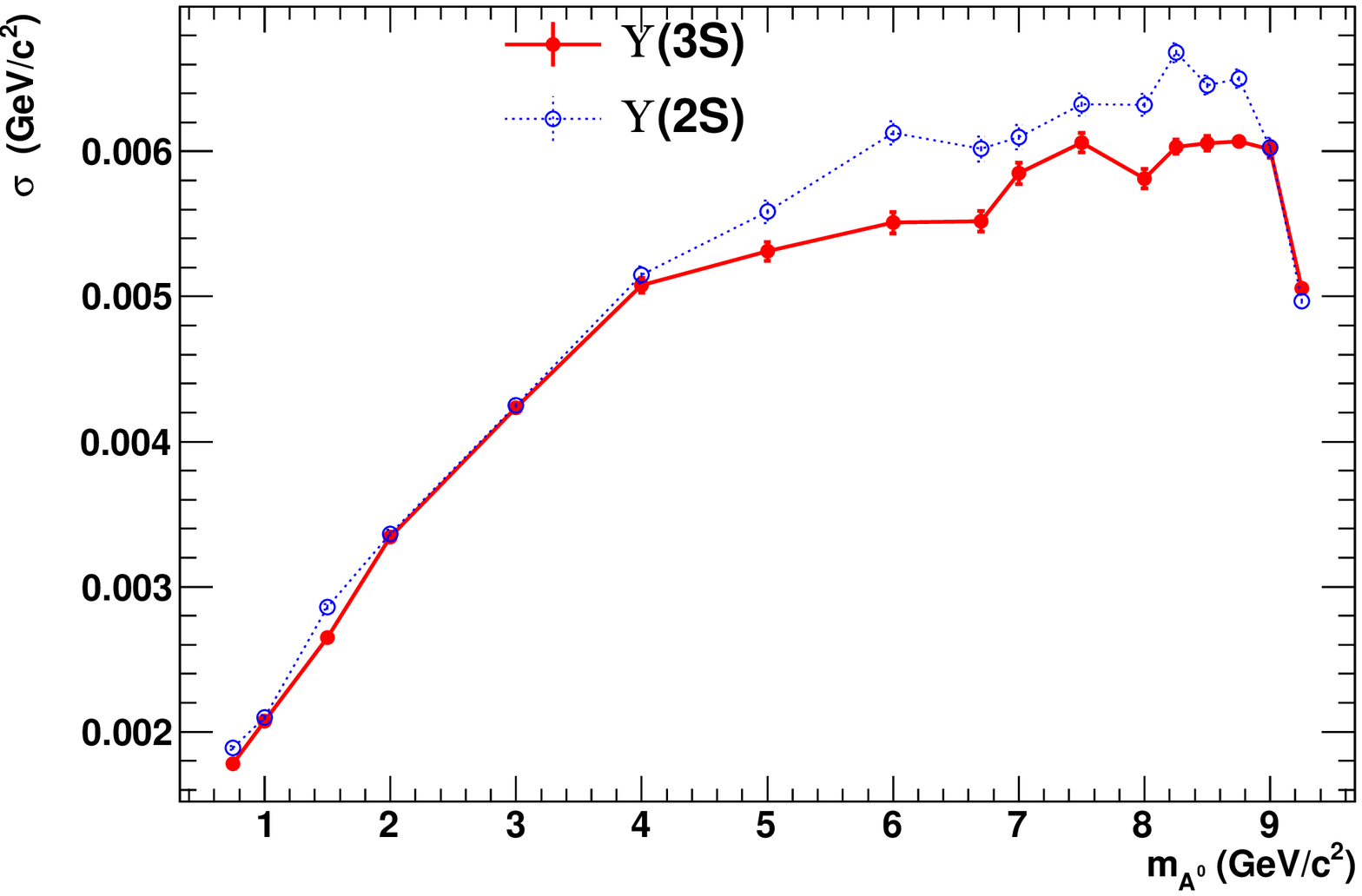}

\smallskip
\centerline{\hfill (a) \hfill \hfill (b) \hfill}
\smallskip

\includegraphics[width=3.0in]{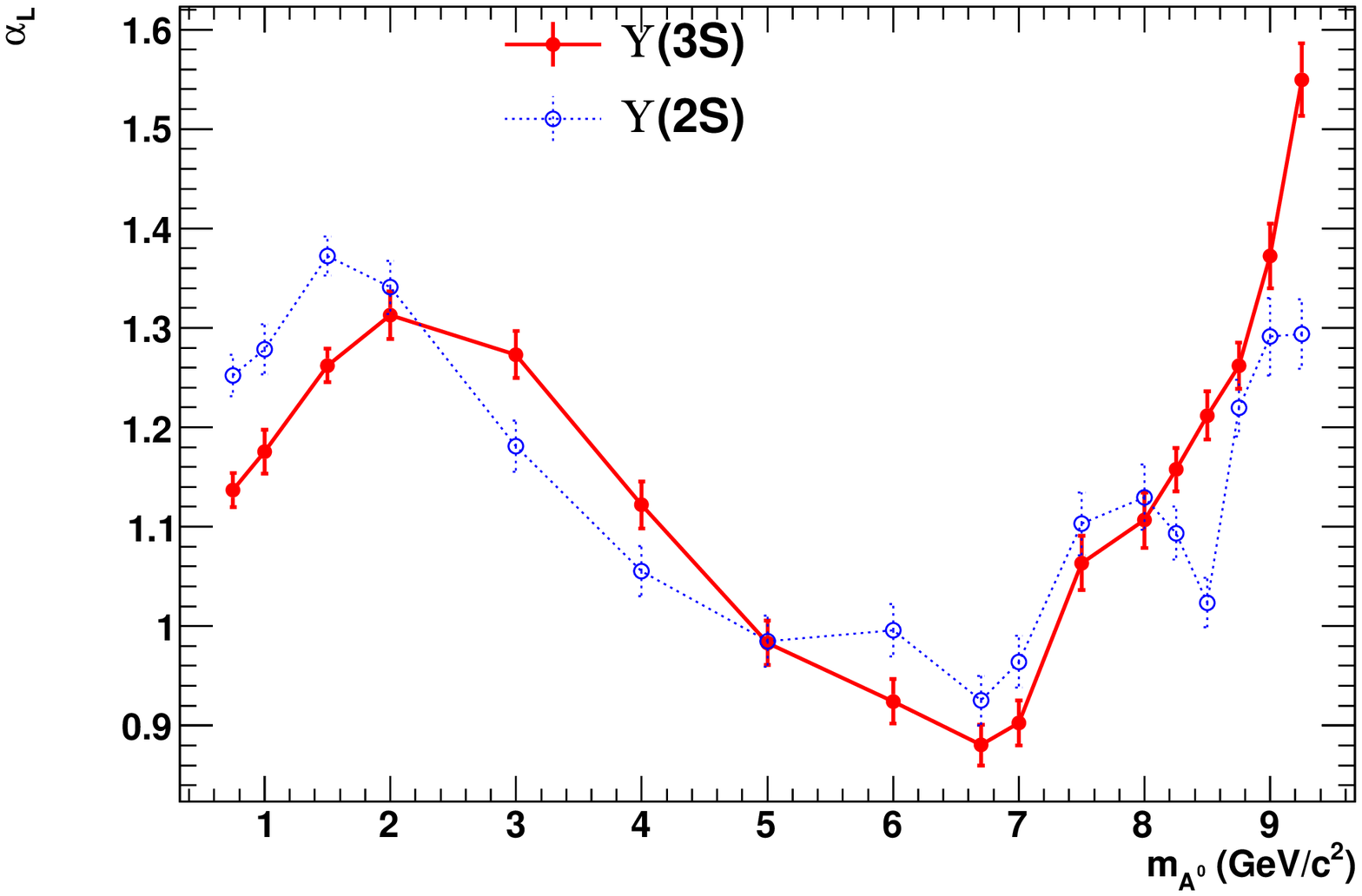}
 \includegraphics[width=3.0in]{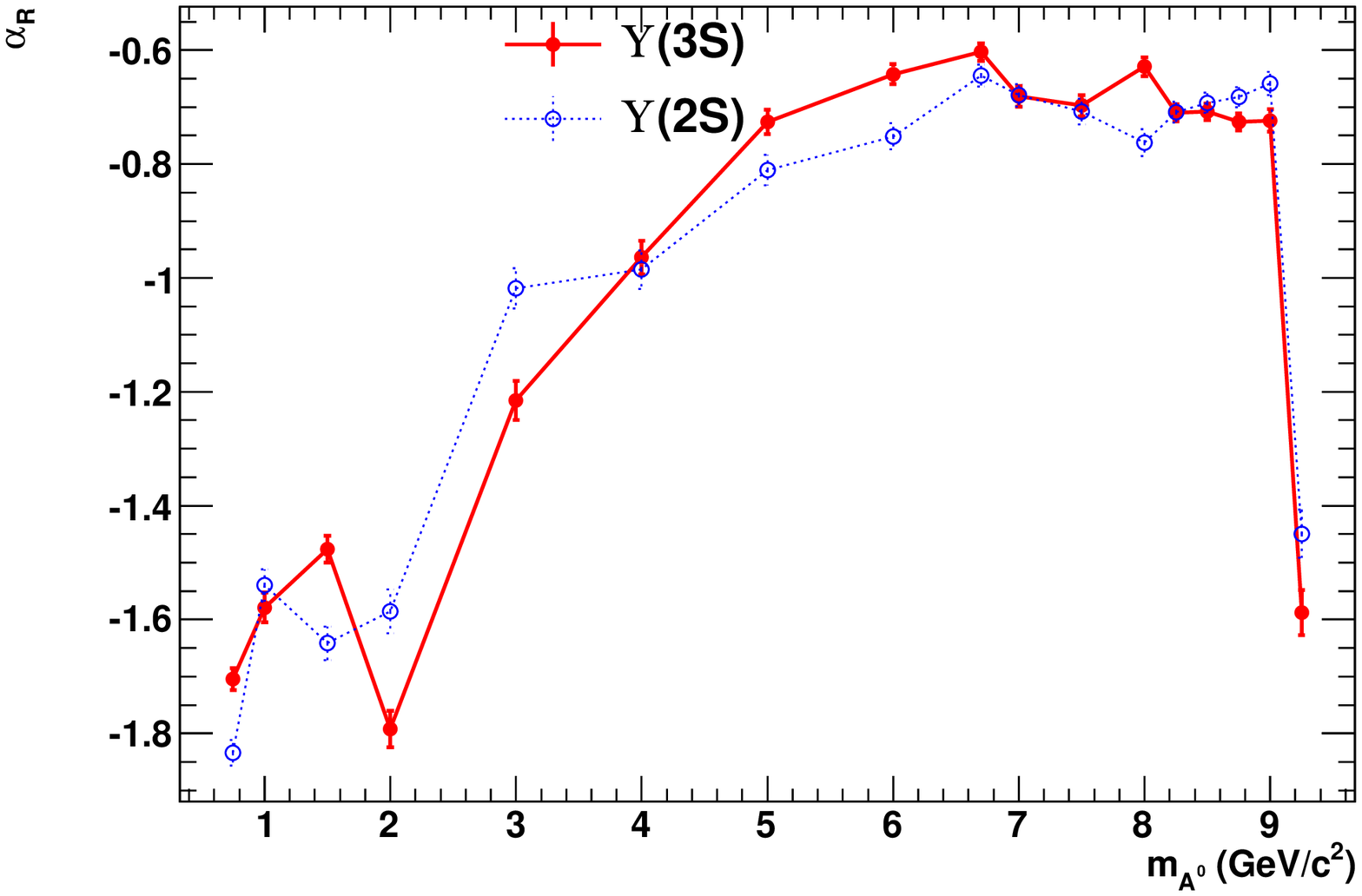}

\smallskip
\centerline{\hfill (c) \hfill \hfill (d) \hfill}
\smallskip
\includegraphics[width=3.0in]{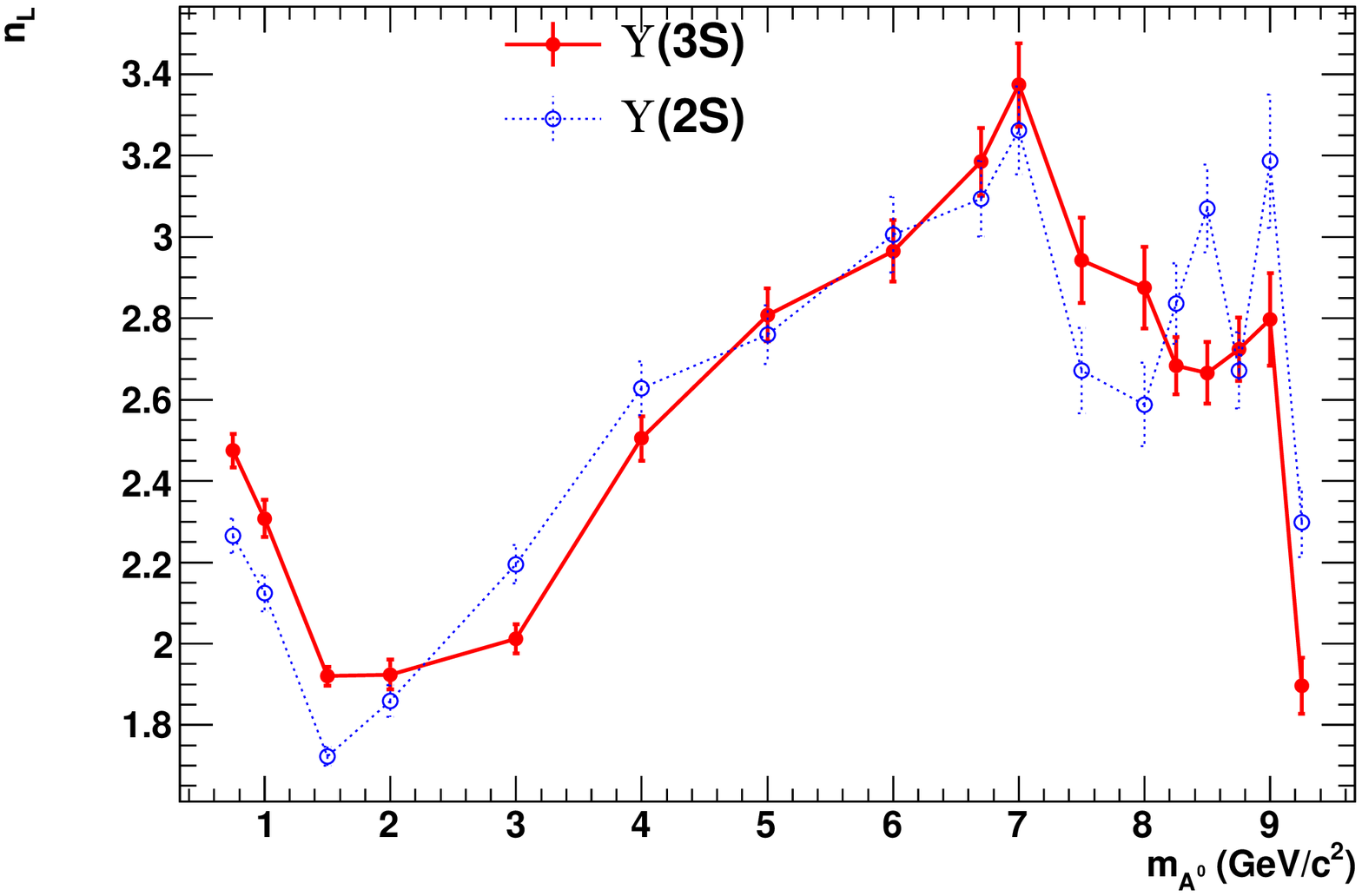}
 \includegraphics[width=3.0in]{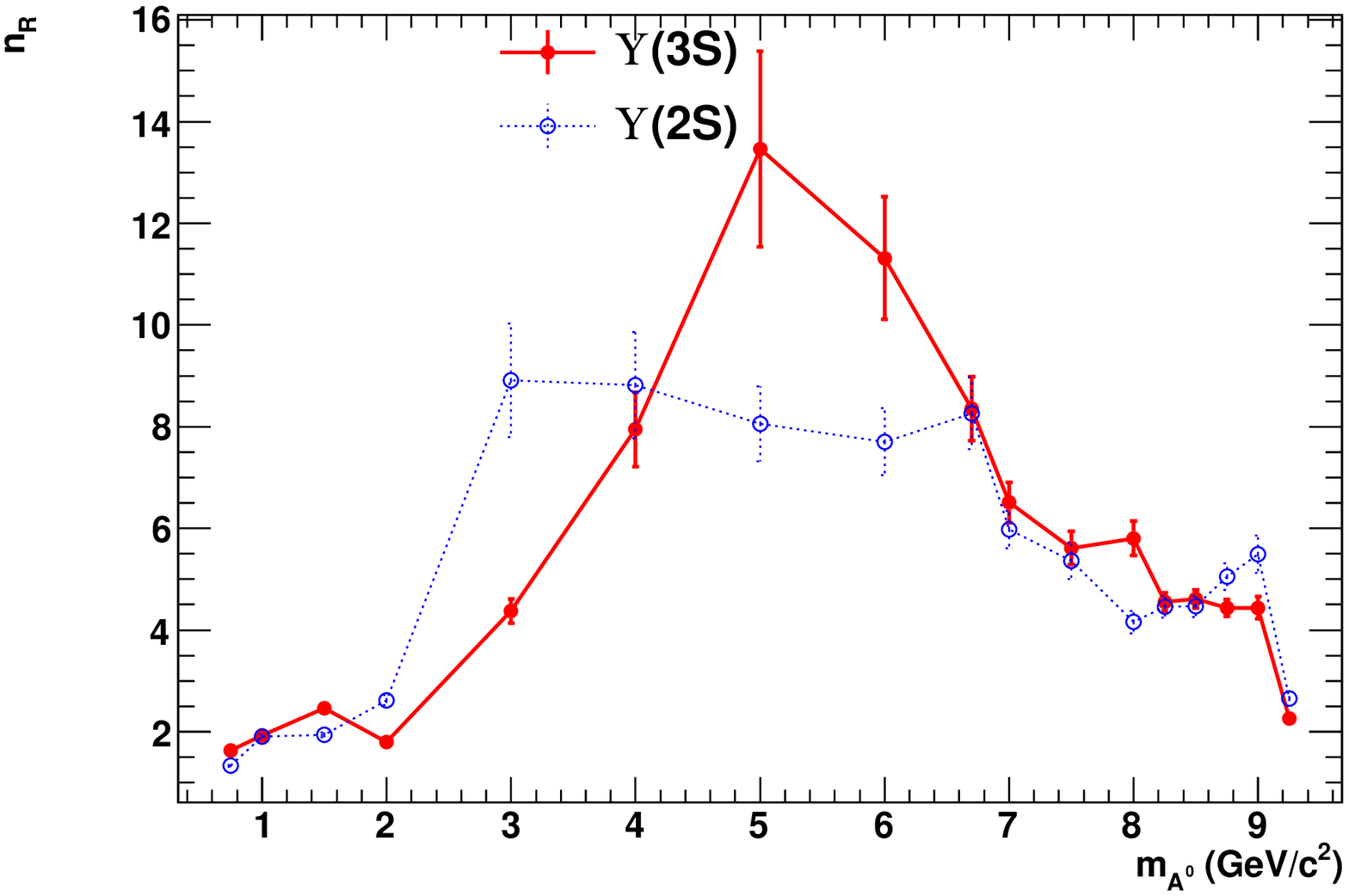}

\smallskip
\centerline{\hfill (e) \hfill \hfill (f) \hfill}
\smallskip

\caption {Parameters of the 1d ML fit to $m_{\rm red}$ distributions for
signal MC for $m_{A^0} > 0.5$ GeV/$c^2$: (a) mean of both CB
functions, (b) width of both CB shapes, (c) cutoff of the \rm{\lq\lq
left\rq\rq} CB, (d) cutoff of the \rm{\lq\lq right \rq\rq} CB, (e)
power of the \rm{\lq\lq left\rq\rq} CB, and (f) power of the
\rm{\lq\lq right\rq\rq} CB. }

\label{fig:Pdfparget0.5}
\end{figure}

\begin{figure}
\centering
 \includegraphics[width=6.0in]{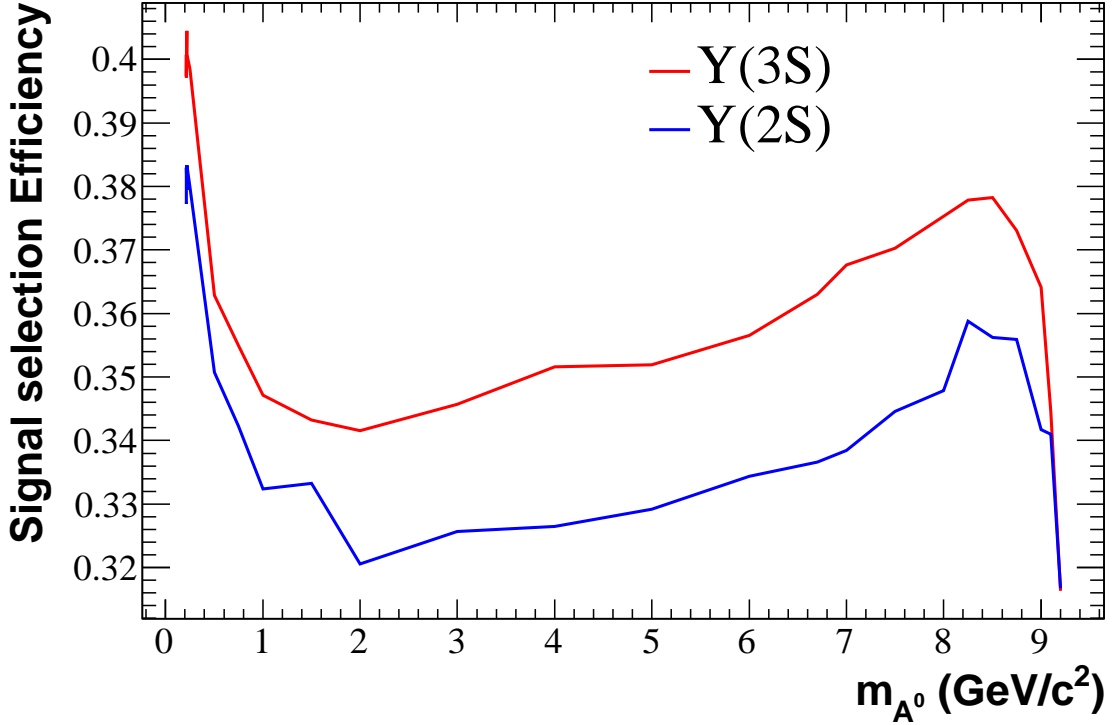}
\caption {The signal selection efficiency as a function of $m_{A^0}$.}
\label{fig:Eff}
\end{figure}

The fit to the $m_{\rm red}$ distributions for the signal MC for the
selected mass points are shown in Figure~\ref{fig:SigPDF}. Rest of the
other plots are shown in Appendix~\ref{AppendixA} in
Figure~\ref{fig:SigPDFY2S1} --~\ref{fig:SigPDFY2S3} for $\Upsilon(2S)$
and in Figure~\ref{fig:SigPDFY3S1} --~\ref{fig:SigPDFY3S2} for
$\Upsilon(3S)$ dataset. The summary of the PDF parameters for both datasets
are shown in Figure~\ref{fig:Pdfparles0.5} -- ~\ref{fig:Pdfparget0.5}. Figure~\ref{fig:Eff} shows the signal selection efficiency as a function $m_{A^0}$ for both the datasets. The PDF parameters of the signal are interpolated linearly from the known $m_{A^0}$ points.

\begin{figure}
\centering
\includegraphics[width=3.0in]{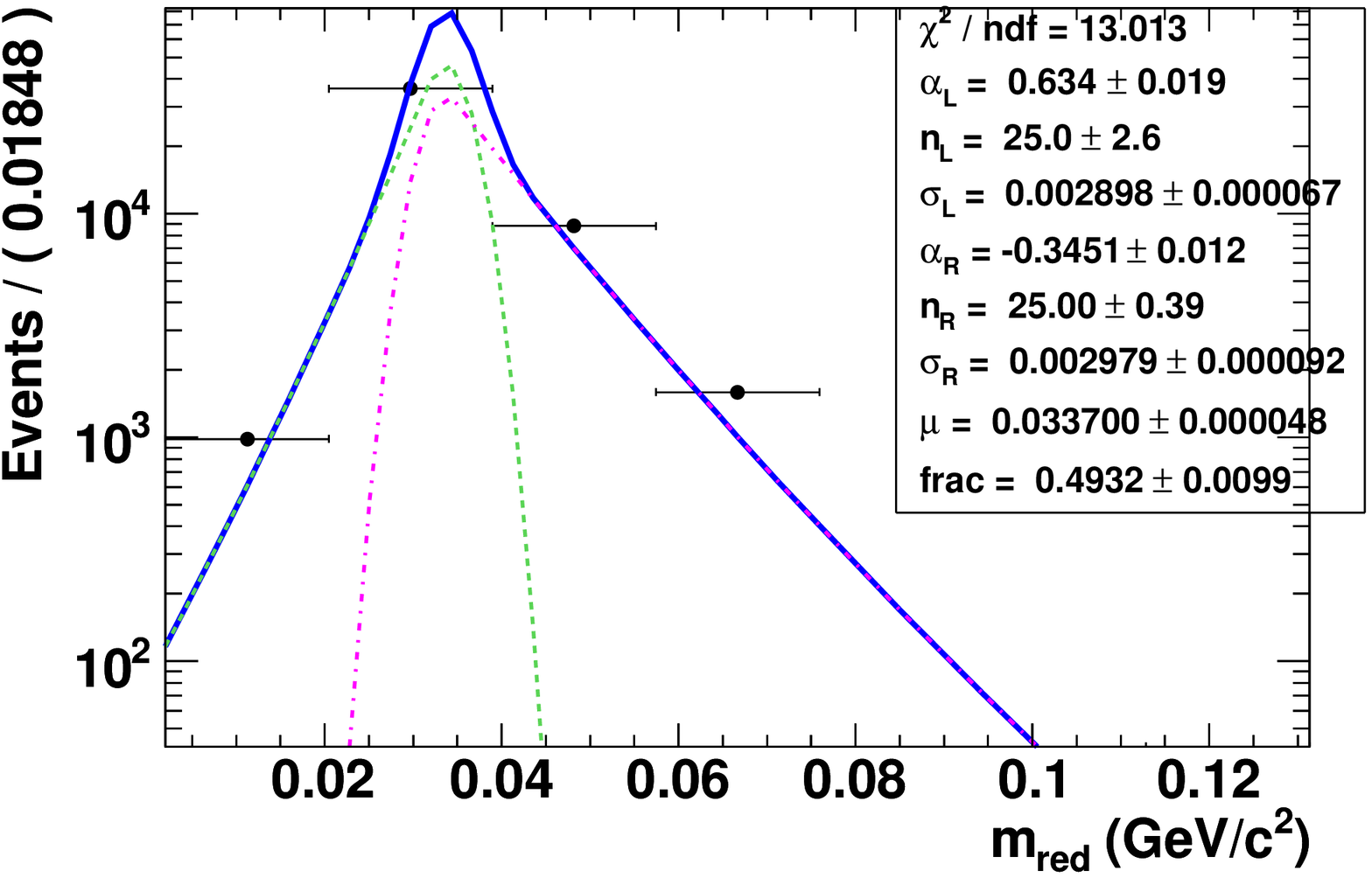}
 \includegraphics[width=3.0in]{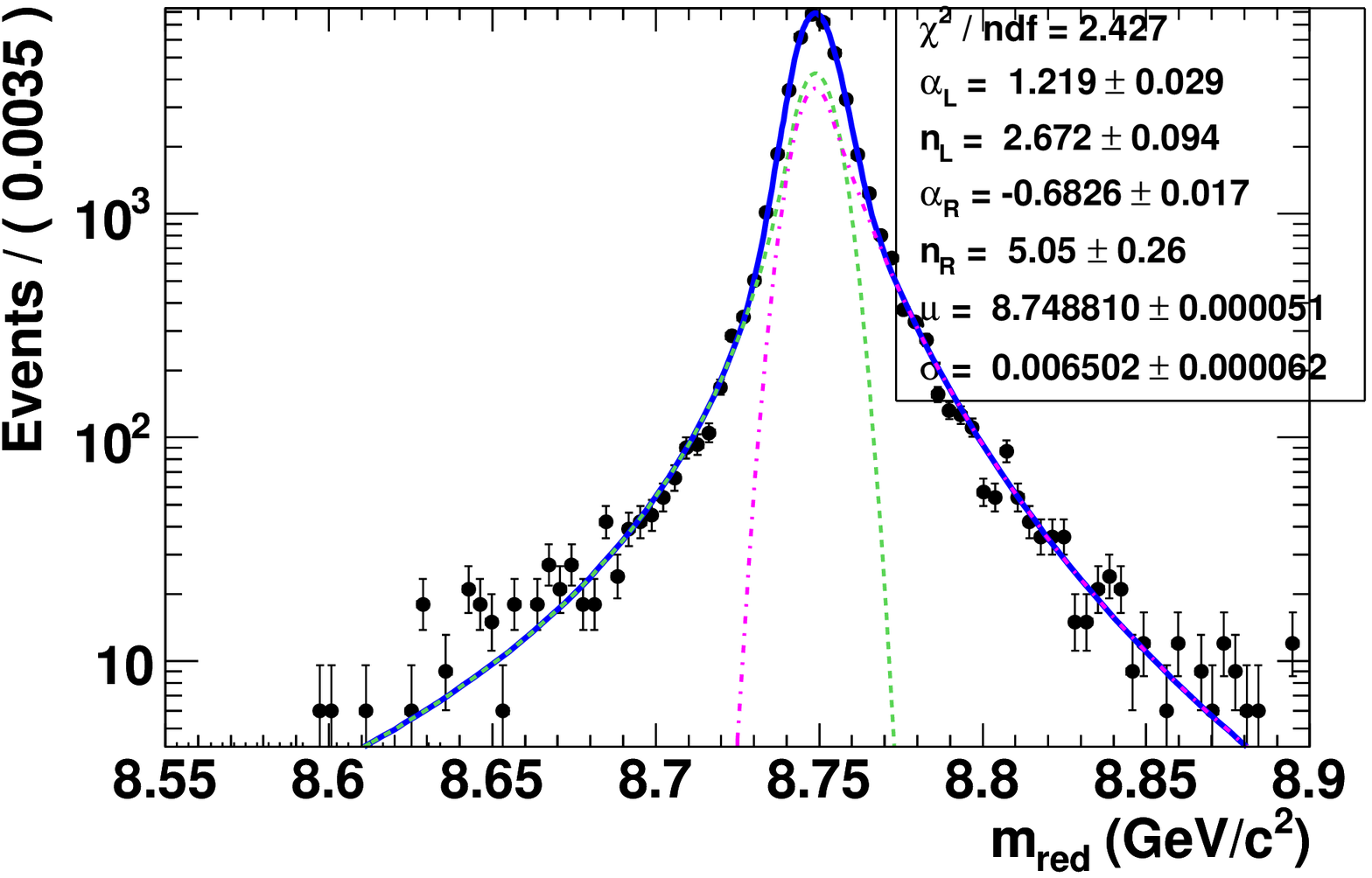}
\smallskip
\centerline{\hfill (j) \hfill \hfill (k) \hfill \hfill (l) \hfill}
\smallskip
\caption {Signal PDFs for the Higgs mass of  (a) $m_{A^0}=0.214$ GeV/$c^2$ and  (b) $m_{A^0}=8.75$ GeV/$c^2$.}
\label{fig:SigPDF}
\end{figure}

\section{Background PDF}
The background PDF in the range of $m_{A^0} \le 1.50$ \gevcc is
modelled using a MC sample of $\Upsilon(2S,3S)\rightarrow \pi^+\pi^-
\Upsilon(1S)$, $\Upsilon(1S) \rightarrow (\g) \mu^+\mu^-$ decays,
which is described by a threshold function 

\begin{equation}
f(m_{\rm red}) \propto [Erf(s(m_{\rm red}-m_0))+1] + exp(\sum_{\ell=0}^1 c_{\ell} m_{\rm red}^{\ell}),
\label{eq:threshold}
\end{equation}

\noindent where $s$ is a threshold parameter and $m_0$ is determined by the kinematic end point of the $m_{\rm red}$ distribution, and $c_{\ell}$ is the coefficient of $\ell^{th}$ order polynomial function. The background PDF is described by a second order Chebyshev polynomial in the range of $1.502 \le m_{A^0} \le 7.10$ \gevcc, and a first order Chebyshev
polynomial for $m_{A^0}>7.10$ \gevcc. The plots of
background PDF near the threshold mass region are shown in
Figure~\ref{fig:threshold} for both $\Upsilon(2S,3S)$ datasets. Rest
of the other background PDFs are shown in the Appendix~\ref{AppendixB}
in Figure~\ref{fig:BackPDFY2S} and ~\ref{fig:BackPDFY3S} for
$\Upsilon(2S)$ and $\Upsilon(3S)$, respectively.

\begin{figure}
\centering
 \includegraphics[width=3.0in]{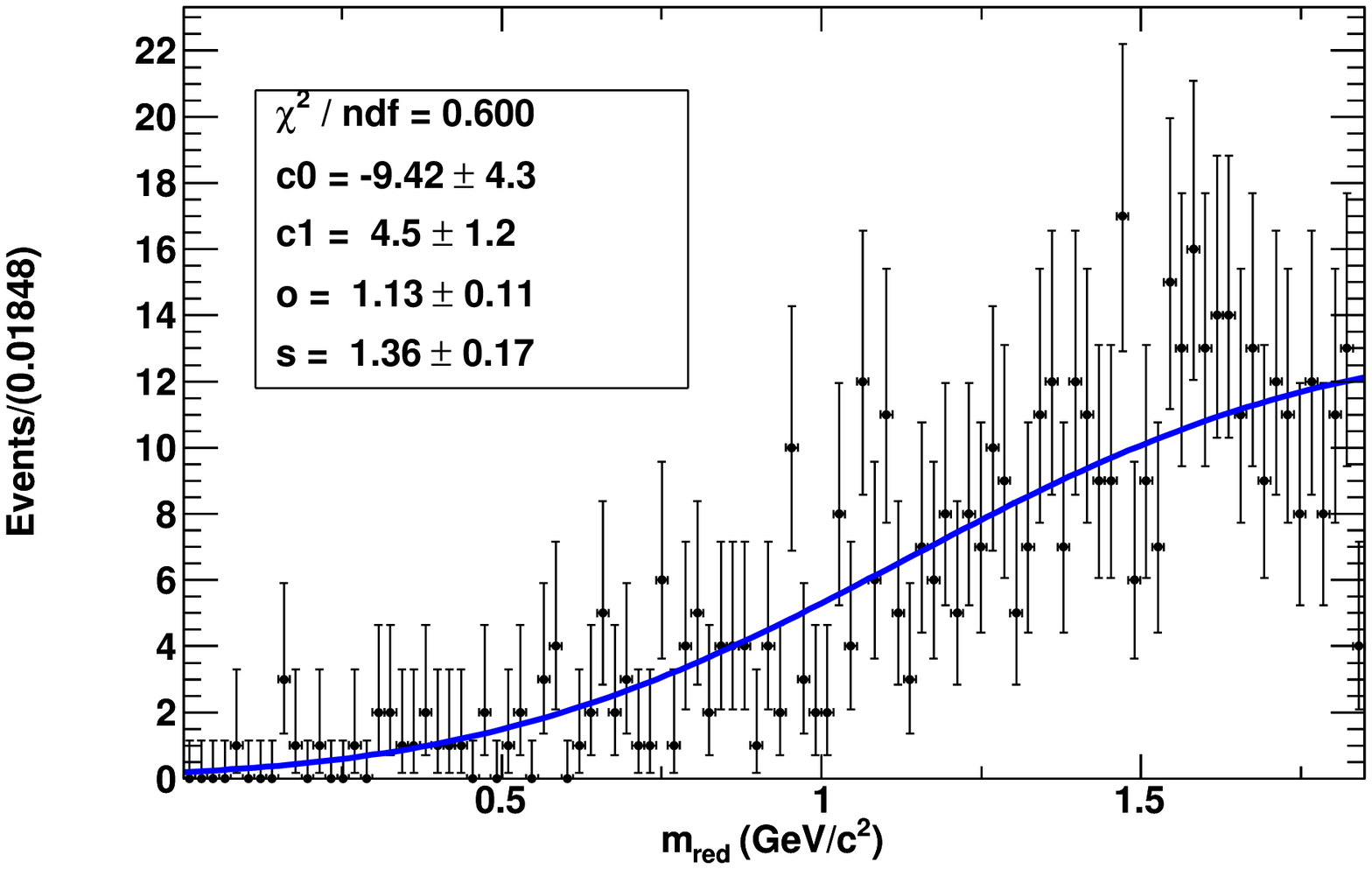} 
\includegraphics[width=3.0in]{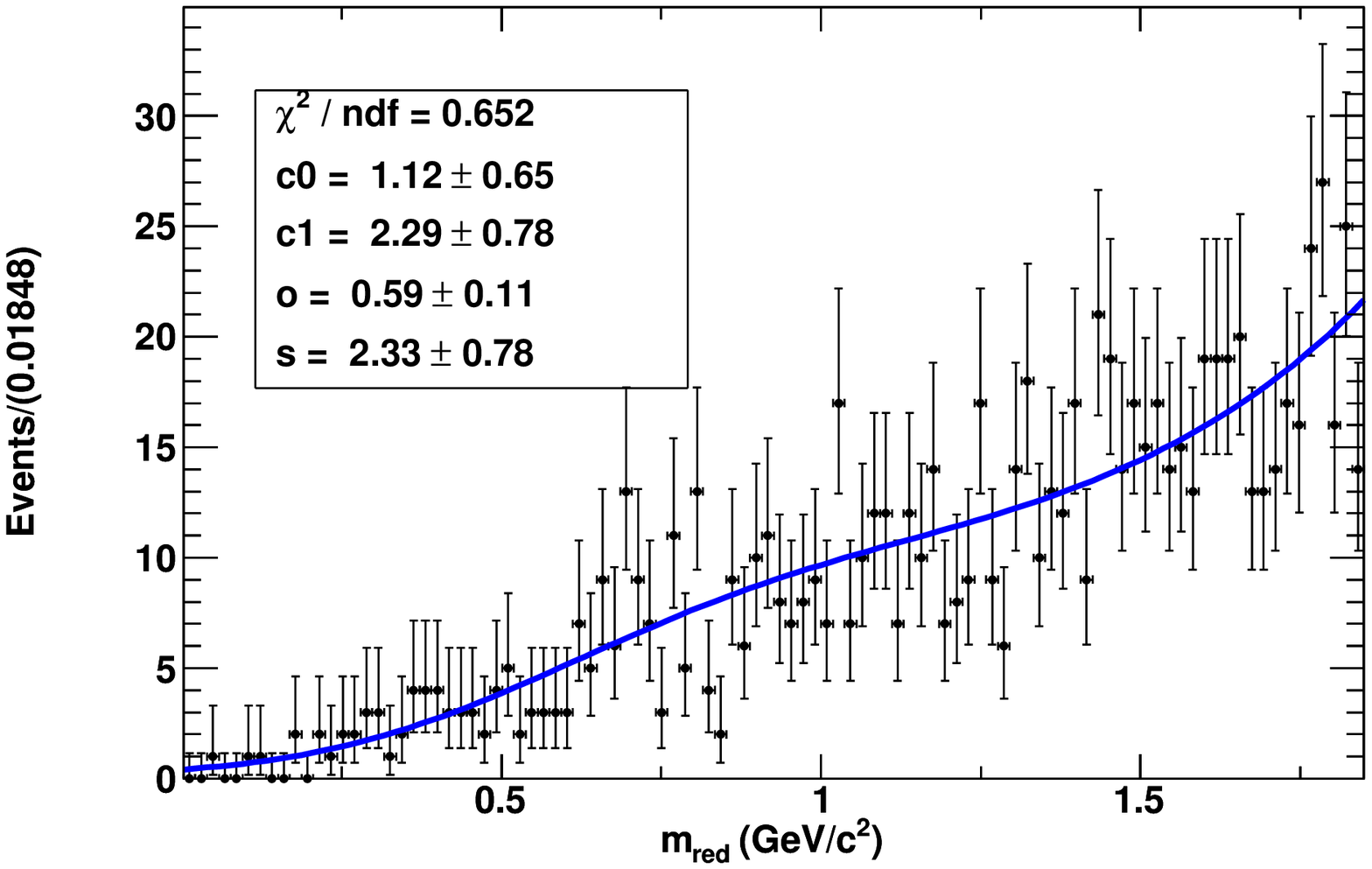}

\caption {The background PDF in the threshold mass region in the
$m_{\rm red}$ distribution. Left plot is for the $\Upsilon(2S)$ dataset
and right plot is for the $\Upsilon(3S)$ dataset.}
\label{fig:threshold} 
\end{figure}

 \subsection{Fit Validation using a  cocktail sample}
\begin{figure}
\centering
\includegraphics[width=3.0in]{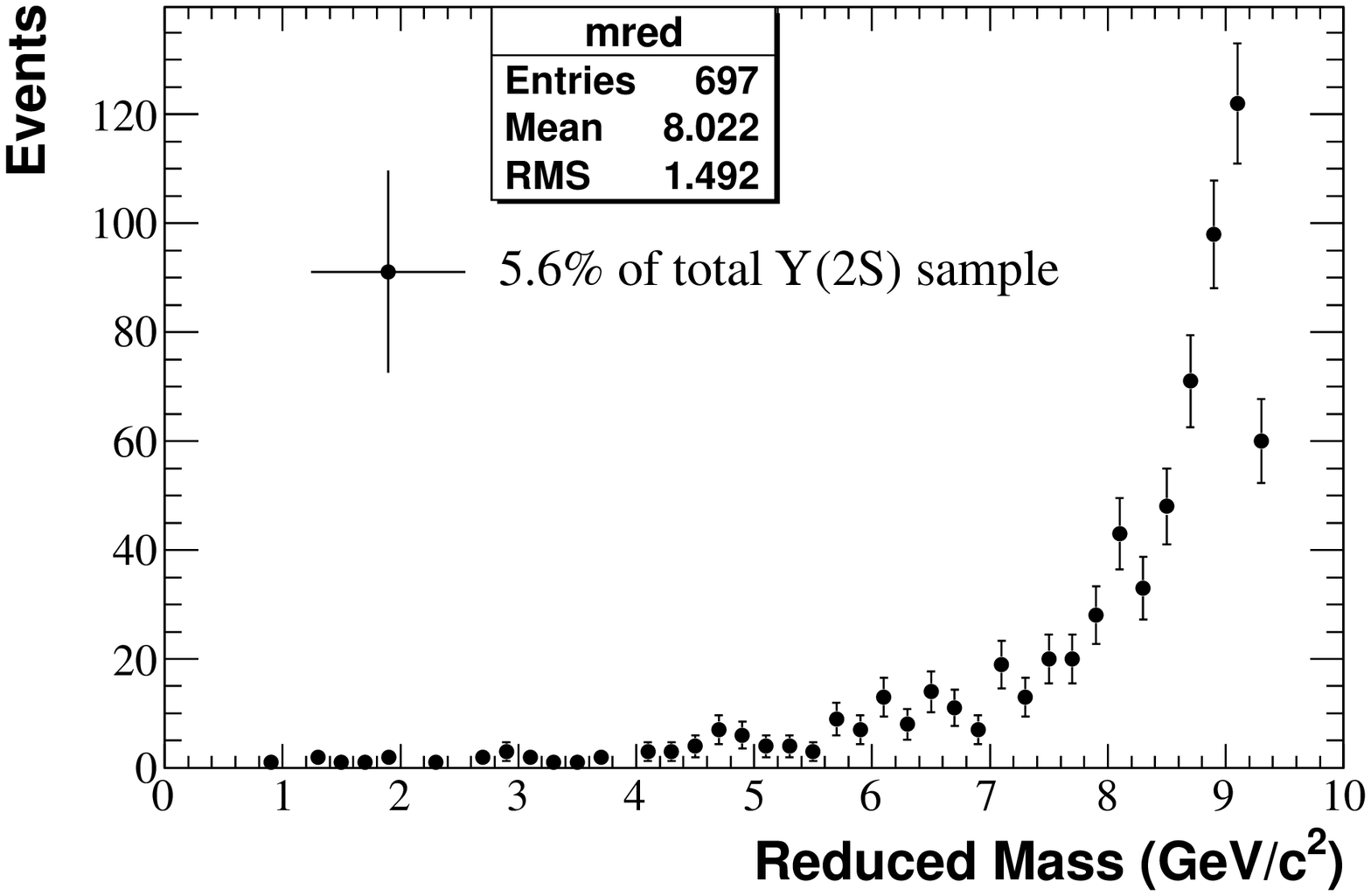} 
\includegraphics[width=3.0in]{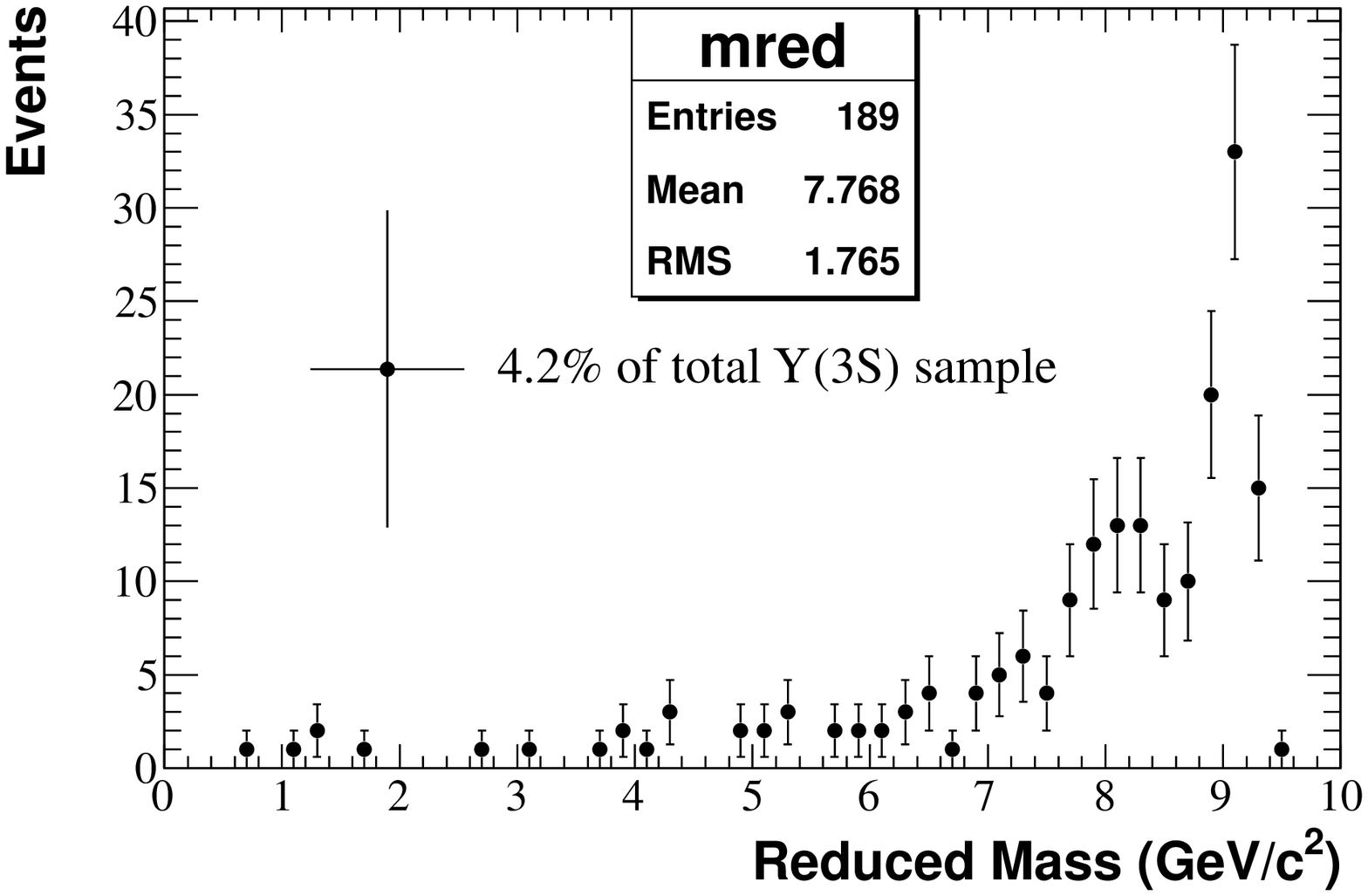} 
\smallskip
\centerline{\hfill (c) \hfill \hfill (d) \hfill}
\smallskip

\includegraphics[width=3.0in]{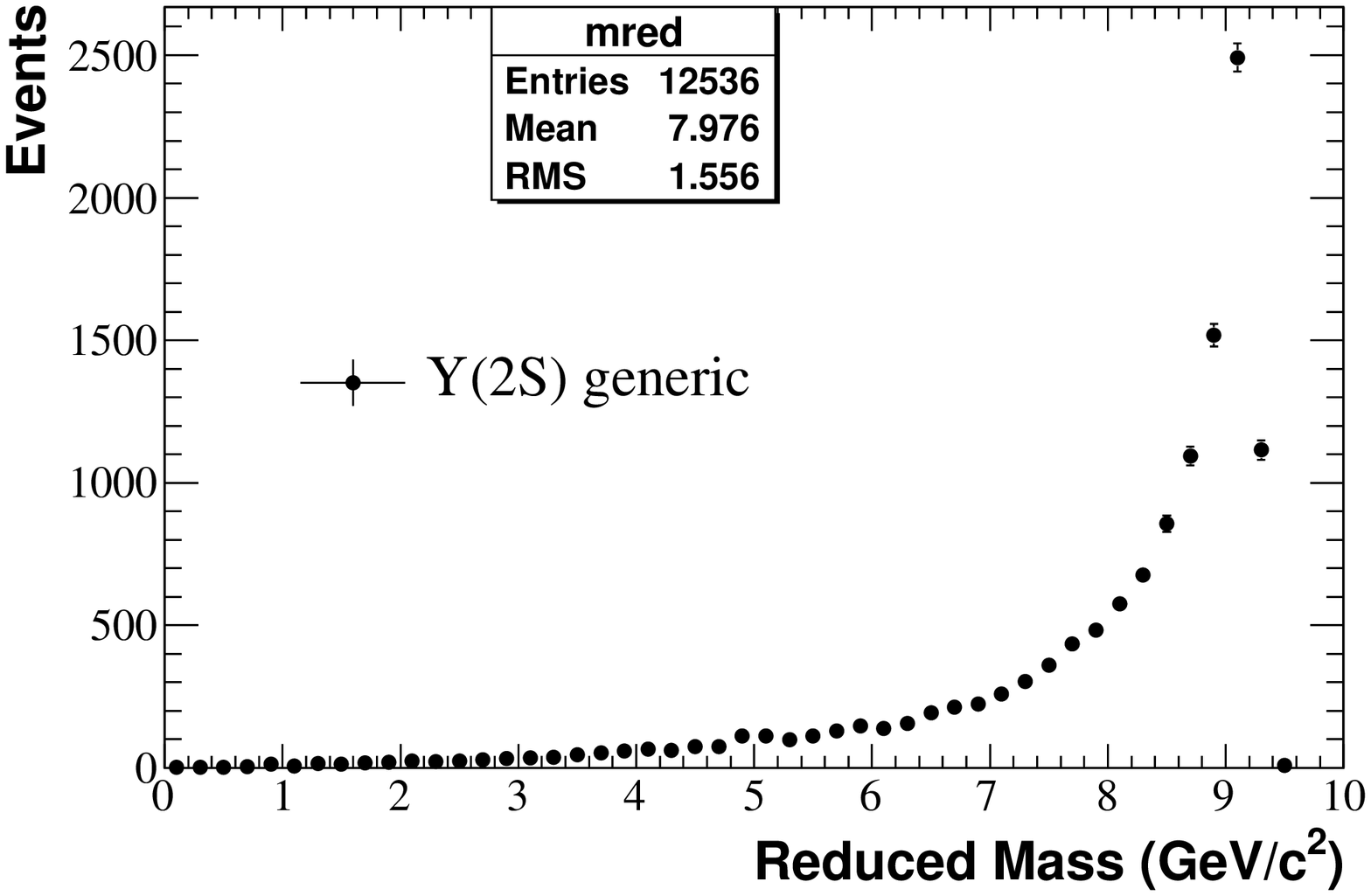} 
\includegraphics[width=3.0in]{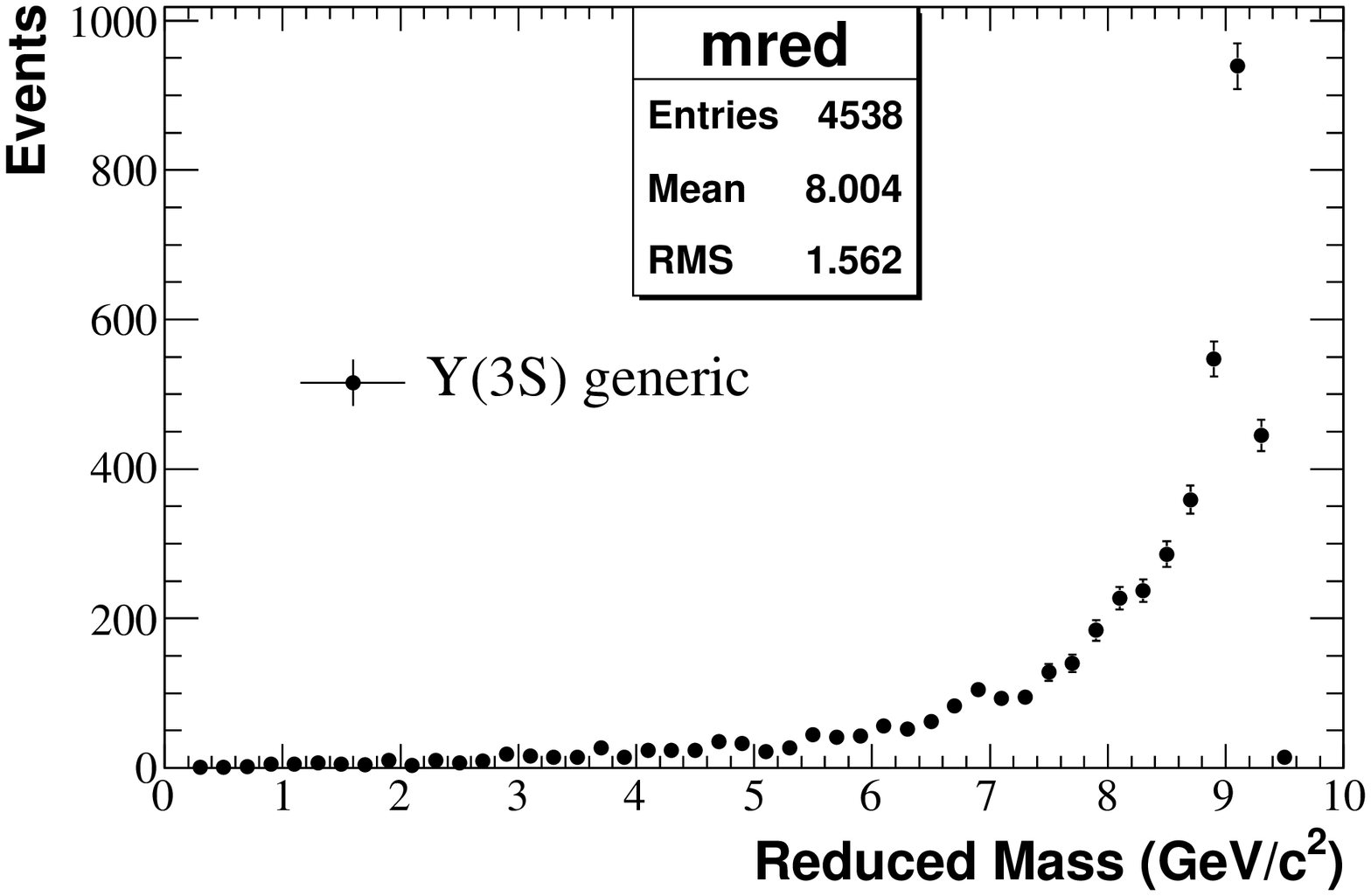} 

\smallskip
\centerline{\hfill (c) \hfill \hfill (d) \hfill}
\smallskip

\caption {$m_{\rm red}$ distribution for $\Upsilon(3S,2S)$ low onpeak and $\Upsilon(3S,2S)$ generic samples. Left plots are for $\Upsilon(2S)$ right plots are for $\Upsilon(3S)$.}
\label{fig:mredonpeak}
\end{figure}

   The validation of the fit procedure is performed using a cocktail
   sample of the $\Upsilon(3S, 2S)$ low onpeak data-sample and $95\%$
   of $\Upsilon(3S, 2S)$ generic MC sample. The cocktail sample contains
   about 4522 events for $\Upsilon(3S)$ and about 12446 events for
   $\Upsilon(2S)$, as  expected in the full data samples.
   Figure~\ref{fig:mredonpeak} shows the reduced mass distribution for
   $\Upsilon(3S,2S)$ low onpeak and $\Upsilon(3S,2S)$ generic samples
   after applying all the selection criteria. As seen in these figures
    the statistics is very limited in the low mass region in both
   the datasets. There are many regions in the $m_{\rm red}$
   distribution where there are no events. The normal ML fit procedure
   gives large negative signal yield in a region of the $m_{\rm red}$
   spectrum, where the statistics is limited. This problem can be
   avoided if we constraint the number of signal and background events
   to be greater or equal to zero. This constraint method works fine
   in the region of limited statistics and ignores the negative
   fluctuation in the datasets but introduces a bias, specially, where
   the statistics is little bit large, but not sufficient to use the
   normal fitting approach. To avoid these difficulties, we impose a
   lower cutoff to the signal yield to ensure that the total signal
   plus background PDF remains non-negative in the integration region
   \cite{PRL88-24}.

\begin{figure}
\centering
\includegraphics[width=3.0in]{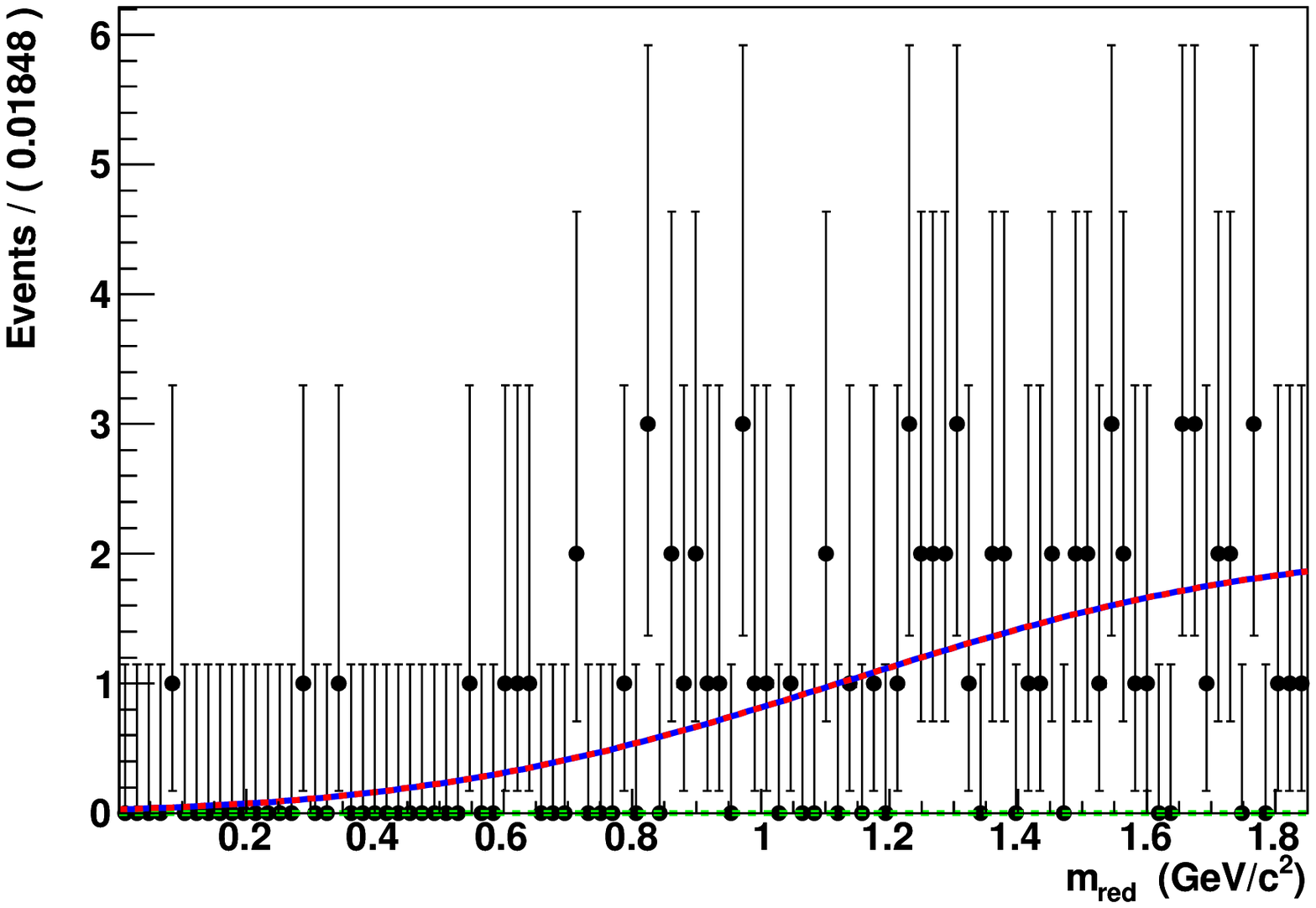}
\includegraphics[width=3.0in]{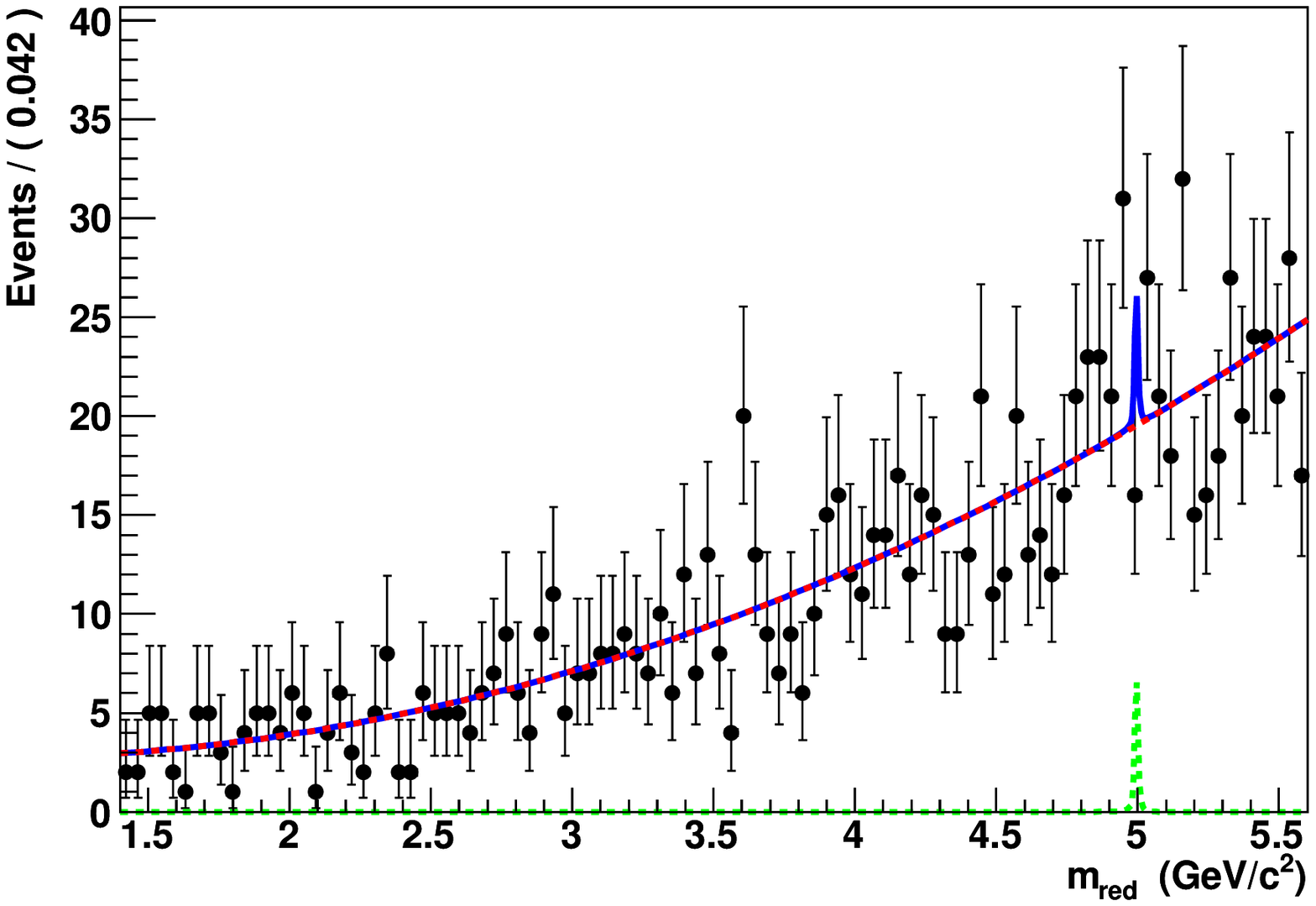}

\includegraphics[width=3.0in]{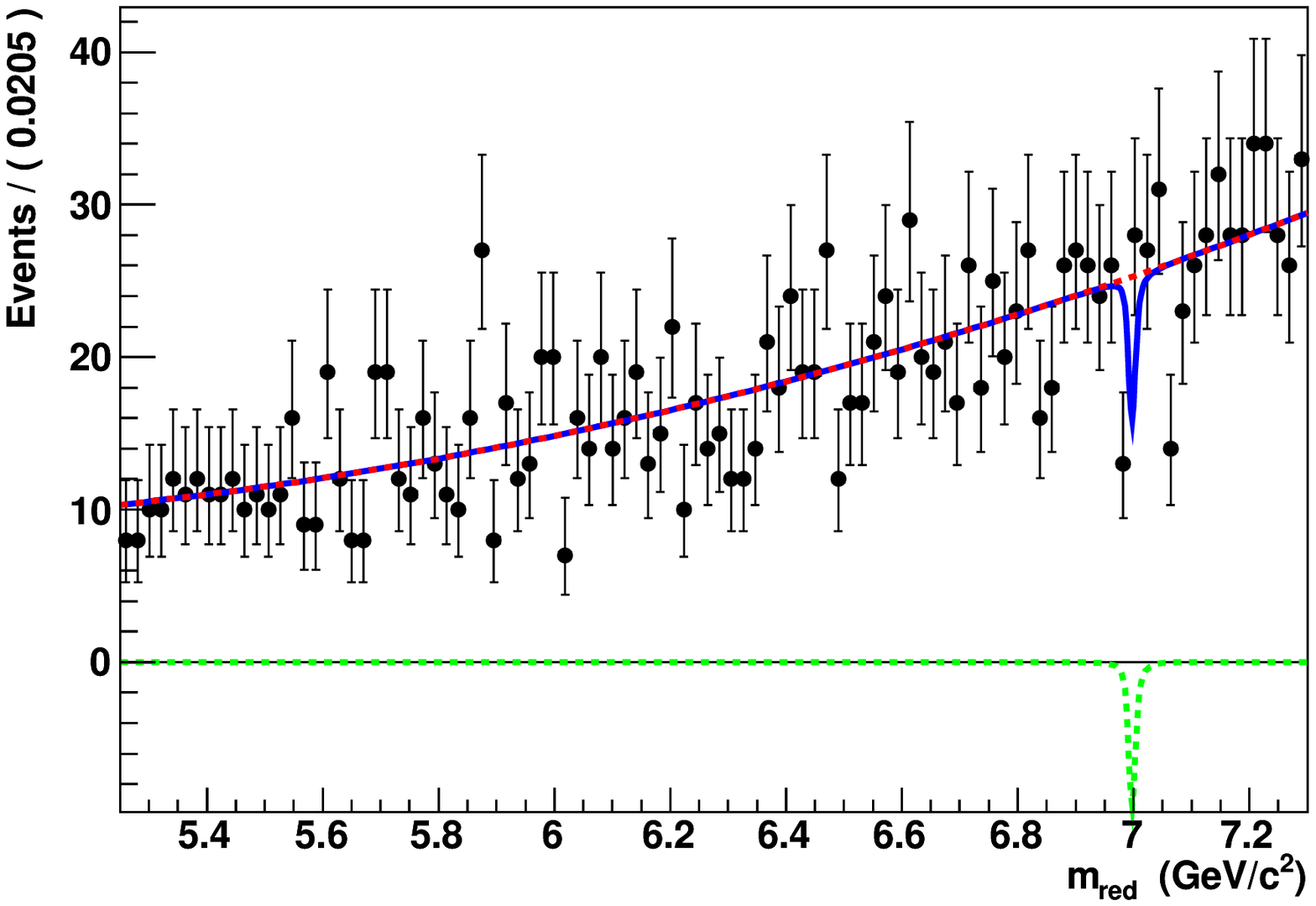} 
\includegraphics[width=3.0in]{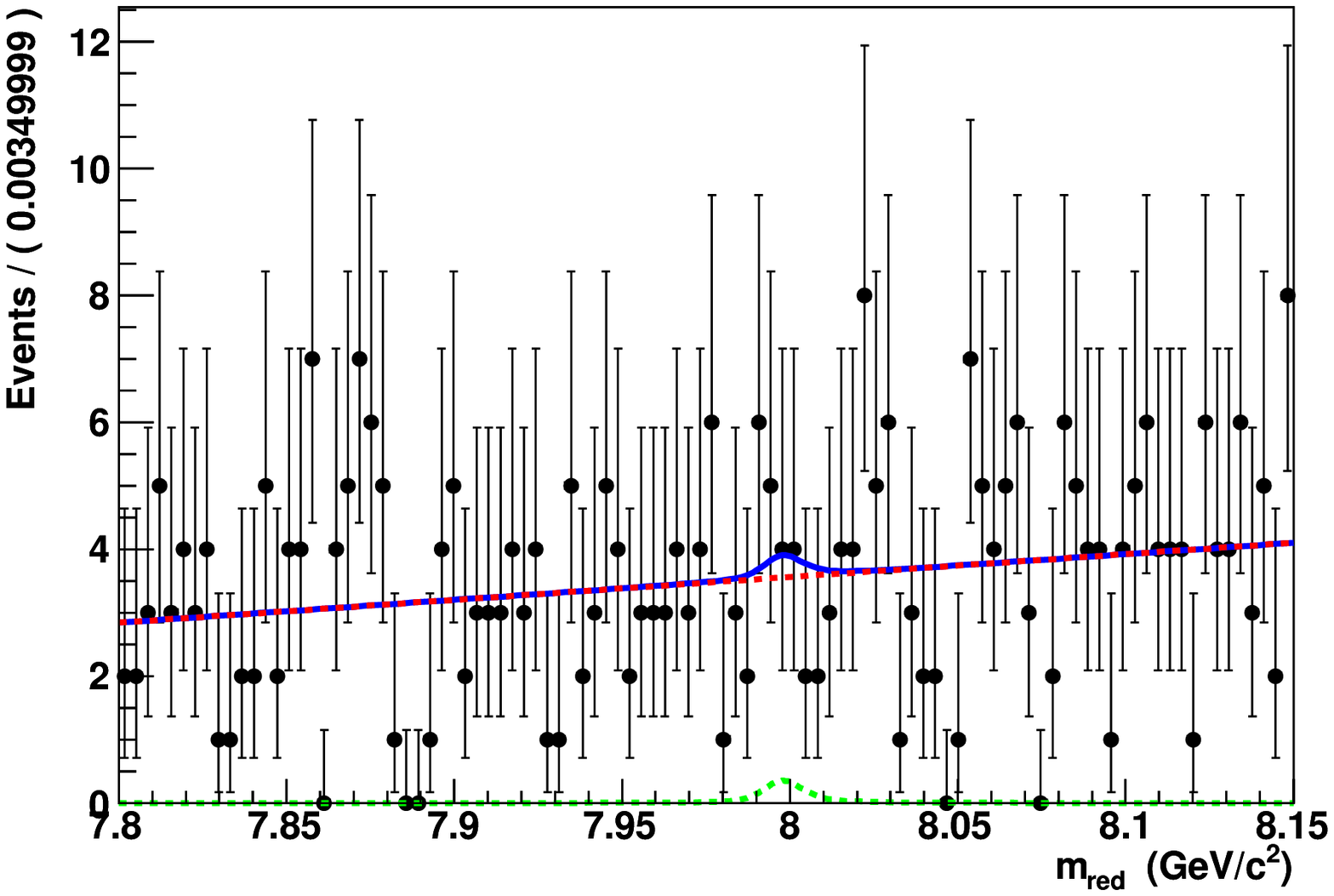}

\includegraphics[width=3.0in]{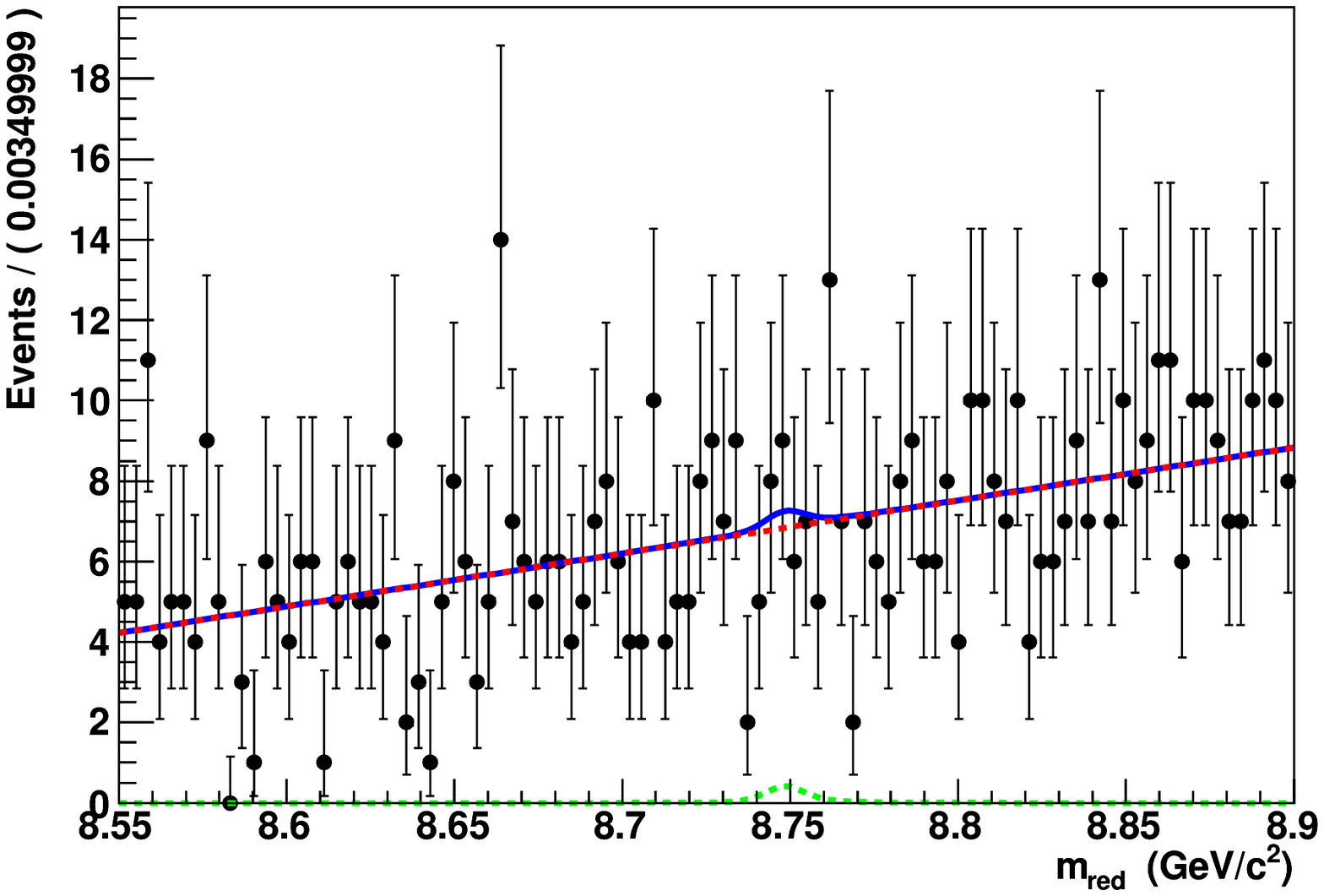} 
\includegraphics[width=3.0in]{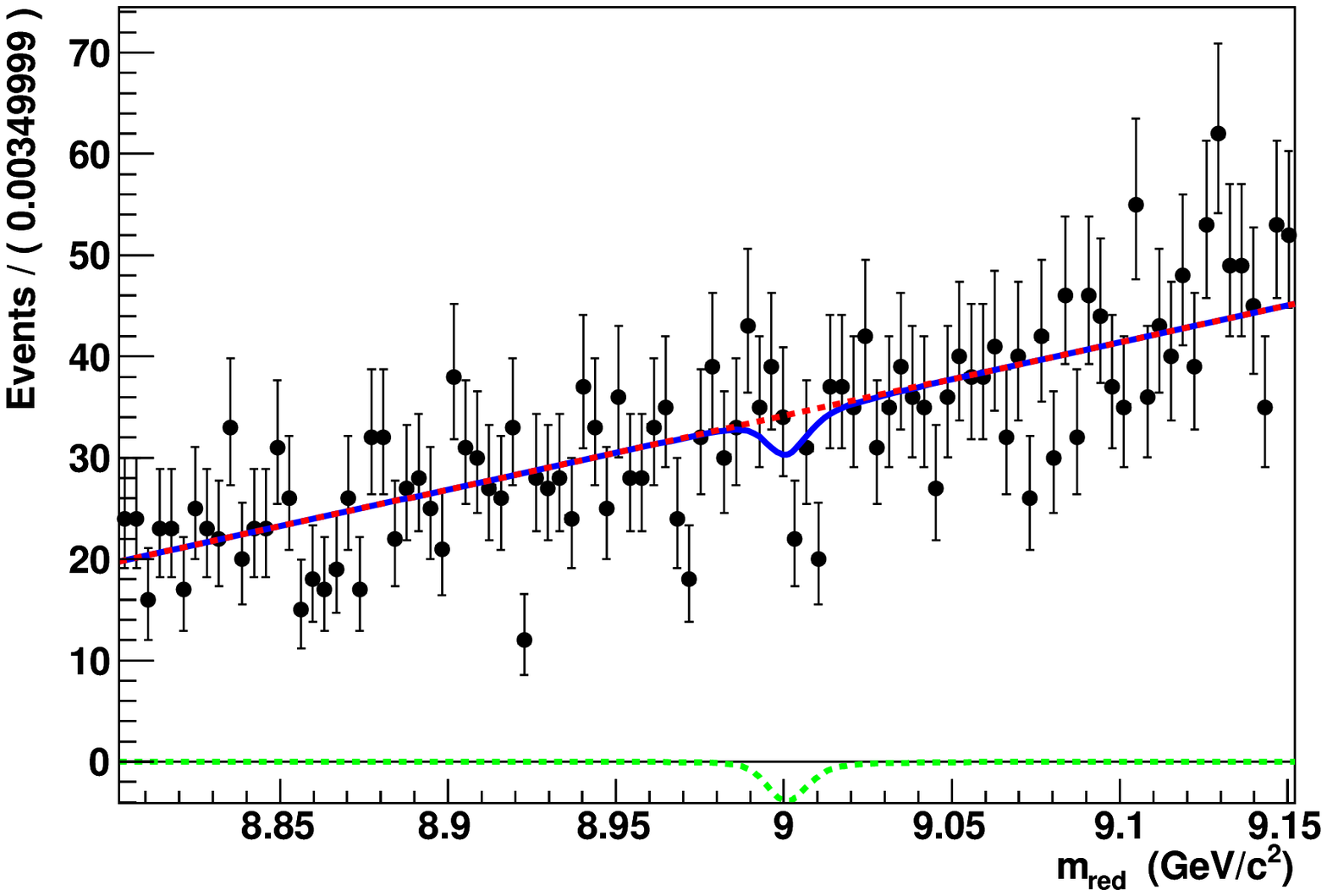}

\caption {Projection plot onto $m_{\rm red}$ distributions at selected $m_{A^0}$ points using the cocktail samples of $\Upsilon(2S,3S)$. The total ML fit is shown in solid blue; the non-peaking background component is shown in dashed green; the signal component is shown in green dashed.} 

\label{fig:Projplot}
\end{figure}

\begin{figure}
\centering
 \includegraphics[width=6.0in]{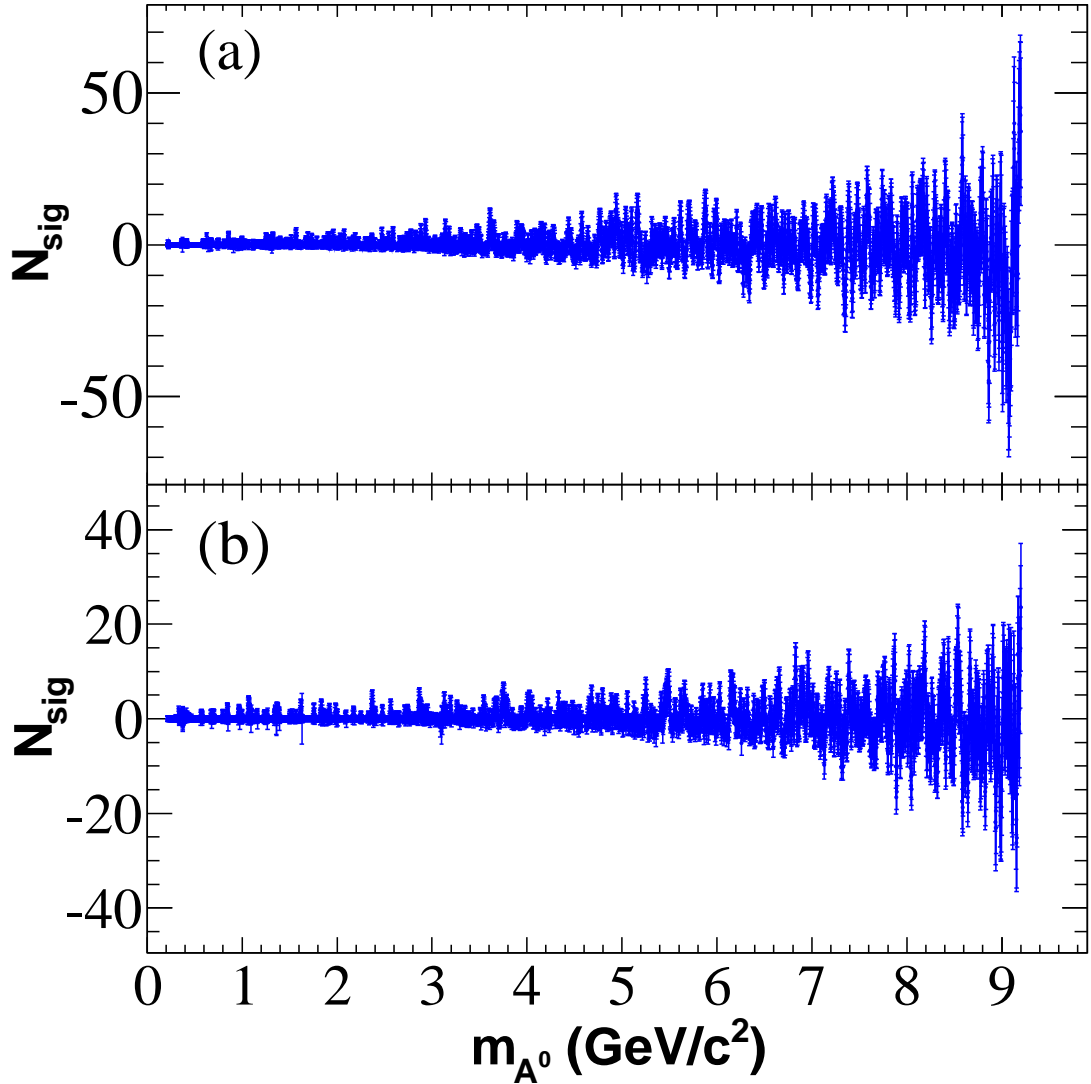}
\caption {The number of signal events ($N_{sig}$) ~as a function of $m_{A^0}$ for (a) the $\Upsilon(2S)$ dataset and (b) the $\Upsilon(3S)$ dataset. These plots are generated using the $\Upsilon(2S,3S)$ cocktail samples.}
\label{fig:Yield}
\end{figure}

\begin{figure}
\centering
 \includegraphics[width=6.0in]{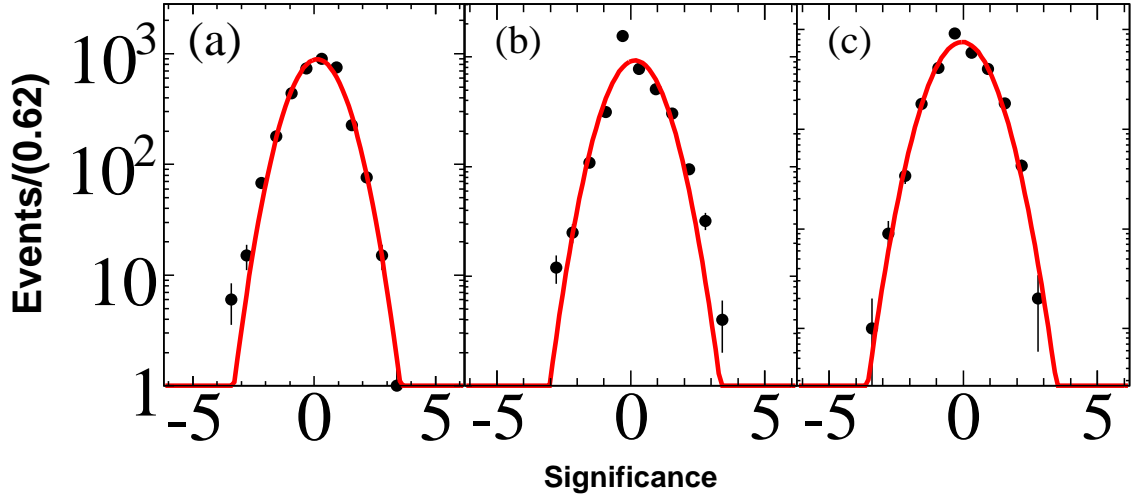}
\caption {The signal significance ($\mathcal{S}$) as a function of $m_{A^0}$ for (a) the $\Upsilon(2S)$ dataset, (b) the $\Upsilon(3S)$ dataset and (c) the combined data of $\Upsilon(2S,3S)$. These plots are generated using the $\Upsilon(2S,3S)$ cocktail samples.} 
\label{fig:Significance}
\end{figure}

    We perform the scan for any possible peaks in the $m_{\rm red}$ distribution from $\Upsilon(3S, 2S)$ cocktail samples in the steps of half of $m_{\rm red}$ resolution, corresponding to 4585 points. The shape of the signal-PDF is fixed while the background-PDF shape, signal and background yields are allowed to float.  The parameters of the signal PDF are interpolated between the known MC points. The representative plots of the 1d ML fit to the $m_{\rm red}$ distributions are shown in Figure~\ref{fig:Projplot} at some selected $m_{A^0}$ points. The signal events ($N_{sig}$) as a function of $m_{A^0}$ are shown in Figure~\ref {fig:Yield}. We also calculate a statistical significance ($\mathcal{S}$)  which is defined as:

 \begin{equation}
\mathcal{S}  = {\rm sign}({N_{sig}} )\sqrt{-2ln(\mathcal{L}_{0}/\mathcal{L}_{max})},
 \end{equation} 

\noindent where $\mathcal{L}_{max}$ is the maximum likelihood value of a fit with a floating signal yield centered at $m_{A^0}$, and $\mathcal{L}_{0}$ is the likelihood value for the null hypothesis. Figure~\ref{fig:Significance} shows the significance distributions for both the $\Upsilon(2S,3S)$ cocktail datasets. The significance  barely deviates more than $3\sigma$ for both the datasets. We also compute the combined significance of the $\Upsilon(2S,3S)$ datasets, which is defined as:

\begin{equation}
     \mathcal{S} = \frac{w_{\Upsilon(2S)}S_{\Upsilon(2S)} + w_{\Upsilon(3S)}S_{\Upsilon(3S)}}{\sqrt{w_{\Upsilon(2S)}^2 + w_{\Upsilon(3S)}^2}},
\end{equation}

\noindent where $S_{\Upsilon(2S,3S)}$ is the significance of the $\Upsilon(2S,3S)$ data-sets, computed at each scanned $m_{A^0}$ points  and $w_{\Upsilon(2S,3S) = 1/\sigma_{N_{sig}}^2}$ is the weight of the  each data-sets.

\subsection{Fit validation using Toy Monte-Carlo}
\label{section:ToyMC} 
We use a large number of toy Monte-Carlo experiments to validate the fit
 procedure further. We first fit the background PDF's to the
 $\Upsilon(3S, 2S)$ cocktail samples. Then, we generate the background
 events according to those PDFs, setting the background yields to the
 number expected in the Run7 $\Upsilon(3S, 2S)$ Onpeak datasets. The
 toy studies are done with different embedded signal events for each
 $m_{A^0}$ points.

\begin{figure}
\centering
 \includegraphics[width=3.0in]{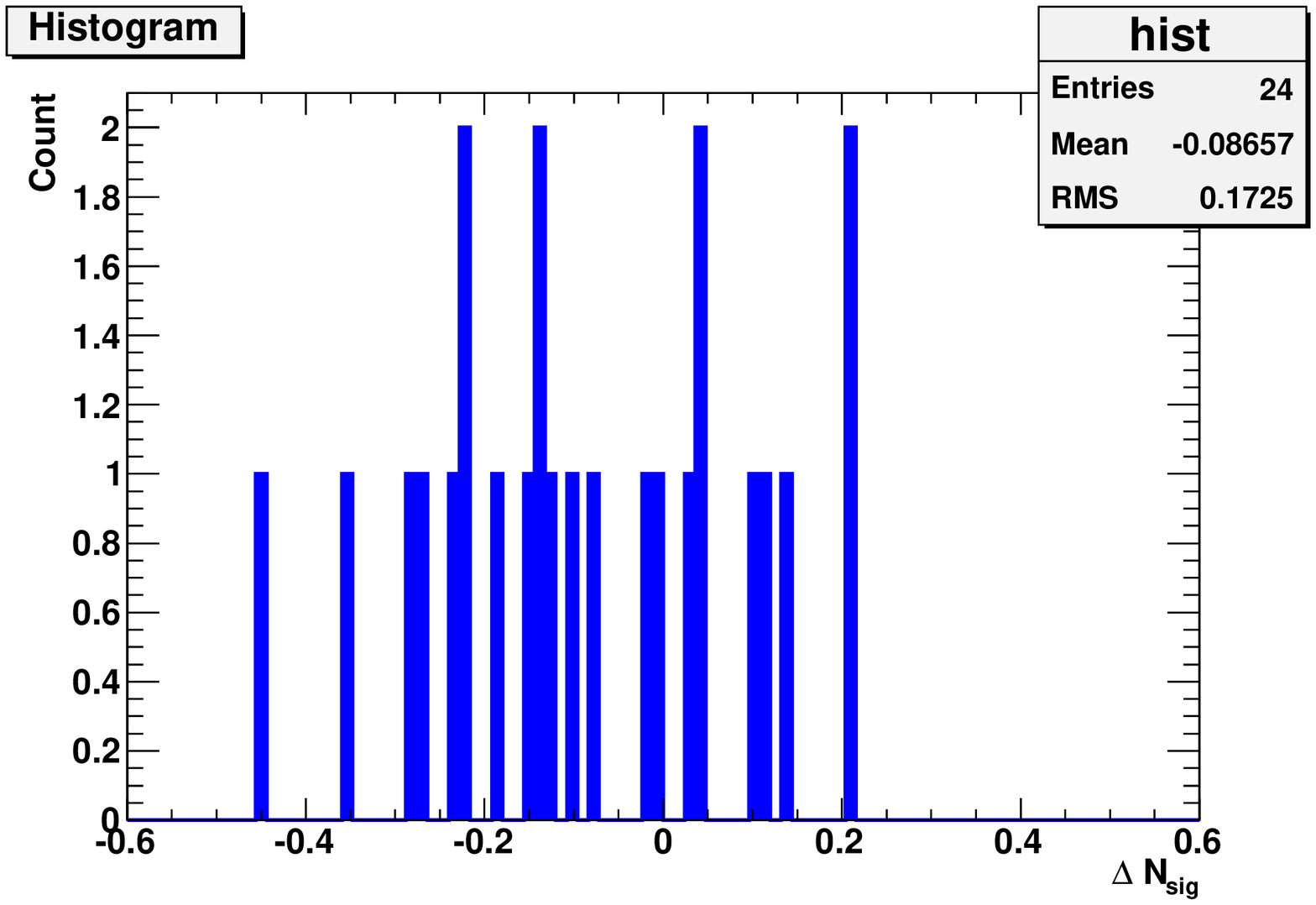}
 \includegraphics[width=3.0in]{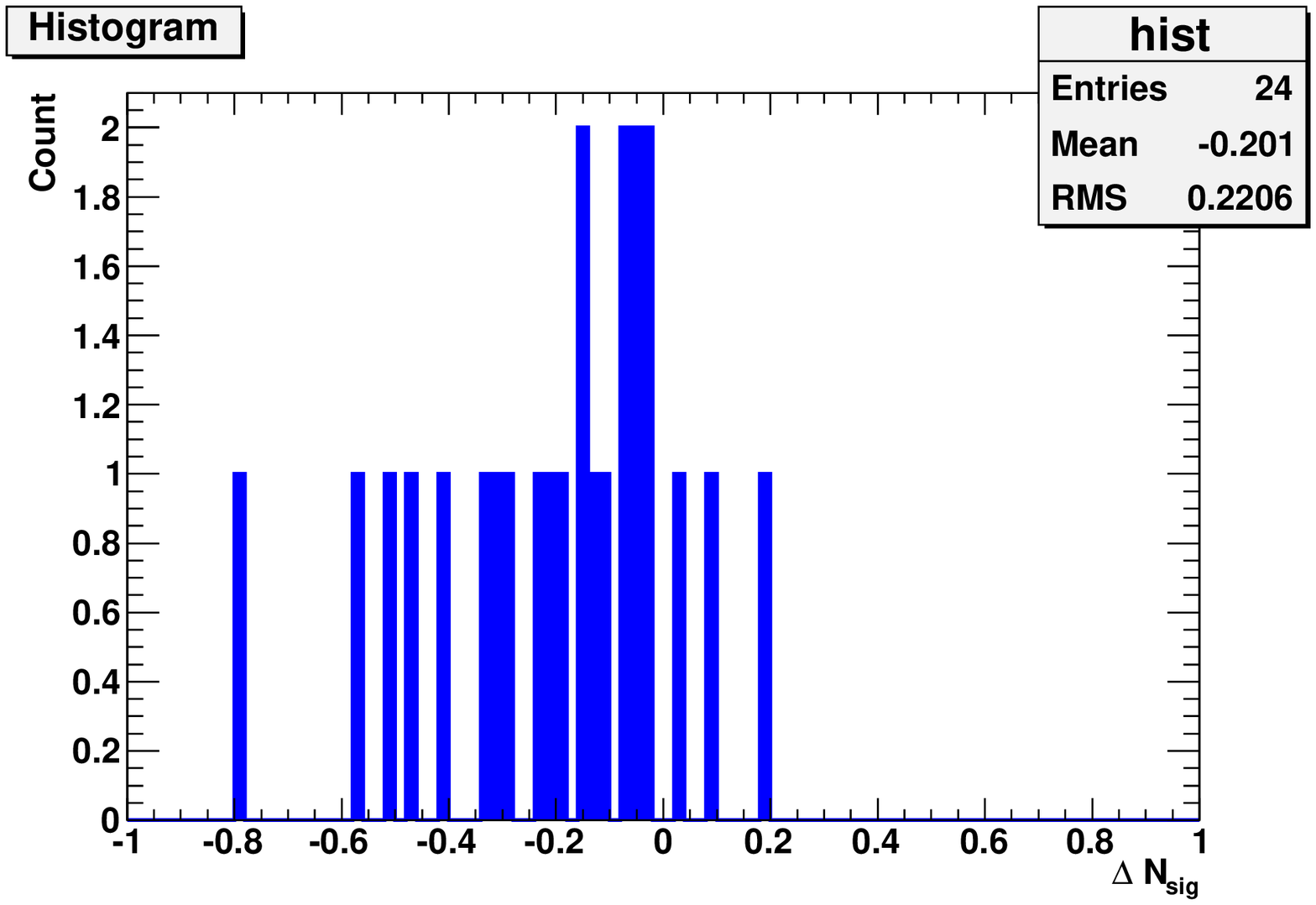}

\caption {Histograms of the intercept of the regression for $\Upsilon(2S)$ (left) and $\Upsilon(3S)$ (right). The fit-residuals as a function of the embedded signal event are fitted by a linear function for each $m_{A^0}$ points and intercept of the regression is accumulated in the Histogram.}

\label{Fig:FitBias}
\end{figure}

            The average fit-residuals (the difference between the
            number of fitted and generated events) as a function of
            embedded signal events for each $m_{A^0}$ are summarized
            in Appendix~\ref{AppendixC} in Figure~\ref{fig:ToyMCY2S1}
            -- ~\ref{fig:ToyMCY2S3} for $\Upsilon(2S)$ and in
            Figure~\ref{fig:ToyMCY3S1} -- ~\ref{fig:ToyMCY3S3} for
            $\Upsilon(3S)$. The fit-residual as a function of embedded
            signal event is fitted by a linear function. We accumulate
            the intercept value of the regression in a histogram for
            all the known mass points for both $\Upsilon(2S)$ and
            $\Upsilon(3S)$ [Figure~\ref{Fig:FitBias}]. Since we do not
            observe any significant bias in the fitting procedure, we
            assign the RMS value of the intercept of the regression as
            a systematic uncertainty.  The RMS value of fit bias
            $(\Delta N_{sig})$ is found to be 0.17 for $\Upsilon(3S)$
            and 0.22 for $\Upsilon(2S)$, which will be considered as
            an additional source of systematic uncertainty for
            $\Upsilon(2S)$ and $\Upsilon(3S)$ datasets.

\section{Unblinding the  $\Upsilon(2S,3S)$ datasets.}
After finalizing all the selection criteria and the ML fitting procedure, including the validation of the analysis, we have  unblinded the $(116.8 \pm 1.0)$ million $\Upsilon(3S)$  events (sum of the \rm{\lq\lq High\rq\rq} and \rm{\lq\lq Medium\rq\rq} samples) and $(92.8 \pm 0.8)$ million $\Upsilon(2S)$ events. 

\begin{figure}
\centering
\includegraphics[width=6.0in]{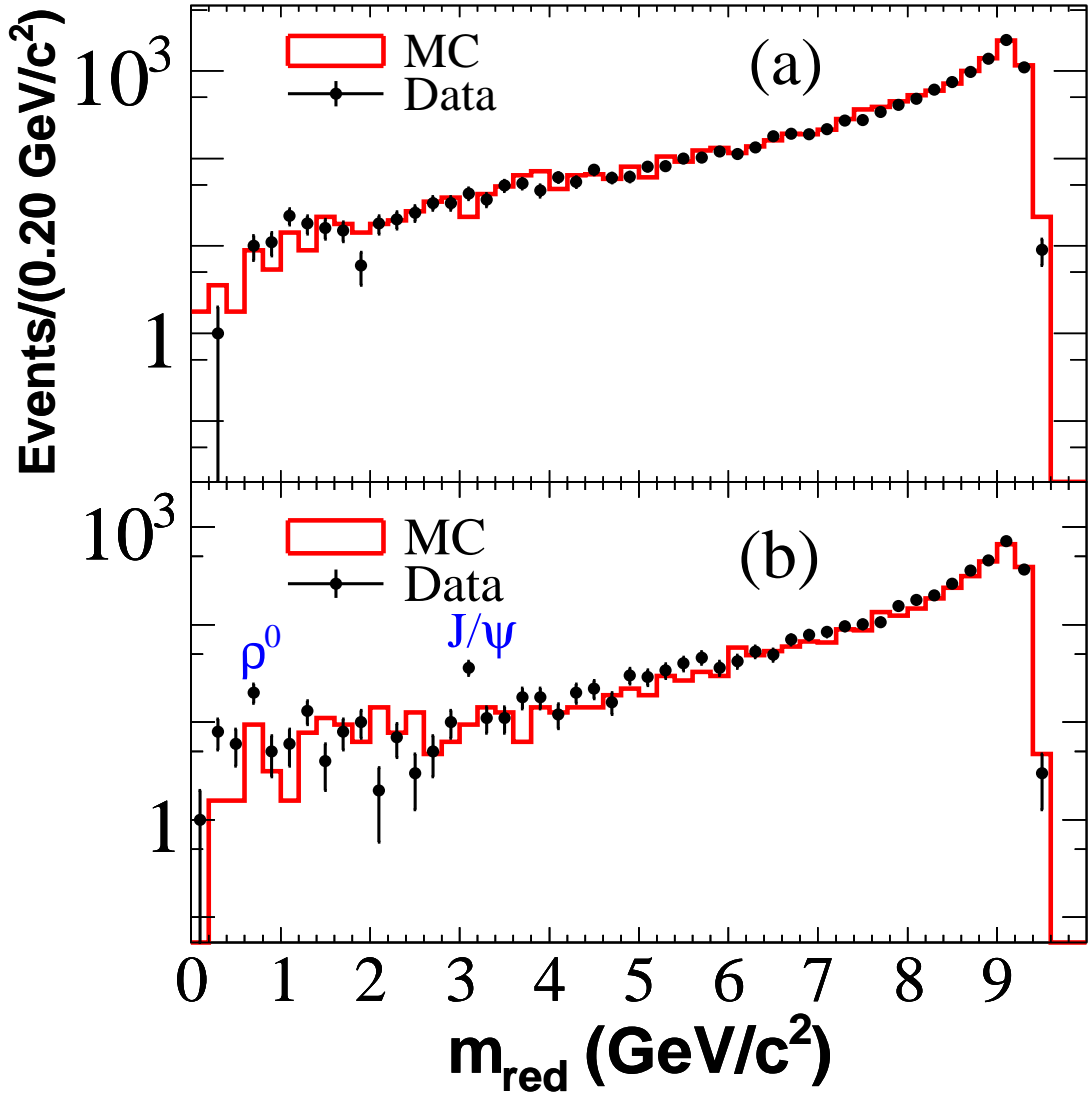}

\caption {The distribution of $m_{\rm red}$  for  (a) the $\Upsilon(2S)$ and (b) the $\Upsilon(3S)$ datasets, together with the remaining background Monte Carlo samples of $\Upsilon(2S.3S) \to \pipi \Upsilon(1S)$, $\Upsilon(1S) \to (\g)\mumu$ decays. The Monte Carlo are normalized to 
the data luminosity. Two peaking 
components corresponding to the $\rho^0$ and \jpsi mesons are observed in the $\Upsilon(3S)$ dataset.}

\label{fig:mred} 
\end{figure}

\begin{figure} 
\centering
\includegraphics[width=6.0in]{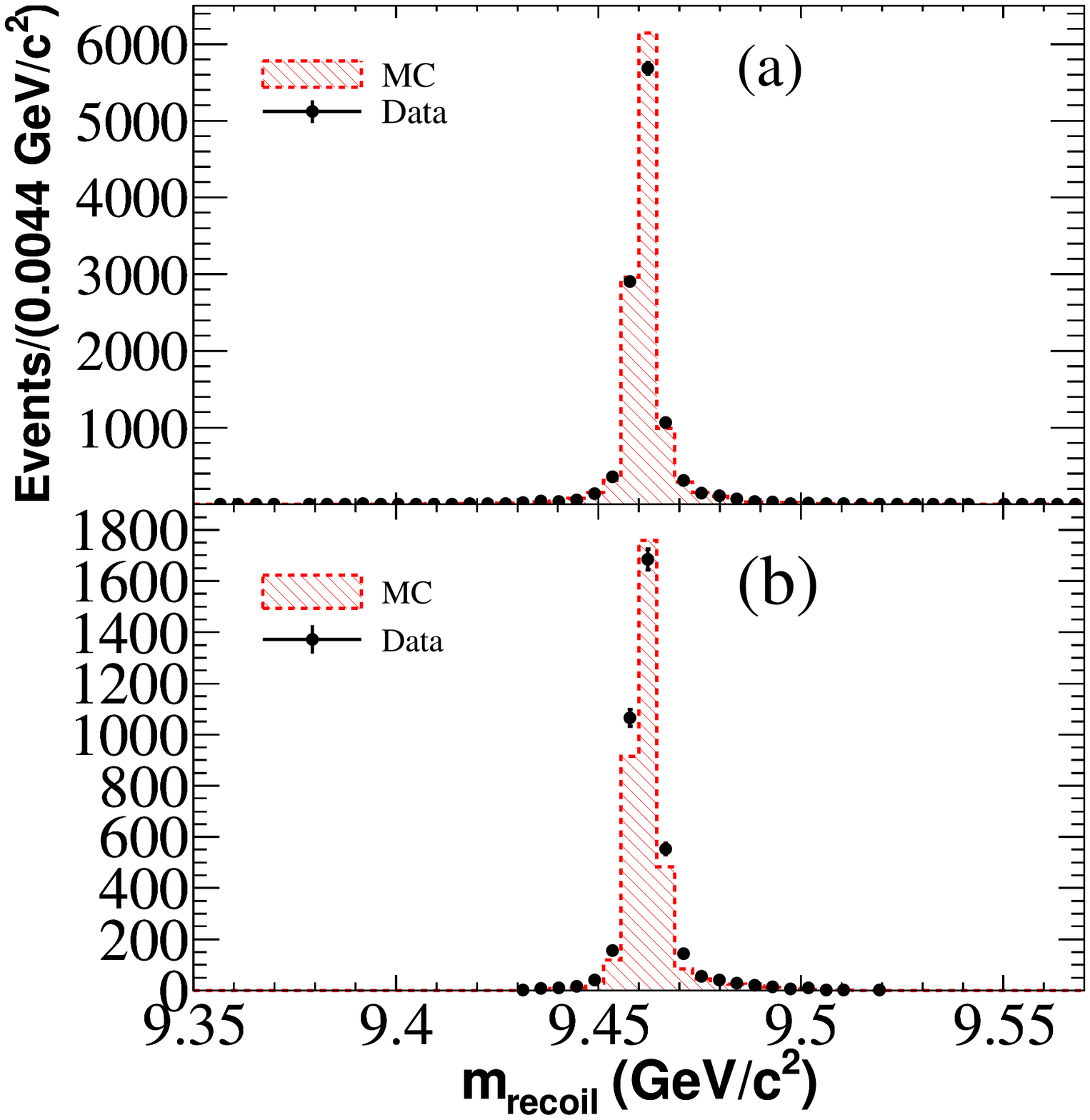}

\caption {The distribution of $m_{\rm recoil}$  for  (a) the $\Upsilon(2S)$ and (b) the $\Upsilon(3S)$ datasets, together with with the remaining  background Monte Carlo samples of $\Upsilon(2S.3S) \to \pipi \Upsilon(1S)$, $\Upsilon(1S) \to (\g)\mumu$ decays. The Monte Carlo are normalized to 
the data luminosity. }

\label{fig:mrec} 
\end{figure}

A total of 11,136 $\Upsilon(2S)$ and 3,857 $\Upsilon(3S)$ candidates are selected by the selection criteria  (mentioned in section~\ref{section:selection}) in the unblinded data samples of $\Upsilon(2S,3S)$. Figure~\ref{fig:mred} and Figure~\ref{fig:mrec} show the distributions of the $m_{\rm red}$ and $m_{\rm recoil}$ together with the remaining background MC samples of  $\Upsilon(2S.3S) \to \pipi \Upsilon(1S)$, $\Upsilon(1S) \to (\g)\mumu$ decays. The MCs are normalized to the data luminosity.

\begin{figure}
\centering
\includegraphics[width=3.0in]{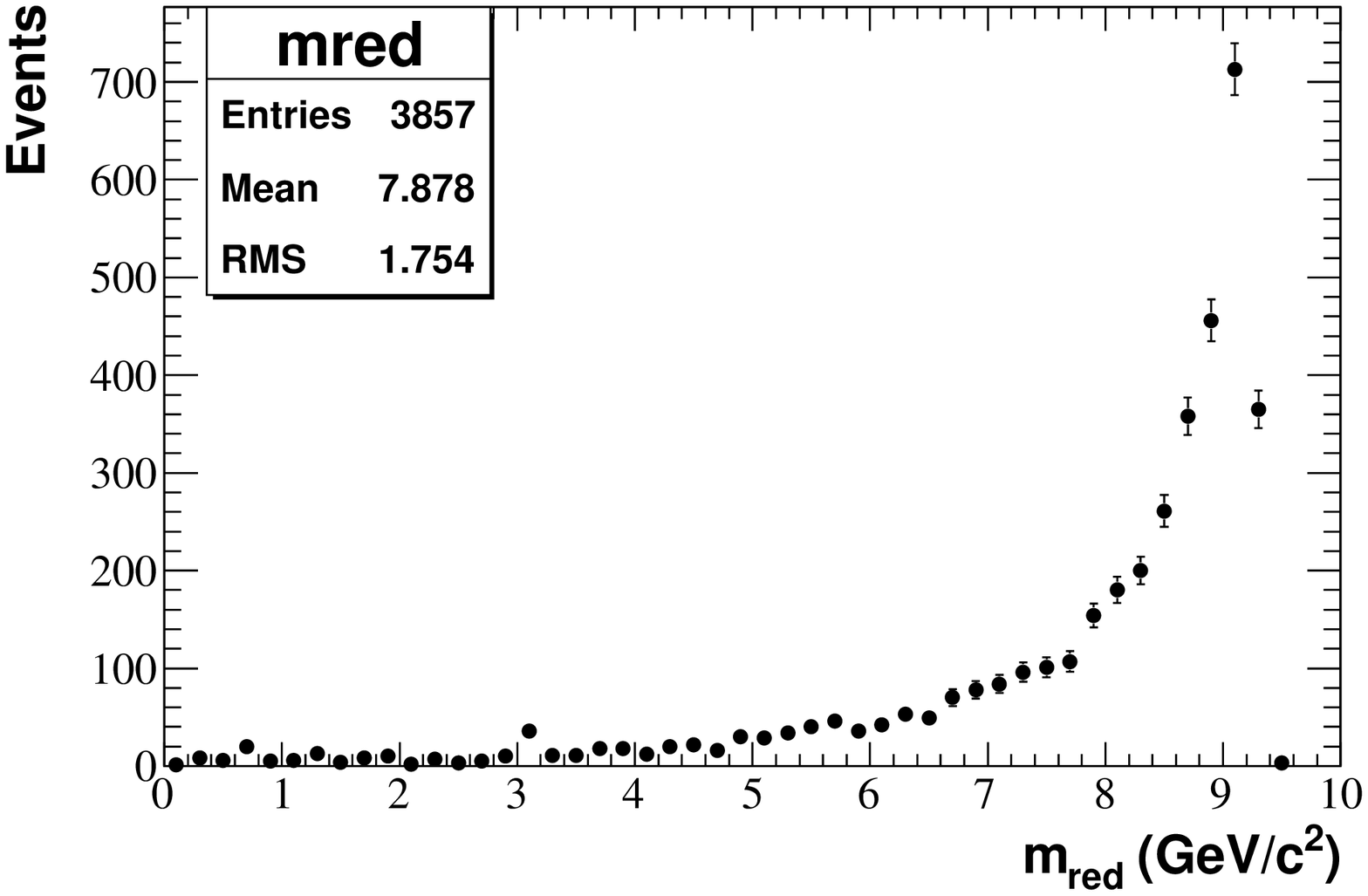}
\includegraphics[width=3.0in]{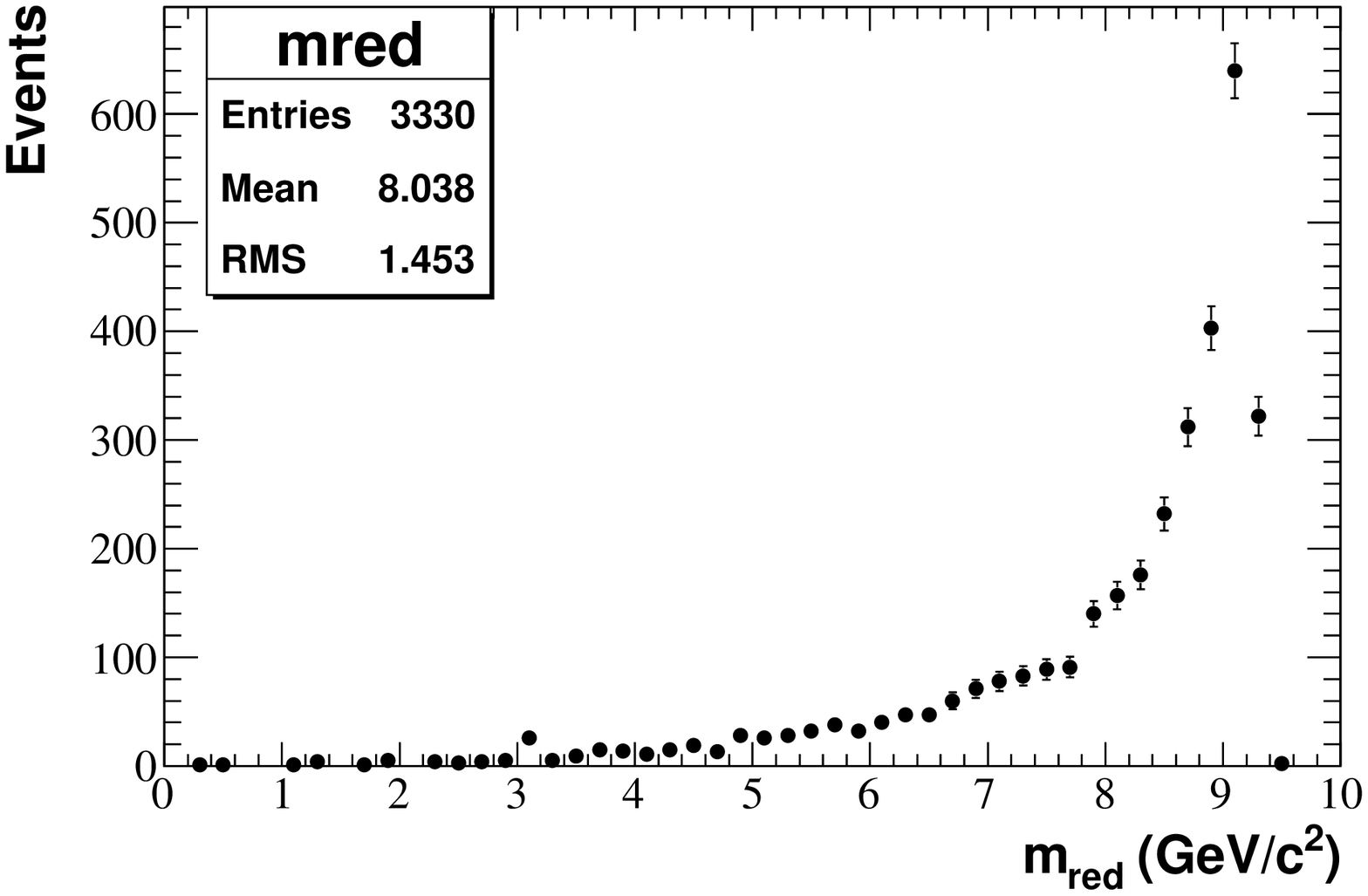}

\caption {The $m_{\rm red}$ distribution for the unblinded $\Upsilon(3S)$ Onpeak data-set after applying all the selection criteria including the OR muon PID cut (left) and AND muon PID cut (right). The first peak disappears after applying the AND muon PID cut but the second peak does not.}

\label{fig:unblindonpeak3S} 
\end{figure}

\begin{figure}
\centering
\includegraphics[width=3.0in]{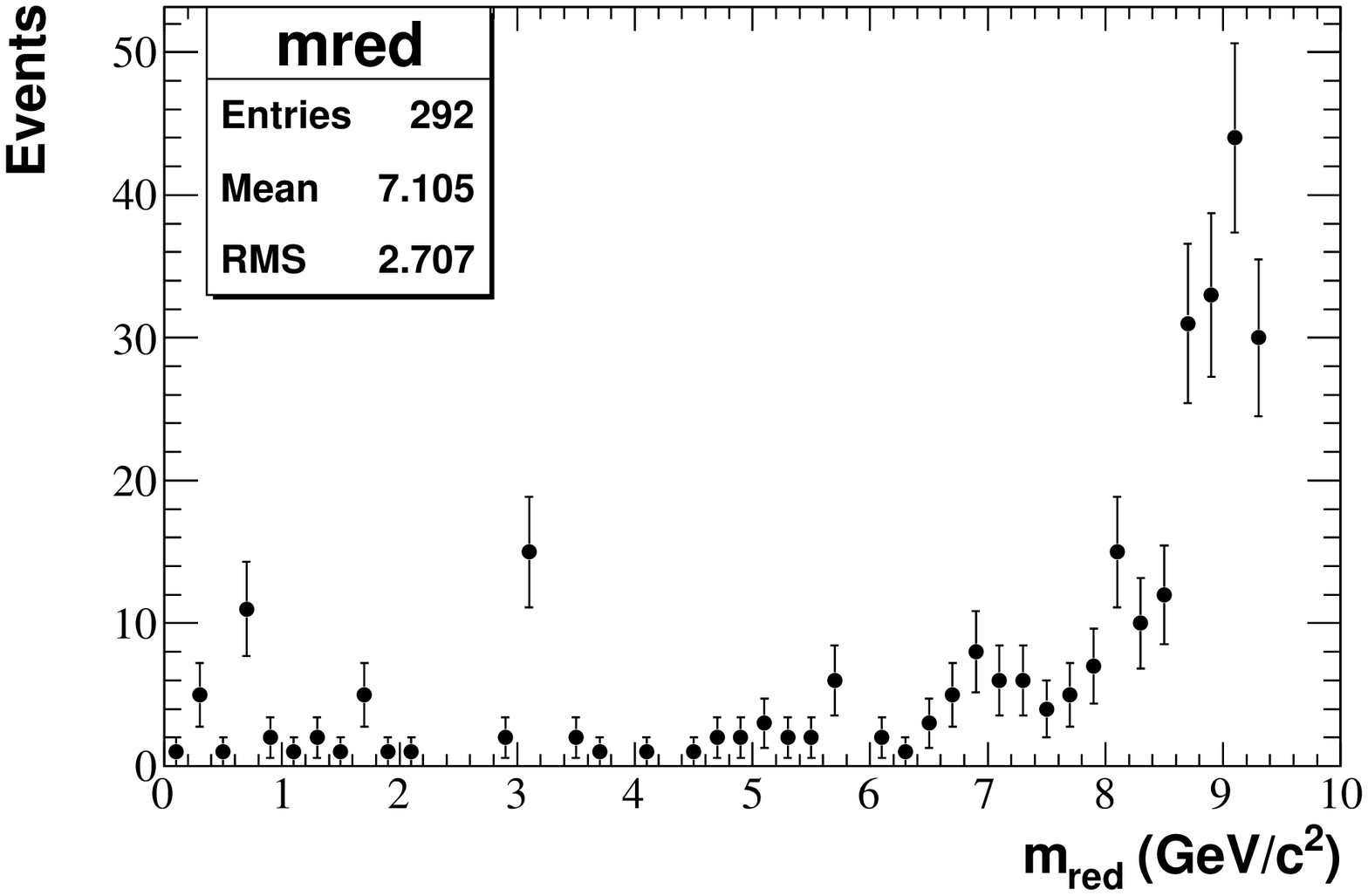}
\includegraphics[width=3.0in]{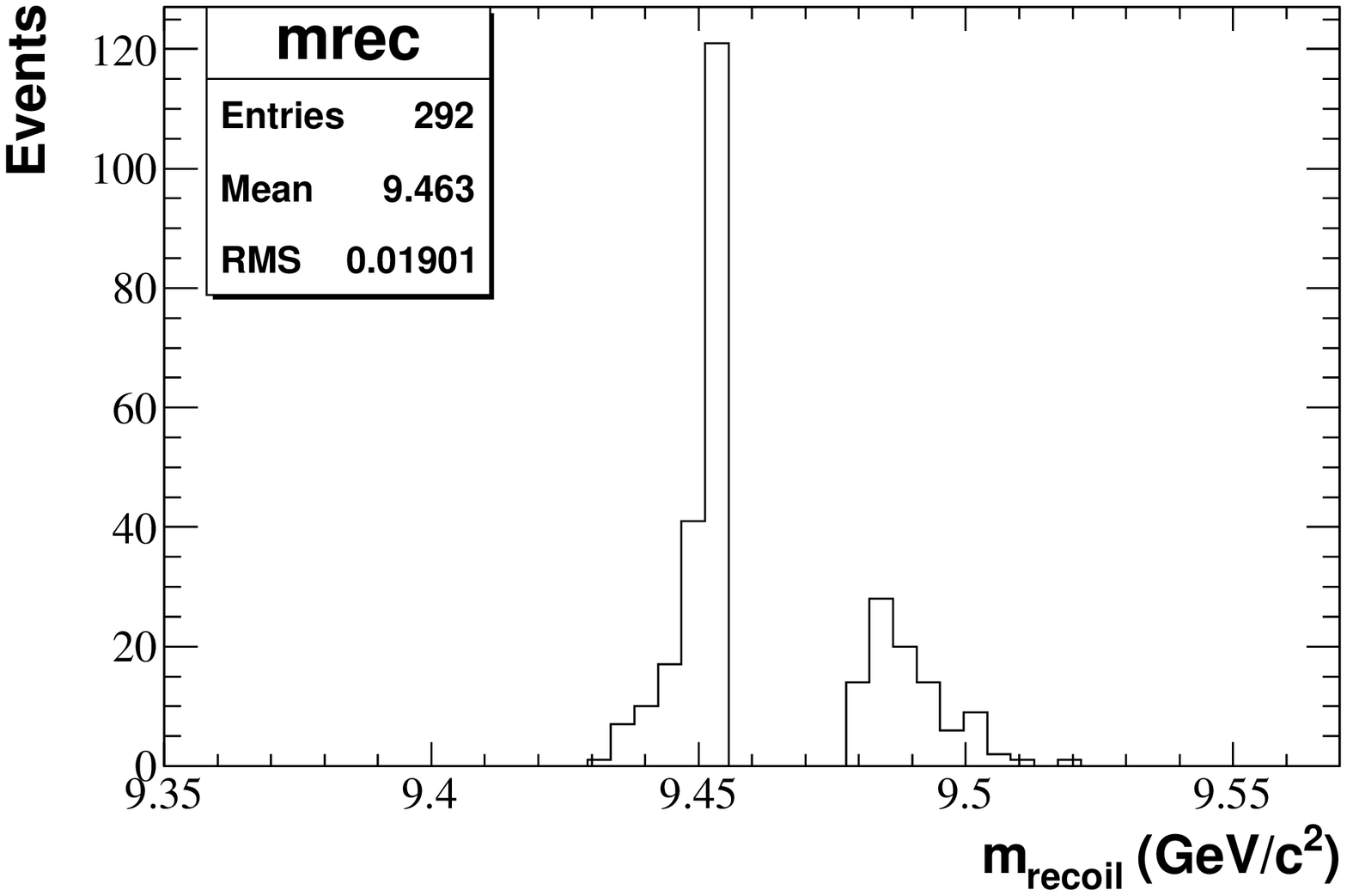}

\caption {The $m_{\rm red}$ distribution (left) and $m_{\rm recoil}$ distribution (right) for the sideband $\Upsilon(3S)$ data of the $m_{\rm recoil}$. We will use sideband of the $m_{\rm recoil}$ distribution in the $\Upsilon(3S)$ Onpeak dataset to model the $\rho^0$ background.}
\label{fig:XOR} 
\end{figure}

Two peaking 
components corresponding to  $\rho^0$ and \jpsi mesons are observed in the $\Upsilon(3S)$ dataset. The $\rho^0$-mesons are mainly produced in initial state radiation events, along with two or more pions, which disappears if we require both candidates to be identified as muons in the $A^0$ reconstruction (Figure~\ref{fig:unblindonpeak3S}) or apply a tighter ($3\sigma$) mass window cut on the $m_{\rm recoil}$ distribution. An enhancement of the $\rho^0$ background is observed outside the signal region of [9.455,9.48] \gevcc in the $m_{\rm recoil}$ distribution of the $\Upsilon(3S)$ dataset (Figure~\ref{fig:XOR}). This data sample is used to model the $\rho^0$ background using a sum of a Gaussian and a constant linear function (Figure~\ref{fig:XOR}). The fixed PDF parameters of the Gaussian function are used to describe the $\rho^0$ peak in the final fit. 

 \begin{figure}
\centering
\includegraphics[width=6.0in]{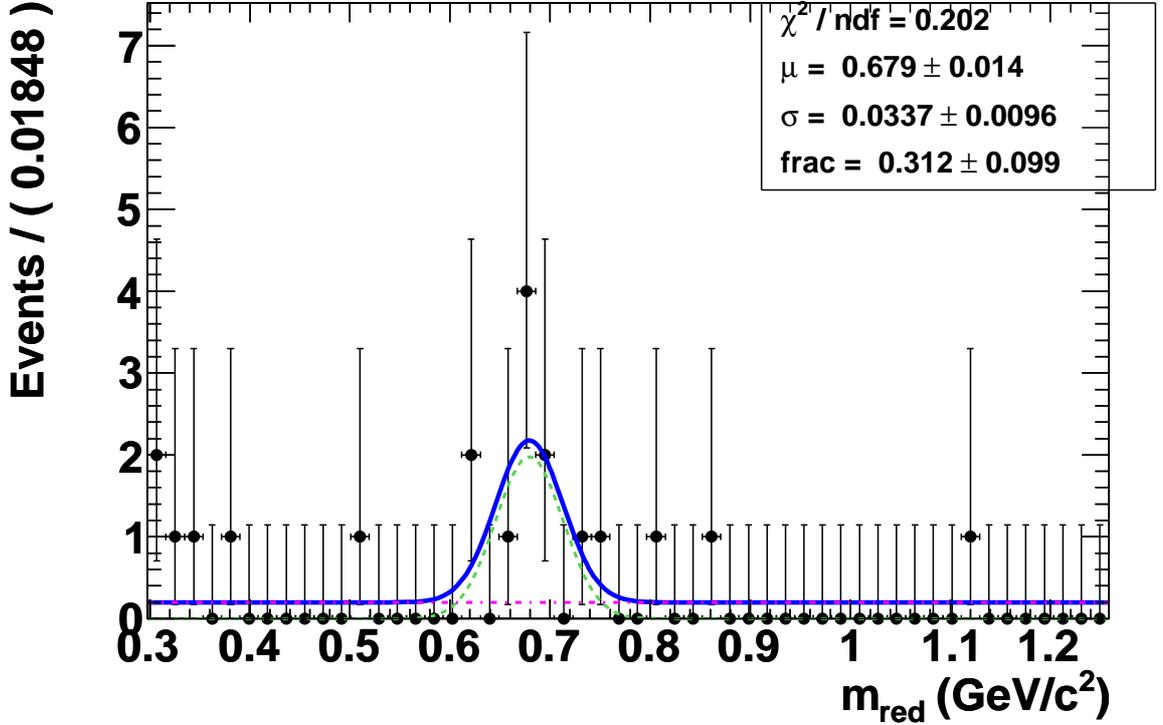}

\caption {The peaking background PDF at the $\rho^0$ mass position. We use the sideband of the $m_{\rm recoil}$ distribution in the $\Upsilon(3S)$ onpeak dataset to model this background.}
\label{fig:peakrho} 
\end{figure}

\begin{figure}
\centering
\includegraphics[width=3.0in]{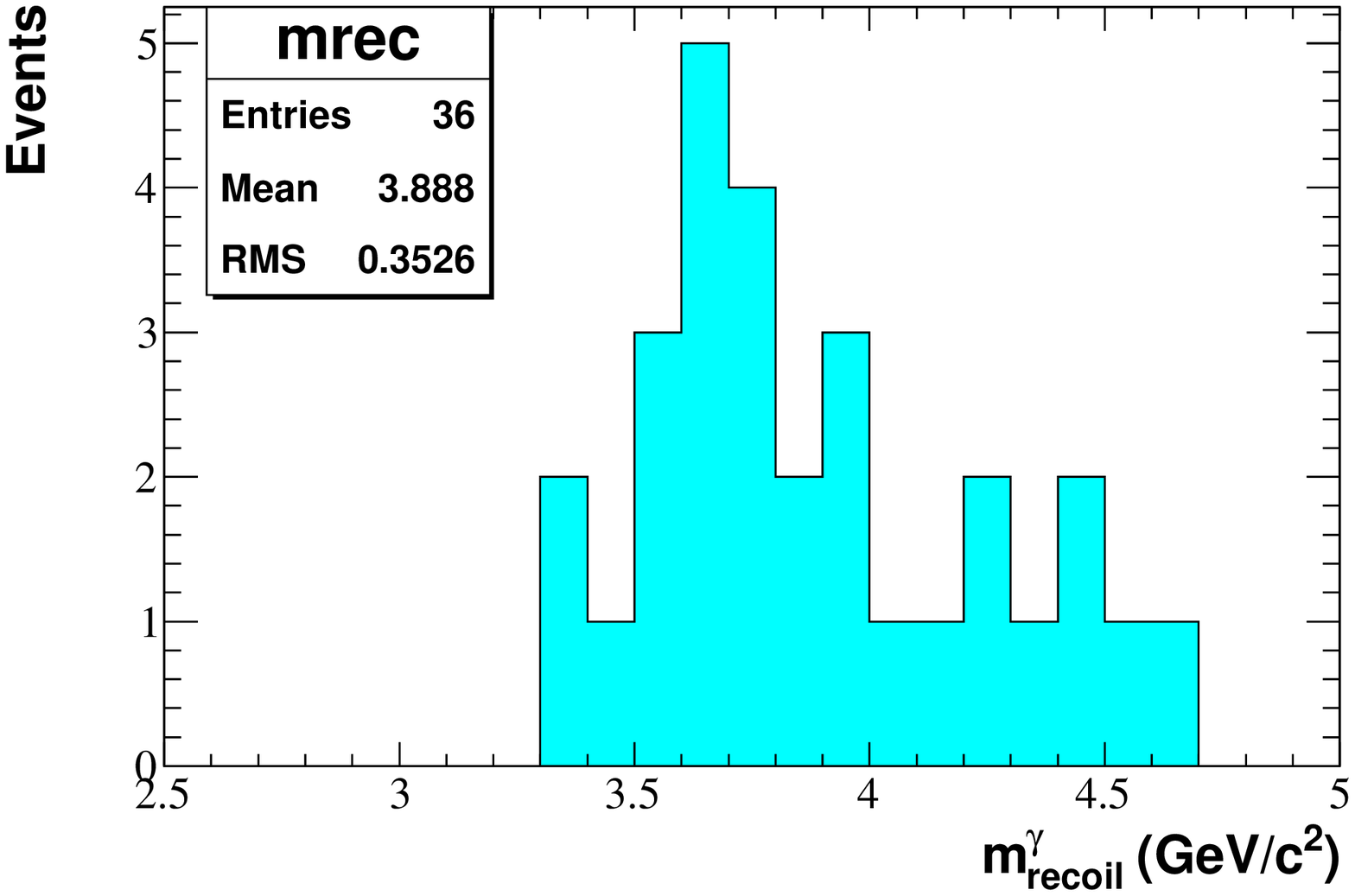}
\includegraphics[width=3.05in]{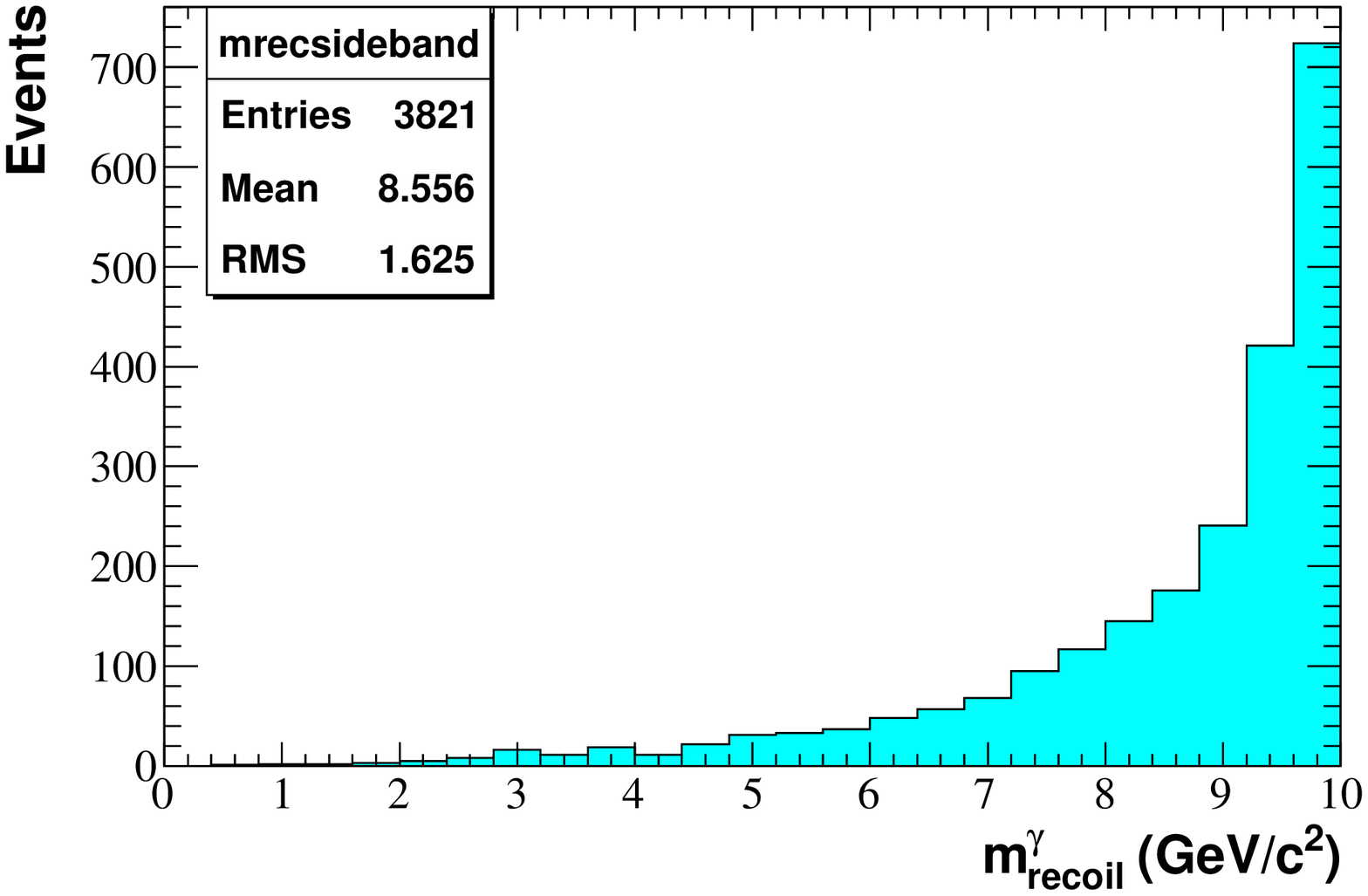}

\caption {The $m_{\rm recoil}^{\gamma}$ distribution for the unblinded $\Upsilon(3S)$ Onpeak data-set after applying all the selection criteria. The left plot shows the  $m_{\rm recoil}^{\gamma}$ distribution in the region of $m_{\rm red} = [3.0,3.2]$ GeV/$c^2$ and the right plot shows the same distribution in the region of  $(m_{\rm red} < 3.0) || (m_{\rm red} > 3.2)$ GeV/$c^2$.} 

\label{fig:gammarecoilmassData} 
\end{figure}

\begin{figure}
\centering
\includegraphics[width=3.0in]{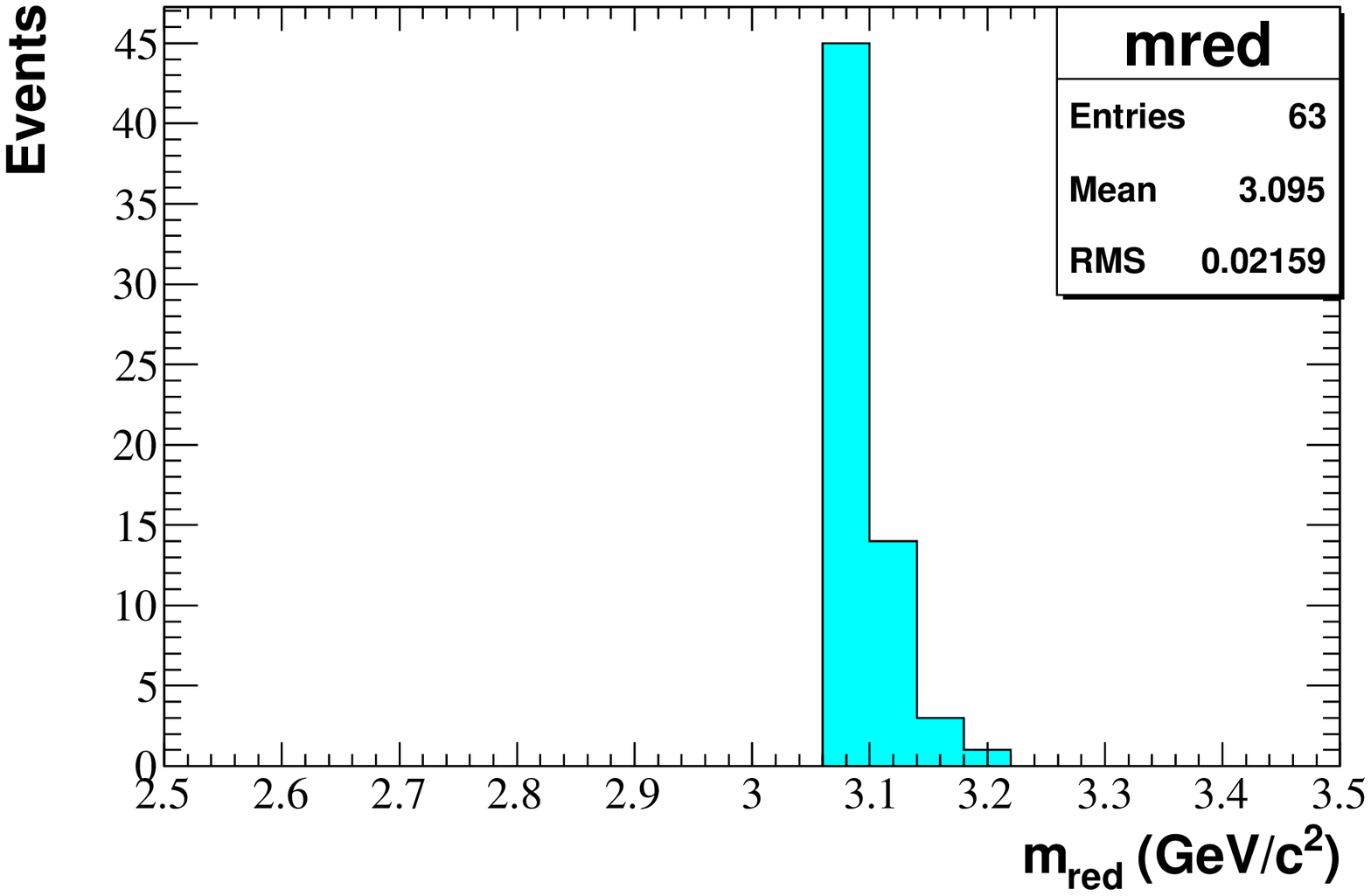}
\includegraphics[width=3.0in]{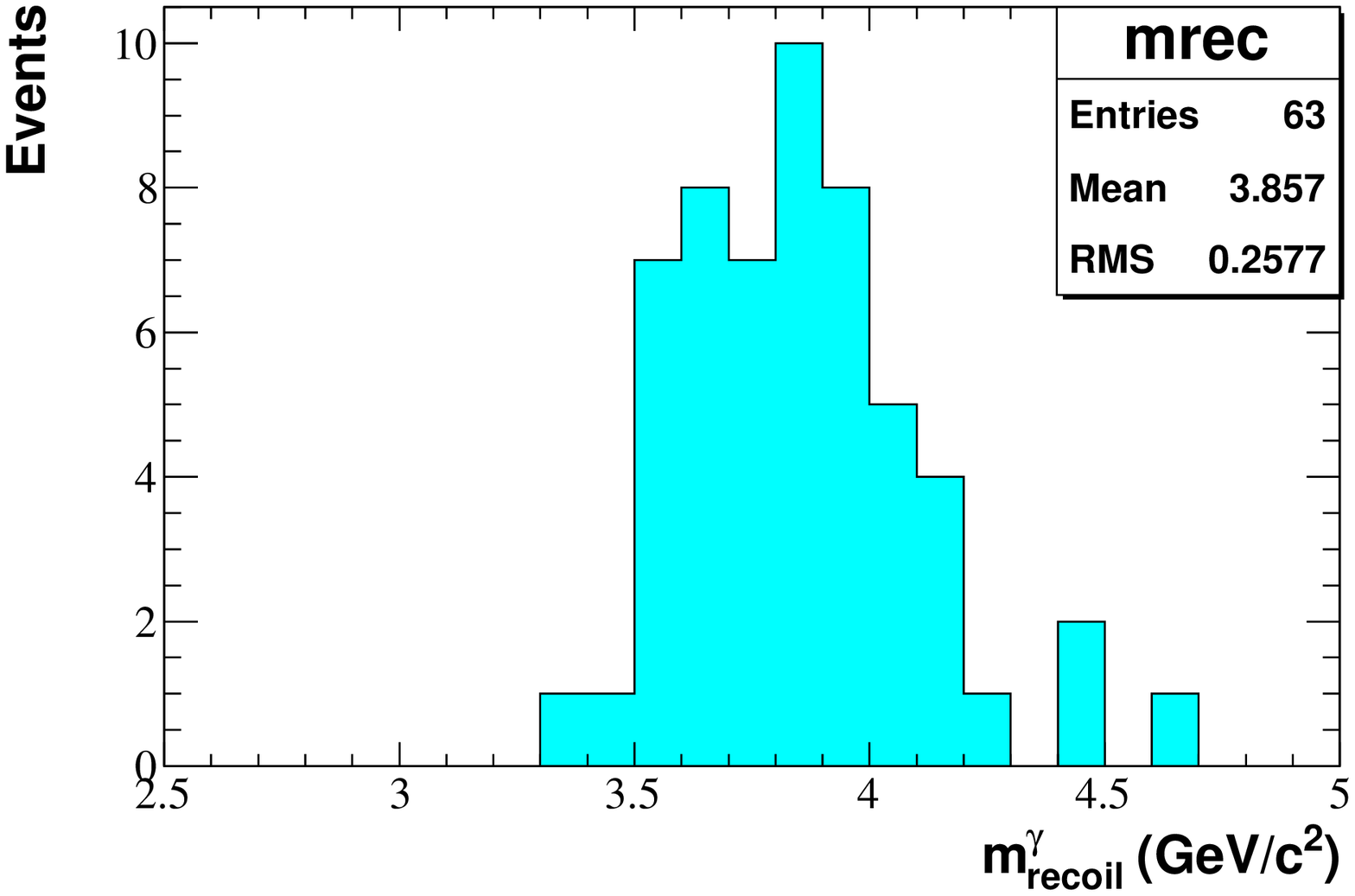}
\caption {The distribution of  $m_{\rm red}$ (left) and $m_{\rm recoil}^{\gamma}$ (right) in the  $\psi(nS)$ generic decays sample after applying all the selection cuts. }
\label{fig:mredSP8922} 
\end{figure}

\begin{figure}
\centering
\includegraphics[width=3.0in]{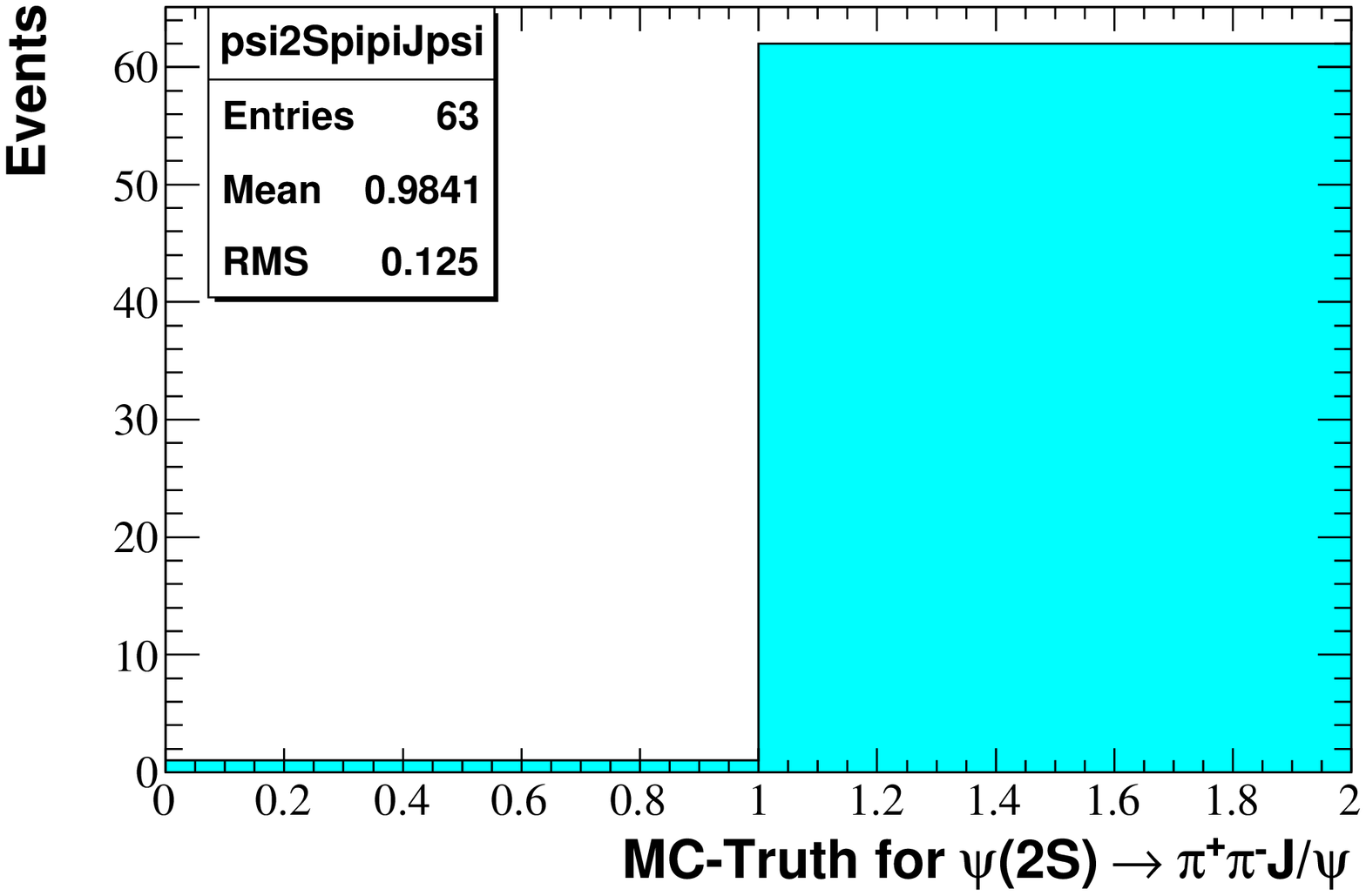}
\includegraphics[width=3.0in]{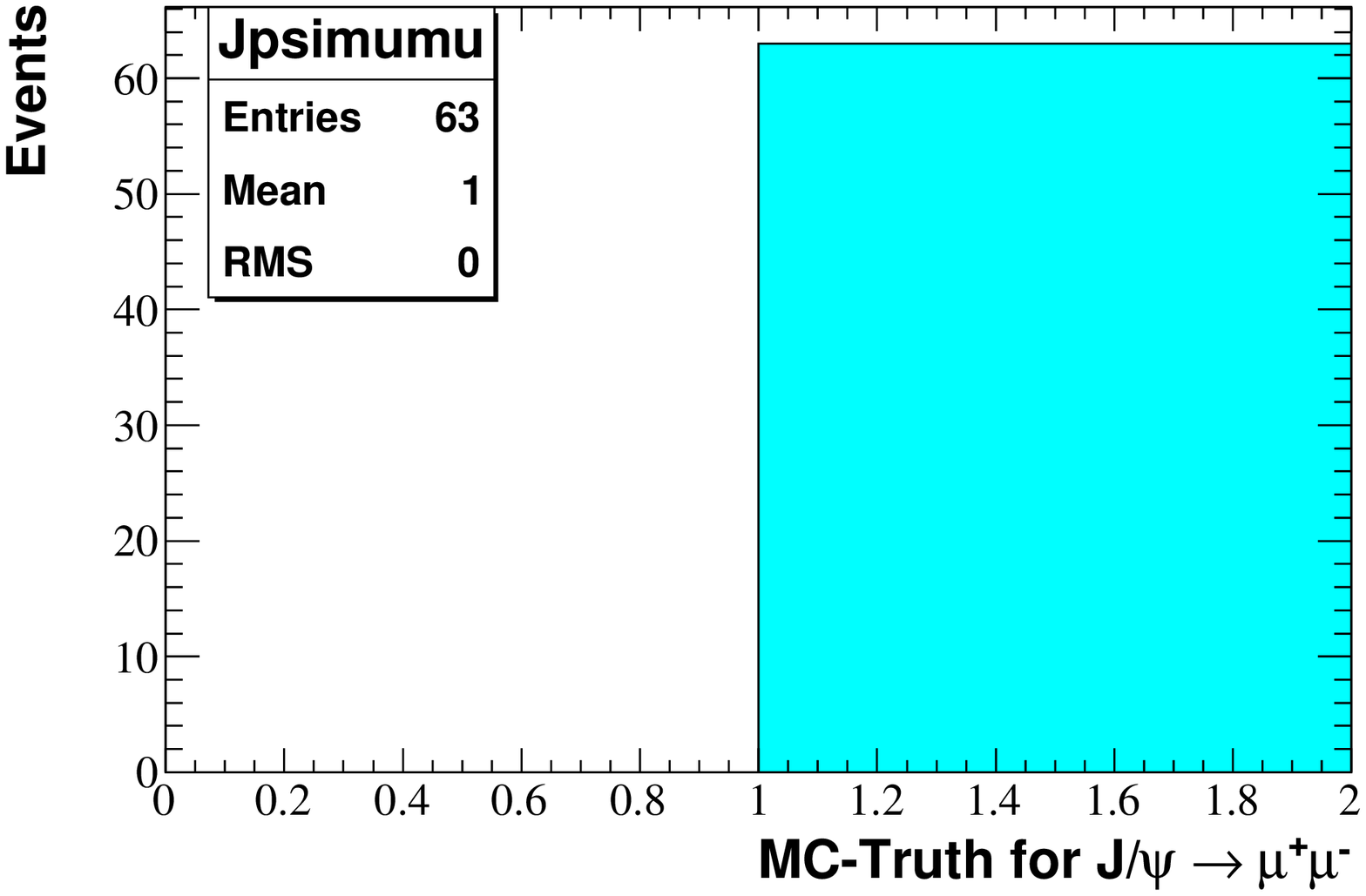}

\caption {MC-Truth Boolean distribution for $\psi(2S) \rightarrow \pi^+\pi^- J/\psi$ decays (left) and for $\jpsi \rightarrow \mu^+\mu^-$ decays (right) in the  $\psi(nS)$ generic decays sample.} 
\label{fig:MCTruthSP8922} 
\end{figure}

\begin{figure}
\centering
\includegraphics[width=6.0in]{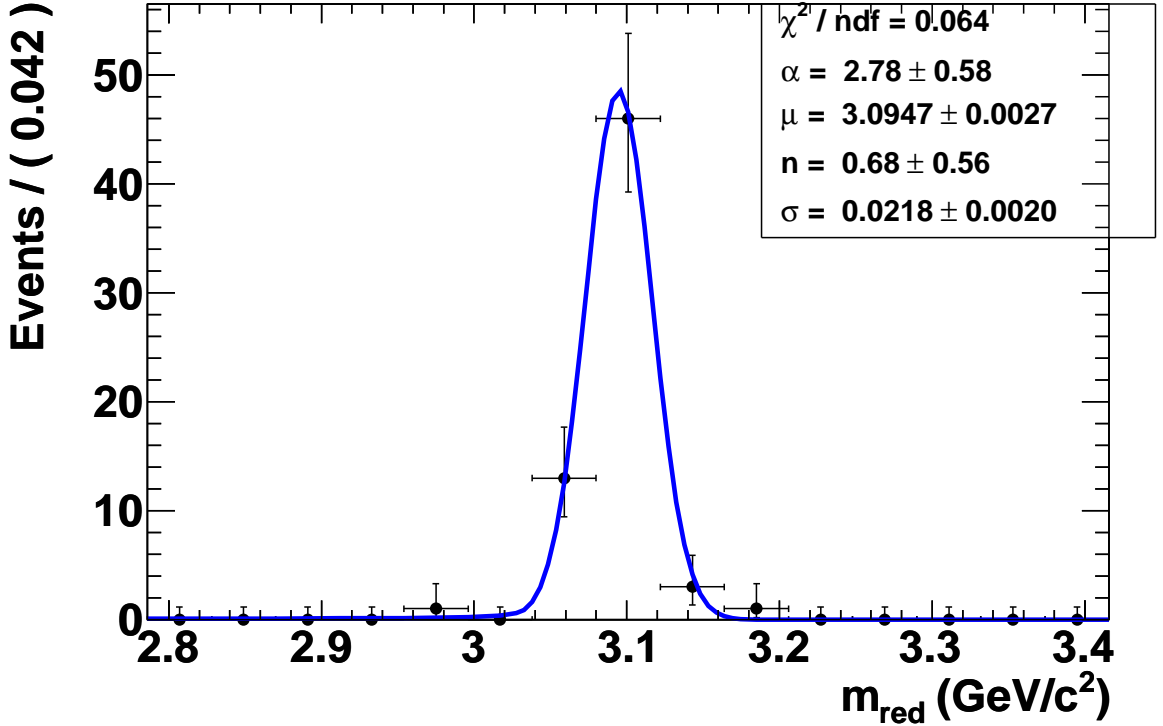} 
\caption {The background PDF for $m_{\rm red}$ at J/$\psi$ mass position. We use a sample of $\psi(nS)$ generic decays to model this background.}
\label{fig:mredPDFSP8922} 
\end{figure}

   To understand the peaking component at the \jpsi mass position in the $\Upsilon(3S)$ dataset, we compute the  mass of the system recoiling against the photon, which is defined as:

\begin{equation}
m_{\rm recoil}^{\gamma} = s - 2\cdot\sqrt{s} \cdot E_{CM}^{\gamma},
\end{equation} 

\noindent where $\sqrt{s}$ is the CM energy of the $e^+e^-$ system and $E_{CM}^{\gamma}$ is the CM energy of the photon. The $m_{\rm recoil}^{\gamma}$ should peak at the mass position of the $X$ resonance, in an ISR decay like $e^+e^- \rightarrow \gamma_{ISR} X$. Figure~\ref{fig:gammarecoilmassData} shows the $m_{\rm recoil}^{\gamma}$ distribution in both $m_{\rm red}$ region of $[3.0-3.2]$ \gevcc as well as the outside of this region using $\Upsilon(3S)$ onpeak dataset. It is clear  that the $m_{\rm recoil}^{\gamma}$ distribution  peaks at $\psi(2S)$ mass position for the $m_{\rm red}$ region of $3.0-3.2$ \gevcc. We have also processed a sample of $e^+e^- \rightarrow \gamma_{ISR} \psi(nS)$ with generic decays of $\psi(nS)$.  The $m_{\rm red}$ distribution of $\psi(nS)$ generic sample at \jpsi mass position is shown in  Figure~\ref{fig:mredSP8922} (left).  The $m_{\rm recoil}^{\gamma}$  distribution of $\psi(nS)$ generic sample is also shown in Figure~\ref{fig:mredSP8922} (right), which peaks at $\psi(2S)$ mass position. Using MC-Truth information of the survived MC events for $\psi(nS)$ generic decays, it is observed that about $95\%$ of the events decay via $\psi(2S) \rightarrow \pi^+\pi^- \jpsi$ and about $99\%$ of the \jpsi events decay via $\mu^+\mu^-$ channel  (Figure~\ref{fig:MCTruthSP8922}). We model the peaking component of the J/$\psi$ background by a CB function using data sample of this $\psi(nS)$ generic decays (Figure~\ref{fig:mredPDFSP8922}). A high statistics data and MC  samples of  $e^+e^- \rightarrow \gamma_{ISR} \psi(2S)$, $\psi(2S) \rightarrow \pi^+\pi^- \jpsi$, $\jpsi
  \rightarrow \mu^+\mu^-$  have also been used to check the resolution of $m_{\rm red}$ distribution at the \jpsi mass peak position. We find a resolution of $(2.014 \pm 0.309) \times 10^{-3}$ \gevcc in the data, compatible with the predictions of the 
 MC of $(2.007 \pm 0.011) \times 10^{-3}$ GeV/$c^2$, which is obtained by applying the mass constraints on the $\psi(2S)$ to improve the resolution
  of the \jpsi. A similar exercise without the mass constraint results in a agreement between data and Monte Carlo as well. However, the resolution of 
 these \jpsi event is not representative of that of the signal, because the kinematic is different.  

\section{Signal yield extraction using the 1d ML fit}
\begin{figure}
\centering
\includegraphics[width=3.0in]{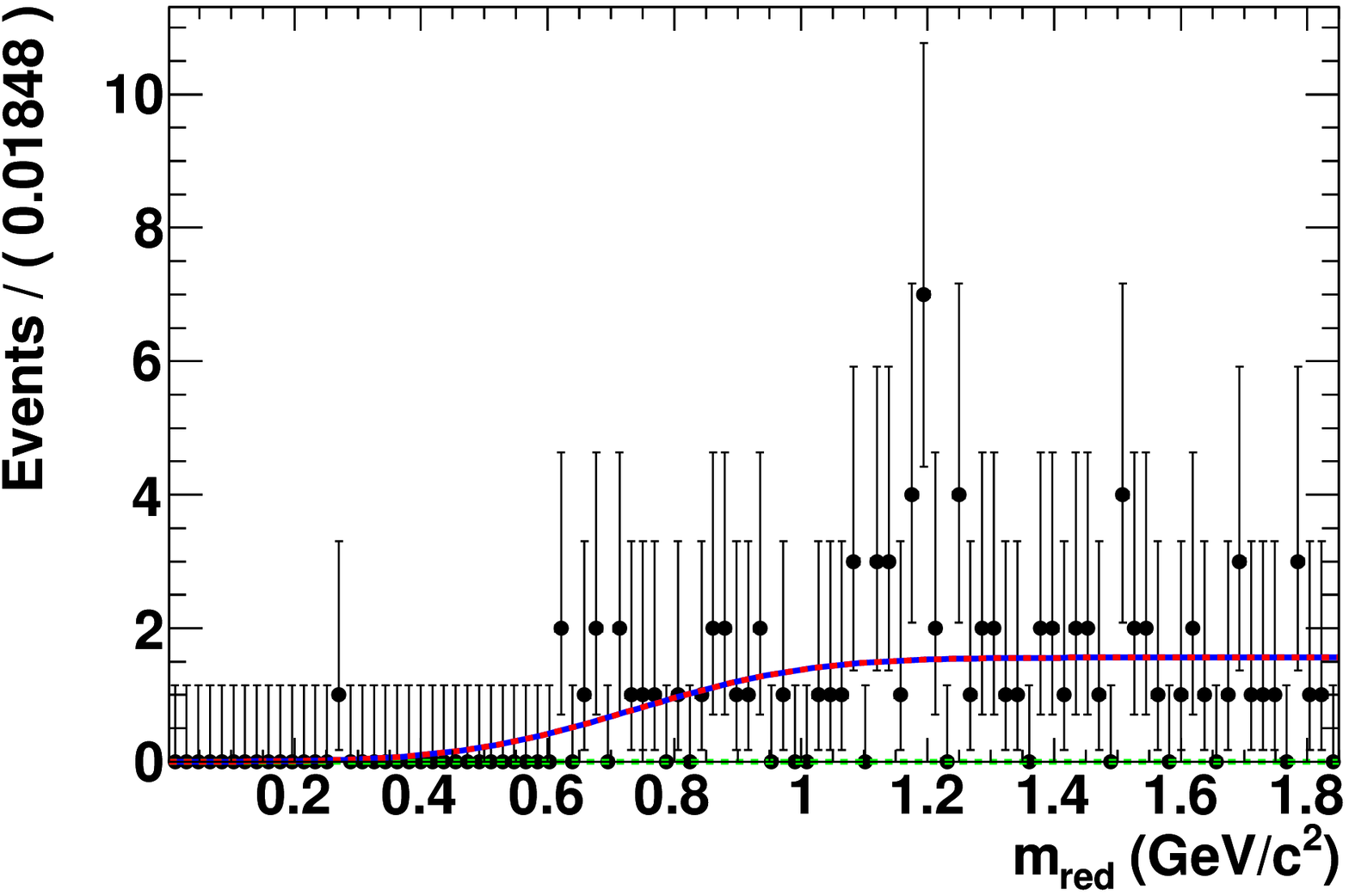}
\includegraphics[width=3.0in]{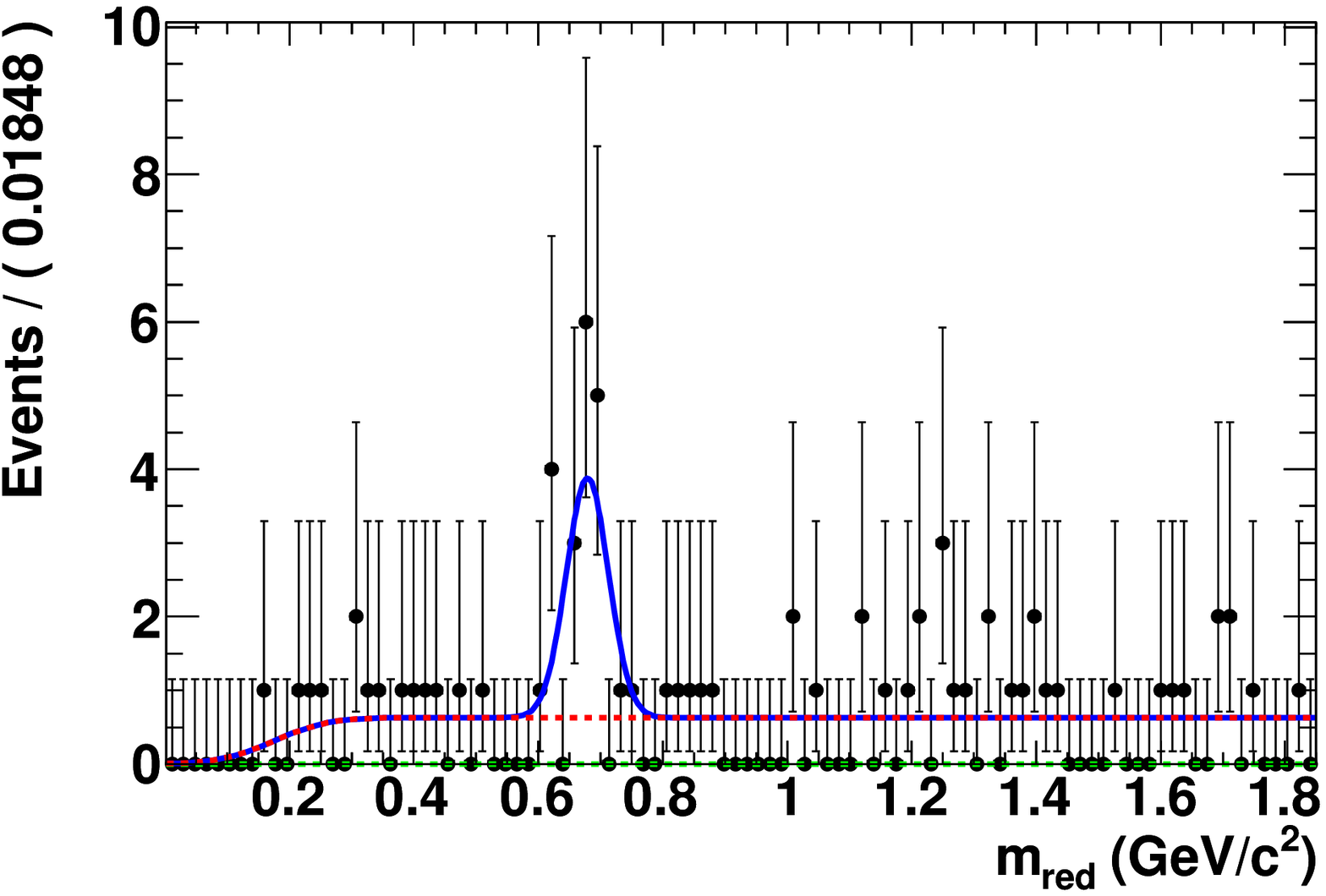}

\smallskip
\centerline{\hfill (a) \hfill \hfill (b) \hfill }
\smallskip

\includegraphics[width=3.0in]{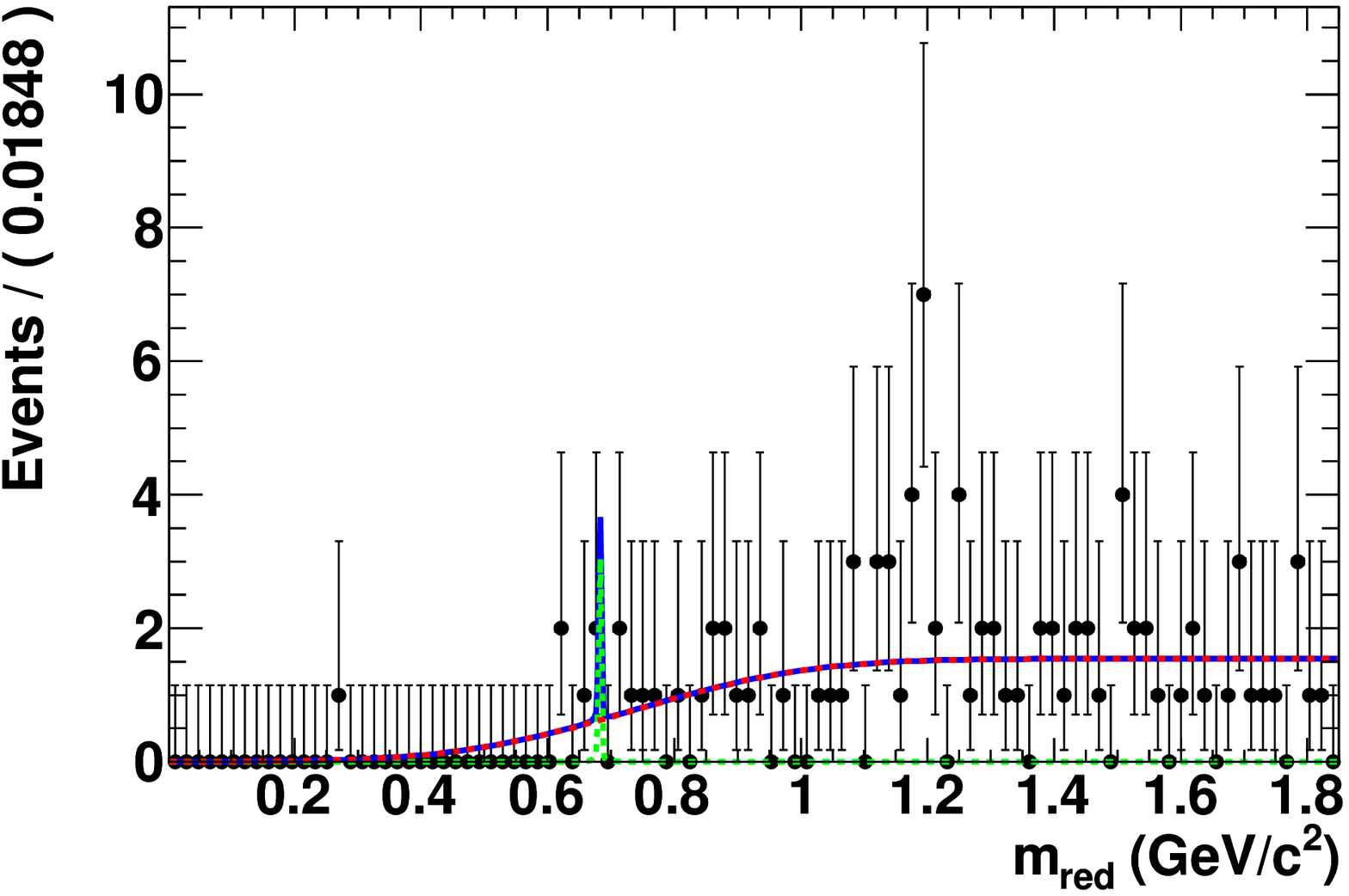}
\includegraphics[width=3.0in]{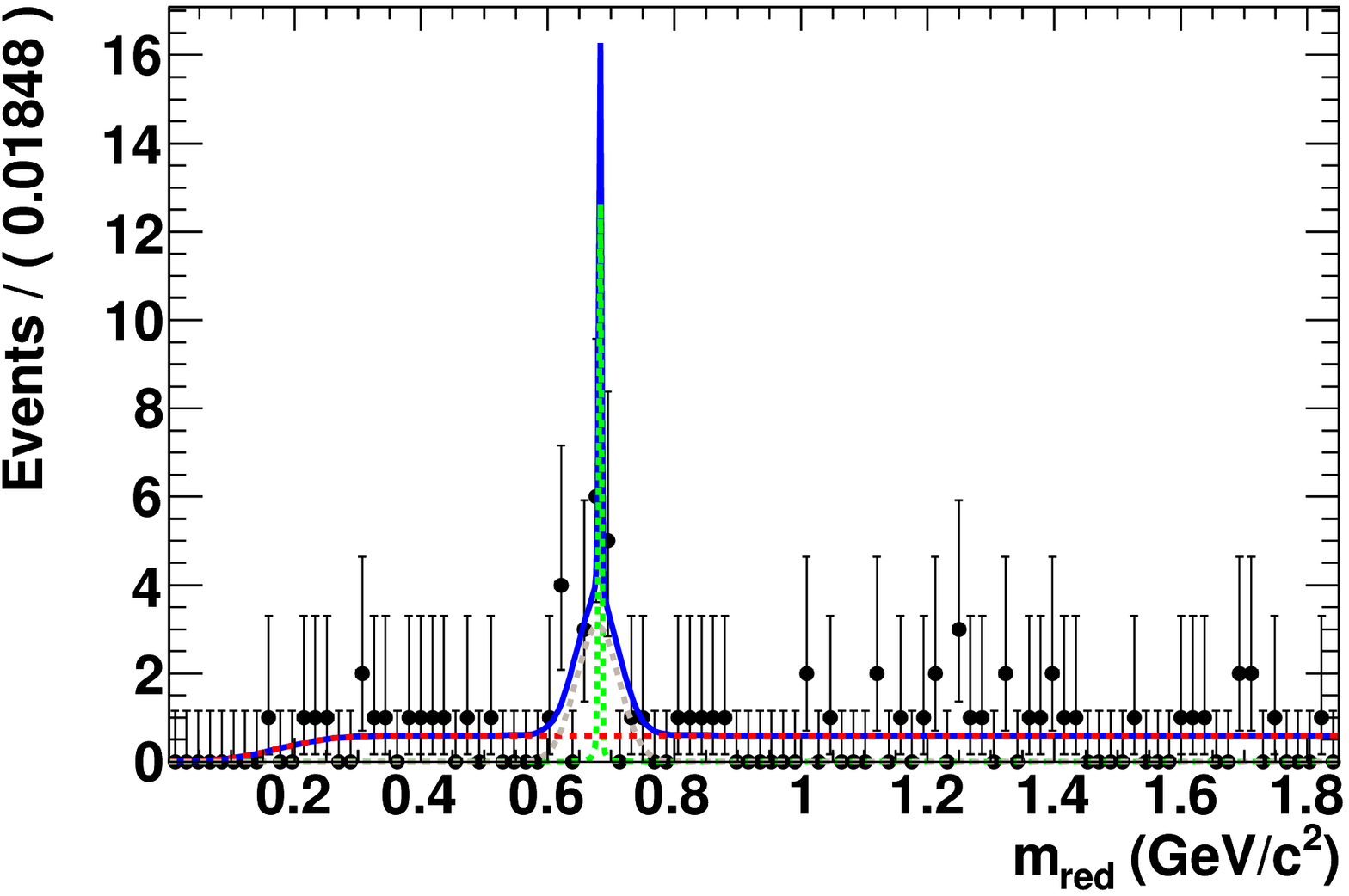}

\smallskip
\centerline{\hfill (c) \hfill \hfill (d) \hfill }
\smallskip

\includegraphics[width=3.0in]{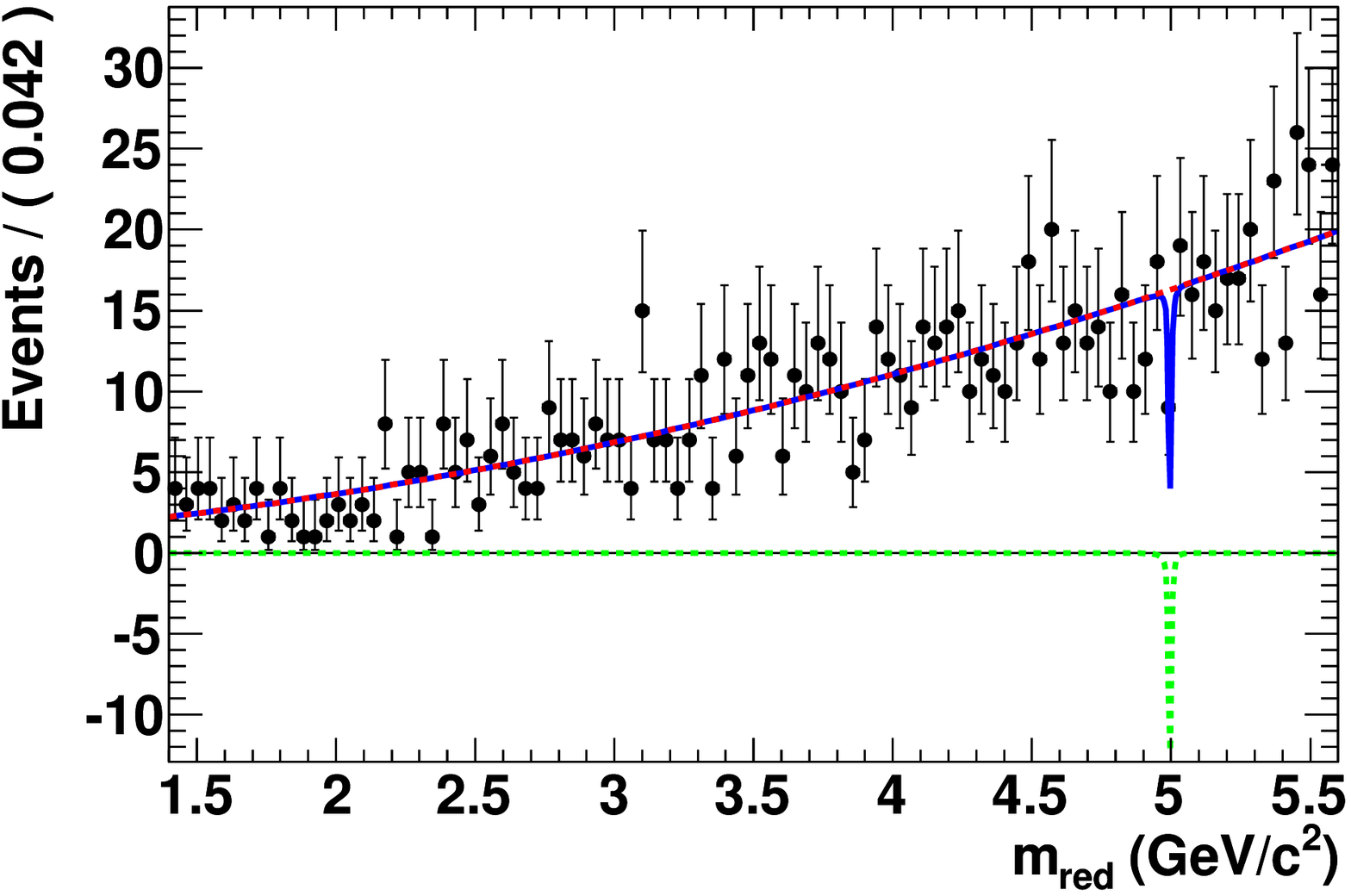}
\includegraphics[width=3.0in]{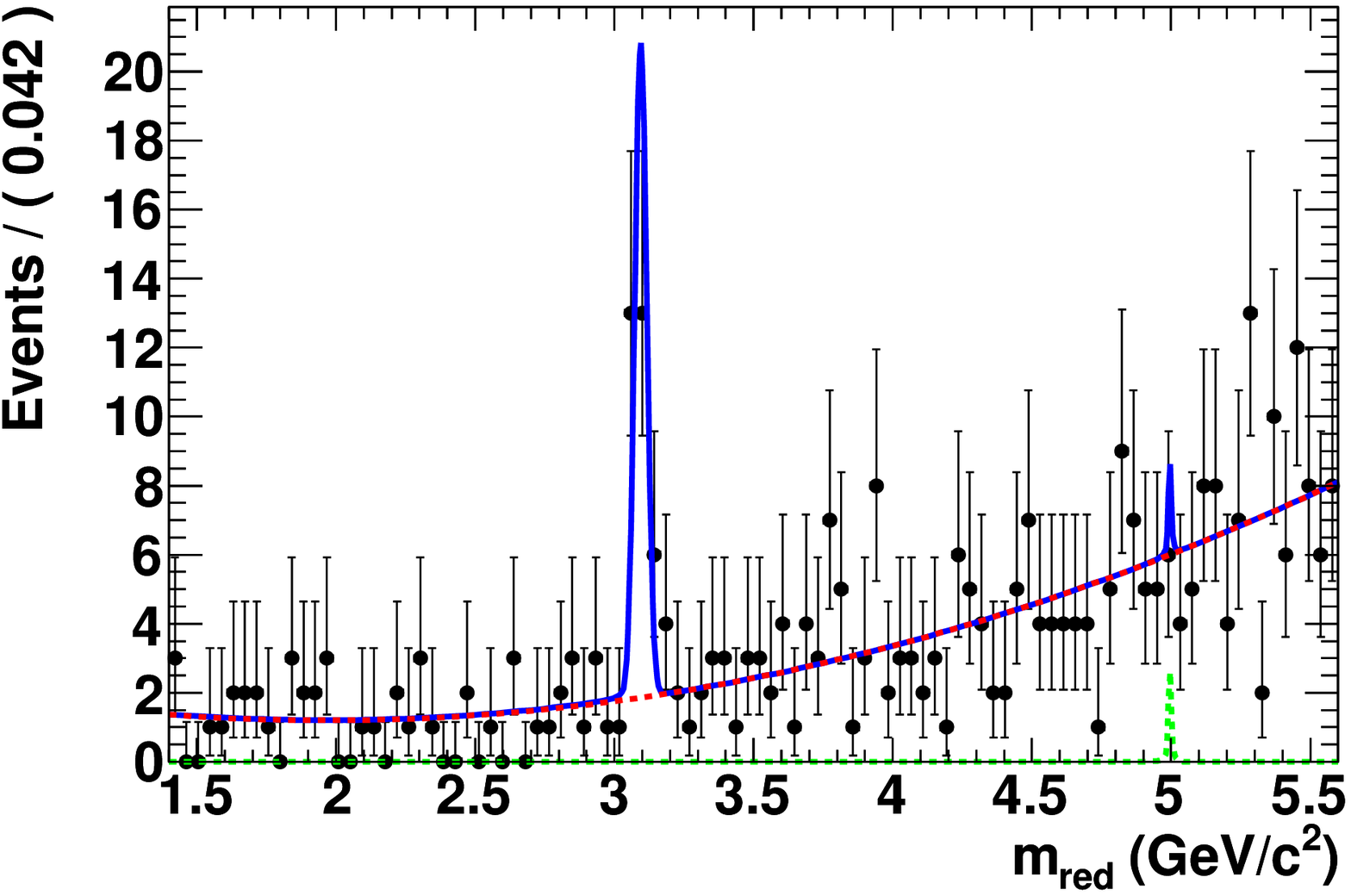}

\smallskip
\centerline{\hfill (e) \hfill \hfill (f) \hfill}
\smallskip

\caption {Result of the likelihood fit to the unblinded $\Upsilon(2S,3S)$ onpeak data samples. Projection plot onto reduce mass distribution for the $m_{A^0}$ of (a, b) $m_{A^0}=0.212$ GeV/$c^2$, (c, d) $m_{A^0}=0.715$ GeV/$c^2$   and (e, f) $m_{A^0}=5.0$ GeV/$c^2$. Left plots are for $\Upsilon(2S)$ and right plots are for $\Upsilon(3S)$ data sample. The total ML fit is shown in solid blue; the non-peaking background component is shown in dashed magenta; the signal component is shown in green dashed. The peaking components of $\rho^0$ and $J/\psi$ resonances are modelled by a  Gaussian  and a CB function, respectively in the $\Upsilon(3S)$ data-set.} 

\label{fig:Newlowproj}
\end{figure}

\begin{figure}
\centering
\includegraphics[width=3.0in]{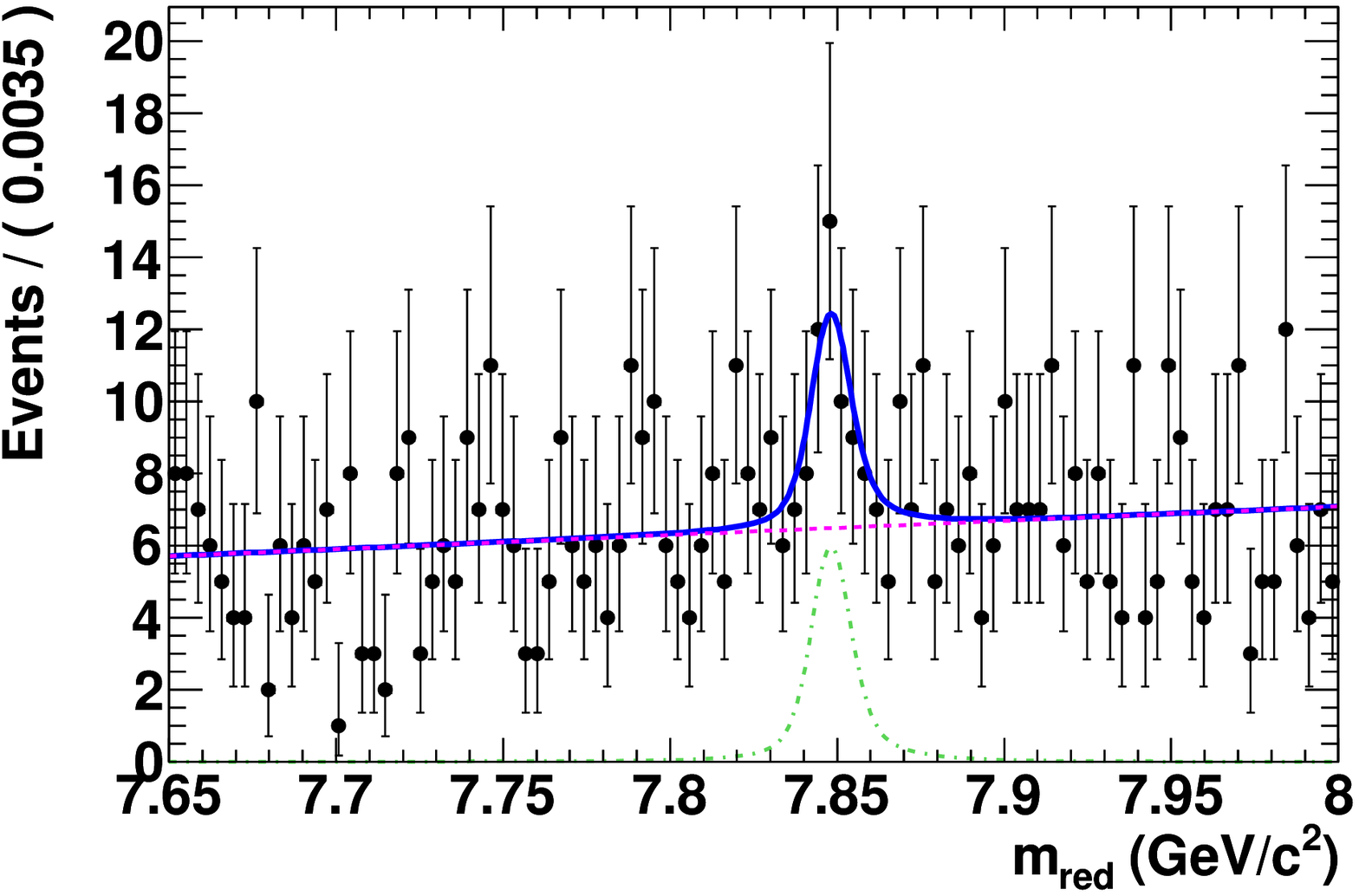}
\includegraphics[width=3.0in]{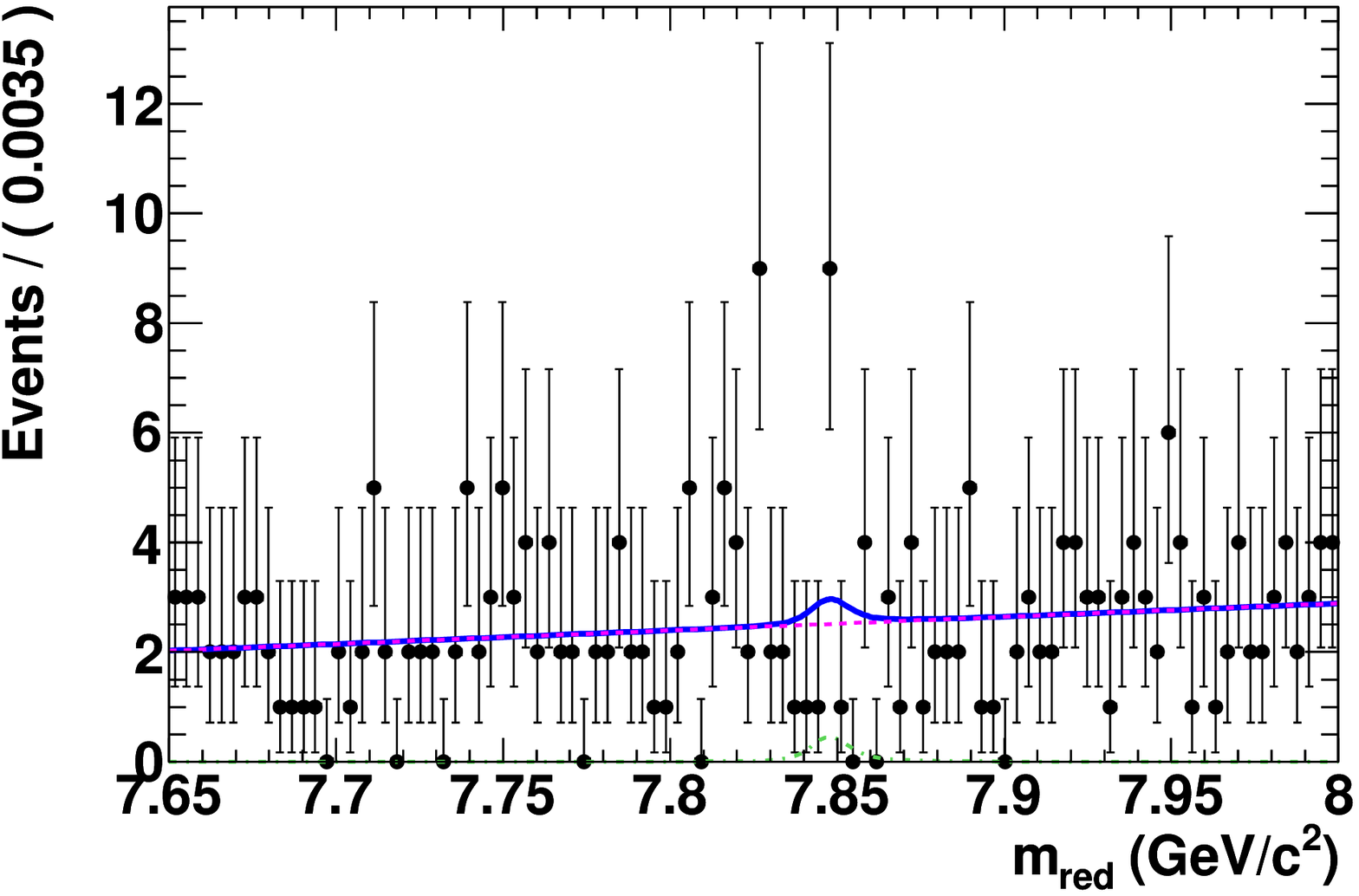}

\smallskip
\centerline{\hfill (a) \hfill \hfill (b) \hfill }
\smallskip

\includegraphics[width=3.0in]{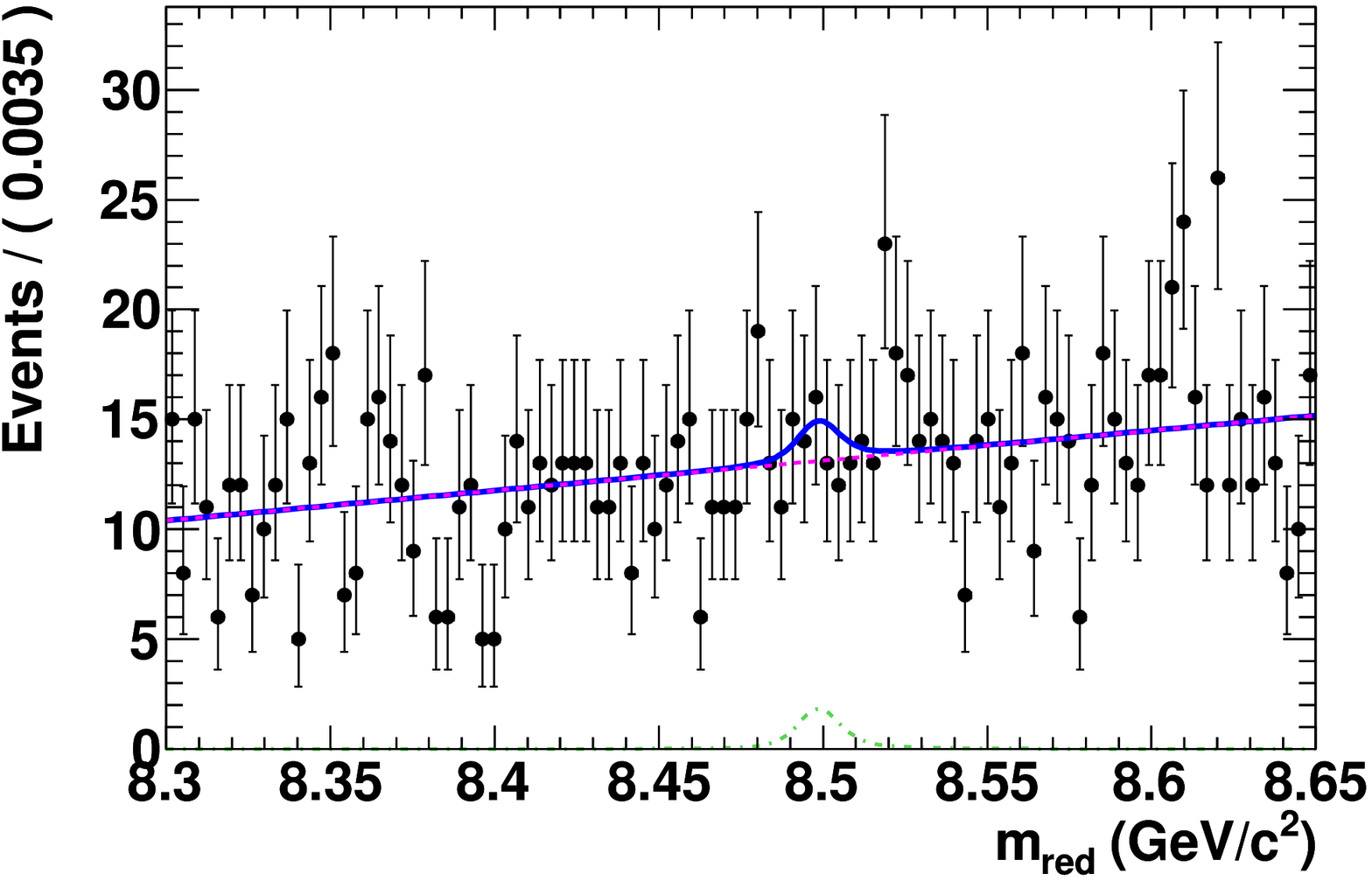}
\includegraphics[width=3.0in]{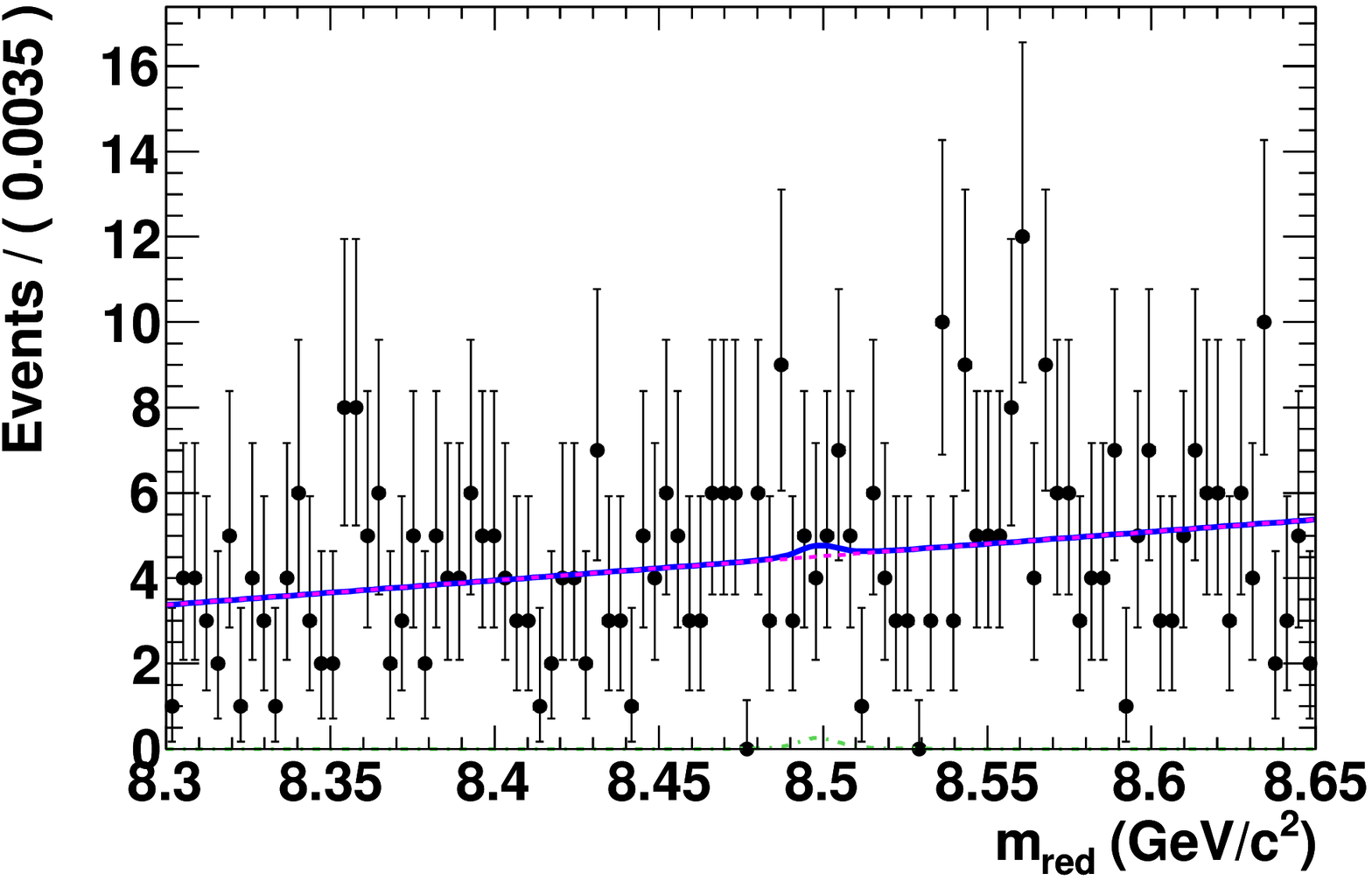}

\smallskip
\centerline{\hfill (c) \hfill \hfill (d) \hfill }
\smallskip

\includegraphics[width=3.0in]{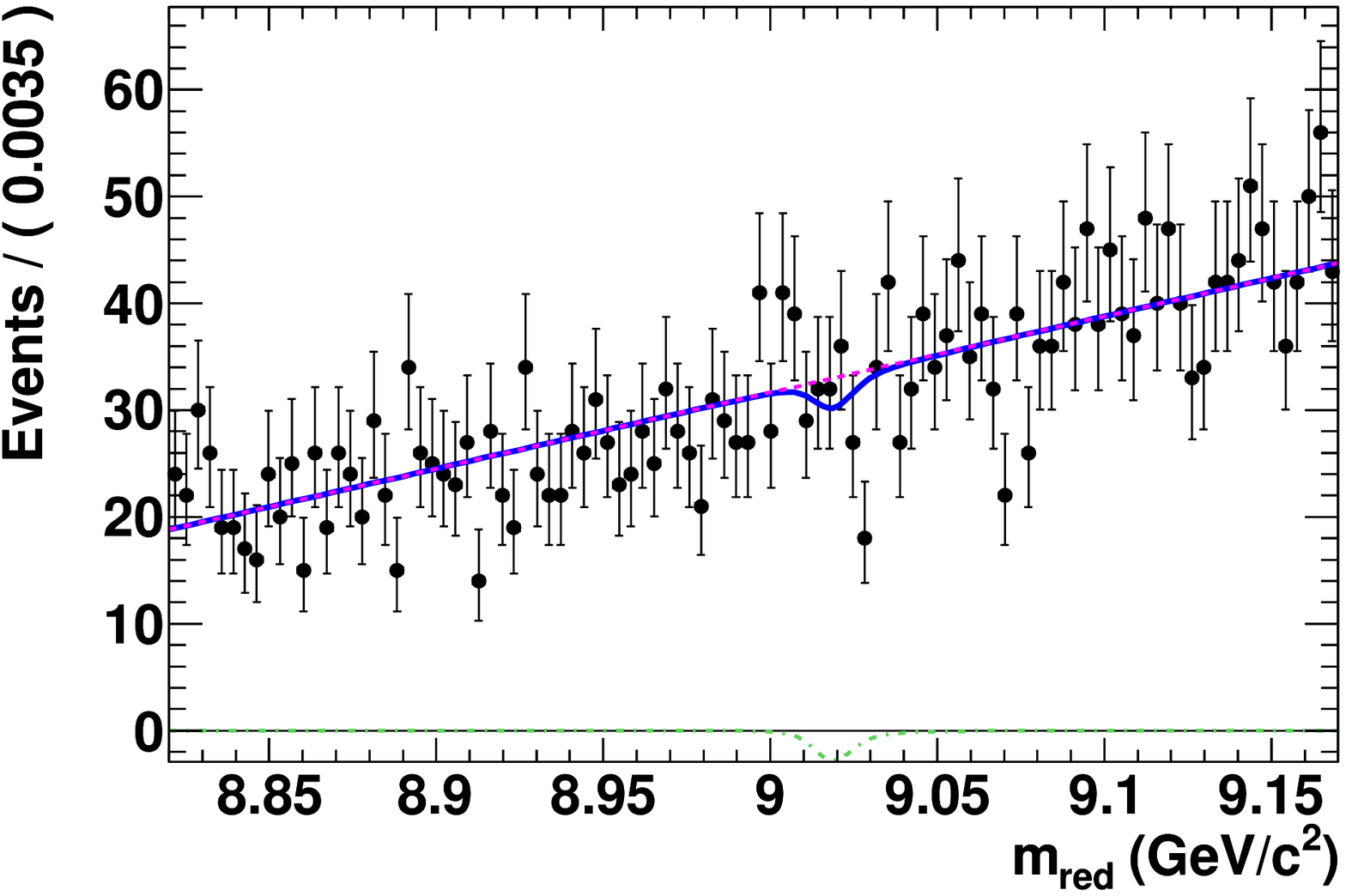}
\includegraphics[width=3.0in]{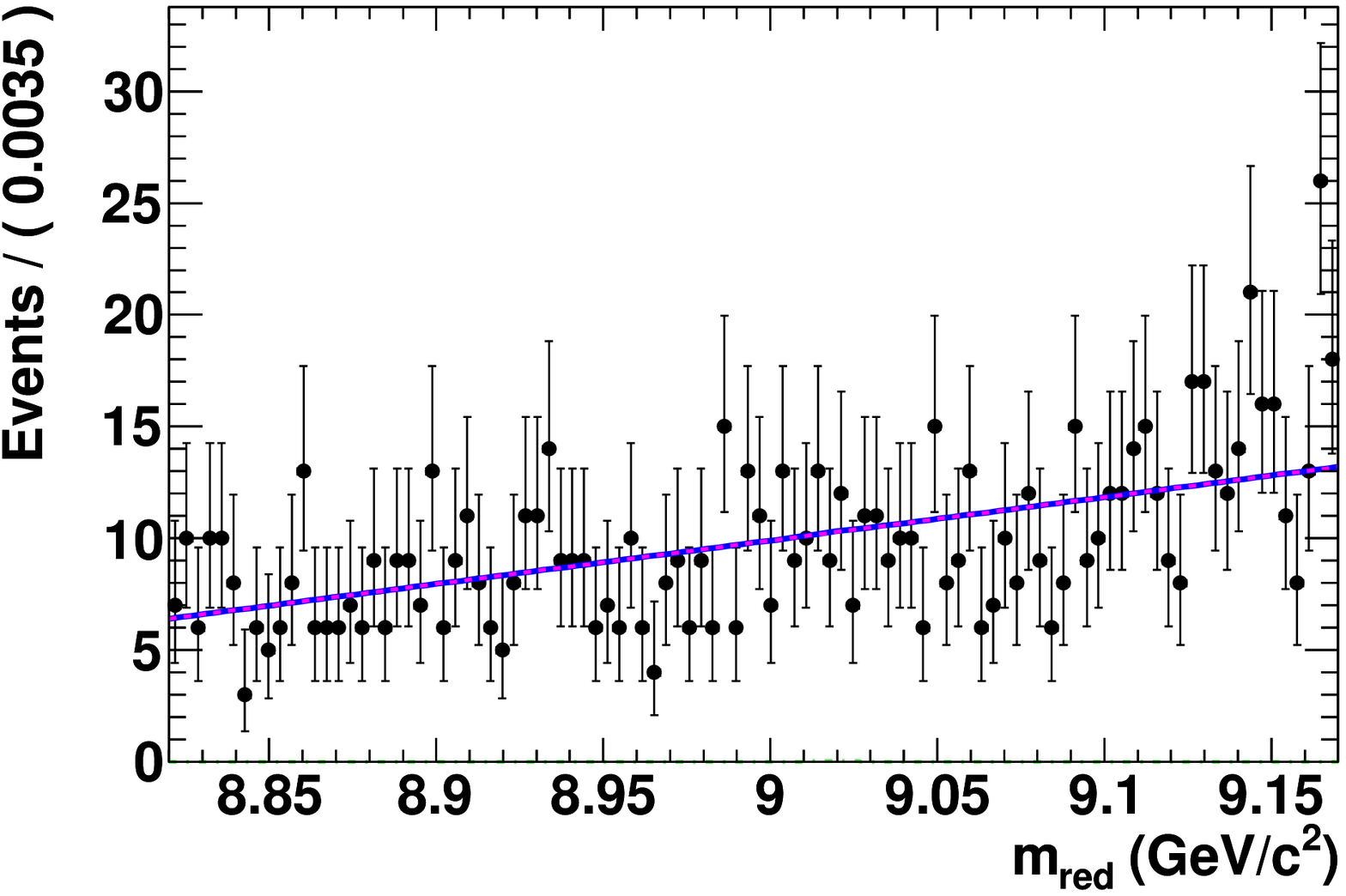}

\smallskip
\centerline{\hfill (e) \hfill \hfill (f) \hfill}
\smallskip

\caption {Result of the likelihood fit to the unblinded $\Upsilon(3S,2S)$ onpeaks data samples. Projection plot onto reduce mass distribution for the $m_{A^0}$ of (a, b) $m_{A^0}=7.85$ GeV/$c^2$  (c, d) $m_{A^0}=8.5$ GeV/$c^2$  and (e, f)  $m_{A^0}=9.02$ GeV/$c^2$. Left plots are for $\Upsilon(2S)$ and right plots are for $\Upsilon(3S)$ data sample. The total ML fit is shown in solid blue; the non-peaking background component is shown in dashed magenta; the signal component is shown in green dashed.} 

\label{fig:highproj}
\end{figure}

\begin{figure}
\centering
\includegraphics[width=5.0in]{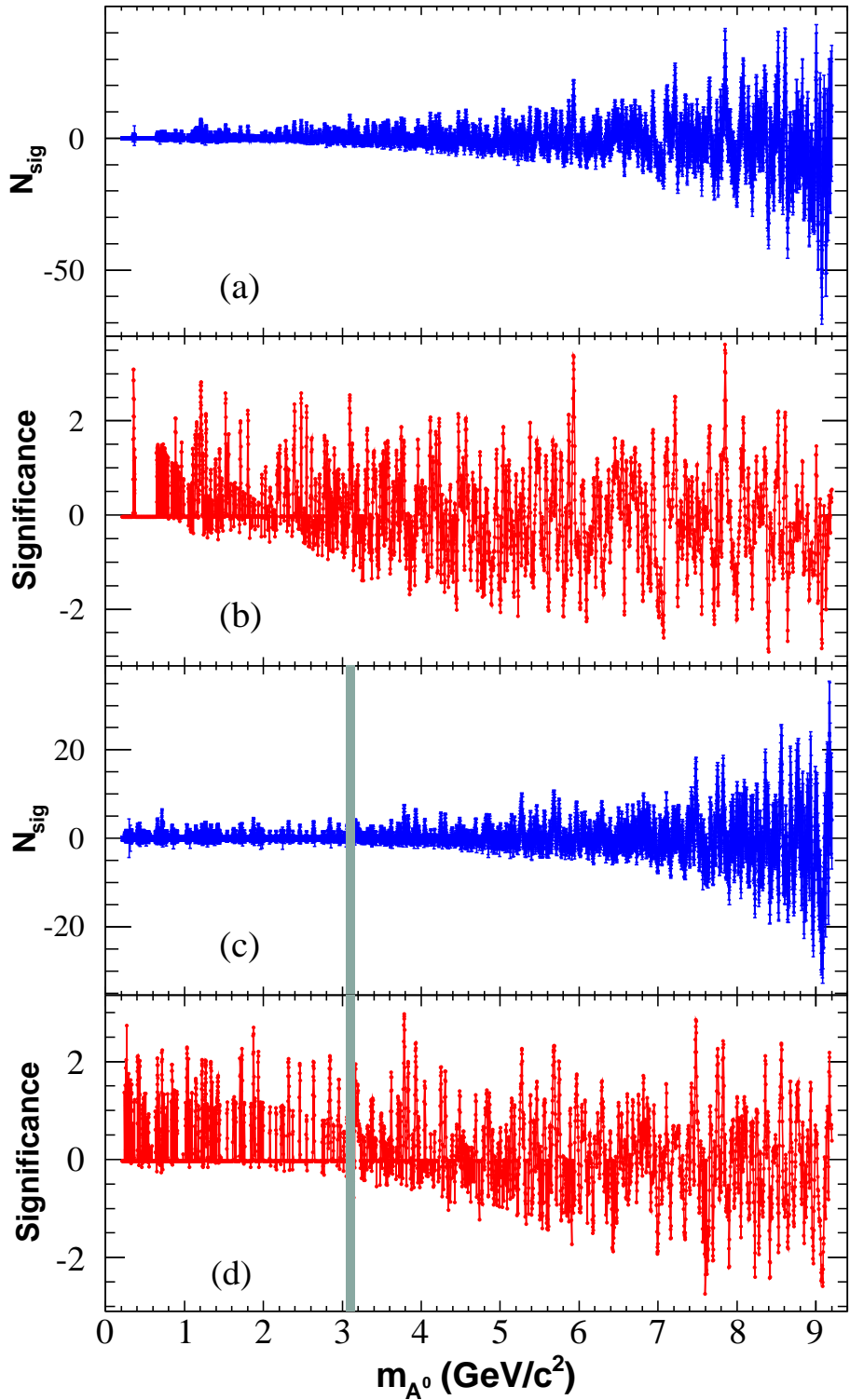}

\caption {The number of signal events and signal significance as a function of $m_{A^0}$ for (a,b) $\Upsilon(2S)$ and (c,d) for $\Upsilon(3S)$. The shaded area shows the region of the \jpsi resonance, excluded from the search in the $\Upsilon(3S)$ dataset.} 

\label{fig:Newnsig} 
\end{figure}

\begin{figure}
\centering
\includegraphics[width=6.0in]{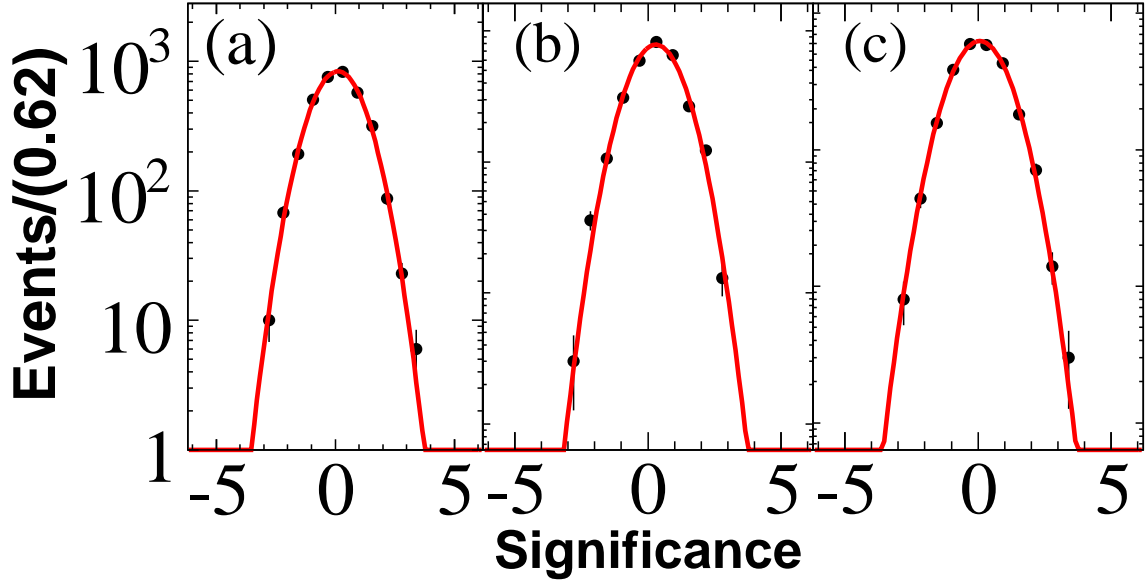}

\caption { Histogram of the signal significance $\mathcal{S}$  with statistical error for (a) $\Upsilon(2S)$ fit, (b) $\Upsilon(3S)$ fit and (c)  the combined data of $\Upsilon(2S)$ and $\Upsilon(3S)$. The $\mathcal{S}$ is excluded in the range of $-0.04 < \mathcal{S} < 0$.  The overlaid curve shows the standard normal distribution expected in the absence of signal.} 

\label{fig:Newsignif} 
\end{figure}

 We perform  the likelihood scan for any  possible peaks in the $m_{\rm red}$ distribution using the unblinded $\Upsilon(3S,2S)$ onpeak data-set in the steps of half of $m_{\rm red}$ resolution, corresponding to 4585 $m_{A^0}$ points. The \jpsi mass region in the $\Upsilon(3S)$ dataset, defined as $3.045 \le m_{\rm red} \le 3.162$ \gevcc, is excluded from the search due to large background from $\jpsi \rightarrow \mu^+\mu^-$ decays. The projection plots for selected mass points are shown in Figure~\ref{fig:Newlowproj} and ~\ref{fig:highproj}. Figure~\ref{fig:Newnsig}~   shows the number of signal events as well as signal significance for the $\Upsilon(3S,2S) \rightarrow \pi^+\pi^- \Upsilon(1S)$; $\Upsilon(1S) \rightarrow \gamma A^0$; $A^0 \rightarrow \mu^+\mu^-$ decay as a function of $m_{A^0}$.  Figure~\ref{fig:Newsignif} shows the distribution of signal significance ($\mathcal{S}$), where $\mathcal{S}$ is excluded in the range of $-0.04 < \mathcal{S} < 0$. The significance is expected 
to follow a normal distribution with $\mu =0$ and $\sigma = 1$ for the pure background hypothesis. The largest values of significance 
are found to be 3.62 (2.96) in the $\Upsilon(2S)$ ($\Upsilon(3S)$) dataset, and 3.24 for the combined $\Upsilon(2S,3S)$ dataset.

\section{Trial factor study: true significance observation}
\label{sect:trial_factor}
When we search for a narrow resonance for $A^0$  at unknown mass points over a broad range of background, special care must be exercised in evaluating the true significance of observing a local excess of events.  The log-likelihood ratio method is used to compute the significance of any positive signal observation. Since we need to  scan the $m_{\rm red}$ distribution of the $\Upsilon(2S,3S)$ onpeaks datasets at 4585 $m_{A^0}$ points, we should expect at least a few statistical fluctuations at the level of $\mathcal{S} \approx 3$, even for the null hypothesis. Hence, we need to determine the probability for the background fluctuation to a particular value of $\mathcal{S}$ anywhere in a given $m_{A^0}$ range. 

  We generate toy Monte-Calro data according to the PDFs using the background only hypothesis. We then scan the toy data in the same way as was done for the $\Upsilon(2S,3S)$ onpeak data-sets for all 4585 $m_{A^0}$ points, and pick up one of the maximum value of significance $S_{max}$ from these 4585 $m_{A^0}$ points.  We repeat this process about $5000$ times and accumulate the $S_{max}$ value each time in a histogram. We also compute the $S_{max}^{comb}  = (S_{max}^{\Upsilon(2S)} + S_{max}^{\Upsilon(3S)})/2$ for the combined $\Upsilon(2S,3S)$ datasets.

\begin{figure}
\centering
 \includegraphics[width=3.0in]{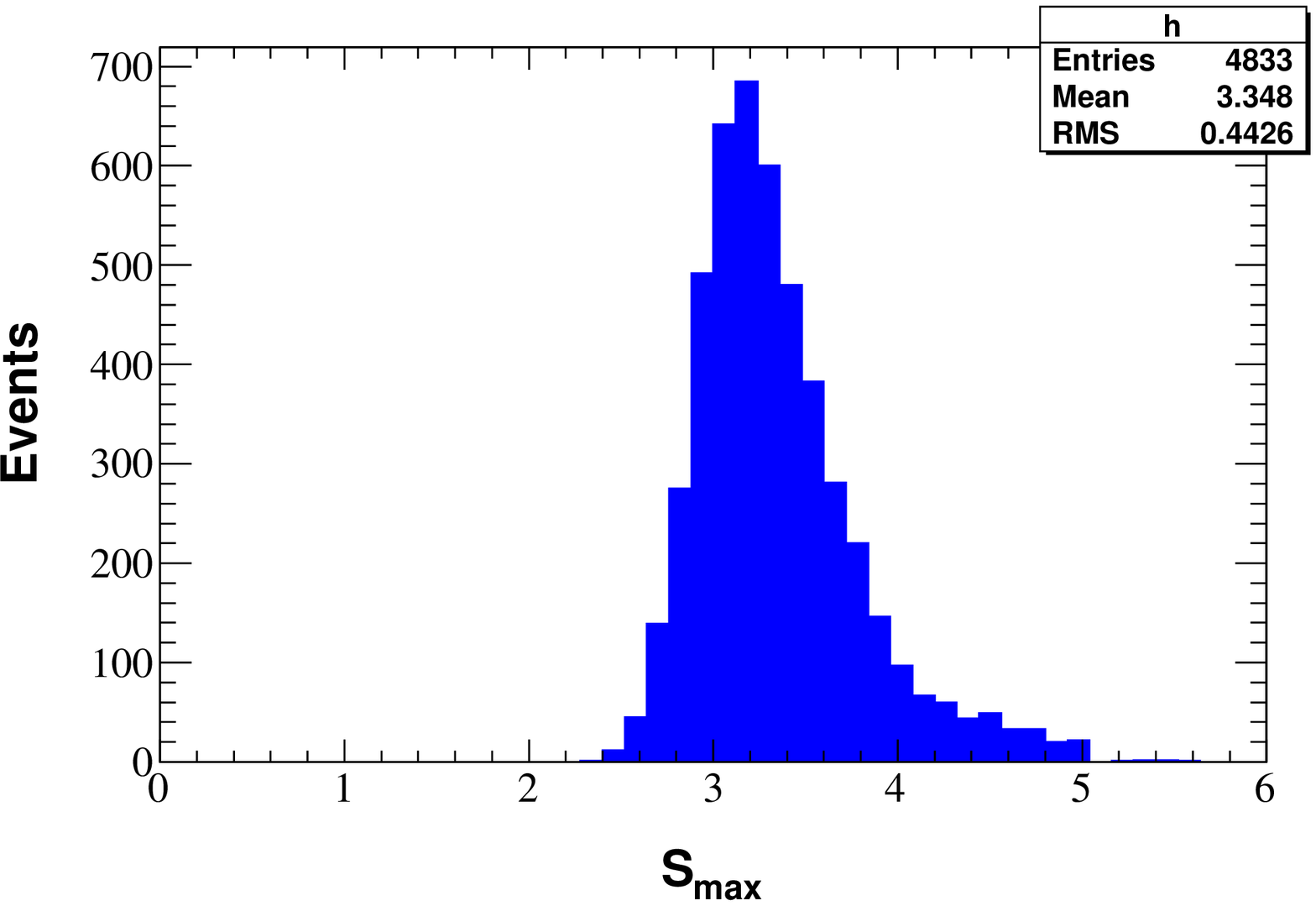}
 \includegraphics[width=3.0in]{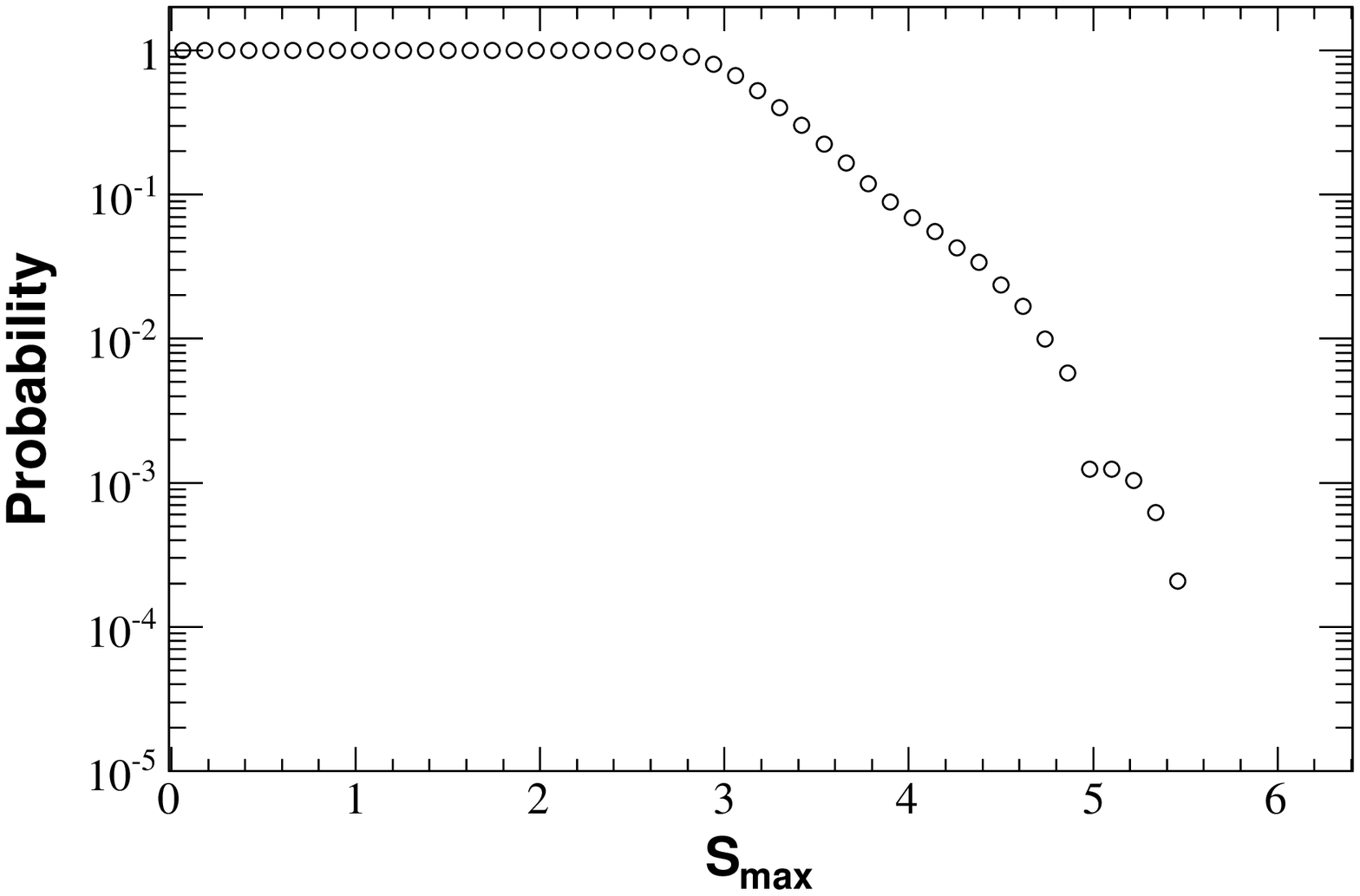}

\caption { Histogram of the $\mathcal{S}_{max}$  (left) and its cumulative distribution  (right) for the $\Upsilon(2S)$ dataset.} 

\label{fig:signifY2Stoy} 
\end{figure}

\begin{figure}
\centering
 \includegraphics[width=3.0in]{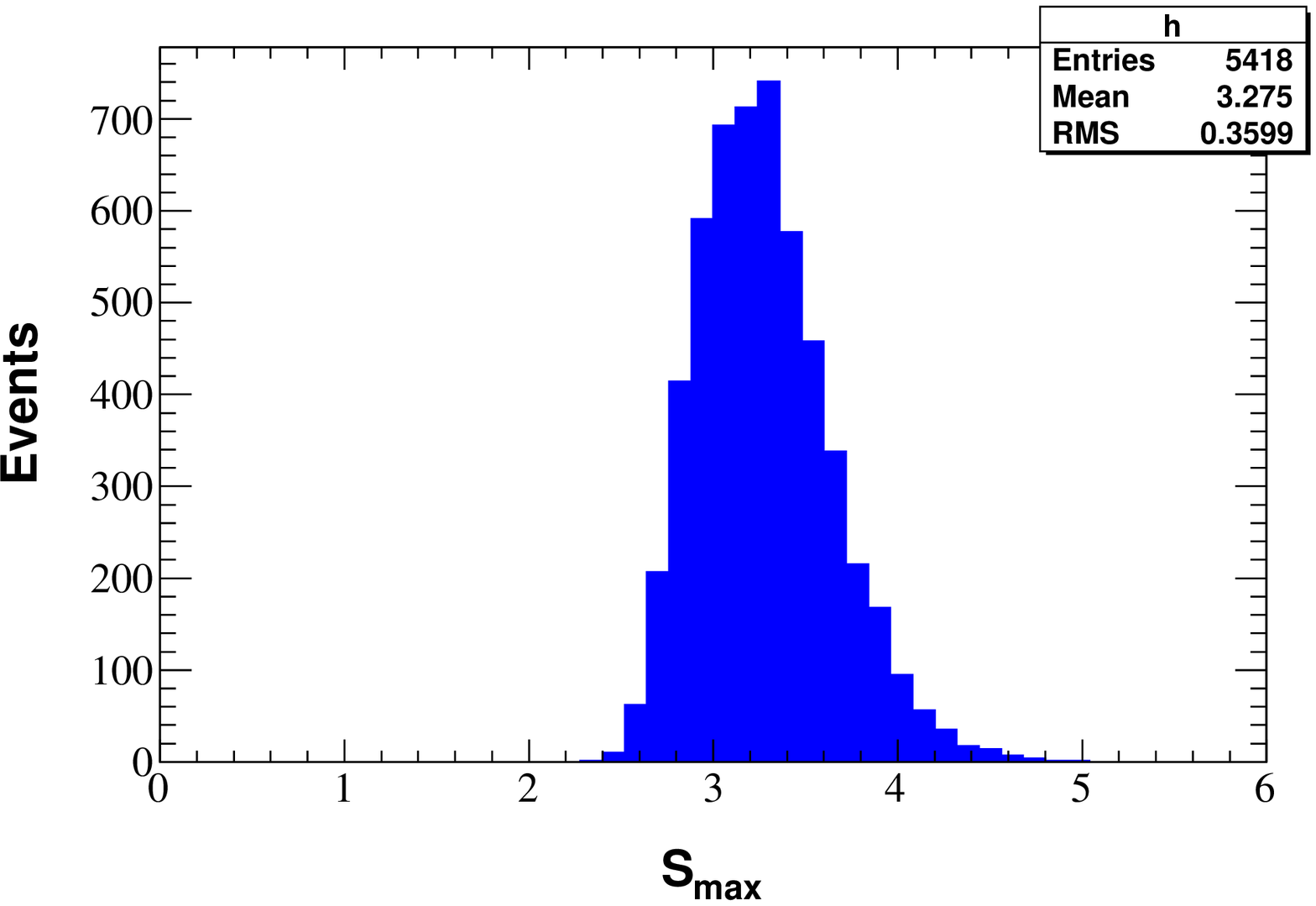}
 \includegraphics[width=3.0in]{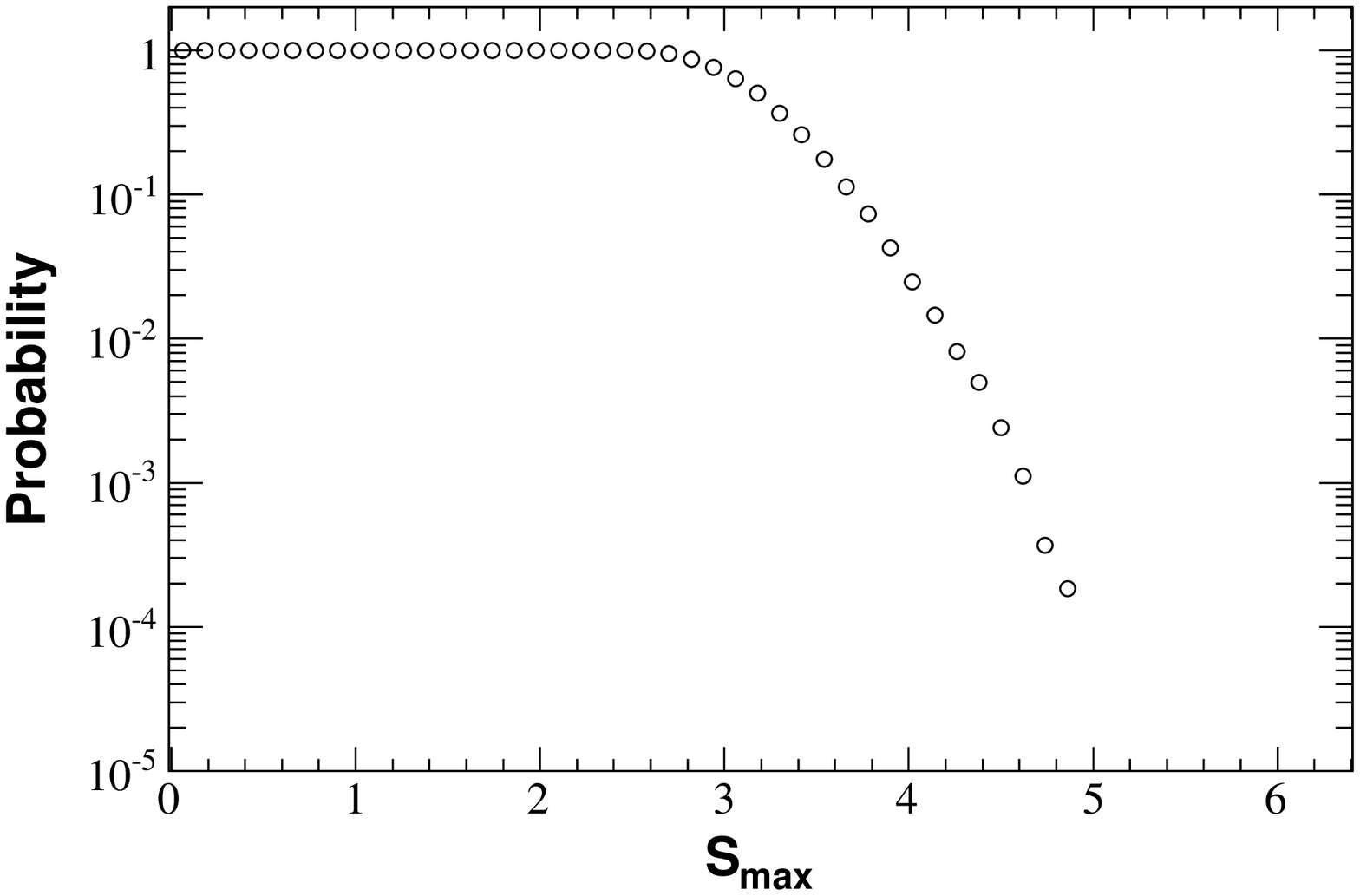}

\caption { Histogram of the $\mathcal{S}_{max}$  (left) and its cumulative distribution  (right) for the $\Upsilon(3S)$ dataset.} 

\label{fig:signifY3Stoy} 
\end{figure}

\begin{figure}
\centering
\includegraphics[width=3.0in]{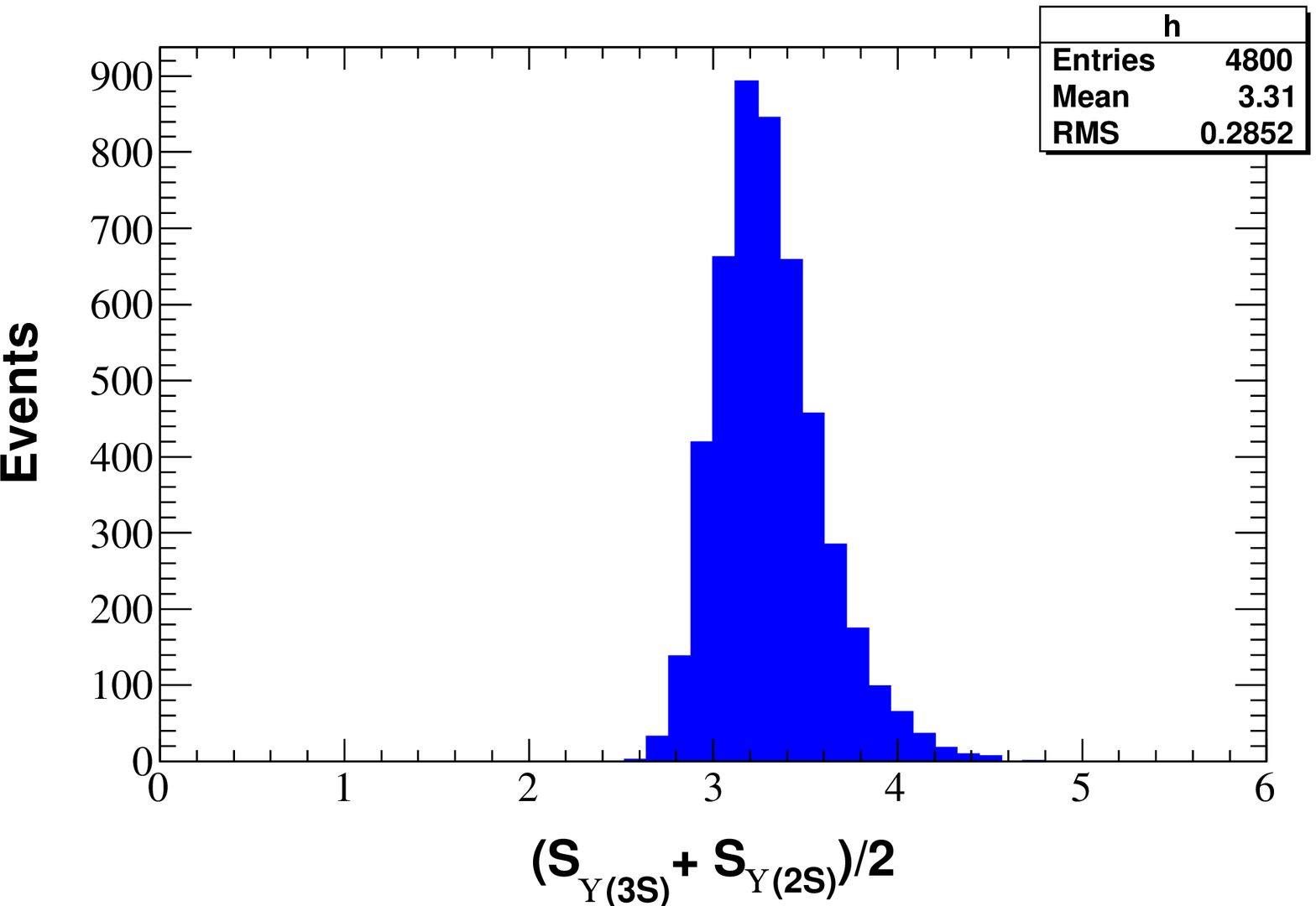} 
 \includegraphics[width=3.0in]{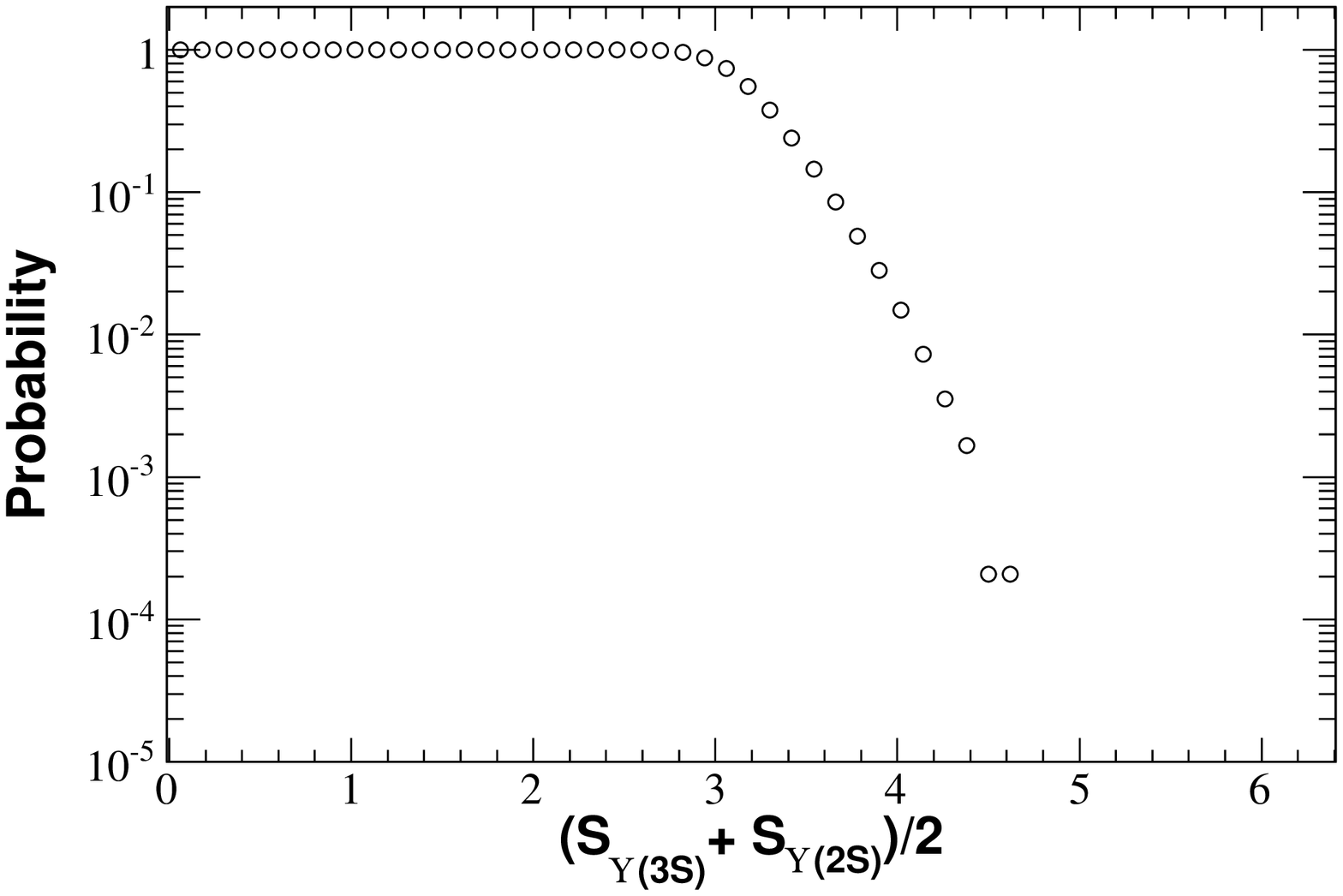}

\caption { Histogram of the $\mathcal{S}_{max}$  (left) and its cumulative distribution  (right) for the  combined dataset of $\Upsilon(2S,3S)$.} 

\label{fig:signifcomb} 
\end{figure}

We compute the inverse cumulative distribution (also called p-value) of $\mathcal{S}_{max}$ while integrating the PDF of $\mathcal{S}_{max}$ from $\mathcal{S}_{max}$ to $\infty$. The histograms of $\mathcal{S}_{max}$ and its inverse cumulative distribution for $\Upsilon(2S)$, $\Upsilon(3S)$ and combined $\Upsilon(2S,3S)$ datasets are  shown in Figure~\ref{fig:signifY2Stoy}, ~\ref{fig:signifY3Stoy} and  ~\ref{fig:signifcomb}, respectively. The p-value is the probability of a test statistics that describes the chance that a pure background would fluctuate to a signal peak with the significance $\mathcal{S}_{max}$. If the null hypothesis is correct, the p-value is uniformly distributed between zero and one. To express a given value of probability in terms of standard deviations ($\sigma$), a convention is adopted for  one sided Gaussian value of  $p= 1.35 \times 10^{-3}$ for $3\sigma$ and  $p= 2.865 \times 10^{-7}$ for $5\sigma$. 
 
We  estimate the probability to observe a fluctuation of $\mathcal{S}_{max} \ge 3.62$ ($18.1\%$)  in the $\Upsilon(2S)$ ($\Upsilon(3S)$) data-set to be $18.1\%$ ($66.2\%$),  and $\mathcal{S}_{max} \ge 3.24$ in the combined $\Upsilon(2S,3S)$ data-set to be $46.5\%$ based upon this trial factor study. Hence we interpret the observed local excess of events at several mass points in both the datasets as a mere background fluctuations.

\section{Chapter Summary}
In this chapter we have described the ML fit procedures used to extract the signal yield from the data. We have developed the signal and background PDFs using the signal MC samples generated at 26 $m_{A^0}$ points and the combined background MC. The fit validations are performed using a cocktail sample as well as a large number of Toy  MC experiments with different embedded signal events at selected $m_{A^0}$ points. The signal yields are extracted using the unblinded data of $\Upsilon(2S,3S)$. A trial factor study is also performed, which shows that there is no evidence for the di-muon decay of the $A^0$ in the radiative decays of the $\Upsilon(1S)$ in the $\Upsilon(2S,3S)$ data samples. The next chapter will describe the possible sources of systematic uncertainties for this analysis.   

 




















\chapter{Systematic Uncertainties}
\label{Chapter5}
This chapter describes the sources of the systematic uncertainty which we consider in this analysis. Two kinds of systematic uncertainties are identified, which are additive and multiplicative systematics. The additive systematics reduce the significance of any observed peak and does not scale with the number of reconstructed events. It arises from the uncertainty on the PDF parameters and the fit bias. The multiplicative systematics do not change the significance of any observed peak and scales with the number of reconstructed events. The primary contributions to the multiplicative systematic uncertainties come from the RF classifier selection, muon-ID, photon-selection, tracking and $\Upsilon(2S,3S)$ kinematic fit $\chi^2$.

\section {PDF systematics}
 The dominant contribution to the additive systematic uncertainty  comes from the uncertainties in the extracted signal yield ($N_{sig}$), which are primarily due to uncertainties in the PDF shapes. We evaluate the PDF systematic uncertainties after unblinding the Run7 $\Upsilon(3S, 2S)$ onpeak datasets by varying each parameter by its statistical error and observing the change in the fitted signal yield $\delta = \Delta N_{sig}$. The total systematic uncertainty in the signal yield is given by $\delta_{Tot} = \sqrt{ \vec{\delta^T}C\vec{\delta}}$, where $\vec{\delta} = < \delta_1 ...\delta_N >$ and C is the parameter correlation matrix, giving a systematic uncertainty in the signal yield.  The $\delta_{Tot}$ value is found to be very small for most of the $m_{A^0}$ points and it varies from (0.00 -- 0.62) events for the $\Upsilon(2S)$ dataset and (0.04 -- 0.58) events for $\Upsilon(3S)$ dataset. 

\section{Fit Bias} 
 We perform a study of fit bias on the signal yield with a large number of Toy MC experiments as mentioned in section~\ref{section:ToyMC}. The biases are consistent with zero and their average uncertainty is taken as a systematic uncertainty.   

\section{ Systematic uncertainty for Particle ID}
\begin{figure}
\centering
 \includegraphics[width=2.9in]{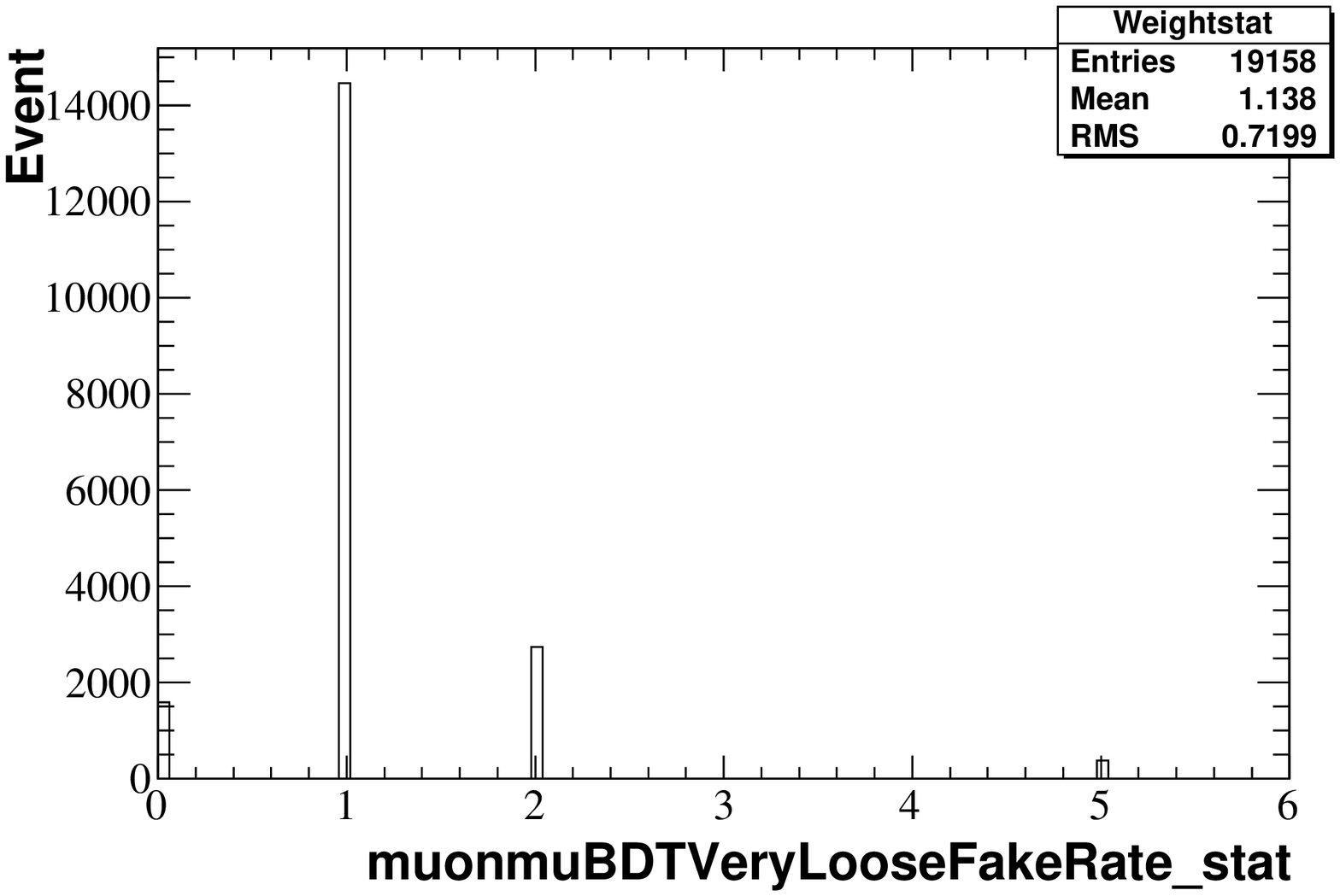}
 \includegraphics[width=3.1in]{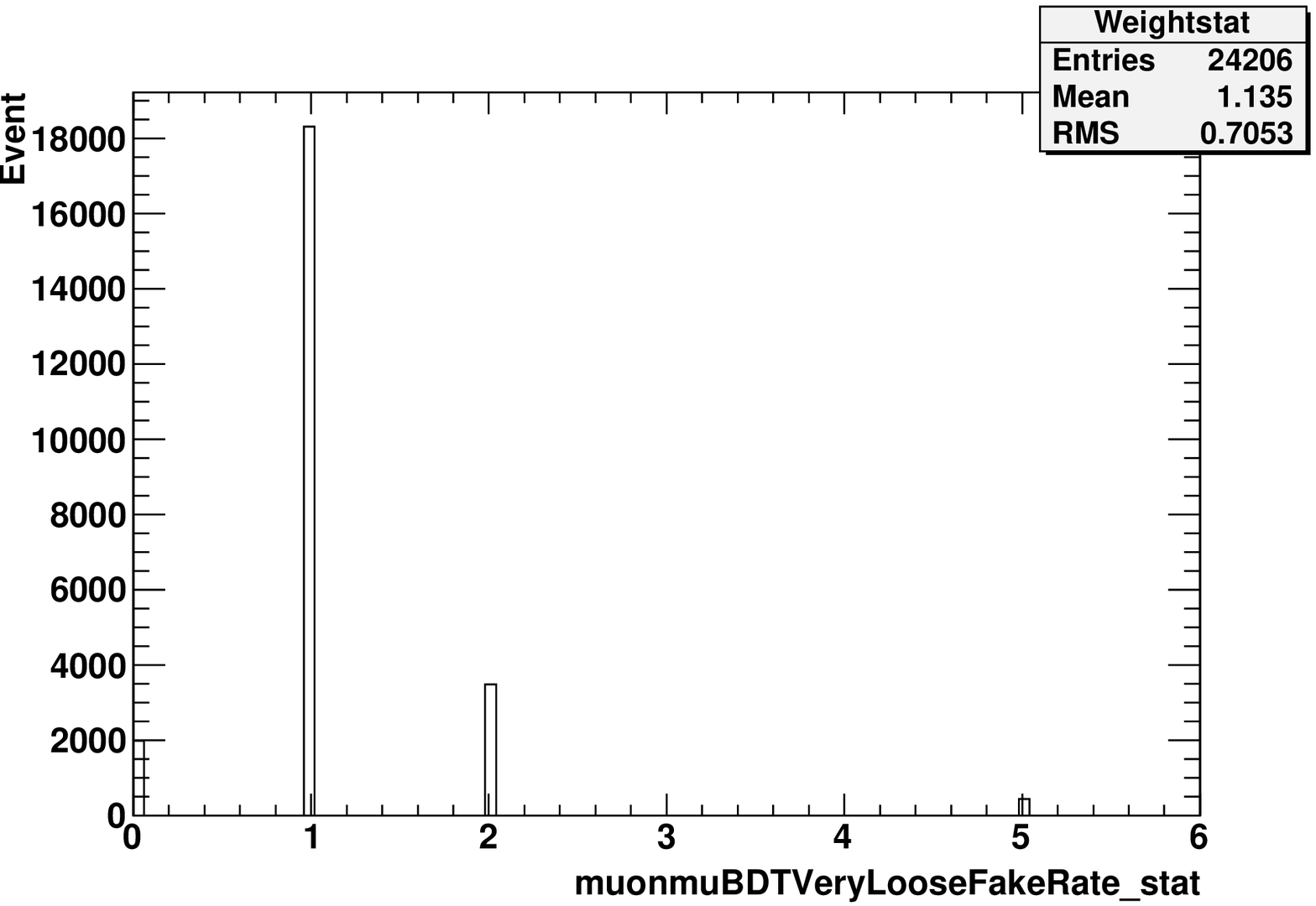}

\caption {PID-weight status distribution for OR-muon PID selection cut for $\Upsilon(2S)$ (left) $\Upsilon(3S)$ (right).} 
\label{Fig:PID_status}       
\end{figure}

The systematic uncertainty for muon PID selection cuts is evaluated by a standard PID weighting recipe developed by the PID group in \babar\ experiment \cite{PID}. This recipe creates a map that assigns a weight of each selected track, where weight is the ratio of efficiency in data and MC. Weight comes from the PID tables, which include the central value of weight, statistical uncertainty of weight, and status of given charged tracks. We first apply all the optimal selection cuts (excluding muon ID cut) to the signal MC sample in the mass range of 0.212 -- 9.46 GeV/$c^2$. Then we check the status of charged tracks after applying the OR-muon PID (BDTMuon1IDFakeRate $||$ BDTMuon2IDFakeRate) selection cut. Figure~\ref {Fig:PID_status} shows PID-weight status distribution for muon PID selection cut for both $\Upsilon(3S)$ and $\Upsilon(2S)$.

\begin{figure}
\centering
 \includegraphics[width=2.9in]{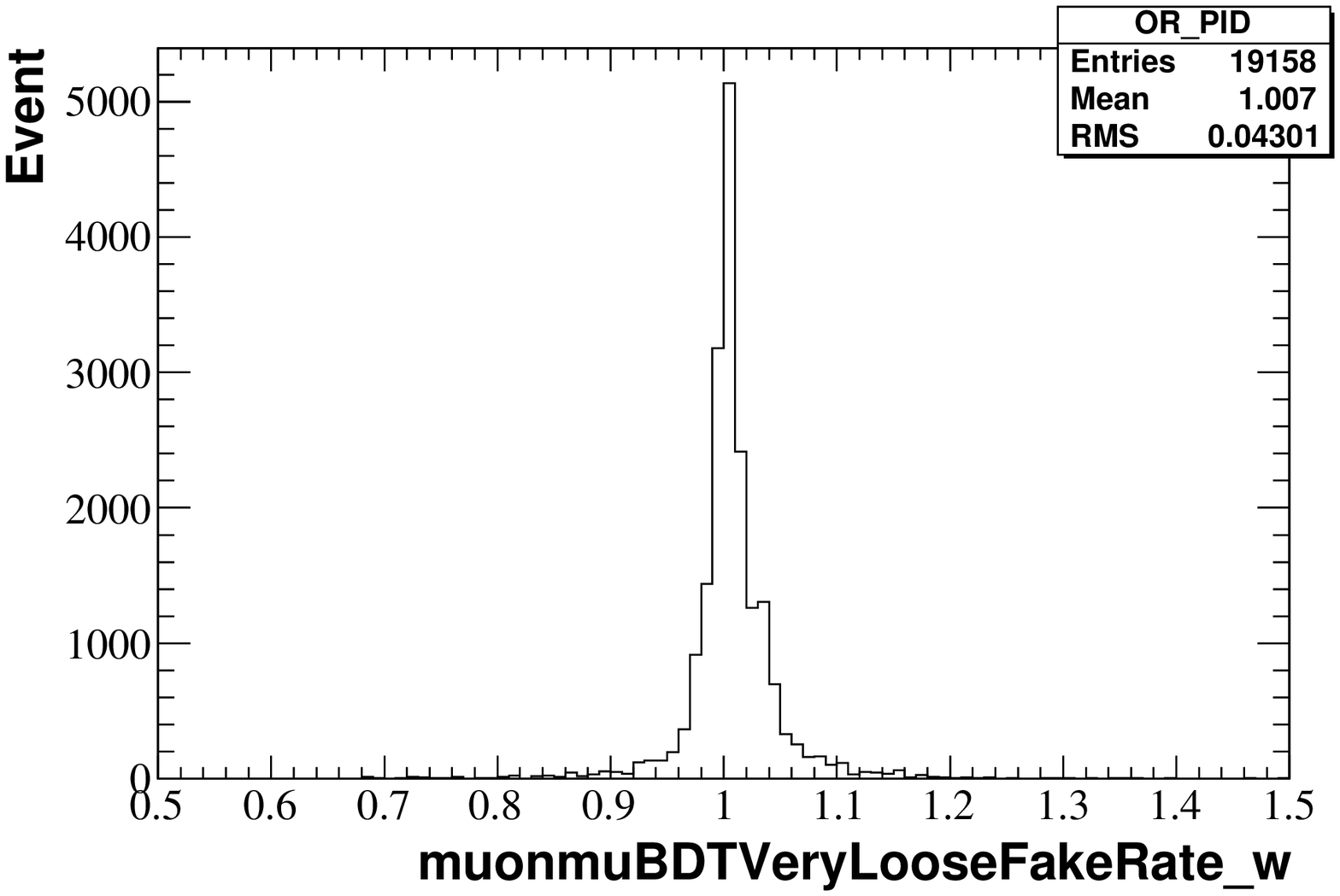}
 \includegraphics[width=3.1in]{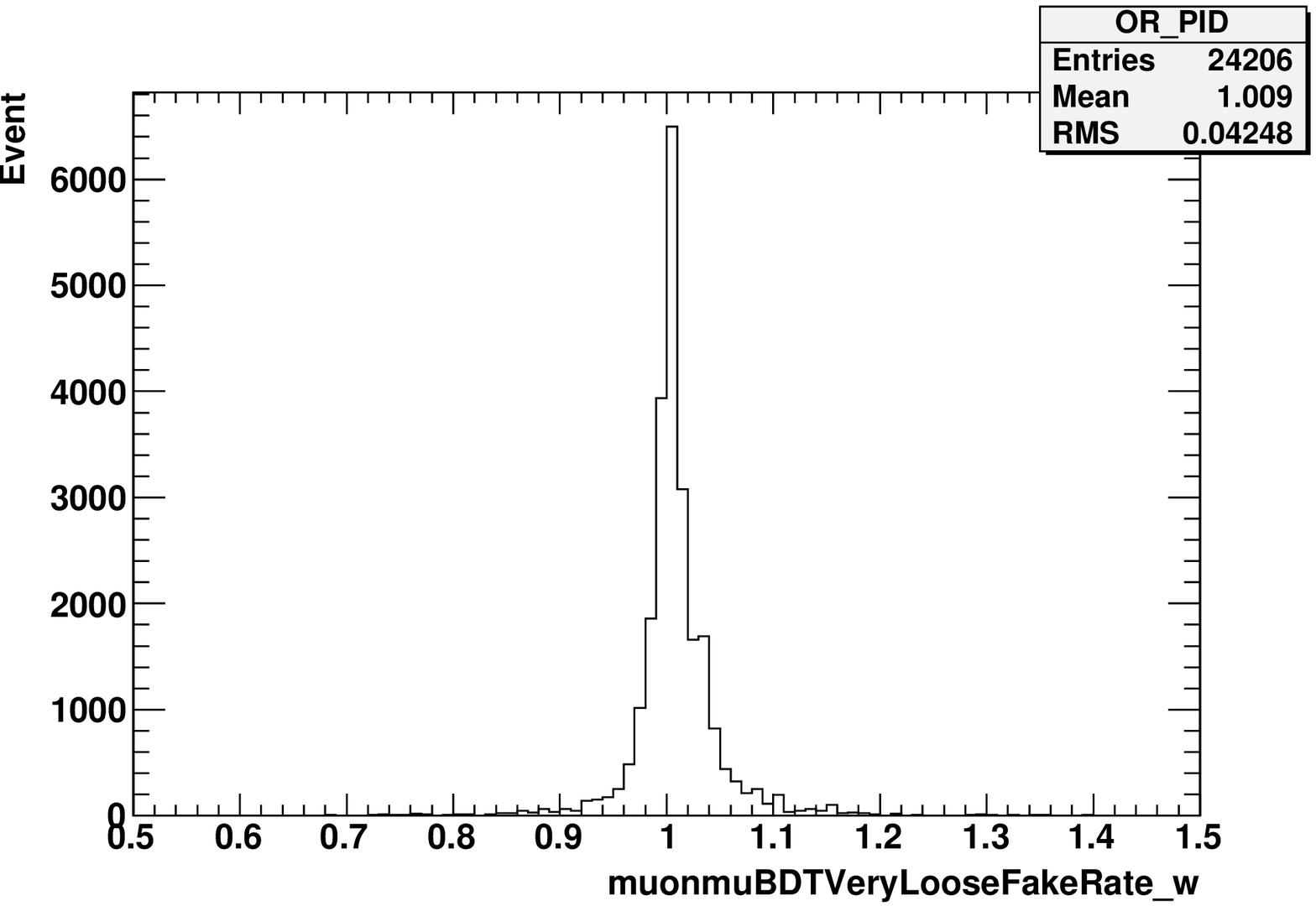}

\caption {New generated PID$\_$weight distribution (mean value of each Gaussian random number) for OR-muon PID selection cut for $\Upsilon(2S)$ (left) and $\Upsilon(3S)$ (right).} 
\label{Fig:PID_modified}       
\end{figure}

 We construct a new table by generating 10000 Gaussian random number with mean = PID$\_$weight, and sigma = PID$\_$weighterr for the PIDWeight$\_$status = 1, 2 and 3, where 1 means the PID efficiency of data and MC are well measured, 2 means the PID efficiency in data and MC are poorly measured due to limited statistics, but still ok, and 3 means the PID efficiency in MC is zero, but upper limit in the weight is compatible with that of a lower momentum bin, from which the PID$\_$weight and PID$\_$weighterr have been taken.  For other PIDWeight$\_$status, we have used Gaussian mean = 1 and sigma = 0. Figure~\ref{Fig:PID_modified} shows the new generated PID$\_$weight distribution (mean value of each Gaussian random number) for OR-muon PID selection cut for both $\Upsilon(3S)$ and $\Upsilon(2S)$. The systematic uncertainty on the muon PID efficiency is taken as the RMS value of the Gaussian. We find that the systematic uncertainty on the muon PID is $4.30\%$ ($4.25\%$) for $\Upsilon(2S)$ ($\Upsilon(3S)$).    

\section{ Systematic uncertainty for the charged tracks}
\begin{figure}
\centering
 \includegraphics[width=3.0in]{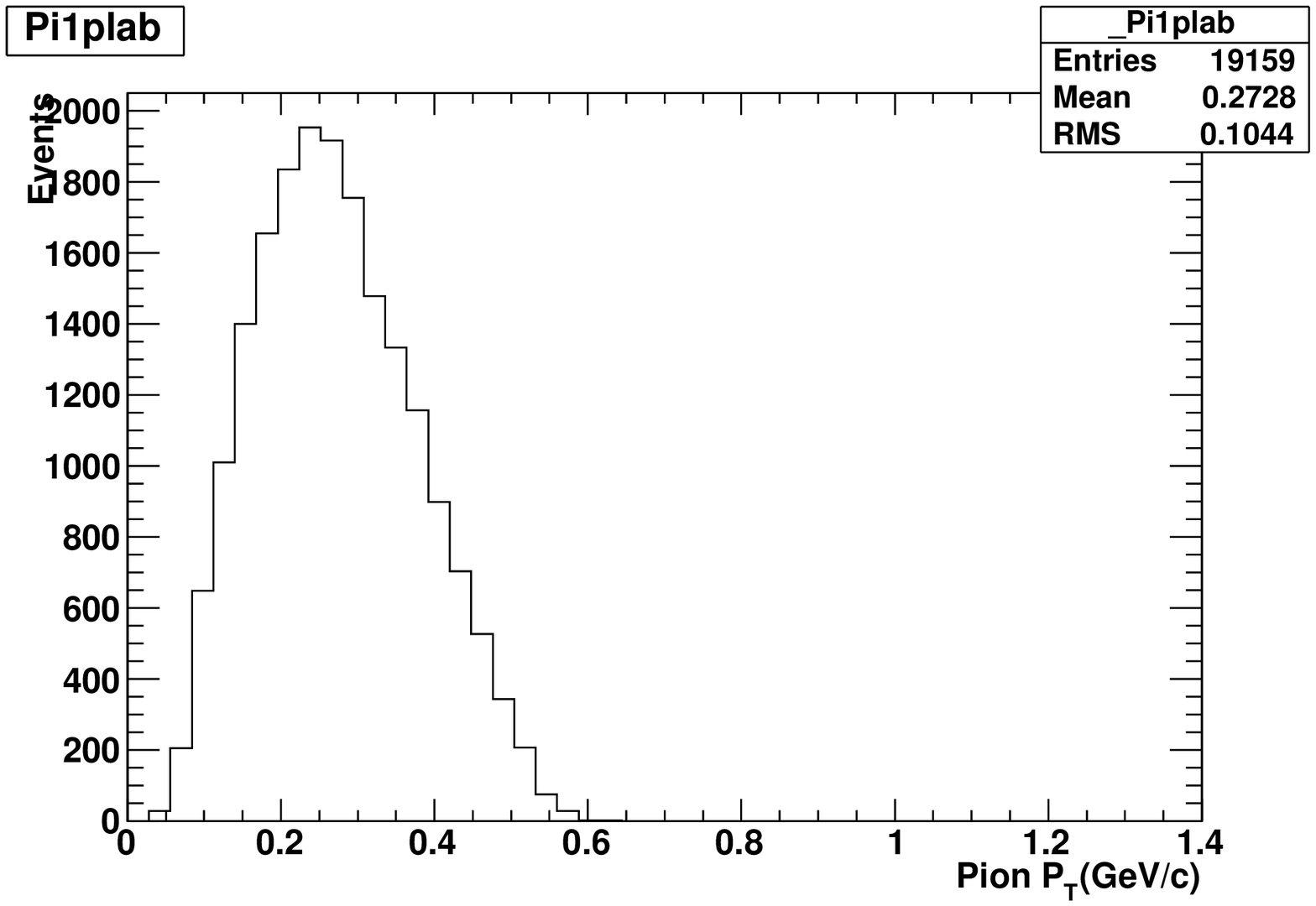}
 \includegraphics[width=3.0in]{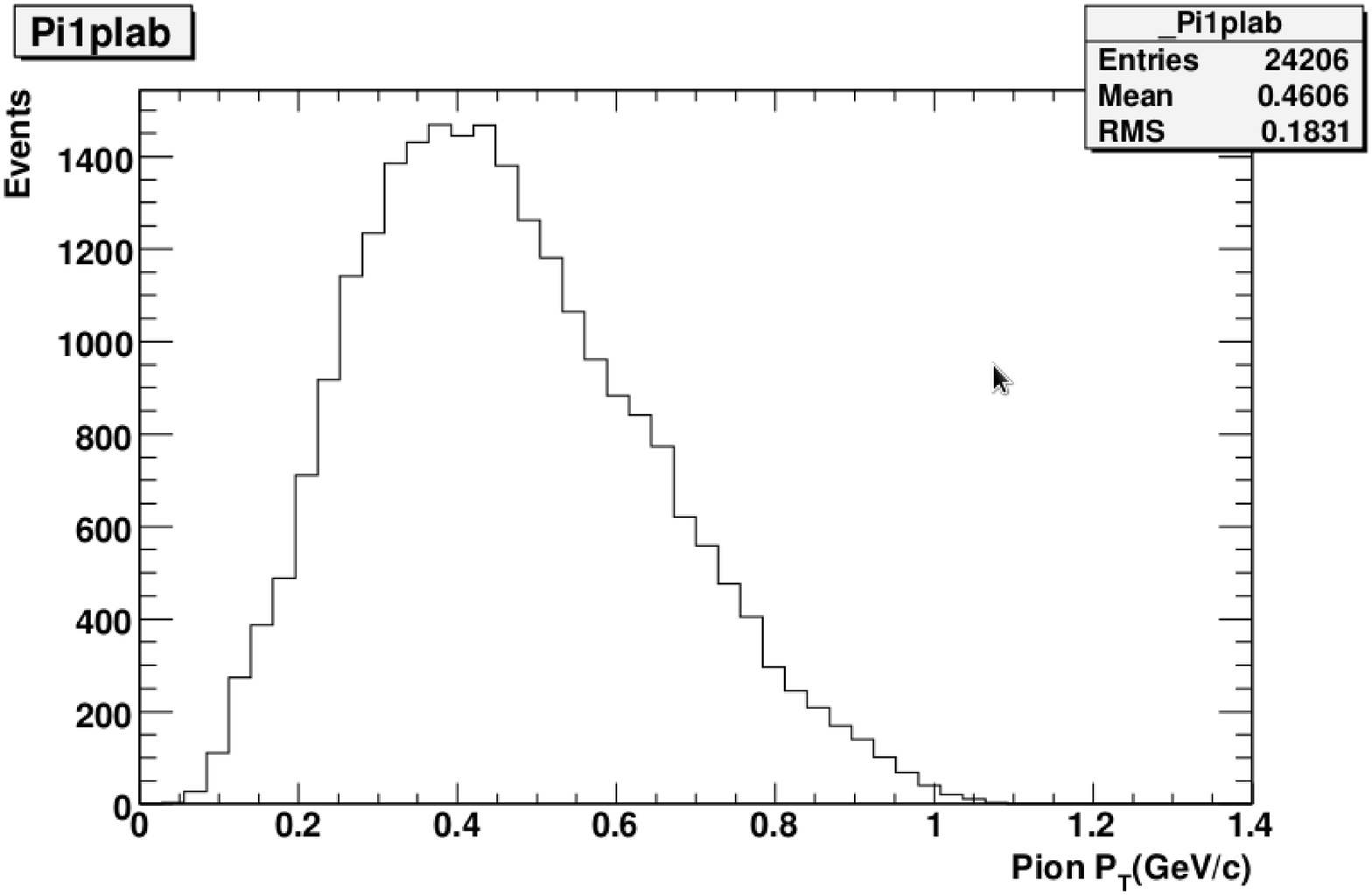}

\smallskip
\centerline{\hfill (a) \hfill \hfill (b) \hfill }
\smallskip
 
 \includegraphics[width=3.0in]{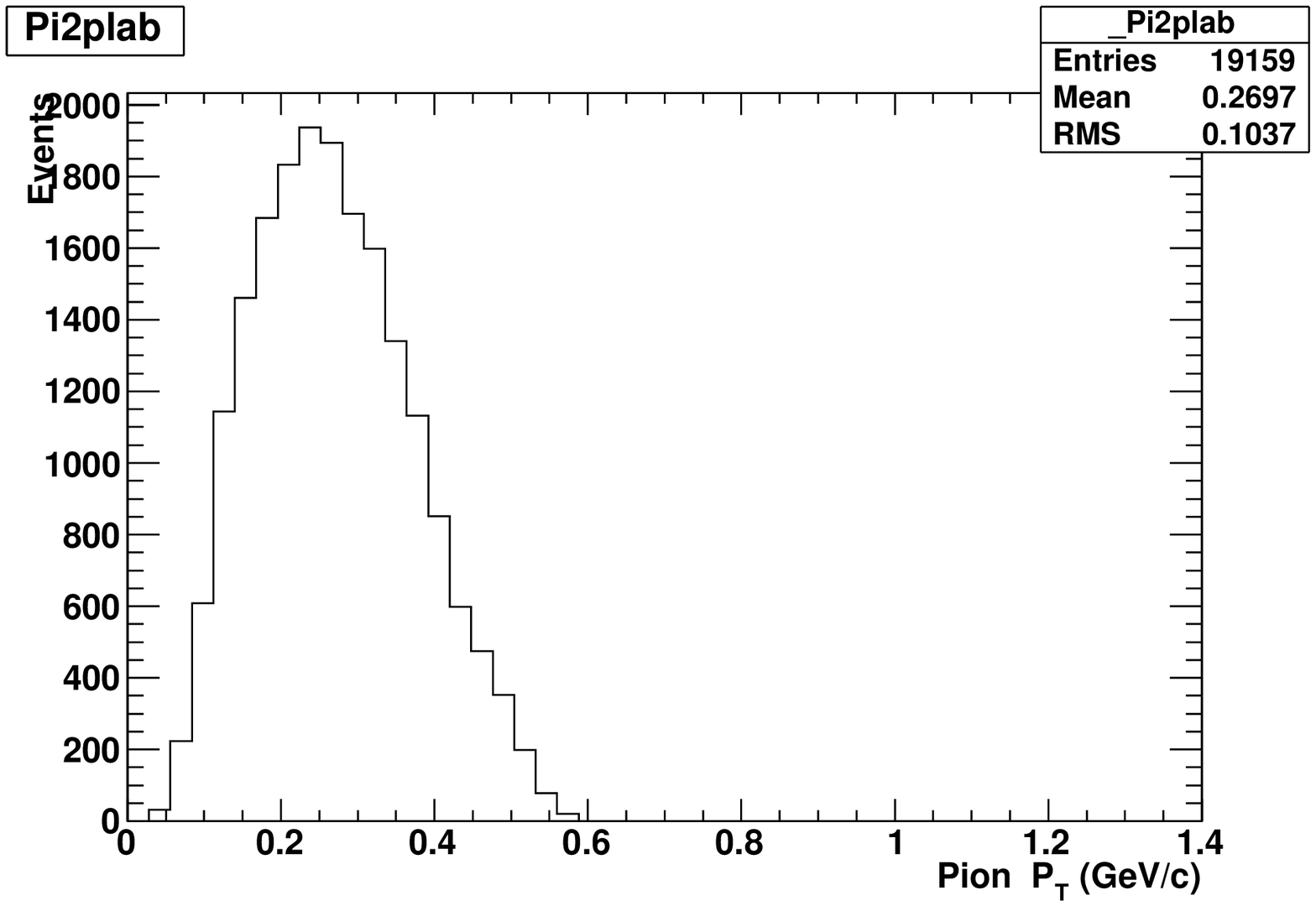}
\includegraphics[width=3.0in]{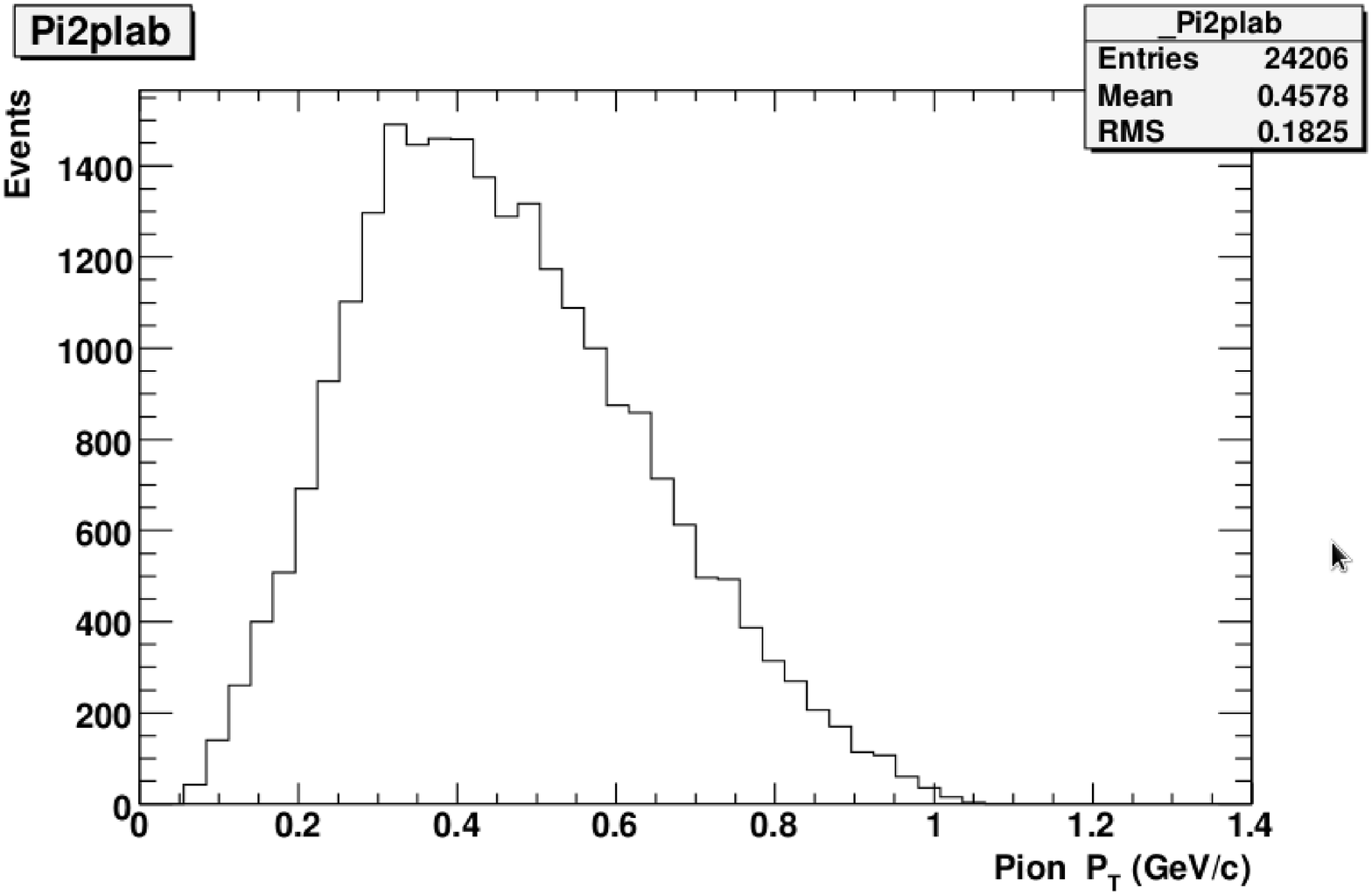}

\smallskip
\centerline{\hfill (a) \hfill \hfill (b) \hfill }
\smallskip

\caption {Pion transverse momentum distribution for both pions in signal MC. Left plots show for $\Upsilon(2S)$ and right plots show for $\Upsilon(3S)$.} 
\label{Fig:Pions}
\end{figure}

The systematic uncertainties for the four charged tracks are taken from  \cite{	arXiv:1207.2849}, which results in a systematic uncertainty of $1.74\%$ for the two highly energetic muon tracks. The systematic uncertainty for the  pions with $p_{T} < 180$ MeV/c is taken  from the soft-pion study and for tracks with $P_{T} > 180$ GeV/c, the systematic uncertainty is taken from the Tau31 study as discussed in \cite{arXiv:1207.2849}. Figure~\ref{Fig:Pions} shows that around (20.6$\%$) ($4\%$)  signal MC events lie in the range of $p_{T} < 180$ MeV/c (for both pions) and $96\%$ (79.4$\%$) signal MC events  lie in the range of $p_{T} > 180$ MeV/c for $\Upsilon(2S)$ ($\Upsilon(3S)$) dataset. So we evaluate  the uncertainty due to the reconstruction of  both pions to be 1.99$\%$  ($1.76\%$) and the total systematic uncertainty for the four tracks to be  $3.73\%$ ($3.5\%$) for $\Upsilon(2S)$ ($\Upsilon(3S)$).

\section{Systematic uncertainty for $\Upsilon(2S,3S)$ kinematic fit $\chi^2$}
\begin{figure}
\centering
 \includegraphics[width=3.0in]{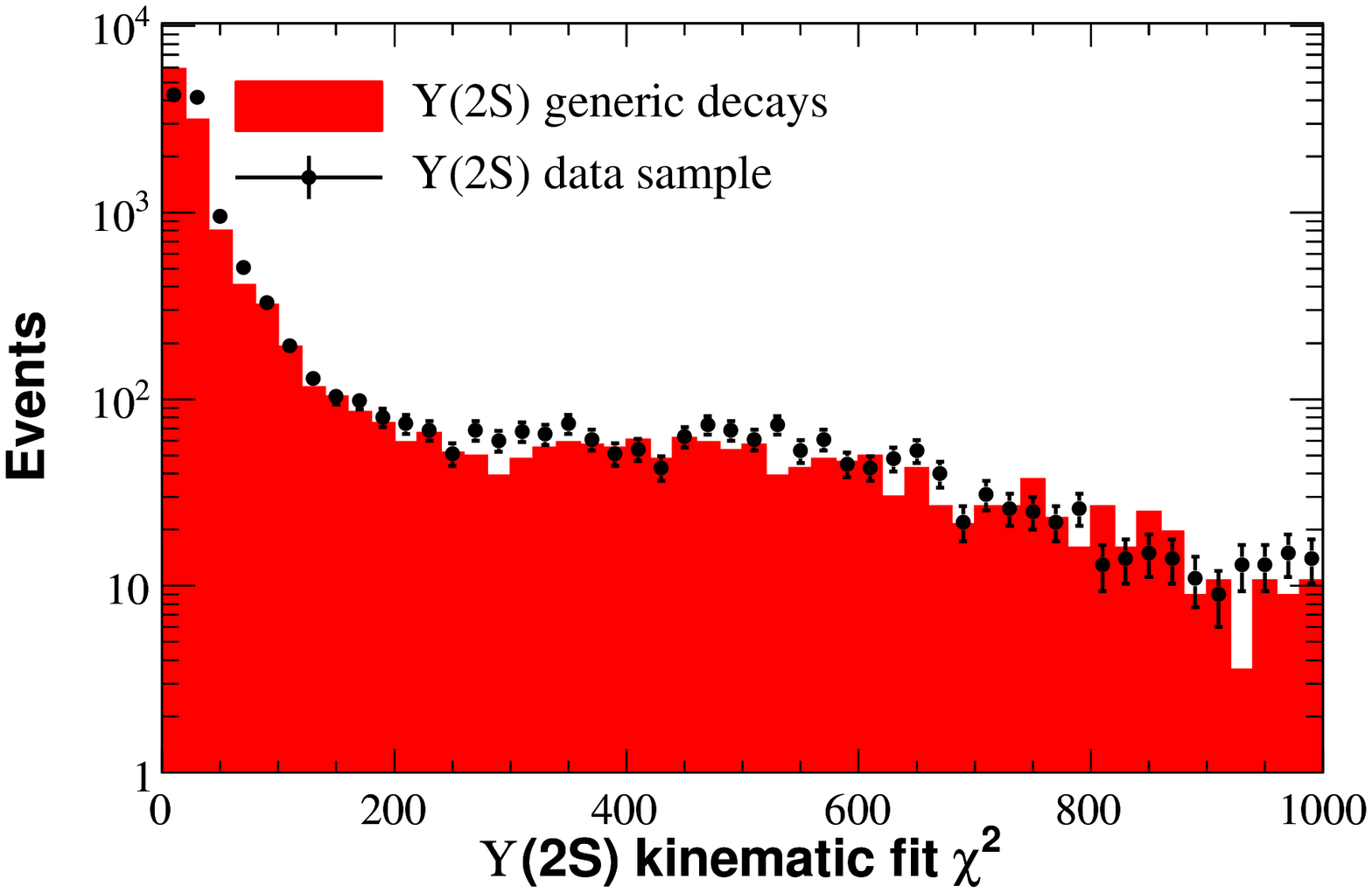}
 \includegraphics[width=3.0in]{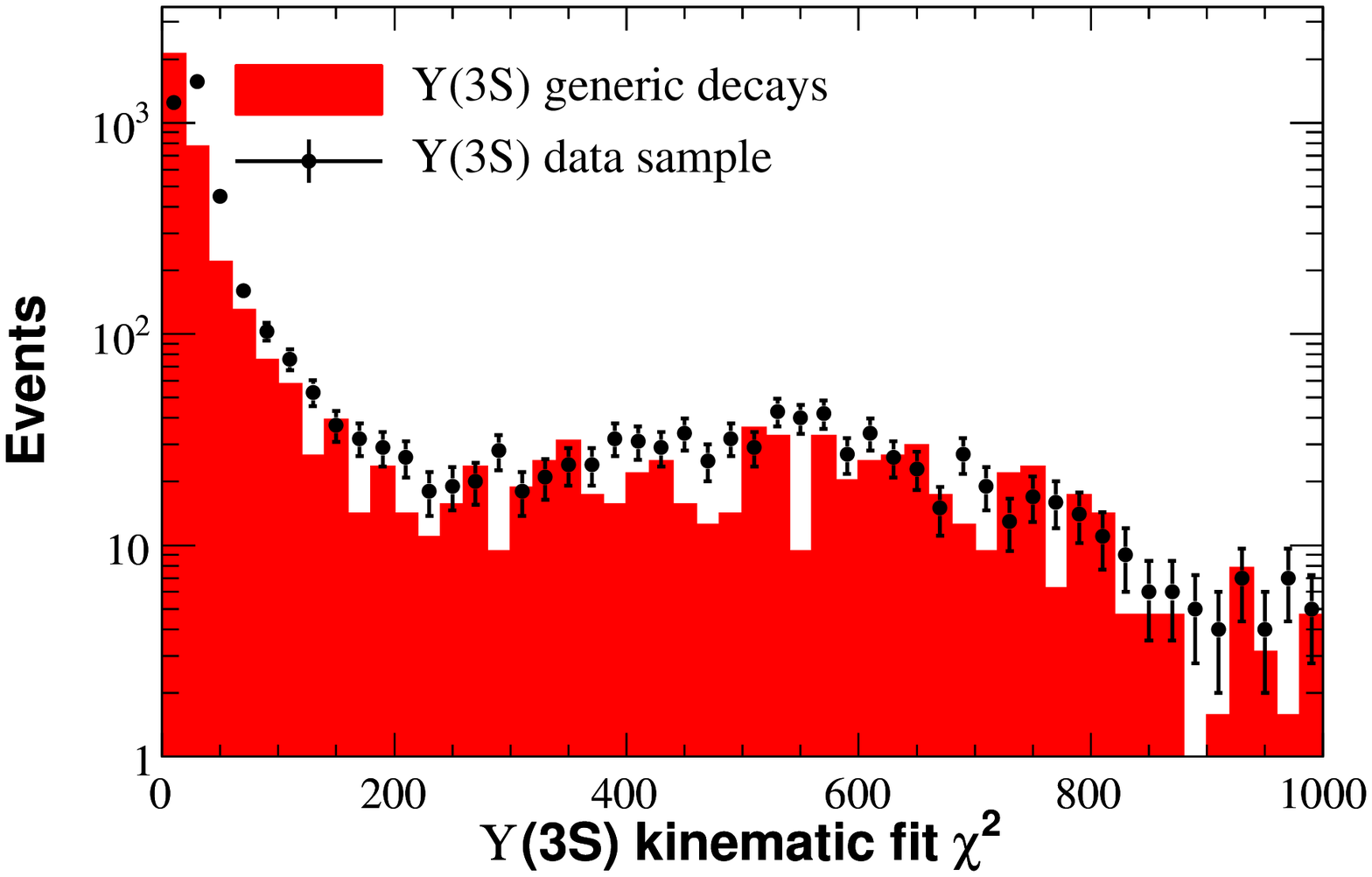}

\caption {$\Upsilon(nS)$ kinematic fit $\chi^2$ distributions after applying all the optimal selection cuts except $\Upsilon(nS)$ kinematic fit $\chi^2$ cut  for $\Upsilon(2S)$ (left) and $\Upsilon(3S)$ (right).} 
\label{Fig:Ynschi2sys}       
\end{figure}

  We use the test sample of $\Upsilon(3S, 2S)$ generic MC and the $\Upsilon(3S, 2S)$  onpeak data samples to evaluate the systematic  uncertainties for $\Upsilon(3S, 2S)$ kinematic fit $\chi2$ after unblinding the data samples. We first apply all the optimal selection cuts to the $\Upsilon(3S, 2S)$ generic MC and the $\Upsilon(3S, 2S)$ onpeak data samples except $\Upsilon(3S, 2S)$ kinematic fit $\chi^2$ cut. We then apply $\Upsilon(3S, 2S)$ kinematic fit $\chi^2$ to the both data and MC to calculate the systematic uncertainties. Figure~\ref{Fig:Ynschi2sys} shows the $\Upsilon(3S)$ kinematic fit $\chi^2$ distributions  for both $\Upsilon(3S)$ and $\Upsilon(2S)$. The relative number of events for both data and MC after applying the kinematic fit $\chi^2$ cut for both $\Upsilon(3S)$ and $\Upsilon(2S)$ are summarized in Table\ref{table:Ynschi2}. The systematic uncertainty due to the $\Upsilon(3S, 2S)$ kinematic fit $\chi^2$ are found to be 1.52$\%$ and 2.96$\%$  for the $\Upsilon(2S)$ and $\Upsilon(3S)$, respectively.

\begin{table}

\centering
\begin{tabular}{|l |r|r|r|r|}
\hline
&\multicolumn{2}{c|} {$\Upsilon(2S)$}   &\multicolumn{2}{c|}  {$\Upsilon(3S)$} \\
\cline{2-5}
\multicolumn{1}{|c|}{Selection cuts }     &  Data                & MC     &  Data                & MC \\
\hline 
Pre-selection cuts                         & 13264               & 7508   & 4706                & 2682               \\
\hline
\parbox{4.0cm} {$\chi_{\Upsilon(3S, 2S)}^2$ $<$ 300}  & 11136     & 6402            & 3857             & 2267               \\     
\hline
Efficiency         & $0.840 \pm 0.0032$     & $0.853 \pm 0.0041$        & $0.820 \pm 0.0056$  &$0.845 \pm 0.007$   \\
\hline 
\end{tabular} 
\caption{The relative number of events in data and MC after applying the $\Upsilon(3S, 2S)$ kinematic fit $\chi^2$  cuts.}
\label{table:Ynschi2}
\end{table} 

\section{ Systematic uncertainty for {\bf $\mathcal{B}(\Upsilon(2S,3S) \to \pipi \Upsilon(1S)$}}
The uncertainties on the branching fractions $\mathcal{B}(\Upsilon(2S,3S) \to \pipi \Upsilon(1S)$ are $2.2\%$ and $2.3\%$ for $\Upsilon(2S)$ and $\Upsilon(3S)$ datasets, respectively, which are taken from the PDG \cite{PDG}

\section{ Systematic uncertainty for RF-selection}
\begin{figure}
\centering
 \includegraphics[width=3.0in]{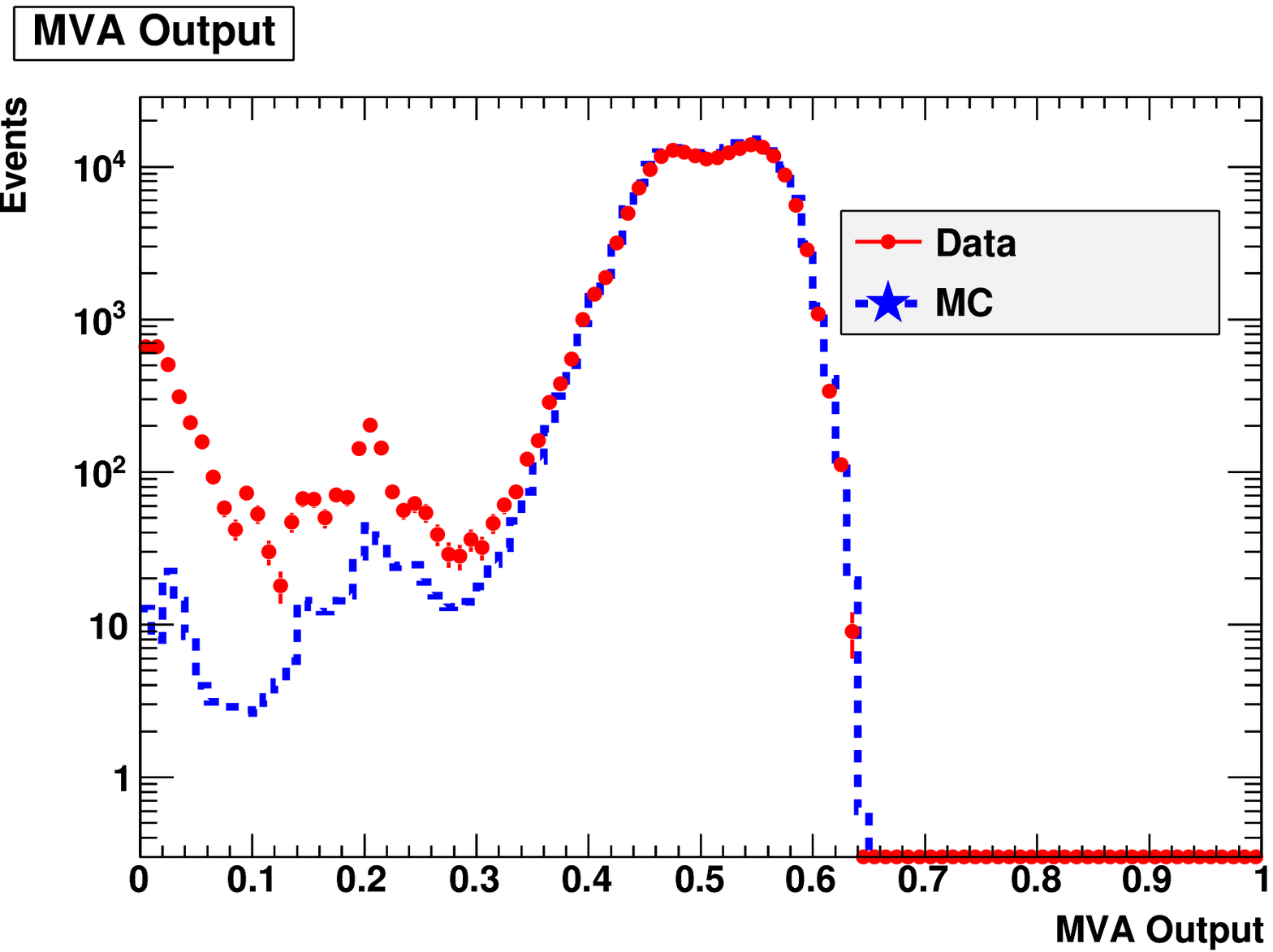}
 \includegraphics[width=3.0in]{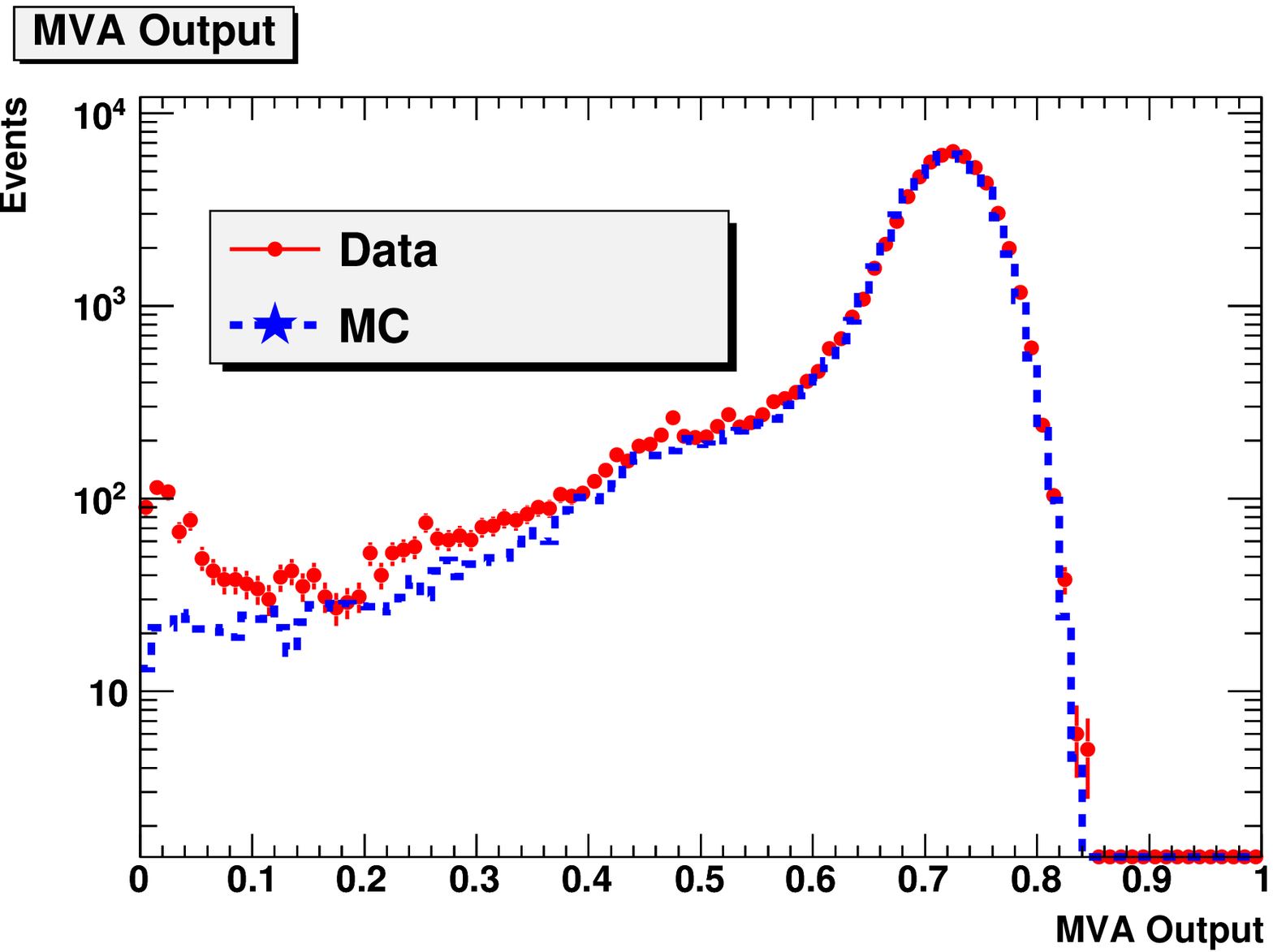}

\caption {Comparison of dipion RF  discriminant output between data and Monte-Carlo control samples for $\Upsilon(2S)$ (left) and  $\Upsilon(3S)$ (right).}

\label{Fig:Dipions}
\end{figure}

We study the systematic uncertainties for pion related variables using a control data and MC samples of $\Upsilon(3S, 2S) \rightarrow \pi^+\pi^- \Upsilon(1S); \Upsilon(1S) \rightarrow \mu^+ \mu^-$ as mentioned in section~\ref{section:mrec-study}.  Figure~\ref{Fig:Dipions} shows the data and MC comparison of the output of RF for both $\Upsilon(2S)$ and $\Upsilon(3S)$ datasets and the relative number of events for both data and MC after applying RF  cut are summarized in Table~\ref{table:Dipions}. Based on the relative difference in the efficiencies of the RF cut on the data and MC, we assign a systematic uncertainty of $2.21\%$ for the $\Upsilon(2S)$ dataset and $2.16\%$ for the $\Upsilon(3S)$ dataset.

\begin{table}
\centering
\begin{tabular}{|l |r|r|r|r|}
\hline
&\multicolumn{2}{c|} {$\Upsilon(2S)$}   &\multicolumn{2}{c|} {$\Upsilon(3S)$} \\
\cline{2-5}
\multicolumn{1}{|c|}{Selection cuts }     &  Data                & MC     &  Data                & MC \\
\hline 
Pre-selection cut  & 189644  & 1655701   & 66147               & 169913                \\
\hline
\parbox{4.0cm} { RF $>$ 0.388 (0.568) for  $\Upsilon(2S)$ ($\Upsilon(3S)$)}  & 183663  & 1642292    & 60116   & 157834              \\     
\hline
Efficiency        & $0.970 \pm 0.0004$ & $0.992 \pm 0.00007$        & $0.909 \pm 0.0011$  &$0.929 \pm 0.0006$  \\
\hline 

\end{tabular} 
\caption{The relative number of events in data and MC after applying the RF  cut.}
\label{table:Dipions}
\end{table}

\section{ Systematic uncertainty due to photon selection}
The systematic uncertainty related to the photon selection is measured using an $e^+e^- \rightarrow \gamma \gamma$ sample in which 
one of the photon converts into an $e^+e^-$ pair in the detector material \cite{Sanchez:2010bm}. The relative selection efficiencies of the photon selection variables are summarized in Table~\ref{table:PhotonEff} and we assign a systematic uncertainty of $1.96\%$ for the photon related variables in the $\Upsilon(3S)$ dataset. Since the photon selection criteria are similar in both $\Upsilon(3S)$ and $\Upsilon(2S)$ datasets, we  use the same systematic uncertainty value for $\Upsilon(2S)$.     

\begin{table}

\centering
\begin{tabular}{|c |r|r|}

	\hline
\hline
\multicolumn{3}{|c|}{For $\Upsilon(3S)$ dataset} \\
\hline
\hline 
Selection cuts & Efficiency $\%$ (Data) & Efficiency $\%$ (MC) \\
\hline
e2Mag $<$ 0.2  & $95.42 \pm 0.108$      & $95.74 \pm 0.063$ \\
\hline
Lateral moment [0.06, 0.74] & $97.83 \pm 0.075$  & $99.13 \pm 0.029$ \\
\hline
Zernika-42 moment $<$ 0.1 & $98.14 \pm 0.070$  & $99.19 \pm 0.028$ \\
\hline
Total Efficiency & $92.78 \pm 0.134$  & $94.63 \pm 0.070$ \\
\hline

\end{tabular}
 \caption{The relative efficiencies in data and MC after applying the photon related variables.}
\label{table:PhotonEff}
\end{table}

\section{ Systematic uncertainty for $\Upsilon(nS)$ counting}
   The systematic uncertainty for $\Upsilon(3S)$ counting has been studied using the on-resonance and off-resonance samples of $\Upsilon(3S)$ data, and the MC samples \cite{Hearty}. The number of $\Upsilon(3S)$ events passing a set of selection criteria in an on-resonance sample is equal to the total number of hadronic events selected less the number of non-$\Upsilon(3S)$  events. The number of non-$\Upsilon(3S)$ events can be expressed in terms of the production cross section, the efficiency to pass the cuts, and the luminosity. The off-resonance sample is used to separate the number of non-$\Upsilon(3S)$ events. A sample of $\epem \to \g\g$ events is also used to provide a relative luminosity normalization between  between resonant and non-resonant $\Upsilon(3S)$ samples. This study quotes a systematic uncertainty of $0.86\%$.  We also use this systematic uncertainty value for $\Upsilon(2S)$.

\subsection{Final systematic uncertainties}
\label{section:syst}
  Table~\ref{table:Systematic} summarizes the final systematic uncertainties and their sources for both $\Upsilon(3S)$ and $\Upsilon(2S)$, which will be incorporated to evaluate the branching ratio or upper limit of B.R. in the analysis.

\begin{table}[htb]
\centering
\begin{tabular}{|l |r|r|}
	\hline
&\multicolumn{2}{c|} {Uncertainty} \\
\cline{2-3}
 \multicolumn{1}{|c|}{Source}  &  $\Upsilon(2S)$    & $\Upsilon(3S)$  \\
\hline
\hline
\multicolumn{3}{|c|}{Additive systematic uncertainties (events) } \\
\hline
\hline 
$N_{s}$ PDF          & (0.00 -- 0.62)    & (0.04 -- 0.58)        \\
\hline
 Fit Bias            & 0.22              & 0.17    \\
\hline
Total                & (0.22 -- 0.66)    & (0.18 -- 0.60)  \\
\hline
\hline
\multicolumn{3}{|c|}{Multiplicative systematic uncertainties ($\%$)} \\
\hline
\hline
Muon-ID              & 4.30              & 4.25     \\
\hline
Charged tracks       & 3.73              & 3.50  \\
\hline
$\Upsilon(nS)$ kinematic fit $\chi^2$  & 1.52  & 2.96\\
\hline
$\mathcal{B}(\Upsilon(nS) \rightarrow \pi^+\pi^- \Upsilon(1S))$  & 2.20  & 2.30   \\
\hline
RF selection                 & 2.21  & 2.16  \\
\hline

Photon efficiency              & 1.96  & 1.96  \\
\hline
$N_{\Upsilon(nS)}$            & 0.86   & 0.86   \\
\hline 
Total                        &   7.00     &   7.32    \\  
\hline 
\end{tabular} 
\caption{Systematic uncertainties and their sources.}
\label{table:Systematic}
\end{table} 

\section {Chapter Summary}
 In this chapter, we  have summarized the possible sources of the systematic uncertainties for this analysis. These systematic uncertainties are included by convolving the likelihood curve with a Gaussian of width $\sigma_{syst}$, which is used to compute the $90\%$ confidence level (C.L.) upper limit using the Bayesian approach with an uniform prior. The details are discussed in the next chapter.

 

\chapter{Results and Conclusion}
\label{Chapter6}
This chapter  presents the $90\%$ C.L. Bayesian upper limits on the product branching fraction of $\mathcal{B}(\Upsilon(1S) \to \g A^0) \times \mathcal{B}(A^0 \to \mu^+\mu^-)$ as well as the effective Yukawa coupling of the b-quark to the $A^0$ as a function of $m_{A^0}$, which are calculated in the absence of any signal events. Finally, we present the summary and conclusion of this dissertation.

\section{Upper-limit}
\begin{figure}
\centering
\includegraphics[width=3.0in]{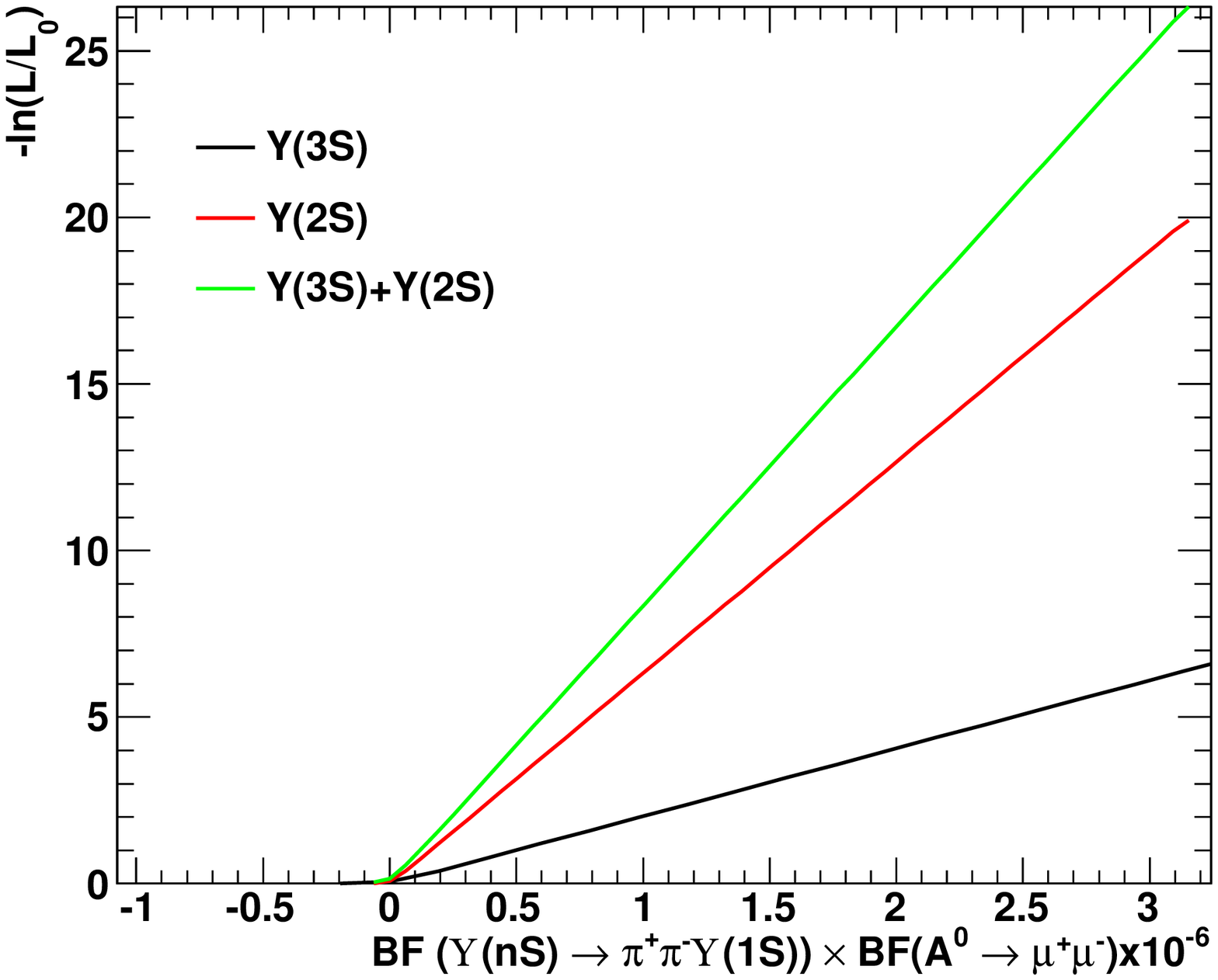}
\includegraphics[width=3.0in]{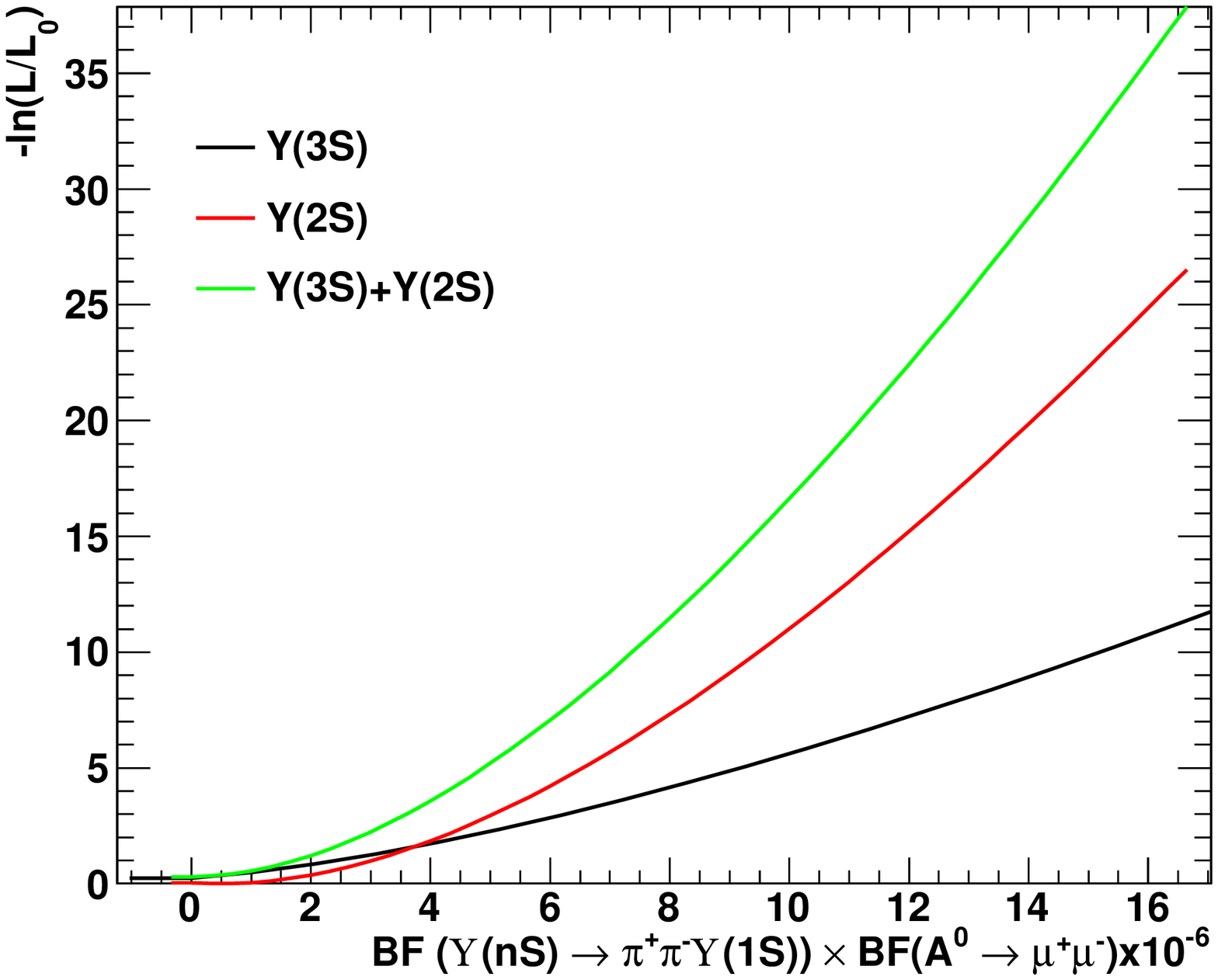}

\caption {The likelihood function as a function of branching fraction (B.F.) for  the Higgs mass of (a) $m_{A^0} = 0.212$ \gevcc and  (b) $m_{A^0} = 8.21$ \gevcc.}

\label{fig:NewNLL} 
\end{figure}

\begin{figure}
\centering
\includegraphics[width=6.0in]{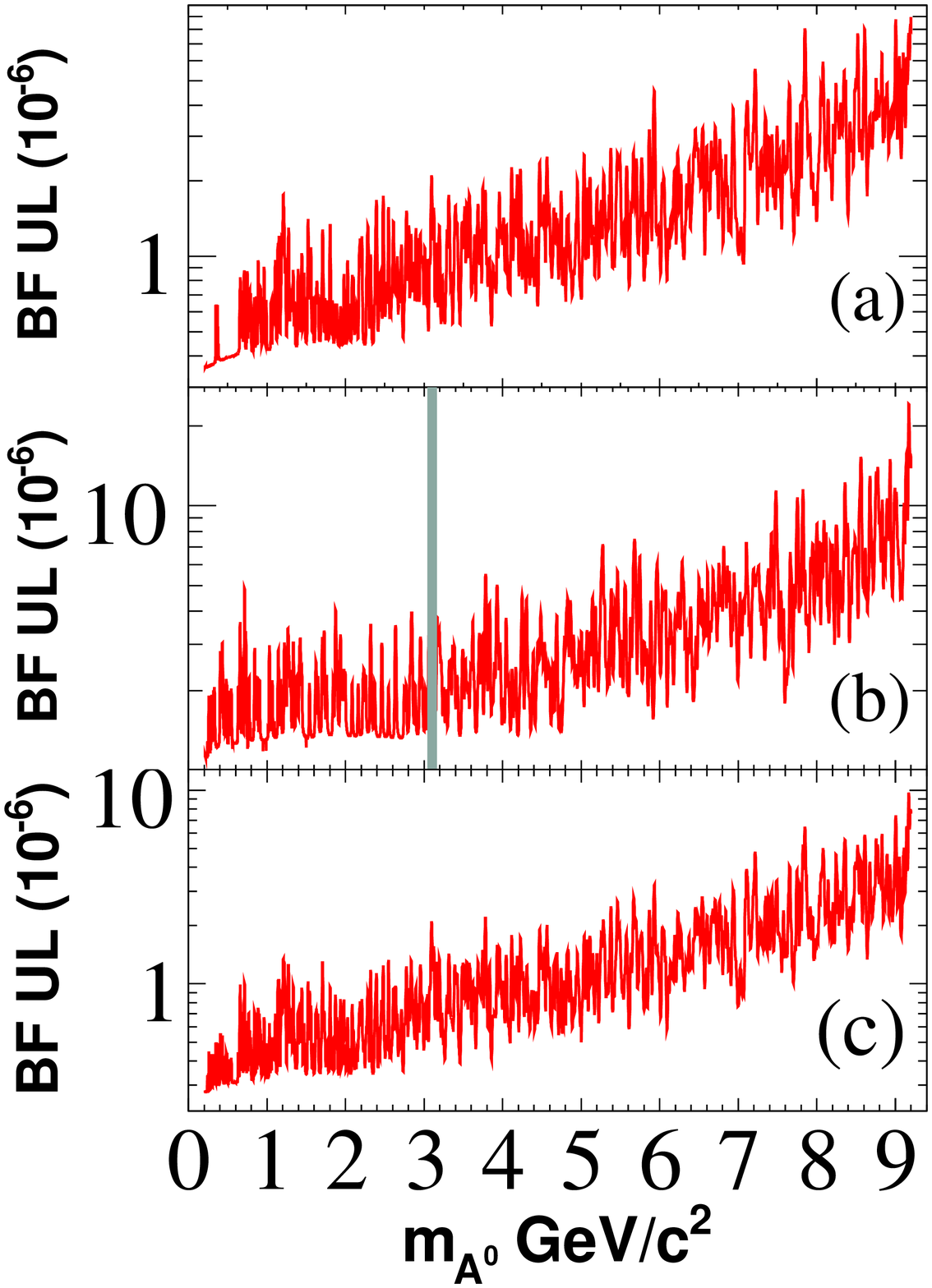}

\caption {The $90\%$ C.L. upper limit on the product of branching fractions 
$\mathcal{B}(\Upsilon(1S) \rightarrow \gamma A^0) \times \mathcal{B}(A^0 \rightarrow \mu^+\mu^-)$ for (a) the $\Upsilon(2S)$ dataset, (b) the $\Upsilon(3S)$ dataset and   (c) 
the combined $\Upsilon(2S,3S)$ dataset. The shaded area shows the region of the \jpsi resonance, excluded from the search in the $\Upsilon(3S)$ dataset.} 

\label{fig:Newullog} 
\end{figure}

 As discussed in section~\ref{sect:trial_factor}, the trial factor study shows that we find  no evidence of signal for the di-muon decay of a light \CP-odd scalar particle in the radiative $\Upsilon(1S)$ decays in the $\Upsilon(3S,2S)$ samples. In the absence of any significant signal yield, we calculate the $90\%$ C.L upper limit on the product branching fraction $\mathcal{B}(\Upsilon(1S) \rightarrow \gamma A^0) \times B(A^0 \rightarrow \mu^+\mu^-)$ as a function of $m_{A^0}$, including the systematic uncertainties. The systematic uncertainty is included by convolving the likelihood curve with a Gaussian of width $\sigma_{syst}$. A convolution is an integral that blends one function with another  producing new function that is typically viewed as modified version of the original functions. Mathematically, the convolution of the two functions $f$ and $g$ over an infinite range is given by:
\begin{equation}
h(x) = \int\limits_{-\infty}^\infty f(x-y)g(x)dx
\end{equation}
 
\noindent where $h(x)$ is the modified version of original functions $f$ and $g$ after the convolution. We plot the negative log likelihood (NLL) as a function of branching fraction ($BF$) and integrate it from zero upward until we find an integral which yields $90\%$  of the total integral (above zero) under the likelihood curve to compute the $90\%$ confidence level Bayesian upper limits. The $BF$ is defined as:

\begin{equation}
BF = \frac{N_{sig}}{\epsilon \cdot \mathcal{B} \cdot N_{\Upsilon(nS)}}
 \end{equation}    
 
\noindent where $N_{sig}$ is the number of the fitted signal yield, $\epsilon$ is the signal selection efficiency,  $\mathcal{B}$ is the branching fraction of $\Upsilon(2S, 3S) \rightarrow \pi^+\pi^-\Upsilon(1S)$ transitions, and $N_{\Upsilon(nS)}$  is the number of $\Upsilon(2S, 3S)$ mesons used in this analysis. For combining the results of the $\Upsilon(2S,3S)$ datasets, we add the log of the $\Upsilon(2S, 3S)$ likelihoods. Figure~\ref{fig:NewNLL} shows the likelihood function as a function of $BF$ at selected mass points for $\Upsilon(2S)$, $\Upsilon(3S)$ and combined data of $\Upsilon(2S,3S)$. The correlated and uncorrelated systematic uncertainties are taken into account for combining the two datasets. The systematic uncertainties of $\Upsilon(nS)$ counting, photon efficiency, tracking and PID are considered as correlated systematic uncertainties and rest of the  systematic uncertainties discussed Table~\ref{table:Systematic} are considered as uncorrelated systematic uncertainties. The \jpsi mass region in the $\Upsilon(3S)$ dataset, defined as $3.045 \le m_{\rm red} \le 3.162$ \gevcc, is excluded from the search due to a large background from $\jpsi \rightarrow \mu^+\mu^-$.  Figure~\ref{fig:Newullog}  shows the $90\%$ C.L upper limits on $\mathcal{B}(\Upsilon(1S) \rightarrow \gamma A^0) \times B(A^0 \rightarrow \mu^+\mu^-)$ as a function of $m_{A^0}$ .
The limits vary between $(0.37 - 8.97)\times 10^{-6}$ for the $\Upsilon(2S)$ dataset, 
$(1.13 - 24.2)\times 10^{-6}$ for the $\Upsilon(3S)$ dataset, and $(0.28 - 9.7)\times 10^{-6}$ for the combined $\Upsilon(2S,3S)$ dataset.

\begin{figure}
\centering
\includegraphics[width=3.0in]{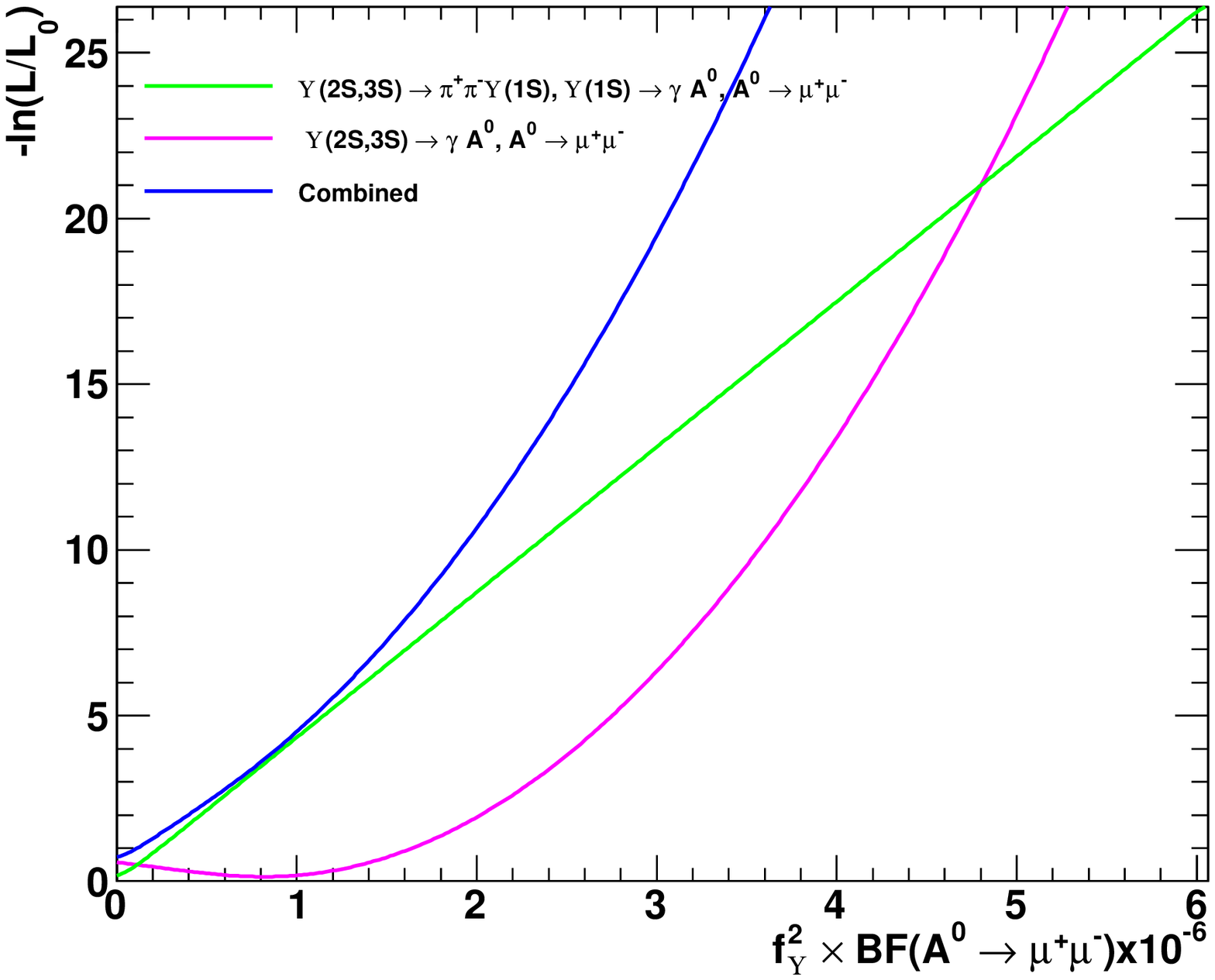}
\includegraphics[width=3.0in]{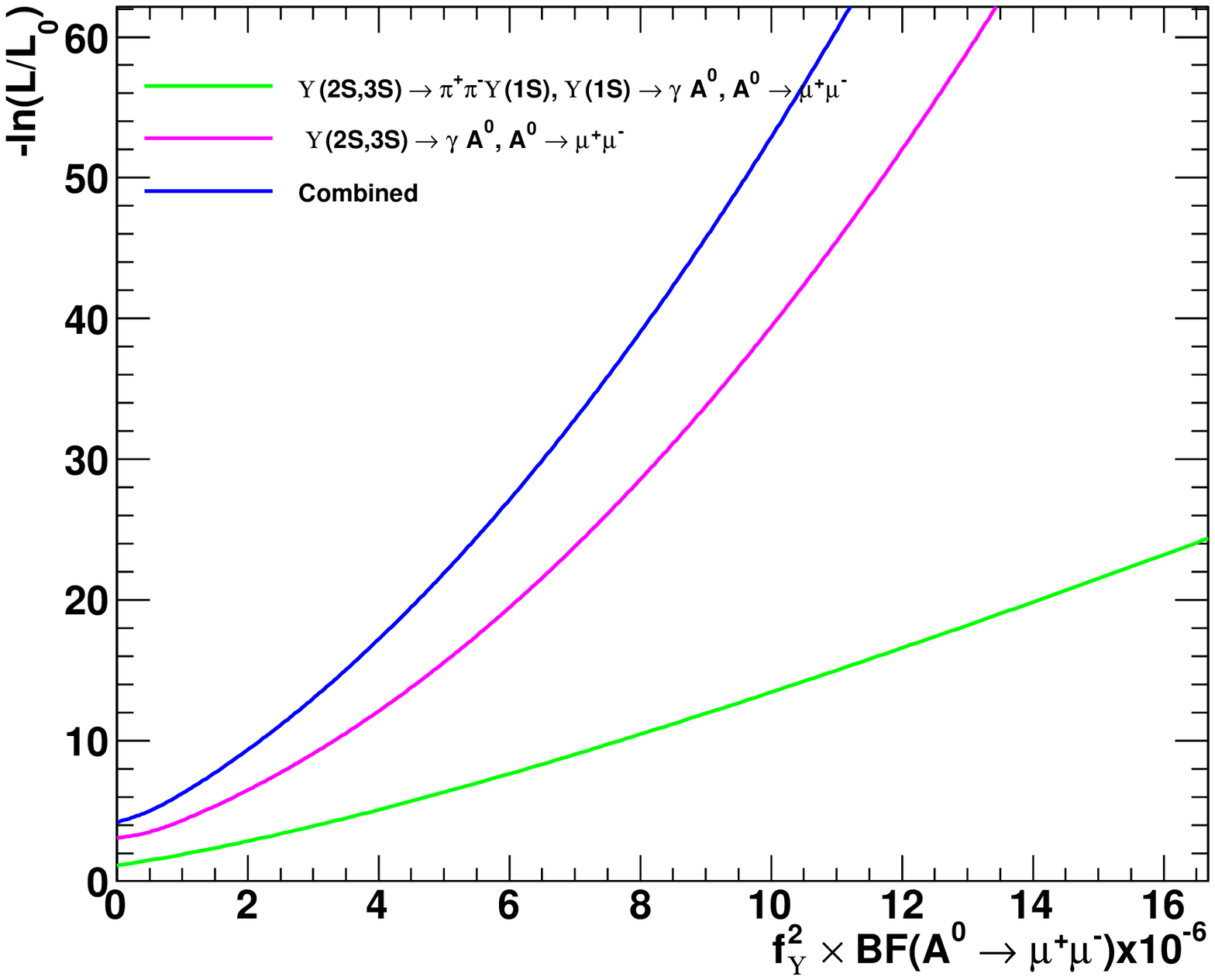}

\caption {The likelihood function as a function of effective Yukawa coupling of \b-quarks to the $A^0$ ($f_{\Upsilon}$) for  the Higgs mass of (a) $m_{A^0} = 0.214$ \gevcc and  (b) $m_{A^0} = 5.60$ \gevcc.} 

\label{fig:FyNLL} 
\end{figure}

\begin{figure}
\centering
\includegraphics[width=6.0in]{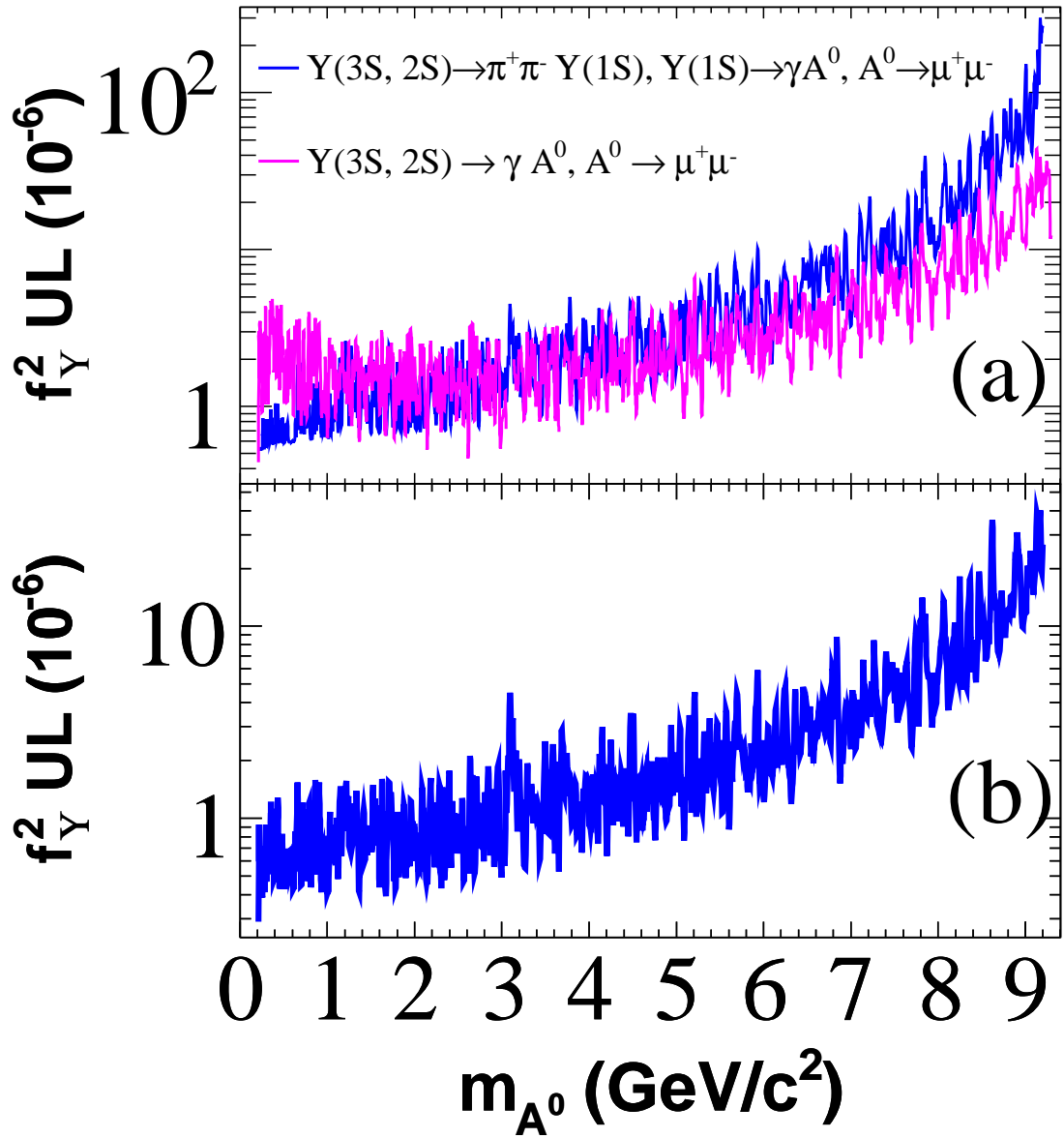}

\caption {The $90\%$ C.L. upper limit on the effective Yukawa coupling ($f_{\Upsilon}$)  of bound  \b-quarks to the $A^0$, $f_{\Upsilon}^2 \times \mathcal{B}(A^0 \rightarrow \mu^+\mu^-)$,  for (a) this and  previous \babar\ measurement of $\Upsilon(2S,3S) \rightarrow \gamma A^0$, $A^0 \rightarrow \mu^+\mu^-$ \cite{Aubert:2009cp}, and (b) the combined  limit.} 

\label{fig:Newfy} 
\end{figure}

 The branching fractions of $\mathcal{B}(\Upsilon(nS) \to \g A^0)$ $(n=1,2,3)$ are related to the effective Yukawa coupling ($f_{\Upsilon}$) of the \b-quark to the $A^0$ via Equation~\ref{Eq:fycoupling}. The value of $f_{\Upsilon}$  incorporates the $m_{A^0}$ dependent QCD and relativistic corrections to $\mathcal{B}(\Upsilon(nS) \to \g A^0)$ \cite{P_Nason}, as well as the leptonic width of $\Upsilon(nS) \to l^+l^-$ \cite{Upsilonwidth}. These corrections are as large as $30\%$ to first order in strong coupling constant ($\alpha_S$), but have comparable uncertainties \cite{Beneke}. The 90$\%$ C.L. upper limits on $f_{\Upsilon}^2 \times \mathcal{B}(A^0 \to \mumu)$ for  combined $\Upsilon(2S,3S)$ datasets range from $0.54\times 10^{-6}$ to $2.99 \times 10^{-4}$ depending upon the mass of $A^0$, which is shown in  Figure~\ref{fig:Newfy}(a). For comparison, the results from previous \babar\ measurements of $\Upsilon(2S,3S) \to \g A^0$, $\mathcal{B}(A^0 \to \mumu)$ \cite{Aubert:2009cp} are also shown. We combine our results with previous \babar\ measurements \cite{Aubert:2009cp}, taking into account both correlated and uncorrelated uncertainties. Figure~\ref{fig:FyNLL} shows the likelihood function as a function of  $f_{\Upsilon}$ at selected $m_{A^0}$ points for the combined data of $\Upsilon(2S,3S)$, previous \babar\ measurements \cite{Aubert:2009cp} and combination of these two measurements. The combined upper limits on $f_{\Upsilon}^2 \times \mathcal{B}(A^0 \to \mumu)$  for these two measurements vary in the range of $(0.29 - 40)\times 10^{-6}$  for $m_{A^0} \le 9.2$ \gevcc (Figure~\ref{fig:Newfy}(b)). 

\section{Summary and Conclusion}
  This thesis describes a search for di-muon decays of a low-mass Higgs boson in the fully reconstructed decay chain of $\Upsilon(2S,3S) \to \pipi\Upsilon(1S)$, $\Upsilon(1S) \to \g A^0$, $A^0 \to \mumu$. The $\Upsilon(1S)$ sample is selected by tagging the pion pair in the $\Upsilon(2S,3S) \rightarrow \pipi \Upsilon(1S)$ transitions, using a data sample of $(92.8 \pm 0.8)\times 10^6$ $\Upsilon(2S)$ and  $(116.8\pm 1.0) \times 10^6$ $\Upsilon(3S)$ mesons collected with the \babar\ detector at the PEP-II asymmetric-energy \epem\ collider located at SLAC National Accelerator Laboratory. The $A^0$ is assumed to be a scalar or pseudoscalar particle with a negligible decay width compared to the experimental resolution \cite{Fullana}. We find no evidence for $A^0$ production and set $90\%$ confidence level (C.L.) upper limits on the product branching fraction  $\mathcal{B}(\Upsilon(1S) \to \g A^0) \times \mathcal{B}(A^0 \to \mumu)$ in the range of $(0.28 - 9.72)\times 10^{-6}$ for   $0.212 \le m_{A^0} \le 9.20$ \gevcc. These results improve the current best limits by a factor of 2--3 for $m_{A^0}< 1.2$ \gevcc and are comparable to the previous \babar\ result  \cite{Aubert:2009cp} in the mass range of $1.20 < m_{A^0} < 3.6$ \gevcc. Within this range, our limits rule out substantial amount of the parameter space allowed by the light Higgs \cite{PRD76-051105} and axion \cite{nomura} model. We also combine our results with previous \babar\ results of $\Upsilon(2S,3S) \to \g A^0$, $A^0 \to \mumu$ to set limits on the effective coupling ($f_{\Upsilon}$) of the \b-quarks to the $A^0$, $f_{\Upsilon}^2 \times \mathcal{B}(A^0 \to \mumu)$, at the level of  $(0.29 - 40.18)\times 10^{-6}$ for $0.212 \le m_{A^0} \le 9.2$ \gevcc. The combined limits on the product $f_{\Upsilon}^2 \times \mathcal{B}(A^0 \to \mumu)$ are the most stringent to date, and significantly constrain the theoretical Models. A high luminosity  \epem asymmetric  energy Super-B factory and International Linear Collider (ILC) experiments can significantly improve the searches of these low-mass scalar particles, difficult to explore by the LHC,  and elucidate the structure of the new Physics.


\appendix 






















\addtocontents{toc}{\vspace{1em}}
\addcontentsline{toc}{chapter}{Appendices}
\begin{figure}
\chapter{Signal PDFs}
\label{AppendixA}
\section{Signal PDFs for $\Upsilon(2S)$}
\centering
 \includegraphics[width=2.0in]{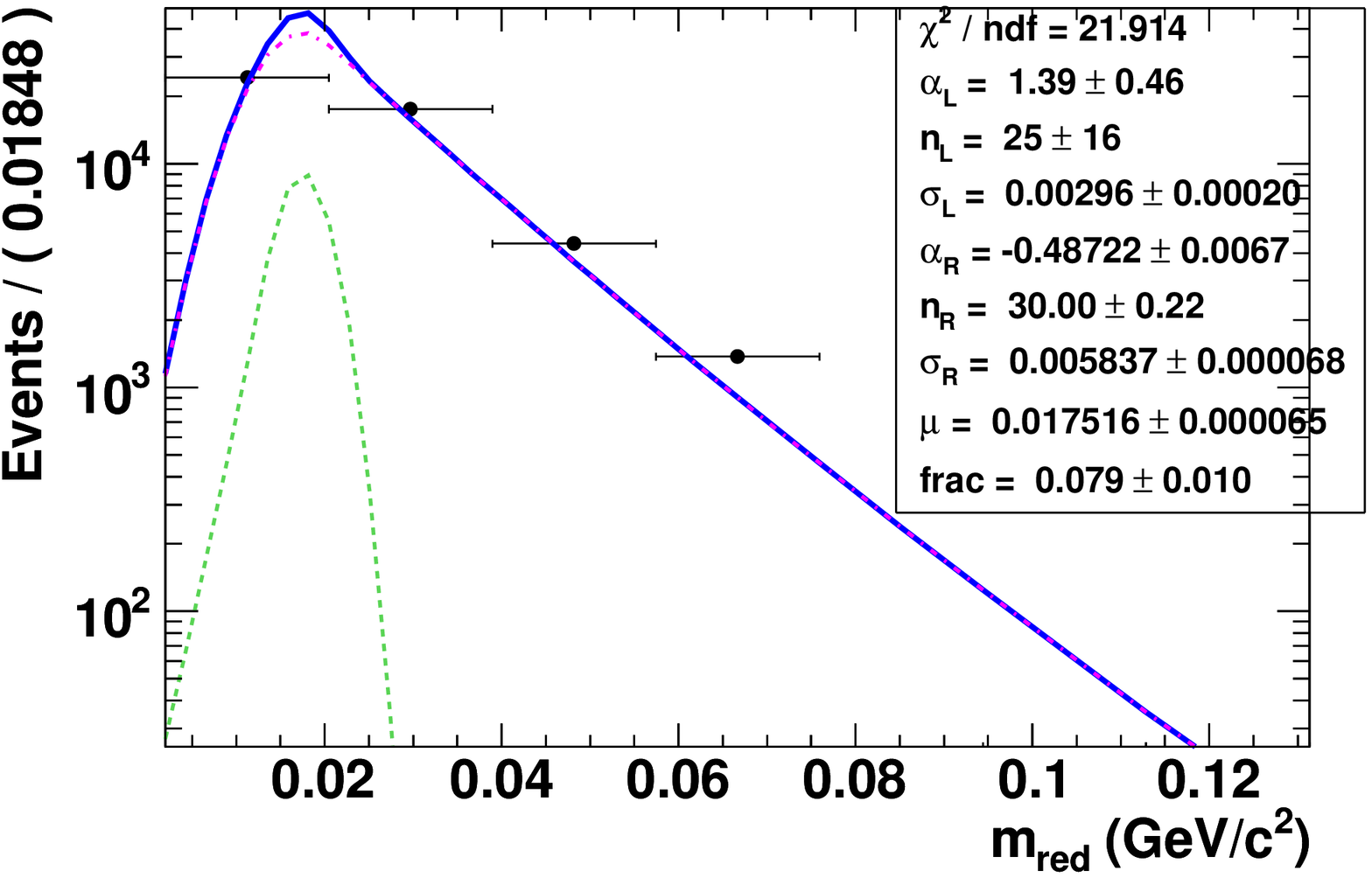}
\includegraphics[width=2.0in]{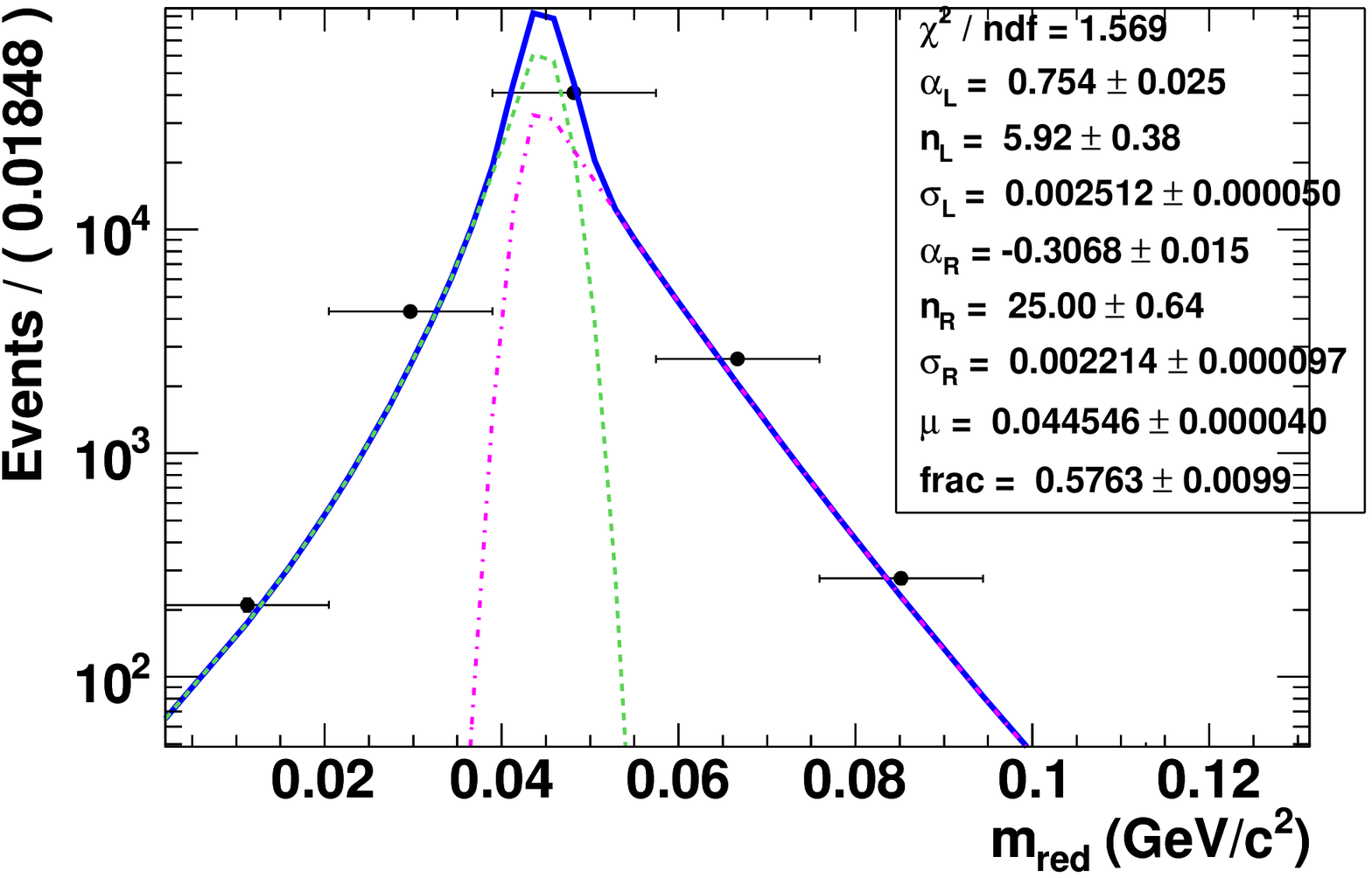}
 \includegraphics[width=2.0in]{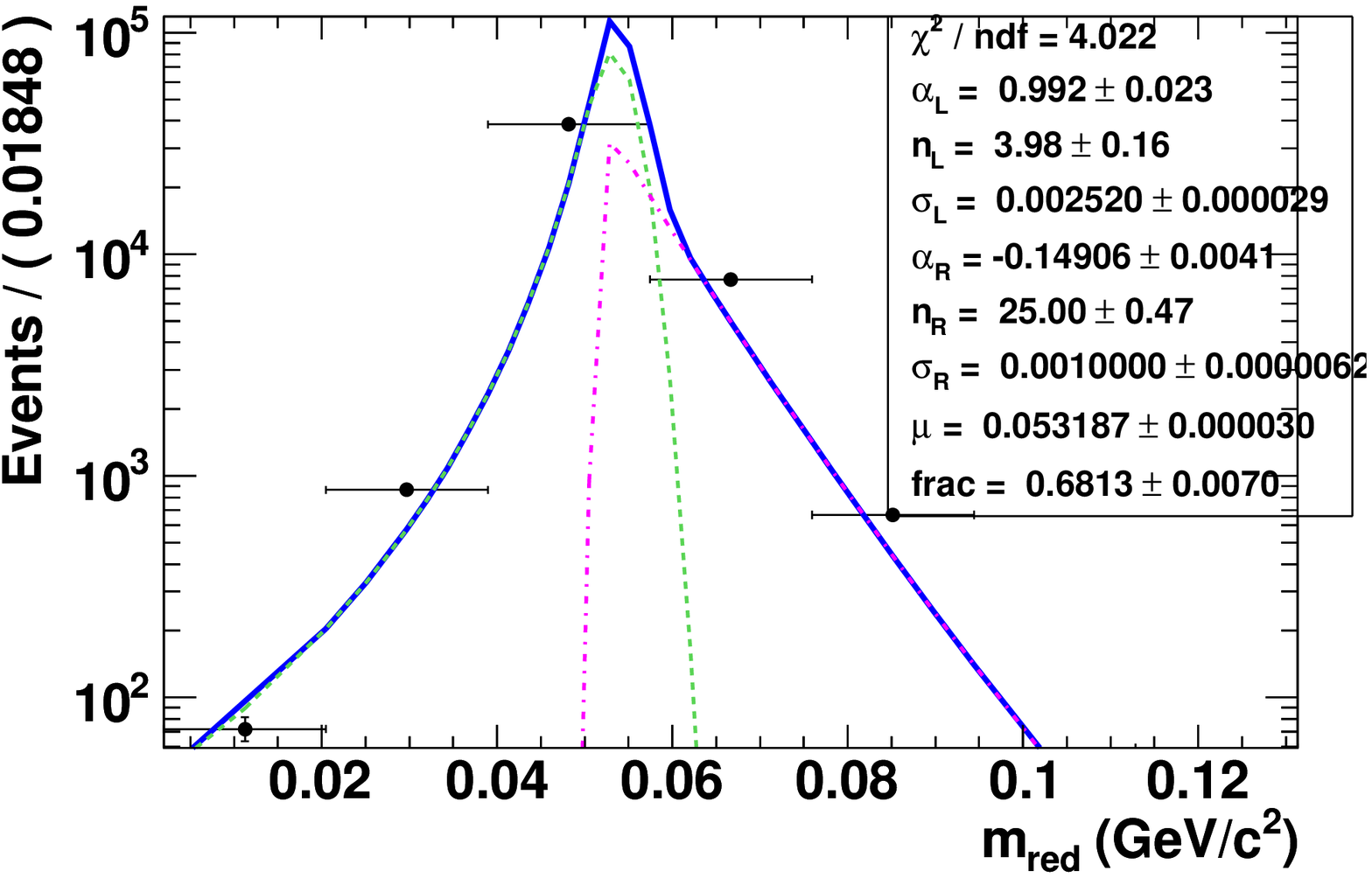}

\smallskip
\centerline{\hfill (a) \hfill \hfill (b) \hfill \hfill (c) \hfill}
\smallskip

\includegraphics[width=2.0in]{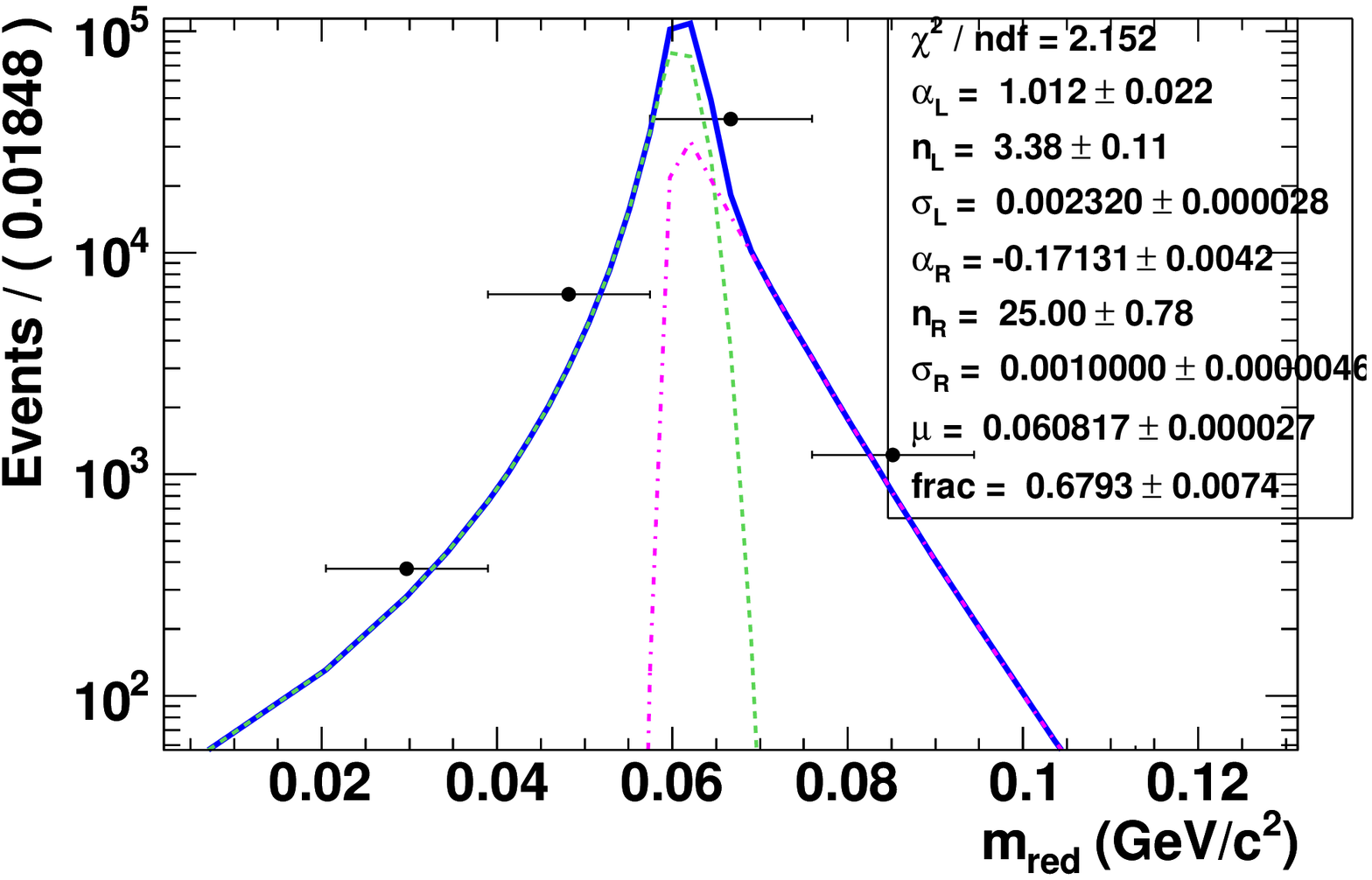}
\includegraphics[width=2.0in]{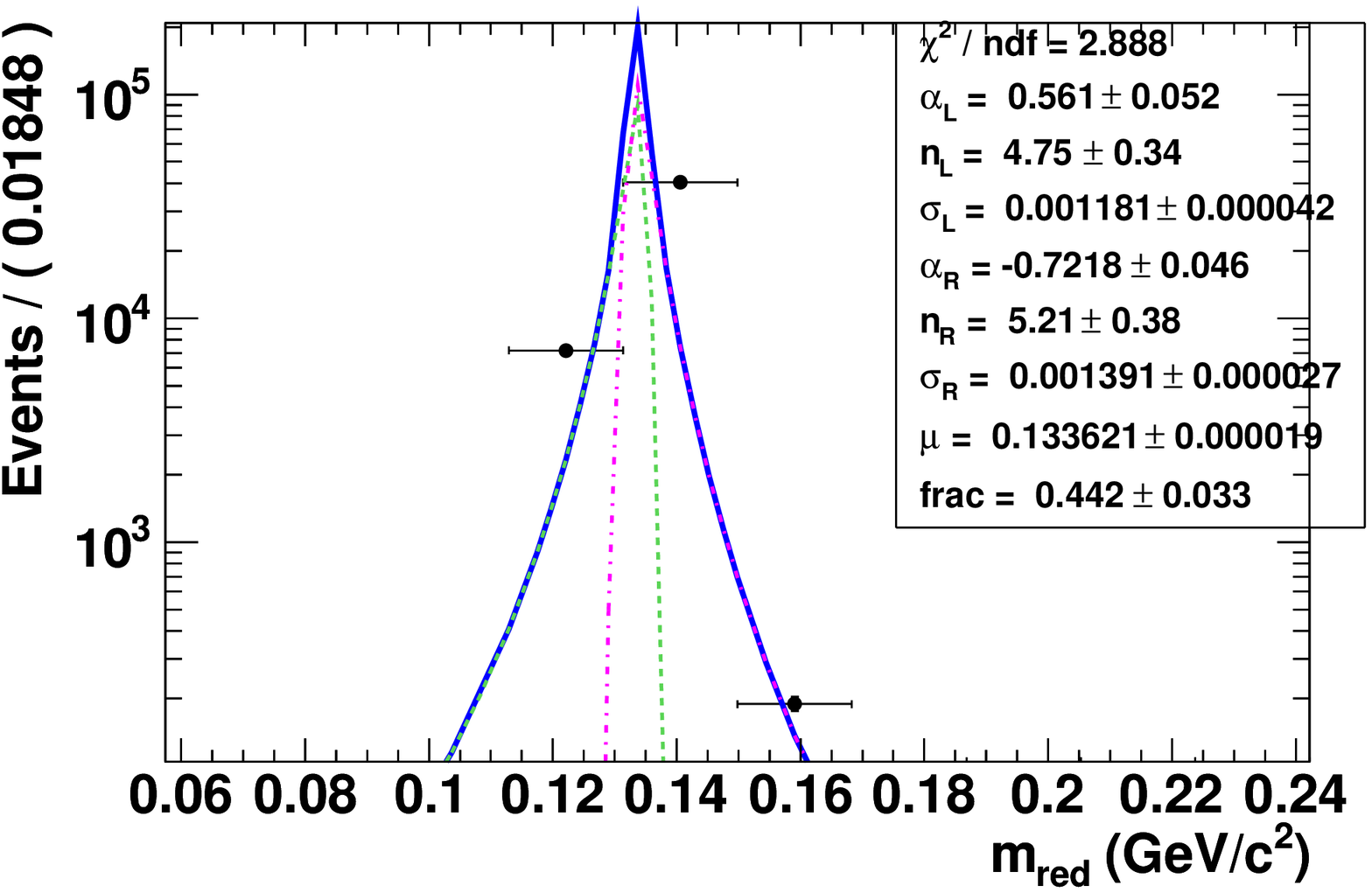}
 \includegraphics[width=2.0in]{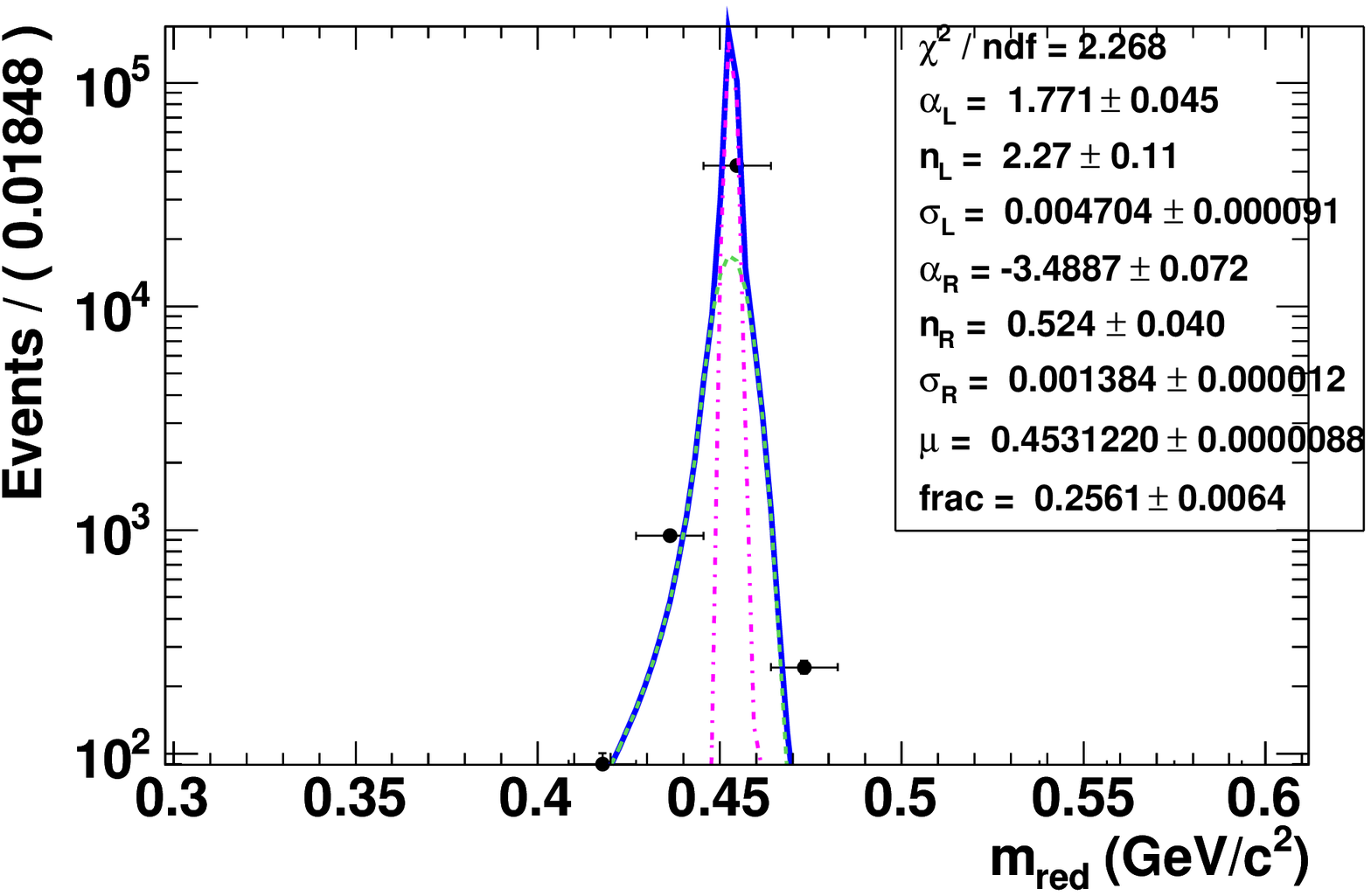}

\smallskip
\centerline{\hfill (d) \hfill \hfill (e) \hfill \hfill (f) \hfill}
\smallskip

\includegraphics[width=2.0in]{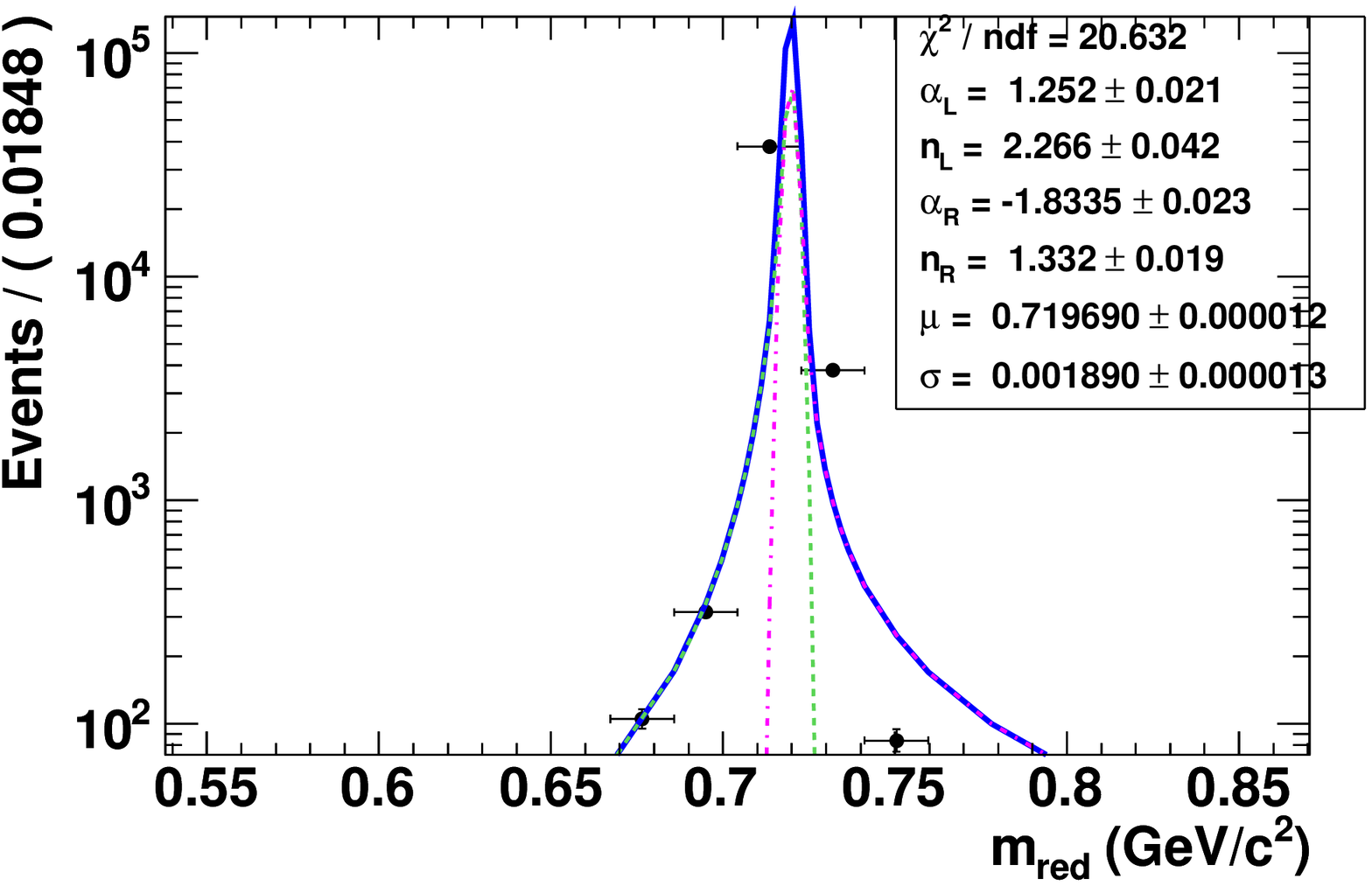}
\includegraphics[width=2.0in]{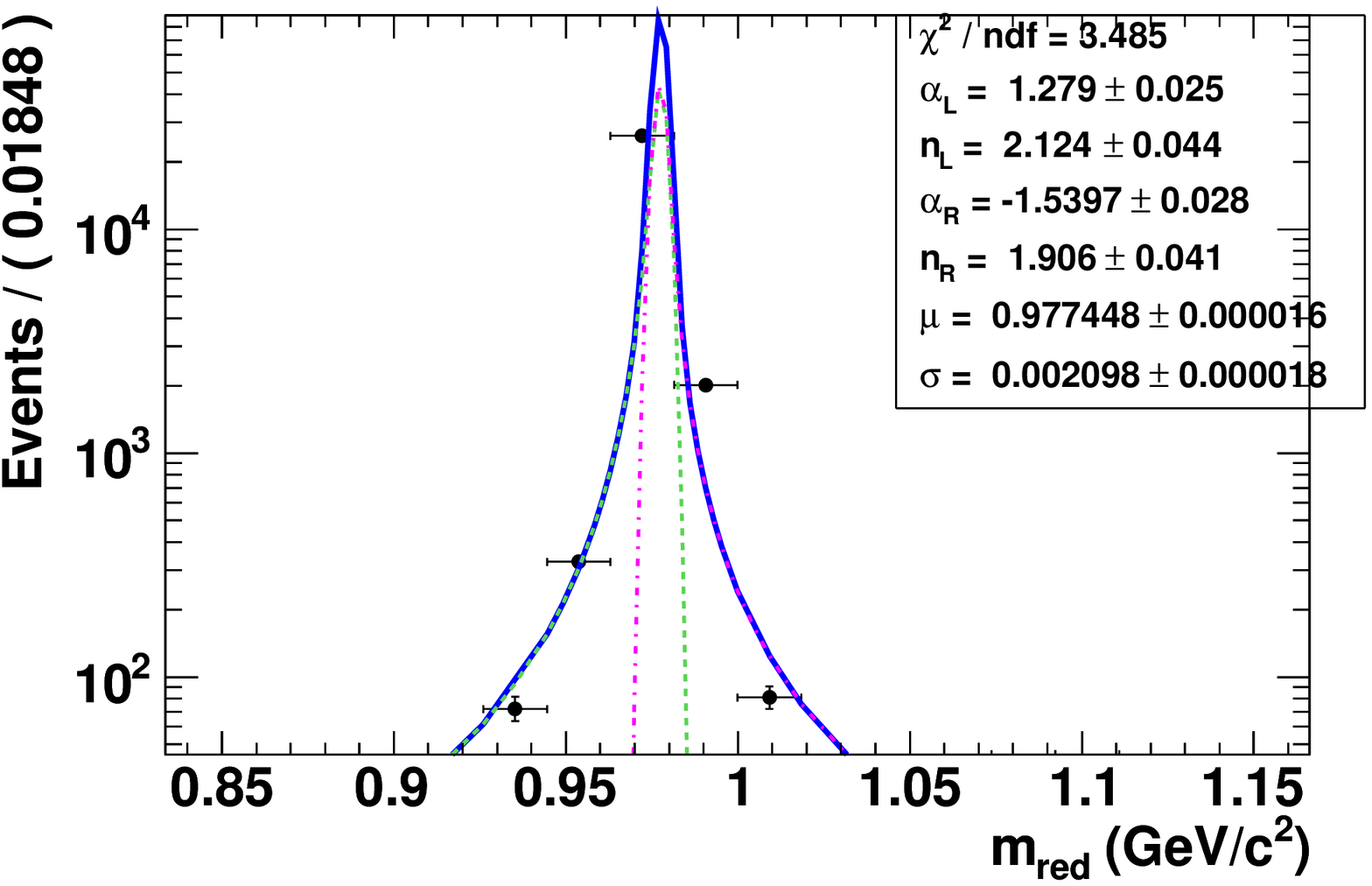}
 \includegraphics[width=2.0in]{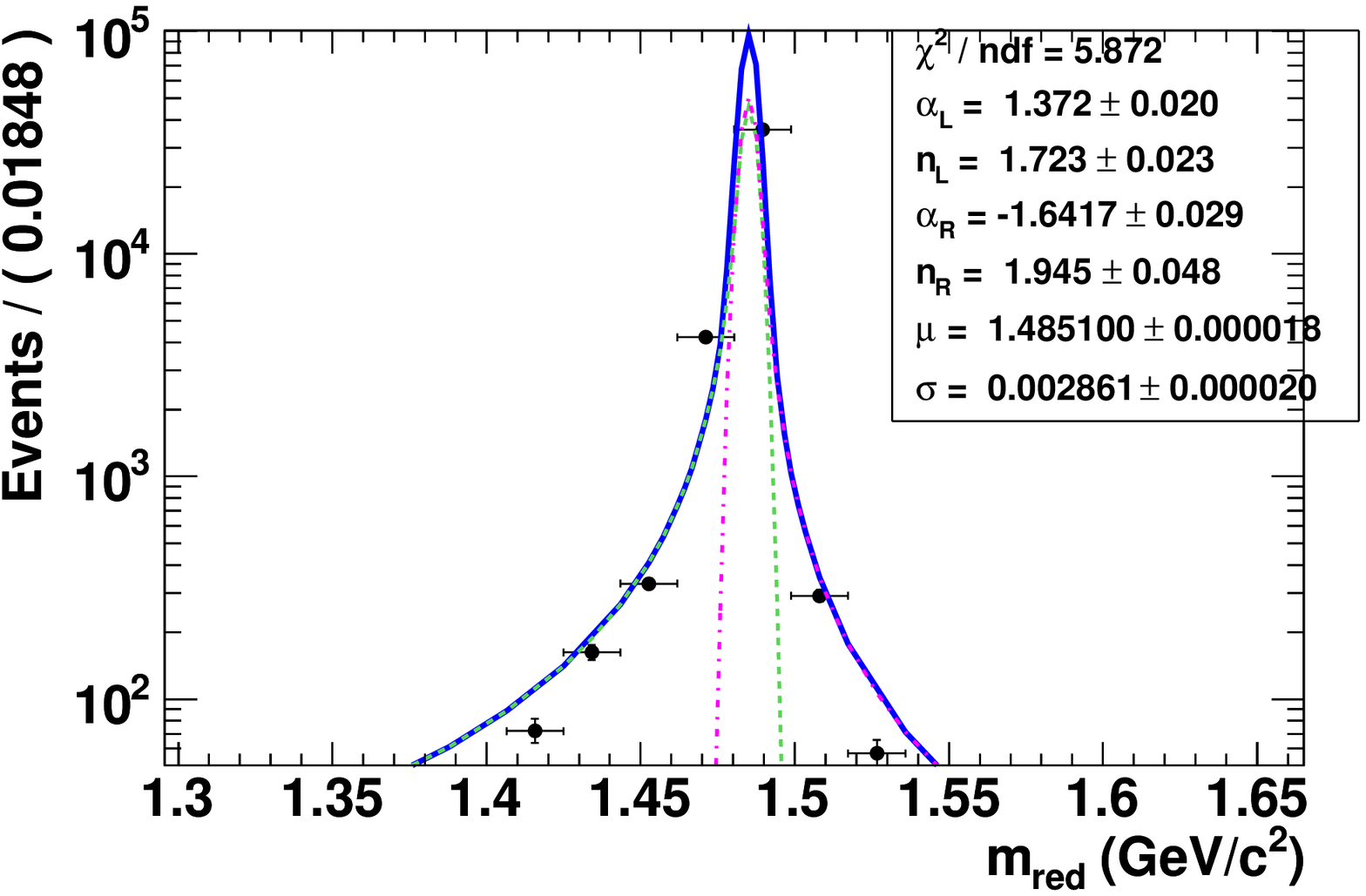}

\smallskip
\centerline{\hfill (g) \hfill \hfill (h) \hfill \hfill (i) \hfill}
\smallskip

\caption {Signal PDFs for the Higgs mass of (a) $m_{A^0}=0.212$ GeV/$c^2$  (b) $m_{A^0}=0.216$ GeV/$c^2$ (c) $m_{A^0}=0.218$ GeV/$c^2$ (d) $m_{A^0}=0.220$ GeV/$c^2$ (e) $m_{A^0}=0.250$ GeV/$c^2$ (f) $m_{A^0}=0.500$ GeV/$c^2$  (g) $m_{A^0}=0.75$ GeV/$c^2$ (h) $m_{A^0}=1.0$ GeV/$c^2$  and (i) $m_{A^0}= 1.5$ GeV/$c^2$.}

\label{fig:SigPDFY2S1}
\end{figure}

\begin{figure}
\centering
\includegraphics[width=2.0in]{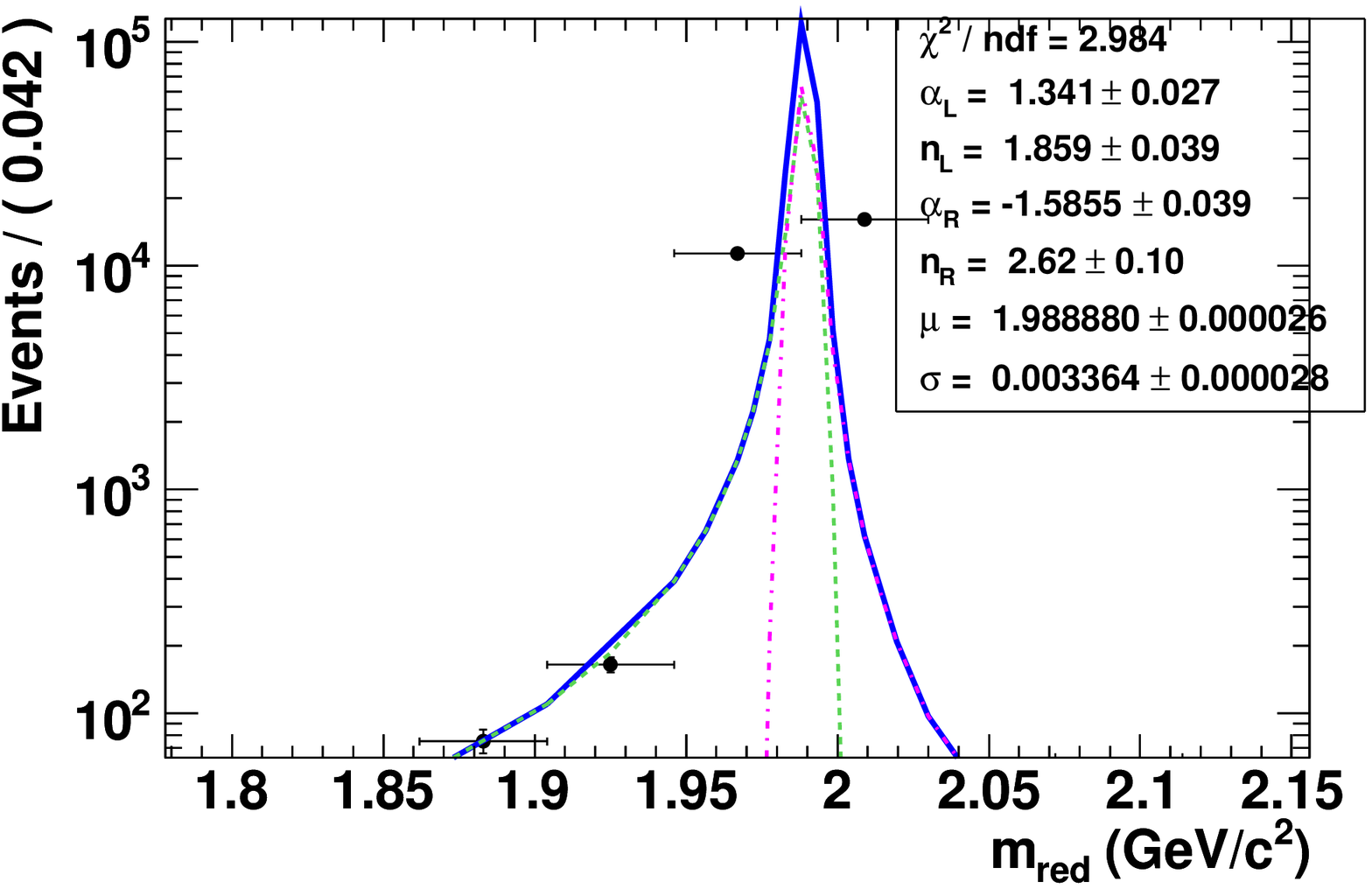}
\includegraphics[width=2.0in]{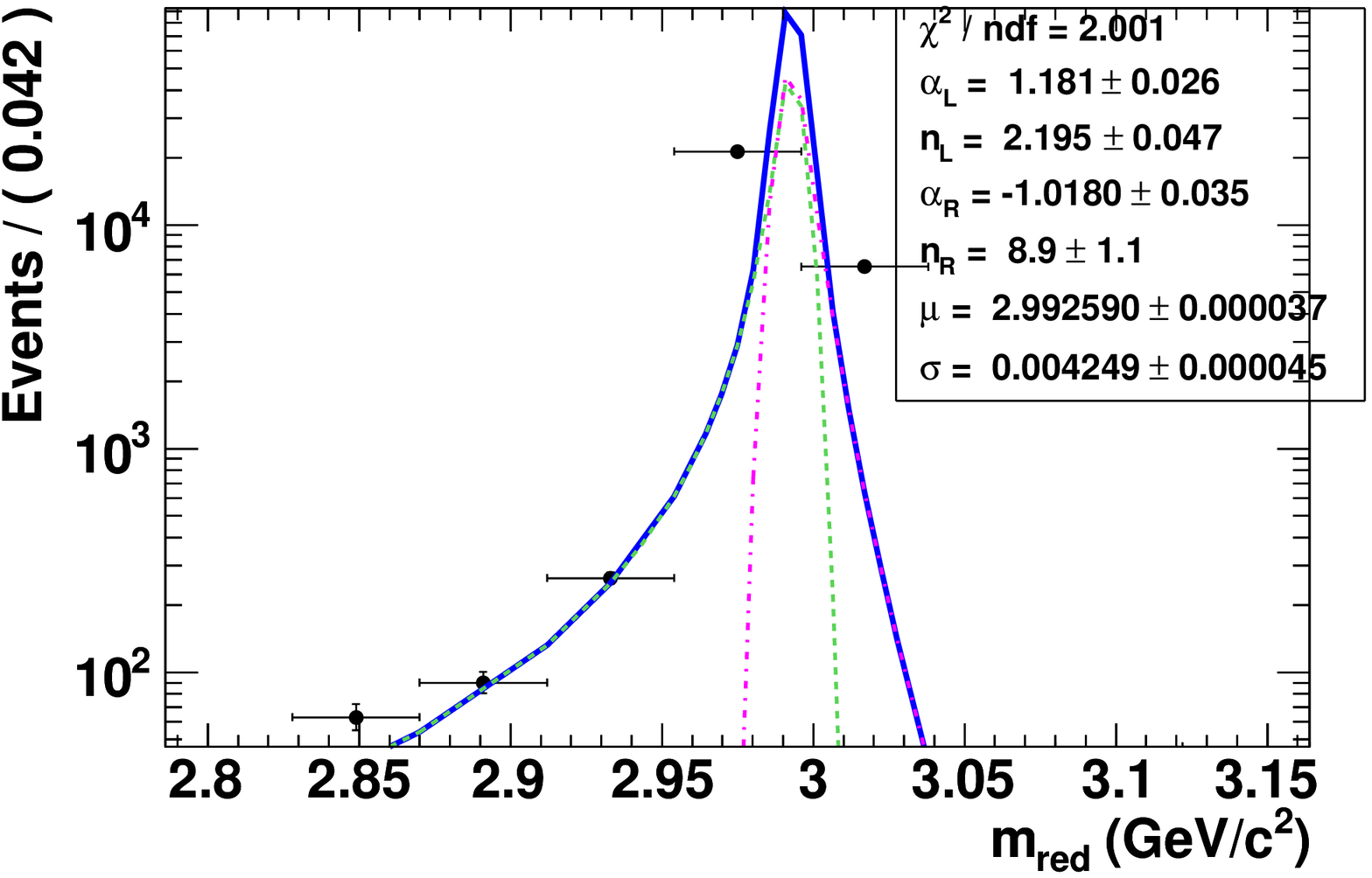}
 \includegraphics[width=2.0in]{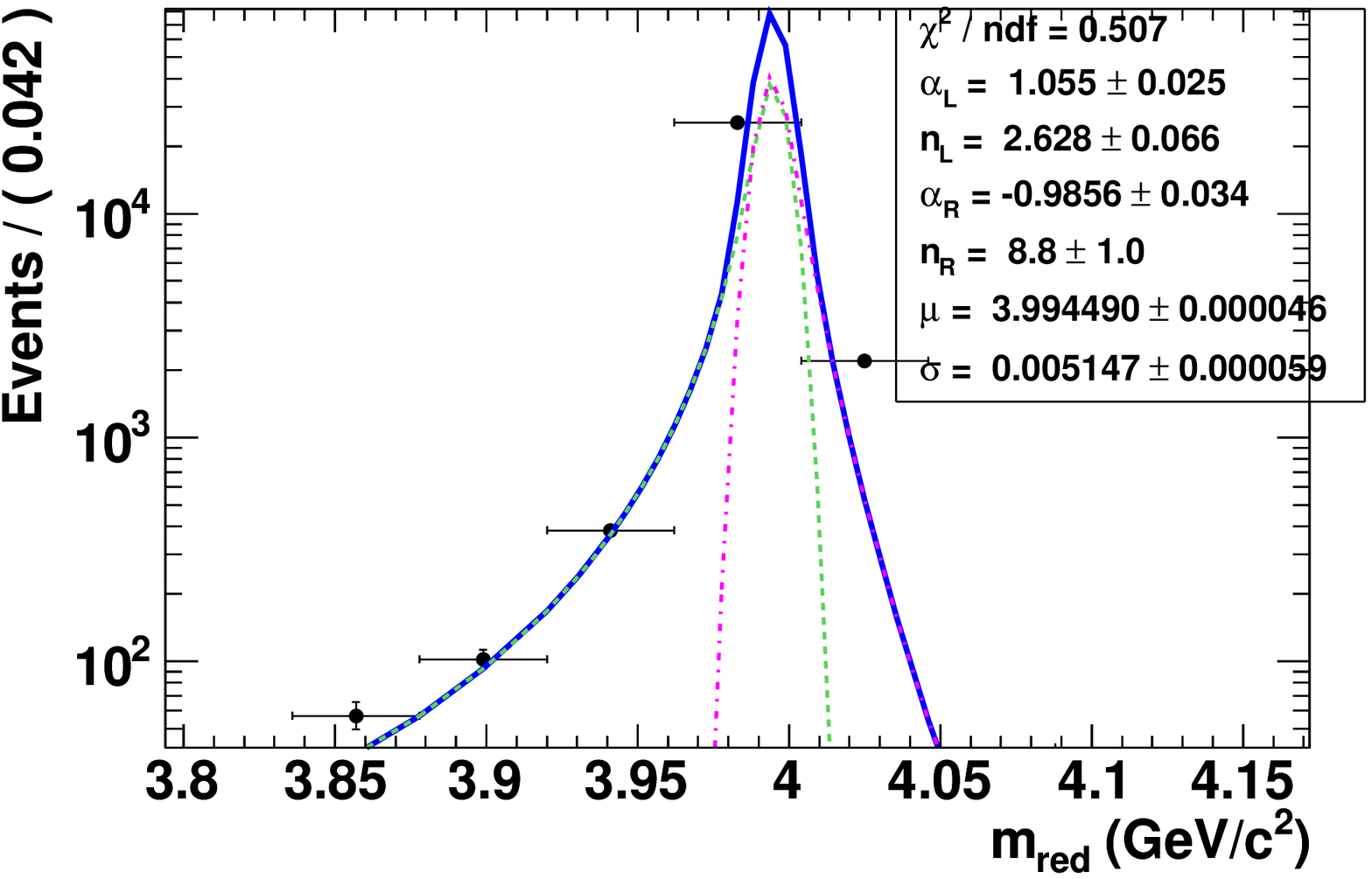}

\smallskip
\centerline{\hfill (a) \hfill \hfill (b) \hfill \hfill (c) \hfill}
\smallskip

\includegraphics[width=2.0in]{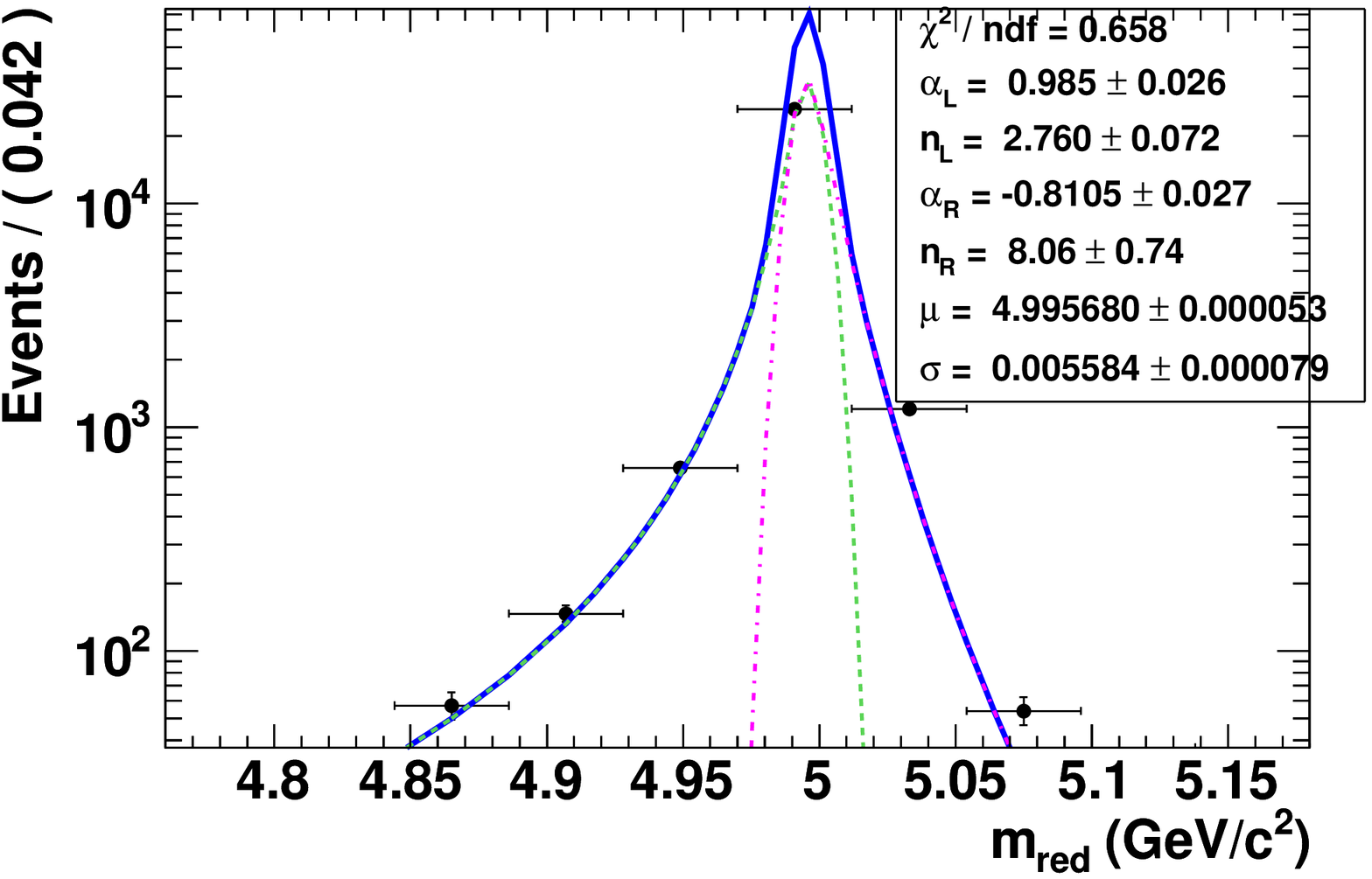}
\includegraphics[width=2.0in]{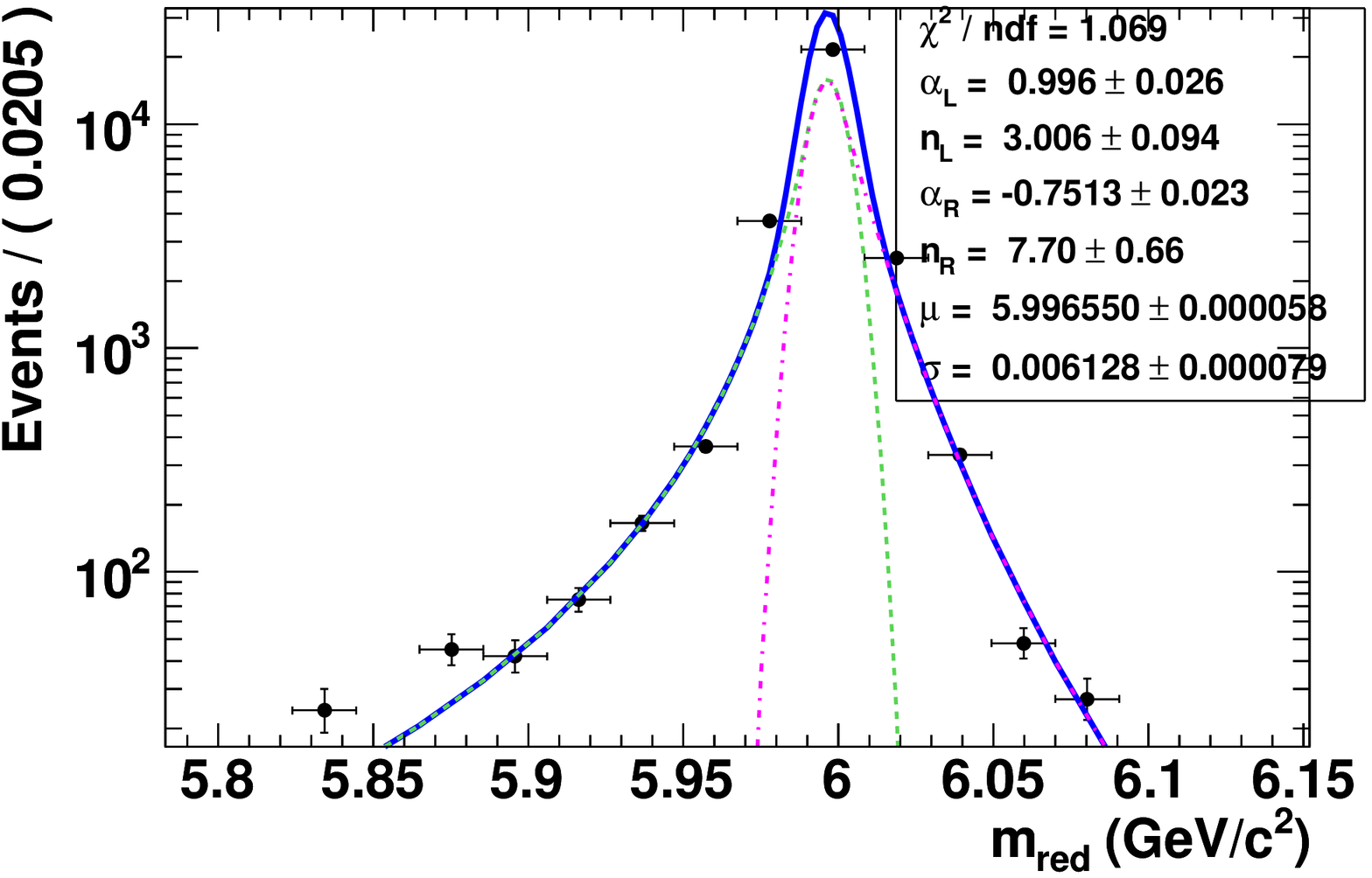}
 \includegraphics[width=2.0in]{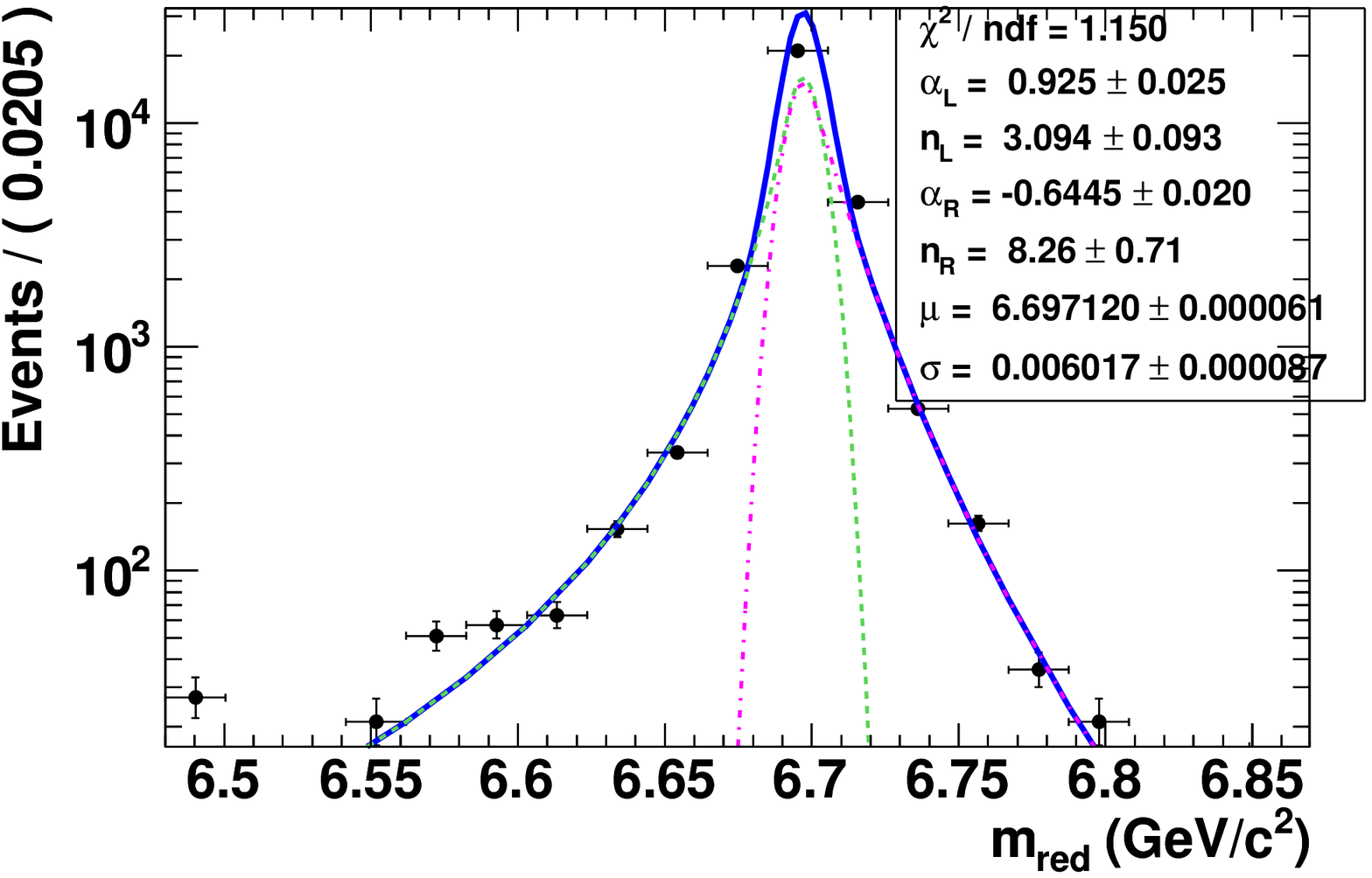}

 \smallskip
\centerline{\hfill (d) \hfill \hfill (e) \hfill \hfill (f) \hfill}
\smallskip

\includegraphics[width=2.0in]{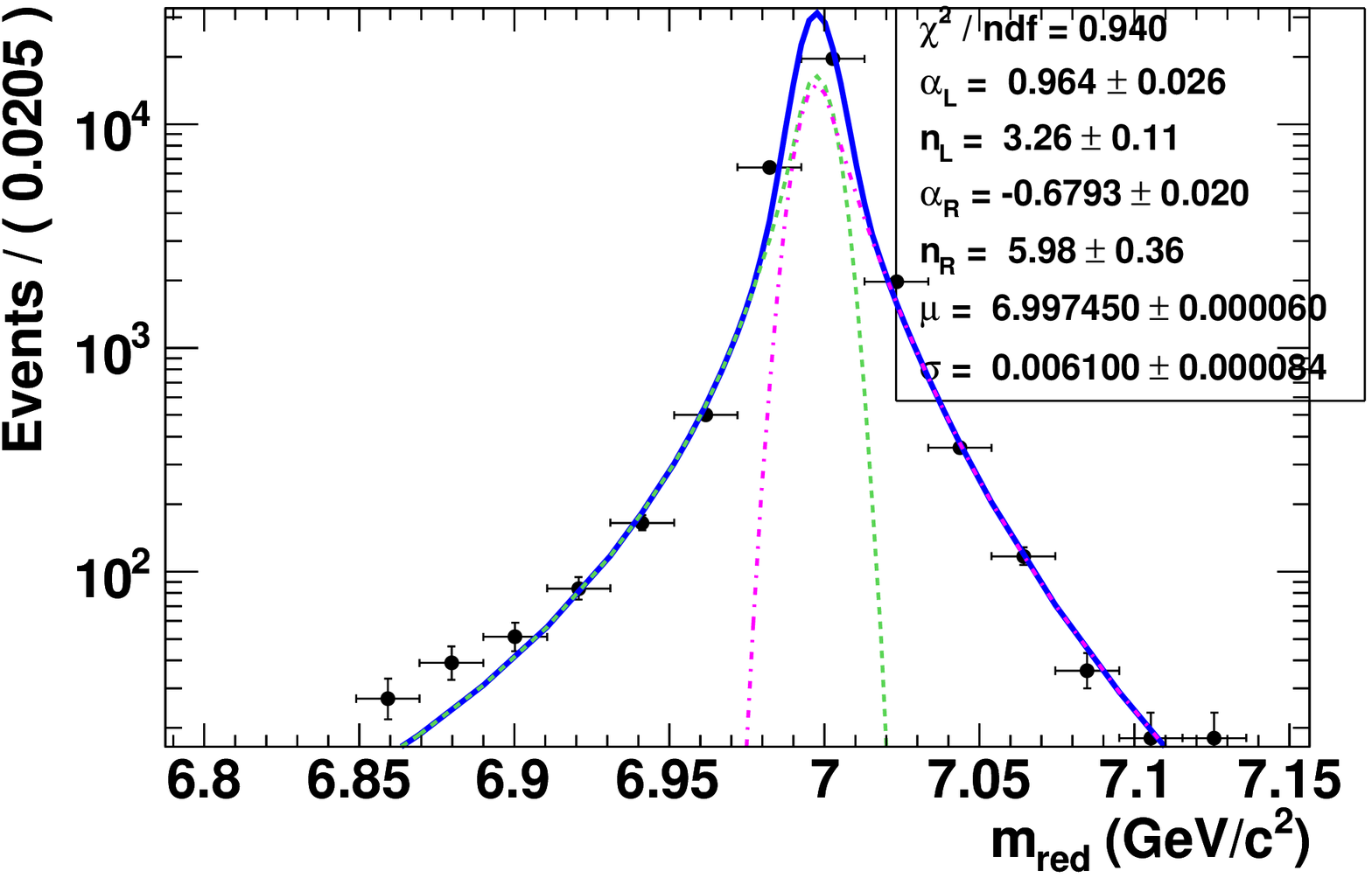}
\includegraphics[width=2.0in]{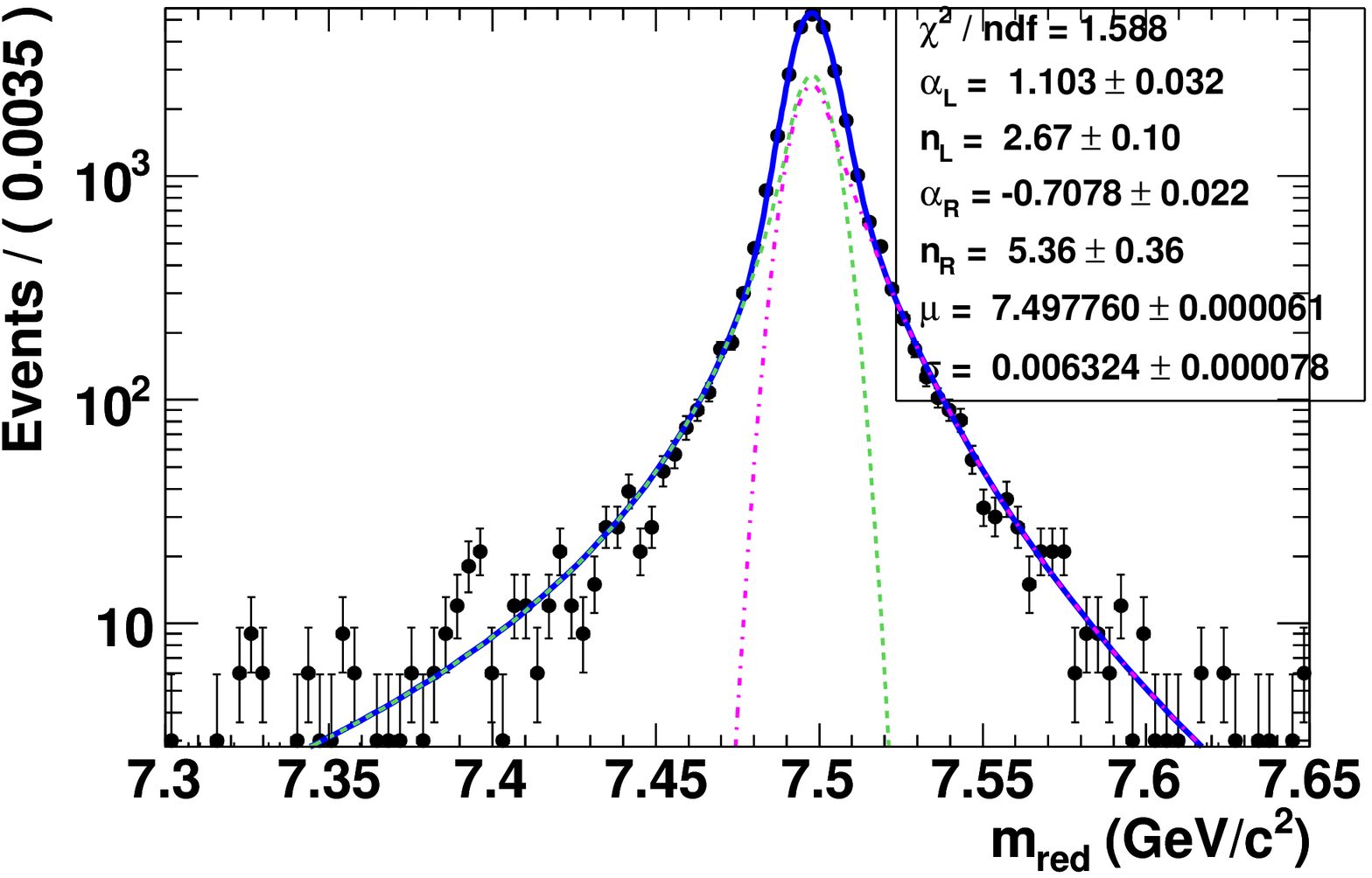}
\includegraphics[width=2.0in]{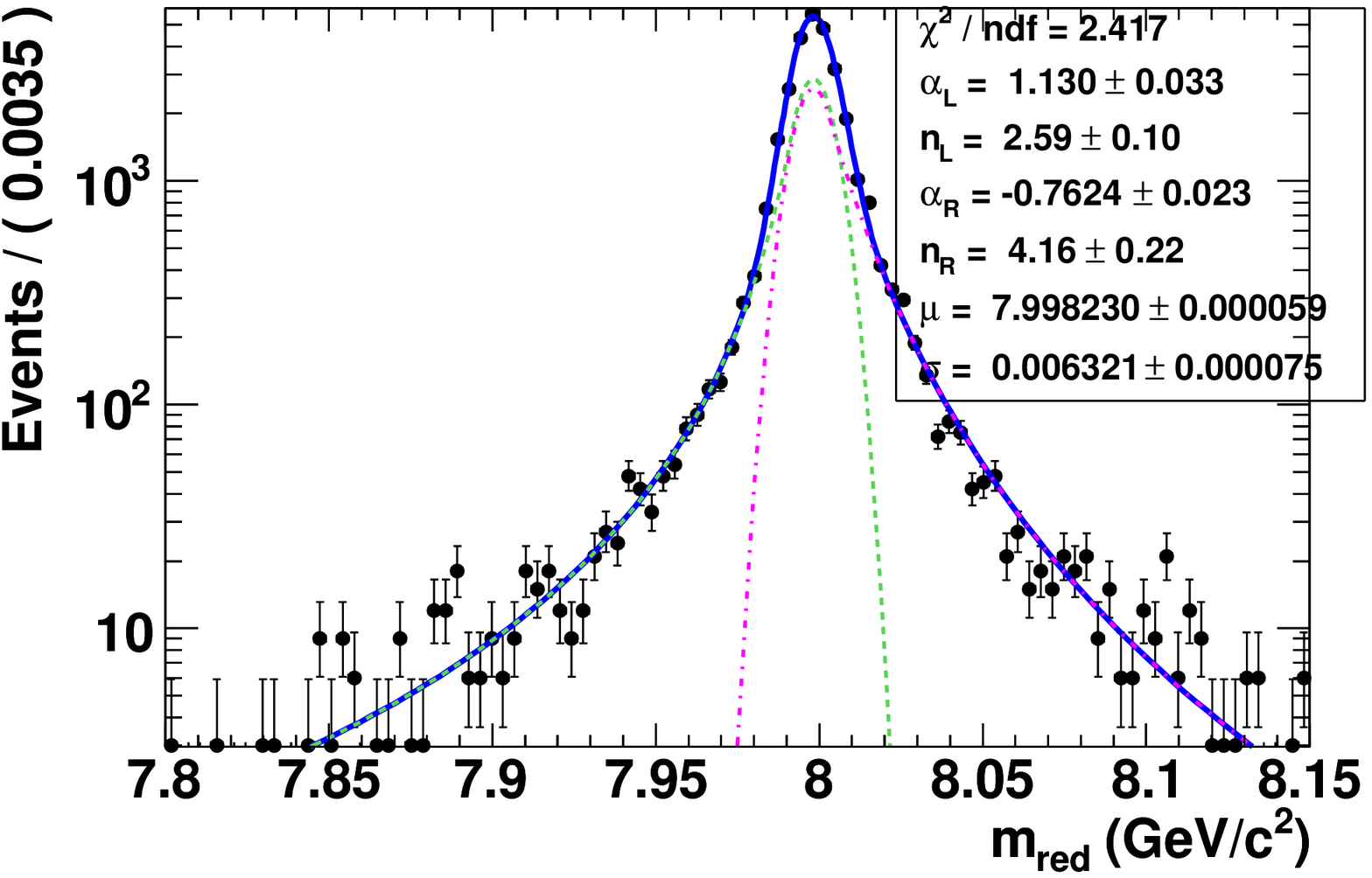}

\smallskip
\centerline{\hfill (g) \hfill \hfill (h) \hfill \hfill (i) \hfill}
\smallskip

\caption {Signal PDFs for the Higgs mass of  (a) $m_{A^0}=2.0$ GeV/$c^2$  (b) $m_{A^0}=3.0$ GeV/$c^2$  (c) $m_{A^0}=4.0$ GeV/$c^2$ (d) $m_{A^0}=5.0$ GeV/$c^2$ (e) $m_{A^0}=6.0$ GeV/$c^2$  (f) $m_{A^0}=6.7$ GeV/$c^2$ (g) $m_{A^0}=7.0$ GeV/$c^2$  (h) $m_{A^0}=7.5$ GeV/$c^2$ and (i) $m_{A^0}=8.0$ GeV/$c^2$.}

\label{fig:SigPDFY2S2}
\end{figure}

\begin{figure}
\centering
 \includegraphics[width=2.0in]{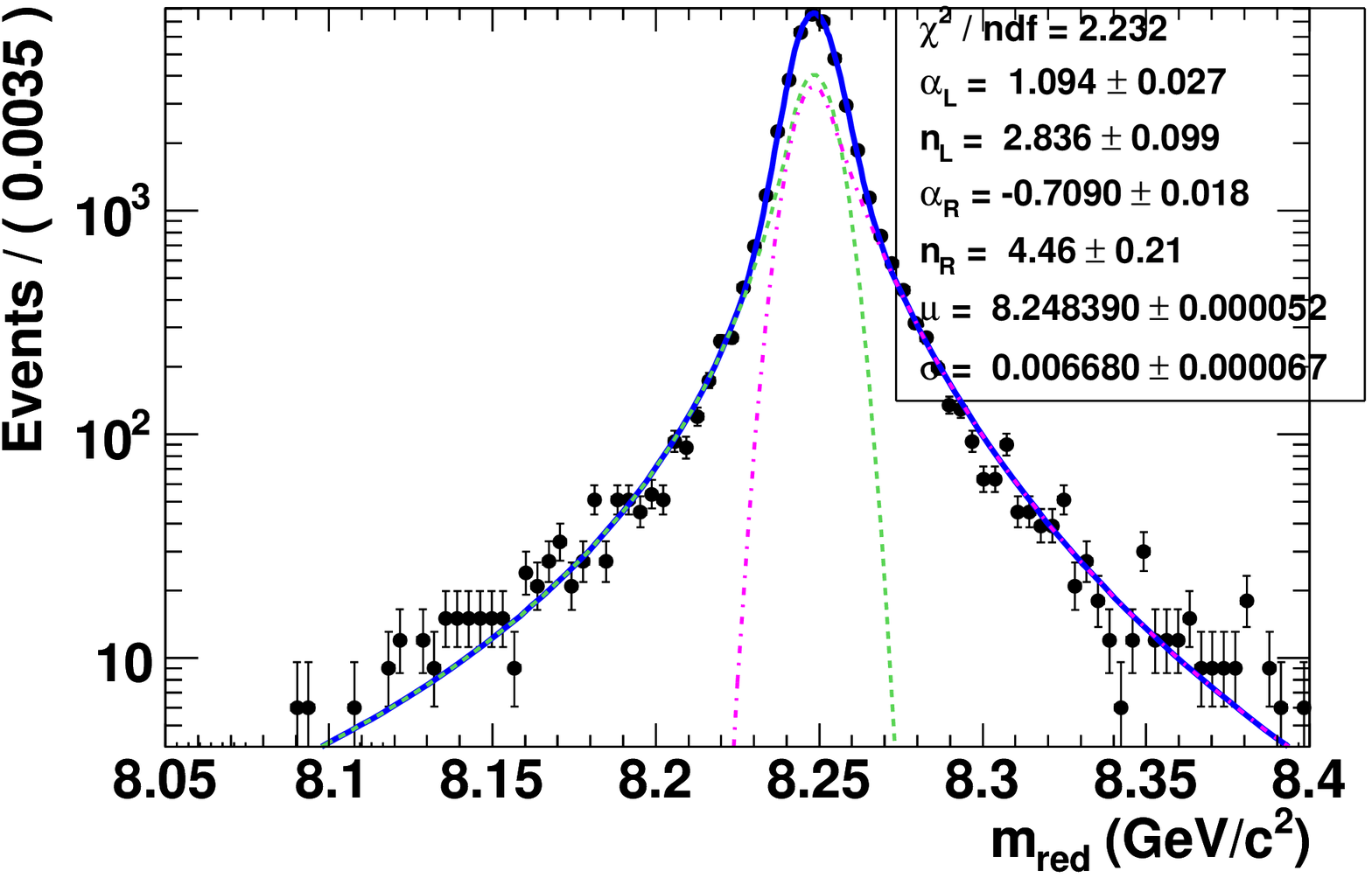}
\includegraphics[width=2.0in]{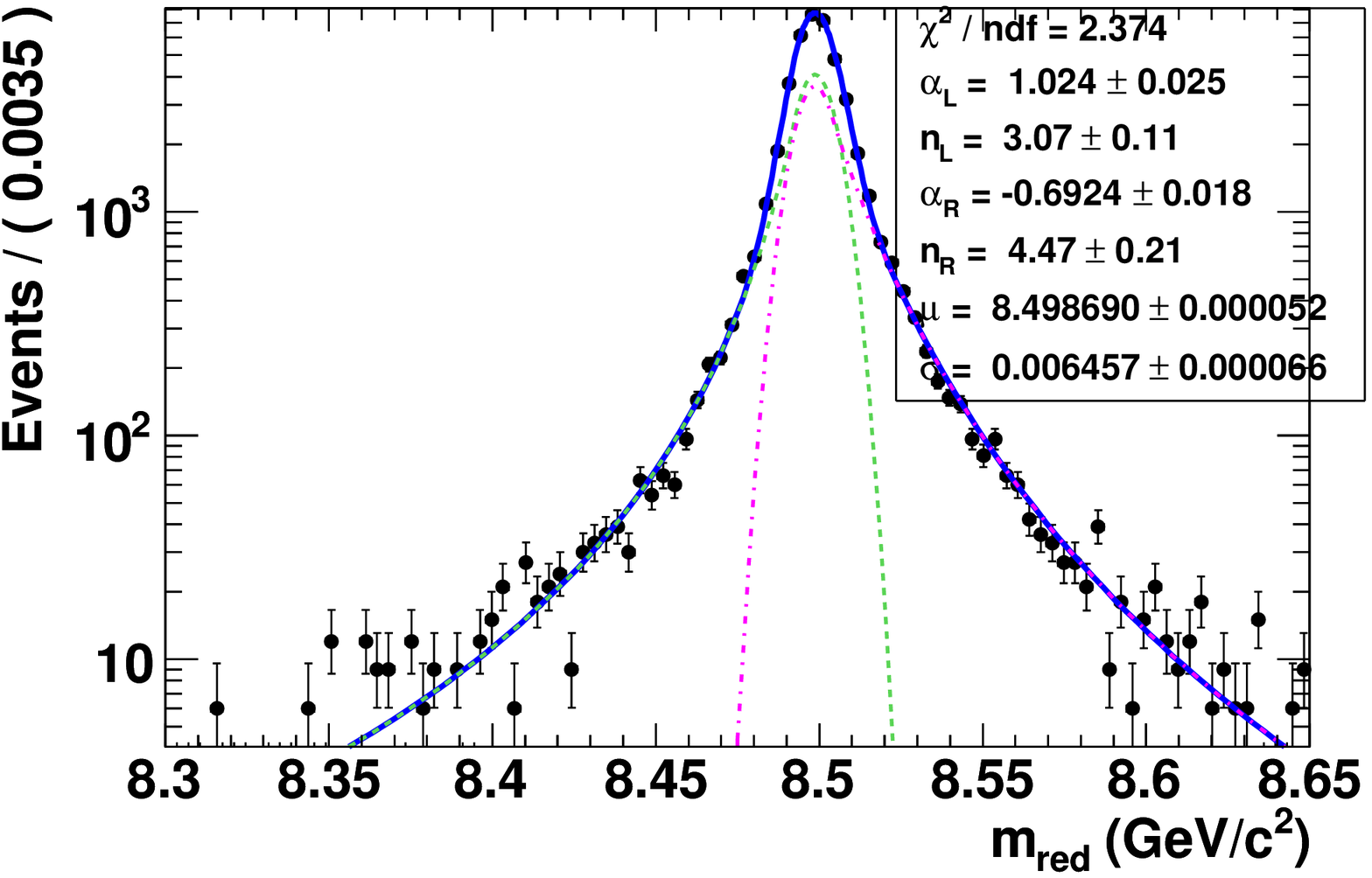}
\includegraphics[width=2.0in]{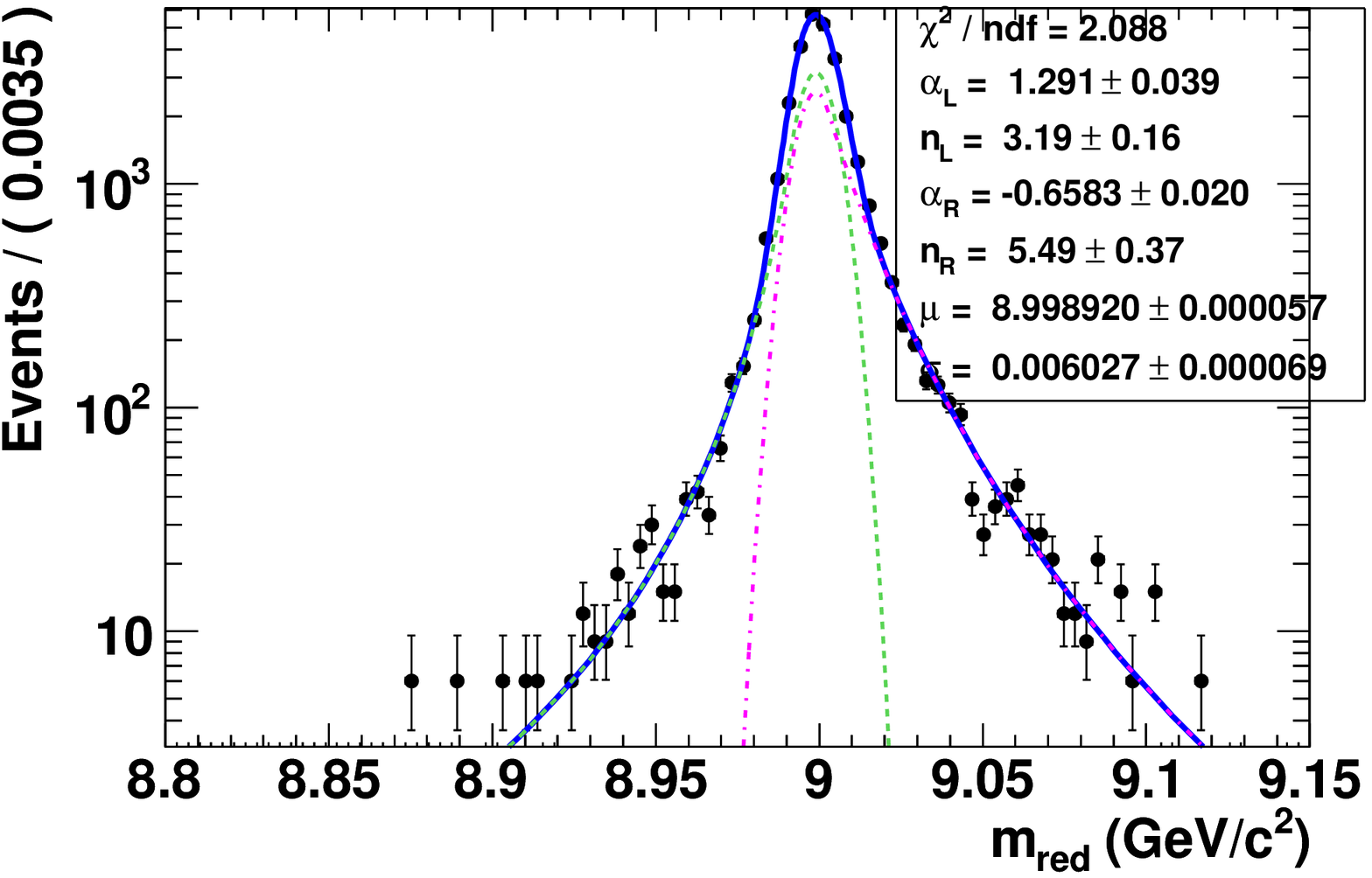}

\smallskip
\centerline{\hfill (a) \hfill \hfill (b) \hfill \hfill (c) \hfill}
\smallskip

 \includegraphics[width=2.0in]{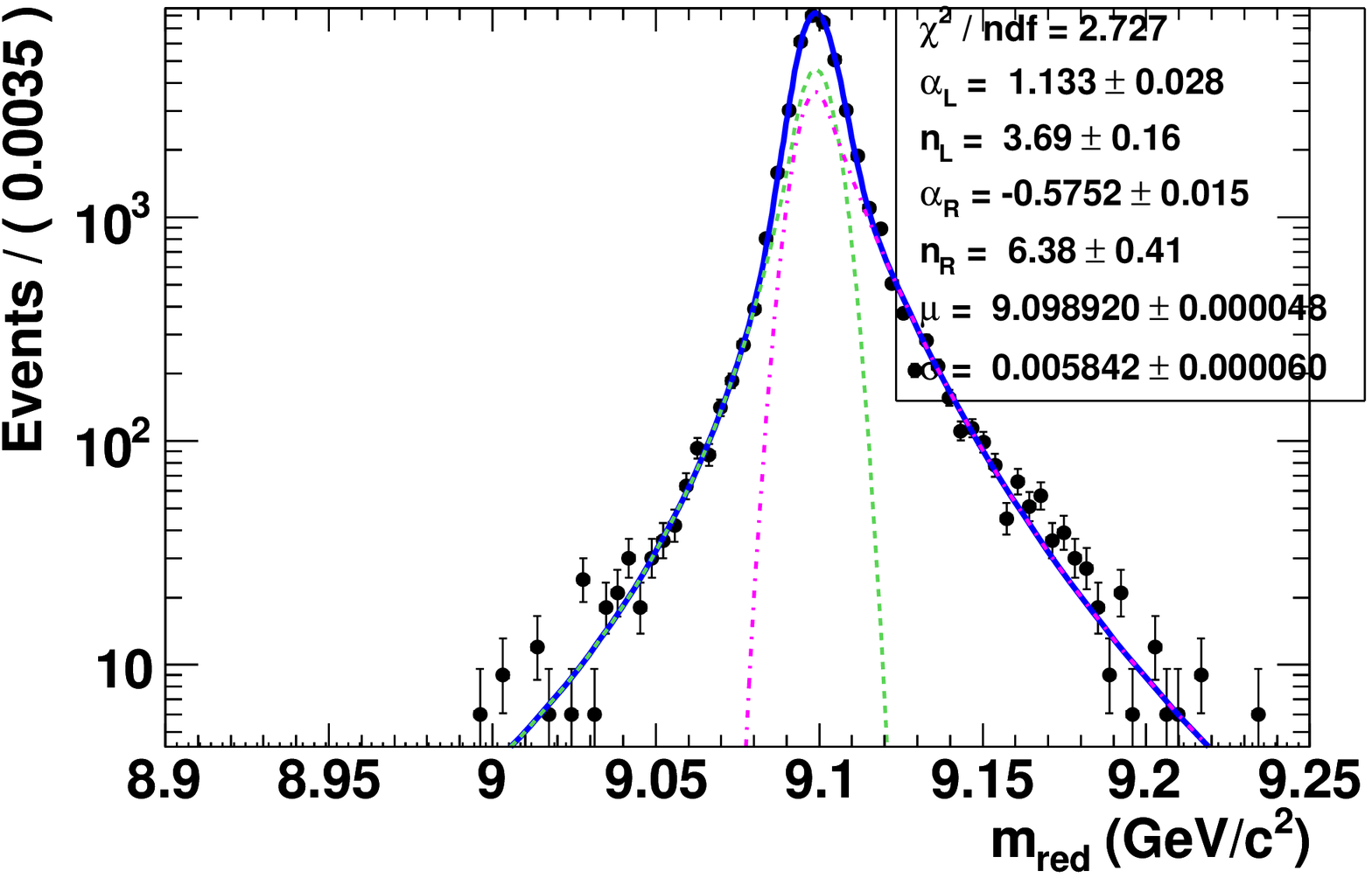}
\includegraphics[width=2.0in]{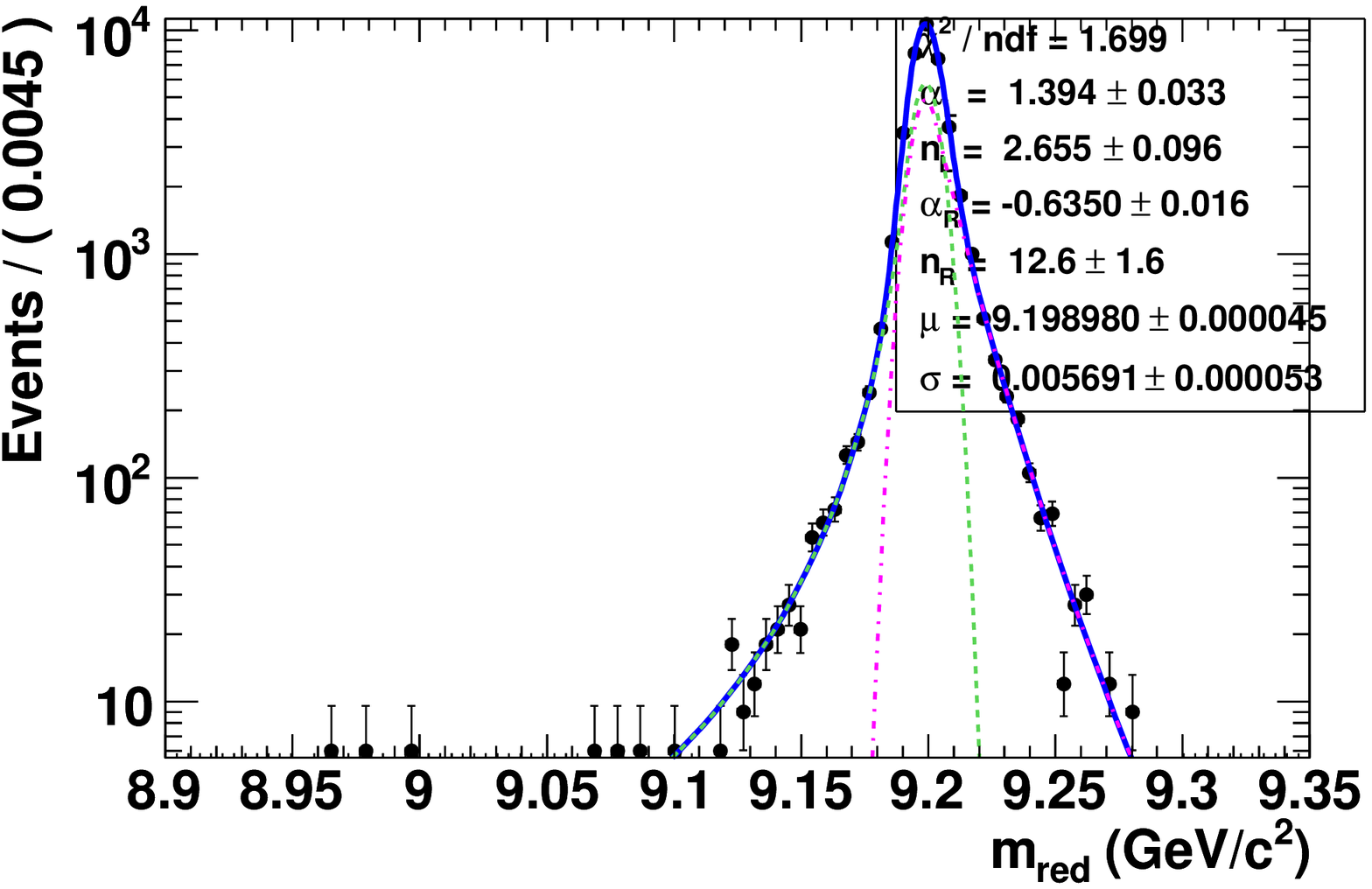}
\includegraphics[width=2.0in]{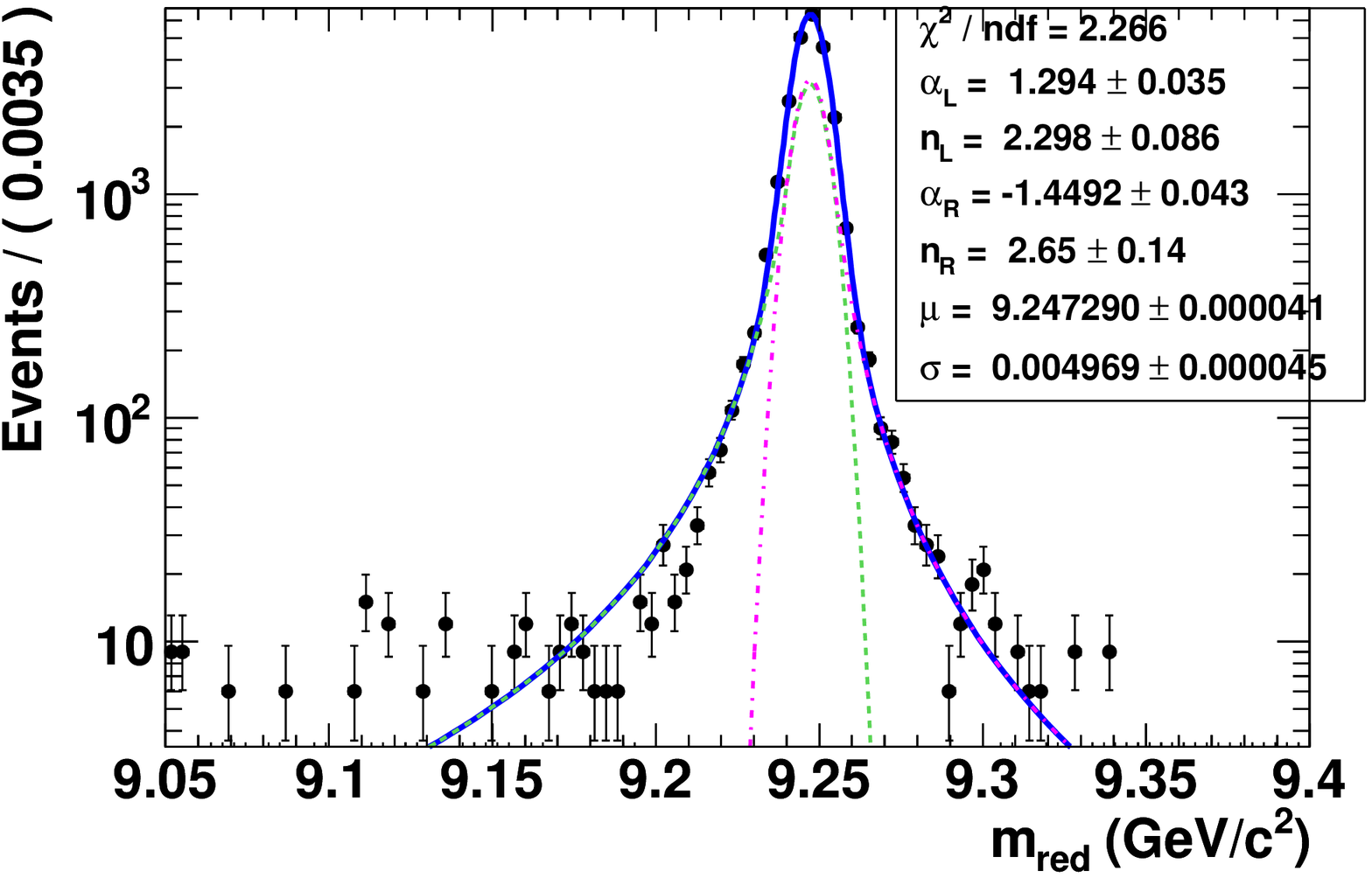}
\smallskip
\centerline{\hfill (d) \hfill \hfill (e) \hfill \hfill (f) \hfill}
\smallskip

\caption {Signal PDFs for the Higgs mass of  (a) $m_{A^0}=8.25$ GeV/$c^2$ (b) $m_{A^0}=8.50$ GeV/$c^2$  (c) $m_{A^0}=9.0$ GeV/$c^2$ (d) $m_{A^0}=9.10$ GeV/$c^2$  (e) $m_{A^0}=9.20$ GeV/$c^2$ and (f) $m_{A^0}=9.25$ GeV/$c^2$.}

\label{fig:SigPDFY2S3}
\end{figure}

\begin{figure}
\section{Signal PDFs for $\Upsilon(3S)$}
\centering
\includegraphics[width=2.0in]{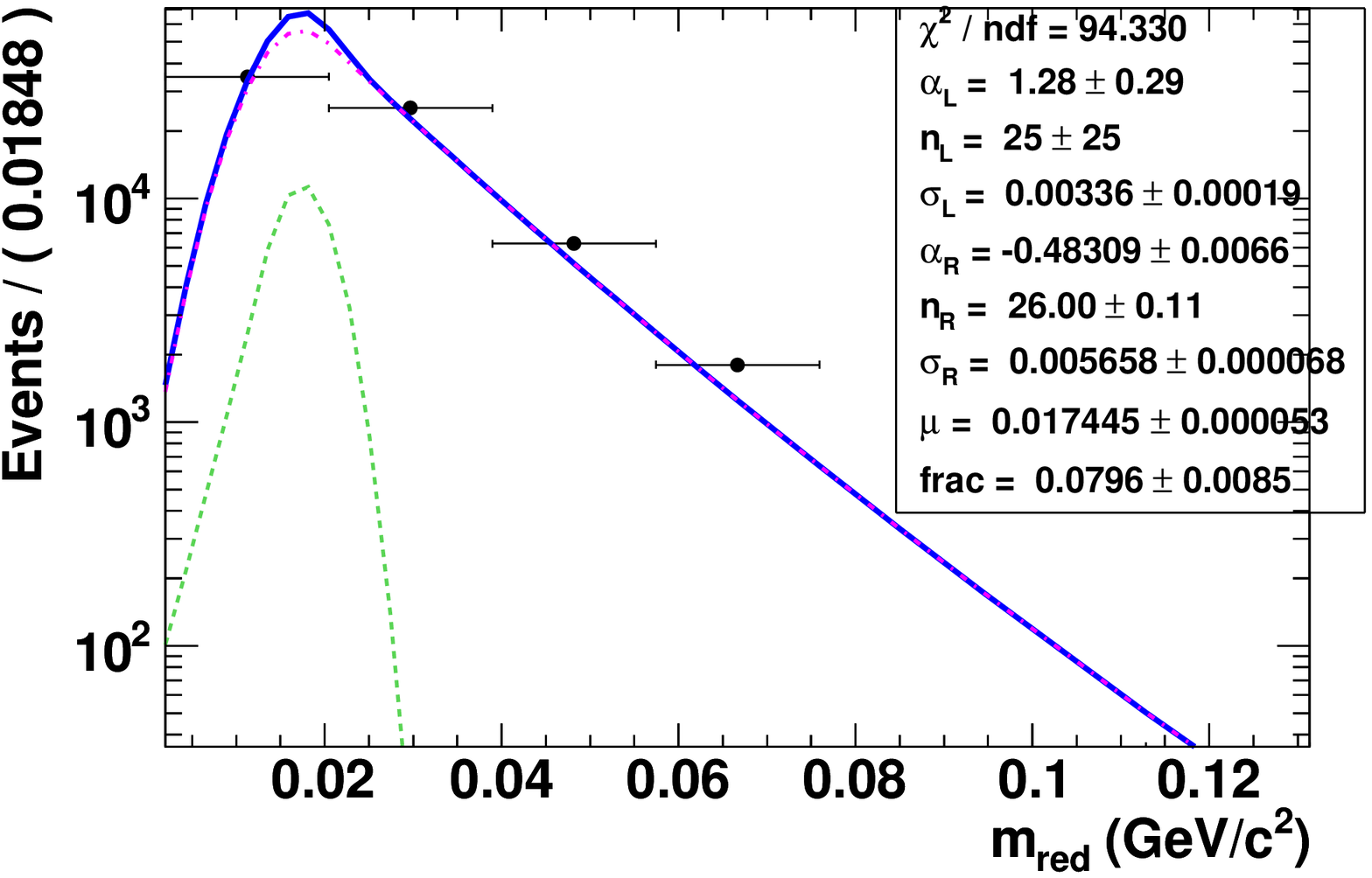}
\includegraphics[width=2.0in]{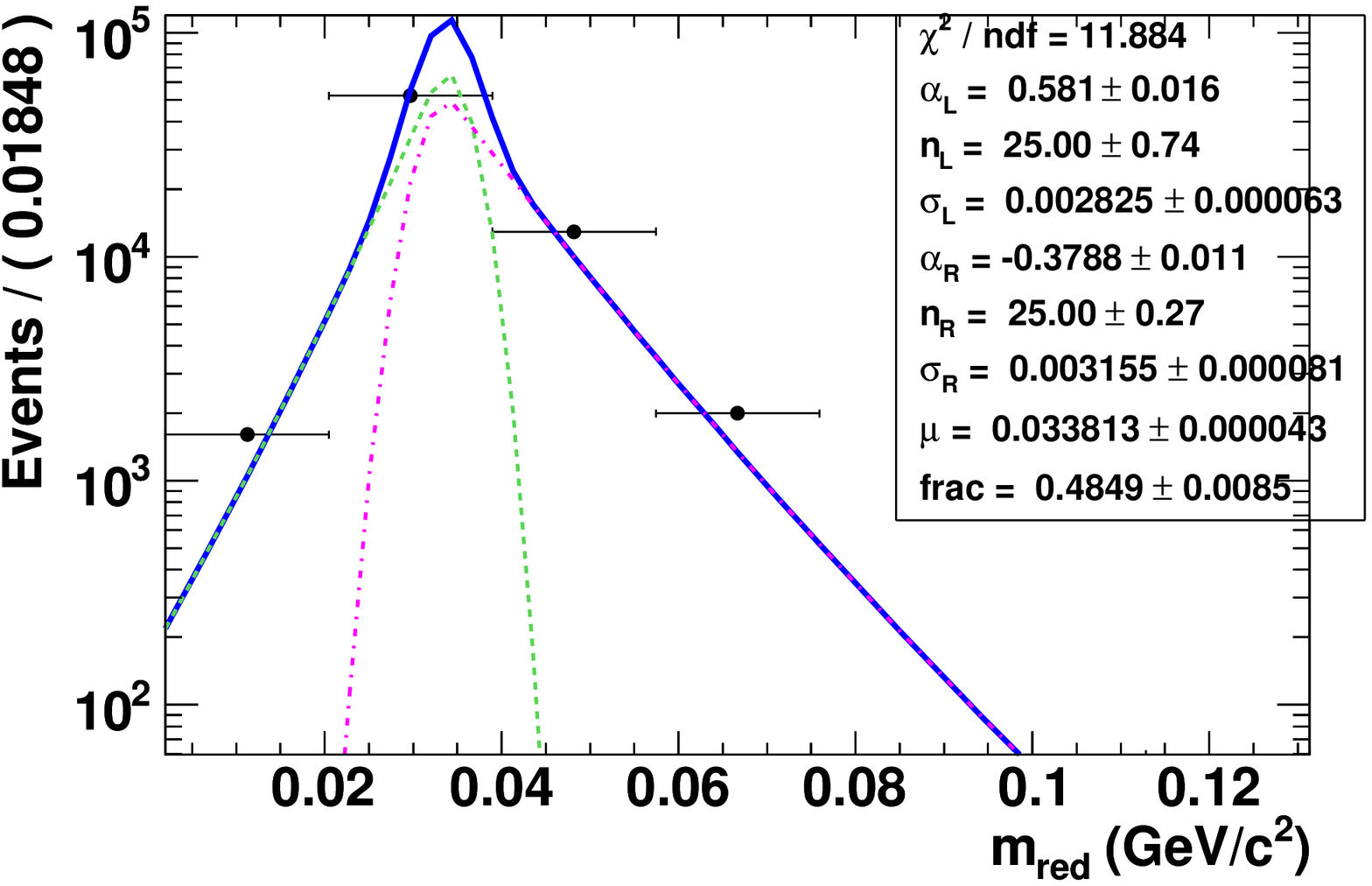}
\includegraphics[width=2.0in]{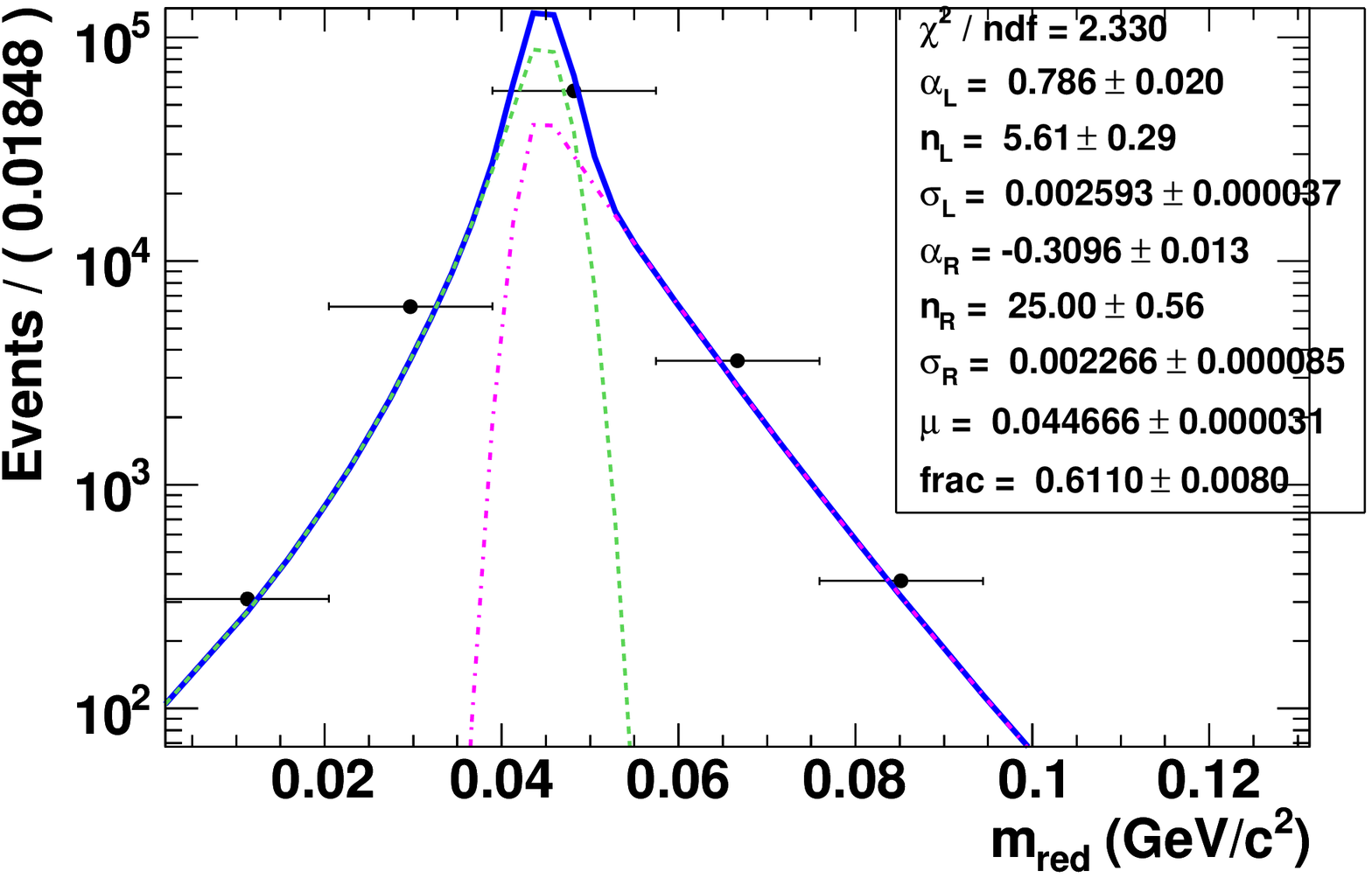}

\smallskip
\centerline{\hfill (a) \hfill \hfill (b) \hfill \hfill (c) \hfill}
\smallskip

\includegraphics[width=2.0in]{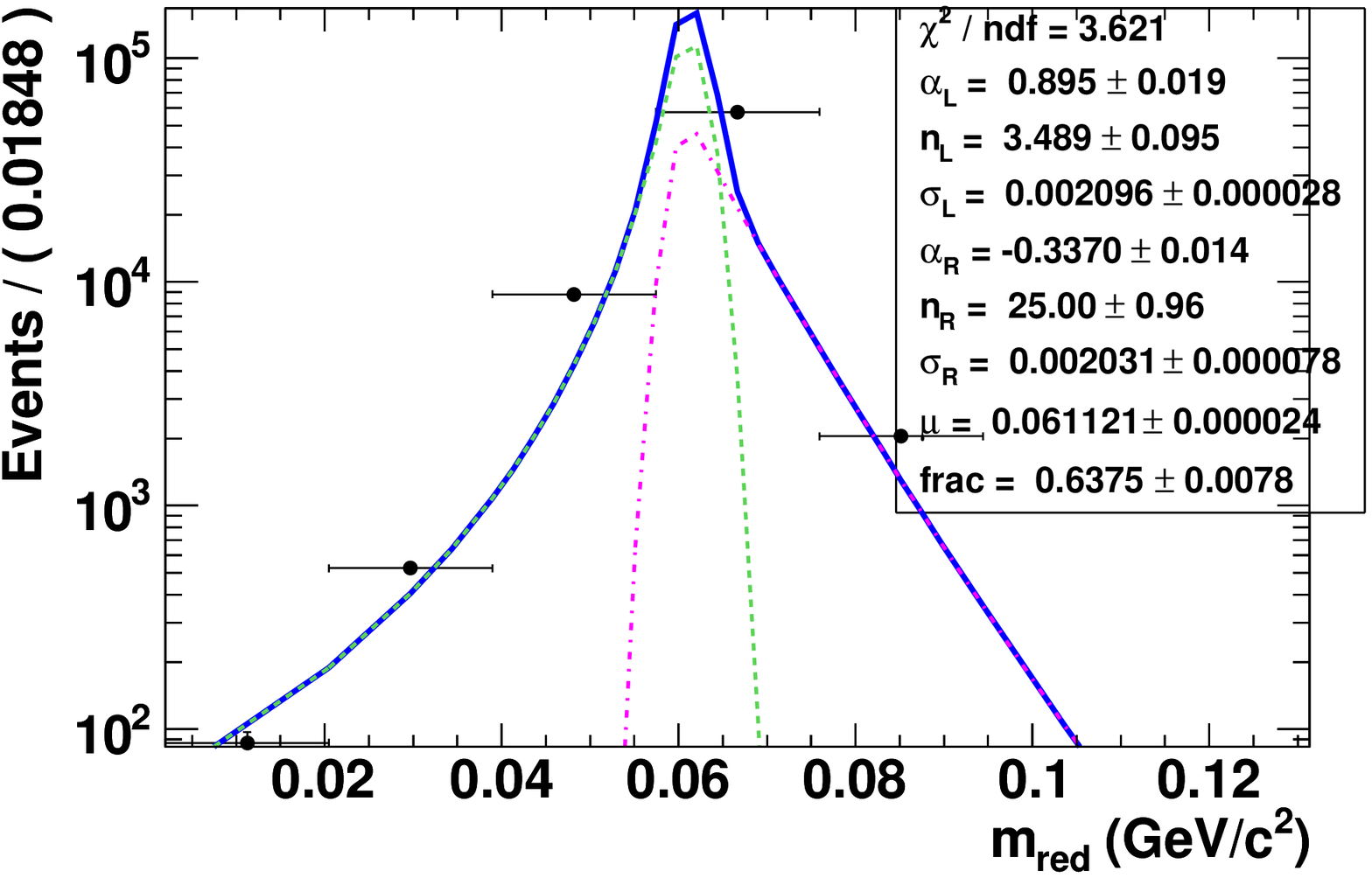}
\includegraphics[width=2.0in]{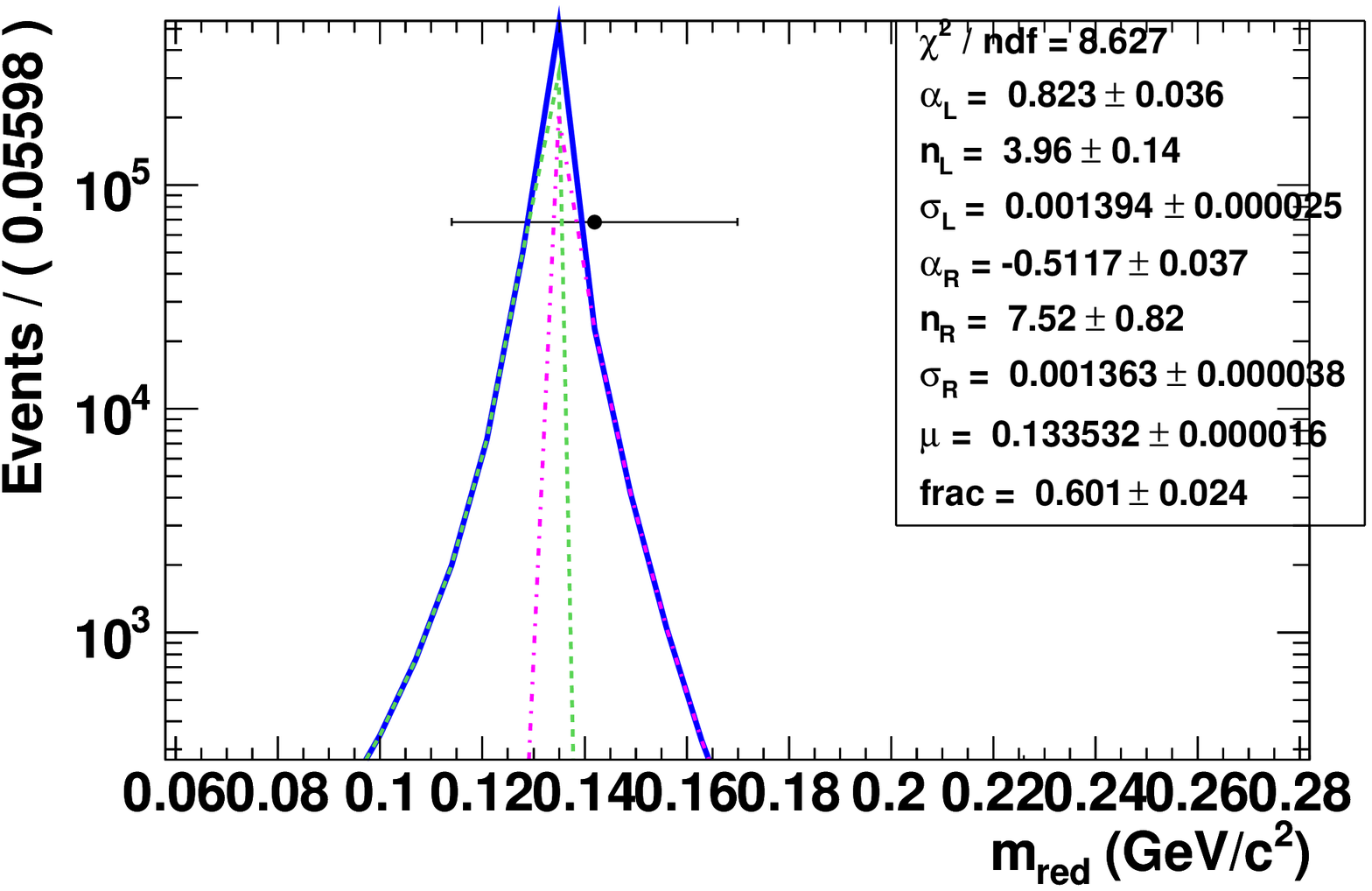}

\smallskip
\centerline{\hfill (d) \hfill \hfill (e) \hfill \hfill (f) \hfill}
\smallskip

\includegraphics[width=2.0in]{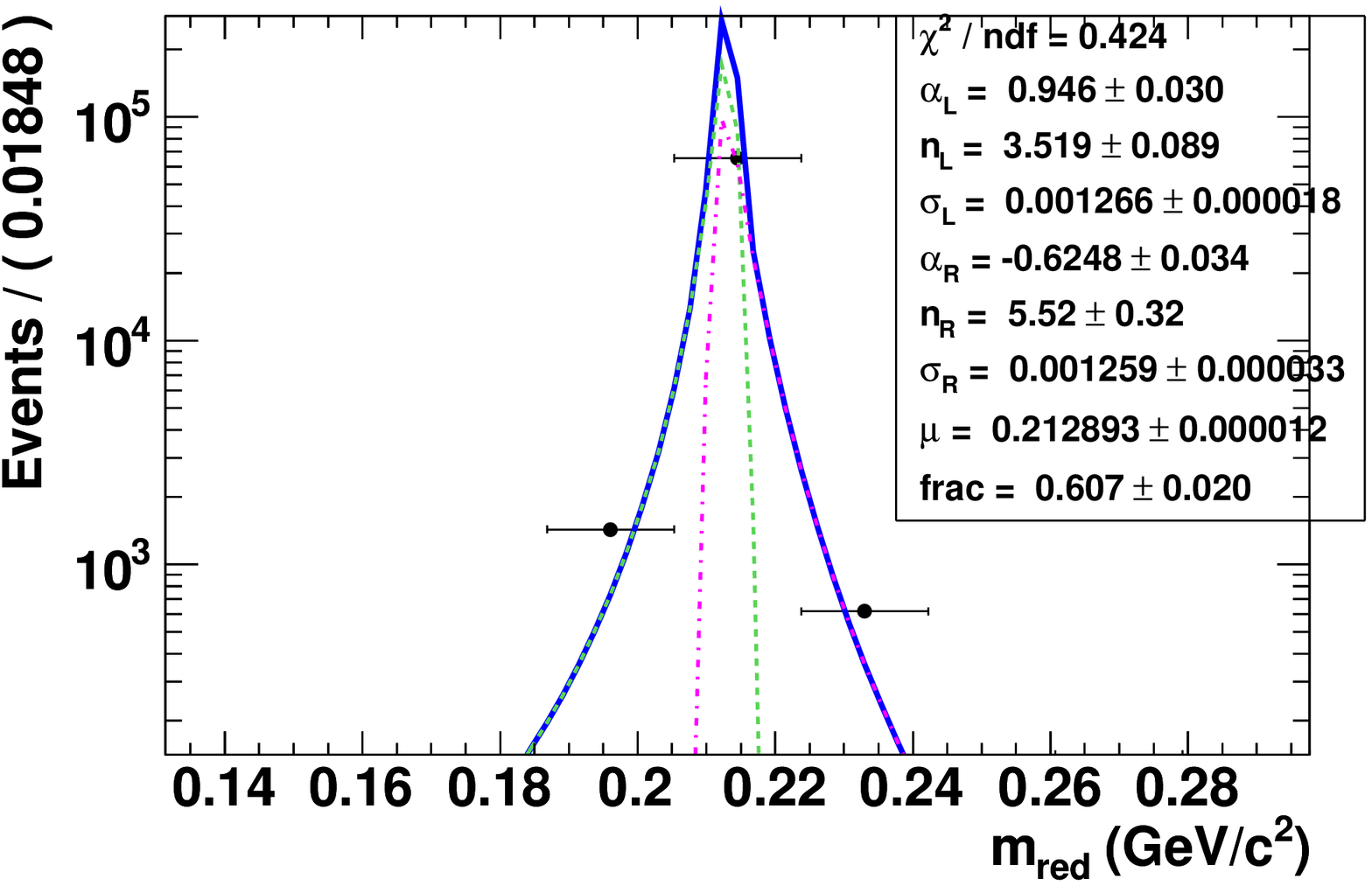}
\includegraphics[width=2.0in]{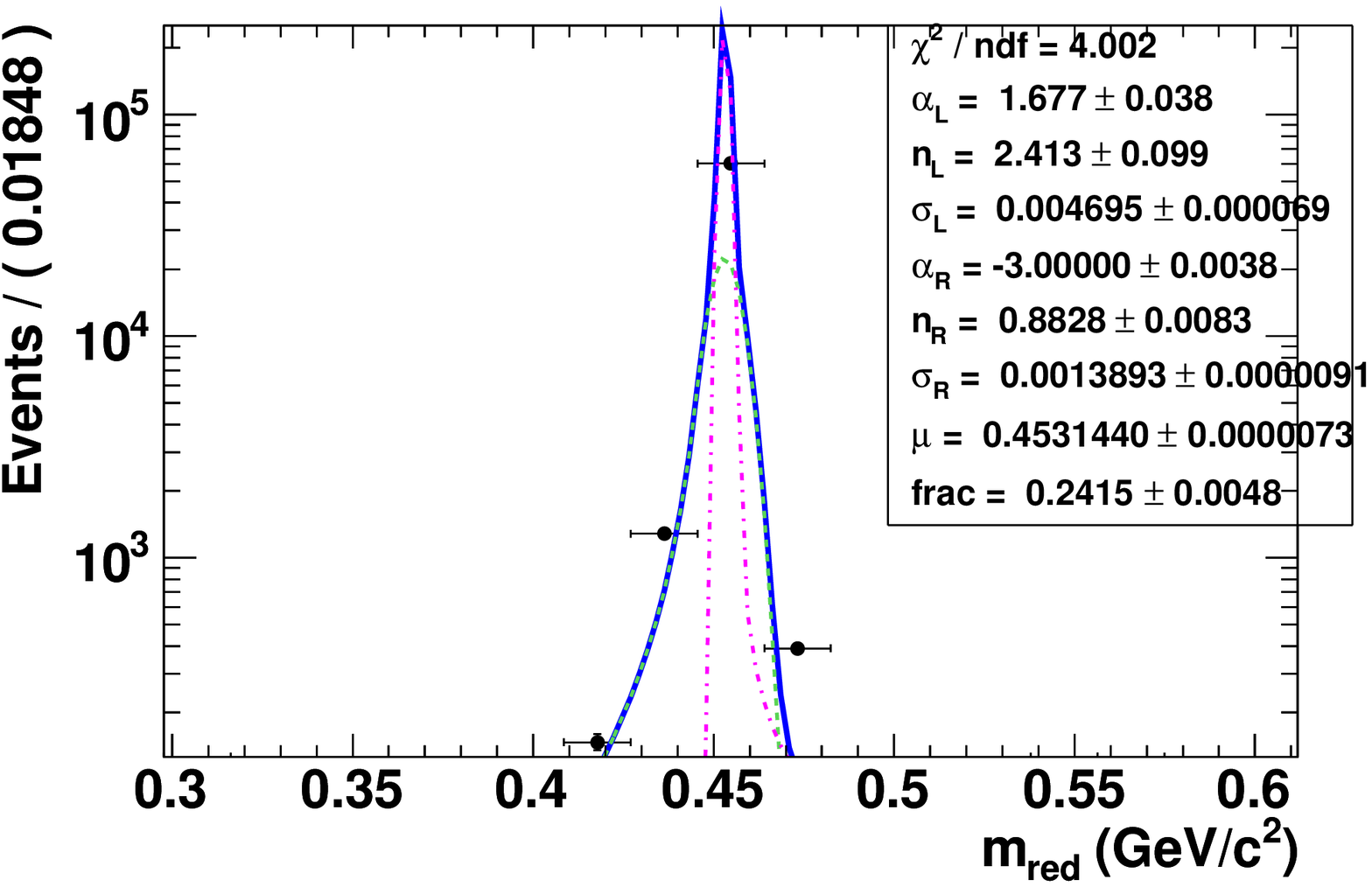}
\includegraphics[width=2.0in]{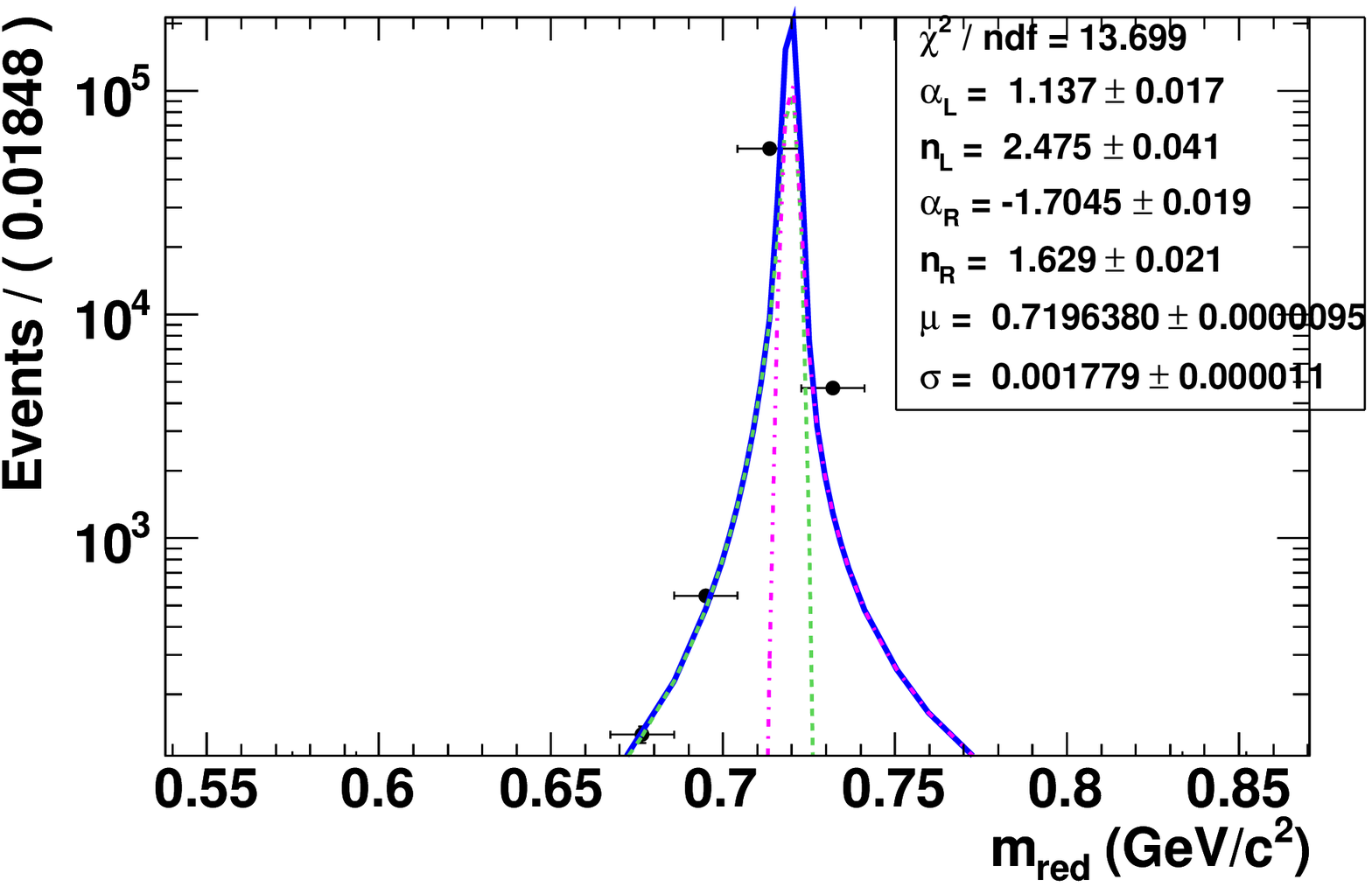}

\smallskip
\centerline{\hfill (g) \hfill \hfill (h) \hfill \hfill (i) \hfill}
\smallskip

\caption {Signal PDFs for the Higgs mass of (a) $m_{A^0}=0.212$ GeV/$c^2$  (b) $m_{A^0}=0.214$ GeV/$c^2$ (c) $m_{A^0}=0.216$ GeV/$c^2$ (d) $m_{A^0}=0.218$ GeV/$c^2$ (e) $m_{A^0}=0.220$ GeV/$c^2$ (f) $m_{A^0}=0.250$ GeV/$c^2$  (g) $m_{A^0}=0.300$ GeV/$c^2$ (h) $m_{A^0}=0.50$ GeV/$c^2$ and (i) $m_{A^0}=0.75$ GeV/$c^2$.}

\label{fig:SigPDFY3S1}
\end{figure}

\begin{figure}
\centering
 \includegraphics[width=2.0in]{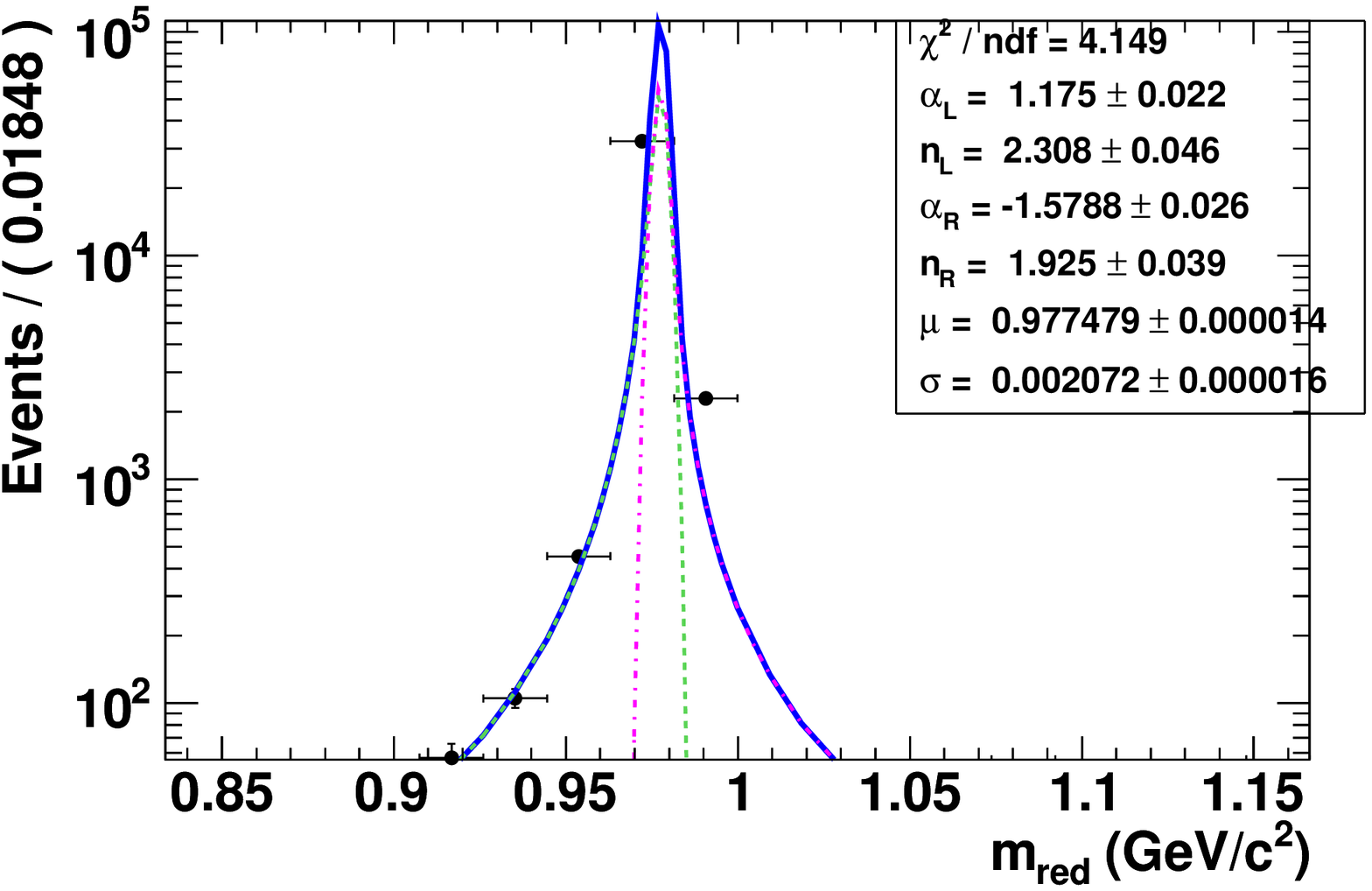}
\includegraphics[width=2.0in]{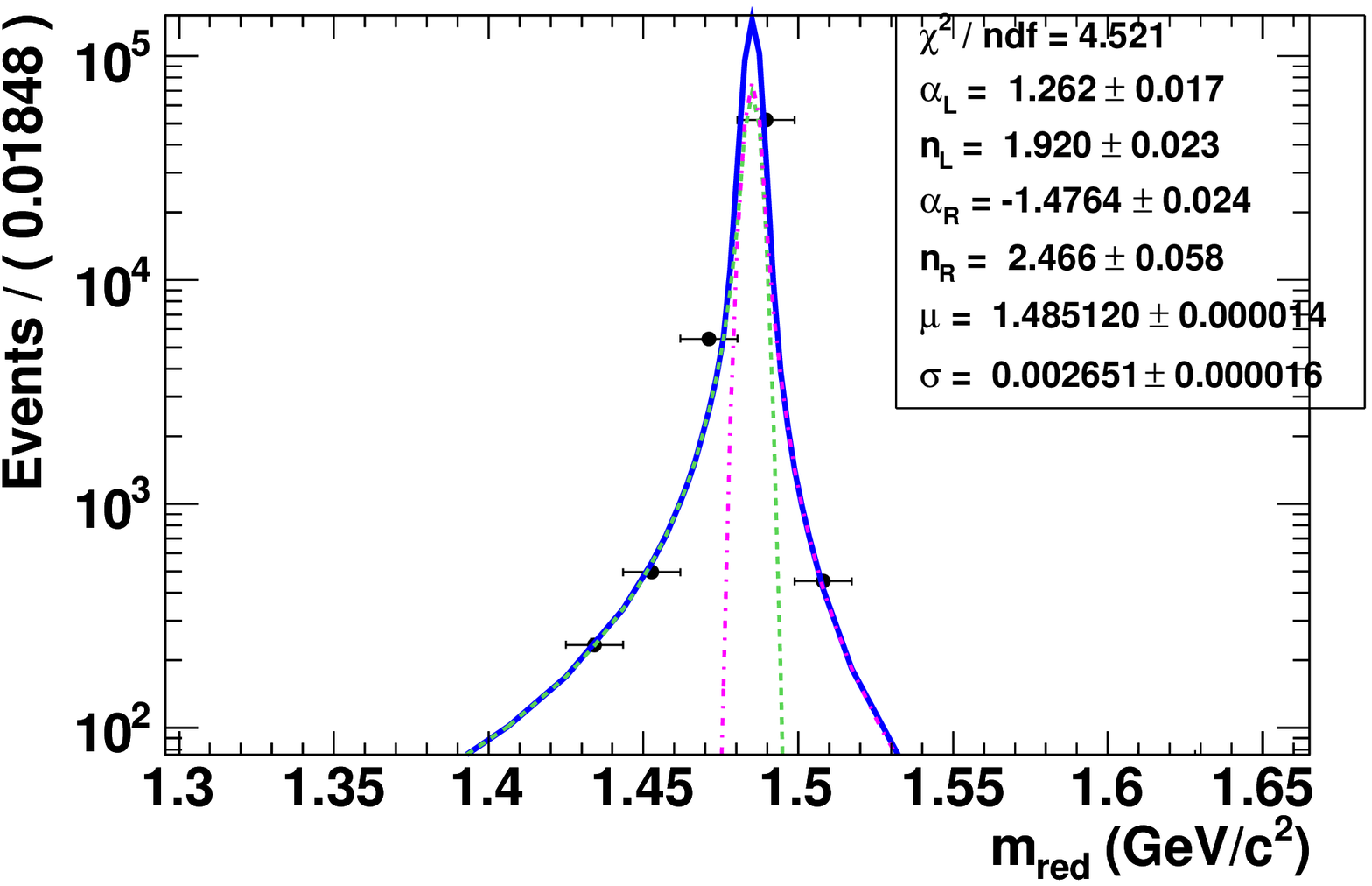}
\includegraphics[width=2.0in]{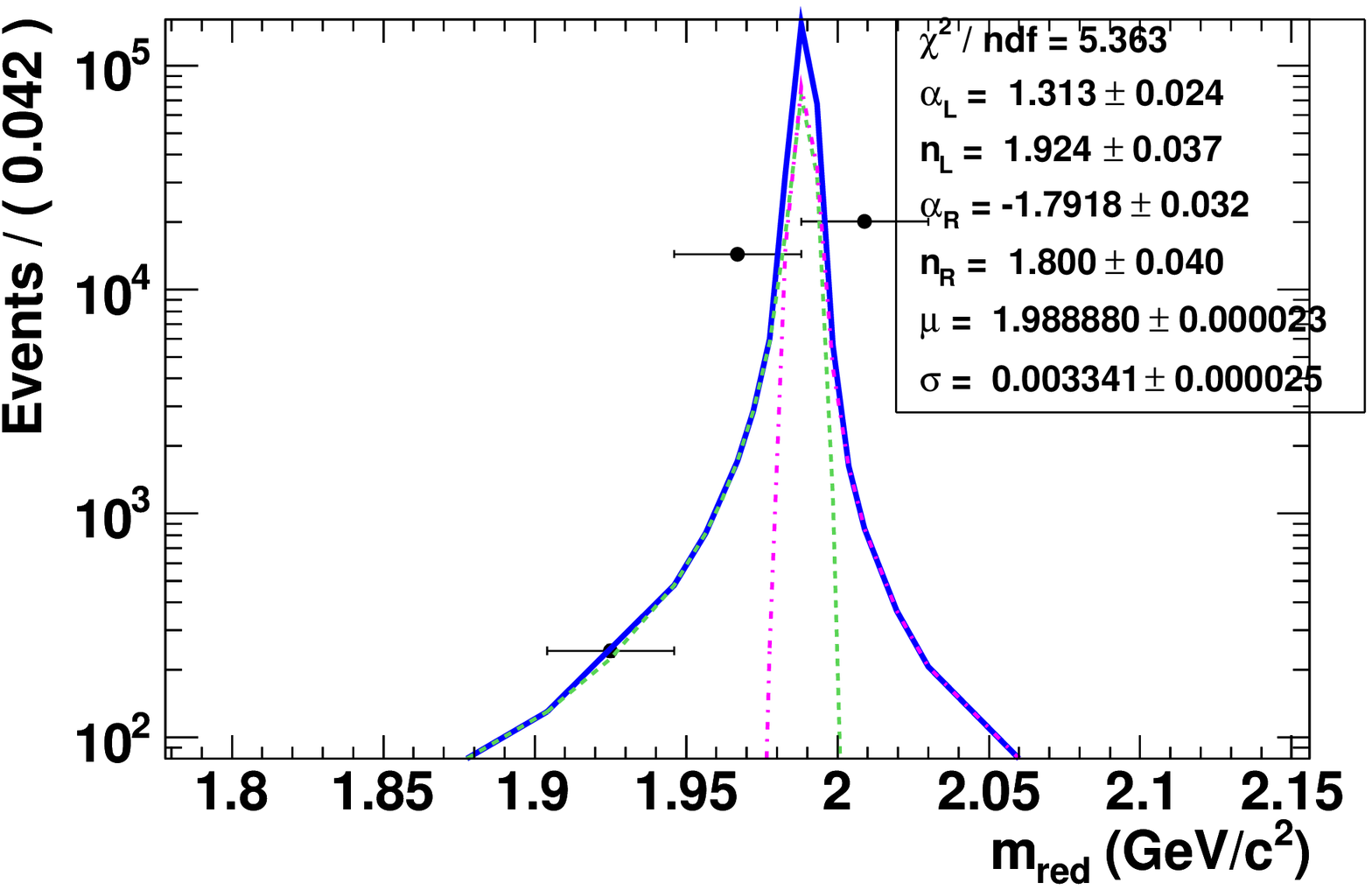}

\smallskip
\centerline{\hfill (a) \hfill \hfill (b) \hfill \hfill (c) \hfill}
\smallskip

 \includegraphics[width=2.0in]{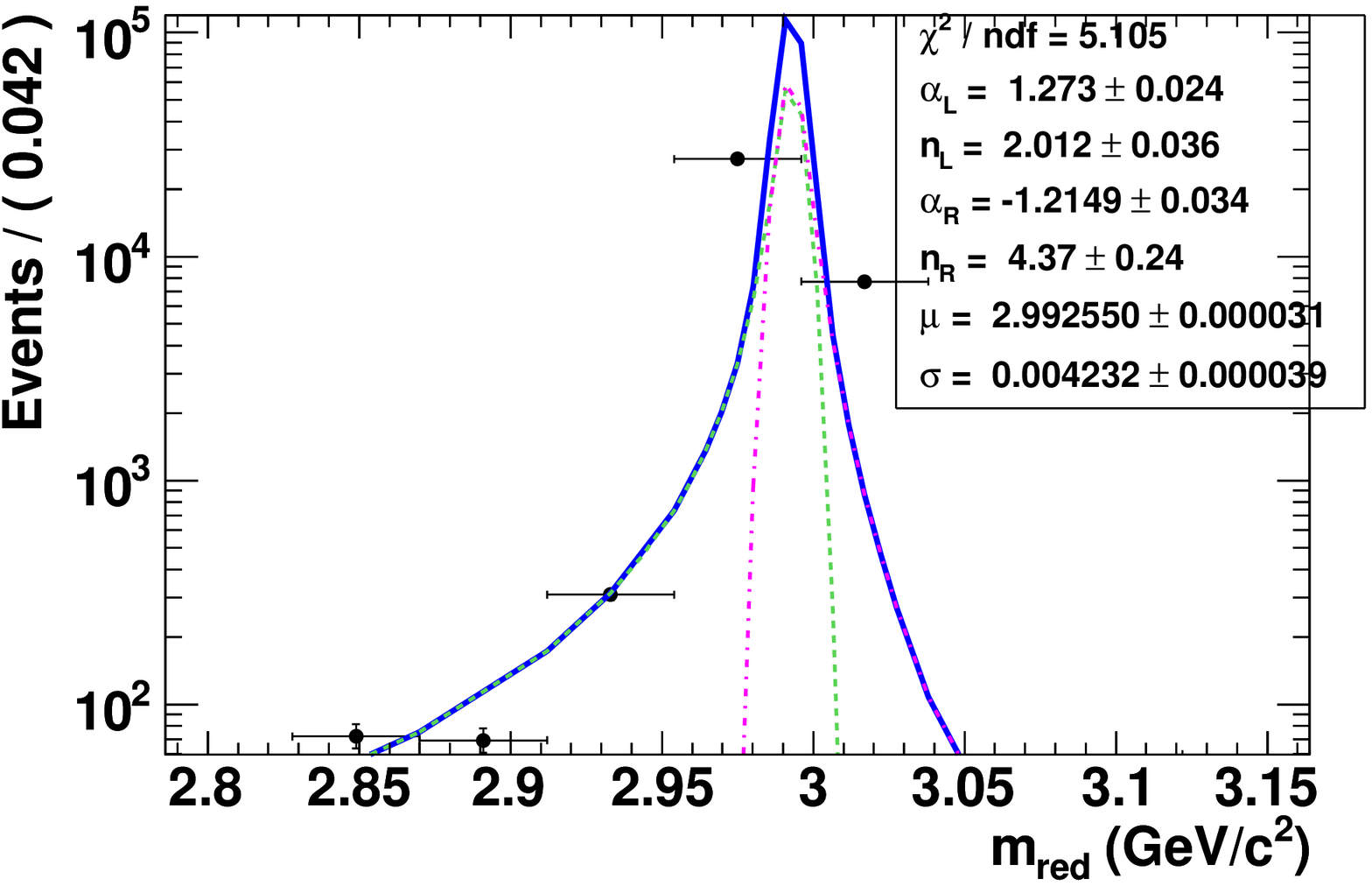}
\includegraphics[width=2.0in]{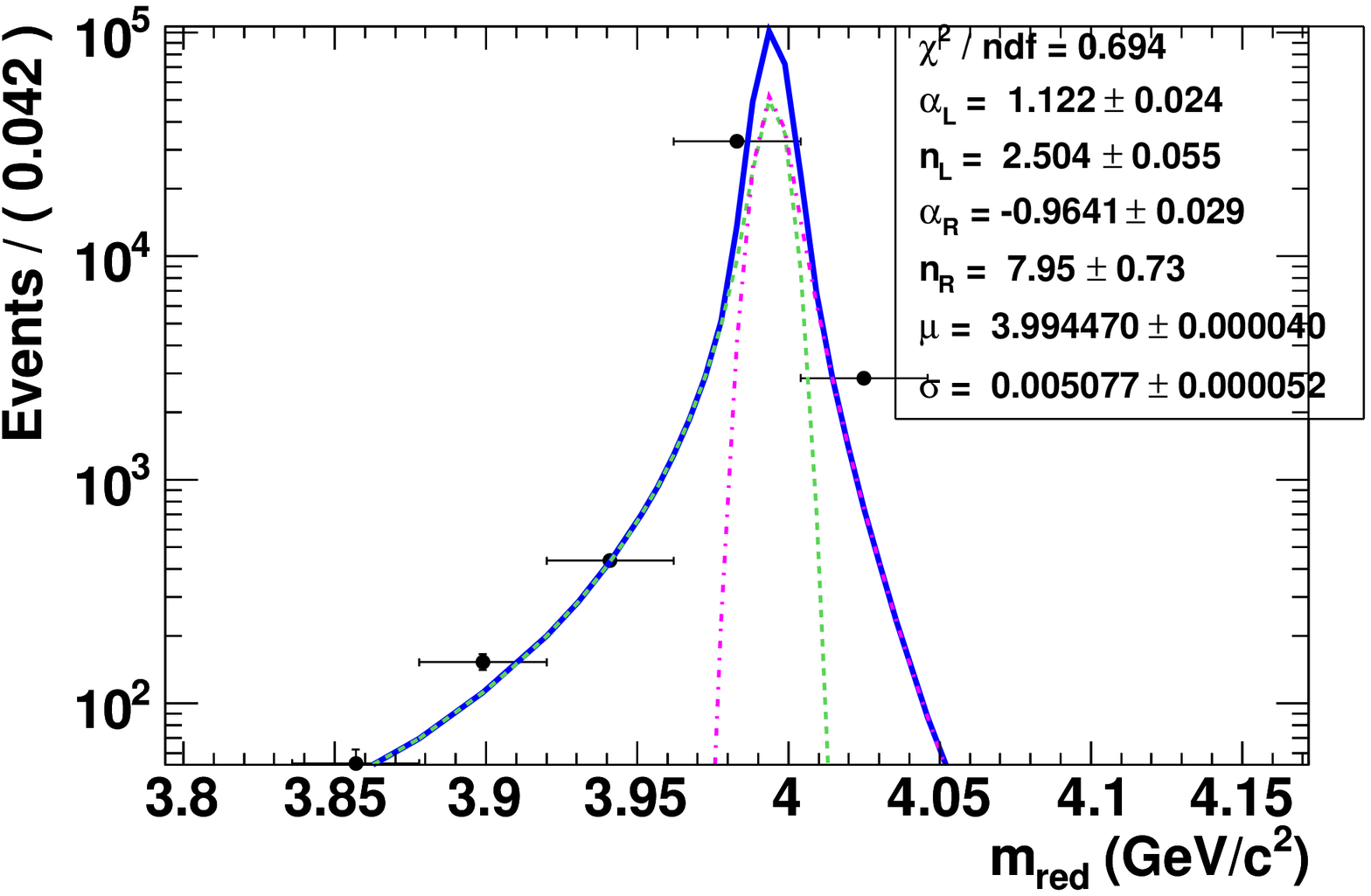}
\includegraphics[width=2.0in]{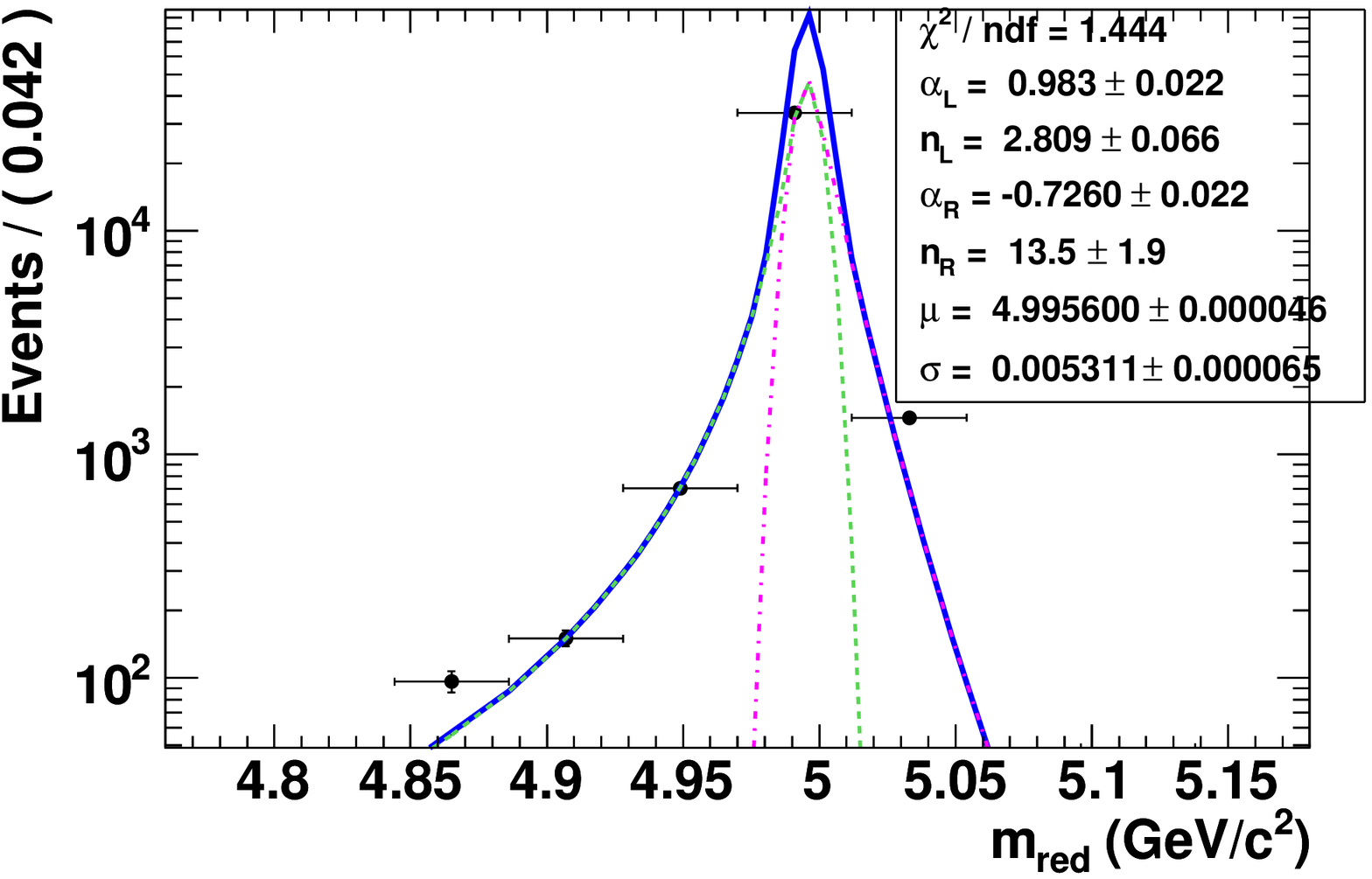}

\smallskip
\centerline{\hfill (d) \hfill \hfill (e) \hfill \hfill (f) \hfill}
\smallskip

\includegraphics[width=2.0in]{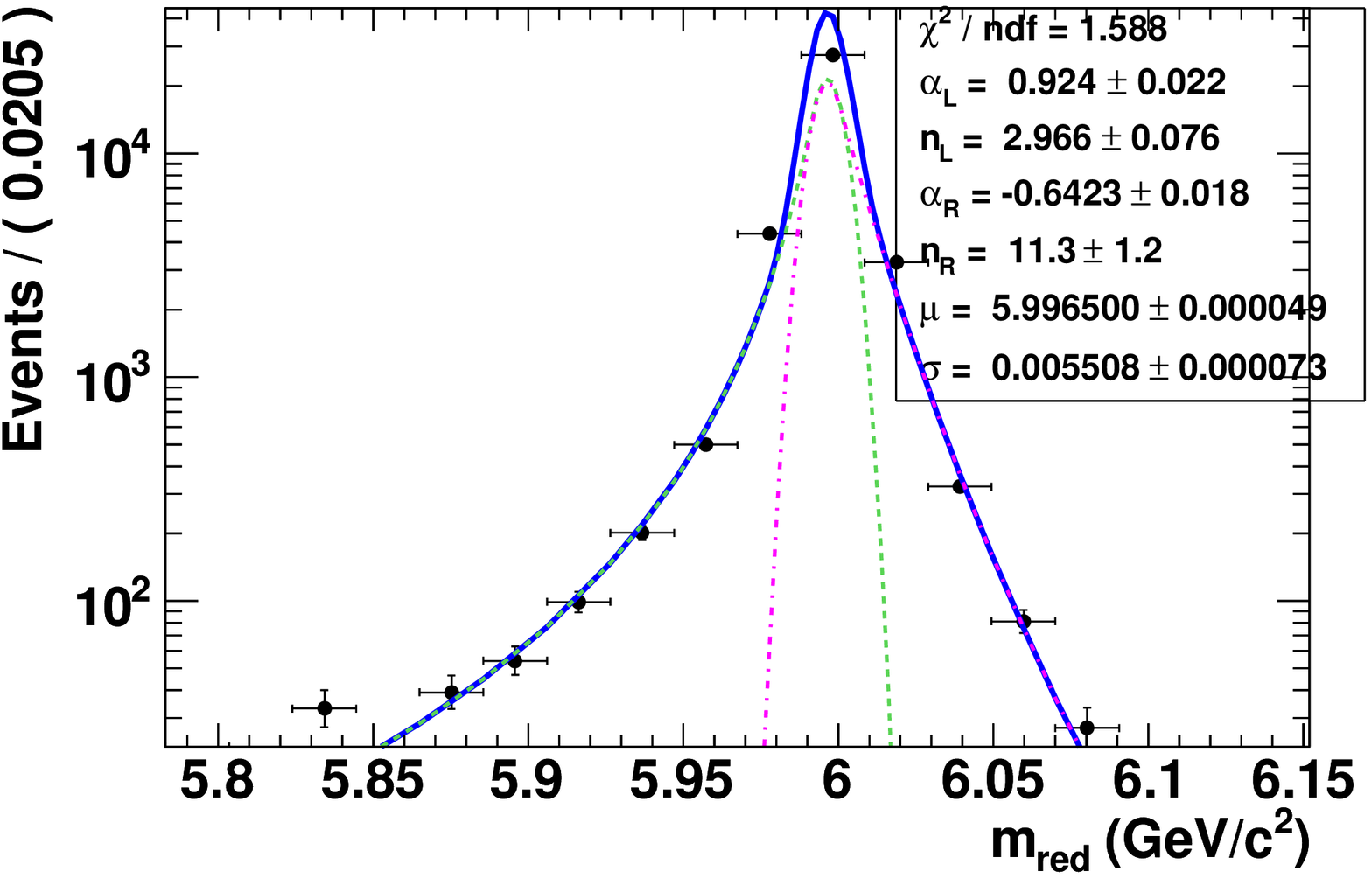}
\includegraphics[width=2.0in]{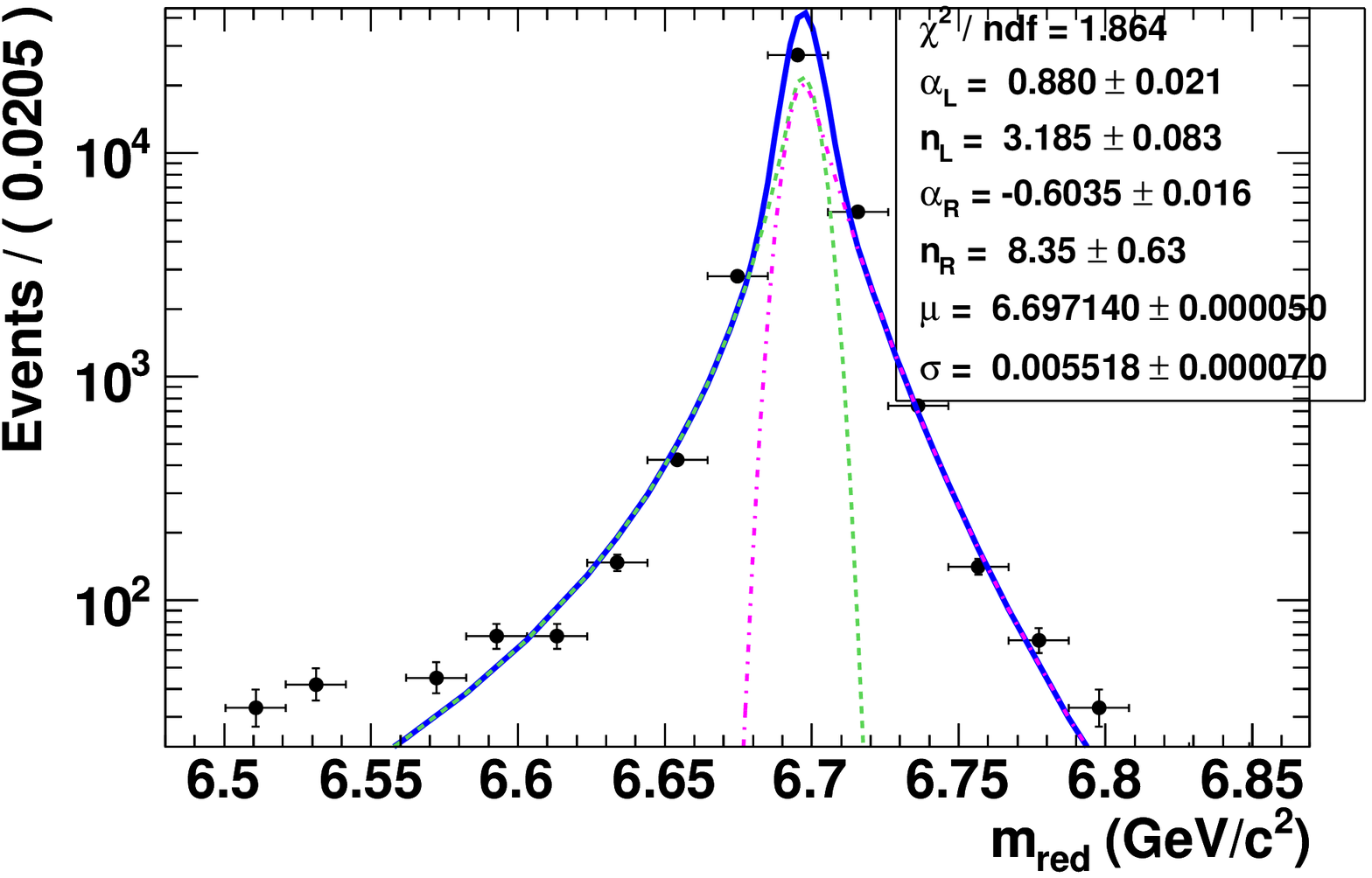}
\includegraphics[width=2.0in]{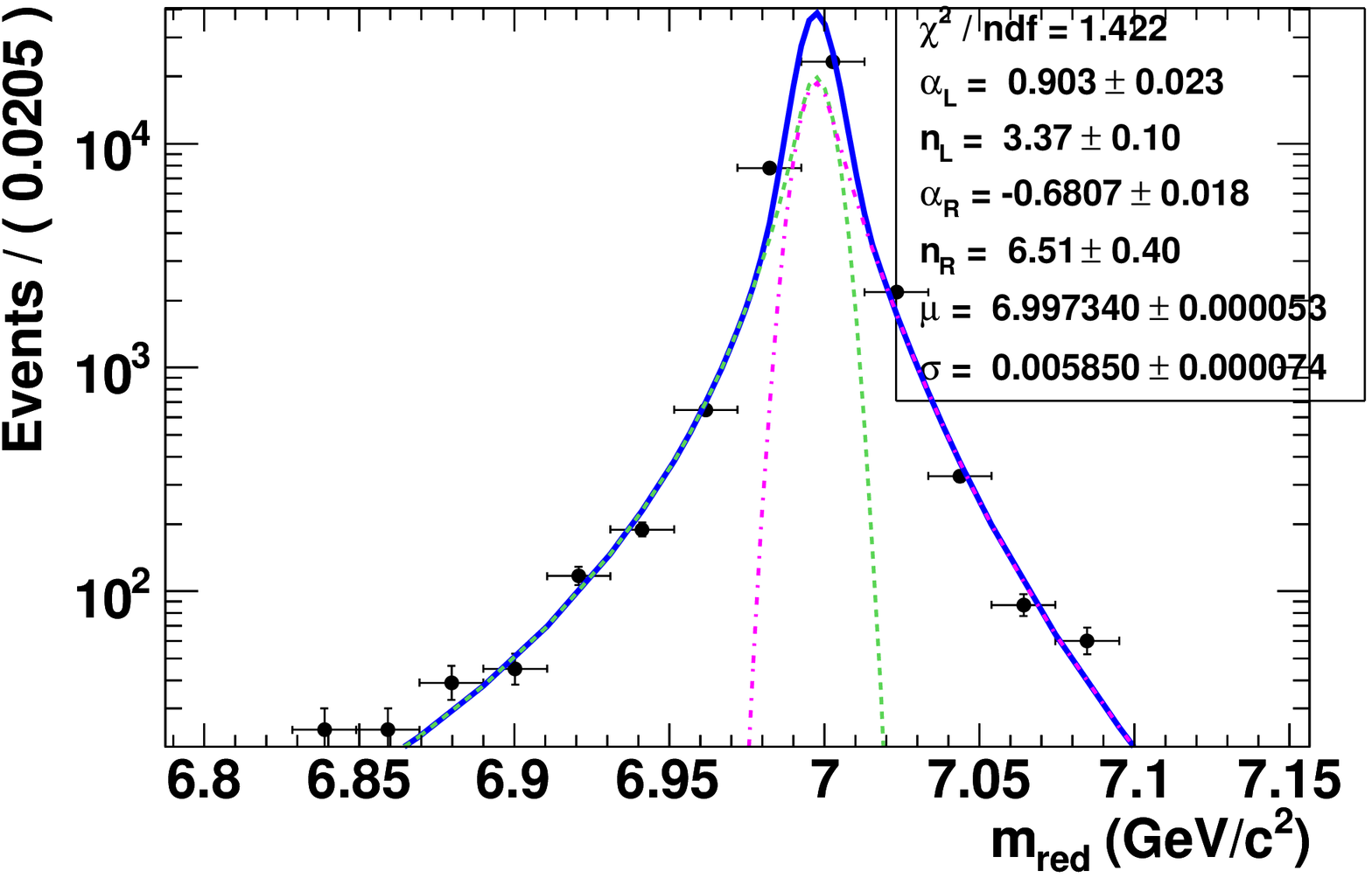}

\smallskip
\centerline{\hfill (g) \hfill \hfill (h) \hfill \hfill (i) \hfill}
\smallskip

\caption {Signal PDFs for the Higgs mass of (a) $m_{A^0}=1.0$ GeV/$c^2$,  (b) $m_{A^0}=1.5$ GeV/$c^2$ (c) $m_{A^0}=2.0$ GeV/$c^2$ (d) $m_{A^0}=3.0$ GeV/$c^2$ (e) $m_{A^0}=4.0$ GeV/$c^2$  (f) $m_{A^0}=5.0$ GeV/$c^2$ (g) $m_{A^0}=6.0$ GeV/$c^2$  (h) $m_{A^0}=6.7$ GeV/$c^2$ and (i) $m_{A^0}=7.0$ GeV/$c^2$.}

\label{fig:SigPDFY3S2}
\end{figure}

\begin{figure}
\centering
 \includegraphics[width=2.0in]{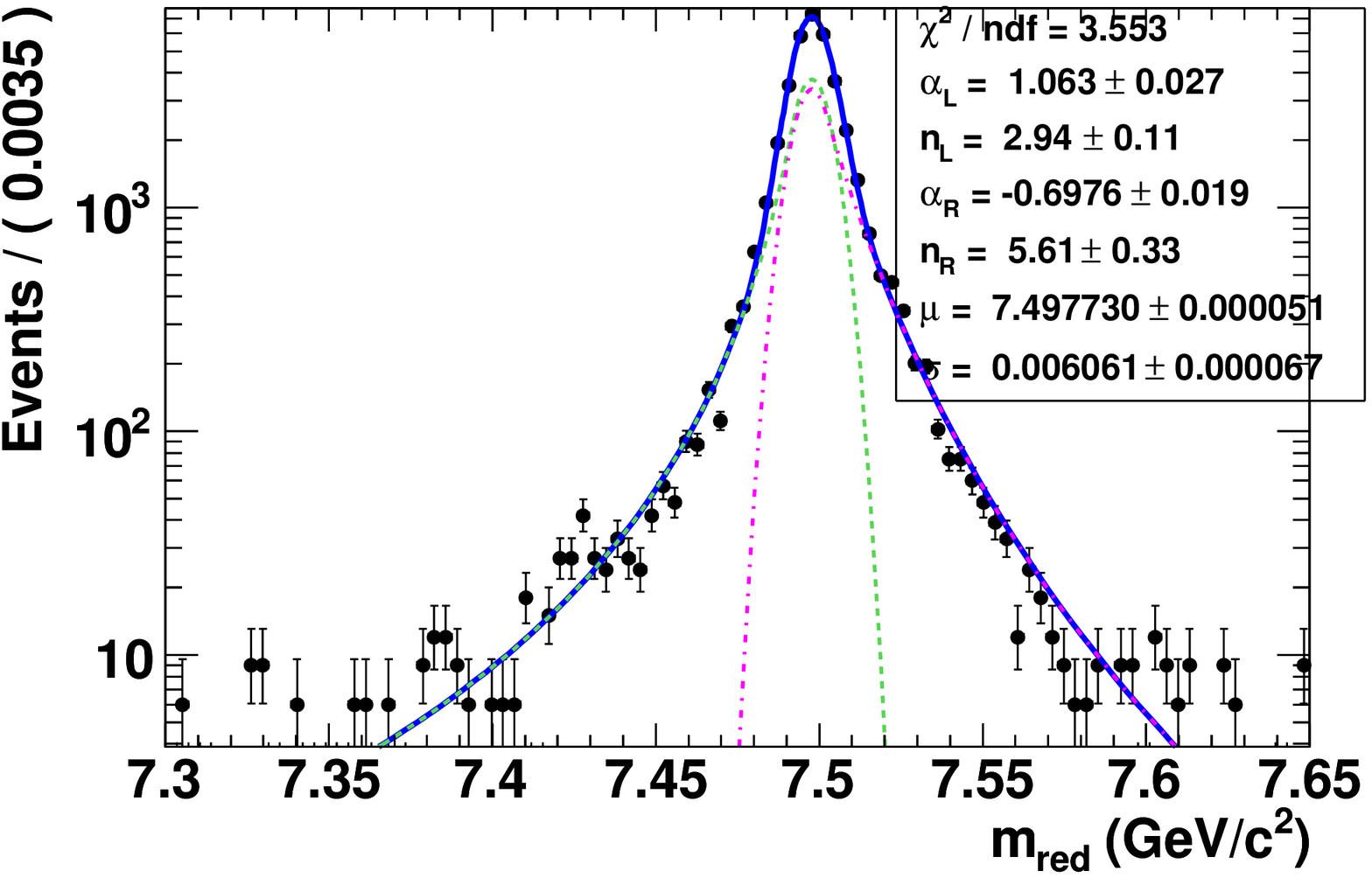}
\includegraphics[width=2.0in]{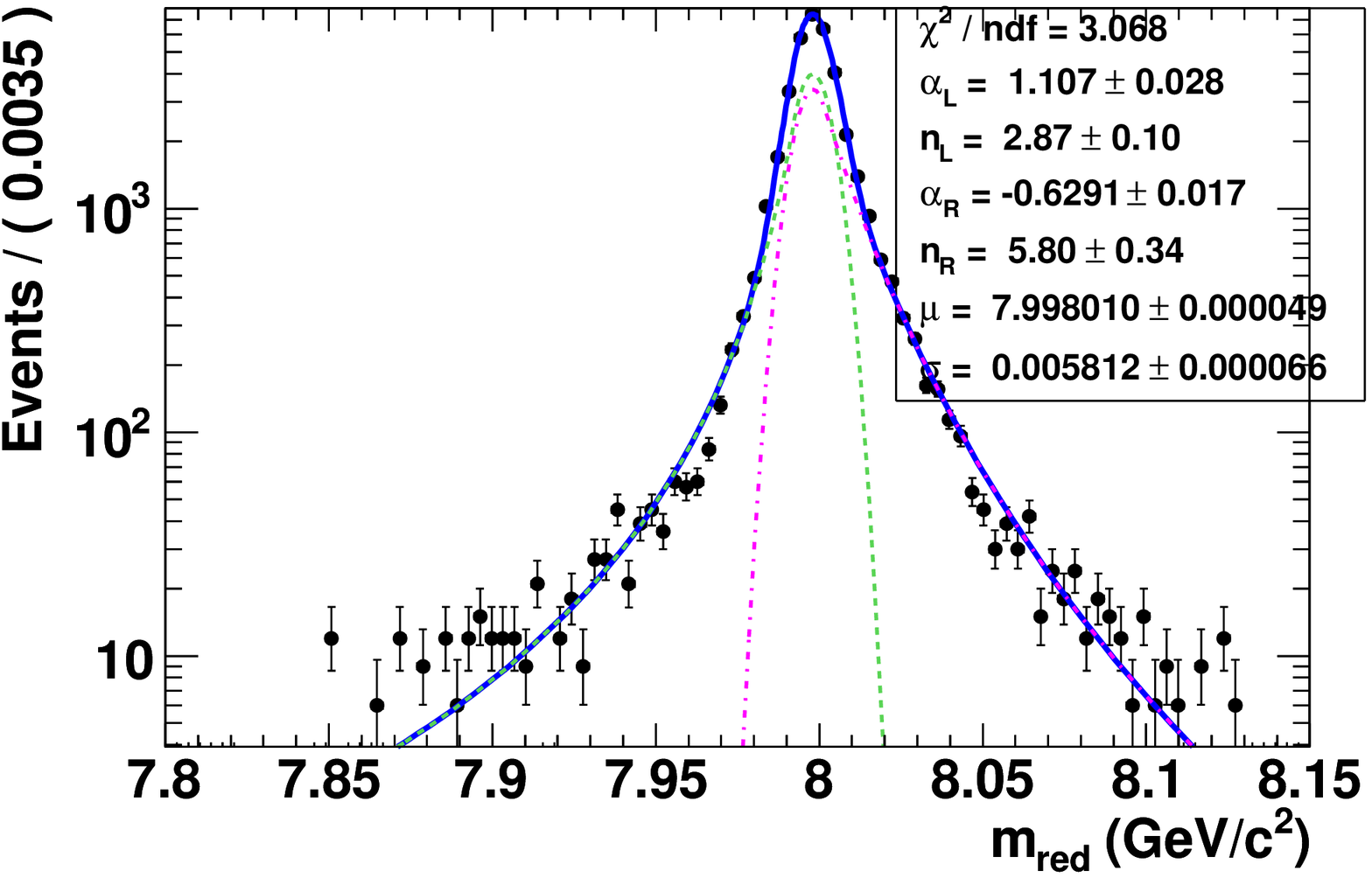}
\includegraphics[width=2.0in]{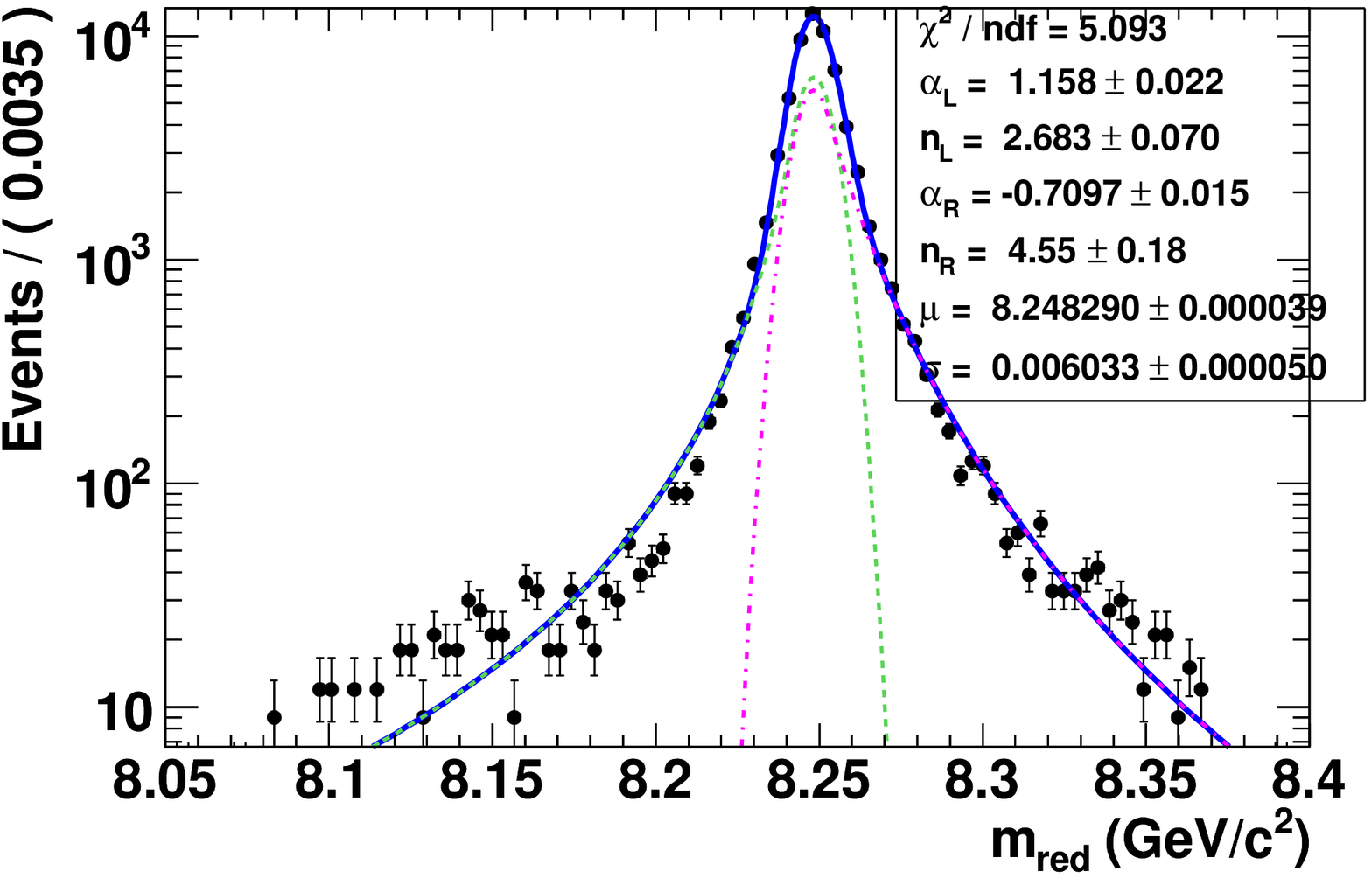}

\smallskip
\centerline{\hfill (a) \hfill \hfill (b) \hfill \hfill (c) \hfill}
\smallskip

 \includegraphics[width=2.0in]{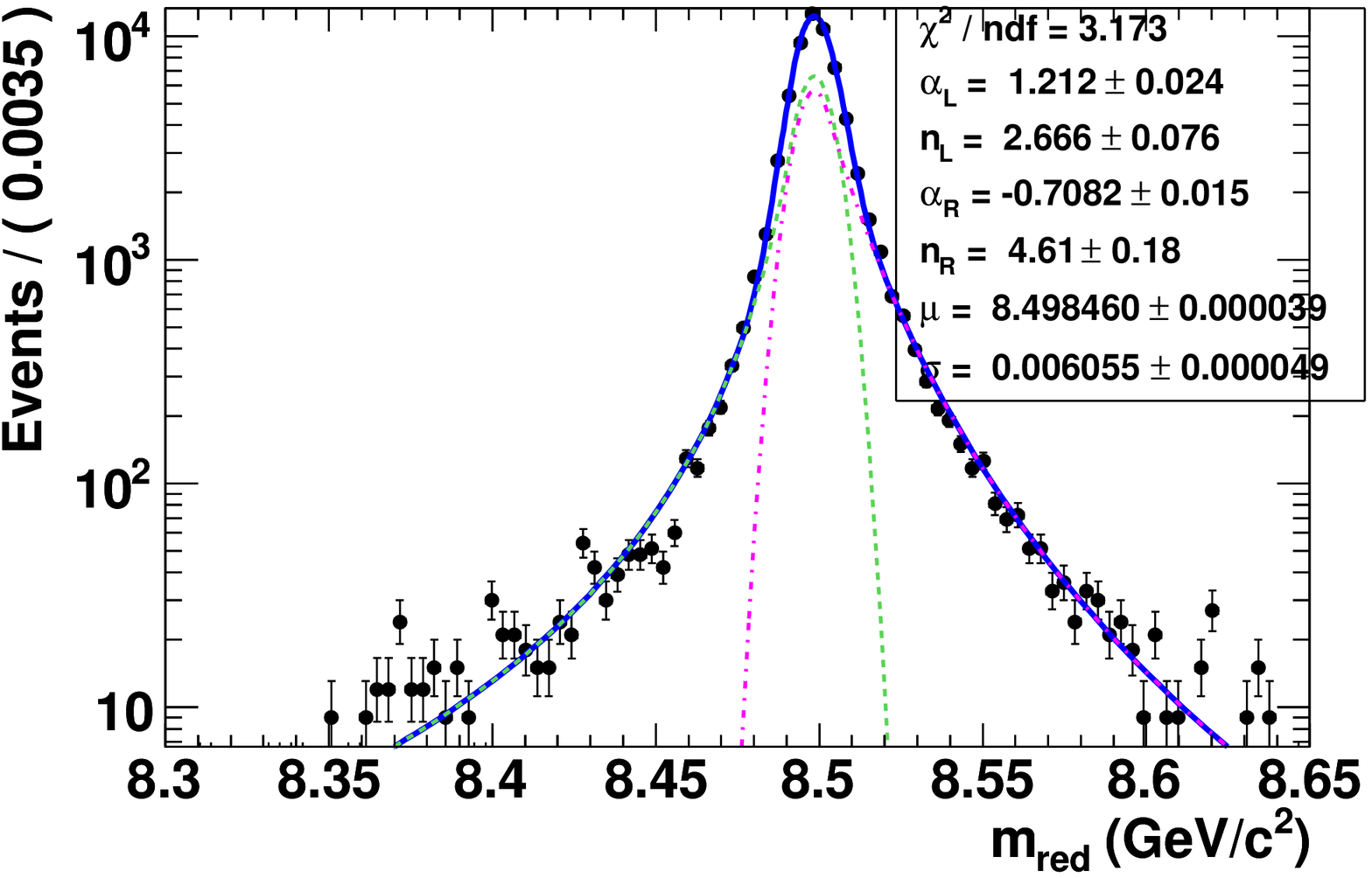}
\includegraphics[width=2.0in]{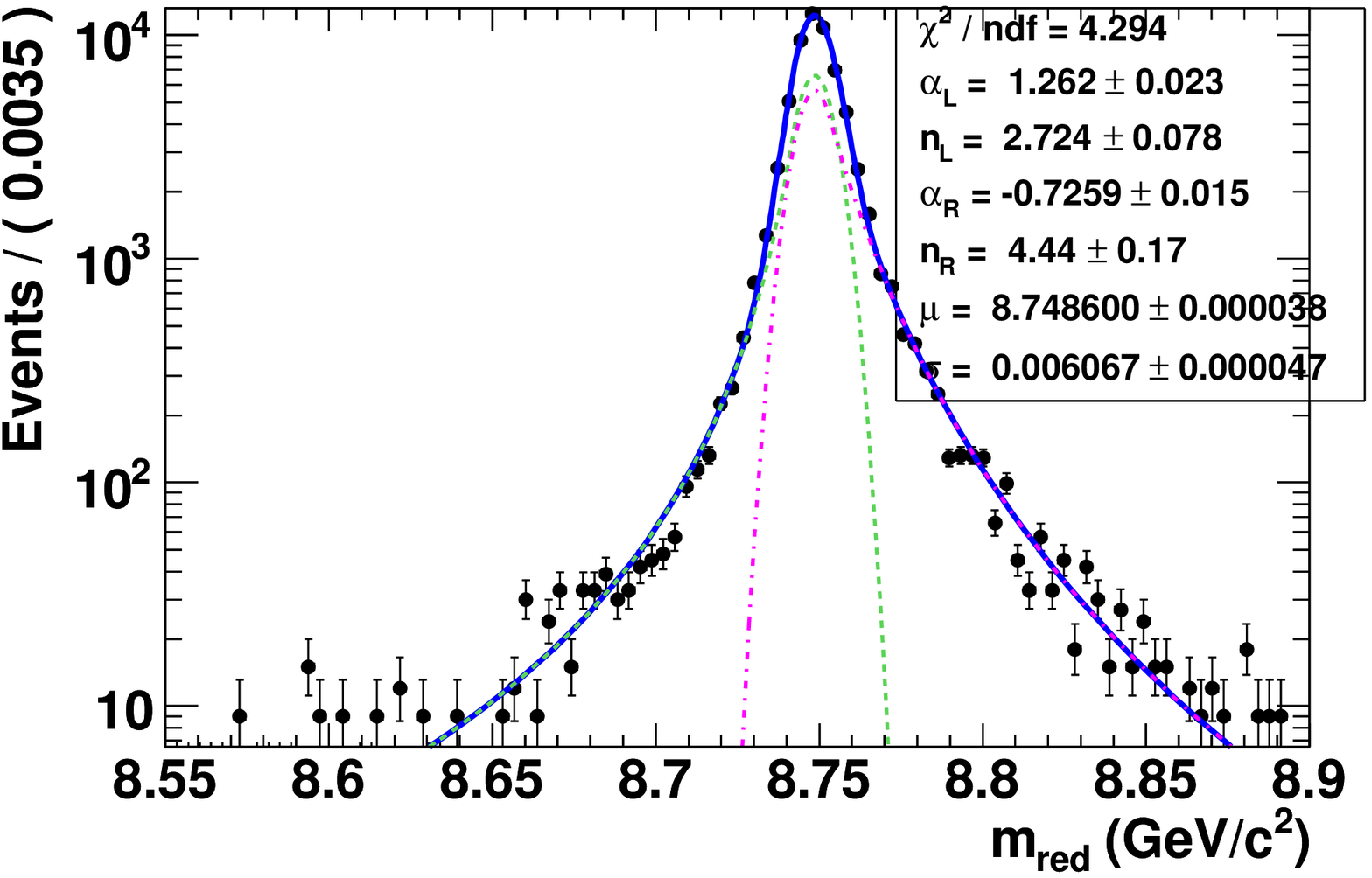}
\includegraphics[width=2.0in]{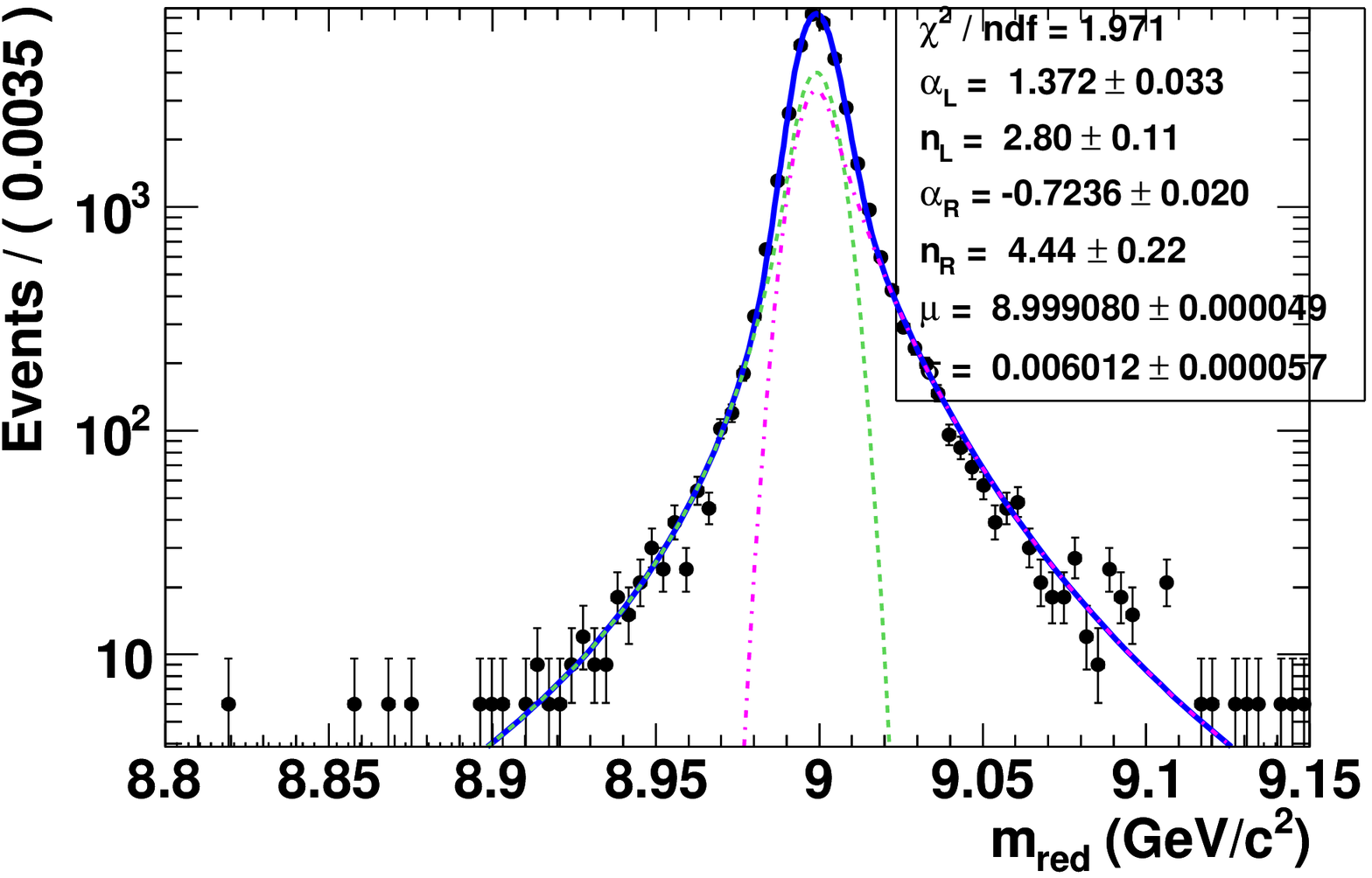}

\smallskip
\centerline{\hfill (d) \hfill \hfill (e) \hfill \hfill (f) \hfill}
\smallskip

\caption {Signal PDFs for reduced mass distribution for the Higgs mass of (a) $m_{A^0}=7.5$ GeV/$c^2$ (b) $m_{A^0}=8.0$ GeV/$c^2$ (c) $m_{A^0}=8.25$ GeV/$c^2$  (d) $m_{A^0}=8.50$ GeV/$c^2$ (e) $m_{A^0}=8.75$ GeV/$c^2$ and  (f) $m_{A^0}=9.0$ GeV/$c^2$. }

\label{fig:SigPDFY3S3}
\end{figure}
























\addtocontents{toc}{\vspace{1em}}
\addcontentsline{toc}{chapter}{Appendices}
\begin{figure}
\chapter{Background PDFs}
\label{AppendixB}
\section{Background PDFs for $\Upsilon(2S)$}
\centering
 \includegraphics[width=2.0in]{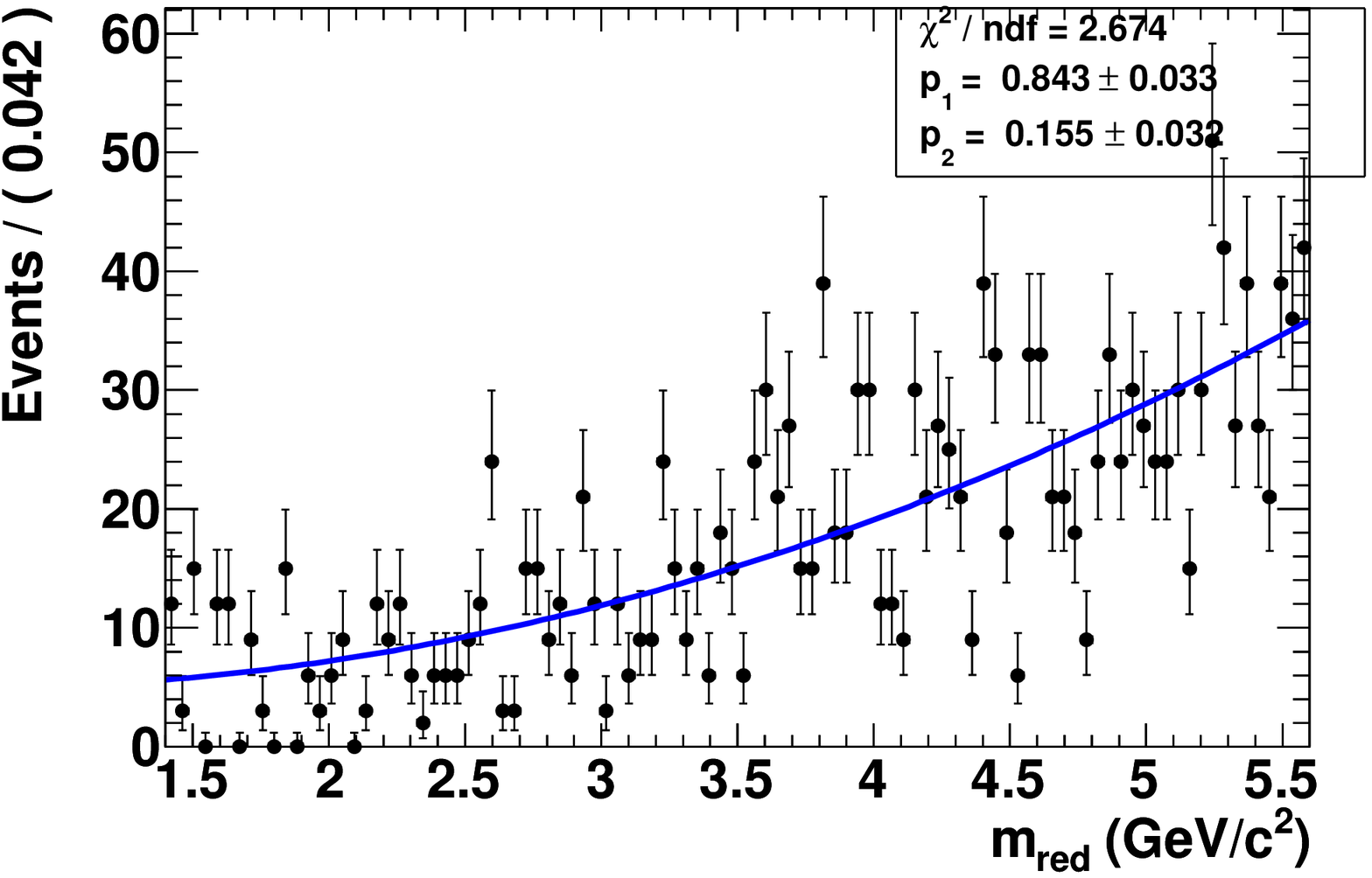}
\includegraphics[width=2.0in]{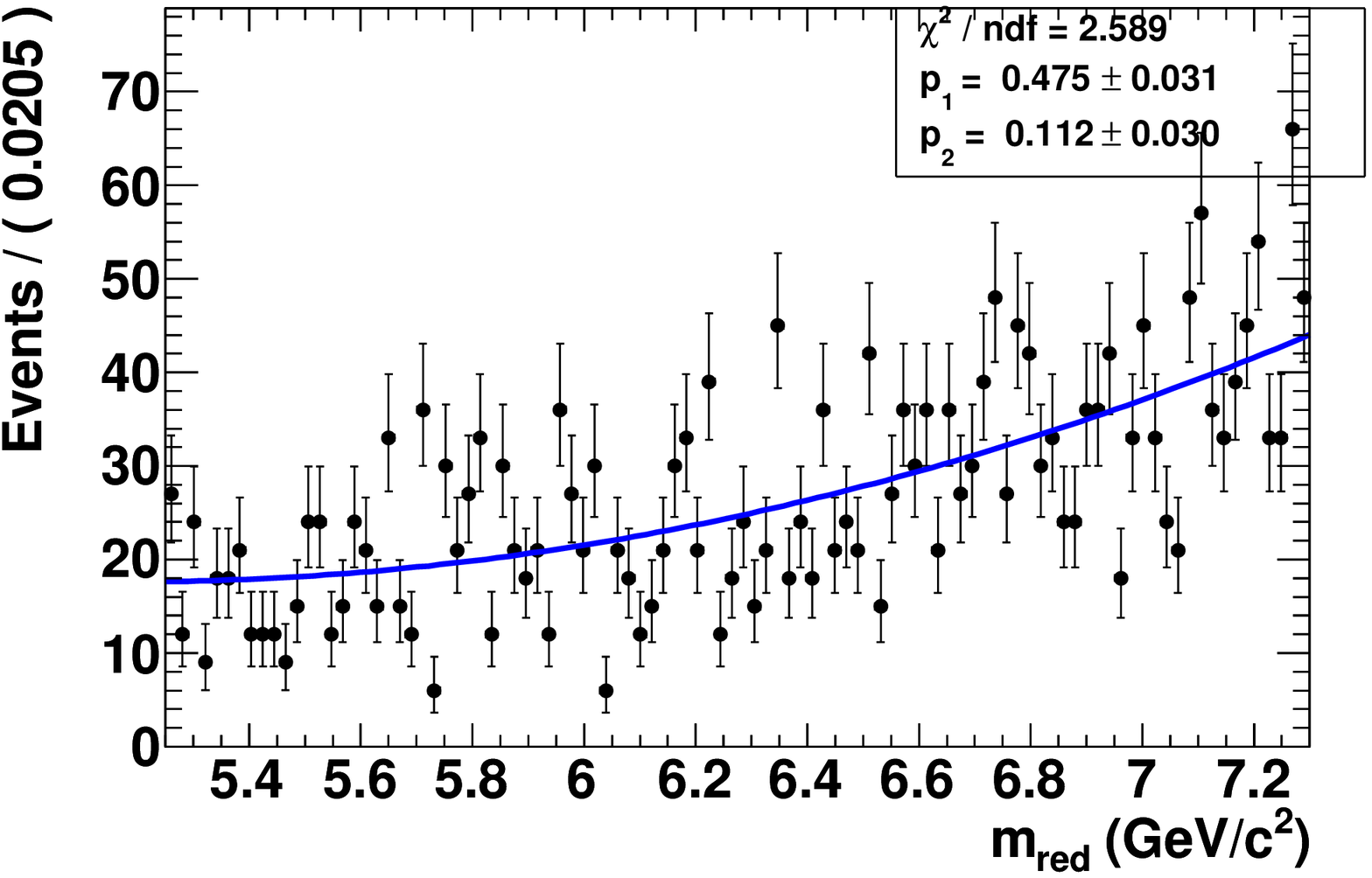}
 \includegraphics[width=2.0in]{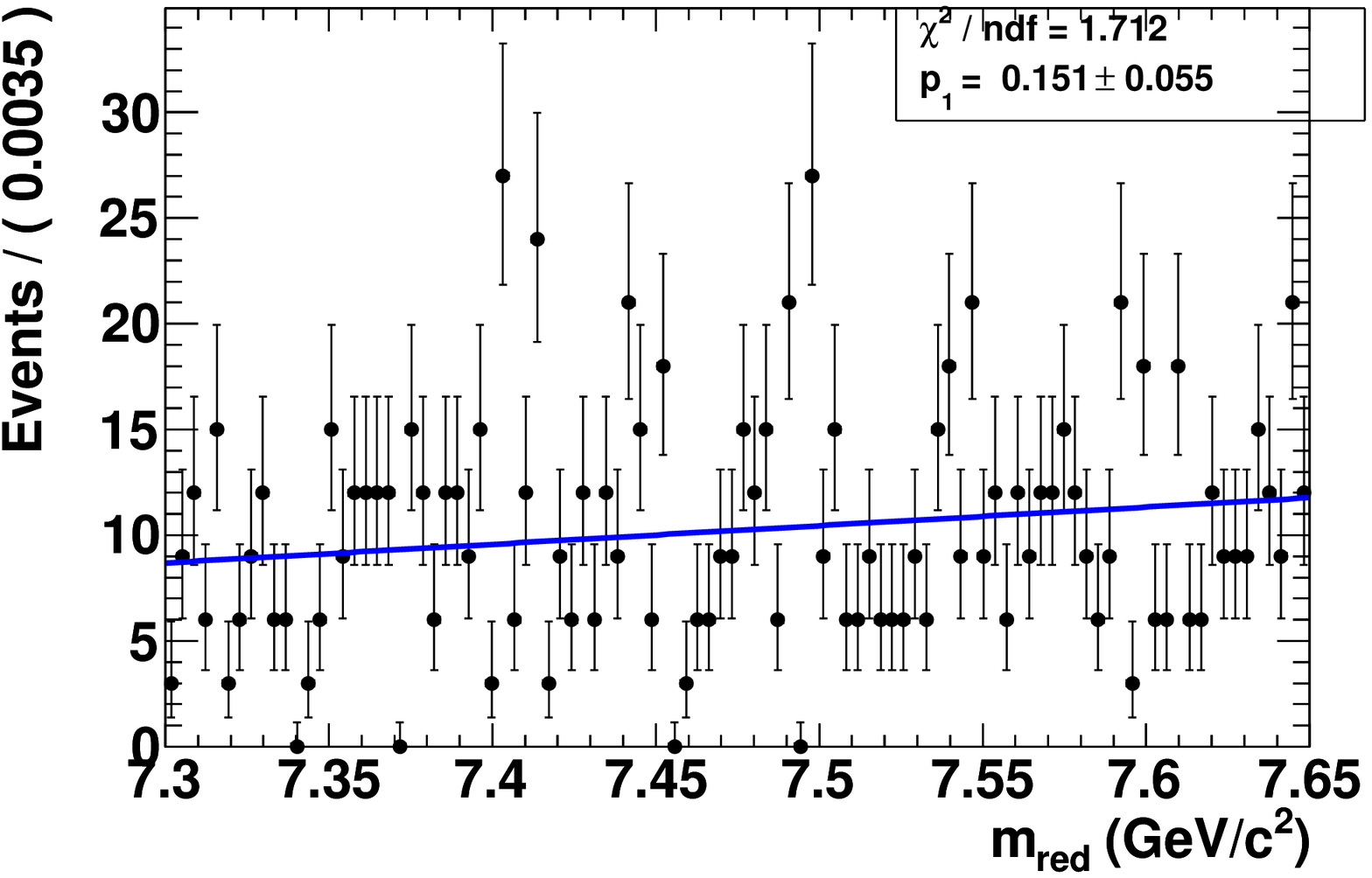}

\smallskip
\centerline{\hfill (a) \hfill \hfill (b) \hfill \hfill (c) \hfill}
\smallskip

\includegraphics[width=2.0in]{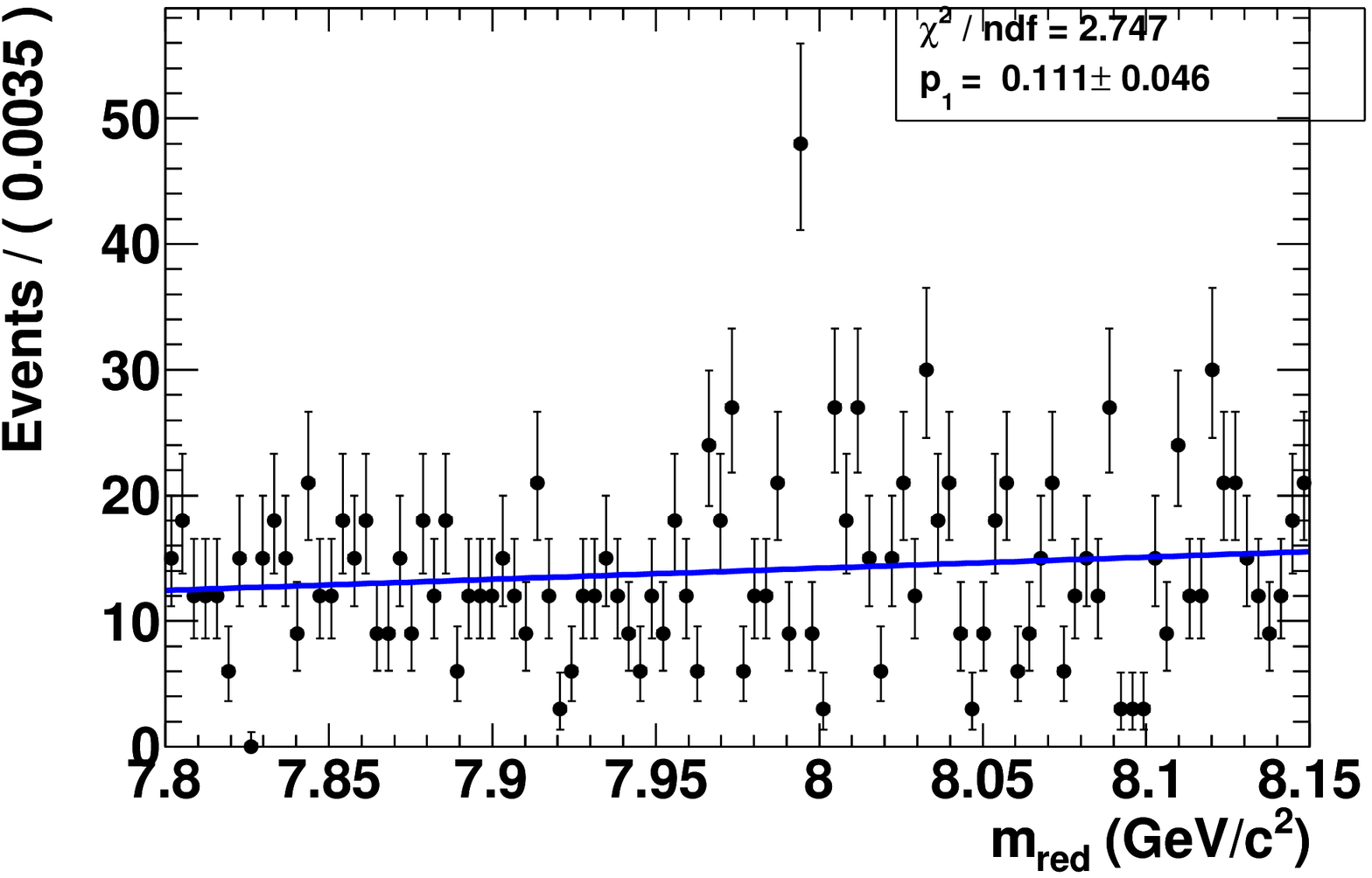}
\includegraphics[width=2.0in]{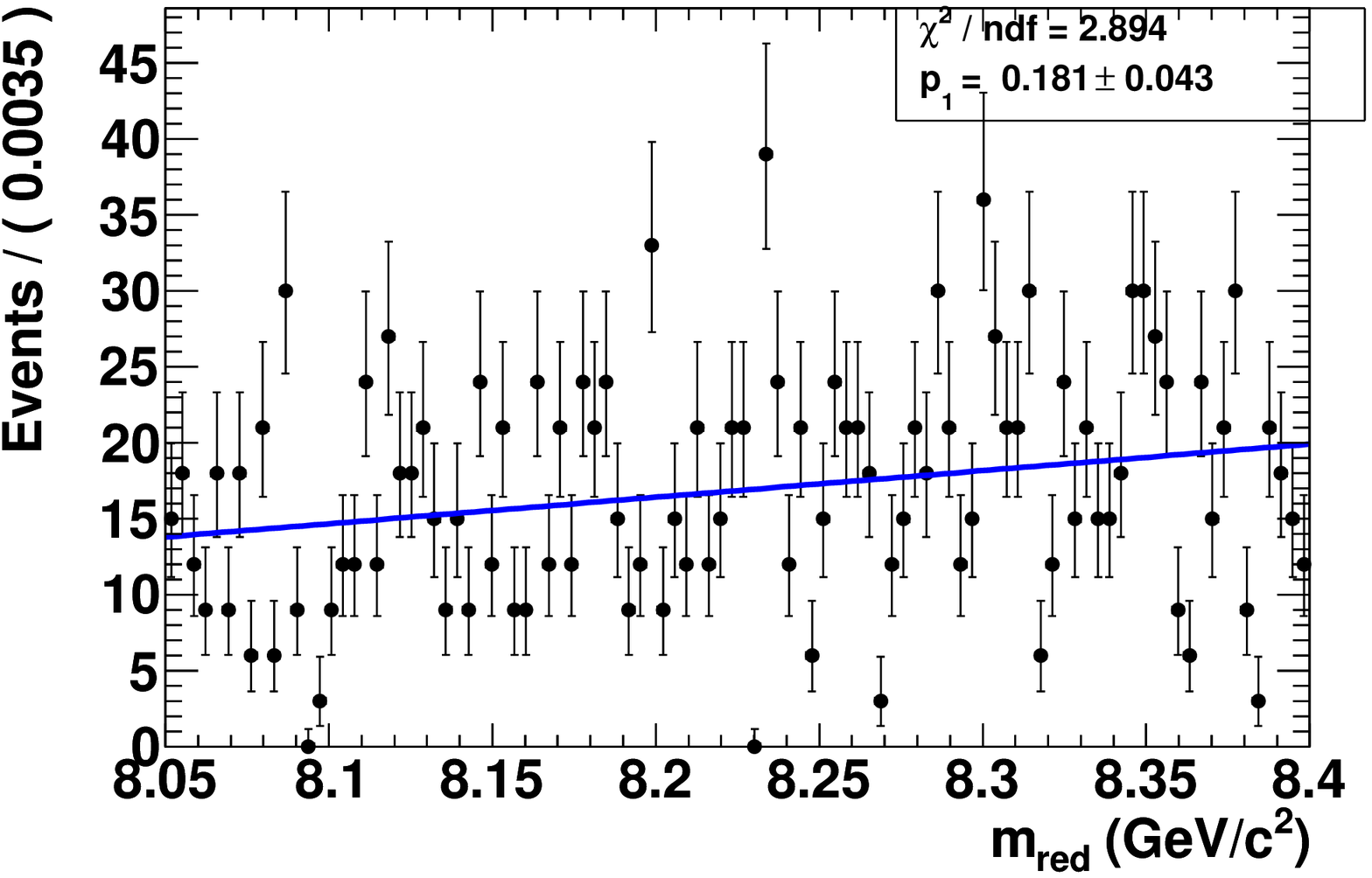}
 \includegraphics[width=2.0in]{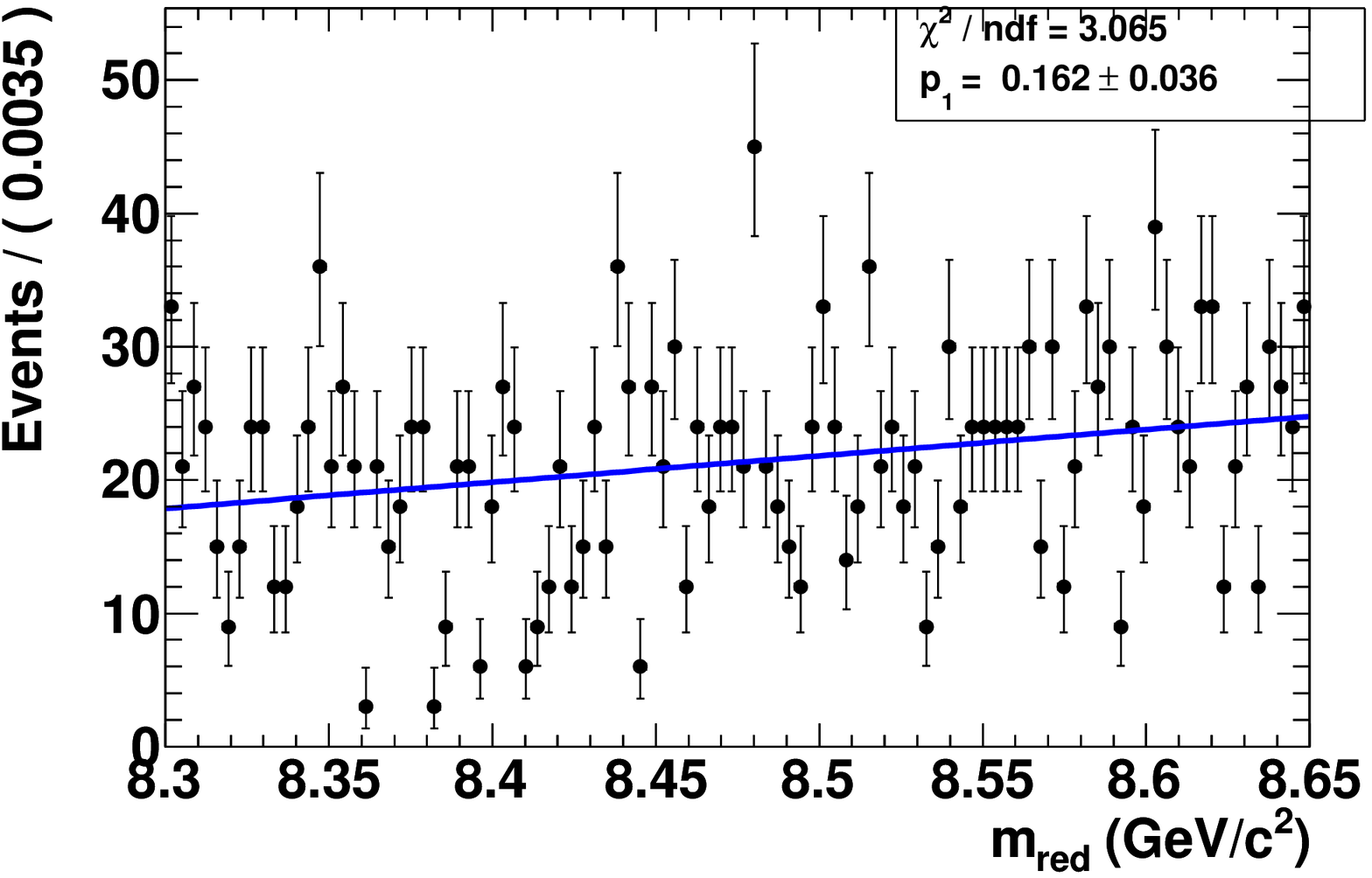}

\smallskip
\centerline{\hfill (d) \hfill \hfill (e) \hfill \hfill (f) \hfill}
\smallskip

\includegraphics[width=2.0in]{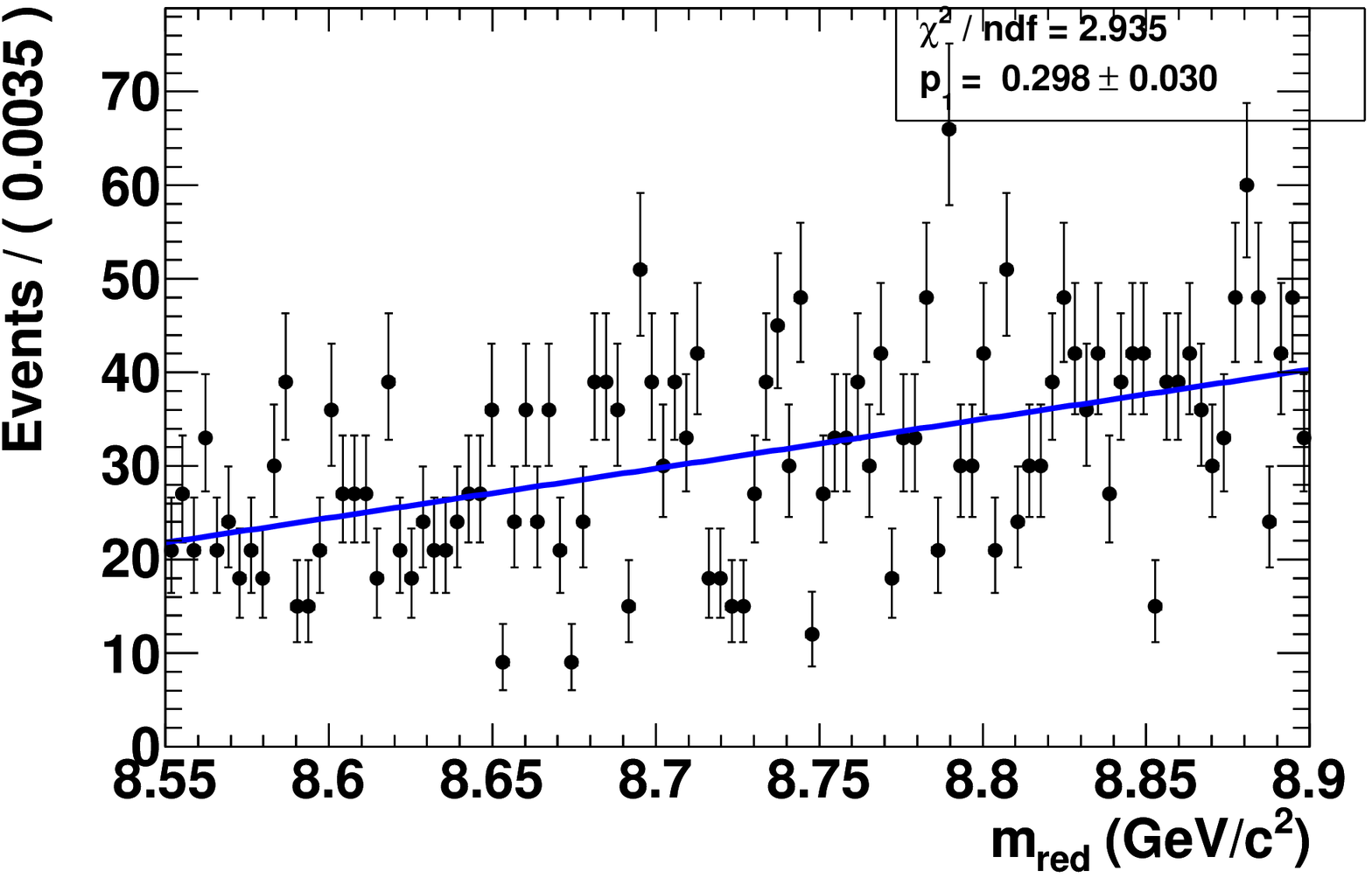}
\includegraphics[width=2.0in]{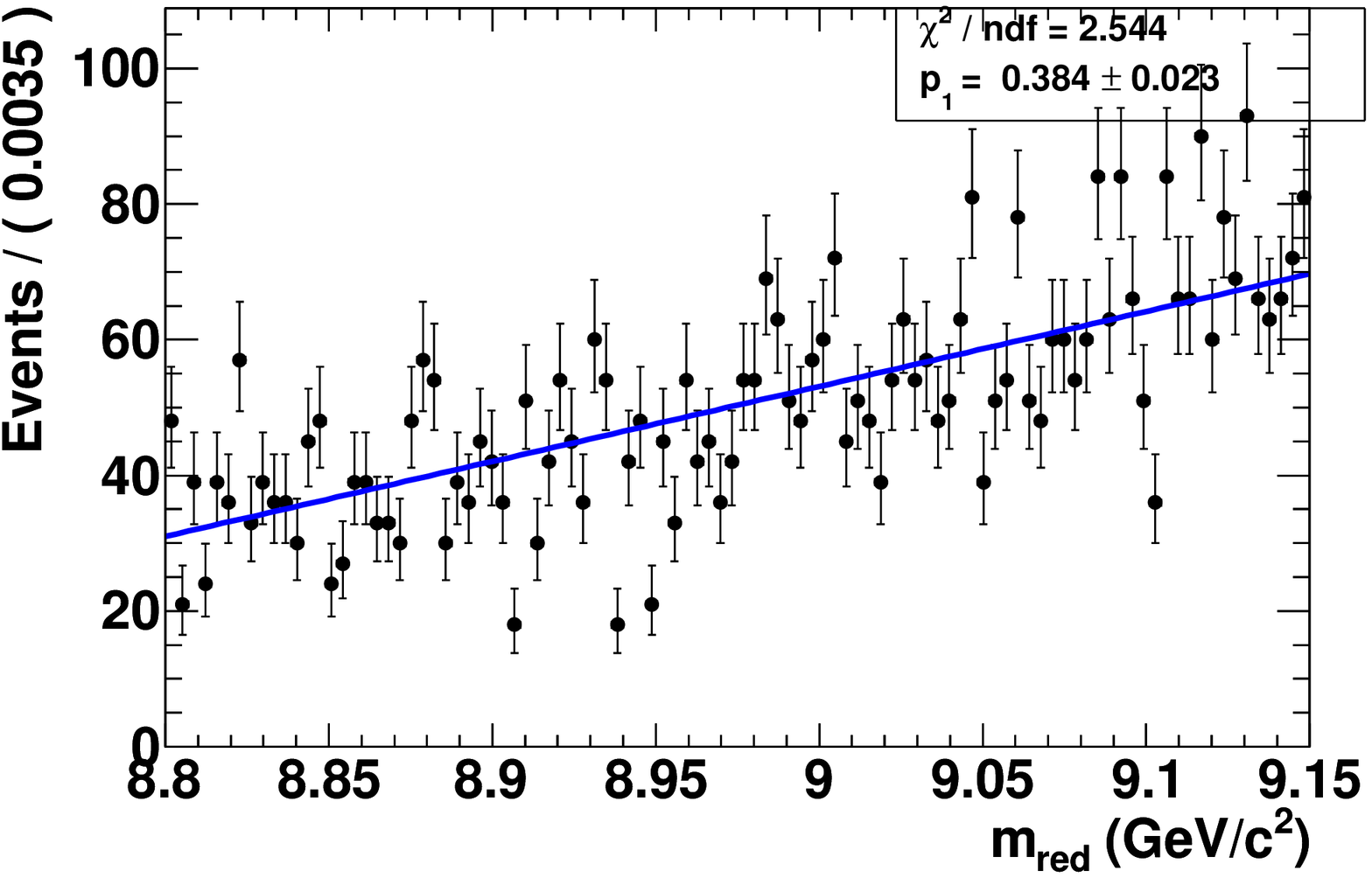}
 \includegraphics[width=2.0in]{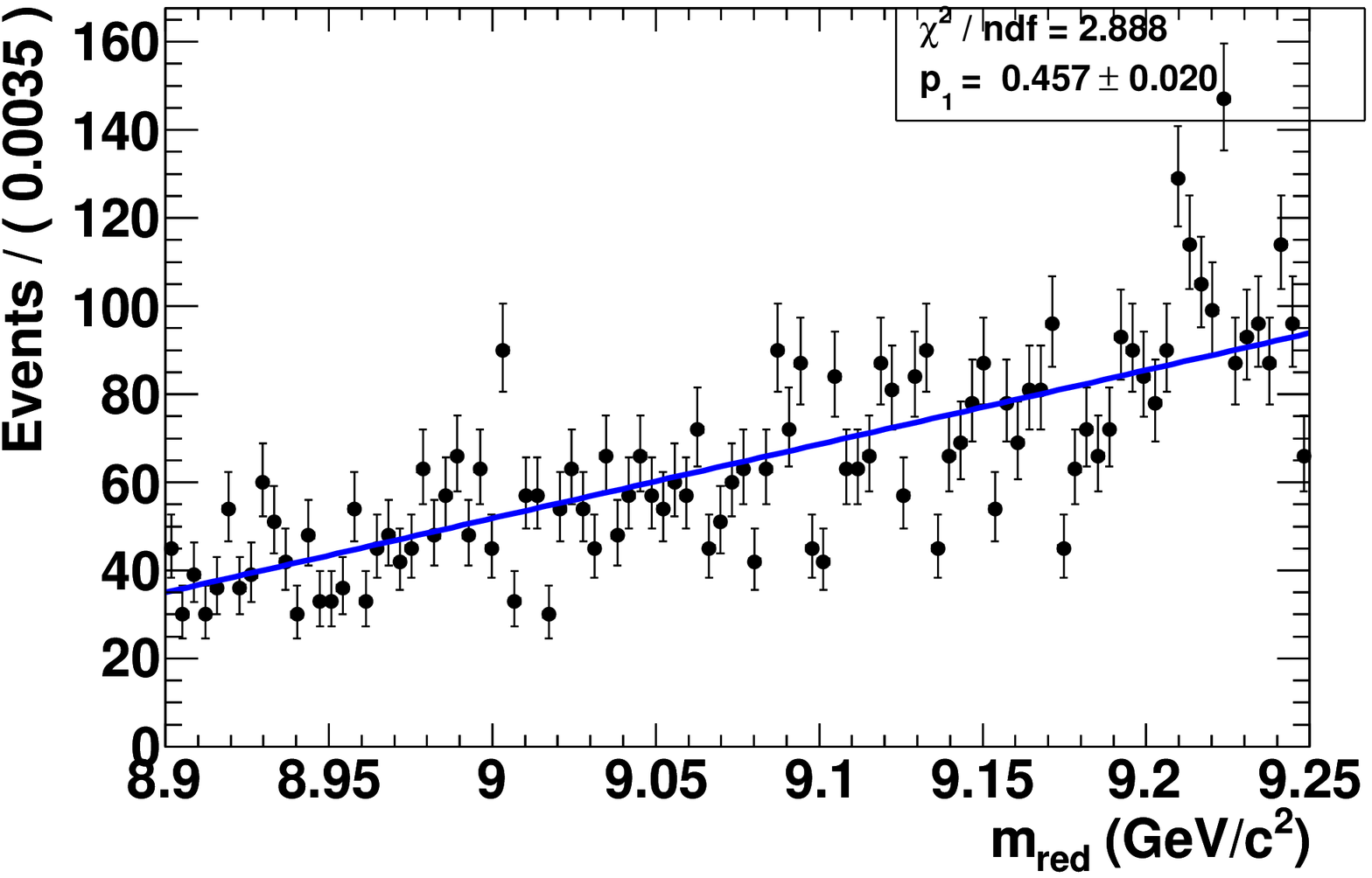}

\smallskip
\centerline{\hfill (g) \hfill \hfill (h) \hfill \hfill (i) \hfill}
\smallskip

\caption {Background PDFs for $m_{\rm red}$ distribution for the Higgs mass of (a) $1.502 \le m_{A^0} \le 5.50$ GeV/$c^2$ (b) $5.25 \le m_{A^0} \le 7.3 $ GeV/$c^2$  (c) $m_{A^0}=7.5$ GeV/$c^2$  (d) $m_{A^0}=8.0$ GeV/$c^2$ (e) $m_{A^0}=8.25$ GeV/$c^2$ (f) $m_{A^0}=8.5$ GeV/$c^2$ (g) $m_{A^0}=8.75$ GeV/$c^2$ (h) $m_{A^0}=9.0$ GeV/$c^2$ and (i) $m_{A^0}=9.1$ GeV/$c^2$.}

\label{fig:BackPDFY2S}
\end{figure}

\begin{figure}
\section{Background PDFs for $\Upsilon(3S)$}
\centering
 \includegraphics[width=2.0in]{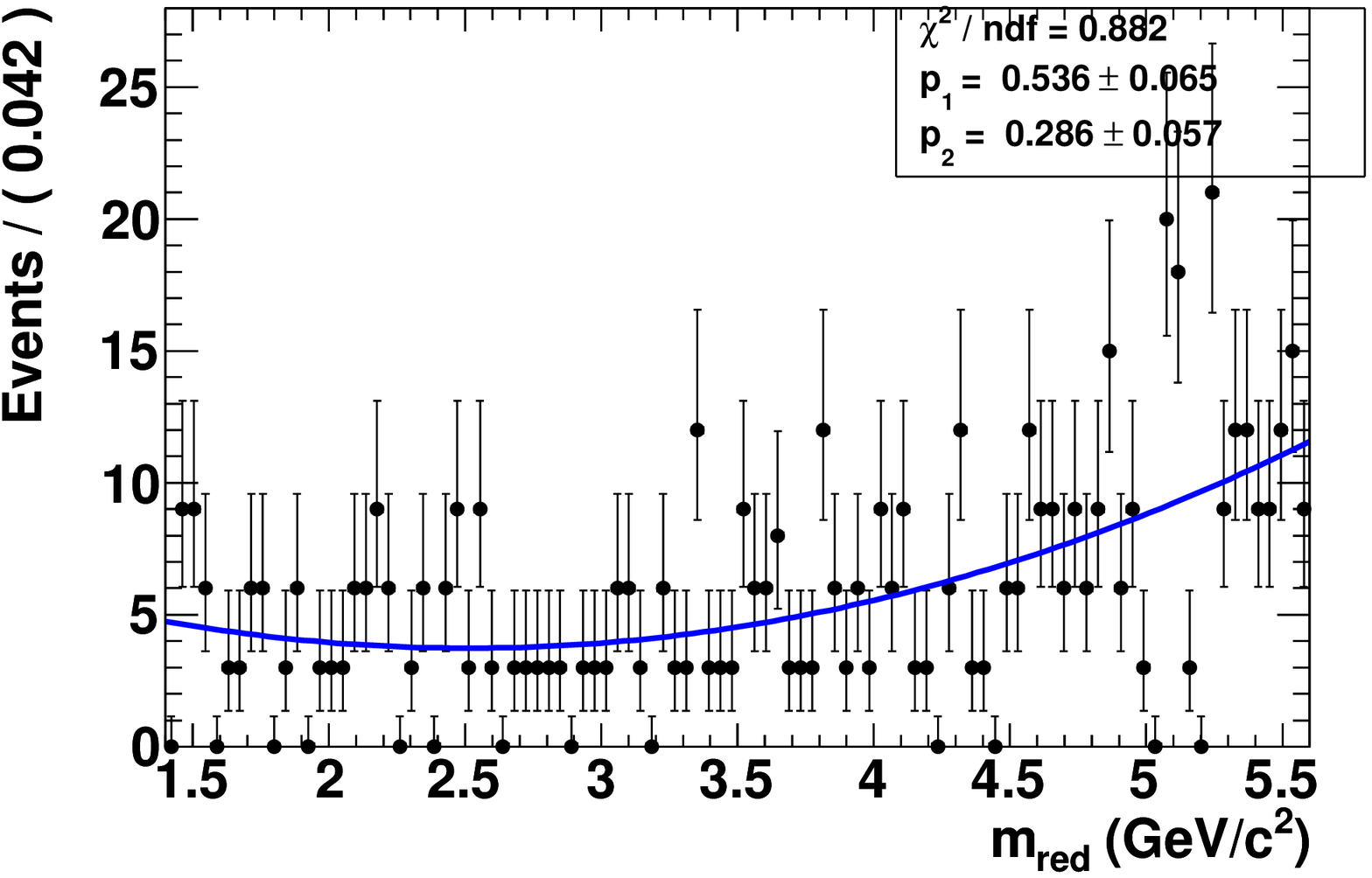}
\includegraphics[width=2.0in]{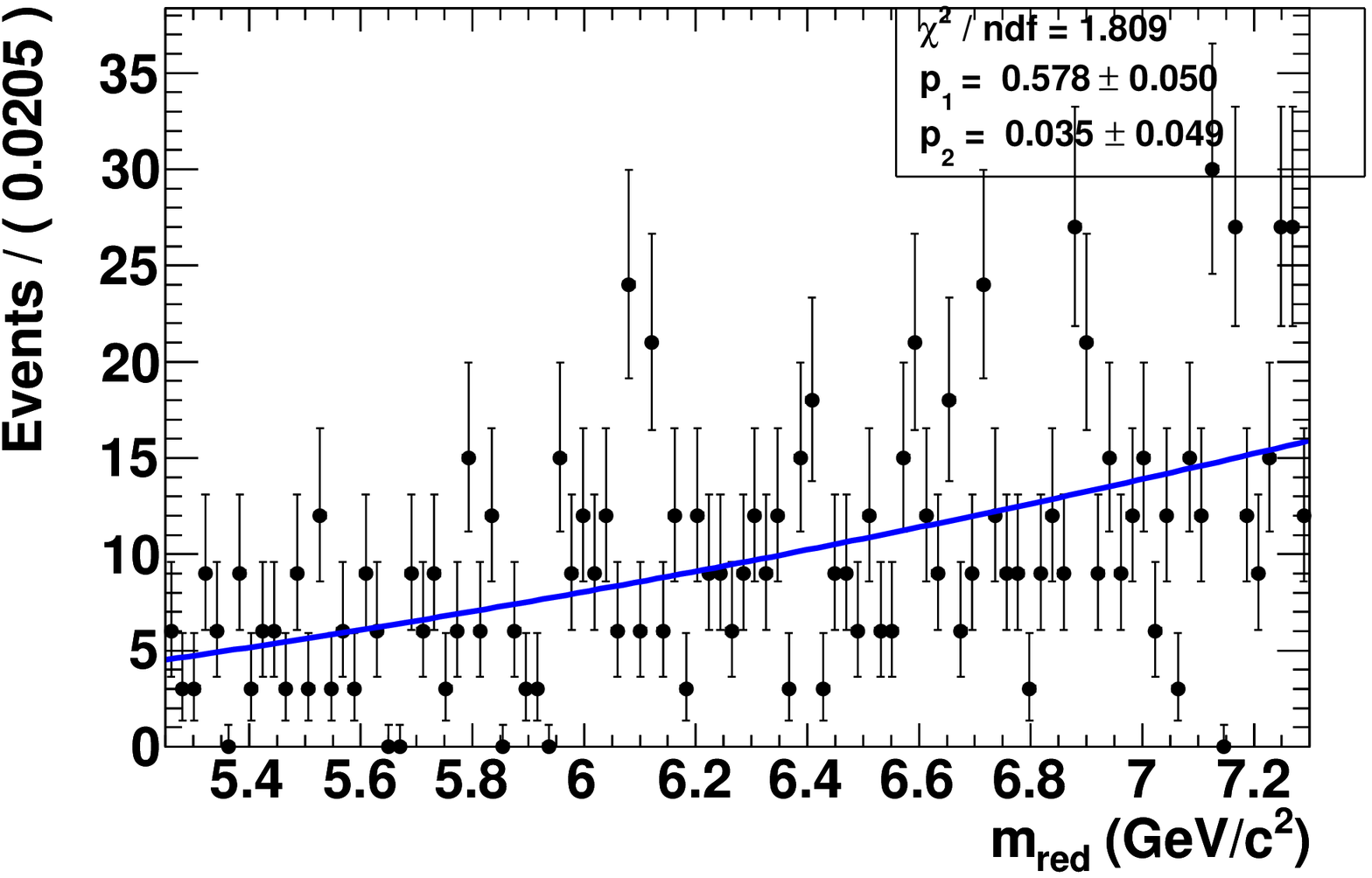}
 \includegraphics[width=2.0in]{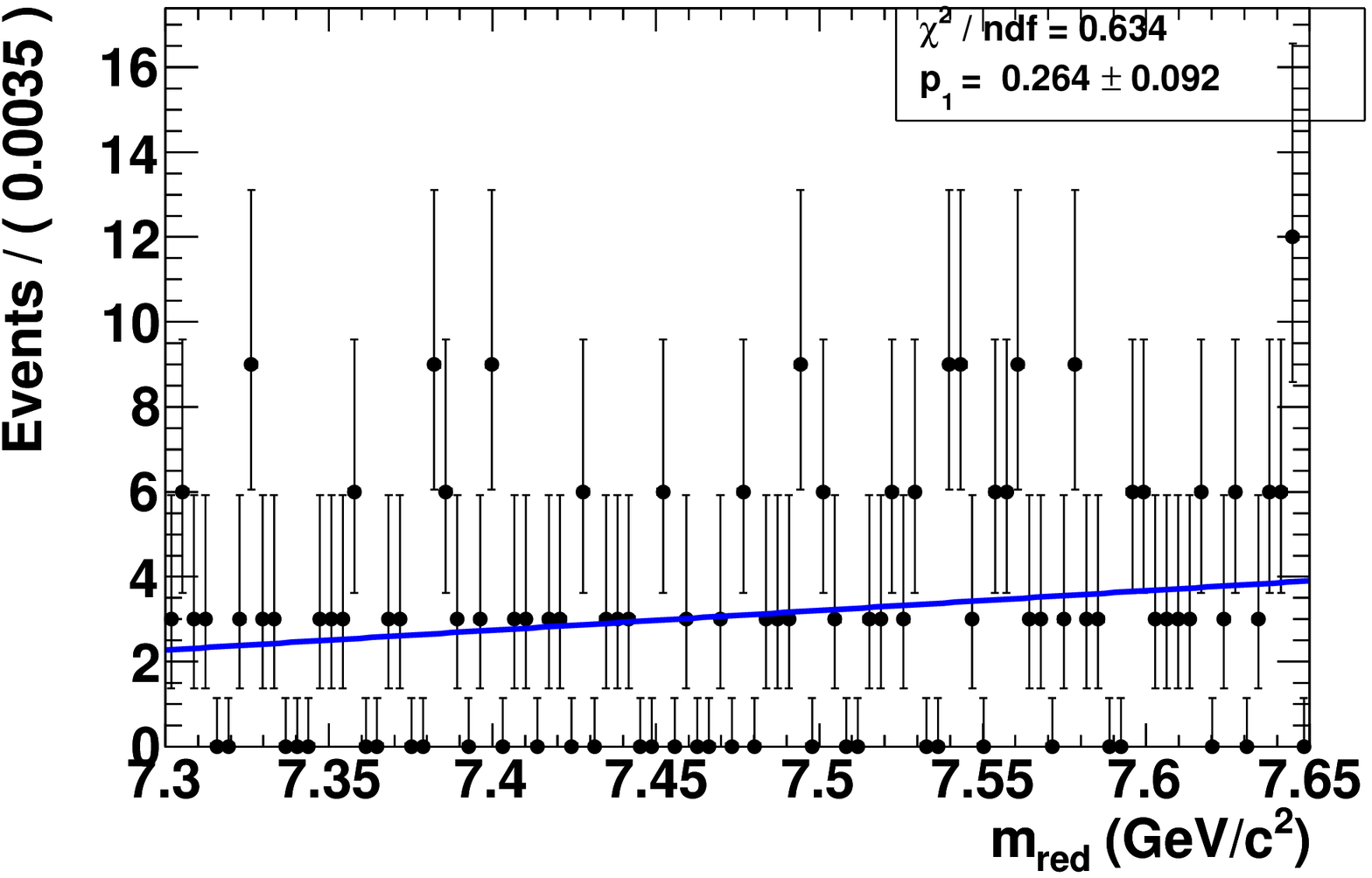}

\smallskip
\centerline{\hfill (a) \hfill \hfill (b) \hfill \hfill (c) \hfill}
\smallskip

\includegraphics[width=2.0in]{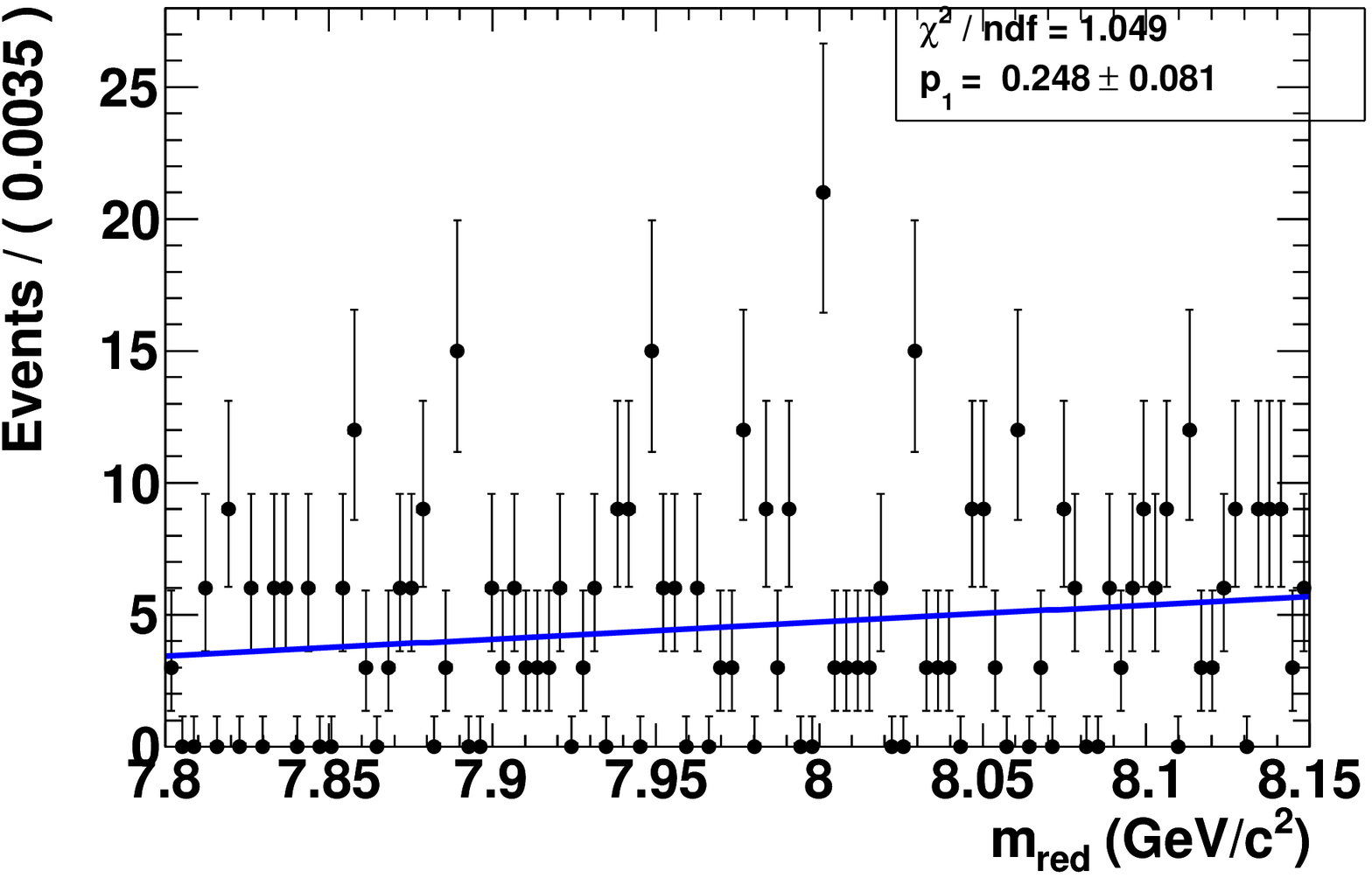}
\includegraphics[width=2.0in]{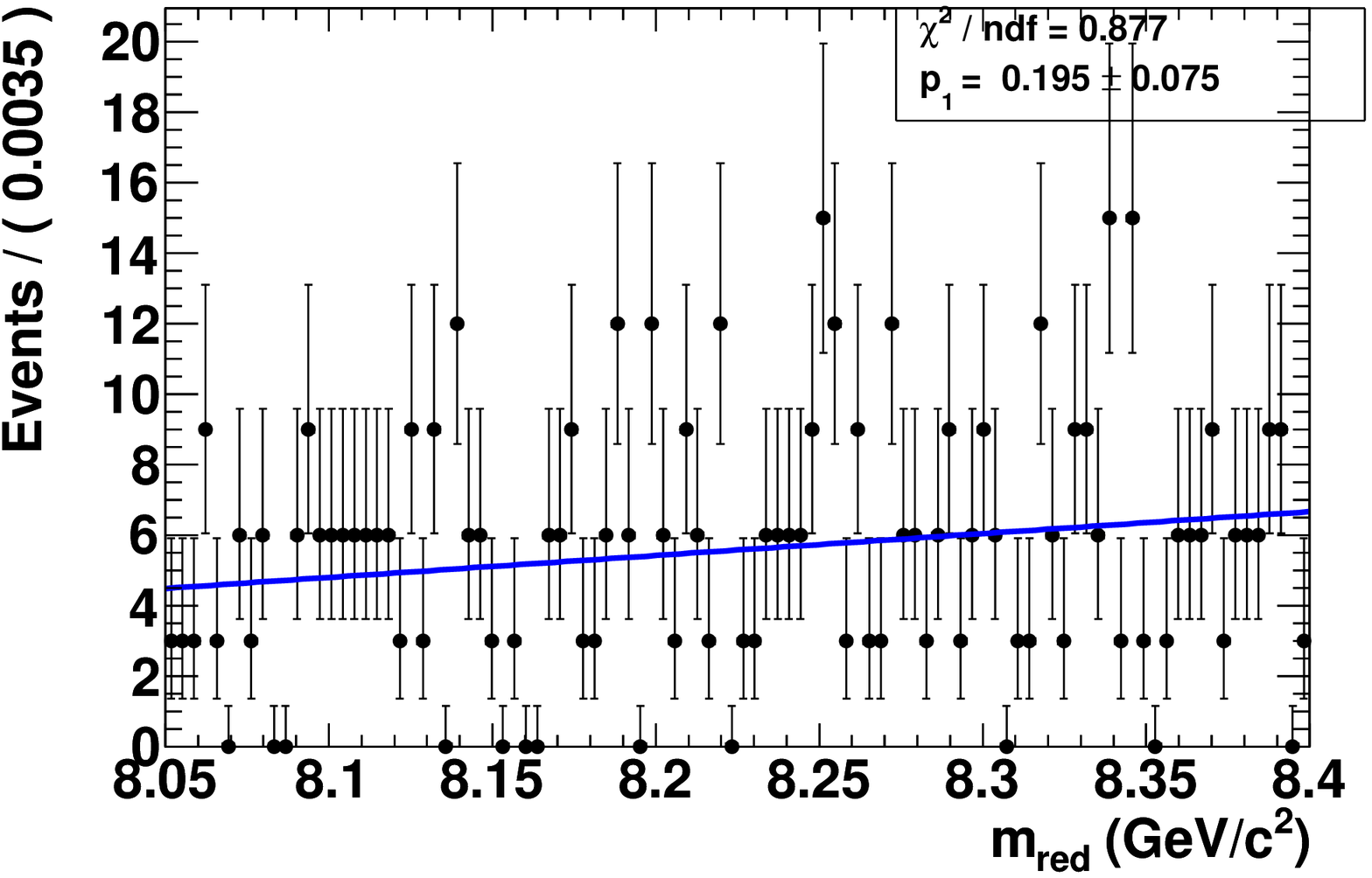}
 \includegraphics[width=2.0in]{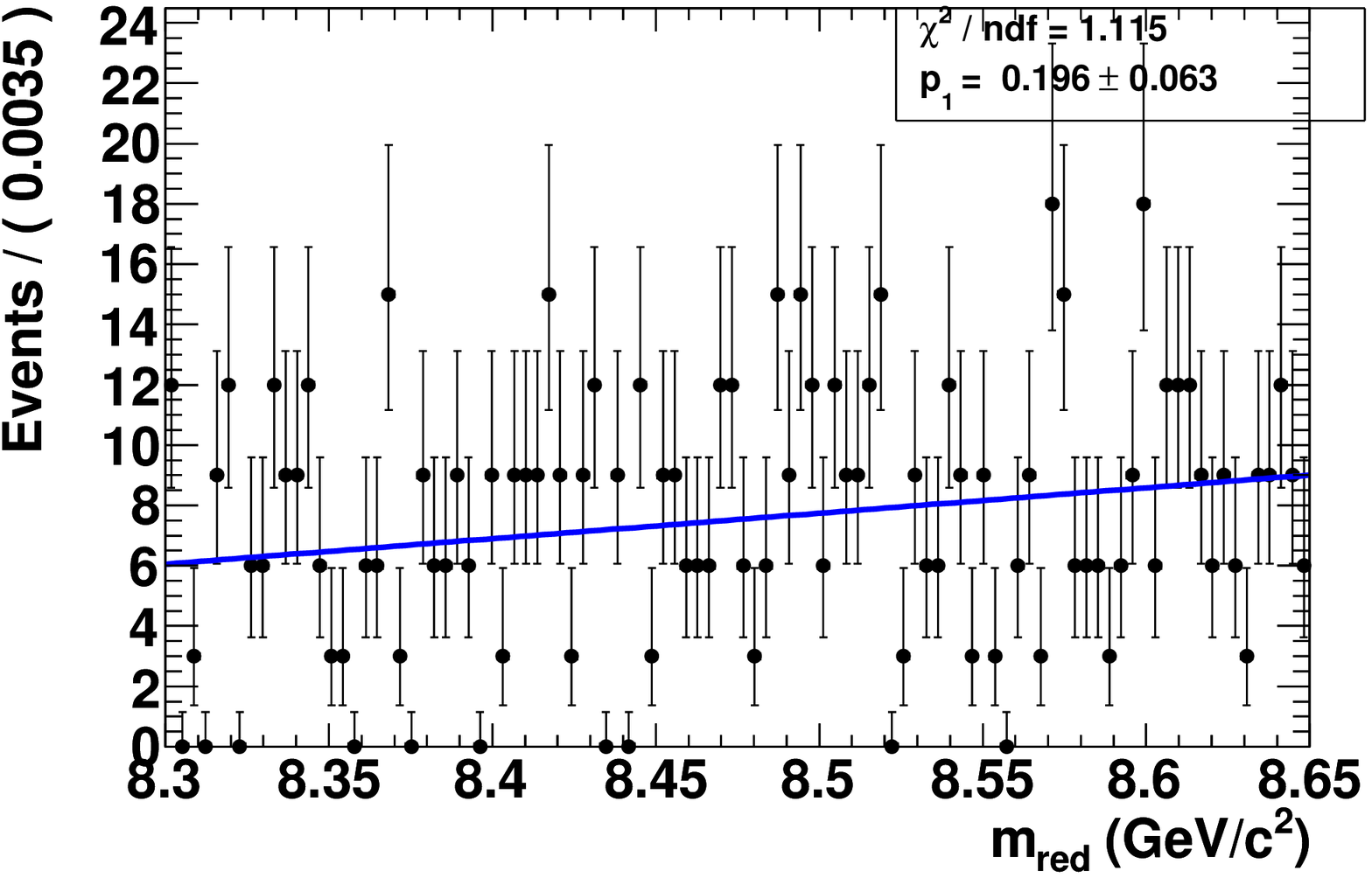}

\smallskip
\centerline{\hfill (d) \hfill \hfill (e) \hfill \hfill (f) \hfill}
\smallskip

\includegraphics[width=2.0in]{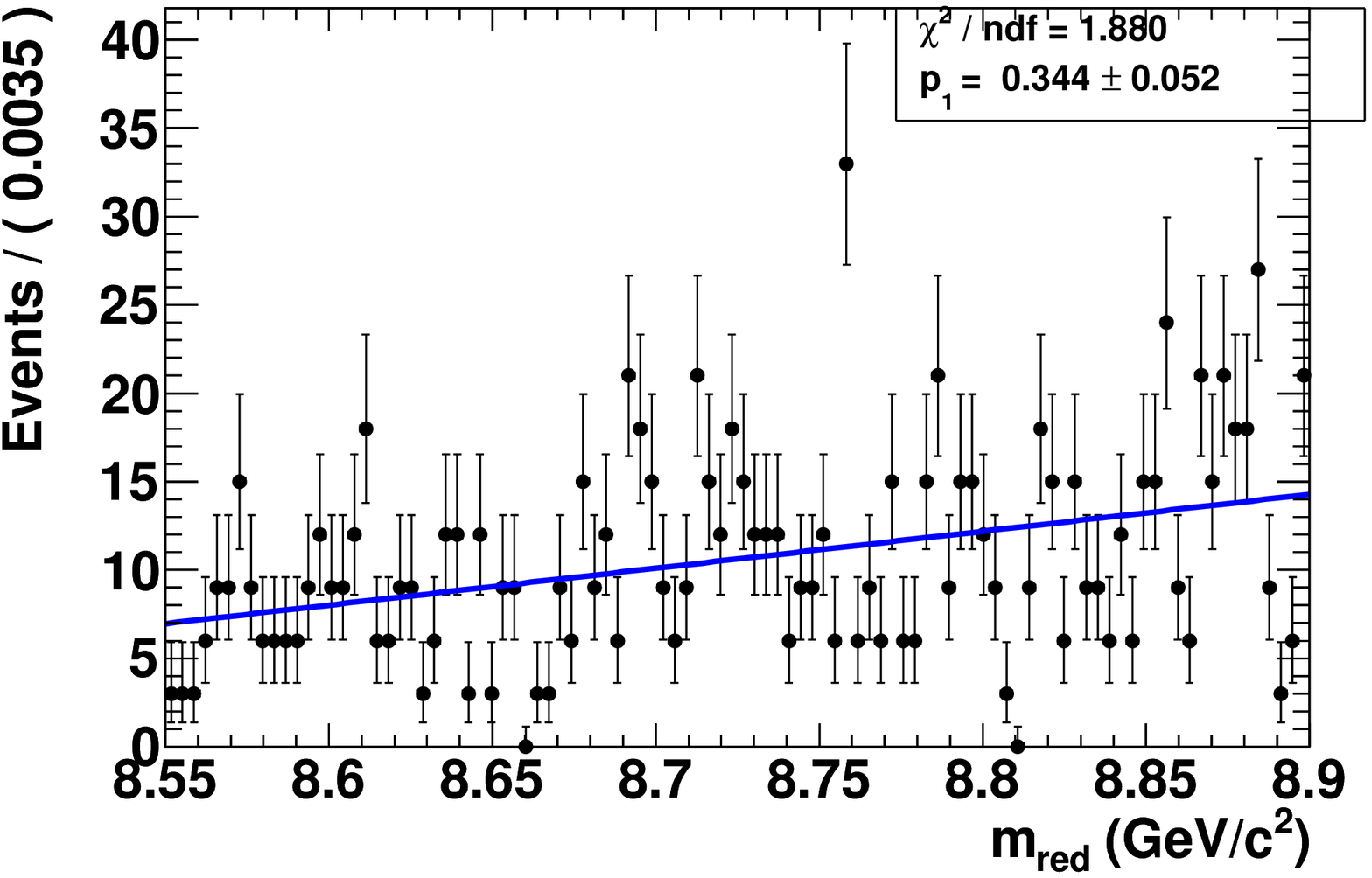}
\includegraphics[width=2.0in]{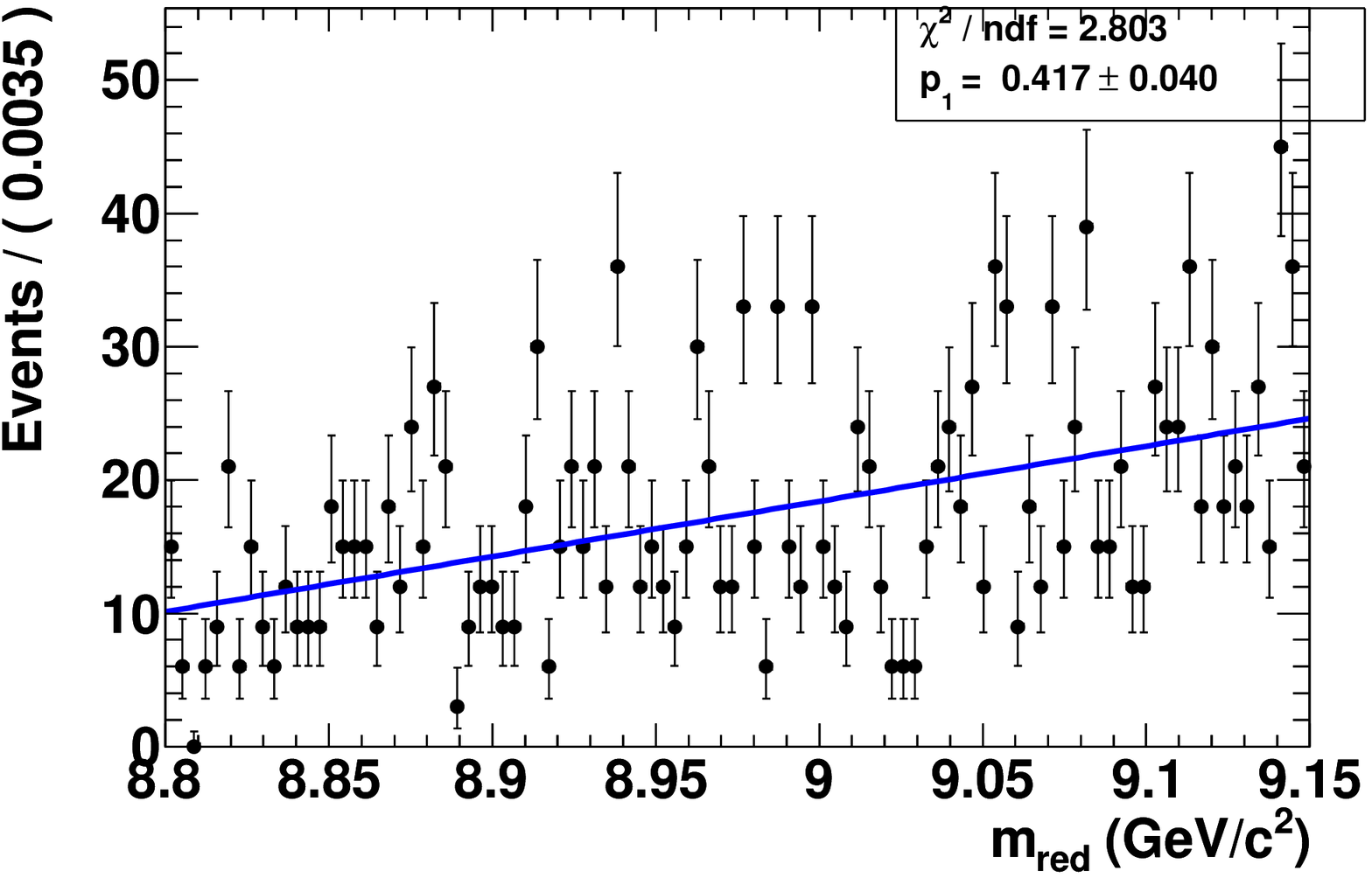}
 
\smallskip
\centerline{\hfill (g) \hfill \hfill (h) \hfill}
\smallskip

\caption {Background PDFs for $m_{\rm red}$ distribution for the Higgs mass of (a) $1.502 \le m_{A^0} \le 5.50$ GeV/$c^2$ (b) $5.25 \le m_{A^0} \le 7.3 $ GeV/$c^2$  (c) $m_{A^0}=7.5$ GeV/$c^2$  (d) $m_{A^0}=8.0$ GeV/$c^2$ (e) $m_{A^0}=8.25$ GeV/$c^2$ (f) $m_{A^0}=8.5$ GeV/$c^2$ (g) $m_{A^0}=8.75$ GeV/$c^2$ and (h) $m_{A^0}=9.0$ GeV/$c^2$.}

\label{fig:BackPDFY3S}
\end{figure}






















\addtocontents{toc}{\vspace{1em}}
\addcontentsline{toc}{chapter}{Appendices}
\begin{figure}
\chapter{Toy Monte Carlo Results}
\label{AppendixC}
\section{For $\Upsilon(2S)$}
\centering
 \includegraphics[width=2.0in]{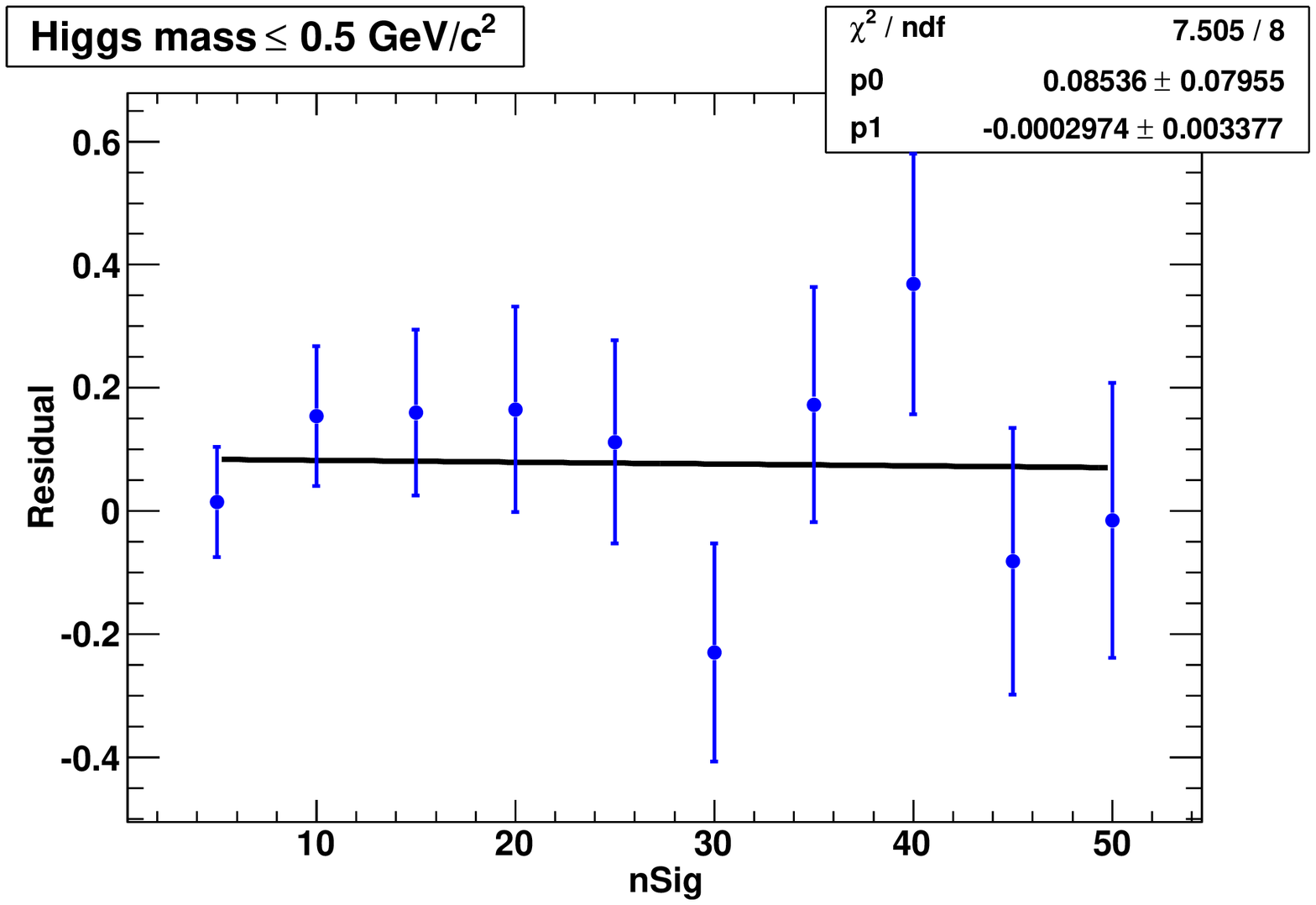}
\includegraphics[width=2.0in]{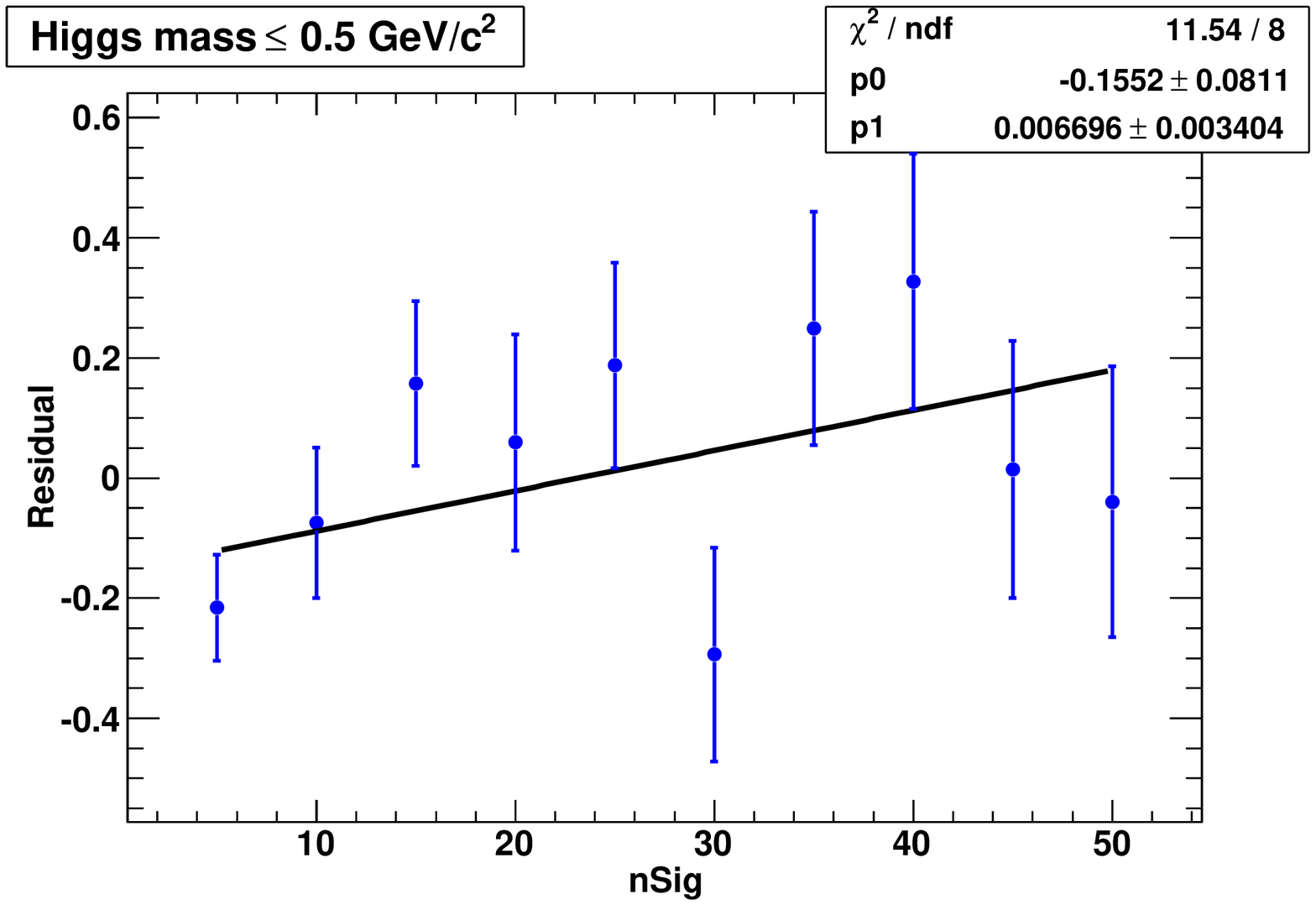}
 \includegraphics[width=2.0in]{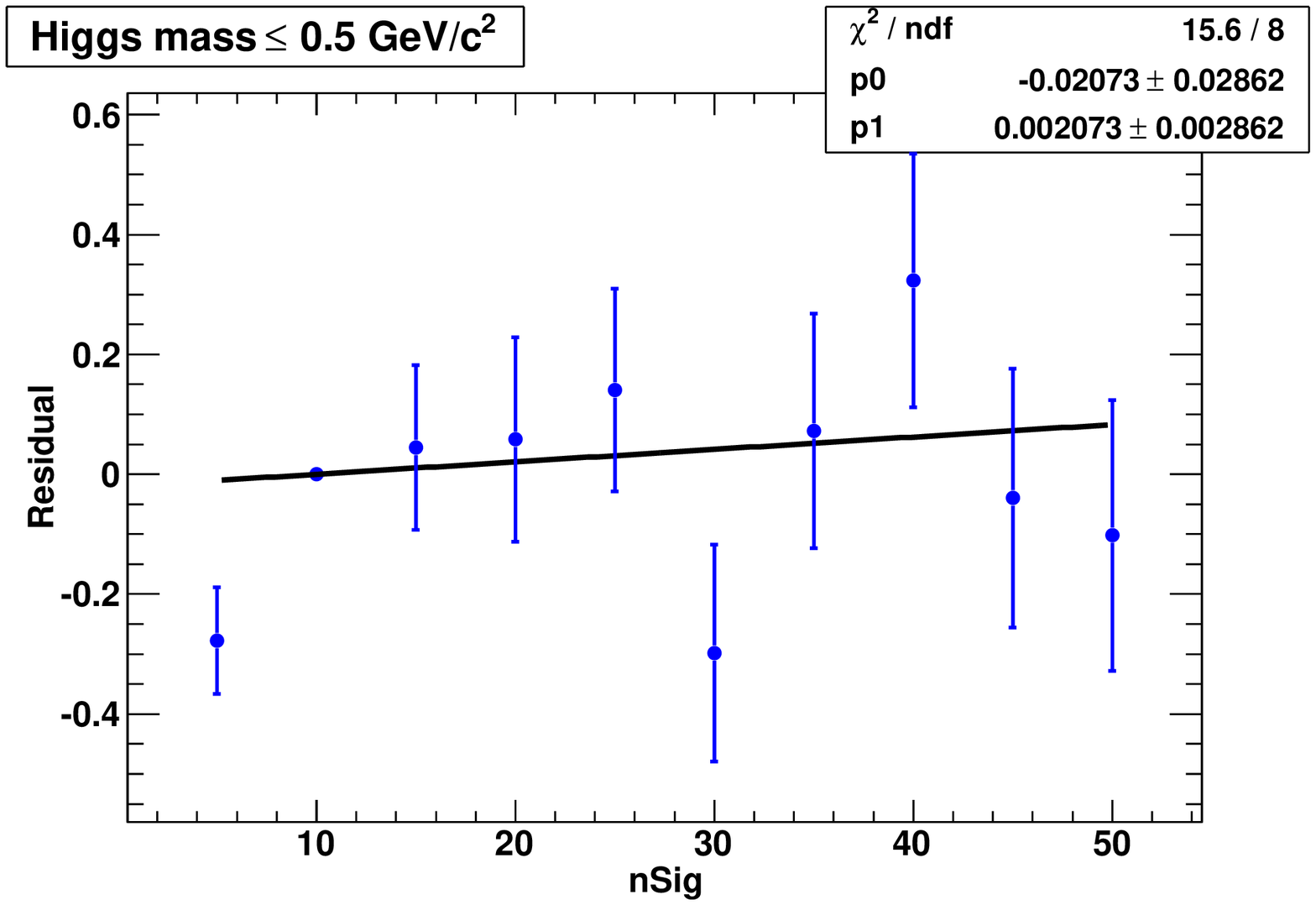}

\smallskip
\centerline{\hfill (a) \hfill \hfill (b) \hfill \hfill (c) \hfill}
\smallskip

 \includegraphics[width=2.0in]{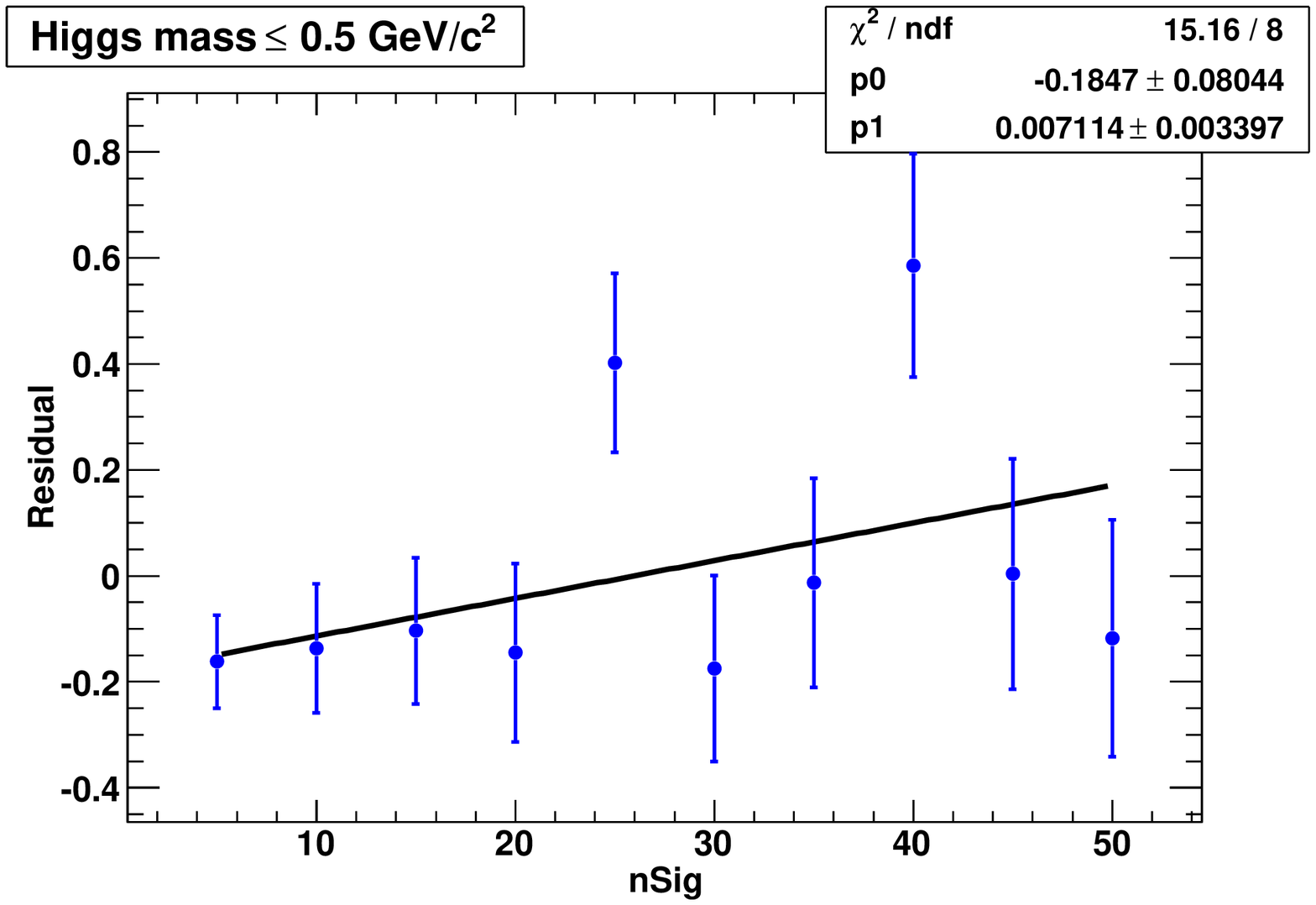}
\includegraphics[width=2.0in]{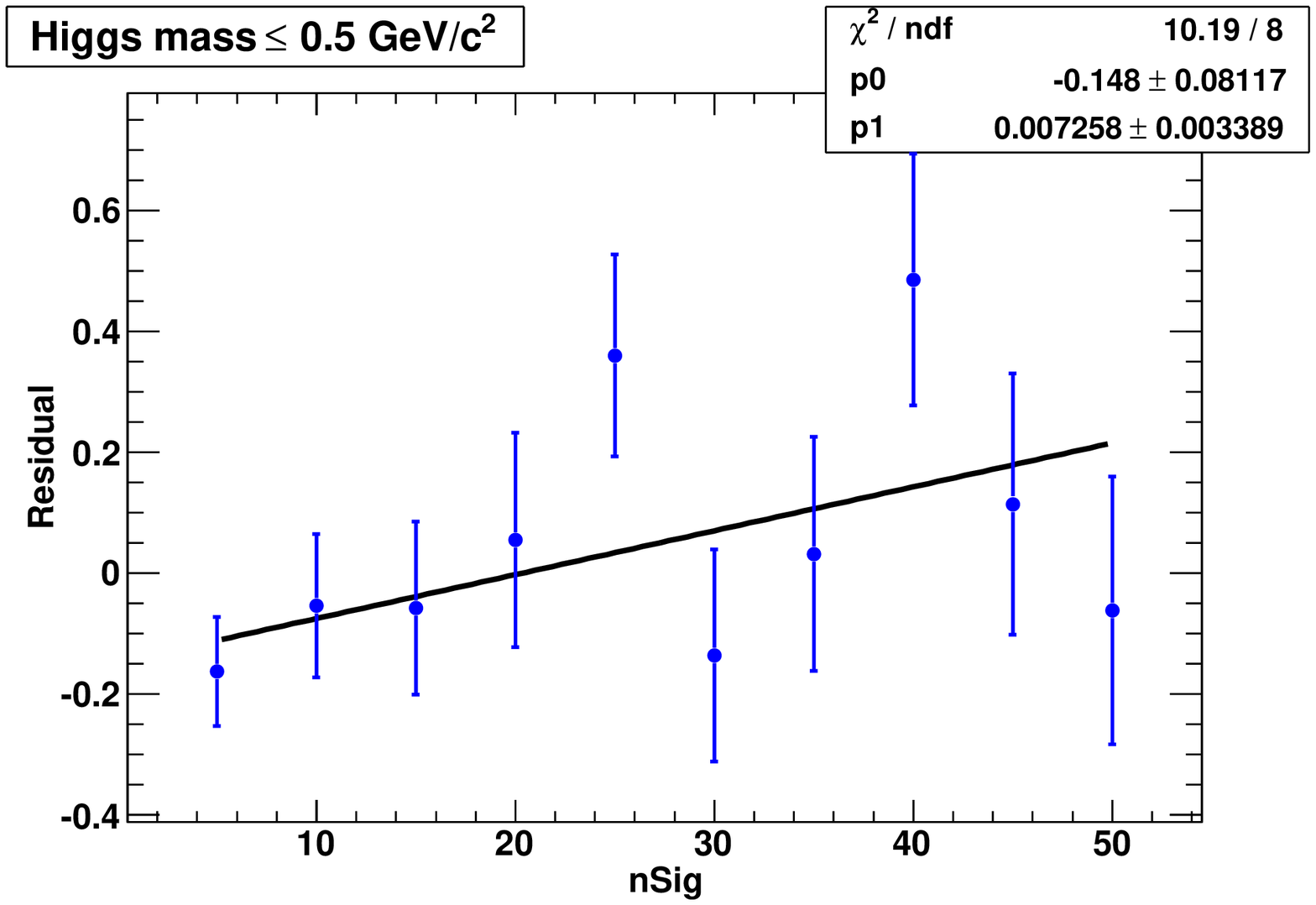}
 \includegraphics[width=2.0in]{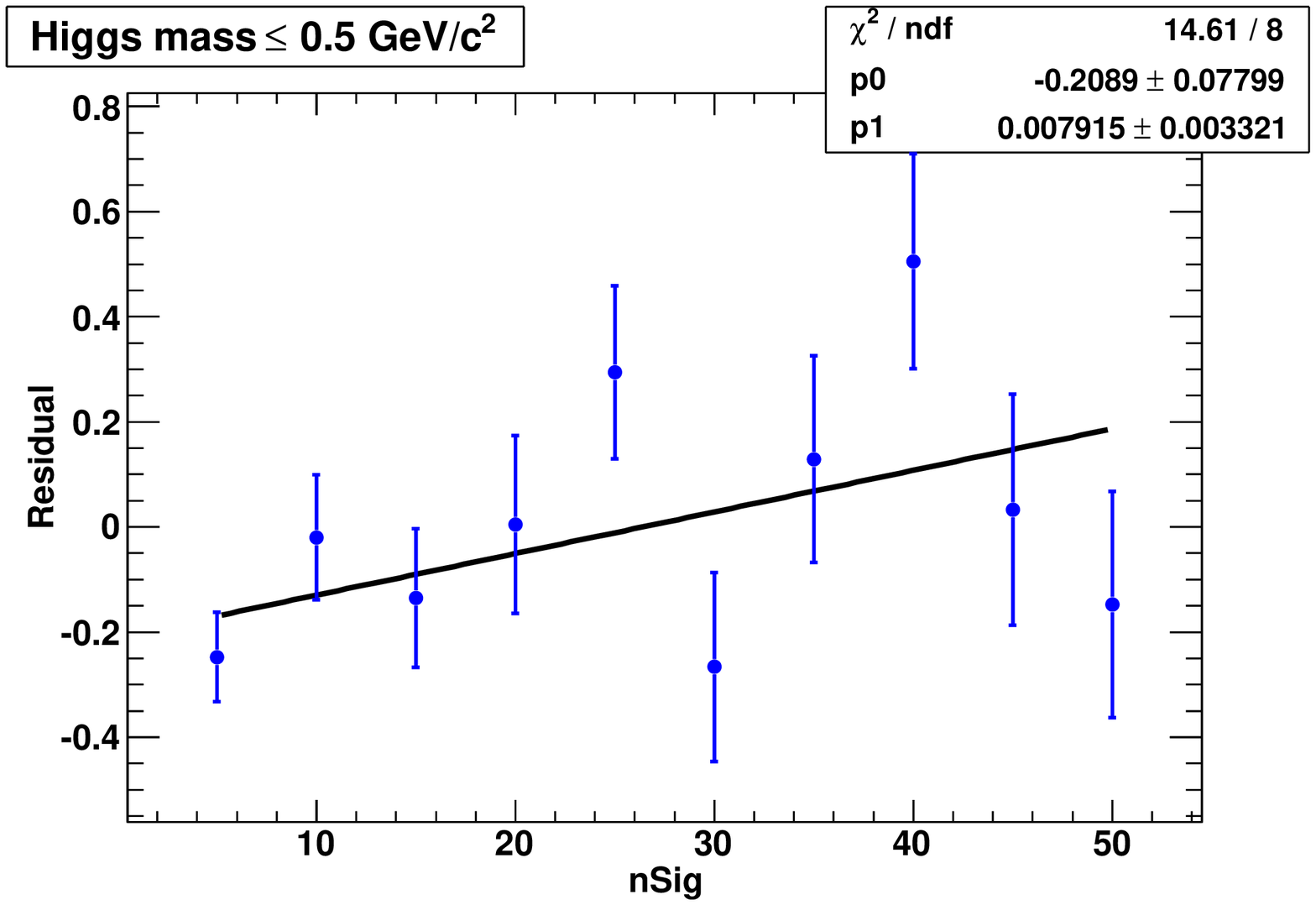}

\smallskip
\centerline{\hfill (d) \hfill \hfill (e) \hfill \hfill (f) \hfill}
\smallskip

 \includegraphics[width=3.0in]{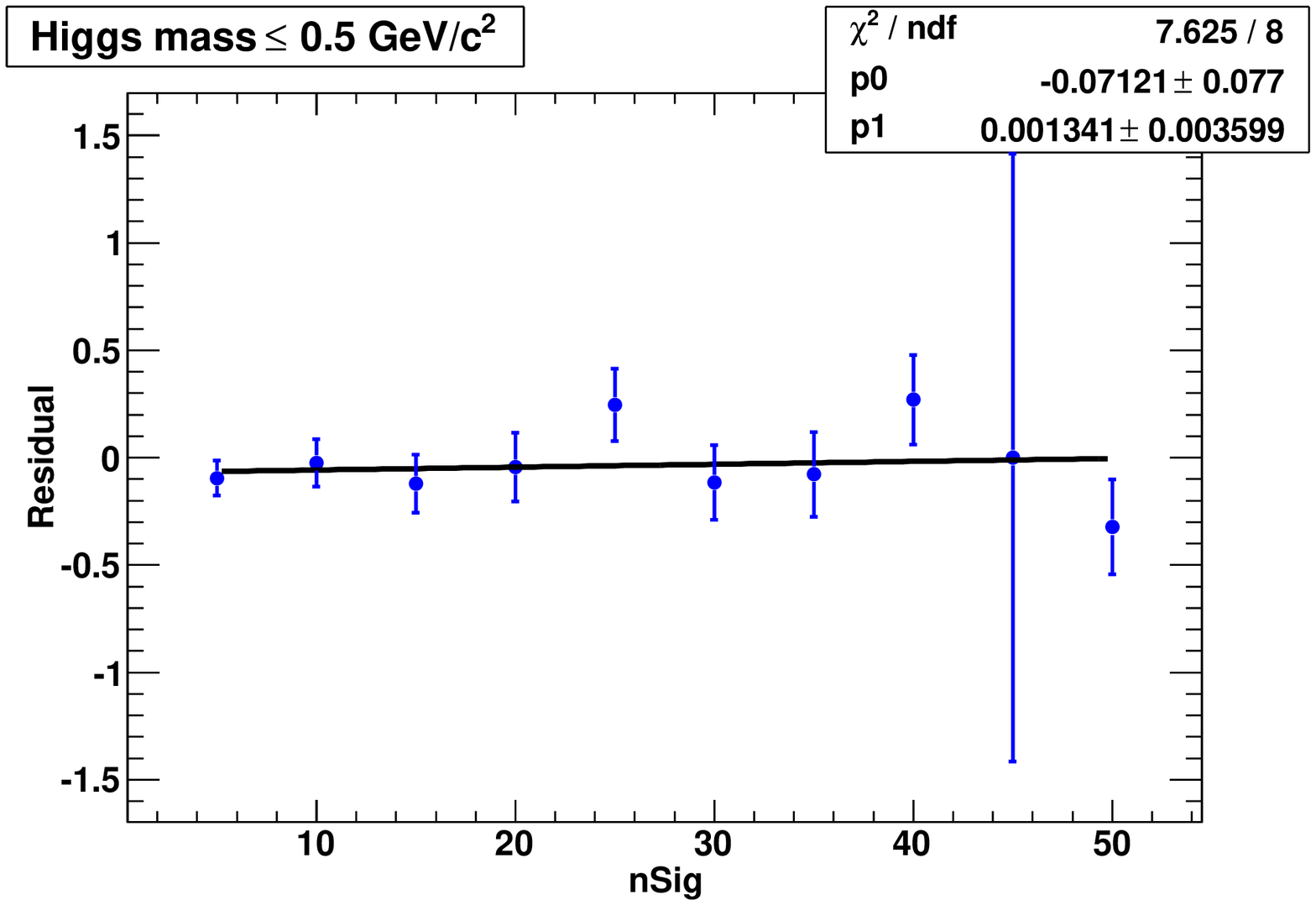}
\includegraphics[width=3.0in]{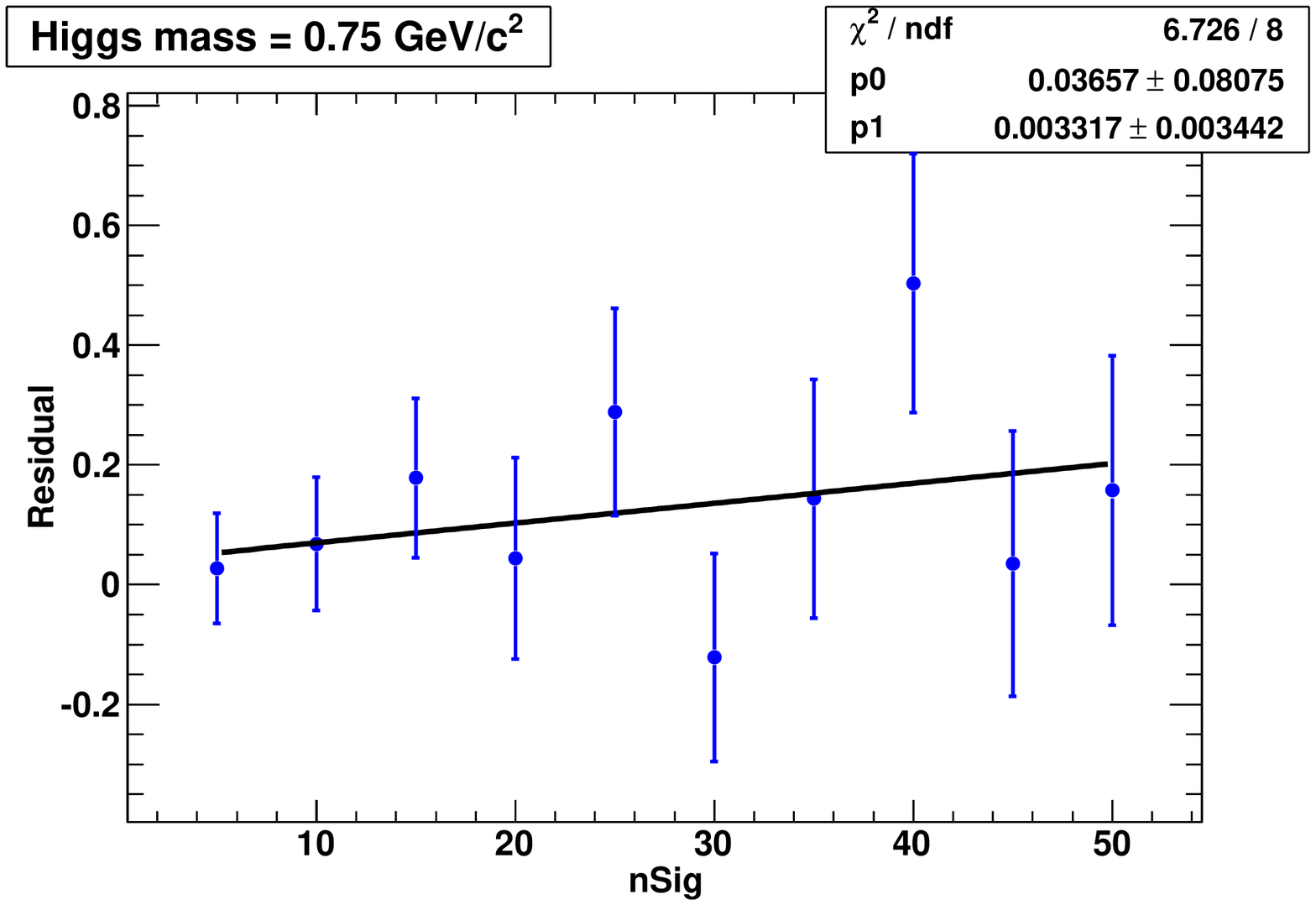}

\smallskip
\centerline{\hfill (g) \hfill \hfill (h) }
\smallskip

\caption {Fit residuals for the Higgs mass of (a) $m_{A^0}=0.212$ GeV/$c^2$  (b) $m_{A^0}=0.214$ GeV/$c^2$ (c) $m_{A^0}=0.216$ GeV/$c^2$ (d) $m_{A^0}=0.218$ GeV/$c^2$ (e) $m_{A^0}=0.220$ GeV/$c^2$ (f) $m_{A^0}=0.250$ GeV/$c^2$  (g) $m_{A^0}=0.50$ GeV/$c^2$ and (h) $m_{A^0}=0.75$ GeV/$c^2$.}

\label{fig:ToyMCY2S1}
\end{figure}

\begin{figure}
\centering
 \includegraphics[width=3.0in]{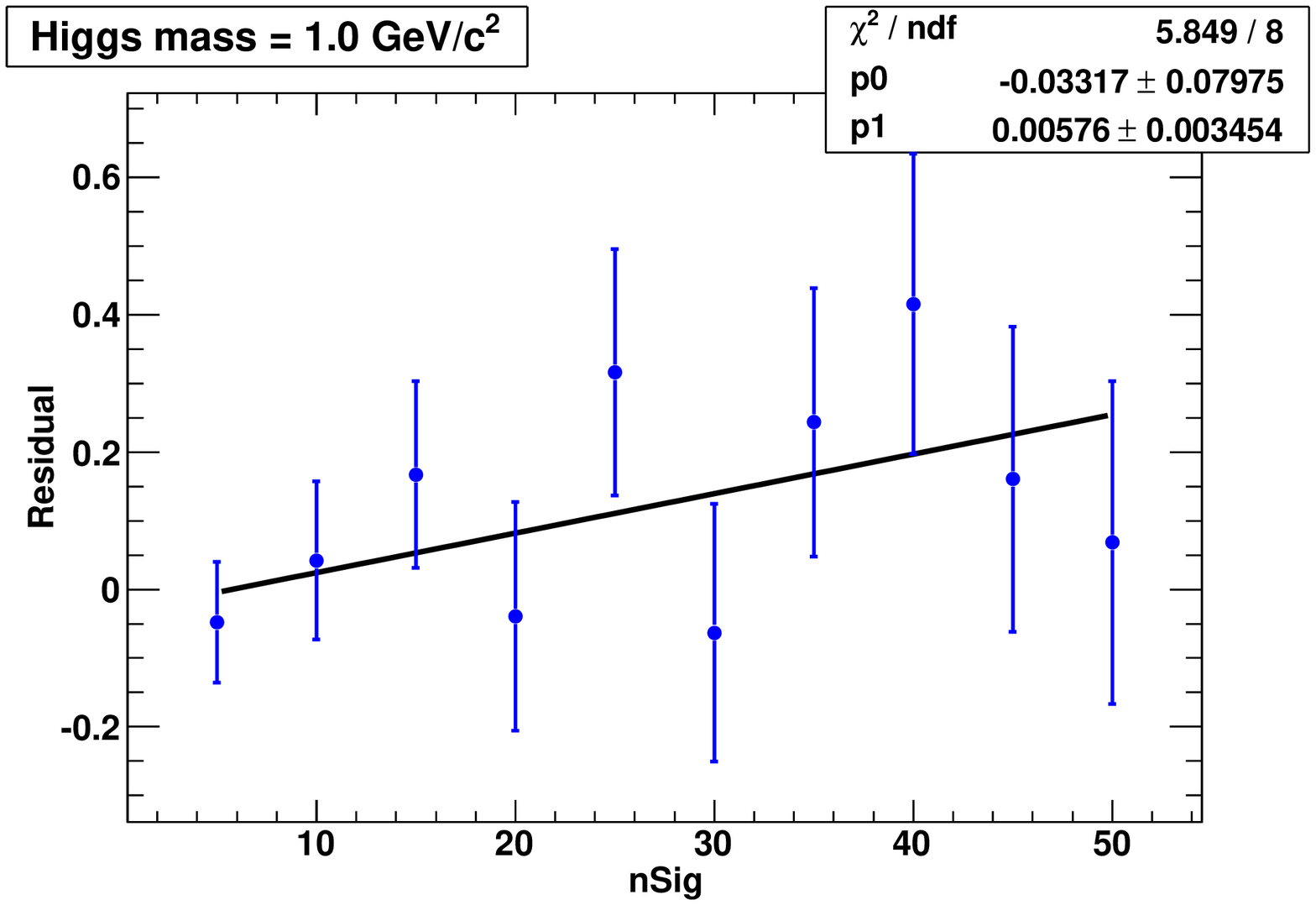}
\includegraphics[width=3.0in]{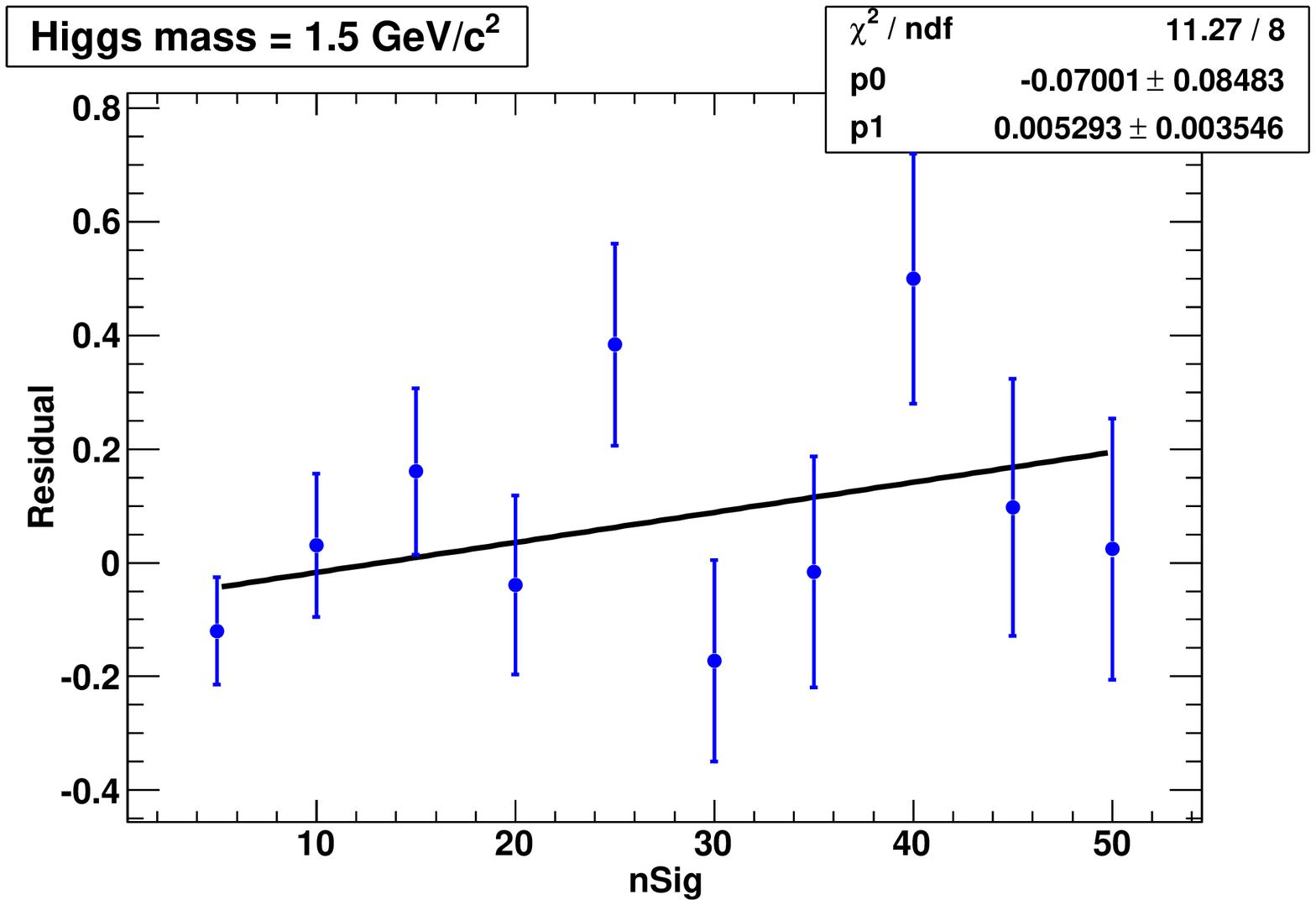}

 \includegraphics[width=3.0in]{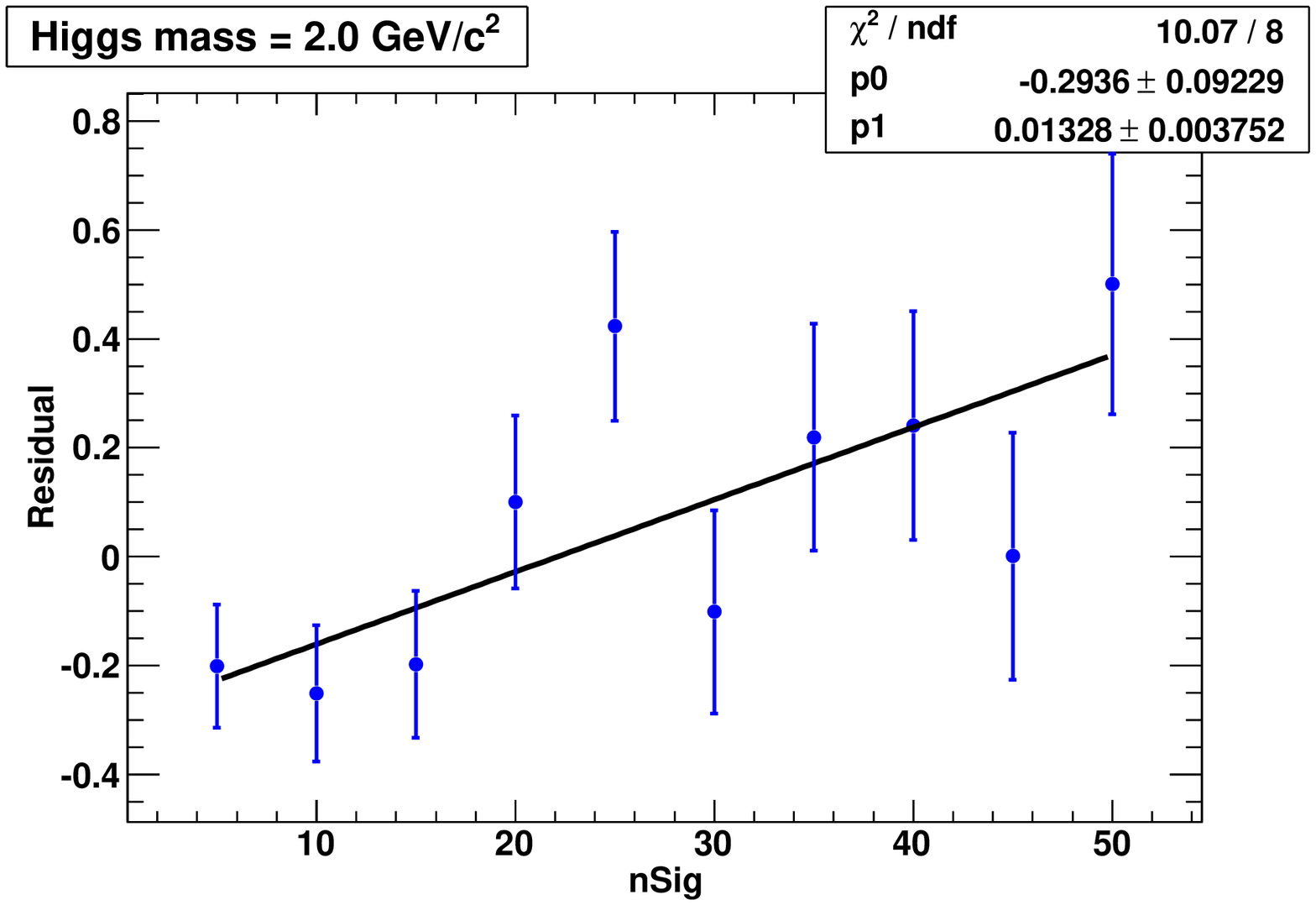}
\includegraphics[width=3.0in]{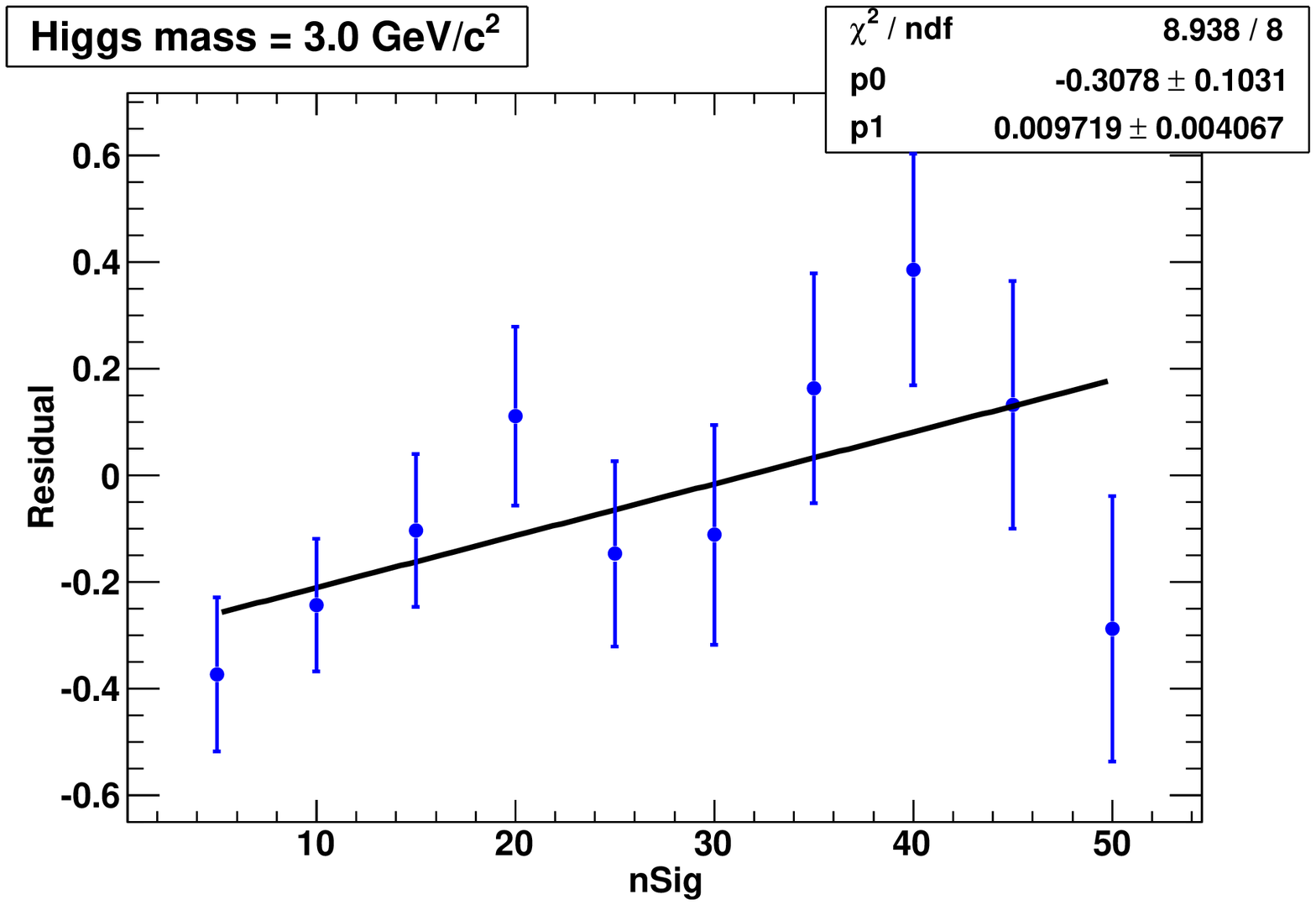}

\includegraphics[width=3.0in]{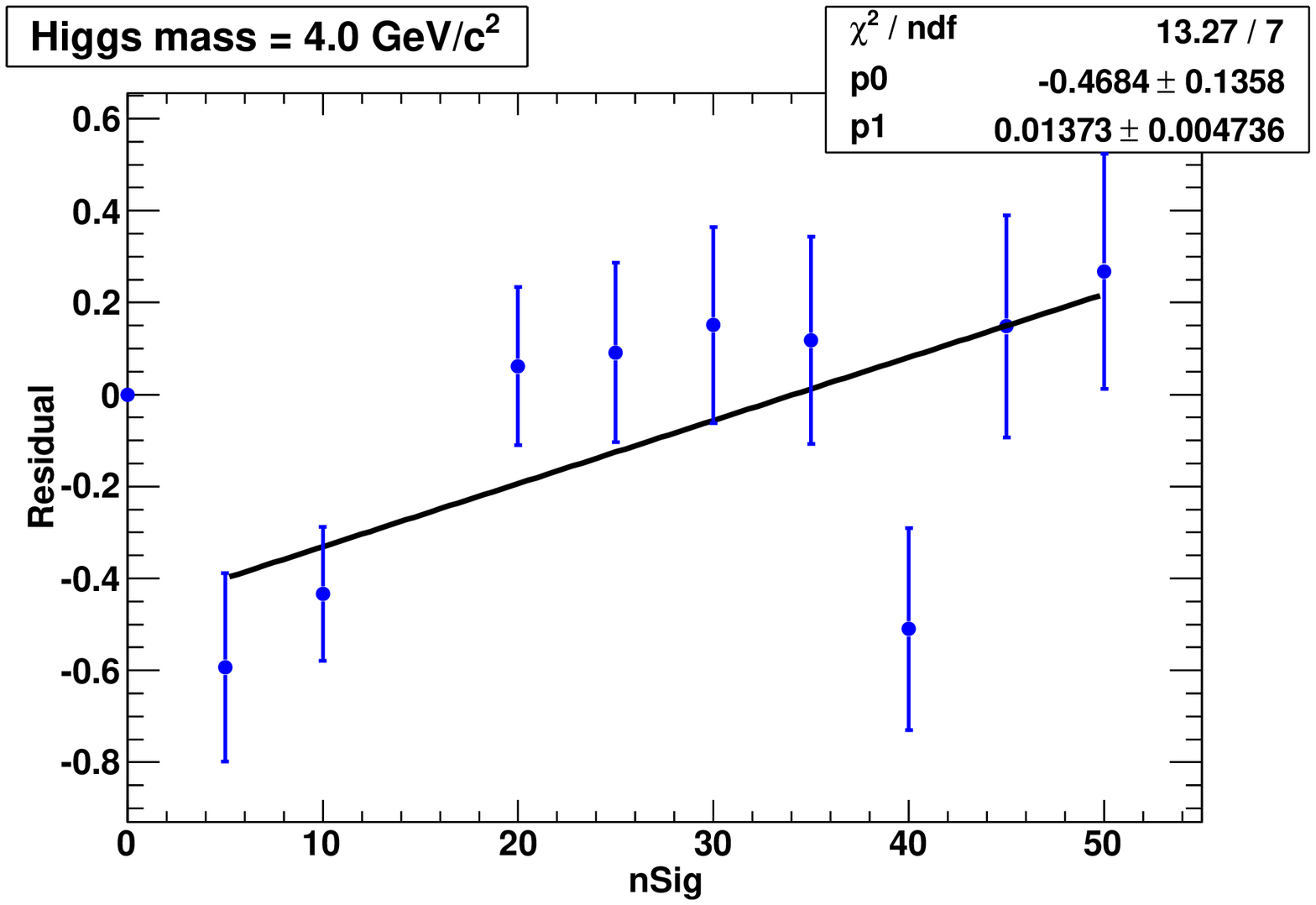}
\includegraphics[width=3.0in]{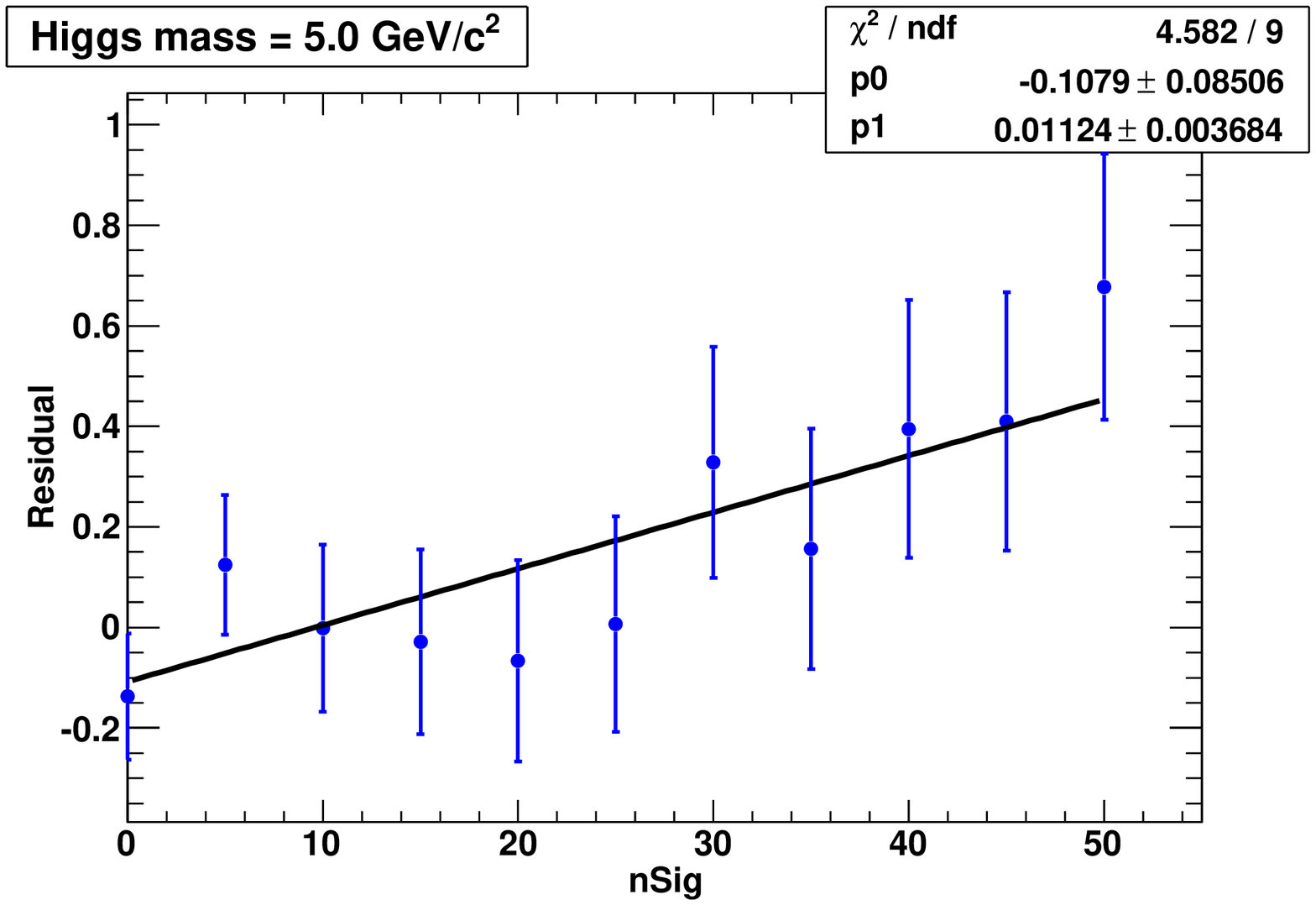}

\includegraphics[width=3.0in]{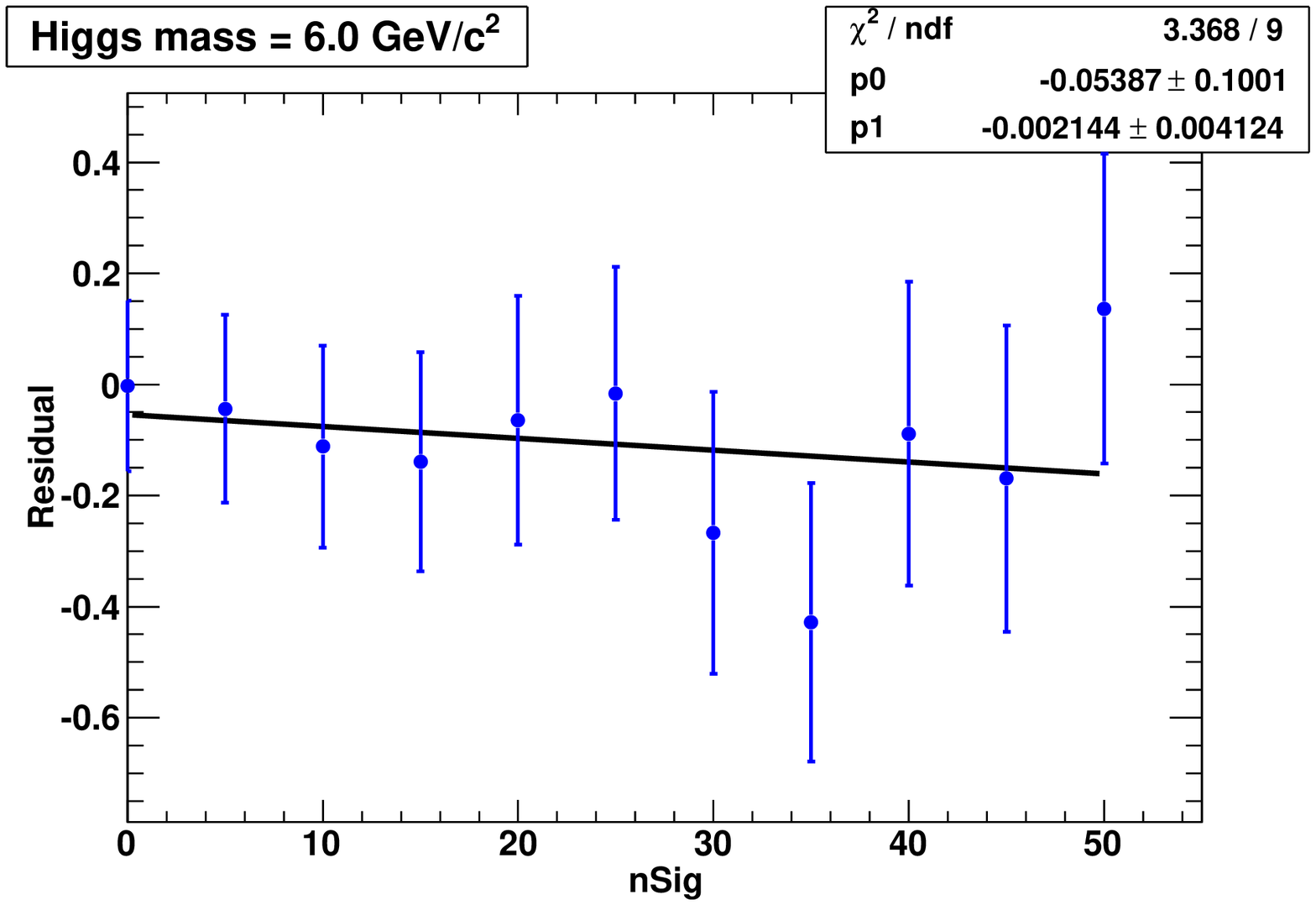}
\includegraphics[width=3.0in]{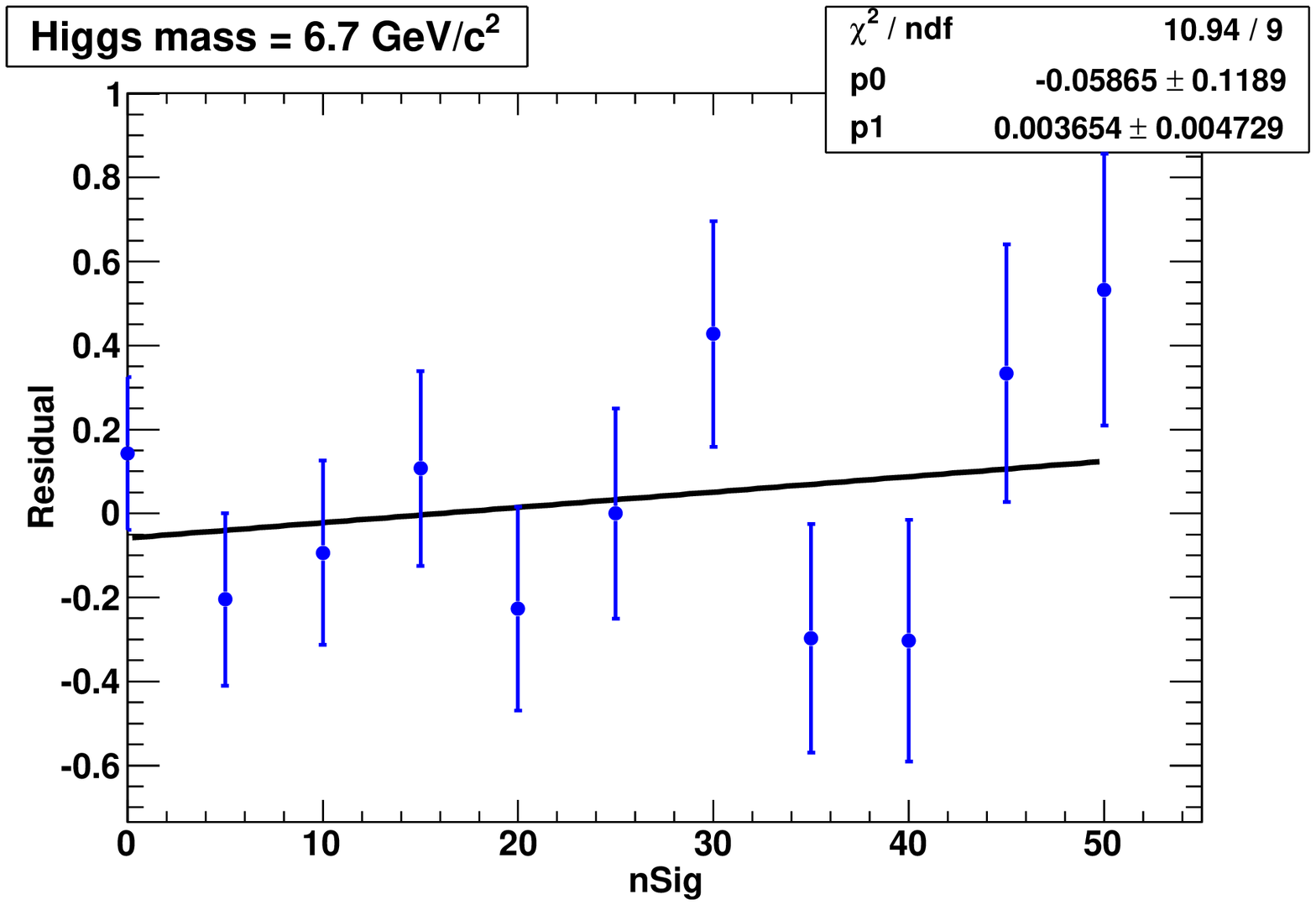}

\caption {Fit residuals for the number of signal events in the toy Monte Carlo experiments generated for each Higgs mass points.}

\label{fig:ToyMCY2S2}
\end{figure}

\begin{figure}
\centering
 \includegraphics[width=3.0in]{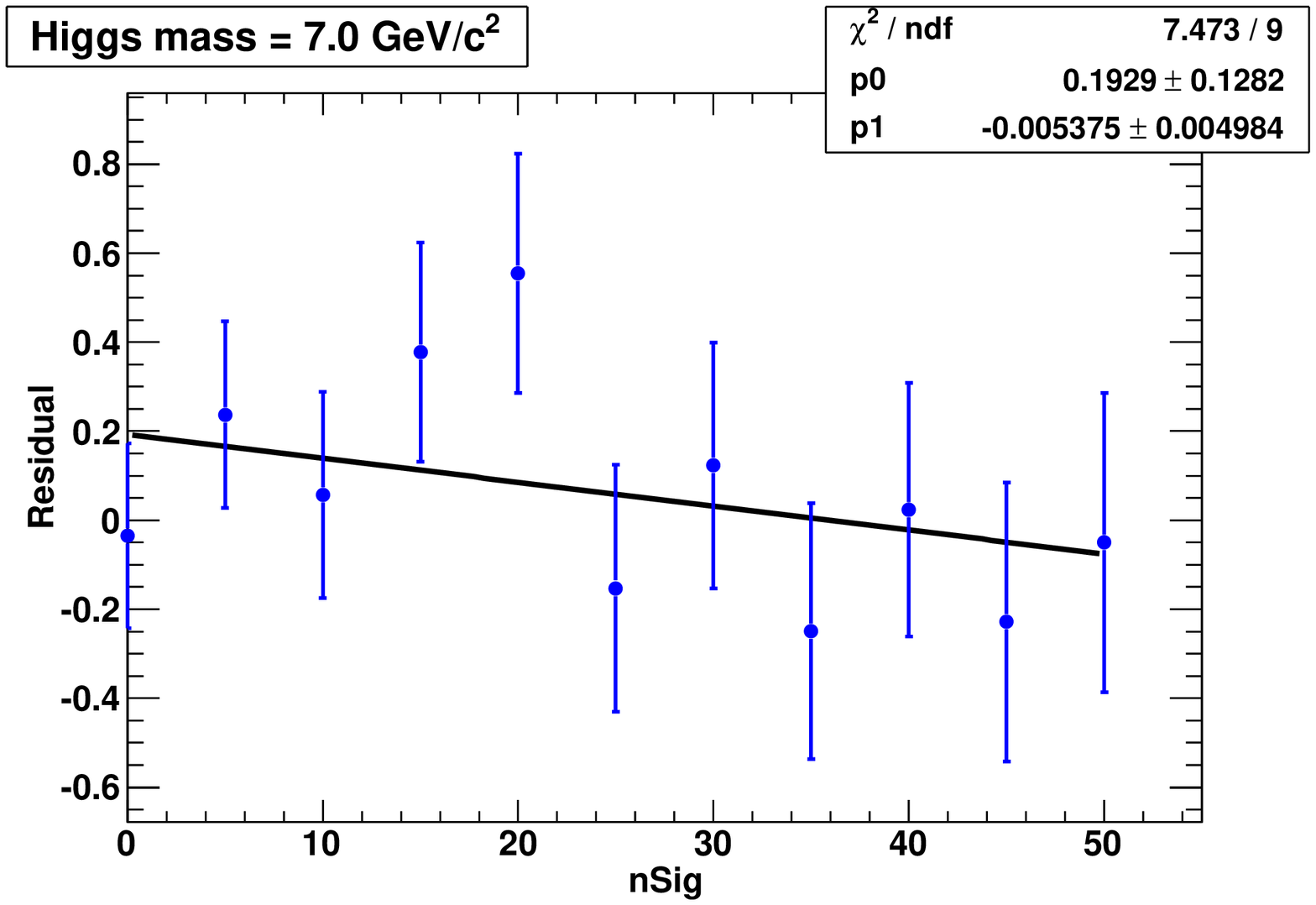}
\includegraphics[width=3.0in]{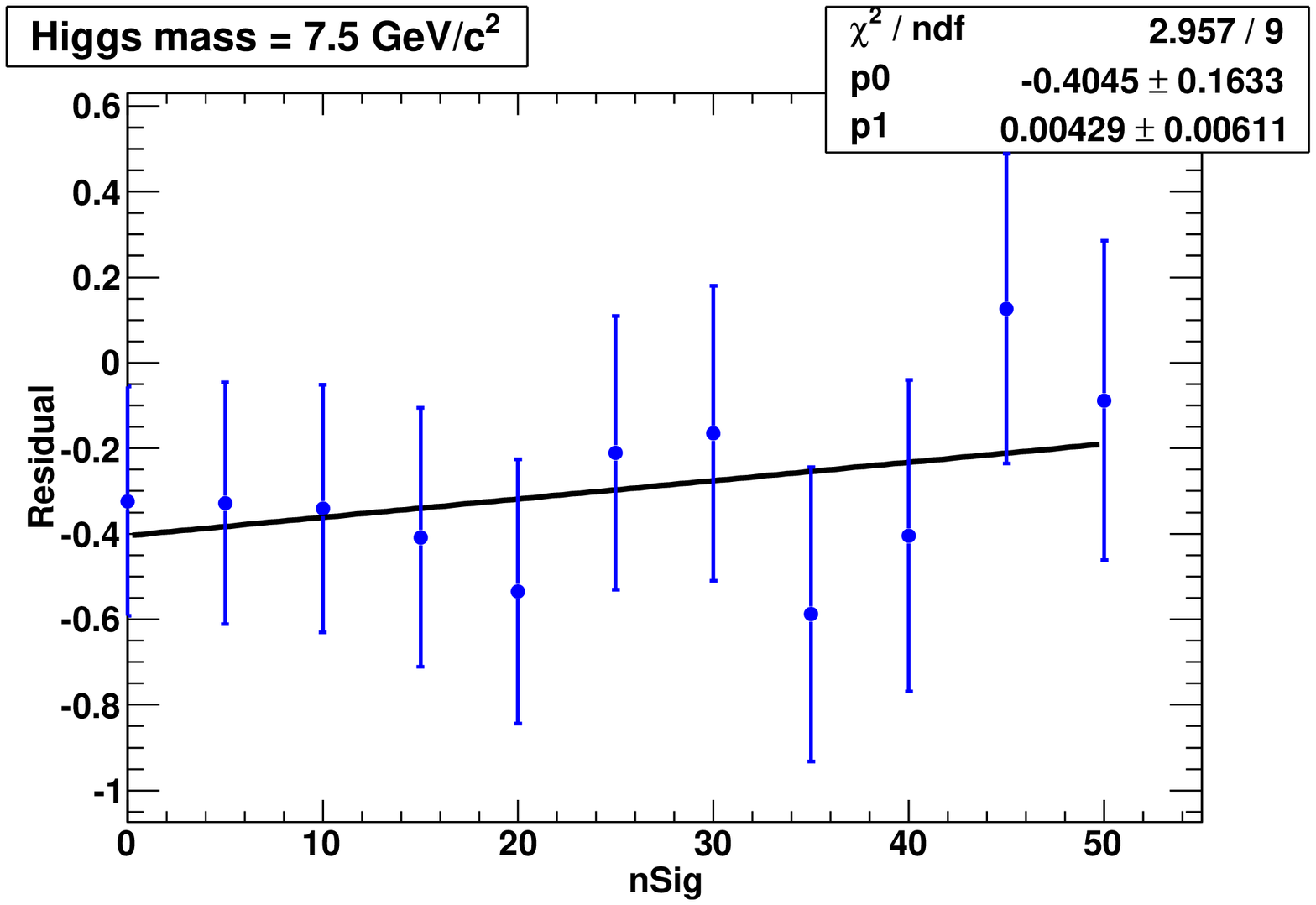}

\includegraphics[width=3.0in]{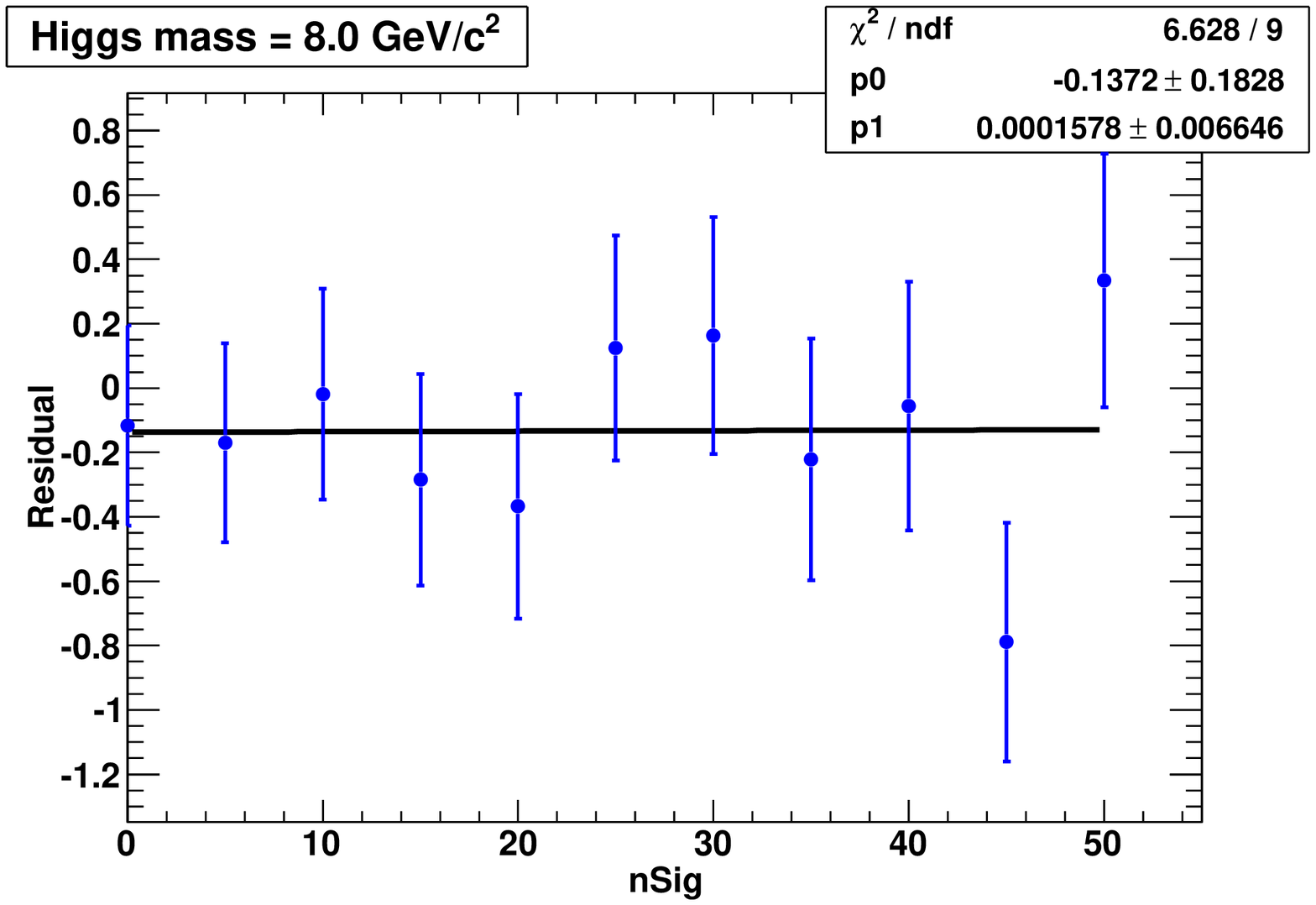}
\includegraphics[width=3.0in]{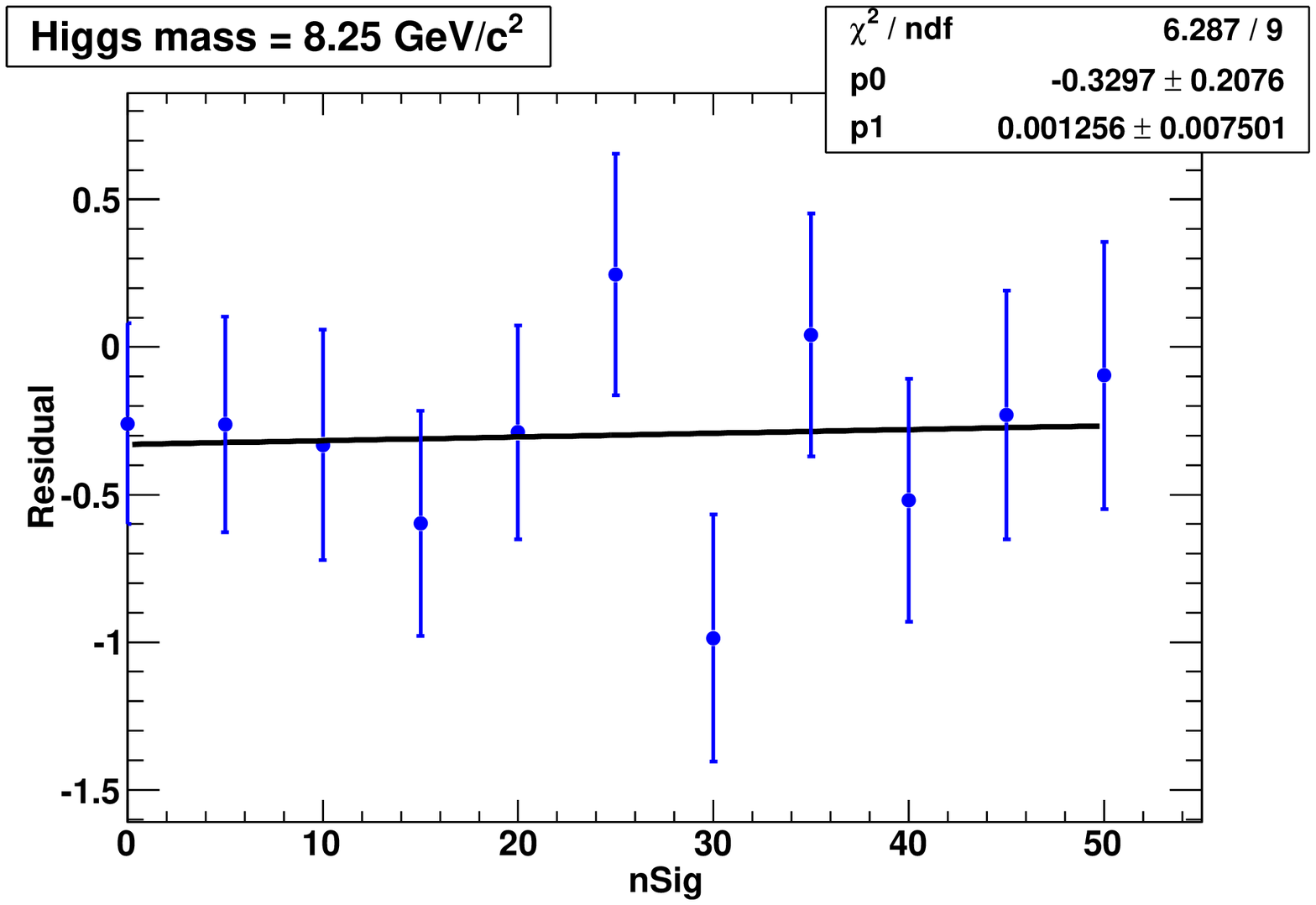}

\includegraphics[width=3.0in]{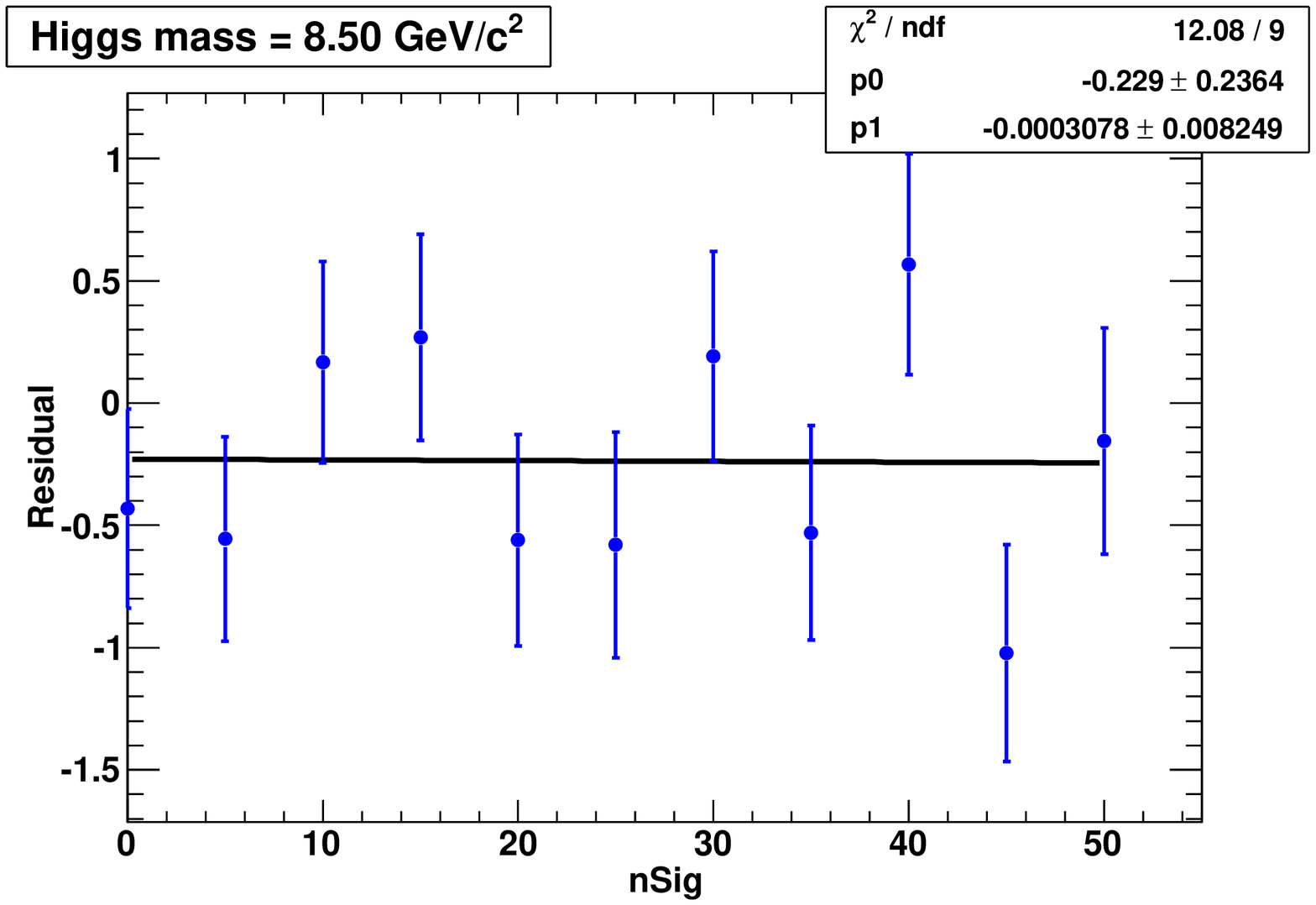}
\includegraphics[width=3.0in]{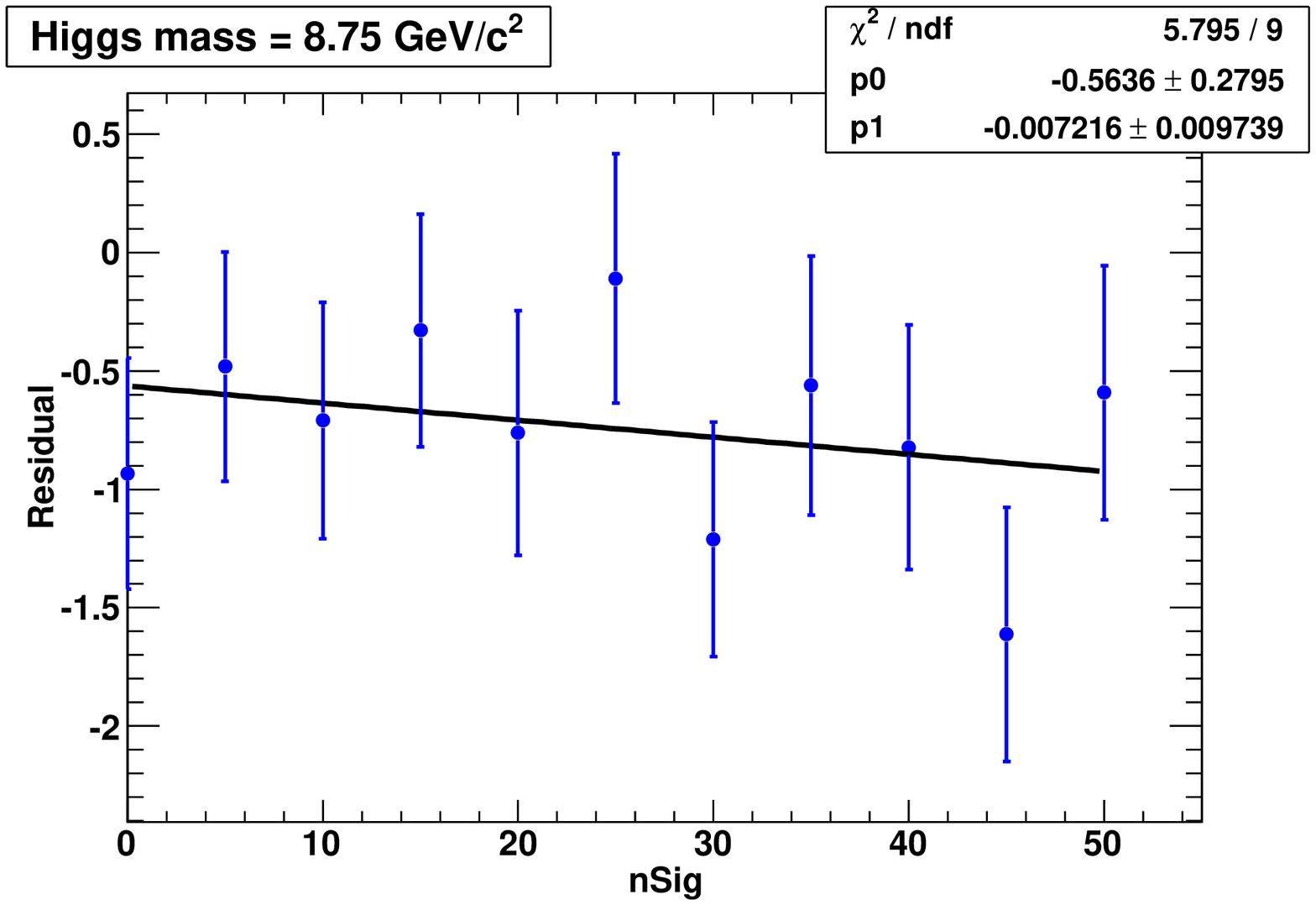}

\includegraphics[width=3.0in]{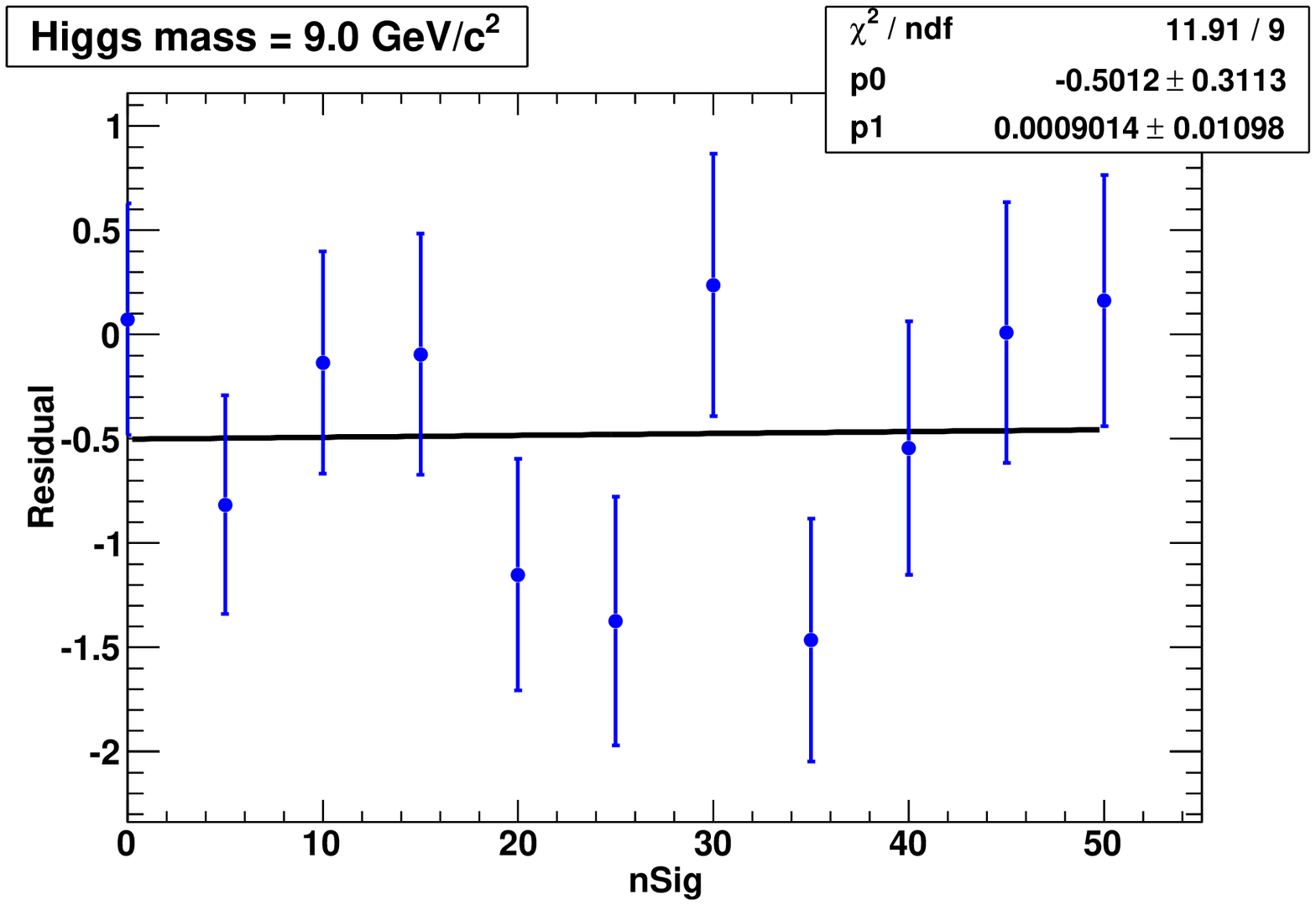}
\includegraphics[width=3.0in]{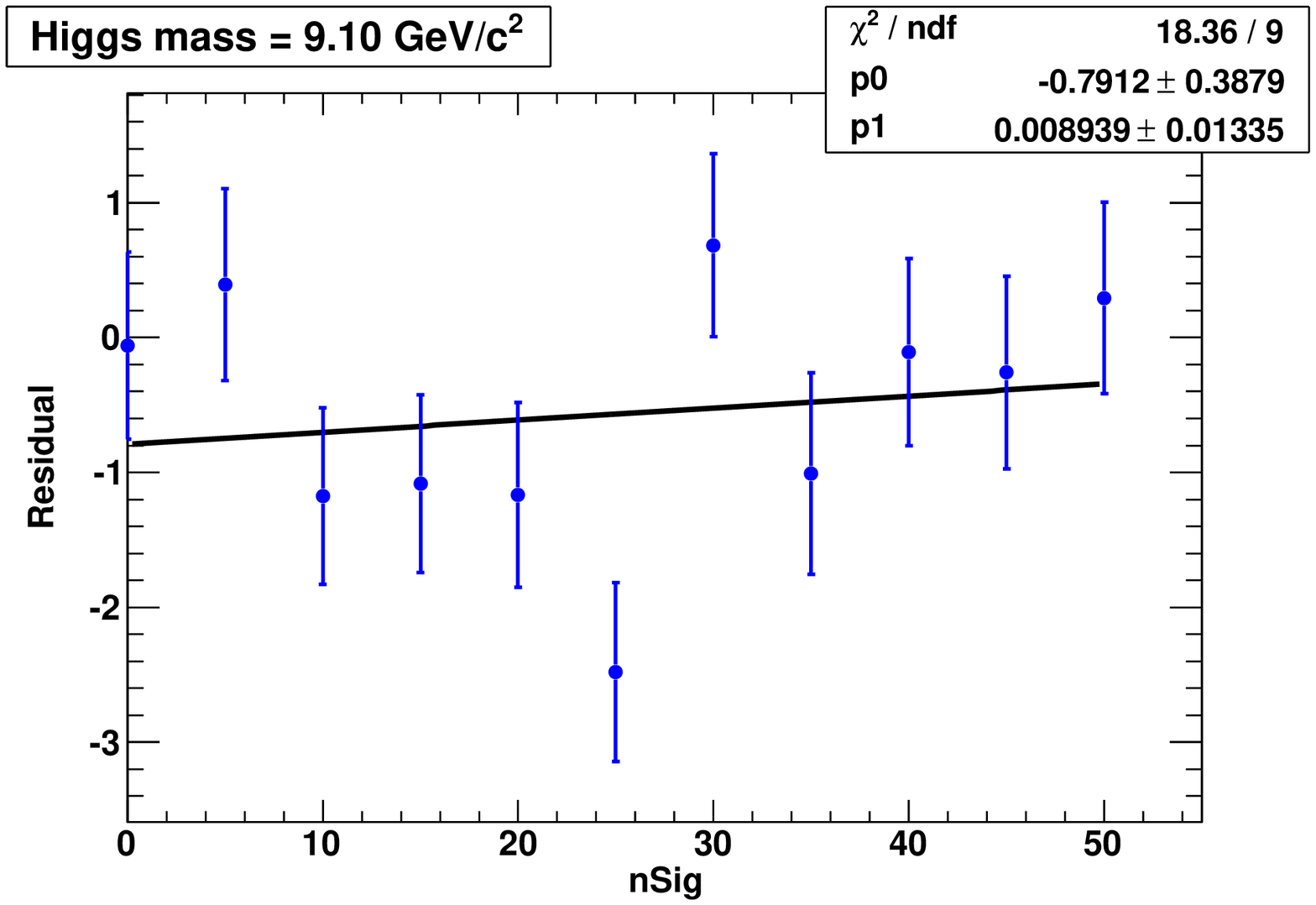}

\caption {Fit residuals for the number of signal events in the toy Monte Carlo experiments generated for each Higgs mass points.}

\label{fig:ToyMCY2S3}
\end{figure}

\begin{figure}
\section{For $\Upsilon(3S)$}
\centering
 \includegraphics[width=2.0in]{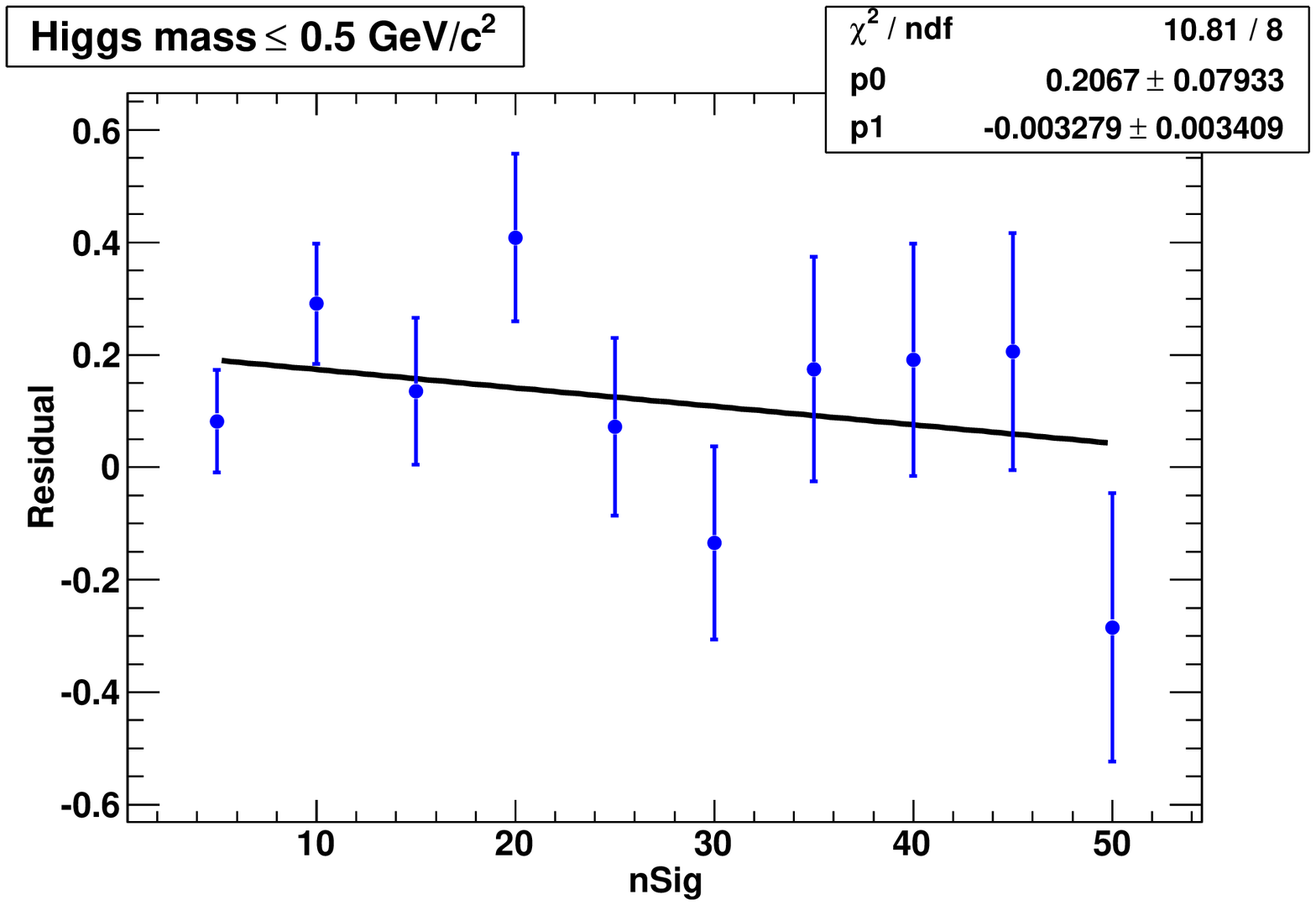}
\includegraphics[width=2.0in]{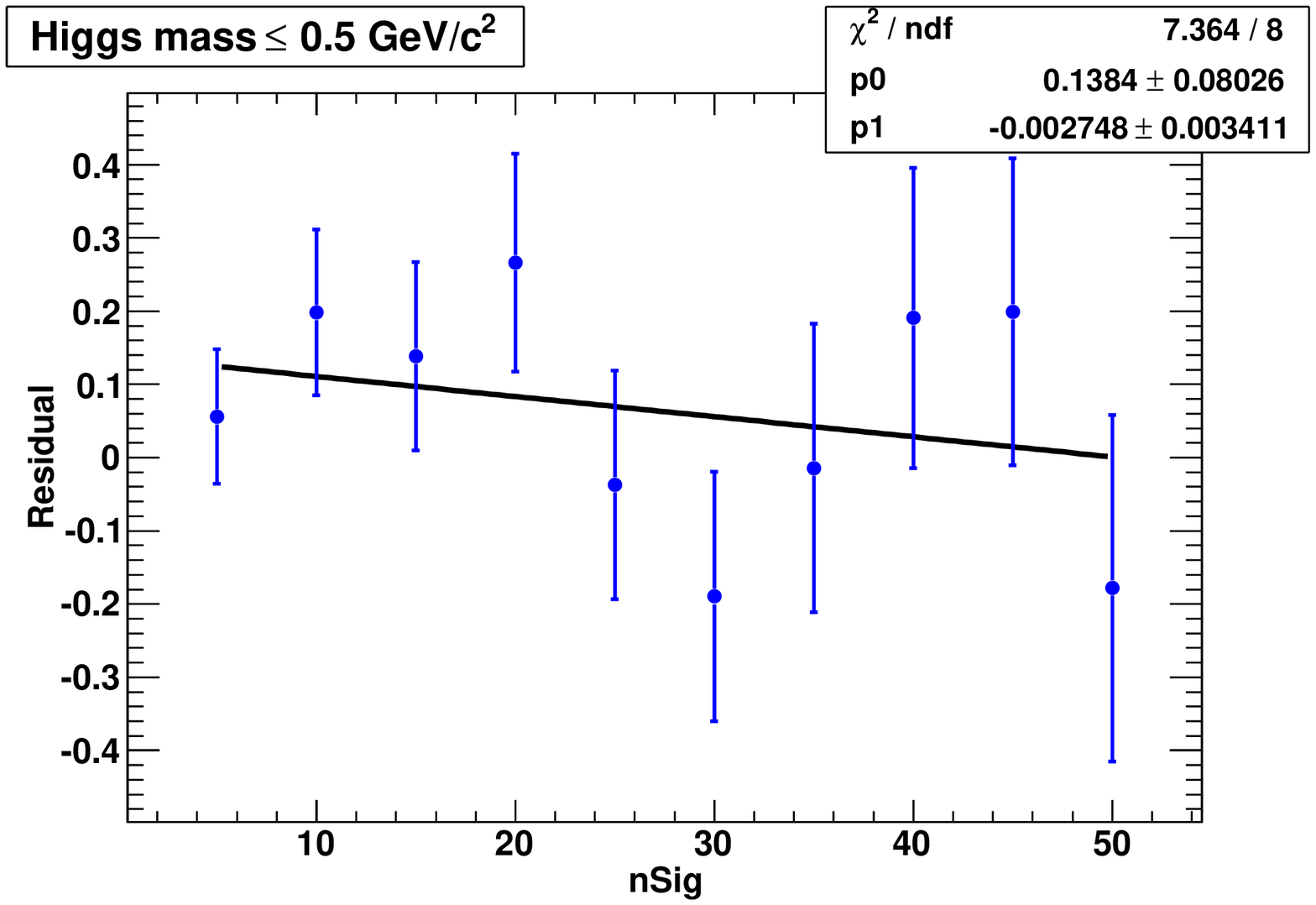}
 \includegraphics[width=2.0in]{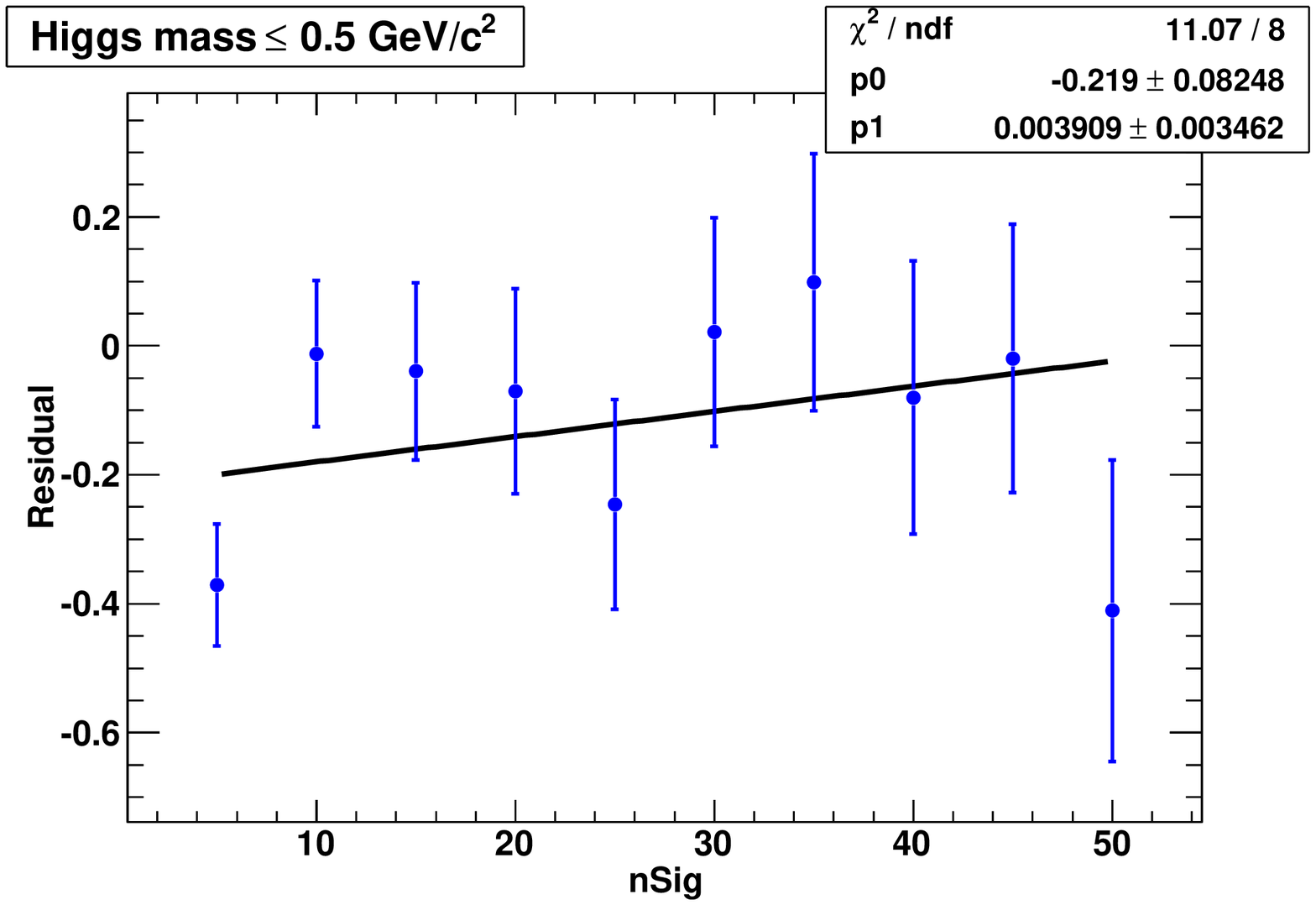}

\smallskip
\centerline{\hfill (a) \hfill \hfill (b) \hfill \hfill (c) \hfill}
\smallskip

 \includegraphics[width=2.0in]{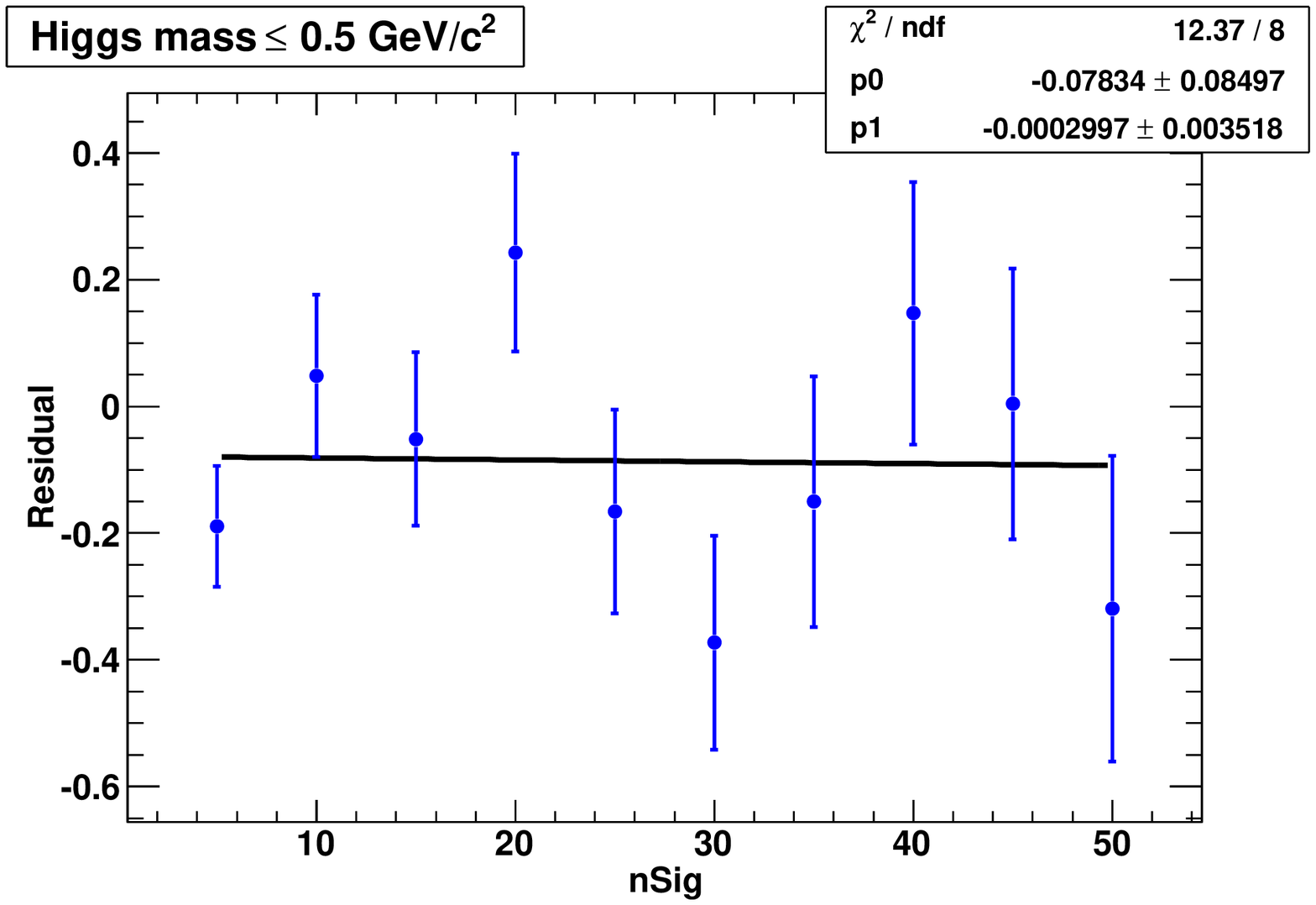}
\includegraphics[width=2.0in]{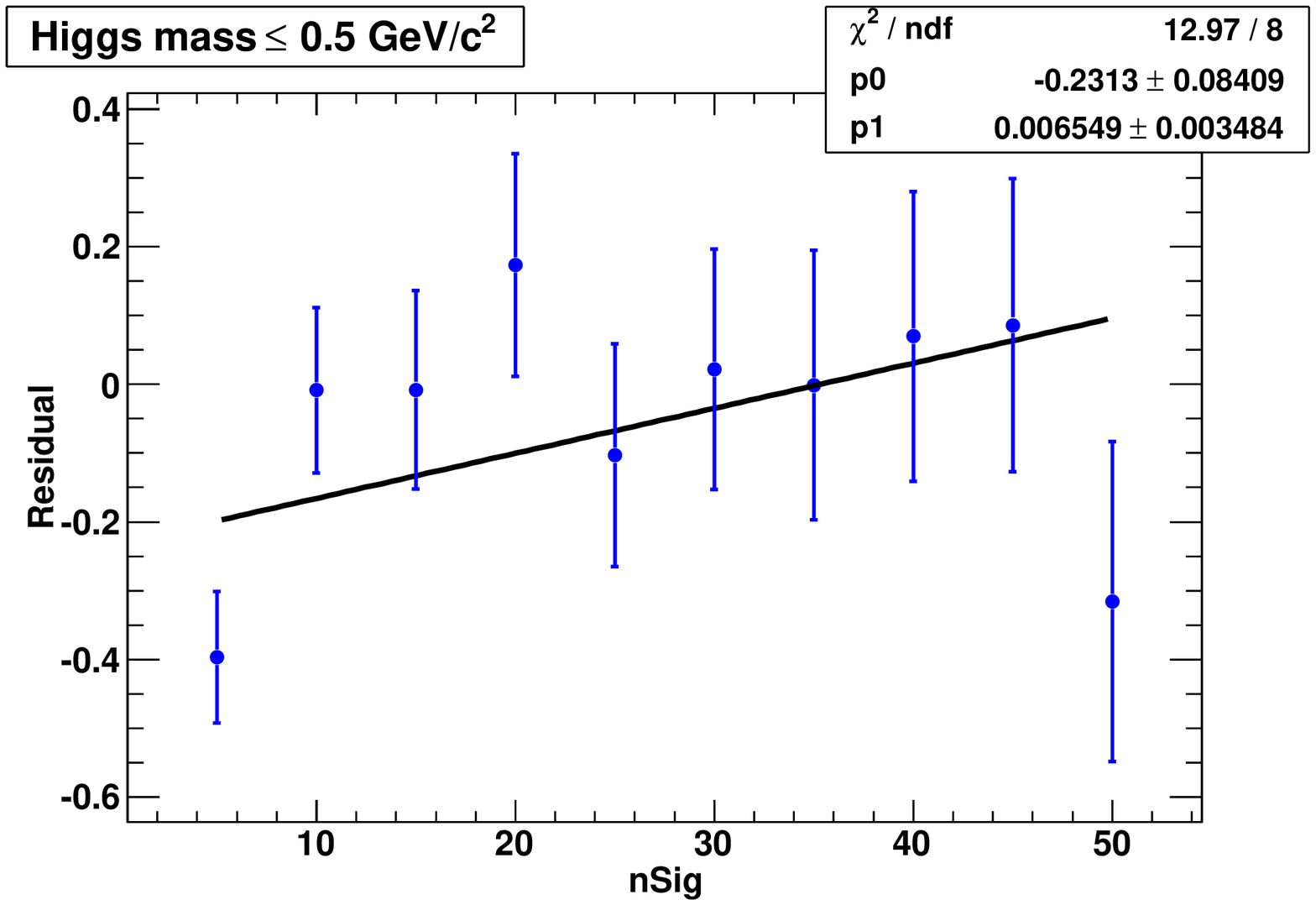}
 \includegraphics[width=2.0in]{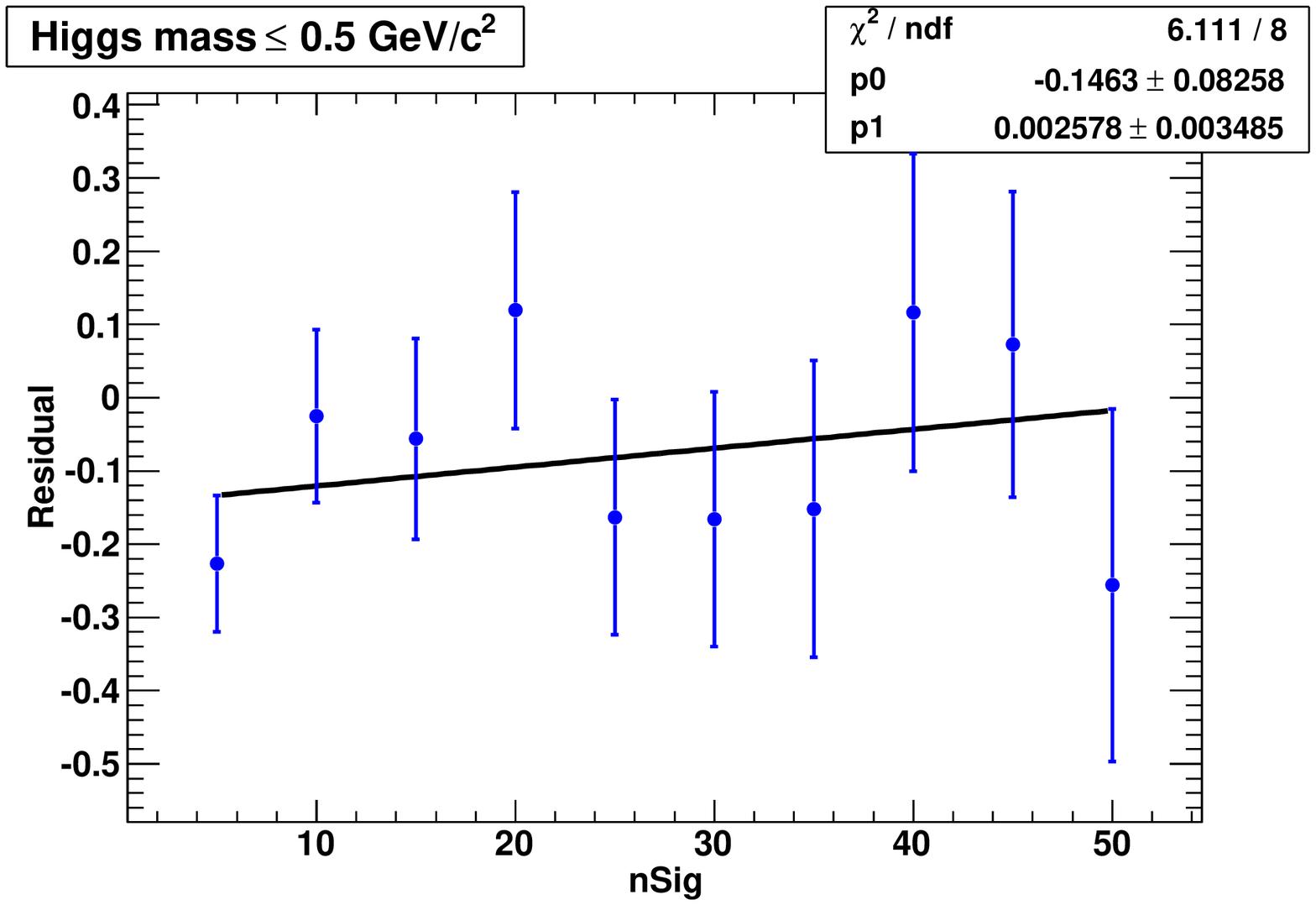}

\smallskip
\centerline{\hfill (d) \hfill \hfill (e) \hfill \hfill (f) \hfill}
\smallskip

 \includegraphics[width=3.0in]{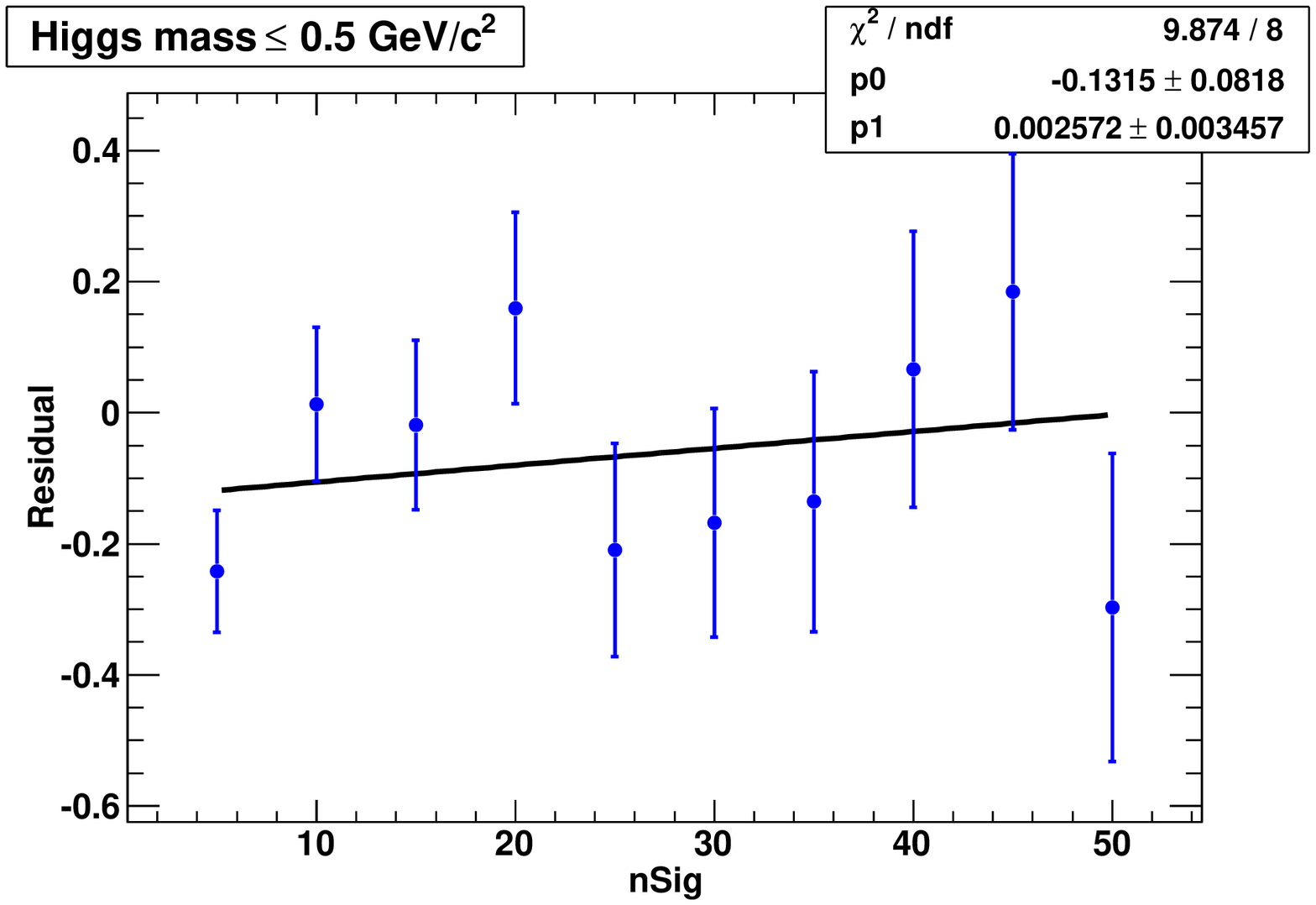}
\includegraphics[width=3.0in]{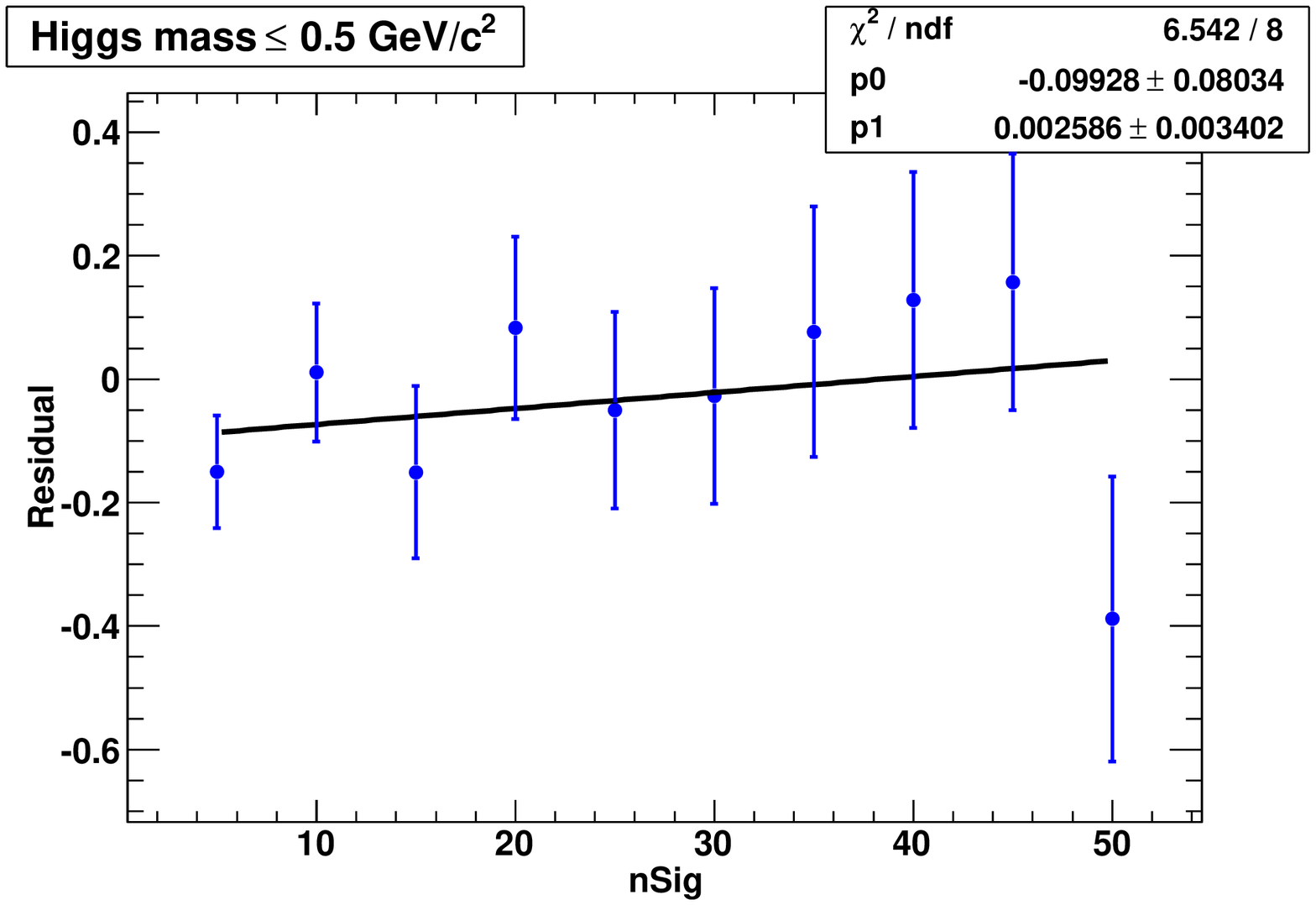}

\smallskip
\centerline{\hfill (g) \hfill \hfill (h) }
\smallskip

\caption {Fit residuals for the Higgs mass of (a) $m_{A^0}=0.212$ GeV/$c^2$  (b) $m_{A^0}=0.214$ GeV/$c^2$ (c) $m_{A^0}=0.216$ GeV/$c^2$ (d) $m_{A^0}=0.218$ GeV/$c^2$ (e) $m_{A^0}=0.220$ GeV/$c^2$ (f) $m_{A^0}=0.250$ GeV/$c^2$  (g) $m_{A^0}=0.300$ GeV/$c^2$ and (h) $m_{A^0}=0.50$ GeV/$c^2$.}

\label{fig:ToyMCY3S1}
\end{figure}

\begin{figure}
\centering
 \includegraphics[width=3.0in]{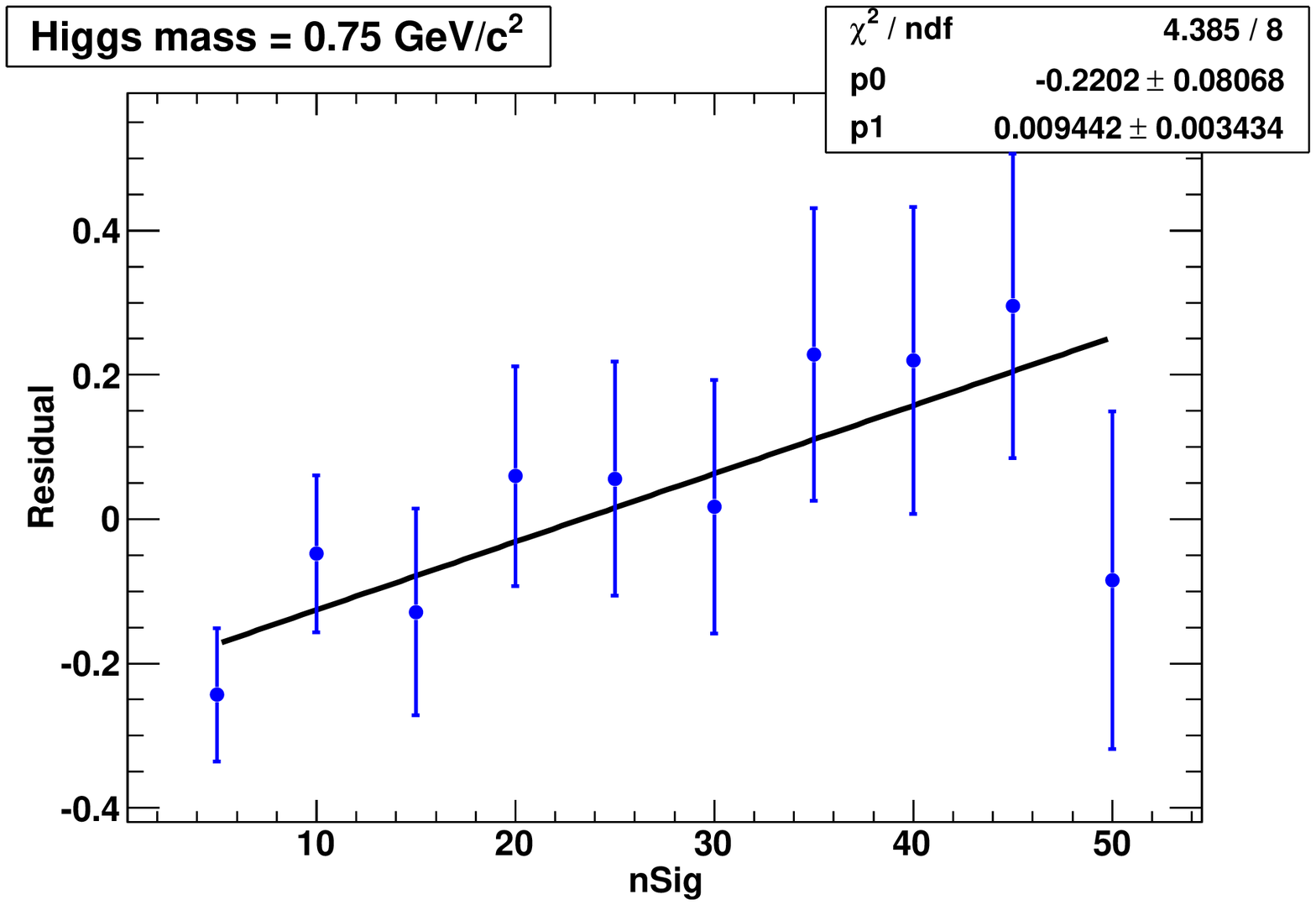}
\includegraphics[width=3.0in]{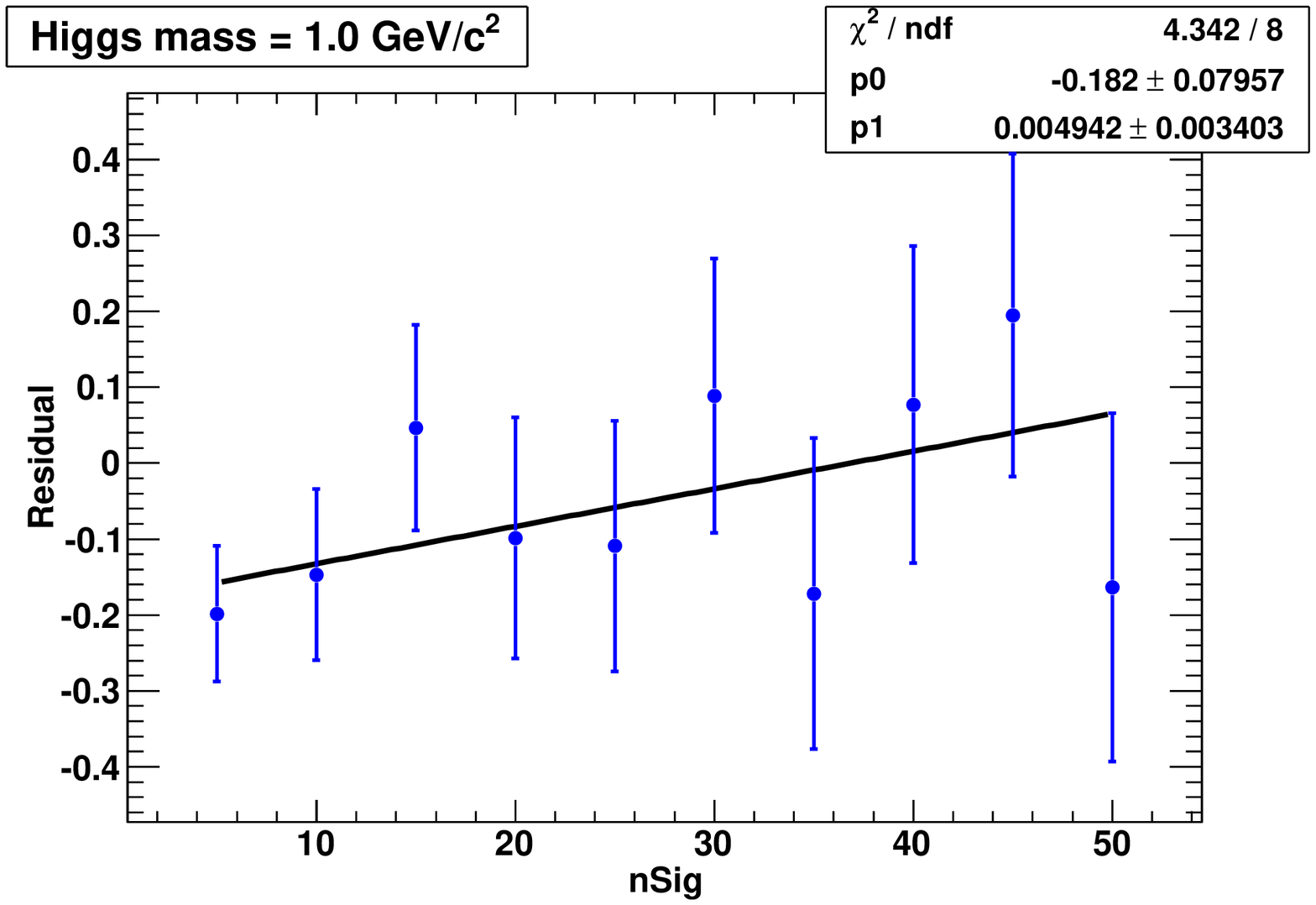}

\includegraphics[width=3.0in]{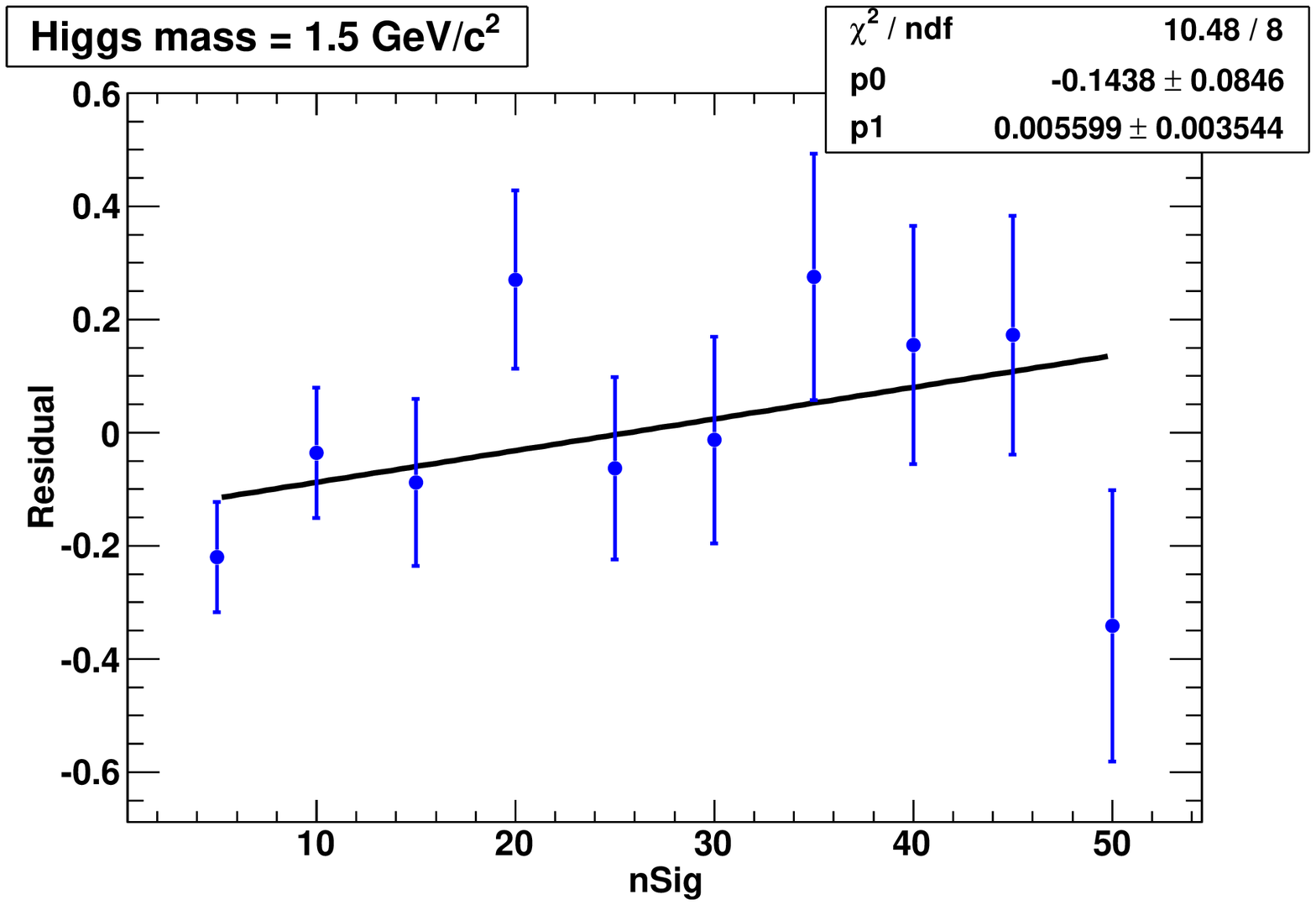}
\includegraphics[width=3.0in]{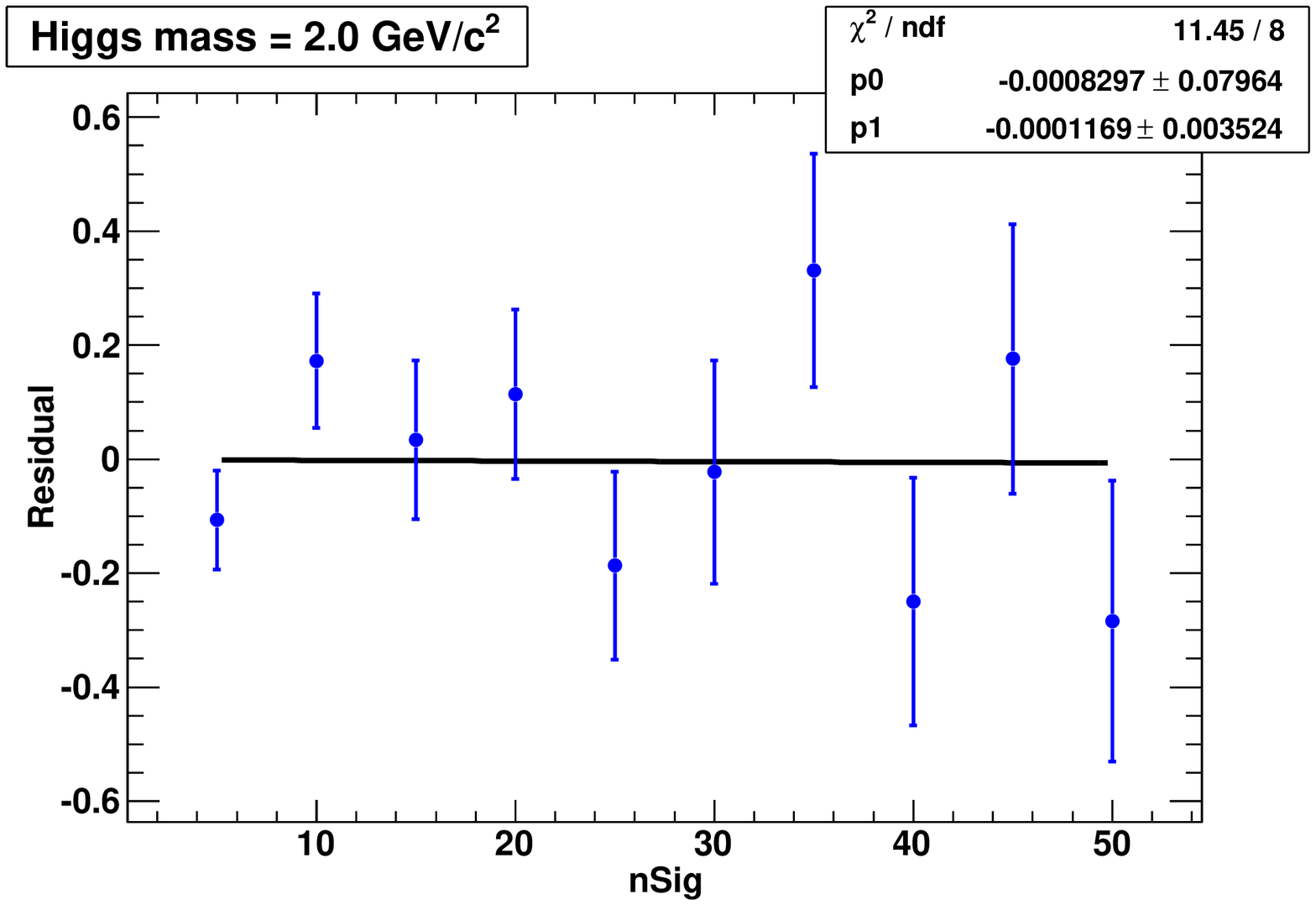}

\includegraphics[width=3.0in]{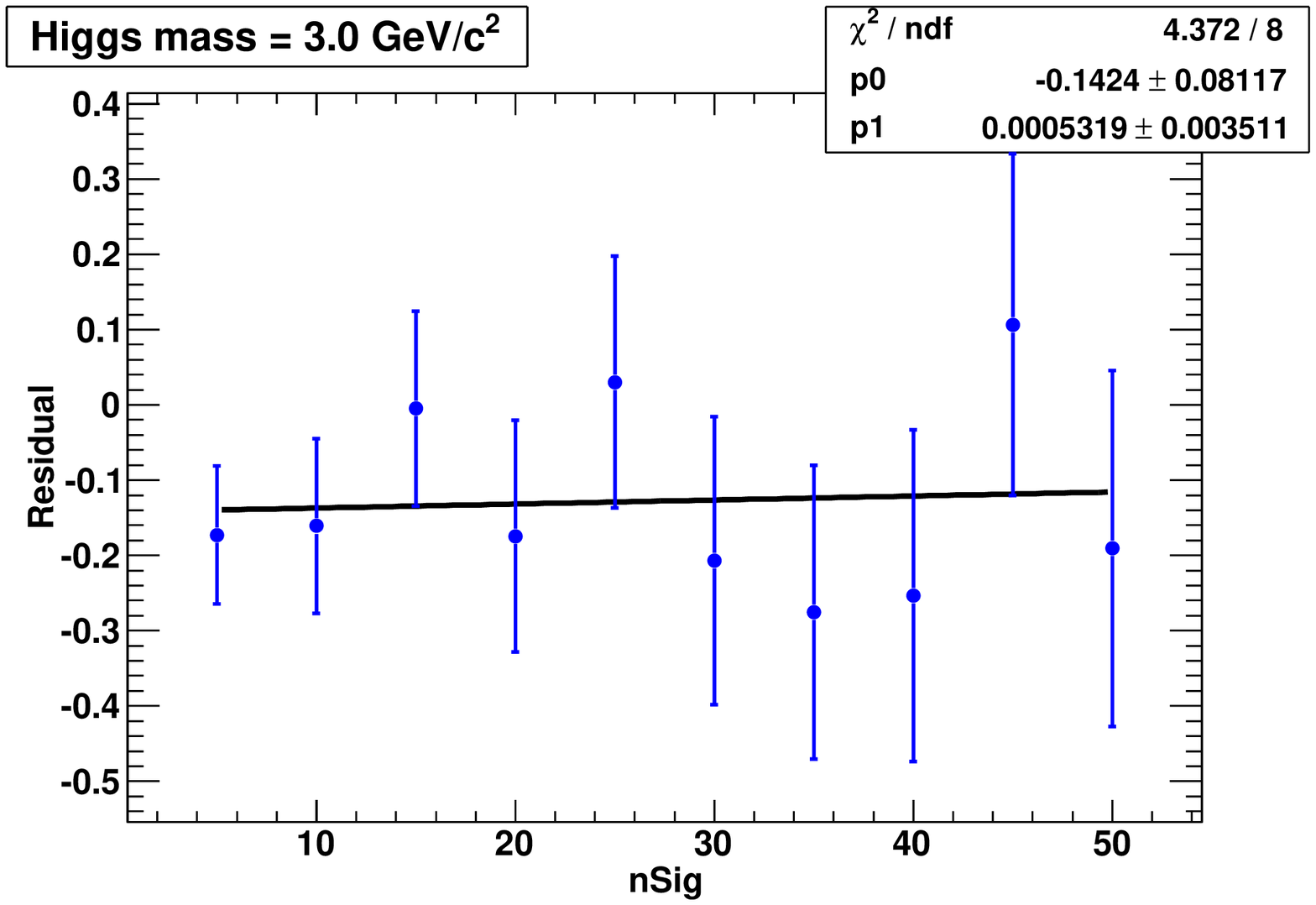}
\includegraphics[width=3.0in]{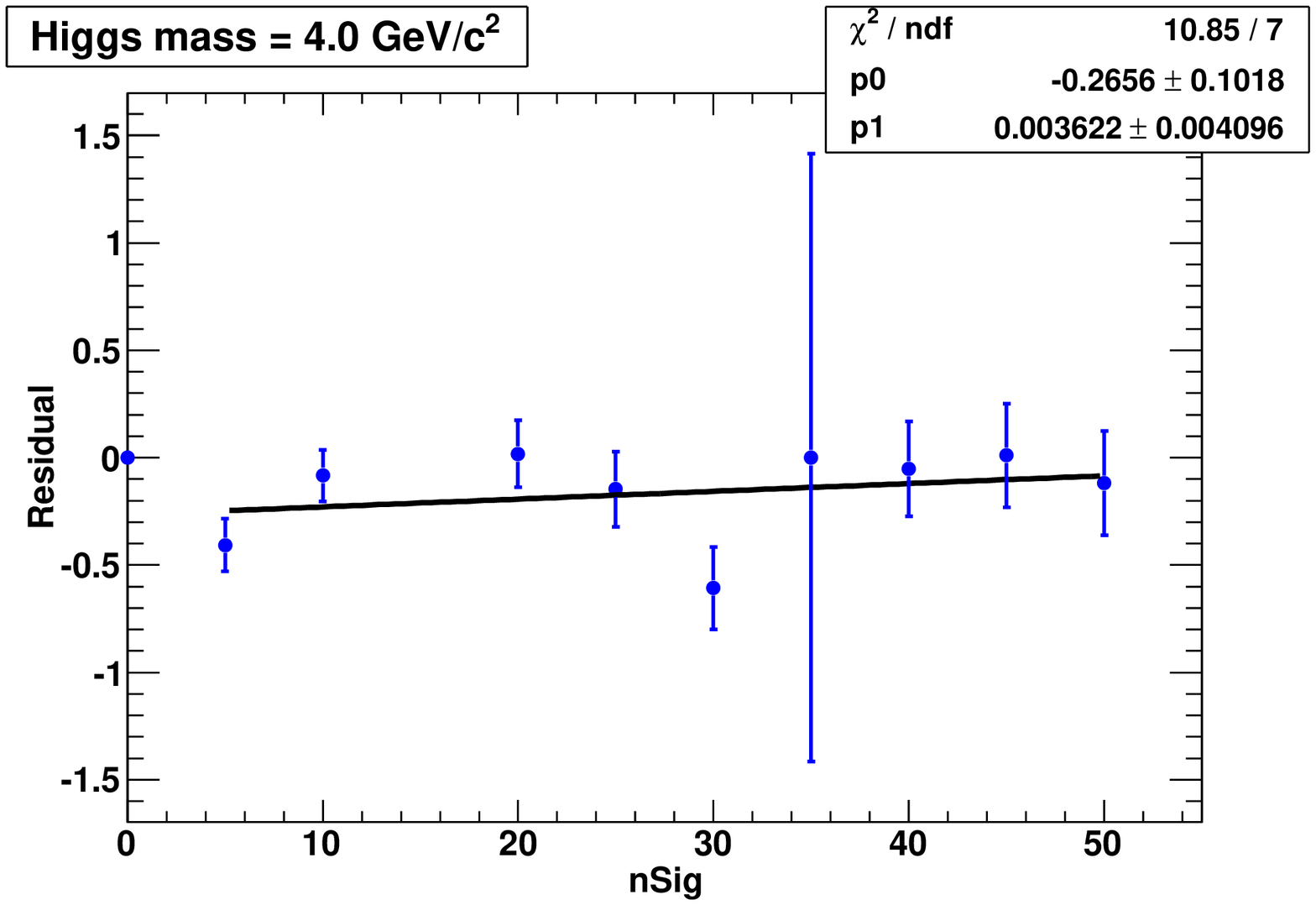}

\includegraphics[width=3.0in]{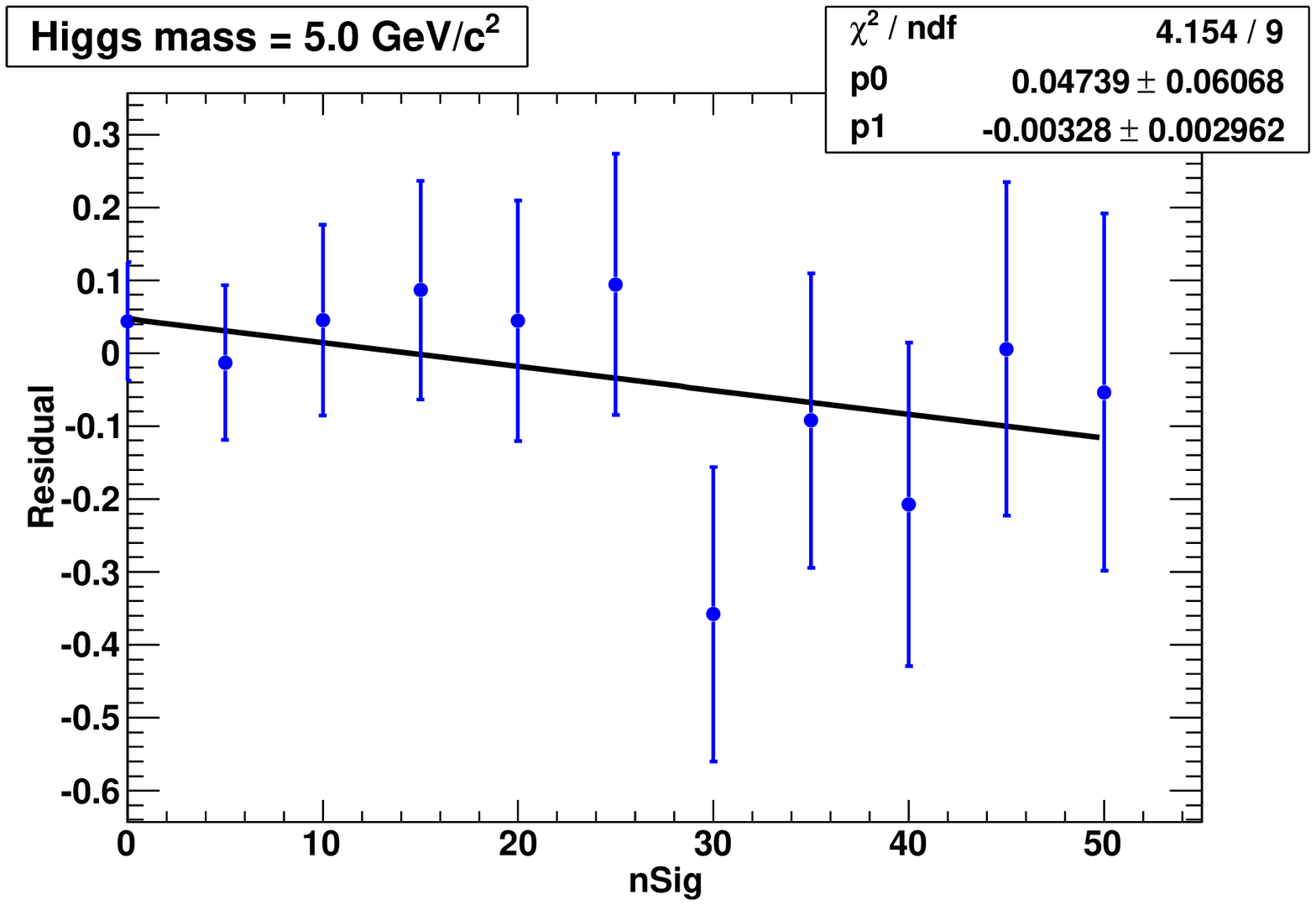}
\includegraphics[width=3.0in]{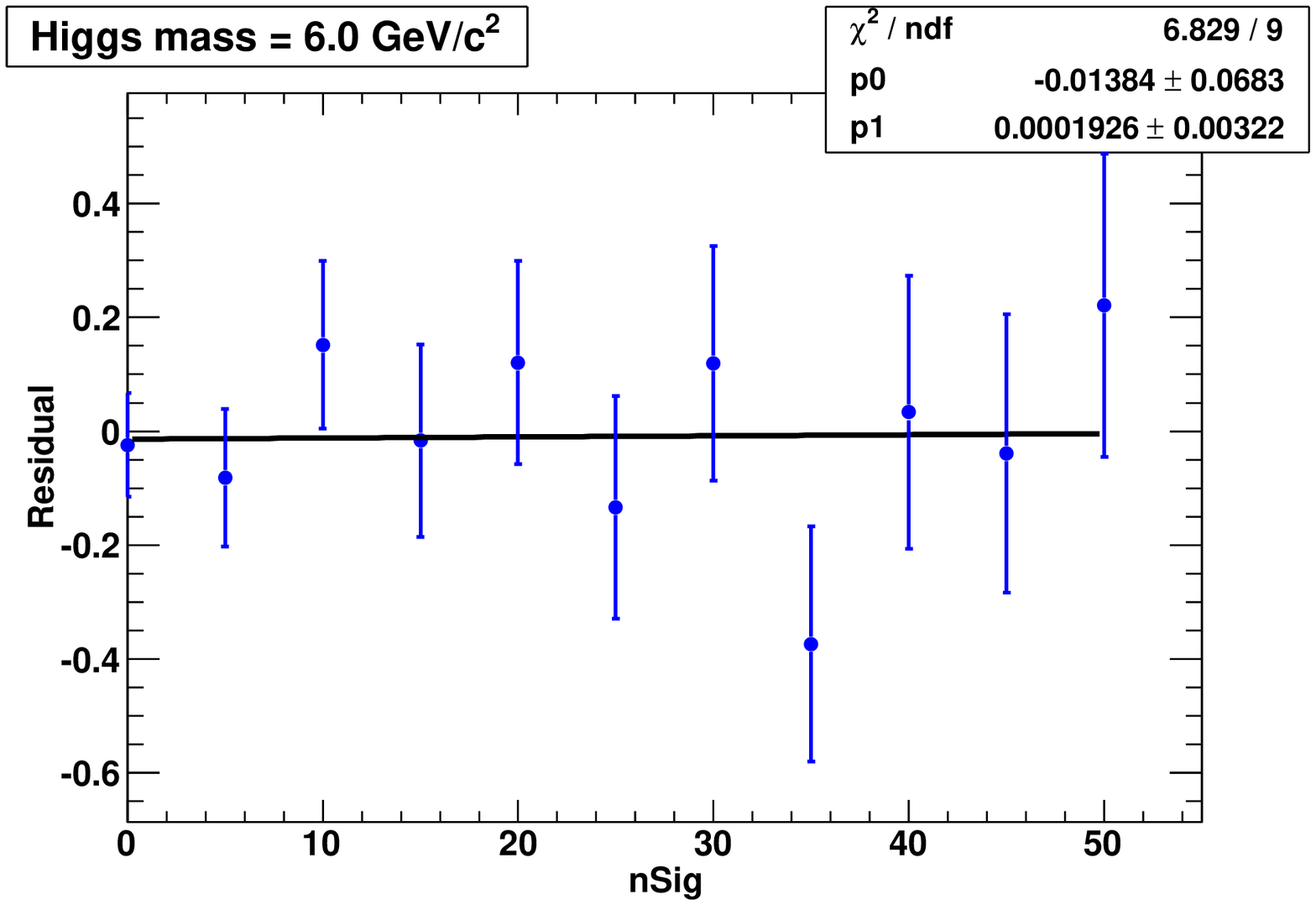}

\caption {Fit residuals for the number of signal events in the toy Monte Carlo experiments generated for each Higgs mass points.}

\label{fig:ToyMCY3S2}
\end{figure}

\begin{figure}
\centering
 \includegraphics[width=3.0in]{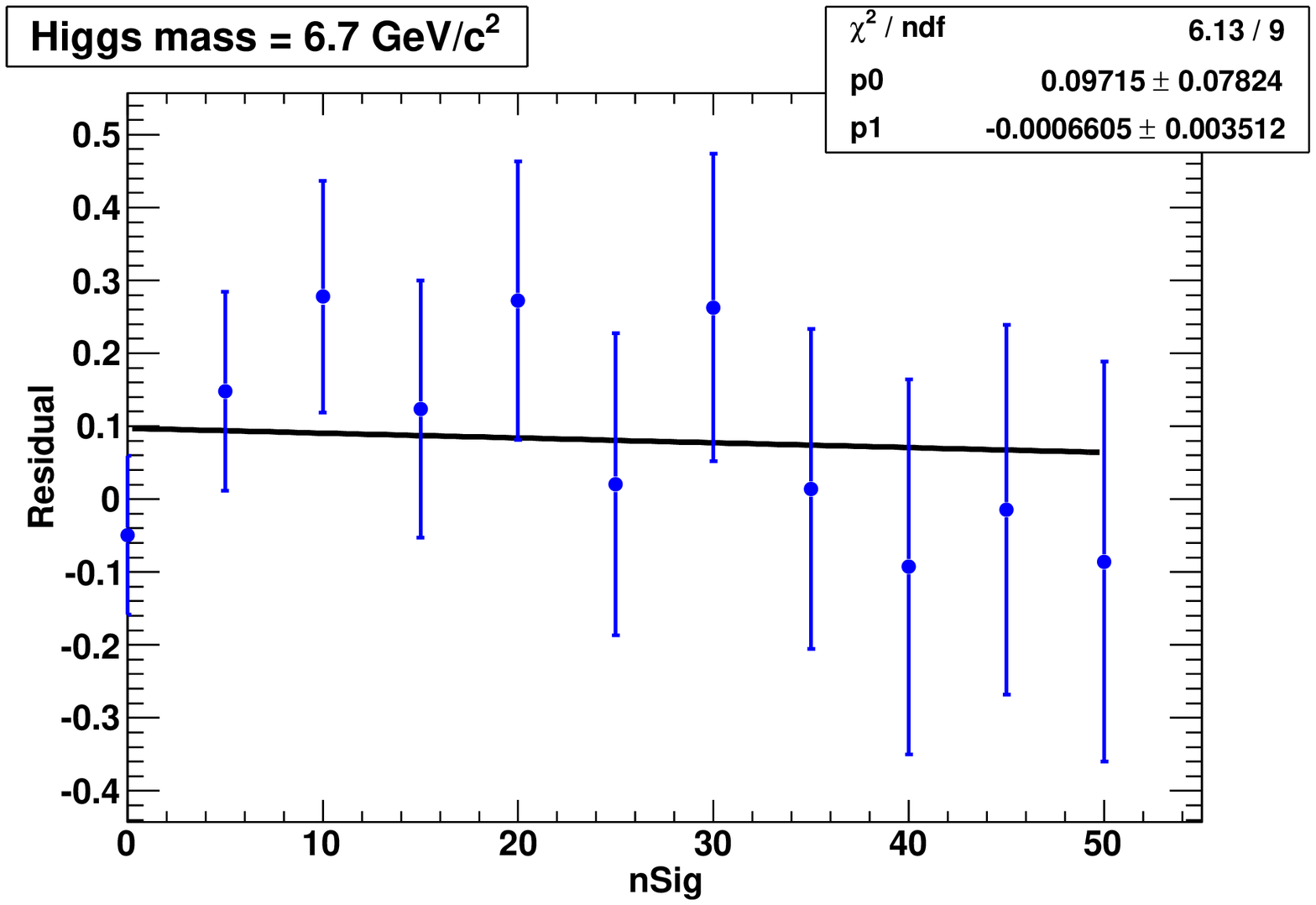}
\includegraphics[width=3.0in]{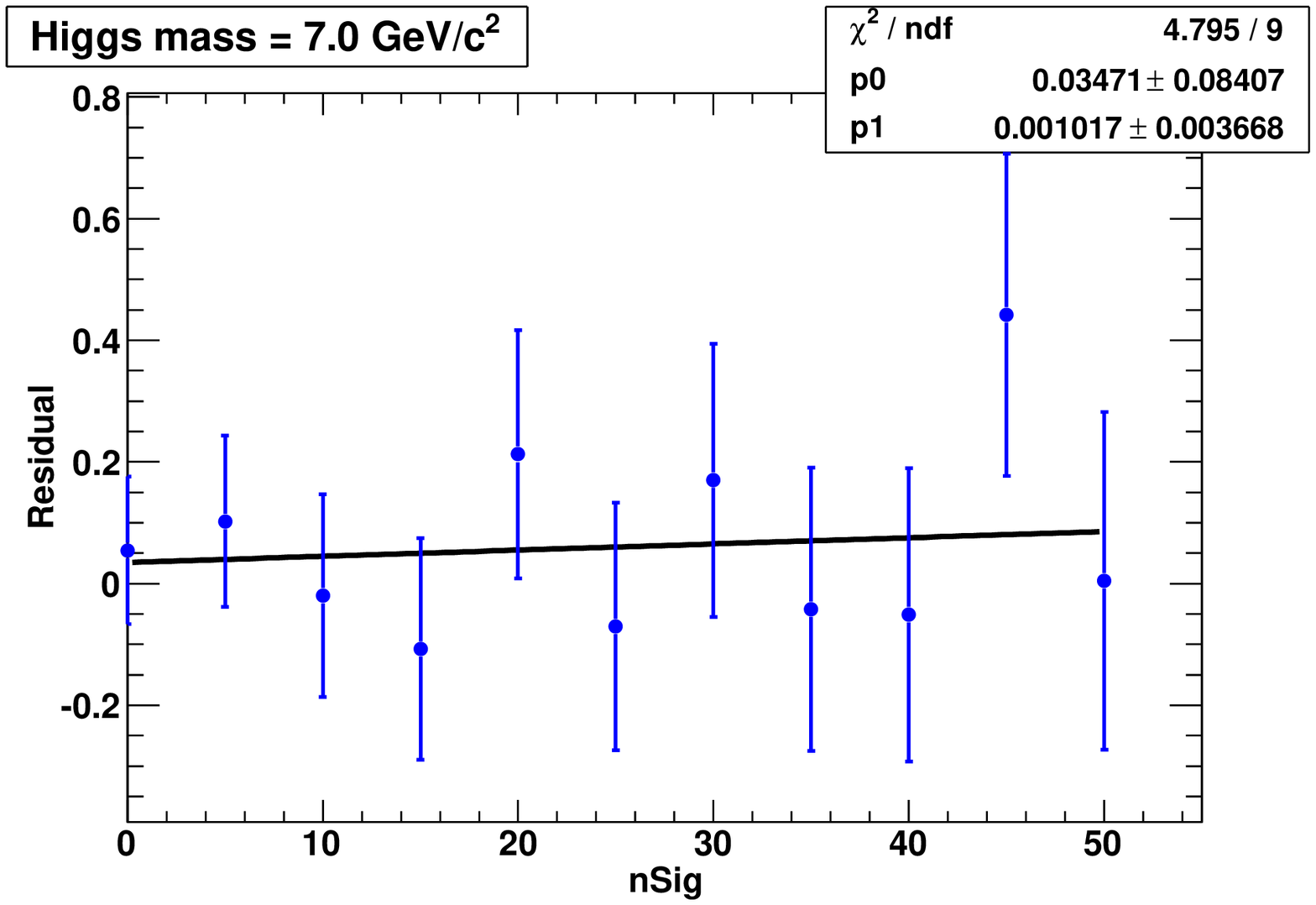}

\includegraphics[width=3.0in]{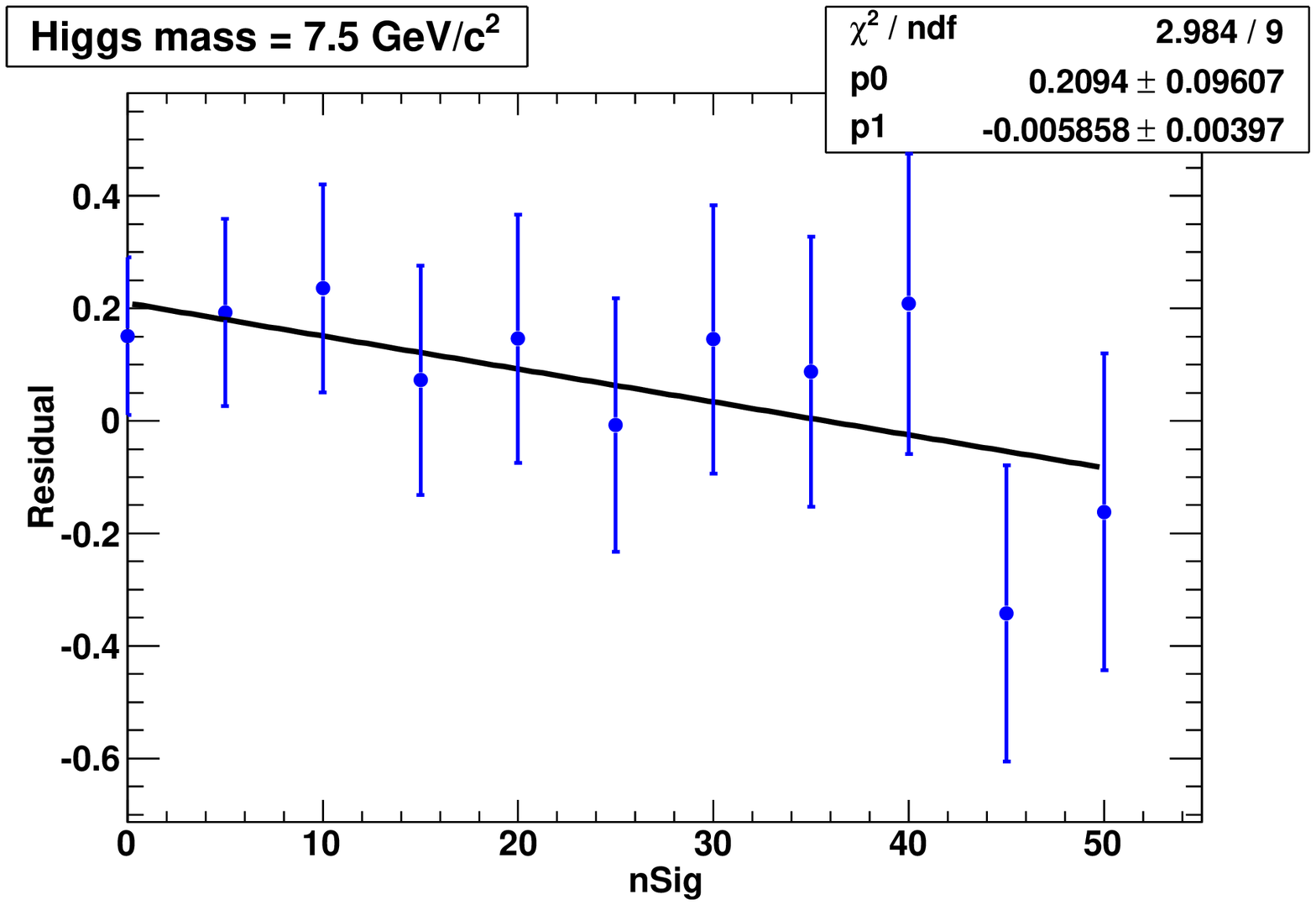}
\includegraphics[width=3.0in]{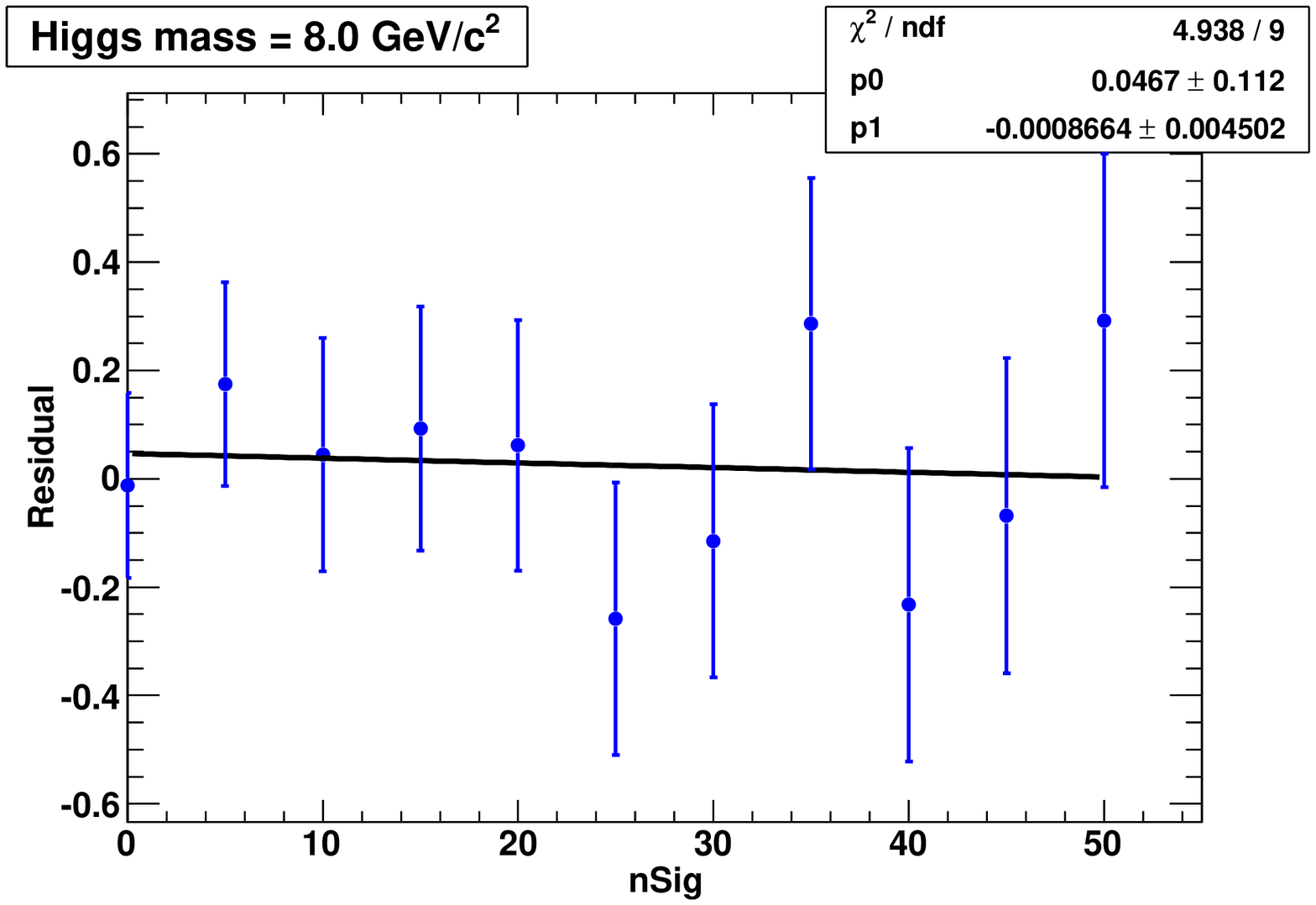}

\includegraphics[width=3.0in]{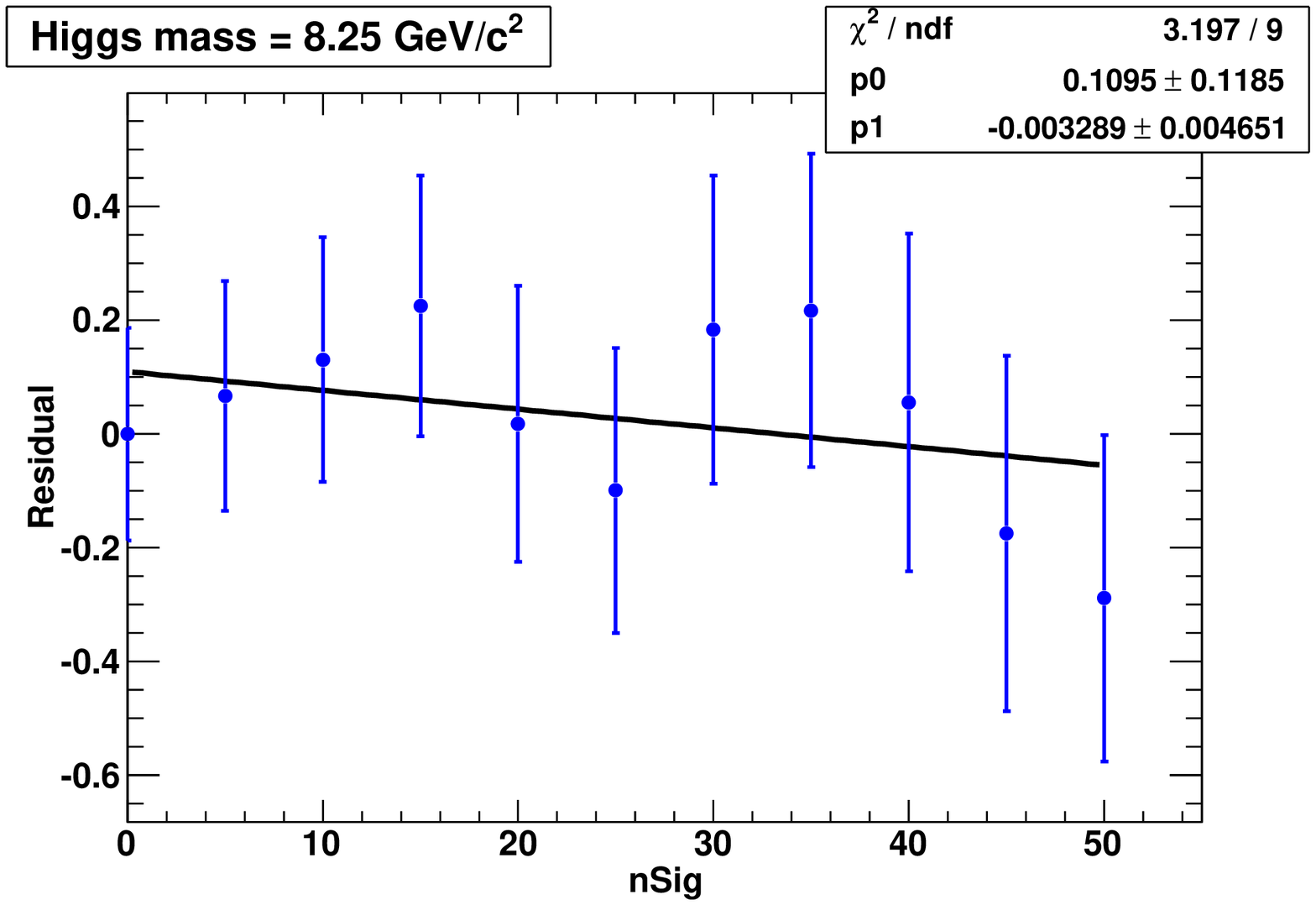}
\includegraphics[width=3.0in]{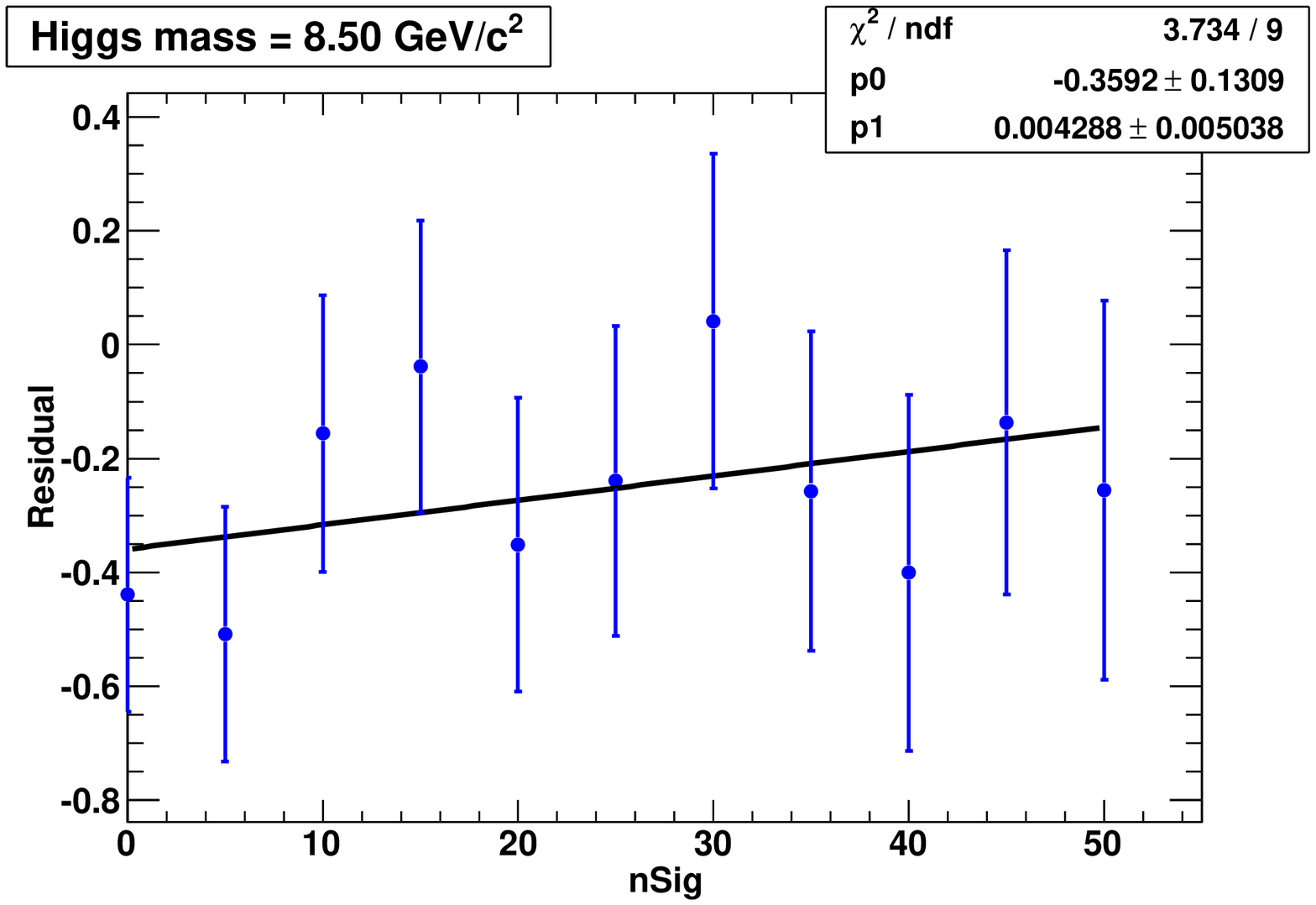}

\includegraphics[width=3.0in]{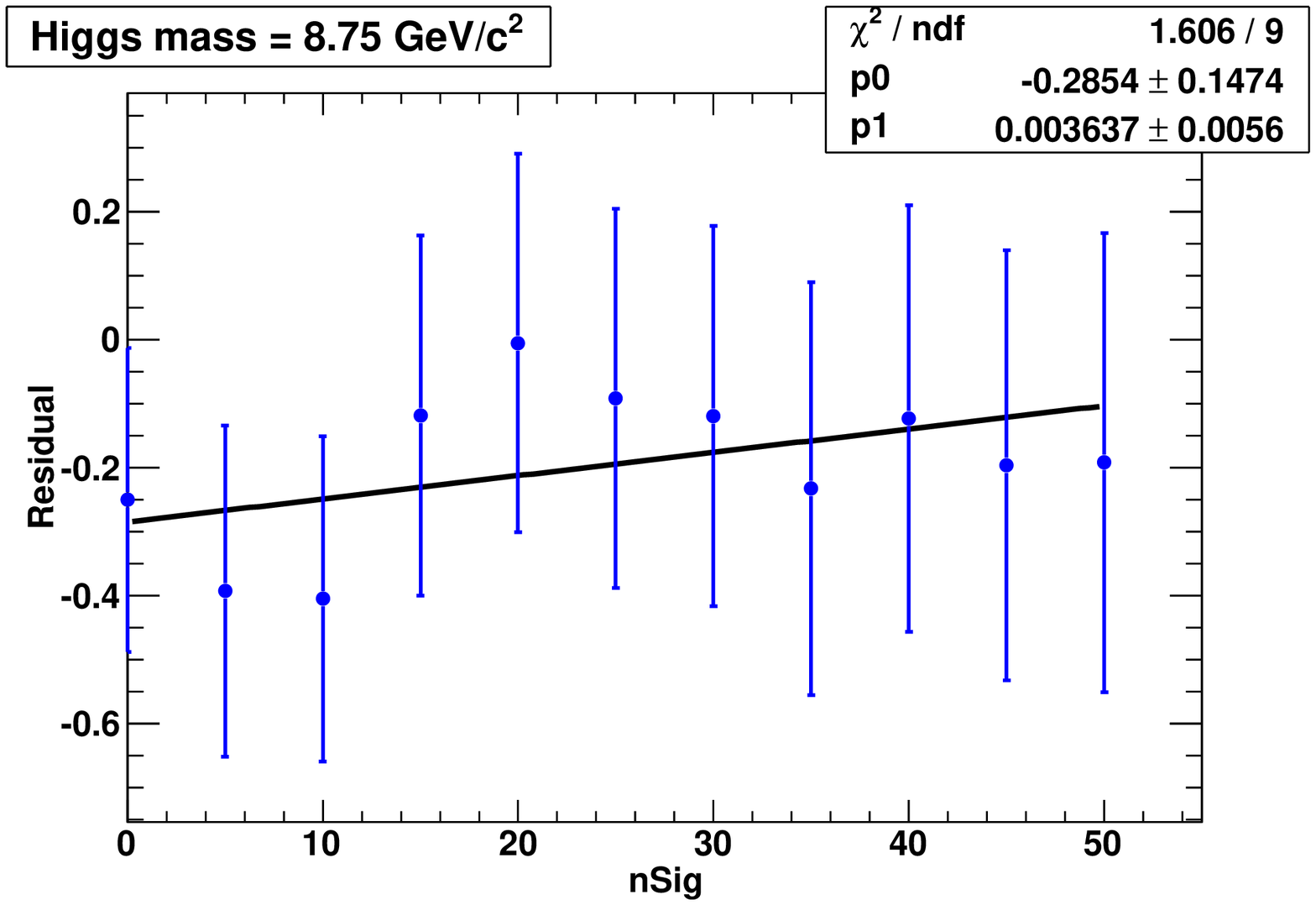}
\includegraphics[width=3.0in]{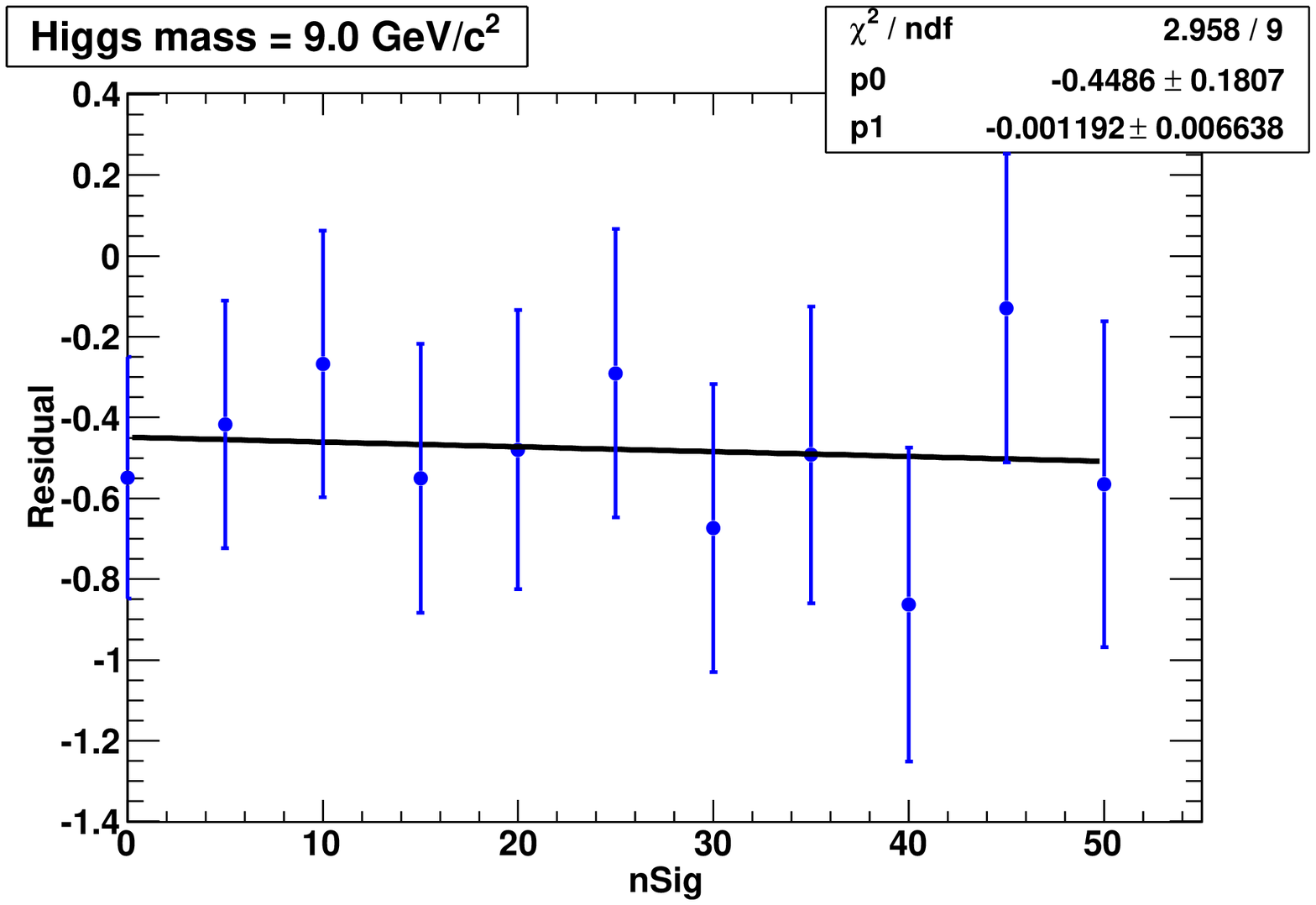}

\caption {Fit residuals for the number of signal events in the toy Monte Carlo experiments generated for each Higgs mass points.}

\label{fig:ToyMCY3S3}
\end{figure}


\addtocontents{toc}{\vspace{1em}}  
\backmatter

\label{Bibliography}
\bibliographystyle{bibliography/aip}  
\bibliography{bibliography/bibliography}  



\end{document}